%% file: main.tex
\documentclass[11pt,a4paper,footinclude=true,headinclude=true]{scrbook}

\usepackage[T1]{fontenc}       
\usepackage{lmodern}
\usepackage{booktabs}
\usepackage[linedheaders,parts,pdfspacing,
    eulerchapternumbers,
    floatperchapter
]{classicthesis}

\usepackage{geometry}
\usepackage[compat=1.1.0]{tikz-feynman}
\usepackage{amsthm}
\usepackage{graphicx} 
\usepackage{hyperref}
\usepackage{amsmath}
\usepackage{amssymb}
\usepackage{amsfonts,amsbsy}
\usepackage{MnSymbol}

\usepackage{slashed}
\usepackage{xcolor}
\usepackage{multirow}
\usepackage{array}
\usepackage{siunitx}

\usepackage[backend=biber, sorting=none, maxbibnames=10, citestyle=numeric-comp]{biblatex}
\DeclareSIUnit\c{c}

\newcommand{\mrm}{\mathrm}
\newcommand{\qhat}{\hat q}
\newcommand{\qperp}{q_\perp}
\newcommand{\Gammael}{\Gamma_{\text{el}}}
\newcommand{\vb}[1]{\mathbf{#1}}
\newcommand{\dd}[2][]{\mathrm d^{#1}{#2}\,} 
\newcommand{\dv}[2][]{\frac{\dd{#1}}{\dd{#2}}}
\newcommand{\pdv}[2][]{\frac{\partial{#1}}{\partial{#2}}}
\newcommand{\re}{Ref.~}
\newcommand{\se}{Section }
\newcommand{\app}{Appendix~}
\newcommand{\eq}{Eq.~}
\newcommand{\tab}{Tab.~}

\newcommand{\Q}{Q_s}
\newcommand{\dA}{d_{\mrm{A}}}
\newcommand{\dR}{d_{\mrm{R}}}
\newcommand{\CA}{C_{\mrm{A}}}
\newcommand{\CR}{C_{\mrm{R}}}
\newcommand{\Cs}{C_{\mrm{s}}}
\newcommand{\NC}{N_\mathrm{c}}
\newcommand{\nf}{n_f} 
\newcommand{\CF}{C_{\mrm{F}}}
\newcommand{\dF}{d_{\mrm{F}}}

\newcommand{\lperp}{\Lambda_\perp} 

\newcommand{\qhatf}{\qhat_{\mathrm{f}}}
\newcommand{\qhatff}{\qhat_{\mathrm{ff}}}

\newcommand{\vecp}{\vec p}
\newcommand{\NOg}{{N_+}} 
\newcommand{\NOq}{{N_-}} 

\renewcommand{\vec}[1]{\mathrm{\mathbf{#1}}}

\newcommand{\pmin}{p_{\mathrm{min}}}
\newcommand{\pmax}{p_{\mathrm{max}}}
\newcommand{\Ejet}{E_{\mathrm{jet}}}

\newcommand{\taubmss}{\tau_{\mathrm{BMSS}}}
\newcommand{\tauT}{\tau_{\mathrm{BMSS}}}
\newcommand{\tauR}{\tau_R}
\newcommand{\tautherm}{\tau_{\mathrm{therm}}}
\newcommand{\ttherm}{t_{\mathrm{therm.}}}
\newcommand{\tauhydro}{\tau_{\mathrm{hydro}}}
\newcommand{\tform}{t^{\mathrm{form}}}

\newcommand{\Conetwo}{\mathcal {C}^{1\leftrightarrow 2}}
\newcommand{\Ctwotwo}{\mathcal{ C}^{2\leftrightarrow 2}}
\newcommand{\Teps}{T_{\varepsilon}}

\newcommand{\thetaqp}{\theta_{qp}}
\newcommand{\thetaqk}{\theta_{qk}}
\newcommand{\phiqk}{\phi_{qk}}
\newcommand{\phiqp}{\phi_{qp}}

\newcommand{\bra}[1]{\left\langle#1\right|}
\newcommand{\ket}[1]{\left|#1\right\rangle}

\newcommand{\Tr}{\mathrm{Tr}\,}

\newcommand{\PiLR}{{\Pi^{00}}}
\newcommand{\tildePiLR}{{\tilde{\Pi}^{00}}}
\newcommand{\PiTR}{{\Pi^{T}}}
\newcommand{\GLR}{{G^{00}}}
\newcommand{\GTR}{{G^{T}}}
\newcommand{\RE}{\mathrm{Re}}
\newcommand{\IM}{\mathrm{Im}}

\newcommand{\Tid}{T_{\mathrm{id}}}
\newcommand{\Tfirst}{T_{\mathrm{1st}}}

\newcommand{\Gret}{G}
\newcommand{\GretHTL}{G^{\mathrm{isoHTL}}}
\newcommand{\epsgoal}{\epsilon^{\mathrm{error goal}}}

\newcommand{\xigauge}{\xi}
\newcommand{\xiscreen}{{\xi_g}}
\newcommand{\xiscreensqr}{{\xi_g^2}}
\newcommand{\xiscreenperp}{{\xi_g^\perp}}
\newcommand{\xiscreenperpsqr}{{(\xi_g^\perp)^2}}
\newcommand{\xianiso}{\xi_0}
\newcommand{\lxi}{{\Lambda_\omega}}
\newcommand{\otild}{{\tilde\omega}}
\newcommand{\ILO}{I^{\mathrm{LO}}}
\newcommand{\INLO}{I^{\mathrm{NLO}}}
\newcommand{\qhatLO}{\qhat^{\mathrm{LO}}}
\newcommand{\qhatNLO}{\qhat^{\mathrm{NLO}}}
\newcommand{\qhattherm}{\qhat^{\mathrm{therm}}}
\newcommand{\qhatftherm}{\qhatf^{\mathrm{therm}}}
\newcommand{\qhatfftherm}{\qhatff^{\mathrm{therm}}}
\newcommand{\qhatffthermimproved}{\qhat_{\mathrm{ff, im}}^{\mathrm{therm}}}
\newcommand{\qhatffimproved}{\qhat_{\mathrm{ff, im}}}
\newcommand{\qhatimproved}{\qhat_{\mathrm{im}}}

\newcommand{\Mhtl}{\mathcal M_{\mathrm{HTL}}}
\newcommand{\Mxi}{{M_\xi}} 
\newcommand{\Mdebyeone}{\mathcal M_{\mathrm{Debye1}}}
\newcommand{\Mdebyetwo}{\mathcal M_{\mathrm{Debye2}}}
\newcommand{\Mdebyethree}{\mathcal M_{\mathrm{Debye3}}}
\newcommand{\Mscreen}{M_{\mathrm{screen}}}
\newcommand{\tildeMscreen}{\tilde M_{\mathrm{screen}}}
\newcommand{\tildeMhtl}{\tilde M_{\mathrm{HTL}}}
\newcommand{\tildeMxi}{{\tilde{M}_\xi}} 
\newcommand{\vbphat}{\hat{\vb p}} 
\newcommand{\qs}{Q_\mathrm{s}}
\newcommand{\phikq}{\phi_{qk}}
\newcommand{\phipq}{\phi_{pq}}

\newcommand{\Ceq}{C_{\mathrm{eq}}}
\newcommand{\Ceqappr}{C_{\mathrm{eq}}^{\mathrm{appr.}}}
\newcommand{\Cisoappr}{C_{\mathrm{iso}}^{\mathrm{appr.}}}
\newcommand{\nmax}{n_{\mathrm{max}}}
\newcommand{\xmax}{x_{\mathrm{max}}}

\newcommand{\LQCD}{\Lambda_{\mathrm{QCD}}}
\newcommand{\Li}{\mathrm{Li}}

\newcommand{\meffg}{m_{\mathrm{eff,g}}}
\newcommand{\meffs}{m_{\mathrm{eff,s}}}

\newcommand{\sign}{\mathrm{sign}}

\newcommand{\fig}{Fig.~}

\makeatletter
\newcolumntype{M}[1]{>{\centering\arraybackslash}m{#1}}
\makeatother

\title{Nonequilibrium QCD in heavy-ion collisions: Kinetic theory and jet modifications during the initial stages}

\author{Florian Lindenbauer}
\date{June 13, 2025}

\begin{document}

\begin{titlepage}
   \begin{center}
       \vspace*{1cm}
        {\Large DISSERTATION}\\
        \vspace{2cm}
       {\LARGE\textbf{Nonequilibrium QCD in heavy-ion collisions:\\Kinetic theory and jet modifications\\during the initial stages}}

       \vspace{1.5cm}

       Ausgeführt zum Zwecke der Erlangung des akademischen Grades eines \\ 
       Doktors der Naturwissenschaften\\ \vspace{0.5cm}
       unter der Leitung von 
        \\
        \vspace{0.5cm}
       {Priv.-Doz.~Dr.~Kirill Boguslavski}\\
       E136\\
       Institut für Theoretische Physik

       \vfill
            
       eingereicht an der Technischen Universität Wien\\
       Fakultät für Physik
            
       \vspace{0.8cm}
        von\\
        \vspace{0.2cm}
       {\Large\textbf{Dipl-Ing.~Florian Lindenbauer, BSc}}\\
       \vspace{0.2cm}
       Matrikelnummer: 01608107
       \\     
       \vspace{2cm}
       
       \begin{tabular}{m{2.5cm} m{5.5cm} m{5.5cm} }
            Wien, & \hrulefill & \hrulefill  \\
            am 13.06.2025& Florian Lindenbauer & Priv.-Doz.~Dr.~Kirill Boguslavski\\
            & (Verfasser) & (Betreuer)\\[1.5cm]
            & \hrulefill & \hrulefill \\
            & Prof.~Dr.~Charles Gale & Prof.~Dr.~Sören Schlichting\\
            & (Gutachter) & (Gutachter)
       \end{tabular}

   \end{center}
\end{titlepage}

\chapter*{Abstract}
\addcontentsline{toc}{section}{Abstract}
\input{100_abstract.tex}

\chapter*{Zusammenfassung}
\addcontentsline{toc}{section}{Zusammenfassung}
\input{150_zusammenfassung.tex}

\chapter*{List of publications}
\addcontentsline{toc}{section}{List of publications}
\input{170_publication-list}
\chapter*{Acknowledgements}
\addcontentsline{toc}{section}{Acknowledgements}
\input{200_acknowledgements.tex}
\tableofcontents

\chapter{Introduction\label{sec:introduction}}
\input{300_introduction.tex}

\chapter{Jet energy loss\label{sec:jet-energy-loss}}
\input{420_energyloss}

\chapter{QCD kinetic theory\label{sec:qcd-kinetic-theory}}
\input{400_theory_background.tex}

\chapter{Momentum broadening of jets\label{sec:momentum-broadening-of-jets}}
\input{450_jet_broadening.tex}

\chapter{Limiting attractors in heavy-ion collisions\label{sec:limiting_attractors}}
\input{470_limiting_attractors.tex}

\chapter{Improving QCD kinetic theory simulations\label{sec:improving-qcd-simulations}}
\input{500_improving_QCD_kinetictheory.tex}

\chapter{Collision kernel and AMY rates out of equilibrium\label{sec:collkern}}
\input{700_collision_kernel.tex}

\chapter{Summary, conclusions and outlook\label{sec:conclusions}}
\input{800_summary.tex}

\appendix
\chapter{QCD and nonequilibrium quantum field theory\label{app:qcd-and-nonequilibrium-qft}}
\input{850_qcd}

\chapter{Numerical and kinematic details on kinetic theory simulations\label{app:numericaldetails}}
\input{950_numerical_details.tex}

\chapter{Large momentum limits of the jet quenching parameter\label{app:large-momentum-limits-qhat}}
\input{908_large_momentum_limits}

\chapter{Numerical details on solving the AMY rate equations\label{app:amyrates-details}}
\input{960_amyrates_numericaldetails}

\chapter{Additional numerical results for bottom-up thermalization\label{sec:additional-results-bottomup}}
\input{900-additional-bottomup-things}

\chapter{Notes on notational conventions here and in the papers\label{sec:notational-differences}}
\input{980_different_notations}

\chapter{Tools used}
\input{985_tools_used}

\printbibliography

\end{document}

%% file: 100_abstract.tex
Heavy-ion collision experiments create a high-temperature plasma that is characterized by a deconfined state of quarks and gluons. It is described by the theory of the strong interaction: Quantum Chromodynamics (QCD). At high energies, QCD admits an effective kinetic description, which allows studying and simulating how the initially far-from-equilibrium plasma reaches thermal equilibrium, where the plasma can be described as a relativistic fluid. While many studies have focused on this fluid phase, the nonequilibrium stages before a fluid picture becomes applicable have recently received increased attention. In particular, they may be experimentally studied using highly energetic particles called jets.

This thesis focuses on how jets are modified by the nonequilibrium quark-gluon plasma.
Its influence on their propagation is typically encoded in a single medium function which is referred to as the dipole cross section.
Its small distance behavior is characterized by the jet quenching parameter $\qhat$, and we obtain its numerical value
throughout the pre-equilibrium stage, finding values comparable in magnitude to the earlier Glasma stage, with also a similar anisotropy.
We also compute the more general elastic collision kernel, obtained by Fourier transforming the dipole cross section. We observe that its small distance behavior
is suppressed during the pre-fluid stages, implying a suppression of jet quenching during the initial stages.
This constitutes an important step to facilitate the understanding of jet-medium interactions during the initial stages in heavy-ion collisions.

Additionally, this thesis improves our understanding of QCD equilibration in heavy-ion collisions. QCD kinetic theory simulations are improved by employing a more realistic (hard thermal loop) screening mechanism to incorporate medium effects, which we compare with simpler screening mechanisms. While the effect on isotropic systems is negligible, an expanding plasma realized in the initial stages of heavy-ion collisions exhibits a significantly reduced maximum anisotropy when using the improved screening prescription. We also quantify its effect on the specific shear viscosity $\eta/s$, finding that its numerical value decreases when using the improved screening prescription. Moreover, we investigate the gluon splitting rates used as input for kinetic theory simulations, which are typically obtained using an isotropic model for the collision kernel. Going beyond that approximation, we find that the splitting rates obtained from the nonequilibrium anisotropic kernel significantly differ from those used in QCD kinetic theory simulations, both in magnitude and qualitative time evolution.
We further identify a novel type of attractor in this thesis, which can be observed in the ratio of the jet quenching parameter, and is obtained by extrapolating to vanishing coupling. This weak coupling limiting attractor is also identified in the pressure ratio.
This improved kinetic theory description and novel limiting attractors contribute towards a more realistic modeling of the nonequilibrium QCD plasma and its equilibration and hydrodynamization process during the initial stages.

%% file: 150_zusammenfassung.tex
In relativistischen Schwerionenkollisionen wird ein Hochtemperaturplasma er\-zeugt, das durch einen Zustand freier Quarks und Gluonen charakterisiert ist. Es wird durch die Theorie der starken Wechselwirkung beschrieben, die Quantenchromodynamik (QCD). 
Bei hohen Energien erlaubt diese eine effektive kinetische Beschreibung, die es ermöglicht, 
die Thermalisierung des anfänglich weit vom Gleich\-ge\-wicht entfernten Plasmas aus Quarks und Gluonen zu simulieren.
In der Nähe des Gleichgewichts kann das Plasma als eine relativistische Flüssigkeit beschrieben werden. Diese hydrodynamische Beschreibung dient als Grundlage vieler Studien zu Schwerionenkollisionen. Jedoch hat in letzter Zeit die vorhergehende Phase außerhalb des Gleichgewichts erhöhte Aufmerksamkeit erhalten, insbesondere, da sie möglicherweise experimentell mittels hochenergetischer Teilchen, genannt Jets, untersucht werden kann.

Diese Arbeit beschäftigt sich mit der Frage, wie Jets durch das Plasma fern des Gleichgewichts modifiziert werden.
Diese Modifikation wird typischerweise durch eine einzige Funktion des Mediums beschrieben, die Dipol-Wirkungsquerschnitt genannt wird.
Dessen Verhalten für kleine Entfernungen wird durch den 
\emph{Jet Quenching Parameter} $\qhat$ beschrieben
und wir berechnen dessen Werte in den frühen Nicht-Gleichgewichtsphasen von Schwerionenkollisionen. Insbesondere finden wir, dass dessen Werte in unseren Simulationen mit jenen Werten dieses Parameters vergleichbar sind, die in der noch früheren Glasma Phase berechnet wurden, sowohl in Größe als auch Richtungsabhängigkeit.
Weiters extrahieren wir den allgemeineren Kollisionskernel durch Fourier-Transformation des Dipol-Wirkungsquerschnitts. Wir finden, dass dessen Werte bei kleinen Abständen kleiner sind als im ther\-mi\-schen Gleichgewicht, was zu einer Unterdrückung von Jet Quenching während der frühen Zeiten führt. Das Extrahieren dieser Größen stellt einen bedeutenden Schritt dar, um die Beschreibung von Jet-Medium Wechselwirkungen in den frühen Phasen von Schwerionenkollisionen besser zu verstehen.

Darüber hinaus verbessert diese Arbeit die zugrundeliegende kinetische Beschreibung durch die Verwendung eines realistischeren Abschirmmechanismus (mittels harter thermischer Schleifen), um die Effekte des Mediums zu berücksichtigen. Diesen vergleichen wir mit einfacheren Abschirmmechanismen. Für isotrope Systeme sind seine Effekte vernachlässigbar, wohingegen ein expandierendes Plasma, das in den frühen Zeiten von Schwerionenkollisionen existiert, eine deutlich geringere Anisotropie aufweist, wenn man den verbesserten Abschirmmechanismus verwendet. Wir quantifizieren diesen Effekt auch durch die Ermittlung der numerischen Werte der spezifischen Viskosität $\eta/s$, deren Wert durch die verbesserte Abschirmvorschrift verringert wird. Schließlich untersuchen wir die Spaltungsraten von Gluonen, die in kinetischen QCD Simulationen auf Basis einer isotropen Näherung verwendet werden. Wir stellen fest, dass sich die Spaltungsraten unter Verwendung des anisotropen Nicht-Gleichgewichts-Kernels deutlich von den genäherten Raten unterscheiden,
und zwar sowohl in der Größe als auch im qualitativen Zeitverhalten.
Weiters identifizieren wir einen neuartigen Attraktor, der für das Verhältnis des Jet Quenching Parameters zwischen verschiedenen Richtungen relevant ist, und über die Extrapolation zu verschwindenden Kopplungen erhalten wird.
Die Weiterentwicklung der kinetischen Theorie und diese neuartigen Attraktoren führen zu einem besseren Verständnis der Thermalisierung des Quark-Gluon Plasmas in Schwerionenkollisionen.

%% file: 170_publication-list.tex
This thesis is based on the following published papers \cite{Boguslavski:2023alu, Boguslavski:2023waw, Boguslavski:2023jvg,Boguslavski:2024kbd} and conference proceedings \cite{Boguslavski:2024pnw}, for which I am the corresponding author, and on a paper in preparation \cite{Altenburger:2025a}:
\begin{itemize}
    \item \cite{Boguslavski:2023alu} Boguslavski, K., Kurkela, A., Lappi, T., Lindenbauer, F., \& Peuron, J. (2024). Jet momentum broadening during initial stages in heavy-ion collisions. Physics Letters B, 850, 138525. \url{https://doi.org/10.1016/j.physletb.2024.138525}

    \item \cite{Boguslavski:2023waw} Boguslavski, K., Kurkela, A., Lappi, T., Lindenbauer, F., \& Peuron, J. (2024). Jet quenching parameter in QCD kinetic theory. Physical Review D, 110(3). \url{https://doi.org/10.1103/physrevd.110.034019}

    \item \cite{Boguslavski:2023jvg} Boguslavski, K., Kurkela, A., Lappi, T., Lindenbauer, F., \& Peuron, J. (2024). Limiting attractors in heavy-ion collisions. Physics Letters B, 852, 138623. \url{https://doi.org/10.1016/j.physletb.2024.138623}

    \item \cite{Boguslavski:2024kbd} Boguslavski, K., \& Lindenbauer, F. (2024). Soft-gluon exchange matters: Isotropic screening in QCD kinetic theory. Physical Review D, 110(7). \url{https://doi.org/10.1103/physrevd.110.074017}
    \item \cite{Boguslavski:2024pnw} Boguslavski, K., Kurkela, A., Lappi, T., Lindenbauer, F., \& Peuron, J. (2024). Limiting attractors in heavy-ion collisions—The interplay between bottom-up and hydrodynamic attractors. EPJ Web of Conferences, 296, 10004. \url{https://doi.org/10.1051/epjconf/202429610004}

    \item Chapter \ref{sec:collkern} is based on a paper in preparation in collaboration with Alois Altenburger and Kirill Boguslavski \cite{Altenburger:2025a}.
\end{itemize}

Other publications not used in this thesis:
\begin{itemize}
    \item \cite{Boguslavski:2023fdm} Boguslavski, K., Kurkela, A., Lappi, T., Lindenbauer, F., \& Peuron, J. (2024). Heavy quark diffusion coefficient in heavy-ion collisions via kinetic theory. Physical Review D, 109(1). \url{https://doi.org/10.1103/physrevd.109.014025} 
    \item \cite{Boguslavski:2023sqn} Peuron, J., Boguslavski, K., Kurkela, A., Lappi, T., \& Lindenbauer, F. (2024). Heavy quark diffusion coefficient during hydrodynamization - non-equilibrium vs. equilibrium. In Proceedings of 11th International Conference on Hard and Electromagnetic Probes of High-Energy Nuclear Collisions — PoS(HardProbes2023) (p. 091). 11th International Conference on Hard and Electromagnetic Probes of High-Energy Nuclear Collisions. Sissa Medialab. \url{https://doi.org/10.22323/1.438.0091}

    \item \cite{Boguslavski:2023zgd} Boguslavski, K., Kurkela, A., Lappi, T., Lindenbauer, F., \& Peuron, J. (2024). Heavy quark momentum diffusion coefficient during hydrodynamization via effective kinetic theory. EPJ Web of Conferences, 296, 09001. \url{https://doi.org/10.1051/epjconf/202429609001}

\end{itemize}

%% file: 200_acknowledgements.tex
During the course of my PhD, I have met many people who I owe gratitude.
First and foremost, I am grateful to my excellent supervisor Kirill Boguslavski, for introducing me to the exciting topic of heavy-ion collisions, and for his help and guidance, both on academic and bureaucratic topics; for saving me from digging too deep into maybe not so relevant aspects, and for motivating me to pursue what seemed interesting to me. Thank you for your support and for believing in me!

Furthermore, I am indebted to all the people I was lucky to collaborate with, who offered wisdom, guidance, and insights, without which this thesis would have never come about. Thank you to Aleksi Kurkela, Tuomas Lappi, and Jarkko Peuron for the collaboration on the projects this thesis is explicitly based on, and additionally to Alois Altenburger, Jo\~ao Barata, Sergio Barrera Cabodevila, Michal Heller, Lucas Hörl, Aleksas Mazeliauskas, Andrey Sadofyev, Michał Spaliński, Adam Takacs and Fabian Zhou for the collaboration on ongoing projects that unfortunately could not be included in this thesis and which have yet to see the light of the arXiv. Thank you not only for sharing ideas about research but also for providing advice and guidance on my journey through academia.

I had the wonderful privilege of being invited to research stays in Santiago de Compostela, Heidelberg, and Stavanger, and I would like to thank Andrey Sadofyev, Aleksas Mazeliauskas, and Aleksi Kurkela for being excellent hosts during my stays there.

I would like to thank fellow PhD students and postdocs that I met at various conferences and with whom I enjoyed enlightening or humorous discussions, dinners, lunches, snacks, or drinks, who made my PhD much more fun! Thank you to 
André,
Bruno,
Carlos,
Clemens,
Dana,
Győző,
Ismail,
Jannis,
János,
Jean,
Jendrik,
Matisse,
Orestis,
Rachel,
Sylwia,
Viktoria,
Victor,
Xiaojian, and
Xoán,
and to the many more I surely failed to mention.

I want to thank my fellow PhD students at the institute, especially Paul, Markus, and Kayran,
for the many shared lunches and discussions, for sharing rooms during various trips, and additionally, for their feedback on early drafts of parts of this thesis.

Finally, and perhaps most significantly, I would like to express my deepest gratitude to my girlfriend Liane for her unwavering support, for being present when it was most needed, and for helpful comments on early and not-so-early drafts of this thesis.
Lastly, I would like to thank my parents for their constant and unconditional support during my studies.

\vspace{3em}
I am a recipient of a DOC Fellowship of the Austrian Academy of Sciences at TU Wien (project 27203). I would also like to acknowledge support by the Austrian Science Fund (FWF) under Grant DOI 10.55776/P34455 and Grant DOI 10.55776/W1252.  

%% file: 300_introduction.tex
\section{The fundamental forces and the Standard Model}
Physics aims to describe, explain, and predict phenomena encountered in nature using mathematical models.
The ultimate goal is to understand and answer fundamental questions about our universe, e.g., \emph{what are the fundamental forces of nature?} or \emph{What forces bind together the matter we are made of?} 
or \emph{Can we probe nonequilibrium effects of the strong interaction using particle colliders?}

Perhaps needless to say, tremendous effort has gone into answering these questions over the past centuries. With the exception of gravity, all fundamental interactions are now conveniently combined in the \emph{Standard Model of particle physics}. It is arguably the most well-tested physical theory today and the cornerstone of modern particle and collider physics, allowing predictions with astonishing precision and accuracy. Despite its compact mathematical formulation, which fits on a simple coffee mug (available for purchase at the European Organization for Nuclear Research (CERN)), 
it describes surprisingly complex dynamics and phenomena.

For instance, the Standard Model includes the \emph{electromagnetic force}, perhaps the most important fundamental force in our everyday lives. It is essential for describing the stability of atoms\footnote{Atoms are important for everyday life despite being banned in Austria per its constitution ("Bundesverfassungsgesetz für ein atomfreies Österreich", \url{https://www.ris.bka.gv.at/GeltendeFassung.wxe?Abfrage=Bundesnormen&Gesetzesnummer=10008058})}, molecules and solids, while also describing light, electricity, and electromagnetic waves.
In the Standard Model, it is neatly combined with the \emph{weak interaction}, which describes how the building blocks of atoms, the protons and neutrons, decay.

The structure of the atomic nucleus itself is governed by the \emph{strong interaction}, the force that binds protons and neutrons together. This interaction describes how protons and neutrons are not fundamental particles, but rather composed of even tinier constituents, the quarks and gluons. This thesis focuses on the mathematical theory of the strong interaction, \emph{quantum chromodynamics} (QCD).

\section{Quantum Chromodynamics and the quark-gluon plasma}
QCD was formulated over 50 years ago \cite{Gross:2022hyw},
and
lives within the mathematical framework of quantum field theory, which combines special relativity with quantum mechanics.
The fundamental ingredients of QCD are quarks and gluons. They have the important property that their interaction strength decreases for increasing energies, a property called \emph{asymptotic freedom}. Thus, at weak coupling---or large energies---one may use perturbation theory to obtain properties of QCD. At lower energies, quarks and gluons are confined in ordinary hadrons, such as protons and neutrons making up most of the visible matter in the universe.
For a QCD plasma in thermal equilibrium, numerical tools such as \emph{Lattice QCD} are available and reveal a crossover phase transition between ordinary nuclear matter and a phase of deconfined quarks and gluons at very high temperatures: the quark-gluon plasma (QGP). This many-body system of quarks and gluons is also predicted to exist shortly after the Big Bang and can be produced experimentally in relativistic heavy-ion collision experiments. How a nonequilibrium QCD plasma equilibrates has been the context of intensive study and remains an important research question \cite{Busza:2018rrf, Achenbach:2023pba}. Moreover, the theoretical and experimental study of this system in and out of equilibrium and its thermalization processes reveals intriguing phenomena and possesses also exciting interdisciplinary connections to other fields of physics, such as condensed matter, cosmology, and ultracold quantum gases \cite{Berges:2020fwq}.

\section{Heavy-ion collisions and thermalization}
As alluded to earlier, the equilibrium and nonequilibrium properties of the quark-gluon plasma can be studied experimentally in relativistic heavy-ion collisions, which are currently performed at the Relativistic Heavy Ion Collider (RHIC, soon to be phased out to make room for the new Electron-Ion Collider (EIC)) at Brookhaven National Laboratory, and at the Large Hadron Collider (LHC) at CERN. There, heavy nuclei such as gold and lead are accelerated nearly to the speed of light and brought to collision. Their decay products are then studied in the detectors, similarly to the study of proton-proton collisions. These heavy-ion collisions create a droplet of QCD matter that is initially far from equilibrium. Analyzing the remnants of these collisions reveals that a deconfined state of matter is created, where quarks and gluons are no longer bound into hadrons, the quark-gluon plasma.

Currently, it is impossible to describe the time evolution of this out-of-equilibrium system using (numerical) first-principle simulations of QCD. Real-time Lattice QCD simulations, for example, suffer from the infamous sign problem \cite{Eisert:2014jea, Gattringer:2016kco, Alexandru:2020wrj}, which prevents real-time simulations or simulations at finite density.
\begin{figure}
    \centering
    \includegraphics[width=\linewidth, clip, trim=0 0.5cm 0 0]{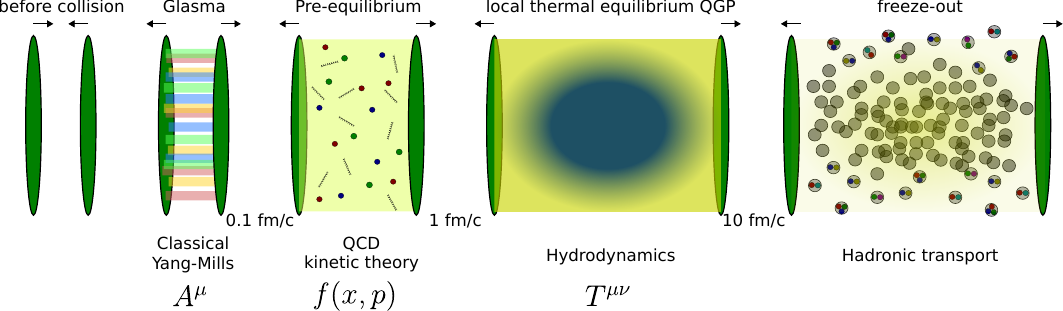}
    \caption{Schematic overview of the initial stages in heavy-ion collisions and their respective effective description.
    }
    \label{fig:schematic-initial-stages}
\end{figure}
Instead, the out-of-equilibrium QCD plasma is often described using a series of effective theories, which are depicted schematically in Fig.~\ref{fig:schematic-initial-stages}. At sufficiently high energies and weak couplings, directly after the collision, the system is dominated by large classical gluon fields and can be described within the Color Glass Condensate \cite{Gelis:2010nm}. The corresponding state (following a specific set of initial conditions) is referred to as the \emph{Glasma}, and its time evolution can be obtained using classical statistical lattice simulations \cite{Krasnitz:1998ns}. After a time of about $\qty{0.1}{\femto\meter/\c}\approx 3\times 10^{-25}\unit{\second}$, a quasiparticle description becomes applicable with gluons initially remaining the dominant degrees of freedom and the---still far from equilibrium---plasma can be described as a weakly-interacting gas of quarks and gluons using an effective kinetic theory \cite{Arnold:2002zm, Ghiglieri:2015ala}, where the time evolution of the system is dictated by a Boltzmann equation
\begin{align}
    \left(\pdv{t}+\vb v\cdot \vb\nabla\right)f(\vb p,\vb x,t)=-\mathcal C[f]
\end{align}
for the particle distribution function $f(\vb p,\vb x,t)$.
Numerical implementations of this QCD kinetic theory allow tracking the system's time evolution towards
equilibrium and
to the later hydrodynamic stage \cite{Kurkela:2014tea, Kurkela:2015qoa, Kurkela:2018oqw, Kurkela:2018xxd, Du:2020dvp, Du:2020zqg},
and 
are also commonly extrapolated to larger coupling values for phenomenological purposes \cite{Kurkela:2018wud, Kurkela:2018vqr, Giacalone:2019ldn, Coquet:2023wjk, Garcia-Montero:2023lrd, Garcia-Montero:2024lbl, Zhou:2024ysb}. 

Common QCD kinetic theory implementations \cite{Kurkela:2014tea, Kurkela:2015qoa, Kurkela:2018oqw, Kurkela:2018xxd, Du:2020dvp, Du:2020zqg} use a simple screening approximation to include medium effects in the simulations, approximating internal medium resummed propagators by a simple Debye-like screened propagator. Notably, this isotropic screening approximation neglects the effect of plasma instabilities that are present in anisotropic systems \cite{Mrowczynski:1988dz, Mrowczynski:1993qm, Romatschke:2003ms, Romatschke:2004jh, Kurkela:2011ub, Mrowczynski:2016etf}, but might be dampened by collisions \cite{Schenke:2006xu, Zhao:2023mrz}. However, they have been shown not to have very dramatic effects on the earliest stages \cite{Berges:2013eia, Berges:2013fga}.

The nonequilibrium plasma hydrodynamizes when a hydrodynamic description becomes applicable. This description using relativistic hydrodynamics \cite{Rezzolla:2013dea, Gale:2013da, Romatschke:2017ejr, Heinz:2024jwu} has been successfully used to understand many experimentally observed collective phenomena. For example, when off-central collisions lead to an initial almond-shaped overlap region of the colliding nuclei, hydrodynamics predicts an elliptical-shaped particle distribution observed in the detectors, known and observed as \emph{elliptic flow} \cite{Elfner:2022iae}. Hydrodynamics is a macroscopic description where microscopic properties only enter via effective transport parameters. As such, hydrodynamics is applicable for both weakly and strongly coupled systems. However, it is typically formulated as a (gradient) expansion around local thermal equilibrium, limiting its applicability to systems close to equilibrium, while kinetic theory provides a genuine description of the microscopic dynamics of the system valid also out of equilibrium.

When the temperature of the fluid drops below a critical ``freeze-out'' temperature, the hydrodynamic evolution is stopped and the energy density is converted back into hadrons which are then measured in the detectors (after a subsequent stage of hadronic interactions) \cite{Elfner:2022iae}.

It should be noted that while hydrodynamics as an effective macroscopic theory is applicable for all values of the coupling constant, kinetic theory and the Glasma stage rely on a weak coupling picture. There is a parallel effort to investigate the features of strongly coupled effective models of QCD, which can be achieved using the holographic principle and the gauge/gravity duality \cite{Maldacena:1997re, Witten:1998qj, Gubser:1998bc}. 
There, one uses the duality of a strongly coupled gauge theory to a weakly coupled gravitational theory, e.g., the AdS/CFT correspondence allows to probe $\mathcal N=4$ supersymmetric Yang-Mills (SYM), which is used as a model for QCD at large temperatures. Heavy-ion collisions are studied in the dual gravitational theory by performing extensive numerical simulations of shock wave collisions \cite{Chesler:2009cy, Chesler:2010bi, vanderSchee:2012qj, vanderSchee:2013pia}.

Despite their different setups, comparing kinetic theory and holographic models of the pre-equilibrium stages reveals astonishing similarities \cite{Keegan:2015avk, Kurkela:2019set}, such as the approach to a hydrodynamic attractor \cite{Heller:2015dha}.
This attractor refers to the property that, for various initial conditions, the system's time evolution follows a universal curve characterized by a reduced number of parameters \cite{Soloviev:2021lhs, Jankowski:2023fdz}. These attractors have been observed in various formulations of hydrodynamics \cite{Kurkela:2019set, Strickland:2017kux}, kinetic theory at weak couplings \cite{Kurkela:2015qoa, Heller:2016rtz, Kurkela:2018xxd, Giacalone:2019ldn, Du:2020zqg, Almaalol:2020rnu, Du:2022bel, Nugara:2024net, Rajagopal:2024lou}, and holographic models at strong couplings \cite{Heller:2012km, Heller:2016gbp, Keegan:2015avk, Kurkela:2019set}.
In particular, this identified universal behavior in certain observables allows making predictions from the pre-hydrodynamic stages despite incomplete information about the initial conditions \cite{Giacalone:2019ldn, Garcia-Montero:2023lrd, Garcia-Montero:2024lbl}. The notion of attractors in heavy-ion collisions and other theoretical setups has been the extensive subject of recent studies \cite{Mazeliauskas:2018yef, Du:2022bel, Heller:2023mah, DeLescluze:2025jqx, Berges:2025ccd, Brewer:2019oha, Brewer:2022vkq, Rajagopal:2024lou, Nugara:2024net, Mazeliauskas:2025jyi}.

Recently, there has been an increased interest to also consider the pre-hydrody\-namic evolution and search for possible experimental observables of this stage \cite{Kurkela:2018wud, Andres:2019eus, Gale:2021emg, Coquet:2023wjk, Avramescu:2024xts}. This is thought to be even more important for light ion collisions and smaller collision systems which spend a larger fraction of their time
before hydrodynamics is applicable. In particular, the upcoming lighter ions collision experiments, such as the planned oxygen-oxygen run at the LHC, will provide an environment in which nonequilibrium effects become more important \cite{Citron:2018lsq, Brewer:2021kiv}, making it now an ideal time to study nonequilibrium and pre-hydro\-dy\-na\-mic effects in heavy-ion collisions.

\section{Jet quenching}
\begin{figure}
    \centering
    \includegraphics[width=0.5\linewidth]{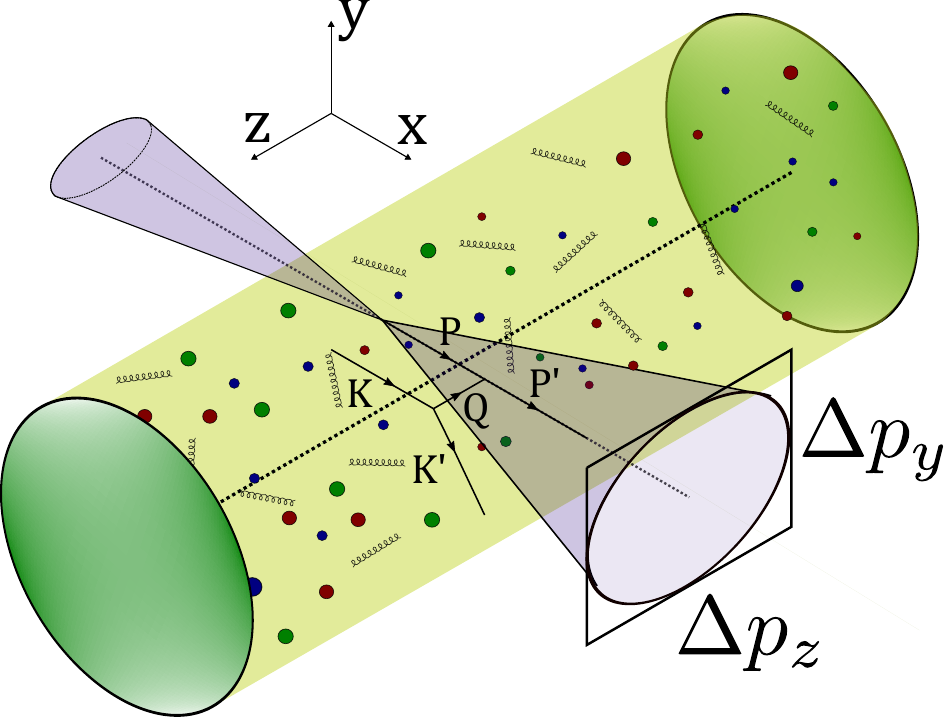}
    \caption{Schematic sketch of jet modification when traversing the quark-gluon plasma generated in a heavy-ion collision. Figure reused from \cite{Boguslavski:2023waw}.}
    \label{fig:jet-quenching-illustration}
\end{figure}
Experimentally accessible probes with potential contributions from the pre-hydro\-dynamic nonequilibrium stages are jets. They are observed as a collection of collimated particles with large momenta and are experimentally reconstructed using jet clustering algorithms \cite{Qin:2015srf, Apolinario:2022vzg}.

Generally, they originate from a highly energetic parton (quark or gluon) created in the initial collision, and, thus, 
interact with the QCD plasma during all stages of its evolution. Hence, they may carry imprints of the nonequilibrium initial stages. 
Experimental evidence suggests that jets lose energy while traversing the plasma, which is referred to as \emph{jet quenching} and is considered to be strong evidence for the formation of the quark-gluon plasma.

Experimentally, jet observables are typically compared to a corresponding observable measured in proton-proton collisions. For example, the nuclear modification factor is obtained by normalizing the yields or cross sections measured in nuclear collisions to their values in proton-proton collisions scaled by the number of binary collisions. If it is equal to one, there are no nuclear effects, i.e., the observable can be obtained as a superposition of proton-proton collisions, and is not modified by the medium generated in heavy-ion collisions.

Jet energy loss and jet quenching are dominated by the process of inelastic gluon emissions.
Without going into much more detail (see Chapter \ref{sec:jet-energy-loss}), the probability for medium-induced gluon radiation and the gluon spectrum depend on an effective propagator in a potential given by the \emph{dipole cross section} \cite{Zakharov:1996fv, Zakharov:1997uu, Zakharov:1998sv, Baier:1996kr, Baier:1996sk, Baier:2000mf}
\begin{align}
    C(\vb x)=\int\frac{\dd[2]{\vb q_\perp}}{(2\pi)^2}C(\vb q_\perp)\left(1-e^{i\vb x\cdot\vb q_\perp}\right). \label{eq:intro-dipole-crosssection-fouriertrafo}
\end{align}
It depends on the elastic collision kernel $C(\vb q_\perp)$ describing the probability per time of a jet particle to receive a momentum kick with transverse momentum $\vb q_\perp$. Different jet quenching formalisms \cite{Zakharov:1996fv, Zakharov:1997uu, Zakharov:1998sv, Baier:1996kr, Baier:1996sk, Baier:2000mf, Gyulassy:1999zd, Gyulassy:2000er, Gyulassy:2003mc, Wiedemann:2000za, Salgado:2003gb, Arnold:2002ja, Djordjevic:2008iz, Arnold:2008iy, Caron-Huot:2010qjx, Blaizot:2012fh} (see Ref.~\cite{Armesto:2011ht} for a comparison) differ in how they approximate the propagator or dipole cross section. In the \emph{harmonic oscillator approximation}, the small distance (small-$|\vb x|$) behavior of the dipole cross section is encoded in the jet quenching parameter $\qhat$,
\begin{align}
    C(\vb x)=\frac{1}{4}\qhat \vb x^2+\mathcal O(\vb x^4).\label{eq:intro-Cb-expansion}
\end{align}
This parameter has the physical interpretation of quantifying the transverse momentum broadening per unit time,
\begin{align}
    \qhat = \dv[\langle p_\perp^2\rangle]{t}=\int\frac{\dd[2]{\vb q_\perp}}{(2\pi)^2}\vb q_\perp^2C(\vb q_\perp).\label{eq:intro-qhat-definition}
\end{align}
Importantly, all the medium information is contained in the dipole cross section $C(\vb x)$ or, in the small-$|\vb x|$ limit, in the jet quenching parameter $\qhat$. They have been calculated analytically for a weakly-coupled plasma using perturbative QCD in thermal equilibrium at leading \cite{Aurenche:2002pd, Arnold:2008vd} and next-to-leading order \cite{Caron-Huot:2008zna} in the coupling. Further thermal results exist also for strongly coupled systems using AdS/CFT \cite{Liu:2006ug, Liu:2006he, Armesto:2006zv, DEramo:2010wup, Zhang:2012jd}, Lattice QCD \cite{Kumar:2020wvb}, dimensionally reduced electrostatic QCD \cite{Panero:2013pla, Moore:2021jwe}, and quasiparticle models \cite{Grishmanovskii:2022tpb, Song:2022wil, Grishmanovskii:2024gag}. 
There exist also extractions of the jet quenching parameter $\qhat$ from experimental data \cite{JET:2013cls, JETSCAPE:2021ehl, Xie:2022ght, JETSCAPE:2024cqe}.
In thermal equilibrium, the evolution of jets has been studied using kinetic theory \cite{He:2015pra, Schlichting:2020lef, Mehtar-Tani:2022zwf, Zhou:2024ysb}.
For the radiation spectrum, formalisms have been developed to go beyond the small $x$ limit of the dipole cross section by expanding in logarithms \cite{Arnold:2008zu, Arnold:2008vd, Mehtar-Tani:2019tvy, Mehtar-Tani:2019ygg, Barata:2021wuf} or using the full kernel to obtain the gluon spectrum and splitting rate \cite{Caron-Huot:2010qjx, Andres:2020vxs, Moore:2021jwe, Schlichting:2021idr, Andres:2023jao, Yazdi:2022bru, Modarresi-Yazdi:2024vfh}.
While these results have been obtained in equilibrium, there has been significant theoretical progress towards describing jet quenching also in inhomogeneous, anisotropic and flowing systems \cite{Romatschke:2004au, Romatschke:2006bb, Dumitru:2007rp, Hauksson:2021okc, Sadofyev:2021ohn, Andres:2022ndd, Barata:2022krd, 
Barata:2023qds, Kuzmin:2023hko, Barata:2024xwy}.

While many phenomenological models ignore or drastically simplify the time evolution of the plasma during these initial stages \cite{Bass:2008rv, Schenke:2009gb, Zapp:2012ak, Casalderrey-Solana:2014bpa, Cao:2017hhk, Huss:2020dwe, Huss:2020whe, Andres:2019eus, Zigic:2019sth, Andres:2022bql, JET:2013cls, JETSCAPE:2021ehl, Yazdi:2022bru, JETSCAPE:2022jer, Shi:2022rja, JETSCAPE:2023hqn, JETSCAPE:2024cqe}, it has been argued that the pre-equilibrium evolution is important for the correct description of experimental observables \cite{Andres:2019eus, Andres:2022bql}.
In particular, the jet quenching parameter $\qhat$ has recently been extracted during the Glasma stage during the earliest times of the nonequilibrium evolution \cite{Ipp:2020mjc, Ipp:2020nfu, Carrington:2020sww, Carrington:2021dvw, Carrington:2022bnv, Avramescu:2023qvv}, and its value was found to be surprisingly large while other studies find $\qhat$ should be negligible at early times to be compatible with experimental data \cite{Andres:2019eus}. However, different treatments of jet quenching during the initial stages reach partly different conclusions \cite{Andres:2019eus, Zigic:2019sth, Andres:2022bql, JETSCAPE:2023hqn}. This clearly highlights the need for a better theoretical understanding of jet quenching during the initial stages.

\begin{figure}
    \centering
    \includegraphics[width=0.5\linewidth]{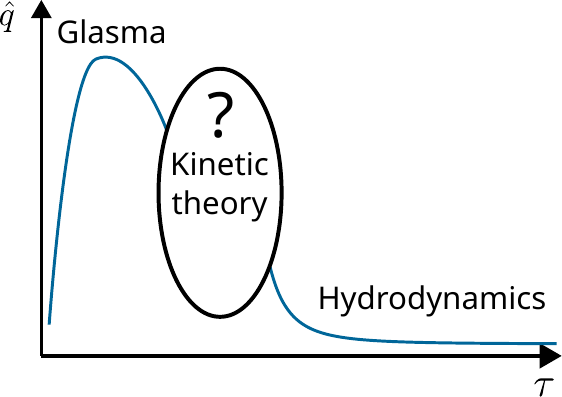}
    \caption{Schematic evolution of the jet quenching parameter $\qhat$: Its evolution has been obtained in the early Glasma stage and in the later hydrodynamic stage. Figure reused from \cite{Boguslavski:2023alu}.}
    \label{fig:qhat-schematic-evolution-intro}
\end{figure}

Although numerical simulations of the equilibration process in heavy-ion collisions using QCD kinetic theory have been performed \cite{Kurkela:2014tea, Kurkela:2015qoa, Kurkela:2018oqw, Kurkela:2018xxd, Du:2020dvp, Du:2020zqg}, the value of the jet quenching parameter $\qhat$ or the form of the collision kernel $C(\vb q_\perp)$ during the pre-equilibrium stages have not been established, as is sketched in Fig.~\ref{fig:qhat-schematic-evolution-intro}. This will be the main focus of this thesis. We will discuss the extraction of the jet quenching parameter $\qhat$ in Chapter \ref{sec:momentum-broadening-of-jets}, and argue that its anisotropy ratio follows a novel type of attractor, which is referred to as \textit{limiting attractor}, in Chapter \ref{sec:limiting_attractors}. Finally, we study the more general collision kernel $C(\vb q_\perp)$ in Chapter \ref{sec:collkern}.

Another important topic of this thesis is the question of how to properly include medium effects, both for obtaining the jet quenching parameter $\qhat$ and for QCD kinetic theory simulations. In the past, simple approximations have been used for the medium-resummed
matrix elements in the collision kernels. The effect of these assumptions has never been fully studied and will be the topic of Chapter~\ref{sec:improving-qcd-simulations}.

\newpage
\section{Guiding questions}
This thesis aims to answer the following questions, which are grouped in two main topics:
\begin{enumerate}
    \item Improving our understanding of jet-medium interactions, particularly during the initial stages
    \begin{enumerate}
    \item\label{question:qhat-definition} How can the jet quenching parameter $\qhat$ be obtained in QCD kinetic theory for a general, possibly anisotropic, distribution function $f(\vb p)$?

    \item\label{question:qhat-evolution} What is the time evolution of the jet quenching parameter $\qhat(\tau)$ during the hydrodynamization process in heavy-ion collisions and is its numerical value close to the large values reported during the Glasma stage?

    \item \label{question:qhat-ratio-attractor}Do the jet quenching parameter or its directional values exhibit universal dynamics? Does it follow a hydrodynamic attractor?
    
    \item\label{question:collisionkernel} How different is the collision kernel $C(\vb q_\perp)$ from its thermal and small-$x$ counterparts during the bottom-up evolution from its thermal counterpart? Can we identify specific features from the anisotropic plasma background?

    \end{enumerate}
    \item Improving our understanding of QCD equilibration
    \begin{enumerate}
    \item\label{question:screening-approximations} What is the impact of typically employed simplified screening prescriptions in QCD kinetic theory simulations, both on the jet quenching parameter and the time evolution itself?

    \item\label{question:other-attractors} Do late-time attractors beyond the hydrodynamic attractors exist? For which observables are they relevant?

    \item\label{question:splittingrates}
    How do the splitting rates obtained from the nonequilibrium anisotropic collision kernel $C(\vb q_\perp)$ compare to those employed in QCD kinetic theory simulations using an isotropic approximation?
    \end{enumerate}
\end{enumerate}

\section{Outline}
This thesis is organized as follows.
Chapter \ref{sec:jet-energy-loss} starts with a discussion on how an energetic parton loses energy in a quark-gluon plasma, and how different energy loss formalisms depend on the collision kernel $C(\vb q_\perp)$.
The following Chapter \ref{sec:qcd-kinetic-theory} contains an introduction to QCD kinetic theory, discusses all the relevant ingredients, and how one might, in principle, derive it from QCD itself.
It also discusses how QCD thermalizes in isotropic and expanding systems and how the QCD kinetic theory simulations employed in this thesis are performed.

Chapter \ref{sec:momentum-broadening-of-jets} addresses Questions \ref{question:qhat-definition}, \ref{question:qhat-evolution} and the parts of Question \ref{question:screening-approximations} regarding $\qhat$. It first discusses how the jet quenching parameter $\qhat$ can be obtained in QCD kinetic theory. Then $\qhat$ is calculated using a scaled thermal distribution and an effectively two-dimensional distribution as toy models to model features of the bottom-up equilibration process. Chapter \ref{sec:momentum-broadening-of-jets} concludes with obtaining the jet quenching parameter $\qhat$ during the hydrodynamization process between the Glasma and hydrodynamic stage.

The following Chapter \ref{sec:limiting_attractors} addresses Questions \ref{question:qhat-ratio-attractor} and \ref{question:other-attractors}. There, it is shown that the ratio of the jet quenching parameters cannot be well described using the typical hydrodynamic attractor time scale. Instead, the new concept of \emph{limiting attractors} is introduced. 

Question \ref{question:screening-approximations} is addressed in Chapter \ref{sec:improving-qcd-simulations}, where the screening prescription is discussed in more detail, in particular its gauge invariance. Additionally, numerical studies using different screening prescriptions are performed, both in isotropic and expanding systems. 
In this chapter, also the effects of different screening prescriptions in the background evolution on the jet quenching parameter $\qhat$ are studied.

Finally, Chapter \ref{sec:collkern} is concerned with Questions \ref{question:collisionkernel} and \ref{question:splittingrates}. First, the collision kernel is obtained during the hydrodynamization process in heavy-ion collision. Then, its Fourier transformed quantity, the dipole cross section, is computed and used as input to calculate the splitting rates using a novel method employed here for the first time.

A summary and conclusion are provided in Chapter \ref{sec:conclusions}.

In Appendix \ref{app:qcd-and-nonequilibrium-qft}, aspects of QCD and nonequilibrium quantum field theories are discussed. Appendix \ref{app:numericaldetails} discusses numerical details on performing QCD kinetic theory simulations.
Appendix \ref{app:large-momentum-limits-qhat} describes the limits of the jet quenching parameter $\qhat$ for large jet energies or large transverse momentum cutoffs. In Appendix \ref{app:amyrates-details}, numerical details on solving the anisotropic AMY rate equations are given. Finally, Appendix \ref{sec:additional-results-bottomup} provides an overview of additional plots and results of QCD kinetic theory simulations of the hydrodynamization process. Appendix \ref{sec:notational-differences} discusses notational differences between this thesis and the publications this thesis is based on.

\section{Conventions}
During this thesis, we will adopt the usual convention of setting the fundamental constants $c=\hbar=1$, which simplifies our expressions and allows measuring every dimensionful quantity in units of energy.
Throughout this thesis, we will employ the mostly-plus metric convention for the Minkowski metric $\eta_{\mu\nu}$, i.e.,
\begin{align}
    \eta_{\mu\nu}=\begin{pmatrix}
        -1&0&0&0\\
        0&1&0&0\\
        0&0&1&0\\
        0&0&0&1
    \end{pmatrix},\label{eq:metric-minkowski}
\end{align}
such that the scalar product of two vectors $P^\mu=(P^0,\vb p)$ and $K^\mu=(K^0,\vb k)$ is $P\cdot K=-P^0K^0 + \vb p\cdot \vb k$. Four-vectors in Minkowski space are denoted by uppercase letters, e.g., $P$. Lowercase boldfaced letters, e.g., $\vb p$, denote three-dimensional (or sometimes also two-dimensional) Euclidean vectors, whose magnitude is given by lowercase non-bold symbols, i.e., $p=|\vb p|$. For instance, a light-like $P^2=0$ vector can be represented as $P^\mu = (p, \vb p)$.

%% file: 420_energyloss.tex
\begin{figure}[t]
    \centering
    \includegraphics[width=0.5\linewidth]{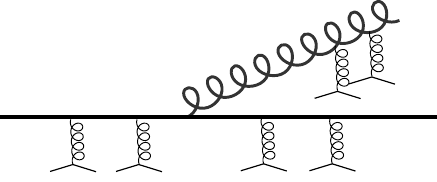}
    \caption{Inelastic gluon emission: A parton receives momentum kicks while traveling through the quark-gluon plasma and emits a gluon.}
    \label{fig:gluon-emission}
\end{figure}
In heavy-ion collisions, highly energetic particles are created, move through the quark-gluon plasma and cascade into a series of lower energetic particles, which then hadronize and are measured as jets in the detectors. Measuring their modification in the presence of a medium may yield insights into the nonequilibrium properties of QCD. However, a first principle calculation of this process is difficult. From a perturbative point of view, the dominant process for energy loss is inelastic gluon emission, where an energetic quark or gluon emits another gluon, as depicted in Fig.~\ref{fig:gluon-emission}. Therefore, jet energy loss is typically modeled as a sequence of single gluon emissions, neglecting the interference of two subsequent emissions for simplicity. In this chapter, we discuss how the rate of a jet particle emitting a single gluon can be calculated. Although a \emph{jet} is experimentally only well-defined through jet clustering algorithms, we will often refer to the single energetic particle we are considering as a \emph{jet}.

\section{\label{sec:lpm}How does an energetic gluon lose energy in a medium}
In a vacuum, a $1\to2$ splitting process is kinematically forbidden. In a medium, however, the quark in Fig.~\ref{fig:gluon-emission} can receive elastic momentum kicks (as depicted by the gluon exchange lines) and be brought slightly off-shell, which opens the phase space for a splitting process.
Describing such a single-splitting process is surprisingly complicated. To illustrate the physical picture, we will start by making simple estimates for the relevant processes, which can be found, e.g., in the review \cite{Schlichting:2019abc}.

In the simplest case, the rate for gluon emission is proportional to the rate of receiving a momentum kick $\Gammael$. This can be estimated using Eq.~\eqref{eq:intro-qhat-definition}, $\qhat=\dv[\langle p_\perp^2\rangle]{t}$. Integrating this, we obtain that during a time interval $\Delta t$, a jet acquires the transverse momentum $\Delta p_\perp^2\sim m^2\sim \qhat\Delta t$, where $m$ is a typical momentum transfer usually taken to be of the order of the screening mass. The rate $\Gammael\sim 1/\Delta t$ is then
\begin{align}
    \Gammael \sim \frac{\qhat}{m^2}.
\end{align}
Assuming the emission process is triggered by these elastic collision, we obtain the \emph{Bethe-Heitler} rate 
\begin{align}
    \dv[\Gamma^{\mathrm{BH}}]{z}\sim \alpha_s P_{g\to g}(z)\frac{\qhat}{m^2}, \label{eq:rate-bethe-heitler}
\end{align}
where the splitting function is given by
\begin{align}
    P_{g\to g}(z)=\CA\frac{1+z^4+(1-z)^4}{z(1-z)}, \quad \overset{z\ll 1}{\longrightarrow}\quad P_{g\to g}^{\mathrm{soft}}(z)\approx \frac{2\CA}{z},
\end{align}
where $z$ is the energy fraction of the emitted gluon, which we approximated for the case of soft gluon emissions $z\ll 1$. The constant $\CA=\NC$ is a group constant, see Appendix \ref{sec:qcd-lagrangian} and Eq.~\eqref{eq:group-constants}.
The Bethe-Heitler rate is only relevant if the formation time of the splitting process $\tform$ is much smaller than the rate at which the elastic collisions occur that trigger the splitting process, i.e., $\tform\ll 1/\Gammael$. To begin with, one may easily estimate the formation time by considering the time it takes for the wave packet of the emitted gluon with size $\Delta x_\perp\sim 1/k_\perp$ not to overlap with the mother gluon \cite{Kurkela:2011ti}. Taking the transverse velocity to be $v_\perp=k_\perp/k$, we obtain $\tform\sim \Delta x_\perp/v_\perp= k/k_\perp^2$, where we can then use the definition of the jet quenching parameter $\qhat$ from Eq.~\eqref{eq:intro-qhat-definition} to arrive at
\begin{align}
    \tform \sim \sqrt{\frac{\omega}{\qhat}},\label{eq:formation-time}
\end{align}
where $\omega=k$ is the energy of the emitted gluon (and, in general, the smallest energy of all participating partons). Thus, the Bethe-Heitler rate is valid for emitted energies
\begin{align}
    \omega\ll \frac{m^4}{\qhat}.
\end{align}
If this condition is not met, the individual emissions would overlap and one needs to take quantum mechanical interference effects into account. This is known as LPM suppression (first considered for QED in Refs.~\cite{Landau:1953ivy, Migdal:1956tc}, and obtained for QCD in \cite{Zakharov:1996fv, Zakharov:1998sv, Baier:1996sk, Baier:1996kr}).
The rate can then be estimated similarly as in Eq.~\eqref{eq:rate-bethe-heitler}, but using the formation time \eqref{eq:formation-time},
\begin{align}
    \dv[\Gamma^{\mathrm{LPM}}]{z}\sim \alpha_s P^{\mathrm{soft}}_{g\to g}(z)\sqrt{\frac{\qhat}{\omega}}.\label{eq:rate-lpm}
\end{align}

For its treatment in QCD, slightly different formalisms have been developed by different authors.\footnote{Ref.~\cite{Arnold:2008iy} provides a convenient overview of how the different conventions and formalisms are related.}

\section{AMY rate equations\label{sec:amy-rate-equation}}
In the limit of an infinite and time-independent medium, the rates for (nearly collinear) gluon emission or quark-antiquark creation were obtained by Arnold, Moore, and Yaffe \cite{Arnold:2002ja, Arnold:2002zm}, and are given by\footnote{With a slightly different convention of $\dv[\Gamma]{z}=\frac{(2\pi)^3}{\nu_g |\vb p|}\gamma \left(p; zp, (1-z)p\right)$.}
\begin{align}
    \gamma^{q}_{qg}(p;p',k)&=\gamma^{\bar q}_{\bar q g}(p; p', k)=\frac{p'^2+p^2}{p'^2p^2k^3}\mathcal F^{\hat n}_q(p,p',k),\\
    \gamma^g_{q\bar q}(p; p', k)&=\frac{k^2+p'^2}{p'^3p^3k^3}\mathcal F^{\hat n}_q(k,-p',p),\\
    \gamma^{g}_{gg}(p; p',k)&=\frac{p'^4+p^4+k^4}{p'^3p^3k^3}\mathcal F^{\hat n}_g(p,p',k),
\end{align}
where
\begin{align}
    \mathcal F^{\hat n}_s(p,p,k)=\frac{d_s C_s\alpha_s}{2(2\pi)^3}\int\frac{\dd[2]{\vb h}}{(2\pi)^2}2\vb h\cdot \RE \vb F_s^{\hat n}(\vb h; p',p,k),
\end{align}
where $C_s$ is the quadratic Casimir group constant corresponding to species $s$, given by Eq.~\eqref{eq:group-constants}.
Furthermore, $\vb F$ is the solution to the integral equation
\begin{align}
    \begin{split}
    2\vb h&=i\delta E(\vb h; p',p,k)\vb F_s^{\hat n}(\vb h; p', p,k)+\int\frac{\dd[4]{Q}}{(2\pi)^4}2\pi \delta(v_{\hat n}\cdot Q)v_{\hat n}^\mu v_{\hat n}^\nu g^2\llangle A_\mu(Q)[A_\nu(Q)]^*\rrangle \\
    &\qquad\times\Bigg\{(C_s-\frac{\CA}{2})[\vb F_s^{\hat n}(\vb h; p',p,k)-\vb F_s^{\hat n}(\vb h-k\vb q_\perp; p',pk)]\\
    &\qquad\qquad+\frac{\CA}{2}[\vb F_s^{\hat n}(\vb h; p',p,k)-\vb F_s^{\hat n}(\vb h+p'\vb q_\perp; p',p,k)]\\
    &\qquad\qquad+\frac{\CA}{2}[\vb F_s^{\hat n}(\vb h; p',p,k)-\vb F_s^{\hat n}(\vb h-p\vb q_\perp; p',p,k)]\Bigg\}.
    \end{split} \label{eq:amy-integralequation-long}
\end{align}
The vector $\vb h$ is a two-dimensional vector in the plane transverse to the direction of the splitting particles, $\vb {\hat n}$.
The double brackets $\llangle A_\mu(Q)[A_\nu(Q)]^*\rrangle$ denote the mean square fluctuations of the background gauge fields and is given by the (Fourier transformed) Wightman correlator \eqref{eq:wightman-correlator-gluons}.
The energy difference $\delta E$ is given by
\begin{align}
    \delta E(\vb h; p',p,k)=\frac{\meffg^2}{2k}+\frac{\meffs^2}{2p}-\frac{\meffs^2}{2p'}+\frac{\vb h^2}{2pkp'}, \label{eq:energy-difference}
\end{align}
and can be obtained by taking the particle momenta in Fig.~\ref{fig:gluon-emission} to be
\begin{align}
    (P')^\mu=(E_{\vb p'}, p', p_\perp+k_\perp,0), && K^\mu=(E_{\vb k},p,k_\perp,0), && P^\mu =(E_{\vb p}, p, p_\perp,0)
\end{align}
for a splitting $P'\to P + K$. Enforcing momentum conservation and expanding the energy for large $p'=p+k$, $E_{\vb p}=\sqrt{p^2+p_\perp^2+\meffs^2}\approx p + \frac{\meffs^2+p_\perp^2}{2p}$, which leads to Eq.~\eqref{eq:energy-difference} when defining $\vb h=z\vb k_\perp-(1-z)\vb p_\perp$, where $z=p/p'$. The effective masses encode medium-dependent corrections to the dispersion relation of the particles and are given explicitly in Eqs.~\eqref{eq:debyemass-general} and \eqref{eq:effective-mass-quark} in Section \ref{sec:qcd-kinetic-theory}.

The four-dimensional integral can be split into one part along the splitting and temporal direction, and another part containing the transverse integral (along $\vb q_\perp$),
\begin{align}
    \begin{split}
    2\vb h&=i\delta E(\vb h; p',p,k)\vb F_s^{\hat n}(\vb h; p', p,k)+\int\frac{\dd[2]{\vb q_\perp}}{(2\pi)^2}\bar C(\vb q_\perp)\\
    &\times\Bigg\{(C_s-\frac{\CA}{2})[\vb F^{\hat n}(\vb h; p',p,k)-\vb F_s^{\hat n}(\vb h-k\vb q_\perp; p',pk)]\\
    &+\frac{\CA}{2}[\vb F_s^{\hat n}(\vb h; p',p,k)-\vb F_s^{\hat n}(\vb h+p'\vb q_\perp; p',p,k)]\\
    &+\frac{\CA}{2}[\vb F_s^{\hat n}(\vb h; p',p,k)-\vb F_s^{\hat n}(\vb h-p\vb q_\perp; p',p,k)]\Bigg\},
    \end{split}
\end{align}
where the function \begin{align}
    \bar C(\vb q_\perp)=C(\vb q_\perp)/\CR=g^2\int\frac{\dd Q^0\dd Q^\parallel}{(2\pi)^2}2\pi\delta(v_{\hat n}\cdot Q)v_{\hat n}^\mu {v_{\hat n}}^\nu\llangle A_\mu(Q)[A_\nu(Q)]^*\rrangle \label{eq:collisionkernel-formula-from-correlator}
\end{align}
is the collision kernel $C(\vb q_\perp)$ stripped of its color factor $\CR$ and encodes the broadening of hard particles during the splitting process. It can be represented as a Wightman correlator of the gluon field generated by the hard particles moving through the plasma. In an isotropic plasma, it can be evaluated analytically (see Appendix \ref{app:sum-rule}).

This integral equation \eqref{eq:amy-integralequation-long} is typically solved by going to impact parameter space, where the before-mentioned dipole cross section or potential $C(\vb x)$ naturally appears (see Eq.~\eqref{eq:intro-dipole-crosssection-fouriertrafo}),
\begin{align}
    \bar C(\vb x)=\int\frac{\dd[2]{\vb q_\perp}}{(2\pi)^2}(1-e^{i\vb x\cdot \vb q_\perp})\bar C(\vb q_\perp).\label{eq:Cbar-fouriertrafo-collisionkernel}
\end{align}
We will discuss in Chapter \ref{sec:collkern} (and Appendix \ref{app:amyrates-details}) how to obtain the splitting rate $\gamma$ numerically for an anisotropic collision kernel $C(\vb q_\perp)$. If the incoming particle is highly energetic, only the small-$|\vb x|$ behavior of this dipole cross section $C(\vb x)$ is important, and the rate can be obtained by an expansion in logarithms \cite{Arnold:2008zu, Arnold:2008vd}.

These rate equations are used to study jet quenching in the \texttt{MARTINI} framework \cite{Schenke:2009gb}, and will be a fundamental ingredient for the effective kinetic description of QCD \cite{Arnold:2002zm}, which we will introduce and discuss in the next Chapter \ref{sec:qcd-kinetic-theory}.

\section{Medium-induced radiation spectrum in a finite medium}
\subsection{Energy spectrum}
In a medium with finite length and also time varying properties, the spectrum is usually written in a seemingly different form (see, e.g., \cite{Zakharov:1997uu, Arnold:2008iy, Caron-Huot:2010qjx}),
\begin{align}
    \omega\dv[I]{\omega}=\frac{\alpha_s z P_{s\to g(z)}}{[z(1-z)E]^2}\RE\int_0^\infty\dd{t_1}\int_{t_1}^\infty\dd{t_2}\partial_{\vb x}\cdot \partial_{\vb y}\left\{\mathcal K(\vb y,t_2; \vb x, t_1)-\mathcal K_{\mathrm{vac}}(\vb y,t_2;\vb x,t_1)\right\}, \label{eq:spectrum-energy}
\end{align}
where $\mathcal K$ is the Green's function for a two-dimensional Schrödinger equation
\begin{align}
    i\partial_t\psi(\vb x,t)=\left(\delta E(\vb p_x)-i\Gamma_3(\vb x,t)\right)\psi(\vb x,t),\label{eq:2D-schroedinger-equation}
\end{align}
with initial condition $\mathcal K(\vb x,t; \vb y,t)=\delta^{(2)}(\vb x-\vb y)$. That this spectrum \eqref{eq:spectrum-energy} is related to the rates obtained in the previous section can be seen, e.g., that in the integral equation \eqref{eq:amy-integralequation-long} the energy difference $\delta E$ from Eq.~\eqref{eq:energy-difference} appears as well. The potential $\Gamma_3$ is also reminiscent of the terms in the integral equation,
\begin{align}
    \Gamma_3(\vb x,t)=\frac{\CA}{2}\bar C(\vb x, t)+\left(\Cs - \frac{\CA}{2}\right)\bar C(z \vb x,t)+\frac{\CA}{2}\bar C\left((1-z)\vb x,t\right),
\end{align}
where the function $\bar C(\vb x,t)$, as in Eq.~\eqref{eq:Cbar-fouriertrafo-collisionkernel}, is related to the collision kernel $\bar C(\vb q_\perp,t)$.
For highly-energetic partons, we have seen before in the qualitative discussion that the LPM regime is that of relevance. There, the results depended crucially on the jet quenching parameter $\qhat$, which determines the small distance behavior (see Eq.~\eqref{eq:intro-Cb-expansion}) of this dipole cross section $C(\vb b,t)$, which we will discuss in more detail in Section \ref{sec:smalldistance-dipolecrosssection}.

\subsection{Medium-induced radiation spectrum differential in transverse momentum}
One may also, instead of just giving the energy spectrum of emitted gluons, consider explicitly gluons with different transverse momenta, i.e., become differential in $k_\perp$. Doing that for a general gluon energy fraction $z$ becomes rather convoluted (see, e.g., Ref.~\cite{Apolinario:2014csa}), and we will stick to the case of soft emissions $z\ll 1$ here. In that case, the spectrum can be written as (see, e.g., \cite{Wiedemann:2000za})
\begin{align}\label{eq:spectrum}
\frac{\dd I}{\dd\omega \dd[2]{\vb k}}=\frac{\bar \alpha}{2\pi\omega^3}\RE\!\int_{t_0}^\infty\!\dd{t_2}\!\int_{t_0}^{t_2}\!\dd{t_1}\int_{\vb x}e^{-i\vb k\cdot\vb x}\mathcal P(\vb x,\infty; t_2)\partial_{\vb x}\cdot \partial_{\vb y} \mathcal K(\vb x,t_2; \vb y, t_1)_{\vb y = 0}-\frac{8\bar \alpha \pi}{\vb k^2\omega}.
\end{align}

The functions $\mathcal P$ and $\mathcal K$ describe the broadening of the emitted gluon and the splitting process, respectively. Again, the medium enters these quantities via the dipole cross section $C(\vb x, t)$,
\begin{align}
    \partial_t \mathcal P(\vb x, t; t_1)&=-C(\vb x,t)\mathcal P(\vb x, t; t_1),\\
    \left(i\partial_t-\frac{\partial_{\vb x}^2}{2\omega}+iC(\vb x,t)\right)\mathcal K(\vb x,t; \vb y, t_1)&=i\delta^{(2)}(\vb x - \vb y)\delta(t_1-t).\label{eq:propagator-in-2d-schroedinger-equation-smallx}
\end{align}
Indeed, the equation for the Green's function $\mathcal K$ here can be obtained by taking the $z\ll 1$ limit of the previous section and is the same function that appears in Eq.~\eqref{eq:spectrum-energy}.
As discussed before, for a highly energetic jet, the small-$|\vb x|$ behavior of the dipole cross section is most important and can be expressed (to a first approximation) as 
\begin{align}
    C(\vb x,t)=\frac{1}{4}\qhat(t)\vb x^2+ \dots \,\,,\label{eq:Cx_harmonic-approximation-first}
\end{align}
where the parameter $\qhat$ is the jet quenching parameter from Eq.~\eqref{eq:intro-Cb-expansion} and \eqref{eq:intro-qhat-definition}. Because this parameter characterizes the small distance behavior of the dipole cross section, it is an important ingredient for jet energy loss and jet quenching calculations. This parameter will also be the central object of Chapter \ref{sec:momentum-broadening-of-jets}.

This quadratic approximation is also known as the \emph{harmonic approximation} or the \emph{harmonic oscillator approximation}, 
because in the approximation \eqref{eq:Cx_harmonic-approximation-first}, the two-dimensional Schrödinger equation \eqref{eq:2D-schroedinger-equation} (and Eq.~\eqref{eq:propagator-in-2d-schroedinger-equation-smallx}) reduces to that of a (time-dependent) harmonic oscillator. This allows for an analytic solution.
More physically motivated,  this approximation is also referred to as the \emph{multiple soft scattering approximation}, because the jet quenching parameter $\qhat$ describes the momentum broadening due to many (soft) scatterings with the medium (see Eq.~\eqref{eq:intro-qhat-definition}).

We will further discuss this small distance behavior in more detail in Section~\ref{sec:smalldistance-dipolecrosssection}.

%% file: 400_theory_background.tex
In this chapter, we discuss the assumptions and equations of the effective kinetic theory description of QCD. We discuss how medium effects enter and how they are typically approximated. In Section \ref{sec:kinetic-theory-derivation}, we discuss how a kinetic description can be derived from the underlying quantum field theory, and derive the expression for the elastic scattering rate. We discuss how QCD kinetic theory has been used to study QCD equilibration in Section \ref{sec:known-things-about-qcd-equilibration}. Finally, we conclude this chapter in Section \ref{sec:performing-qcd-kinetic-theory-simulations} by discussing and giving more details on how to perform numerical simulations of QCD kinetic theory.

\section{Validity of QCD kinetic theory}
At sufficiently weak coupling (which, in physical terms for QCD, means sufficiently high energies), QCD admits an effective kinetic description, first formulated by Arnold, Moore, and Yaffe \cite{Arnold:2002zm}. Generally, a kinetic description is valid when the mean free path between collisions (or interactions) is much larger than the duration of a scattering/collision process, which is treated as instantaneous. Additionally, the quantum mechanical wave packets corresponding to the particles must be well describable by classical particles, meaning that the De Broglie wavelength must be much smaller than the mean free path.

In the weak-coupling picture, this is the case: The extent of typical excitations with momenta $P\sim T$ in a thermal system is $\sim 1/T$, while the mean free path is $1/(g^2T)$ for small-angle scatterings and $1/(g^4T)$ for large angle scatterings. Since the formation time for a gluon emission $1/(g^2T)$ is of similar order as the mean free path between small-angle scatterings, it is required to take the quantum mechanical interference between different splitting processes into account, leading to the LPM rate \eqref{eq:rate-lpm} with mean free path\footnote{This can be seen by using the simple estimate $\qhat\sim g^4 T^3$ (see later Chapter \ref{sec:momentum-broadening-of-jets}) in the LPM rate $\dd{\Gamma^{\mathrm{LPM}}}/\dd{z}\sim g^2\sqrt{\hat q/\omega}$ from Eq.~\eqref{eq:rate-lpm}.} $1/(g^4T)$.

While the estimates in the previous paragraph were given for a plasma in thermal equilibrium, the kinetic description is also valid out of equilibrium, given that certain conditions are fulfilled \cite{Arnold:2002zm}. We assume that all particle masses are negligible with respect to thermal (or effective) masses $\sim gT$, which should again be much smaller than the momenta of ``relevant'' physical excitations, constituting a \emph{scale separation} between the ``relevant'' excitations and medium-dependent corrections to dispersive relations or to the inverse Debye screening length. 
Further required assumptions are that the distribution functions $f_s(t,\vb p,\vb x)$ do not vary significantly over spacetime regions of the formation time $\tform$ of near-collinear processes (see Eq.~\eqref{eq:formation-time}) for the relevant excitations. Furthermore, the distribution functions should not vary significantly with $\mathcal O(\meffs)\sim gT$ changes in momentum, and the distribution function should not be non-perturbatively large, $f_s\ll 1/\alpha_s$.

\section{The equations of QCD kinetic theory\label{sec:equations-of-qcd-kinetic-theory}}
The Boltzmann equations for the effective kinetic theory description of QCD have been formulated in Ref.~\cite{Arnold:2002zm}, where all leading-order relevant scattering processes are taken into account. While next-to-leading order (NLO) corrections have also been obtained \cite{Ghiglieri:2015ala}, this thesis focuses on the leading order description.

In a kinetic description, the fundamental quantity is the distribution function 
\begin{align}
    f_s(t,\vb p,\vb x)=\frac{(2\pi)^3}{\nu_s}\frac{\dd{N_s}}{\dd[3]{\vb p}\dd[3]{\vb x}},
\end{align}
describing the number of particles $N_s$ of species $s$ per phase-space volume. It is averaged over all $\nu_s$ spin and color states ($\nu_g=2\dA=2(\NC^2-1)$ for gluons and $\nu_q=2\dF=2\NC$ for quarks/antiquarks). Equivalently, we may view the distribution function $f_s(t,\vb p,\vb x)$ as the distribution function for a single species of a particular color and spin state, which is identical for all $\nu_s$ degrees of freedom.

The distribution functions obey the coupled Boltzmann equations
\begin{align}
	\left(\pdv{t}+\vb v\cdot \nabla \right)f_s(t,\vb p, \vb x) =-\Conetwo_s[f_a(t,\vb p,\vb x)]- \Ctwotwo_s[f_a(t,\vb p,\vb x)],\label{eq:boltzmann_equation}
\end{align}
which describe how they change due to elastic ($\Ctwotwo$) and inelastic ($\Conetwo$) collisions.
The collision terms are local in space $\vb x$ and time $t$; therefore, their explicit space-time dependence will be frequently omitted.

The inelastic collision term $\Conetwo$ describes strict collinear splitting (where all momenta are proportional to a unit vector $\vb {\hat n}=\vbphat=\vb p/p$),
and is given by
\begin{align}
	\begin{split}
		\Conetwo_a[f_i(\vb p)]&=\frac{(2\pi)^3}{2p^2\nu_a}\sum_{bc}\int_0^\infty\dd{p'}\dd{k'}\delta(p-p'-k')\gamma^a_{bc}(p;p',k')\\
		&\quad\times\Big\{f_a(\vb p)(1\pm f_b(p'\vbphat))(1\pm f_c(k'\vbphat))-f_b(p'\vbphat)f_c(k'\vbphat)(1\pm f_a(\vb p))\Big\}\\
		&+\frac{(2\pi)^3}{p^2\nu_a}\sum_{bc}\int_0^\infty\dd{p'}\dd{k}\delta(p+k-p')\gamma^c_{ab}(p';p,k)\\
		&\quad\times\Big\{f_a(\vb p)f_b(k\vbphat)(1\pm f_c(p'\vbphat))-f_c(p'\vbphat)(1\pm f_a(k\vbphat))(1\pm f_b(\vb p))\Big\}. \label{eq:c12}
	\end{split}
\end{align}
The upper signs are to be used for gluons (bosons) and the lower signs for quarks (fermions), 
\begin{align}
    (1\pm f_a(\vb p)) = \begin{cases}
        1+f_a(\vb p), & \text{for gluons, i.e., }a = g,\\
        1-f_a(\vb p), & \text{for quarks, i.e., }a = q,\bar q
    \end{cases}\quad,\label{eq:boseenhancement-fermi-blocking}
\end{align}
representing Bose enhancement and Fermi blocking, respectively.
The effective splitting/joining rates $\gamma^a_{bc}(p;p',k)$ depend on the unit vector $\vb {\hat n}$ and interpolate between the Bethe-Heitler and LPM regime. They were already discussed in the last Chapter \ref{sec:lpm}, but are, for completeness, reiterated here. They are given by
\begin{subequations}
\begin{align}
    \gamma^{q}_{qg}(p;p',k)&=\gamma^{\bar q}_{\bar q g}(p; p', k)=\frac{p'^2+p^2}{p'^2p^2k^3}\mathcal F^{\hat n}_q(p,p',k),\\
    \gamma^g_{q\bar q}(p; p', k)&=\frac{k^2+p'^2}{p'^3p^3k^3}\mathcal F^{\hat n}_q(k,-p',p),\\
    \gamma^{g}_{gg}(p; p',k)&=\frac{p'^4+p^4+k^4}{p'^3p^3k^3}\mathcal F^{\hat n}_g(p,p',k),
\end{align}
\end{subequations}
where
\begin{align}
    \mathcal F^{\hat n}_s(p,p,k)=\frac{d_s C_s\alpha_s}{2(2\pi)^3}\int\frac{\dd[2]{\vb h}}{(2\pi)^2}2\vb h\cdot \RE \vb F_s^{\hat n}(\vb h; p',p,k),
\end{align}
and $\vb F$ is the solution to the integral equation
\begin{align}
    \begin{split}
    2\vb h&=i\delta E(\vb h; p',p,k)\vb F_s^{\hat n}(\vb h; p', p,k)+\int\frac{\dd[2]{\vb q_\perp}}{(2\pi)^2}\bar C(\vb q_\perp)\\
    &\times\Bigg\{(C_s-\frac{\CA}{2})[\vb F^{\hat n}(\vb h; p',p,k)-\vb F_s^{\hat n}(\vb h-k\vb q_\perp; p',pk)]\\
    &+\frac{\CA}{2}[\vb F_s^{\hat n}(\vb h; p',p,k)-\vb F_s^{\hat n}(\vb h+p'\vb q_\perp; p',p,k)]\\
    &+\frac{\CA}{2}[\vb F_s^{\hat n}(\vb h; p',p,k)-\vb F_s^{\hat n}(\vb h-p\vb q_\perp; p',p,k)]\Bigg\}.
    \end{split}
\end{align}
The vector $\vb h$ is a two-dimensional vector in the plane transverse to the direction of the splitting particles, $\vb {\hat n}$.
The energy difference $\delta E$ is given by
\begin{align}
    \delta E(\vb h; p',p,k)=\frac{\meffg^2}{2k}+\frac{\meffs^2}{2p}-\frac{\meffs^2}{2p'}+\frac{\vb h^2}{2pkp'}.
\end{align}
The function $\bar C(\vb q_\perp)=C(\vb q_\perp)/\CR$ is the collision kernel stripped of its color factor and encodes the broadening of hard particles during the splitting process. It can be represented as a Wightman correlator (see Eqs.~\eqref{eq:collisionkernel-formula-from-correlator}, and Eq.~\eqref{eq:wightman} in Appendix \ref{app:qcd-and-nonequilibrium-qft} for its definition) of the gluon field generated by the hard particles moving through the plasma.
With an isotropic screening approximation, it can be written as the difference of the retarded transverse and longitudinal hard thermal loop propagators evaluated at $\omega=q_\parallel=0$ (see Appendix \ref{app:sum-rule}),
\begin{align}
    \bar C(\vb q_\perp)=g^2T_\ast\left(\frac{1}{q_\perp^2}-\frac{1}{q_\perp^2+m_D^2}\right),
\end{align}
with the infrared temperature $T_\ast$ given by
\begin{align}
    T_\ast = \frac{\sum_s\nu_s \frac{g^2C_s}{\dA}\int\frac{\dd[3]{\vb p}}{(2\pi)^3}f(\vb p)(1\pm f(\vb p))}{m_D^2}\label{eq:tstar-definition}
\end{align}
and with the nonequilibrium Debye mass
\begin{align}
	m_D^2 =  2\meffg^2=\sum_s4\nu_s\frac{g^2C_s}{\dA}\int\frac{\dd[3]{\vb p}}{(2\pi)^32|\vb p|}f_s(\vb p).\label{eq:debyemass-general}
\end{align}
The effective quark mass of flavor $s$ is given by
\begin{align}
    \meffs^2=2g^2\CF\int\frac{\dd[3]{\vb p}}{(2\pi)^32|\vb p|}\left[2f_g(\vb p)+f_s(\vb p)+f_{\bar s}(\vb p)\right].\label{eq:effective-mass-quark}
\end{align}
In equilibrium, they are given by
\begin{align}
    T_\ast = T, && m_D^2=g^2T^2\left(\frac{\NC}{3}+\frac{\nf}{6}\right), && \meffs^2=\frac{\CF g^2}{4}. \label{eq:equilibriumform-debyemass-tstar-meffs}
\end{align}
The splitting rates $\gamma^a_{bc}$ are symmetric under the exchange of the outgoing particles
\begin{align}
    \gamma^a_{bc}(p; p',k)=\gamma^a_{cb}(p; k,p').
\end{align}

The elastic collision term $\Ctwotwo$ is given by
\begin{align}
	\Ctwotwo_a[f_i(\vb p)]&=\frac{1}{4|\vb{p}|\nu_a}\sum_{bcd}\int_{\vb{kp'k'}}\left|\mathcal M^{ab}_{cd}(\vb{p},\vb{k};\vb{p'}\vb{k'})\right|^2 
	(2\pi)^4\delta^4(P+K-P'-K')\label{eq:c22_first}\\
	&\!\!\!\!\!\times\Big\{f_a(\vb p)f_b(\vb k)\left[1\pm  f_c(\vb p')\right]\left[1\pm f_d(\vb k')\right]
	- f_c(\vb p')f_d(\vb k')\left[1\pm f_a(\vb p)\right]\left[1 \pm  f_b(\vb k)\right]\Big\} \nonumber.
\end{align}
Again, the upper signs are to be used for gluons and the lower signs for quarks, as stated in Eq.~\eqref{eq:boseenhancement-fermi-blocking}.
The Lorentz-invariant integration measure is defined as
\begin{align}
	\int_{\vb k}:=\int\frac{\dd[3]{\vb k}}{(2\pi)^3 2k}.\label{eq:integration-measure}
\end{align}

\begin{table}
\centering
\begin{tabular}{ M{0.2\linewidth}  m{0.65\linewidth} }
\toprule 
\vspace{5pt}
$ab\leftrightarrow cd$ & $\left|\mathcal M^{ab}_{cd}\right|^2/g^4$ \\[5pt] 
\hline 
$q_1q_2\leftrightarrow q_1q_2,$ & \multirow{4}{*}{$8\frac{\dF^2\CF ^2}{\dA}\left(\frac{s^2+u^2}{\underline{t^2}}\right)$}\\ $q_1\bar q_2\leftrightarrow q_1\bar q_2,$ &\\ $\bar q_1 q_2\leftrightarrow \bar q_1 q_2,$ &\\ $\bar q_1\bar q_2\leftrightarrow \bar q_1\bar q_2$ &  \\[12pt] 
 
$q_1q_1\leftrightarrow q_1q_1,$ & \multirow{2}{*}{$8\frac{\dF^2\CF ^2}{\dA}\left(\frac{s^2+u^2}{\underline{t^2}}+\frac{s^2+t^2}{\underline{u^2}}\right)+16 \dF \CF \left(\CF -\frac{\CA }{2}\right)\frac{s^2}{tu}$}\\ $\bar q_1 \bar q_1\leftrightarrow \bar q_1 \bar q_1$ & \\ [12pt]
 
$q_1\bar q_1\leftrightarrow q_1\bar q_1$ &$8\frac{\dF ^2\CF ^2}{\dA}\left(\frac{s^2+u^2}{\underline{t^2}}+\frac{t^2+u^2}{s^2}\right)+16 \dF \CF \left(\CF -\frac{\CA }{2}\right)\frac{u^2}{st}$ \\[12pt]

$q_1\bar q_1\leftrightarrow q_2\bar q_2$ & $8\frac{\dF ^2\CF ^2}{\dA}\left(\frac{t^2+u^2}{s^2}\right)$\\[10pt]
$q_1\bar q_1\leftrightarrow gg$ & $8 \dF  \CF ^2\left(\frac{u}{\underline{\underline{t}}}+\frac{t}{\underline{\underline{u}}}\right) - 8\dF \CF \CA\left(\frac{t^2+u^2}{s^2}\right)$ \\[12pt] 
$q_1 g\leftrightarrow q_1 g,$ & \multirow{2}{*}{$-8\dF \CF ^2\left(\frac{u}{s}+\frac{s}{\underline{\underline{u}}}\right)+8\dF \CF \CA\left(\frac{s^2+u^2}{\underline{t^2}}\right)$}\\
$\bar q_1 g\leftrightarrow \bar q_1 g$ & \\[12pt]

$gg\leftrightarrow gg$ & $16\dA\CA^2\left(3-\frac{su}{\underline{t^2}}-\frac{st}{\underline{u^2}}-\frac{tu}{s^2}\right)$\\[5pt]
\bottomrule 
\end{tabular}
\caption{Matrix elements for the elastic collision term \eqref{eq:c22_first} from \cite{Arnold:2002zm}. Singly-underlined denominators indicate infrared-sensitive contributions from soft-gluon exchange, doubly-underlined denominators from soft-fermion exchange. The constants $\dF$, $\CF$, $\dA$, and $\CA$ are given in Eq.~\eqref{eq:group-constants}.
}
\label{tab:amy_matrix_el}
\end{table}

At leading-order, the required matrix elements are listed in Table \ref{tab:amy_matrix_el}, where $s$, $t$, and $u$ are the usual Lorentz invariant Mandelstam variables, which are given in terms of the incoming ($P$, $K$) and outgoing ($P'$, $K'$) momenta,
\begin{align}
s=-(P+K)^2,&& t=-(P'-P)^2, && u=-(K'-P)^2. \label{eq:Mandelstam-definition}
\end{align}
Recall that
we use the mostly plus metric convention here, see Eq.~\eqref{eq:metric-minkowski}. The Mandelstam variables satisfy the usual relation (for massless particles)
\begin{align}
	s+t+u=0.\label{eq:mandelstam_sum}
\end{align}
The matrix elements for the elastic collision term obey the symmetries
\begin{align}
    \left|\mathcal M^{ab}_{cd}(\vb p,\vb k; \vb p',\vb k')\right|^2=\left|\mathcal M^{ab}_{dc}(\vb p,\vb k;\vb k',\vb p')\right|^2=\left|\mathcal M^{ba}_{cd}(\vb k,\vb p;\vb p',\vb k')\right|^2=\left|\mathcal M^{cd}_{ab}(\vb p',\vb k';\vb p,\vb k)\right|^2,\label{eq:c22-matrixelements-symmetries}
\end{align}
corresponding to switching the outgoing particles ($c\leftrightarrow d$), the incoming particles ($a\leftrightarrow b$), or the incoming with the outgoing particles $(ab)\leftrightarrow(cd)$.

In thermal equilibrium, both elastic and inelastic collision terms identically vanish due to the detailed balance condition. This can be seen easily by including the thermal distribution functions
\begin{align}
    f_\pm(k;T)=\frac{1}{\exp(k/T)\mp 1} \label{eq:thermal-distributionfunctions}
\end{align}
in the collision terms \eqref{eq:c12} and \eqref{eq:c22_first}. The upper sign $f_+$ denotes the Bose-Einstein distribution (relevant for gluons) and $f_-$ the Fermi-Dirac distribution (relevant for quarks and antiquarks). Note that this label in $f_\pm$ is chosen such that the statistical factors in \eqref{eq:boseenhancement-fermi-blocking} become $(1\pm f_\pm(p))$.

\section{Medium effects: Screening of soft modes\label{sec:amy-screening-prescription}}
\subsection{Where screening is needed}
In the inelastic collision term, medium effects enter via the collision kernel $\bar C(\vb q_\perp)$, opening up the phase space for the effective $1\to 2$ splitting and $2\to 1$ merging processes. 

For the elastic $2\leftrightarrow 2$ scattering processes\footnote{This subsection follows the discussion in Ref.~\cite{Boguslavski:2024kbd}.} entering the elastic collision term \eqref{eq:c22_first}, one needs to calculate the $2\to 2$ scattering matrix elements up to the needed order in perturbation theory, in our case up to leading order. However, it is a well-known fact in thermal field theory that for a thermal medium, a na\"ive expansion in loops is not sufficient, i.e., for instance, the leading order expressions need input from an arbitrarily high number of loops. This happens when an internal propagator becomes soft, in which case it has to be resummed using the hard thermal loop effective theory \cite{Braaten:1989mz, Braaten:1990az, Frenkel:1989br}, which has been applied to non-thermal media as well \cite{Mrowczynski:2000ed, Mrowczynski:2004kv, Romatschke:2003ms, Romatschke:2004jh}.

The most straightforward way to compute the scattering matrix elements with medium effects included is to reevaluate the matrix elements and replace all internal propagators with resummed hard thermal loop (HTL) propagators.\footnote{Throughout this thesis, we will keep using the phrase \textit{hard thermal loops} even for systems out of equilibrium, where other authors have used \textit{hard loops} instead.} This is rather cumbersome,\footnote{For instance, even for the vacuum case, where the gluon propagator is given by Eq.~\eqref{eq:gluon-propagator-vacuum} instead of the more complicated HTL propagator \eqref{eq:HTL-propagators}, Peskin and Schroeder note that the cross section for ``gluon-gluon scattering [...] is rather tedious to evaluate'' \cite{Peskin:1995ev}.} and Arnold, Moore, and Yaffe \cite{Arnold:2002zm} propose a different, leading-order equivalent, way to achieve the same effect. One simply needs to replace the underlined terms in Table \ref{tab:amy_matrix_el} by
\begin{subequations}
\begin{align}
	{\frac{(s-u)^2}{\underline{t^2}}}&\to \left|\Gret_{\mu\nu}(P-P')\;(P+P')^\mu(K+K')^\nu\right|^2\label{eq:amy_replacement},\\
    \frac{u}{\underline{\underline{t}}}&\to \frac{4\RE[(P\cdot \mathcal Q)(K\cdot \mathcal Q)^*]+s \mathcal Q\cdot \mathcal Q^*}{|\mathcal Q\cdot \mathcal Q|^2},
\end{align}
\end{subequations}
where the first line is needed for soft-gluon exchanges and the second line is for soft-quark exchanges.
Here, $G_{\mu\nu}(Q)$ denotes the retarded gluon propagator, and $\mathcal Q$ is defined as $\mathcal Q^\mu=P^\mu-P'^\mu-\Pi_{\mathrm{Ret}}^\mu(P-P')$. We will discuss why the retarded propagator needs to be used in more detail in Section \ref{sec:kinetic-theory-derivation}.
Note that when using the vacuum propagator 
\begin{align}
G^0_{\mu\nu}(Q)=\frac{\eta^{\mu\nu}}{Q^2} \label{eq:gluon-propagator-vacuum}
\end{align}
in Eq.~\eqref{eq:amy_replacement}, the right-hand side reduces to the left-hand side, i.e., for no medium modifications, we recover the vacuum case. We discuss the motivation for this replacement and its validity in more detail in the following.

In this thesis, we mainly consider gluons, and in particular, we will only need the prescription for soft-gluon exchange \eqref{eq:amy_replacement}. This can be motivated as follows:
First, note that using relation \eqref{eq:mandelstam_requirement} between the Mandelstam variables, we may rewrite the gluon-gluon matrix element from Table \ref{tab:amy_matrix_el} as 
\begin{align}
	\label{eq:gluonic_matrixel}
	\frac{\left|\mathcal M\right|^2}{4 \lambda^2 \dA} =  9 + \frac{(t-s)^2}{\underline{u^2}} + \frac{(s-u)^2}{\underline{t^2}} + \frac{(u-t)^2}{s^2} \,.
\end{align}
Using the symmetry of exchanging $u$ and $t$ (corresponding to the exchange of the external particles with momenta $\vb p'$ and $\vb k'$), we may rewrite this to
\begin{align}
	\label{eq:gluonic_matrixel_simple}
	\frac{\left|\mathcal M\right|^2}{4 \lambda^2 \dA} =  9 + 2{\frac{(s-u)^2}{\underline{t^2}}} + \frac{(u-t)^2}{s^2} \,.
\end{align}
At tree level, which is relevant for the leading-order matrix elements in Table \ref{tab:amy_matrix_el}, only one internal propagator $G$ appears.
Medium corrections to its free form $G_0$ are conventionally encoded in the self-energy (see Appendix \ref{sec:perturbation-theory-self-energy}), and the full propagator can be schematically written (when $G_0^{-1}=Q^2$) as
\begin{align}
    G(Q)=\frac{1}{Q^2 + \Pi(Q)}.
\end{align}
Thus, the self-energy introduces an effective mass $m_{\mathrm{eff}}$.
In a (thermal) medium, this self-energy is proportional to the screening mass scale $\Pi \sim m_D^2 = \mathcal O(g^2T^2)$ (see Eq.~\eqref{eq:debyemass-general}). 
Therefore, at leading order, medium effects are only relevant for soft momenta $Q\sim\mathcal O(gT)$.

Using the Mandelstam variables \eqref{eq:Mandelstam-definition}, it is easy to see that internal soft momenta correspond to the case\footnote{For the small $u$ region, one may always use the symmetry $t \leftrightarrow u$ to rewrite the matrix element \eqref{eq:gluonic_matrixel_simple} such that only the small $t$ behavior needs to be regulated.} $|t|=|Q^2|\ll s$, where $Q^\mu=(\omega,\vb q)=P'{}^\mu - P^\mu$. This implies that both $\omega, q\ll p,k$ (see Appendix~\ref{app:kinematic-considerations}).  Therefore, the small $t$ behavior is the region of interest for medium modifications, i.e., the region where
\begin{align}
	0 < -t \ll s \approx -u.
\end{align}

At leading order,
the QCD $2\leftrightarrow2$ scattering matrix elements are spin-independent for soft momentum exchange (we show that explicitly in Appendix~\ref{app:scattering-soft-momentum-exchange} for elastic quark-quark and gluon-gluon scattering). Therefore, medium effects can be included as in a theory with fictitious scalar `quarks', which can be regarded as a generalization of scalar QED.
For quark scattering in this fictitious scalar QCD,
the vertex factor is given by $-igt^a(P+K)^\nu$, where $P$ and $K$ are the momenta of the in- and outgoing quark, $g$ is the coupling constant and $t^a$ is a basis element of the $\mathfrak{su}(\NC)$ Lie algebra.
There are no spinor factors for external legs, and thus the squared amplitude (without color and coupling factors) is given by
\begin{align}
	\left|\mathcal M\right|^2\propto\left|(P+P')^\mu(K+K')^\nu G_{\mu\nu}(Q)\right|^2,\label{eq:scalar_quark_result}
\end{align}
which precisely corresponds to the screening prescription \eqref{eq:amy_replacement} (with color factors reinstated such that for no screening the vacuum case is reproduced). 
In \app \ref{app:scattering-soft-momentum-exchange}, it is demonstrated that this simple argument using scalar QCD indeed reproduces the leading order correct prescription for quark and gluon scattering with soft-gluon exchange.

\subsection{Debye-like screening\label{sec:debye-like-screening}}
Equation~\eqref{eq:amy_replacement} is typically implemented in QCD kinetic theory implementations \cite{AbraaoYork:2014hbk, Kurkela:2014tea, Kurkela:2015qoa, Kurkela:2018oqw, Kurkela:2018xxd, Du:2020zqg, Du:2020dvp, kurkela_2023_10409474} in a very simple approximation. Instead of using the (full) hard thermal loop (HTL) propagator (Eq.~\eqref{eq:HTL-propagators} in Appendix \ref{app:qcd-and-nonequilibrium-qft}), one uses a simple (isotropic) Debye-like screened propagator \cite{AbraaoYork:2014hbk},
\begin{align}
	G_{\mu\nu}=\frac{\eta_{\mu\nu}}{Q^2} \frac{q^2}{q^2+\xiscreensqr \, m_D^2},
	\label{eq:simple-isotropic-screening}
\end{align}
which at the level of the matrix elements corresponds to replacing
\begin{align}
	\frac{(s-u)^2}{t^2}\to \frac{(s-u)^2}{t^2} \, \frac{q^4}{(q^2+\xiscreensqr \, m_D^2)^2}.
	\label{eq:simple_replacement}
\end{align}
The constant $\xiscreen=e^{5/6}/\sqrt{8}$ is chosen to approximate the HTL propagator in isotropic systems \cite{AbraaoYork:2014hbk}, or---as we will see in Chapters \ref{sec:momentum-broadening-of-jets} and \ref{sec:improving-qcd-simulations}---for longitudinal momentum broadening. For the soft-fermion diagrams, a similar prescription with a different constant $\xi_q$ is used in the literature.

In Ref.~\cite{AbraaoYork:2014hbk}, the value of $\xiscreen$ is obtained by writing the elastic collision term for an isotropic distribution function as
\begin{align}
	\begin{split}
		\Ctwotwo&=\frac{1}{2^9\pi^5\nu}\int_0^\infty\dd{k}\int_0^{2\pi}\dd{\phiqp}\\
		&\times\int_{-p}^k\dd{\omega}\left\{f(p)f(k)(1+f(p+\omega))(1+f(k-\omega)) - f(p+\omega)f(k-\omega)(1+f(p))(1+f(k))\right\}\\
		&\times\int_{|\omega|}^{\min(2k-\omega,2p+\omega)}\dd{q}\int_0^{2\pi}\dd{\phiqk}\frac{\left|\mathcal M\right|^2}{p^2}.
	\end{split}
\end{align}
Since screening effects are only important for soft internal momenta, $q,\,\omega\ll k,\,p$, we may expand the distribution functions for small $\omega$. The first nonvanishing term is quadratic
in $\omega$ since the matrix element is even. Therefore, we can fix the constant $\xiscreen$ by requiring \cite{AbraaoYork:2014hbk} that in this limit
\begin{align}
	\int_{-\infty}^\infty\dd{\omega}\omega^2\int_{|\omega|}^\infty\dd{q}\int_0^{2\pi}\dd\phi \left(\left|\Mhtl\right|^2-\left|\Mdebyeone\right|^2\right)=0. \label{eq:ximatching-longitduinal-theory}
\end{align}

This isotropic screening prescription neglects the effect of plasma instabilities which would otherwise be present in anisotropic systems \cite{Mrowczynski:1988dz, Mrowczynski:1993qm, Arnold:2003rq, Romatschke:2003ms, Romatschke:2004jh, Kurkela:2011ub}. However, numerical evidence indicates that these instabilities do not play a dominant role at the time scales of interest for kinetic theory simulations \cite{Berges:2013eia, Berges:2013fga} and when a quasiparticle picture has become applicable \cite{Boguslavski:2018beu, Boguslavski:2021buh, Boguslavski:2021kdd}. 

In Chapter \ref{sec:momentum-broadening-of-jets}, we will see that this simple screening prescription does not accurately describe both the longitudinal and transverse broadening of partons and in Chapter \ref{sec:improving-qcd-simulations}, we will compare kinetic theory simulations using this simple Debye-like screened matrix element to simulations where the HTL matrix element is used.

\section{How kinetic theory emerges from quantum field theory\label{sec:kinetic-theory-derivation}}
Before going on to discuss QCD thermalization and how to perform QCD kinetic theory simulations, let us discuss how the Boltzmann equation and, in particular, the collision terms can be obtained from the underlying quantum field theory.

This procedure, in principle, is well understood \cite{kadanoff_baym, Calzetta:1986cq}, but several complications arise when considering QCD.
First, QCD is a gauge theory and one has to deal with the redundancies that come from the gauge symmetry. Secondly, as discussed before, $1\to 2$ particle splittings (see Chapter \ref{sec:jet-energy-loss}) occur at a similar rate as elastic collisions, and thus need to be included in a consistent kinetic description.
We will not consider these complications here in detail, and consider the more easily accessible case of a scalar theory.

We will mainly follow the presentation laid out in Ref.~\cite{Blaizot:2001nr} (similar presentations with slightly different notation and conventions can be found, e.g., in Refs.~\cite{kadanoff_baym, Calzetta:1986cq, Berges:2004pu, Berges:2004yj, Blaizot:1999xk, Rammer_2007}) and do not attempt to provide a very rigorous derivation here. This section should be seen as motivating, which physical approximations are needed to derive kinetic equations and, in particular, motivate why the internal propagators in the matrix elements of the elastic collision terms are retarded ones. For this chapter, we will need some of the formalism of nonequilibrium quantum field theory briefly reviewed in Appendix \ref{app:qcd-and-nonequilibrium-qft}.

A typical technique to derive kinetic theory is to start from the Kadanoff-Baym equations, perform a Wigner transform to phase space and then a gradient expansion, keeping only the lowest-order terms in spatial gradients. Additionally, one performs the quasiparticle approximation, and the Kadanoff-Baym ansatz \cite{kadanoff_baym, Berges:2004pu}, where one identifies the Wightman functions with the phase-space distribution function and the spectral function. The quasiparticle approximation (taking the width of the spectral function to be zero, effectively replacing it with a delta function) means physically approximating the lifetime of the (quasi-)particles to be infinite between collisions, which is needed for any kinetic description to be valid. This implies that the particles are stable between collisions and can only change their momentum state (without the inclusion of force terms in the Boltzmann equation) or decay due to collisions encoded in the collision terms.

\subsection{Kadanoff-Baym equations}
Let us start by discussing the Kadanoff-Baym evolution equations for the propagator, and how these equations arise from the underlying quantum field theory.
For scalar field theory, let us start with the Lagrangian
\begin{align}
    \mathcal L(X)-j(X)\phi(X)=-\frac{1}{2}\partial_\mu\phi(X)\partial^\mu \phi(X)-\frac{m^2}{2}\phi^2(X)-V(\phi(X))-j(X)\phi(X),
\end{align}
where we have included an external source term with the source $j(X)$.
One may derive the equations of motion for the full propagator by first considering the equations of motion (using $\partial^2=\partial_\mu\partial^\mu$)
\begin{align}
    -m^2\phi(X)+\partial^2\phi(X)=\pdv[V]{\phi}+j(X),
\end{align}
and taking the ensemble average $\langle \dots \rangle$ (e.g., by using the path integral formalism, see Eq.~\eqref{eq:general-trace} in Appendix \ref{app:qcd-and-nonequilibrium-qft}). One then obtains
\begin{align}
    (\partial^2-m^2)(-iG(X,Y))+\int\dd[4]{Z}(-i\Pi(X,Z))(-iG(Z,Y))=\delta^{(4)}(X-Y), \label{eq:eom-fullpropagator}
\end{align}
where the full propagator $G(x,y)$ and self-energy $\Pi(x,y)$ can be defined as the functional derivatives
\begin{align}
    -iG(X,Y)=\frac{\delta\langle \phi(X)\rangle}{\delta j(Y)}, && i\Pi(X,Y)=\frac{\delta}{\delta\langle \phi(Y)\rangle}\left\langle \dv[V]{\phi}(X)\right\rangle.
\end{align}
Eq.~\eqref{eq:eom-fullpropagator} has the nice physical interpretation that the difference between the full and free propagator is the self-energy $\Pi$ (see Eq.~\eqref{eq:self-energy-schematically} in Appendix \ref{app:qcd-and-nonequilibrium-qft}).

Eq.~\eqref{eq:eom-fullpropagator} can also be obtained on the Schwinger-Keldysh contour (see Appendix \ref{app:closed-timepath}), and it is useful to decompose the full (time-ordered) propagator into the Wightman functions $G^<$ and $G^>$ (see Appendix \ref{app:closed-timepath}) and perform a similar decomposition of the self-energy (see Eqs.~\eqref{eq:wightman-functions-decomposition} and \eqref{eq:self-energy-decomposition}),
\begin{align}
    G(X,Y)&=\Theta(X^0-Y^0)G^>(X,Y)+\Theta(Y^0- X^0)G^<(X,Y),\\
    \Pi(X,Y)&=-i\Pi^\delta(X)\delta^{(4)}(X-Y)+\Theta(X^0-Y^0)\Pi^>(X,Y)+\Theta(Y^0-X^0)\Pi^<(X,Y).
\end{align}

Inserting this decomposition into \eqref{eq:eom-fullpropagator} and neglecting\footnote{This is justified for initial times in the remote past \cite{Blaizot:2001nr, kadanoff_baym}.} terms depending on the initial time $t_0$, we obtain the Kadanoff-Baym equations
\begin{subequations}\label{eq:kadanoff-baym}
    \begin{align}
    (\partial_X^2-m^2-\Pi^\delta(X))G^>(X,Y)&=i\int_{-\infty}^{\infty}\dd[4]{Z}\left(\Pi^R(X,Z)G^>(Z,Y)+\Pi^>(X,Z)G^A(Z,Y)\right)\label{eq:G>1},\\
    (\partial_X^2-m^2-\Pi^\delta(X))G^<(X,Y)&=i\int_{-\infty}^{\infty}\dd[4]{Z}\left(\Pi^R(X,Z)G^<(Z,Y)+\Pi^<(X,Z)G^A(Z,Y)\right),\\
    (\partial_Y^2-m^2-\Pi^\delta(X))G^>(X,Y)&=i\int_{-\infty}^{\infty}\dd[4]{Z}\left(G^>(X,Z)\Pi^A(Z,Y)+G^R(X,Z)\Pi^>(Z,Y)\right)\label{eq:G>2},\\
    (\partial_Y^2-m^2-\Pi^\delta(X))G^<(X,Y)&=i\int_{-\infty}^{\infty}\dd[4]{Z}\left(G^<(X,Z)\Pi^A(Z,Y)+G^R(X,Z)\Pi^<(Z,Y)\right).
\end{align}
\end{subequations}
These equations are exact, but only useful if we know the exact form of the self-energies. In practice, truncating the self-energy at some point constitutes an approximation here.

\subsection{Wigner transform}
The next step is to perform a Wigner transform
\begin{align}
    f(X,Y)\mapsto \tilde f(\bar X,P),
\end{align}
which consists of a coordinate transformation to relative and central coordinates, 
\begin{align}
    S^\mu = X^\mu-Y^\mu, && \bar X^\mu=\frac{X^\mu+Y^\mu}{2},
\end{align}
and then performing a Fourier transform with respect to the relative coordinate $S$,
\begin{align}
    \tilde f(k,\bar X)=\int\dd[4]{S}e^{-IK\cdot S}f\left(X(S,\bar X),Y(S,\bar X)\right).
\end{align}
Taking the difference of Eqs.~\eqref{eq:G>1} and \eqref{eq:G>2} leads to expressions
\begin{align}
    \partial_X^2-\partial_Y^2=2\partial_S\cdot \partial_{\bar X}, && \Pi^\delta(X)-\Pi^\delta (Y)=-(S\cdot \partial_{\bar X})\Pi^\delta (\bar X).
\end{align}
Importantly, the Wigner transform of a general convolution such as those on the right-hand side of Eqs.~\eqref{eq:kadanoff-baym} is \cite{Rammer_2007}
\begin{align}
\begin{split}
    \int\dd[4]{Z}A(X,Z)B(Z,Y)&\mapsto e^{\frac{i}{2}(\partial^A_{\bar X}\cdot \partial_P^B-\partial^A_P\cdot\partial_{\bar X}^B)}\tilde A(\bar X,P)\tilde B(\bar X,P)\\
    &=\tilde A(\bar X,P)\tilde B(\bar X,P)+\frac{i}{2}\left\{\tilde A(\bar X,P), \tilde B(\bar X,P)\right\}_{\mathrm{PB}}+\dots,
    \end{split}
\end{align}
where $\{A,B\}_{\mathrm{PB}}=\partial_P A\cdot \partial_{\bar X} B-\partial_{\bar X}A\cdot \partial_P B$ denotes the Poisson bracket.
The Boltzmann equation is obtained by next expanding in gradients of $\partial_{\bar X}$. Using identities from Appendix \ref{app:correlator-relations}, we obtain
\begin{align}
    \begin{split}
    \left(-2 K^\mu + \pdv[\RE\tilde \Pi(K,\bar X)]{K_\mu}\right)\pdv[\tilde G^>(K,\bar X)]{\bar X^\mu}-\pdv[\RE\tilde\Pi (K,\bar X)]{\bar X_\mu}\pdv[\tilde G^>(K,\bar X)]{K^\mu}\\
    -\left\{\tilde \Pi^>(K,\bar X),\RE \tilde G^R(K,\bar X)\right\}_{\mathrm{PB}}=-\left(\tilde G^>(K,\bar X)\tilde \Pi^<(K,\bar X)-\tilde\Pi^>(K,\bar X)\tilde G^<(K,\bar X)\right),\label{eq:quantum-kinetic-eq}
    \end{split}
\end{align}
where we have defined $\RE\tilde\Pi(K,\bar X)=\Pi^\delta(\bar X)+\RE\tilde \Pi^R(K,\bar X)$.

\subsection{Quasiparticle approximation}
Using the quasiparticle approximation,
\begin{align}
    \tilde\rho(K,\bar X)=2\pi\sign(K^0)\delta(-K^2-M^2-\RE\tilde\Pi(K,\bar X))g(K)=2\pi\sign(K^0)\delta(K_0^2-E_{\vb k}^2)g(K), \label{eq:quasiparticle-approximation}
\end{align}
we enforce the particles to be stable between collisions, which was one of the requirements for kinetic theory listed in the beginning. With this approximation, one may also neglect the Poisson bracket in Eq.~\eqref{eq:quantum-kinetic-eq} \cite{Blaizot:2001nr}. We have included here an additional function $g(Q)$, which is unity, $g(Q)\equiv 1$, in our case of scalar particles, but for gluons reads $g(Q)=-\eta_{\mu\nu}\delta_{ab}$ and for fermions $g(Q)=\slashed Q + m$.

The distribution function is introduced via 
the Kadanoff-Baym ansatz \cite{kadanoff_baym, Berges:2004pu},
\begin{align}
    \tilde G^<(K,\bar X)=\tilde \rho(K,\bar X)\tilde N(K,\bar X), && \tilde G^>(K,\bar X)=\tilde \rho(K,\bar X)[1+\tilde N(K,\bar X)],\label{eq:kadanoff-baym-ansatz}
\end{align}
which is true in thermal equilibrium when inserting the Bose-Einstein distribution \eqref{eq:thermal-distributionfunctions} \cite{Blaizot:2001nr}, and is used as the definition of the distribution function out of equilibrium. Together with the quasiparticle approximation, the off-shell distribution function $\tilde N(K,\bar X)$ can be brought on-shell, using $\tilde N(K,\bar X)=-(1+\tilde N(-K,\bar X))$ from Eqs.~\eqref{eq:wightmanfunction-negativek} and \eqref{eq:rho-antisymmetric}),
\begin{subequations}\label{eq:bring_f_onshell}
\begin{align}
    \tilde G^<(K,\bar X)&=2\pi \delta(K_0^2-E_{\vb k}^2(\bar X))\left(\Theta(K^0)f(\vb k,\bar X)+\Theta(-K^0)\left(1
+f(-\vb k,X)\right)\right),\\
    \tilde G^>(K,\bar X)&=2\pi\delta(K_0^2-E_{\vb k}^2(\bar X))\left(\Theta(K^0)\left(1+f(\vb k,\bar X)\right)+\Theta(-K^0)f(-\vb k,\bar X)\right),\label{eq:wightmann-G>-distributionfunction}
\end{align}
\end{subequations}
with the on-shell distribution function
\begin{align}
    f(\vb k,\bar X)=\tilde N(K^0=E_{\vb k},\vb k,\bar X).\label{eq:distributionfunction-onshell}
\end{align}
The splitting in Eq.~\eqref{eq:bring_f_onshell} is done to enforce the $K^0$ component in \eqref{eq:distributionfunction-onshell} to be always positive.

We can then integrate Eq.~\eqref{eq:quantum-kinetic-eq} over (positive) $K^0$ to bring the distribution function on-shell and arrive at
\begin{align}
    \begin{split}
    &\left(\partial_{\bar t}+\frac{\vb k}{E_{\vb k}(X)}\cdot \partial_{\bar {\vb x}}-\partial_{\bar {\vb x}} E_{\vb k}(\bar X)\cdot\partial_{\vb k}\right)f(\vb k,\bar X)\\
    &\qquad=-\frac{1}{2E_{\vb k}(\bar X)}\left((1+f(\vb k,\bar X))\Pi_E^<(\vb k,\bar X)-\Pi^>_E(\vb k,\bar X)f(\vb k,\bar X)\right),\label{eq:boltzmannequation-derived}
    \end{split}
\end{align}
which has the correct structure to compare with the Boltzmann equation \eqref{eq:boltzmann_equation}. Note that $\vb v=\vb k/E_{\vb k}$. 
Here, we can identify the gain and loss term, where the loss term is given by $\Pi^>_E(k,X)$ and multiplied by the distribution function $f(\vb k,\bar X)$. Here, the $E$ in $\Pi^>_E(k,X)$ defines the on-shell self-energy, i.e.,
\begin{align}
    \Pi^>_E(\vb k,\bar X)=\tilde \Pi^>(K^0=E_{\vb k},\vb k,\bar X).\label{eq:self-energy-onshell}
\end{align}

To recapitulate, the only approximations to arrive at this equation were the gradient expansion and the quasiparticle approximation, together with setting the initial time $t_0$ in the remote past. Using this, we were able to derive a Boltzmann equation for the distribution function $f(\vb k,\bar X)$. The next step is to evaluate the self-energies $\Pi^>$ and $\Pi ^<$ in terms of the distribution function $f(\vb k,\bar X)$, i.e., to write it in terms of degrees of freedom for which we again may use the quasiparticle approximation.

\subsection{Decay rate\label{sec:decay-rate}}
Let us focus here on the decay rate (or the loss term) given by $\Pi^>_E$. The gain term can be obtained similarly by replacing $>\leftrightarrow<$.

\subsubsection{Cutting rule for the self-energy}
For the self-energy $\tilde\Pi^{\substack{<\\>}}$, there exists a convenient \textit{cutting rule} (see \cite{Caron-Huot:2007zhp,Ghiglieri:2020dpq}),
\begin{align}
\begin{split}\label{eq:cutting-rule}
\tilde \Pi^>(K)&=\sum_n\frac{1}{n!}\left(\prod_{j=1}^n\int\frac{\dd[4]{Q_j}}{(2\pi)^4}\right)(2\pi)^4\delta^4(Q_1+\dots+Q_n-K)\\
&\quad\times\mathcal{M}_{ar\dots r}(P; Q_1,\dots, Q_n)\mathcal M_{ar\dots r}(-P; -Q_1,\dots,-Q_n)\\
&\quad\times \tilde G^>(Q_1)\dots \tilde G^>(Q_n)
\end{split}\\
&=\sum_n\raisebox{-0.5\height}{\begin{tikzpicture}
	\begin{feynman}
		\vertex (a1);
		\node[right=1.8cm of a1, blob] (b1);
		\vertex[right=2.5cm of b1] (c9);
		\vertex[above=0.5cm of c9] (c);
		\vertex[below=1.5cm of c] (c4);
		\vertex[below=2.5cm of c] (c2);
		\vertex[above=1.5cm of c] (c3);
		\vertex[above = 2cm of c] (cut1);
		\vertex[below=3cm of c] (cut2);
		\node[right=2.1cm of c9, blob] (b2);
		\vertex[right=2cm of b2] (a2);
		\diagram* {
			(b1) -- [fermion, edge label= \(K\)] (a1),
			(c3) -- [fermion, edge label'=\(Q_1\)] (b1),
			(c) -- [edge label'=\(Q_2\),fermion] (b1),
			(c2) -- [fermion, edge label = \(Q_n\)] (b1),
			{(c), (c2), (c3)} -- [fermion] (b2),
			(c4) -- [charged scalar, edge label'=\(\vdots\)] (b1),
			(c4) -- [charged scalar, edge label=\(\vdots\)] (b2),
			(b2) -- [fermion] (a2),
			(cut1) -- [scalar] (cut2),
		};
	\end{feynman}
\end{tikzpicture}},
\end{align}
where $\mathcal M_{ar\dots r}$ are \textit{fully retarded amplitudes}, i.e. they represent a matrix element with exactly one $a$-index in the r/a basis (see Appendix \ref{app:closed-timepath}). The sum runs over all the propagators $\tilde G^>(Q_1)\dots$ that are inserted between the two diagrams, or equivalently, over all the lines that are \textit{cut} and then replaced by $\tilde G^>(Q_1)\dots$. For particles with spin or color, the additional indices are summed over (e.g., consider an index in the amplitude $\mathcal M_{ar\dots r}$ to belong to one momentum $Q_i$; then the propagator with $Q_i$ carries the same index and connects it to the corresponding index in the conjugate amplitude $\mathcal M_{ar\dots r}^\ast$).

\subsubsection{Self-energy for pure gluons}
Let us now consider\footnote{A similar discussion was included in my diploma thesis \cite{Lindenbauer:2022}.} the gluon self-energy in a purely gluonic system, which appears in the loss term in the Boltzmann equation \eqref{eq:boltzmannequation-derived}, and use the cutting rule to obtain it.
For that, let us consider a specific term appearing in the self-energy for pure gluons, the term where three internal lines are cut according to the cutting rule \eqref{eq:cutting-rule},
\begin{align}
\delta_{ah}\eta^{\sigma\epsilon}\Pi^>_{\sigma\epsilon}{}^{ah}(P) &\ni \frac{1}{6}\int\frac{\dd[4]{Q_1}\dd[4]{Q_2}\dd[4]{Q_3}}{(2\pi)^{12}}\label{eq:pi>}
\\
&\quad\times \tilde G_{\alpha\mu}^>{}^{gd}(Q_1)\tilde G_{\beta\nu}^>{}^{eb}(Q_2)\tilde G_{\gamma\rho}^>{}^{fc}(Q_3)(2\pi)^4\delta^4(Q_1+Q_2+Q_3-P) \nonumber\\
&\times\left(\raisebox{-0.5\height}{\begin{tikzpicture}
\begin{feynman}
	\vertex (a) {\(\sigma,\, a\)};
	\node[right=2cm of a, blob] (b);
	\vertex[right=2cm of b] (c) {\(\nu,\, b\)};
	\vertex[above=1.5cm of c] (d) {\(\rho,\, c\)};
	\vertex[below=1.5cm of c] (g) {\(\mu,\, d\)};
	\diagram* {
	(g) -- [charged boson, edge label=\(Q_1\)] (b),
	(c) -- [charged boson, edge label=\(Q_2\)] (b),
	(d) -- [charged boson, edge label=\(Q_3\)] (b),
	(b) -- [charged boson, edge label=\(P\)] (a),
	};
\end{feynman}
\end{tikzpicture}}\right)
\left(\raisebox{-0.5\height}{\begin{tikzpicture}
\begin{feynman}
	\vertex (a) {\(\epsilon,\, h\)};
	\node[right=2cm of a, blob] (b);
	\vertex[right=2cm of b] (c) {\(\beta, \, e\)};
	\vertex[above=1.5cm of c] (d) {\(\gamma,\, f\)};
	\vertex[below=1.5cm of c] (g) {\(\alpha,\, g\)};
	\diagram* {
	(g) -- [charged boson, edge label=\(Q_1\)] (b),
	(c) -- [charged boson, edge label=\(Q_2\)] (b),
	(d) -- [charged boson, edge label=\(Q_3\)] (b),
	(b) -- [charged boson, edge label=\(P\)] (a),
	};
\end{feynman}
\end{tikzpicture}}\right)^\ast\nonumber.
\end{align}
The arrows denote (retarded) propagators in the r/a basis (see Appendix \ref{app:closed-timepath}).

Let us bring this expression into a more convenient form. With the Kadanoff-Baym ansatz \eqref{eq:kadanoff-baym-ansatz}, and the quasiparticle approximation \eqref{eq:quasiparticle-approximation} with 
$g(Q)=-\eta_{\mu\nu}\delta_{ab}$ for gluons, the Wightman functions $\tilde G^>$ can be written using \eqref{eq:wightmann-G>-distributionfunction} as
\begin{align}
    \tilde G^>_{\mu\nu}{}^{ab}(Q)=-2\pi\eta_{\mu\nu}\delta^{ab}\frac{1}{2E_{\vb q}}[\delta (Q^0-E_{\vb q})(1+f(\vb q))+\delta(Q^0+E_{\vb q})f(-\vb q)]. \label{eq:decomposition_D>}
\end{align}
Due to rotational symmetry, $E_{\vb q}=E_{-\vb q}$. Inserting this for the gluonic propagators in \eqref{eq:pi>}, we can write the two diagrams as the squared amplitude, 
where we take $\left|\mathcal M\right|^2$ to be summed over all initial and final polarizations and colors.
\begin{align}
\left|\mathcal M\right|^2&=\eta_{\alpha\mu}\eta_{\beta\nu}\eta_{\gamma\rho}\eta_{\sigma\epsilon}\delta_{ah}\delta_{gd}\delta_{eb}\delta_{fc}\nonumber\\
&\quad\times \left(\raisebox{-0.5\height}{\begin{tikzpicture}
\begin{feynman}
	\vertex (a) {\(\sigma,\, a\)};
	\node[right=2cm of a, blob] (b);
	\vertex[right=2cm of b] (c) {\(\nu,\, b\)};
	\vertex[above=1.5cm of c] (d) {\(\rho,\, c\)};
	\vertex[below=1.5cm of c] (g) {\(\mu,\, d\)};
	\diagram* {
	(g) -- [charged boson, edge label=\(Q_1\)] (b),
	(c) -- [charged boson, edge label=\(Q_2\)] (b),
	(d) -- [charged boson, edge label=\(Q_3\)] (b),
	(b) -- [charged boson, edge label=\(P\)] (a),
	};
\end{feynman}
\end{tikzpicture}}\right)
\left(\raisebox{-0.5\height}{\begin{tikzpicture}
\begin{feynman}
	\vertex (a) {\(\epsilon,\, h\)};
	\node[right=2cm of a, blob] (b);
	\vertex[right=2cm of b] (c) {\(\beta, \, e\)};
	\vertex[above=1.5cm of c] (d) {\(\gamma,\, f\)};
	\vertex[below=1.5cm of c] (g) {\(\alpha,\, g\)};
	\diagram* {
	(g) -- [charged boson, edge label=\(Q_1\)] (b),
	(c) -- [charged boson, edge label=\(Q_2\)] (b),
	(d) -- [charged boson, edge label=\(Q_3\)] (b),
	(b) -- [charged boson, edge label=\(P\)] (a),
	};
\end{feynman}
\end{tikzpicture}}\right)^\ast\nonumber \\
&=\left|\raisebox{-0.5\height}{\begin{tikzpicture}
\begin{feynman}
	\vertex (a) {};
	\node[right=2cm of a, blob] (b);
	\vertex[right=2cm of b] (c) {};
	\vertex[above=1.5cm of c] (d) {};
	\vertex[below=1.5cm of c] (g) {};
	\diagram* {
	(g) -- [charged boson, edge label=\(Q_1\)] (b),
	(c) -- [charged boson, edge label=\(Q_2\)] (b),
	(d) -- [charged boson, edge label=\(Q_3\)] (b),
	(b) -- [charged boson, edge label=\(P\)] (a),
	};
\end{feynman}
\end{tikzpicture}}\right|^2,\label{eq:feynman_gg_matrixel}
\end{align}
with the corresponding legs contracted and summed over all colors (and polarizations).

The $Q^0$ integrals in \eqref{eq:pi>} can now be performed, where for every $\tilde D^>$ the delta function removes the $Q^0$ integral and yields two terms according to \eqref{eq:decomposition_D>},
\begin{align}
\begin{split}
\frac{1}{6}&\int\frac{\dd[3]{\vb q_1}\dd[3]{\vb q_2}\dd[3]{\vb q_3}}{(2\pi)^{9} 2E_{\vb q_1}2E_{\vb q_2}2E_{\vb q_3}}(2\pi)^4\delta^3(\vb q_1+\vb q_2+\vb q_3-P)\left|\mathcal M\right|^2\\
&\times \Big\{\delta(E_{\vb q_1}+E_{\vb q_2}+E_{\vb q_3}-E_{\vb p})\left[(1+ f_{\vb q_1})(1+ f_{\vb q_2})(1+ f_{\vb q_3})\right]\\
&\quad + \delta(-E_{\vb q_1}+E_{\vb q_2}+E_{\vb q_3}-E_{\vb p})\left[f_{-\vb q_1}(1+ f_{\vb q_2})(1+ f_{\vb q_3})\right]\\
&\quad + \text{ similar with } \vb q_1\leftrightarrow \vb q_2, \quad \vb q_1\leftrightarrow \vb q_3\\
&\quad + \delta(-E_{\vb q_1}-E_{\vb q_2}+E_{\vb q_3}-E_{\vb p})\left[f_{-\vb q_1}f_{-\vb q_2}(1+ f_{\vb q_3})\right]\\
&\quad + \text{ similar with } \vb q_2\leftrightarrow \vb q_3, \quad \vb q_1\leftrightarrow \vb q_3\\
&\quad \delta(-E_{\vb q_1}-E_{\vb q_2}-E_{\vb q_3}-E_{\vb p})\left[f_{-\vb q_1}f_{-\vb q_2}f_{-\vb q_3}\right]\Big\}
\end{split}\\
\begin{split}
=\frac{1}{6}&\int\frac{\dd[3]{\vb q_1}\dd[3]{\vb q_2}\dd[3]{\vb q_3}}{(2\pi)^{9} 2E_{\vb q_1}2E_{\vb q_2}2E_{\vb q_3}}(2\pi)^4\left|\mathcal M\right|^2\\
&\times \Big\{\left.\delta^4(Q_1+Q_2+Q_3-P)\right|_{(Q_i)^0=E_{\vb q_i}}\left[(1+ f_{\vb q_1})(1+ f_{\vb q_2})(1+ f_{\vb q_3})\right]\\
&\quad + \left.\delta^4(-Q_1+Q_2+Q_3-P)\right|_{(Q_i)^0=E_{\vb q_i}}\left[f_{\vb q_1}(1+ f_{\vb q_2})(1+ f_{\vb q_3})\right]\\
&\quad + \text{ similar with } \vb q_1\leftrightarrow \vb q_2, \quad \vb q_1\leftrightarrow \vb q_3\\
&\quad + \left.\delta^4(-Q_1-Q_2+Q_3-P)\right|_{(Q_i)^0=E_{\vb q_i}}\left[f_{\vb q_1}f_{\vb q_2}(1+ f_{\vb q_3})\right]\\
&\quad + \text{ similar with } \vb q_2\leftrightarrow \vb q_3, \quad \vb q_1\leftrightarrow \vb q_3\\
&\quad \left.\delta^4(-Q_1-Q_2-Q_3-P)\right|_{(Q_i)^0=E_{\vb q_i}}\left[f_{\vb q_1}f_{\vb q_2}f_{\vb q_3}\right]\Big\}.
\end{split}\label{eq:pi_gluons}
\end{align}

For particles with an (effective) mass, energy-momentum conservation can only be fulfilled for the second delta function.

We thus obtain 
\begin{align}
\begin{split}
\delta_{ah}\eta^{\sigma\epsilon}\Pi^>_{\sigma\epsilon}{}^{ah}(P)\overset{3\,\text{lines}}=\frac{1}{2}&\int\frac{\dd[3]{\vb q_1}\dd[3]{\vb q_2}\dd[3]{\vb q_3}}{(2\pi)^{9} 2E_{\vb q_1}2E_{\vb q_2}2E_{\vb q_3}}(2\pi)^4\left|\mathcal M\right|^2\\
&\times \left.\delta^4(-Q_1+Q_2+Q_3-P)\right|_{(Q_i)^0=E_{\vb q_i}}\left[f_{\vb q_1}(1+ f_{\vb q_2})(1+ f_{\vb q_3})\right].
\end{split}\label{eq:pi_gluons_finished}
\end{align}
From the self-energy $\Pi^>$, we can now easily obtain the decay rate of an excitation with momentum $P$ using the Boltzmann equation \eqref{eq:boltzmannequation-derived} with the normalization $1/(2E_{\vb p})$. Taking the trace over all colors and spins in Eq.~\eqref{eq:pi>} leads to the sum of all $\nu_g$ excitations. Additionally, we only considered three cut propagators, which corresponds to elastic processes (2 particles going in, 2 particles going out). Thus, the elastic decay or scattering rate of a particle with momentum $P$ is
\begin{align}
	\Gammael&=\frac{1}{4p\nu_a}\sum_{bcd}\int_{\vec k\vec p'\vec k'}(2\pi)^4\delta^4(P+K-P'-K')\label{eq:decay_rate}\\
	&\quad\times\left|\mathcal M^{ab}_{cd}(\vecp, \vec k;\vec p',\vec k')\right|^2f_b(\vec k)\left[1\pm f_d(\vec k')\right]\left[1\pm f_c(\vec p')\right]. \nonumber
\end{align}
Here, we have used $E_{\vb p}=|\vb p|=p$, and generalized this equation to include also quarks. Eq.~\eqref{eq:decay_rate} describes the decay rate or scattering rate of a particle of species $a$ with momentum $\vb p$ due to elastic collisions $p+k\to p'+k'$. The sum is over all possible scattering processes including particles $b$, $c$ and $d$.

We will use this equation \eqref{eq:decay_rate} in the next Chapter \ref{sec:momentum-broadening-of-jets} as a starting point to obtain the jet quenching parameter $\qhat$.

\subsection{Relation to the Boltzmann equation}
Comparing with the elastic collision kernel in QCD kinetic theory, we recognize that Eq.~\eqref{eq:decay_rate} constitutes exactly the contribution to the elastic collision kernel Eq.~\eqref{eq:c22_first} from the decay rate.
The gain term can be computed in a similar way.

For the inelastic collision kernel \eqref{eq:c12}, evaluating the self-energy using the techniques from the previous sections is more complicated because one has to take into account the physics of LPM suppression. However, using the radiation rate $\gamma$ from Section \ref{sec:amy-rate-equation} as an effective vertex, we can use the cutting rule \eqref{eq:cutting-rule} with two internal cut lines to arrive at \eqref{eq:c12}, similar to the steps taken here.

\section{What we know about QCD equilibration \label{sec:known-things-about-qcd-equilibration}}
As mentioned in the introduction, understanding and simulating how an equilibrated quark-gluon plasma is formed in heavy-ion collisions is an active research area. While for the very earliest times after the collision, classical statistical simulations need to be performed \cite{Berges:2020fwq}, we will focus here on studies and considerations using QCD kinetic theory, which can describe how an over-occupied system (possibly taken from an earlier classical statistical simulation) equilibrates.
While analytic and parametric estimates exist \cite{Baier:2000sb, Kurkela:2011ti, Kurkela:2011ub, Blaizot:2011xf}, numerical studies have been performed using QCD kinetic theory in isotropic systems \cite{Kurkela:2014tea, Kurkela:2015qoa, Fu:2021jhl}, and also including quarks and nonzero chemical potential \cite{Kurkela:2018oqw, Kurkela:2018xxd, Du:2020dvp, Du:2020zqg}.
A review on QCD thermalization can be found, e.g., in \cite{Berges:2020fwq, Schlichting:2019abc}.

\subsection{Thermalization of isotropic systems}
\begin{figure}
    \centering
    \includegraphics[width=\linewidth]{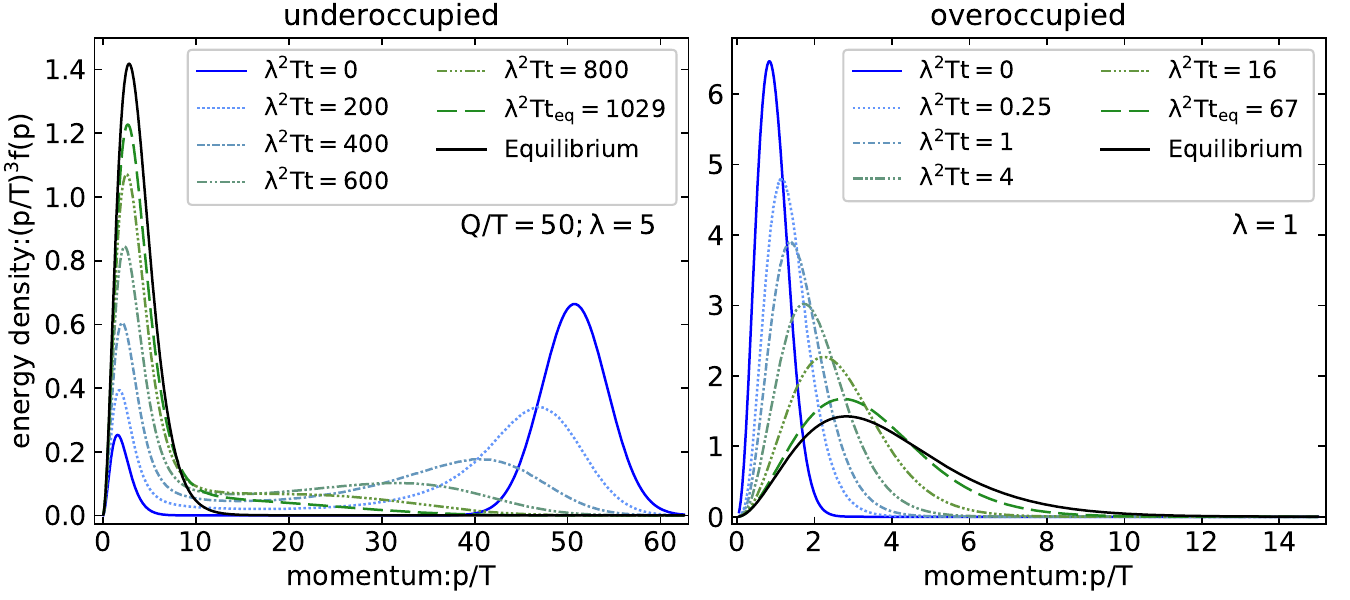}
    \caption{\label{fig:time-evolution-isotropic}
    Thermalization of isotropic nonabelian plasmas at NLO for an under-occupied (left) and over-occupied (right) initial condition.
    Figure from \cite{Fu:2021jhl}.}
\end{figure}
For initially over-occupied systems (right panel of Fig.~\ref{fig:time-evolution-isotropic}), where $f\gg 1$, but still $f\ll 1/\alpha_s$ for kinetic theory to be applicable, the initial energy is carried by a large number of soft particles. There, the initial evolution quickly falls on top of a scaling solution, which is referred to as a \emph{non-thermal fixed point},
\begin{align}
    f(t,p)=(Qt)^\alpha f_S\left((Qt)^\beta \frac{p}{Q}\right).
\end{align}
Conservation laws relate the exponents $\alpha$ and $\beta$. For instance, energy conservation
\begin{align}
    \varepsilon(t)=\int\frac{\dd[3]{\vb p}}{(2\pi)^3}p f(\vb p)=\text{const},
\end{align}
implies the relation
\begin{align}
    \alpha=4\beta.
\end{align}
With a more rigorous analysis considering how the Boltzmann equation behaves at the scaling solution, one obtains further $\alpha=-4/7$, $\beta=-1/7$.

This self-similar evolution breaks down once the condition $f\gg 1$ is no longer satisfied. The thermalization time can be easily obtained by noting that we start with an initial energy density $\varepsilon \sim Q^4/\alpha_s$, which leads to a final temperature $T\sim Q\alpha_s^{-1/4}\gg Q$. The self-similar evolution ends, once the hard scale in the distribution function $Q (Qt)^{-\beta}\sim T$, which leads to
\begin{align}
    \ttherm \sim \frac{1}{\alpha_s^2 T}.
\end{align}
In this regime, classical statistical simulations and kinetic theory simulations are both valid and detailed analysis between both approaches has shown their equivalence \cite{AbraaoYork:2014hbk}.

For initially under-occupied systems (left panel of Fig.~\ref{fig:time-evolution-isotropic}), the total energy is concentrated at a small number $f_0\ll 1$ of very hard particles of momentum $Q$, such that $T\ll Q$. These hard particles emit soft gluons which form a soft thermal bath. After such a bath has formed, the remaining hard particles of momentum $Q$ thermalize by democratic splitting $Q\to Q/2+Q/2$. The formation time $\tform$ of such a splitting is given by Eq.~\eqref{eq:formation-time}, $\tform\sim\sqrt{\omega/\qhat}$. Since this scales with the energy of the emitted gluon $\omega\sim Q$, the formation time is long and the splitting takes place in the LPM regime.
Taking the LPM rate from Eq.~\eqref{eq:rate-lpm}, the number of emitted gluons above a frequency $\omega$ can be estimated to be (see also \cite{Blaizot:2012fh})
\begin{align}
    \Delta N(\omega)=\int_\omega\dd\omega'\int_0^t\dd{t'}\frac{\dd\Gamma}{\dd{\omega}}\sim \alpha_s t\sqrt{\frac{\qhat}{\omega}}.
\end{align}
If we assume that the system thermalizes once the first democratic splitting with $\omega\approx Q/2$ has occurred, we can define the thermalization time when the number of emitted gluons is of order unity, i.e., we consider $\Delta N(Q)\sim 1$ and take $t\sim \ttherm$. We then obtain
\begin{align}
    \ttherm\sim \frac{1}{\alpha_s}\sqrt{\frac{Q}{\qhat}}.
\end{align}
Assuming\footnote{We will
obtain a more precise expression for the jet quenching parameter $\qhat$ in Chapter \ref{sec:momentum-broadening-of-jets}. For now, as a simple estimate, we may take $\qhat=\int\dd[2]{\vb q_\perp} q_\perp^2 \dd\Gammael/\dd[2]{\vb q_\perp}$, which for dimensional reasons is proportional to $T^3$ and we obtain a factor $\alpha_s^2$ from the matrix elements for the scattering rate \eqref{eq:decay_rate} from Table \ref{tab:amy_matrix_el}.
}
$\qhat\sim \alpha_s^2T^3$, we obtain
\begin{align}
    \ttherm \sim \frac{1}{\alpha_s^2 T}\sqrt{\frac{Q}{T}},\label{eq:thermalizationtime-underoccupied}
\end{align}

\subsection{Thermalization of expanding systems\label{sec:thermalization-expanding-systems}}
In expanding and, thus, anisotropic systems, the appearance of plasma instabilities \cite{Mrowczynski:1988dz, Mrowczynski:1993qm, Romatschke:2003ms, Romatschke:2004jh, Arnold:2005vb, Arnold:2005ef, Rebhan:2005re, Romatschke:2006wg, Bodeker:2007fw, Rebhan:2008uj, Ipp:2010uy, Attems:2012js, Hauksson:2020wsm, Hauksson:2021okc} complicates the picture (see also the review \cite{Mrowczynski:2016etf}), but it has been observed in classical simulations that they do not play a significant role at least at the early times where these simulations are valid \cite{Berges:2013fga, Berges:2013eia}. We will, therefore, not consider the effect of these plasma instabilities here.

\begin{figure}
    \centering
    \includegraphics[width=0.5\linewidth]{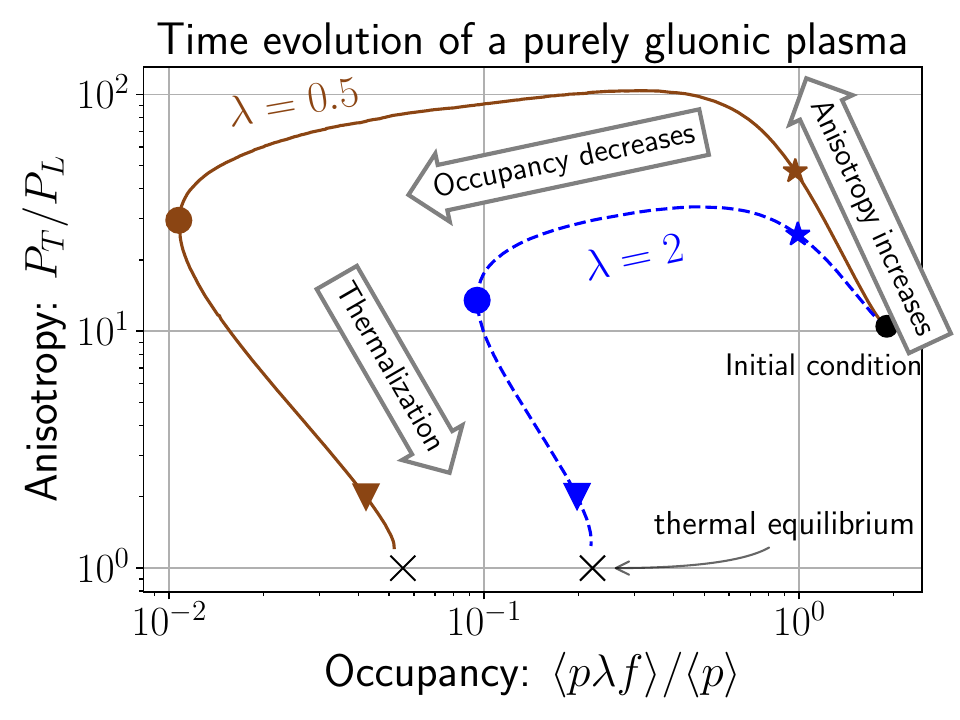}
    \caption{Time evolution of a gluonic plasma undergoing Bjorken expansion with an over-occupied initial condition (black dot) in the anisotropy-occupancy plane. The three stages of bottom-up thermalization are denoted by arrows.}
    \label{fig:bottomup-overview}
\end{figure}

How expanding systems thermalize has been worked out by Baier, Mueller, Schiff, and Son about 25 years ago \cite{Baier:2000sb}.
Starting from an overoccupied initial condition $f\sim 1/\lambda$ which should be obtained from classical statistical simulations of the earlier dynamics, quasiparticles and a kinetic description become applicable at a time scale of around $\tau\sim 1/Q_s$. At this time, the (hard) gluons have momenta of order $Q_s$. While the system is dominated by rapid expansion, the interaction of the hard particles leads to the emission of soft gluons.

When the occupancy drops below $1/\lambda$ (star marker in Fig.~\ref{fig:bottomup-overview}), the plasma becomes under-occupied and a classical description is no longer applicable. While the anisotropy stays roughly constant, a soft bath is built by further branching, which thermalizes first. A significant amount of the total energy is still carried by the remaining small number of hard gluons, which, in the third stage, lose energy through multiple hard branchings, until equilibrium is reached at a time that is parametrically given by
\begin{align}
\tauT=\alpha_s^{-13/5}/Q_s. \label{eq:bottom-up-timescale}   \end{align}

A simple estimate for this thermalization time can be given following Ref.~\cite{Kurkela:2014tea} by noticing that the physical picture for thermalization is as in the under-occupied isotropic case and taking the energy density to be as in a simple free streaming expansion
\begin{align}
    \varepsilon(\tau) \sim \frac{Q_s^3}{\alpha_s \tau}, && T(\tau)\sim \left(\frac{Q_s^3}{\alpha_s \tau}\right)^{1/4}.
\end{align}
Combining this with Eq.~\eqref{eq:thermalizationtime-underoccupied}, we obtain
\begin{align}
    \tautherm =\frac{1}{\alpha_s^2T(\tautherm)}\sqrt{\frac{Q_s}{T(\tautherm)}}=\alpha_s^{-13/5}/Q_s.
\end{align}

The physical picture of bottom-up thermalization is quite robust and is also seen in simulations solving the Boltzmann equation \eqref{eq:boltzmann_equation} in diffusion approximation \cite{BarreraCabodevila:2022jhi, Cabodevila:2023htm, BarreraCabodevila:2025vir}.

\section{Performing QCD kinetic theory simulations\label{sec:performing-qcd-kinetic-theory-simulations}}
This thesis relies heavily on QCD kinetic theory simulations. In this section, all concepts needed for performing these numerical simulations are introduced. The code used for the publications \cite{Boguslavski:2023waw, Boguslavski:2023jvg, Boguslavski:2023alu} is publicly available \cite{kurkela_2023_10409474}. We will discuss how the Boltzmann equation \eqref{eq:boltzmann_equation} is solved, what initial conditions are used, how observables are calculated, and conclude with discussing discretization effects.

\subsection{Symmetries of the distribution function\label{sec:symmetries-f}}
Throughout the thesis, we will consider the distribution function to be a function of only three parameters: $f(t,\vb p,\vb x)=f(\tau,p,\cos\theta_p)$, i.e., the proper time $\tau$, the magnitude of the momentum $p$, and its angle to the beam axis $\theta_p$, which we motivate now.

\begin{enumerate}
    \item 
First, we assume that the system is \emph{boost invariant} \cite{Mueller:1999pi}, i.e., the distribution function only depends on the rapidity difference $\tilde f(\tau, p, y-\eta)$ for momenta $P$ and spacetime $X$,
\begin{subequations}\label{eq:rapidity-definition}
\begin{align}
    P_\mu&=p(\cosh y, \hat{\vb v}_\perp, \sinh y) = (p, \vb p_\perp, p_z), \\
     X_\mu&=(\tau\cosh\eta, \vb x_\perp, \tau\sinh\eta)=(t,\vb x_\perp, z).
\end{align}
\end{subequations}
A boost with rapidity $\eta'$ would shift $y\to y+\eta'$, $\eta\to \eta+\eta'$, thus leaving the distribution function $f$ invariant. It is then enough to consider the system at $z=\eta=0$. 
\item Further, we assume \emph{homogeneity in the transverse plane}, i.e., the distribution function does not depend on $\vb x_\perp$. This is justified for large and homogeneous colliding nuclei.

\item Additionally, we assume rotational symmetry around the beam axis, i.e., in the transverse plane,
\begin{align}
    f(k,\cos\theta_k,\phi_k+\alpha)=f(k,\cos\theta_k,\phi_k), && \forall\alpha\in [0,2\pi),\label{eq:distributionfunction-rotationsymmetry}
\end{align}
which is justified by assuming that the initial particle production is isotropic, with only the longitudinal expansion singling out one preferred direction.

\item Mirror symmetry around $z=0$ in symmetric collisions,
\begin{align}
    f(k,\cos\theta_k,\phi_k)=f(k,-\cos\theta_k,\phi_k) .\label{eq:distributionfunction-mirrorsymmetry}
\end{align}
\end{enumerate}
With all these assumptions, at $z=\eta=0$, the distribution function can be written as
\begin{align}
    f(t,\vb p)=f(\tau, p, p_z)=f(\tau,p,\cos\theta_p).
\end{align}
Mathematically, we employ here a sloppy notation using the same symbol $f$ for $f(\tau,p,p_z)=f(\tau,p,\cos\theta_p)$. Additionally, the time argument will be sometimes suppressed.

At nonzero $z$ or spacetime rapidity $\eta$, the distribution function can be obtained from $f(\tau, p,\cos\theta_p)$ by using \eqref{eq:rapidity-definition},
$\tilde f(\tau, p,y)=f(\tau, p,\tanh y)$ and by boost-invariance,
\begin{align}
    \tilde f(\tau, p, y-\eta)=f(\tau, p,\tanh(y-\eta)).
\end{align}

Using these symmetries, the spatial derivative in the Boltzmann equation \eqref{eq:boltzmann_equation} can be rewritten (at $\eta=z=0$) to \cite{Mueller:1999pi}
\begin{align}
    \vb v\cdot \nabla_x f(\tau, \vb p, z)=-\frac{p_z}{\tau}\pdv{p_z}f(\tau, p, p_z), \label{eq:boost-invariant-expansion-term}
\end{align}
and represents the effective longitudinal Bjorken expansion. Besides that, we will sometimes also consider isotropic distribution functions $f(p)$ without Bjorken expansion, for which $\vb v\cdot\nabla_x f(\tau,\vb p,z)=0$.

\subsection{Boltzmann equation and initial condition\label{sec:boltzmann-equation-and-initial-condition}}
With the approximations mentioned in the previous section, the Boltzmann equation \eqref{eq:boltzmann_equation} for a purely gluonic system is then given by
\begin{align}
    \left(\pdv{\tau}-\frac{p_z}{\tau}\pdv{p_z}\right)f(\tau,\vb p)=-\Conetwo[f(\tau,\vb p)]-\Ctwotwo[f(\tau,\vb p)],\label{eq:actual-boltzmann-equation-to-solve-expanding-system}
\end{align}
with the collision kernels $\Conetwo$ and $\Ctwotwo$ containing inelastic and elastic collisions described in more detail in Section \ref{sec:equations-of-qcd-kinetic-theory}. For the spatial derivative, we have already inserted the boost-invariant expansion term \eqref{eq:boost-invariant-expansion-term}, which is not present for the isotropic nonexpanding simulations also performed in this thesis. If present, this expansion term accounts for the effective longitudinal Bjorken expansion of the system.

\begin{table}
    \centering
    \begin{tabular}{cc} \toprule
         $\xianiso$& $A$\\
         \hline
         $2$& $0.96789$\\
         $4$& $2.05335$\\
         $10$& $5.24171$\\
         \bottomrule
    \end{tabular}
    \caption{Parameter choices for the initial condition \eqref{eq:initial_cond} for different anisotropy parameters $\xianiso$ that leave the initial energy density constant.}
    \label{tab:initconds}
\end{table}

As an initial condition, ideally, we would take at some time, e.g., $\tau_0=1/\Q$ \cite{Baier:2000sb}, the full lattice configuration obtained in a classical statistical simulation of the Glasma, extract a gluon distribution function (perhaps using the methods discussed in Section \ref{sec:kinetic-theory-derivation}, or as, e.g., done in Refs.~\cite{Greif:2017bnr, Greif:2020rhi, Du:2022bel}) and use this as input for our kinetic theory simulation. While performing this matching seems feasible in principle,
for this thesis, we follow the approach and initial conditions from Ref.~\cite{Kurkela:2015qoa}, where a parameterization of the JIMWLK evolved (2+1D) Glasma result from Ref.~\cite{Lappi:2011ju} is used,
\begin{align}  
    f(p_\perp,p_z,\tau{=}1/Q_s)=\frac{2A(\xianiso)\langle p_T\rangle}{\lambda \, p_\xi}\exp\left({-\frac{2p_\xi^2}{3\langle p_T\rangle^2}}\right), 
    \label{eq:initial_cond}
\end{align}
with $p_\xi=\sqrt{p_\perp^2+(\xianiso p_z)^2}$ and $\langle p_T\rangle = 1.8 Q_s$. Different parameter choices for the initial anisotropy parameter $\xianiso$ and the normalization constant $A$ are given in Table \ref{tab:initconds}, which are chosen to keep the initial energy density constant for varying anisotropies.
Here, the $p_z$ dependence is obtained by ``de''-squeezing the transverse momentum $p_\perp \to \sqrt{p_\perp^2+\xi^2p_z^2}$. Recall that the 't Hooft coupling $\lambda$ is defined via
\begin{align}
    \lambda = g^2\NC=4\pi\NC\,\alpha_s\,.
\end{align}

To give results in physical units, one has to choose a value for the saturation momentum $\Q$. For most of this thesis, we will not choose any particular value and discuss the results in units of $\Q$. However, in Section \ref{sec:obtaining-qhat-between-glasma-and-hydro} (which is based on Ref.~\cite{Boguslavski:2023alu}), we will attempt to make a comparison to the value of the jet quenching parameter $\qhat$ during the Glasma stage, for which we need to choose a value of $\Q$. 
There, the saturation momentum $Q_s$ is chosen to reproduce the energy density of the Glasma in Ref.~\cite{Ipp:2020nfu} at initial time $\Q\tau=1$ at the coupling $\lambda=10$, which leads to $\Q=\SI{1.4}{\giga\electronvolt}$. Remarkably, the same value is also obtained in Ref.~\cite{Keegan:2016cpi}, where it was found that precisely this value is needed 
for the EKT setup to be consistent with the later hydrodynamic evolution. We will discuss this matching in more detail in Section \ref{sec:obtaining_qhat_from_kinetic_theory}. Additionally, one could also proceed as in Ref.~\cite{Avramescu:2023qvv}, where coupling and saturation momentum $Q_s$ are related via the one-loop beta function of QCD,
\begin{align}
    \alpha_s(Q_s)=\frac{\lambda(Q_s)}{4\pi\NC}=\frac{1}{\frac{33-3\nf}{12\pi}\ln\frac{Q_s^2}{\LQCD^2}},
\end{align}
to obtain for $\nf=0$ and $\Q=\qty{1.4}{\giga\electronvolt}$ approximately $\lambda\approx 11$ for $\LQCD=\qty{200}{\mega\electronvolt}$, thus leading to a value in the same ballpark. It should be emphasized, however, that this matching is meant to serve as an estimate for which value of $\Q$ to choose, and is needed to provide numerical values in physical units.

\subsection{Observables and Landau matching}
We define the expectation value of an observable $O(\vb p,t)$ for a given distribution function $f(\vb p, t)$ via
\begin{align}
	\langle O(t)\rangle = \frac{\nu_g}{n(t)}\int\frac{\dd[3]{\vb p}}{(2\pi)^3}\, O(\vb p, t) f(\vb p, t), \label{eq:observables}
\end{align}
which is normalized by the particle number density
\begin{align}
	n(t)=\nu_g\int\frac{\dd[3]{\vb p}}{(2\pi)^3}f(\vb p,t),
\end{align}
such that $\langle 1\rangle = 1$.

An important class of such observables 
are the components of the energy-momentum tensor,
\begin{align}
	T^{\mu\nu}=\nu_g\int\frac{\dd[3]{\vb p}}{(2\pi)^3} \frac{p^\mu p^\nu}{|\vb p|}f(\vb p),\label{eq:energy-momentum-tensor-from-f}
\end{align}
whose diagonal entries correspond to the energy density $\varepsilon = T^{00}=n\langle p\rangle$,
\begin{align}
    \varepsilon=\nu_g\int\frac{\dd[3]{\vb p}}{(2\pi)^3}|\vb p|f(\vb p),
\end{align}
transverse pressure $P_T=T^{xx}=T^{yy}$ and longitudinal pressure $P_L = T^{zz}$.
Since QCD kinetic theory is conformal, $T^{\mu\nu}$ must be traceless, which leads to $\varepsilon = 2P_T+P_L$. For an isotropic system, $P_L=P_T=\varepsilon/3$.

We will frequently 
compare the nonequilibrium system (described by the distribution function $f(\vb p,t)$) to a corresponding thermal system. This comparison is not unique. While a thermal system is uniquely identified by its temperature, for a nonthermal system, temperature cannot be uniquely identified. A common way to do that is to use the Landau matching prescription, where the energy density $\varepsilon$ is used to define an effective temperature $\Teps$ as the temperature of the thermal system which has the same energy density as the nonthermal system,
\begin{align}
    \Teps(t) = \left(\frac{30\,\varepsilon(t)}{\nu_g\pi^2}\right)^{1/4}. \label{eq:Landau-matching-condition}
\end{align}
Matching other quantities is also possible, and the best matching prescription may depend on the observable. For instance, in Ref.~\cite{Boguslavski:2023fdm}, several matchings were considered for the \emph{heavy-quark diffusion coefficient}
$\kappa$, and Landau matching \eqref{eq:Landau-matching-condition} provided the best results.

\subsection{Time markers and time scales\label{sec:time-markers-and-scales}}
As discussed above, for plasmas undergoing Bjorken expansion, thermalization occurs in several stages according to the bottom-up picture. To better identify the different stages in the results and plots, 
we introduce time markers that should roughly correspond to the boundaries between the different stages, which we can best see in Fig.~\ref{fig:bottomup-overview}.
The star and circle markers are related to the occupancy $\langle pf \rangle/\langle p\rangle$, with the star placed where it first drops below $1/\lambda$ and the circle placed at its minimum value. Finally, the triangle marker is placed when the pressure ratio $P_T/P_L=2$, and corresponds to a system that has almost isotropized.

\begin{figure}
    \centering
    \includegraphics[width=0.5\linewidth]{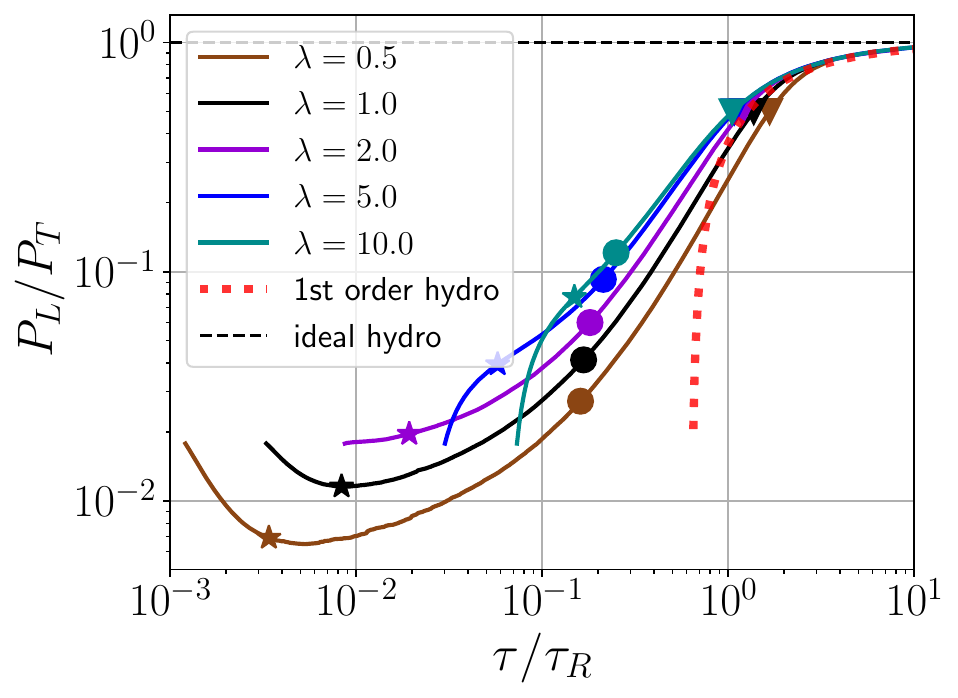}
    \caption{Pressure ratio in QCD kinetic theory simulations with Debye-like screened matrix elements. The first-order hydrodynamic expression \eqref{eq:first-order-hydro-pL/pT} is shown as a red dotted curve and the ideal hydro expectation (unity) as a constant line.}
    \label{fig:pressureratio-approach-hydro}
\end{figure}
We will frequently rescale the time in units of the bottom-up thermalization estimate \eqref{eq:bottom-up-timescale},
\begin{subequations}
\begin{align}
    \taubmss = \alpha_s^{-13/5}/Q_s=\left(\frac{\lambda}{4\pi\NC}\right)^{-13/5}/Q_s\,. \label{eq:taubmss}
\end{align}
Additionally, we will also frequently use the relaxation time,
\begin{align}
    \tau_R(\tau)=\frac{4\pi\eta/s}{\Teps(\tau)},\label{eq:relaxation-time}
\end{align}
\end{subequations}
which depends on the shear viscosity $\eta$ over entropy density $s$, and the effective temperature $\Teps$ defined in Eq.~\eqref{eq:Landau-matching-condition}. It is the only time scale that appears in first-order hydrodynamics. For instance, there the pressure ratio \cite{Romatschke:2017ejr} is given by
\begin{align}
    \frac{P_L}{P_T}=1-\frac{2}{\pi}\frac{\tau}{\tauR}. \label{eq:first-order-hydro-pL/pT}
\end{align}
Note that the dimensionless ratio $\eta/s$ depends on the coupling $\lambda$. It has been calculated from perturbative QCD \cite{Arnold:2000dr, Arnold:2003zc}, but in practice, for using it in QCD kinetic theory simulations, it is extracted numerically \cite{Keegan:2015avk, Kurkela:2018vqr, Kurkela:2018wud, Kurkela:2018oqw}. We will discuss how this ratio can be extracted in more detail in Chapter \ref{sec:improving-qcd-simulations}.

In Fig.~\ref{fig:pressureratio-approach-hydro}, we show the pressure ratio obtained from QCD kinetic theory simulations at various couplings and how it approaches the first order hydrodynamic result \eqref{eq:first-order-hydro-pL/pT}. Additionally, we also show the ideal hydrodynamic estimate, where $P_L=P_T=\varepsilon/3$.
This shows that QCD kinetic theory approaches hydrodynamics at late times. In fact, kinetic theory is more general than hydrodynamics, and the hydrodynamic equations of motion can be obtained from kinetic theory \cite{Israel:1976efz, Denicol:2012cn}.

\subsection{Extrapolating EKT to large couplings\label{sec:ekt_extrapolation_large_couplings}}

Although QCD kinetic theory is only valid at weak couplings, it is often extrapolated to larger values of the coupling, where no first-principle fully dynamical simulation of QCD is possible. 
There, using the Boltzmann equations and QCD kinetic theory should be viewed as a (mathematically well-defined) model for QCD equilibration. Its study using these larger couplings may nevertheless provide useful phenomenological insights into QCD equilibration and hydrodynamization in 
heavy-ion collisions. Since the Debye mass $m_D$ is proportional to the coupling, the na\"ive extrapolation $\lambda\to\infty$ leads to $m_D\to\infty$, and thus the screened $t$-channel small-angle-scattering terms in the matrix element \eqref{eq:usual_screened_matrix_element} vanish and only the $s$-channel and constant parts remain.
Therefore, at sufficiently large couplings, differences in the screening prescription should become less important, albeit in a regime outside of any theoretical control.

\subsection{Numerical details and discretization artifacts\label{sec:discretization-artifacts}}
\begin{figure}
    \centering
    \includegraphics[width=0.5\linewidth]{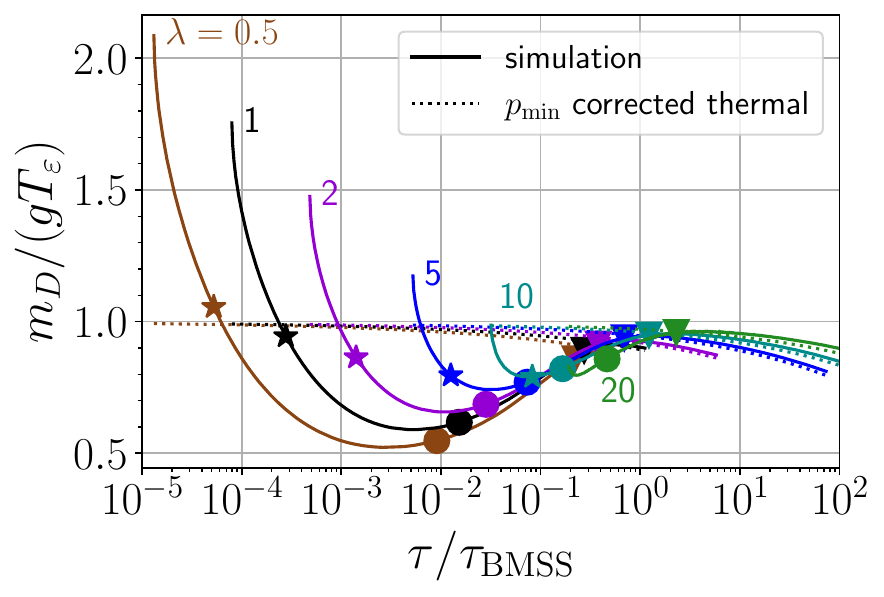}
    \caption{Debye mass over its thermal value $g\Teps$ during kinetic theory simulations of expanding systems for different couplings. As a dotted line, we show the $\pmin$ corrected Debye mass from Eq.~\eqref{eq:debyemass-finitepmin}.}
    \label{fig:debyemass-pmineffects}
\end{figure}
The implementation used in this thesis is based on Refs.~\cite{AbraaoYork:2014hbk, Kurkela:2014tea, Kurkela:2015qoa} and is publicly available \cite{kurkela_2023_10409474}.
The distribution function $f(\tau,\vb p)$ is stored on a finite momentum grid, with $\pmin \leq p \leq  \pmax$. The collision kernels, which consist of several integrals, are evaluated using Monte Carlo methods. The numerical algorithm is constructed in such a way that energy and momentum are exactly conserved. More details on the discretization and on how the collision kernels are evaluated can be found in Appendix \ref{app:numericaldetails}.

Let us now focus on one aspect which is independent of the precise discretization and evaluation of the collision kernels, which is having a nonzero $\pmin$ and finite $\pmax$. This will have the effect that any observable \eqref{eq:observables} will have discretization artifacts. For instance, even for a thermal distribution, the Debye mass $m_D$ obtained from Eq.~\eqref{eq:debyemass-general} for a nonzero $\pmin$ is given by \cite{Boguslavski:2023fdm}
\begin{align}
    m_D^2(\pmin)=\frac{8\lambda}{(2\pi)^2}\int_{\pmin}^\infty \dd{p} p f_+(p)=\frac{2\lambda T\left(T\Li_2\left(e^{-\frac{\pmin}{T}}\right)-\pmin\log\left(1-e^{-\pmin/T}\right)\right)}{\pi^2}.\label{eq:debyemass-finitepmin}
\end{align}

Moreover, if we simulate an expanding system, all characteristic timescales will decrease; thus, discretization effects, and in particular $\pmin>0$ effects, will have increased importance over time. This is demonstrated in Fig.~\ref{fig:debyemass-pmineffects}, where the Debye mass obtained from Eq.~\eqref{eq:debyemass-general} over its thermal value $g\Teps$ is plotted. At late times, the ratio deviates significantly from unity. The dotted lines are the $\pmin$ corrected expression for the Debye mass \eqref{eq:debyemass-finitepmin}, and agree very well with the results from the nonequilibrium simulation. This demonstrates that finite $\pmin$ effects are the leading discretization effect for the Debye mass.

In Appendix \ref{app:numericaldetails}, we discuss in more detail the precise discretization used and how the collision terms are evaluated, including the coordinate system used for the integration in the elastic collision term.

%% file: 450_jet_broadening.tex
The most straightforward aspect of jet-medium interactions to study is jet momentum broadening, in particular, characterized by the jet quenching parameter $\qhat$ from Eq.~\eqref{eq:intro-qhat-definition}. The physical picture is that of a highly energetic parton (e.g., a quark for a quark-jet) moving through the quark-gluon plasma and receiving (random) momentum kicks from the plasma constituents.
This parameter is not only interesting on its own but is also used as input to calculate jet energy loss by medium-induced gluon emission in the harmonic approximation (see Eq.~\eqref{eq:intro-Cb-expansion}). In this chapter, we address how the jet quenching parameter $\hat q$ can be obtained in a nonequilibrium plasma of quarks and gluons. Results are presented for thermal equilibrium and simple (toy) models (scaled thermal distributions and effectively two-dimensional distributions). Finally, we conclude this chapter with the extraction of $\hat q$ using QCD kinetic theory simulations of the initial stages in heavy-ion collisions and compare the extracted values with numerical simulations of the previous Glasma stage.

This chapter is based on Refs.~\cite{Boguslavski:2023alu, Boguslavski:2023waw}.

\section{Obtaining the jet quenching parameter in QCD kinetic theory\label{sec:obtaining_qhat_from_kinetic_theory}}

\subsection{Relation of jet quenching parameter and scattering rate}
Let us consider an energetic parton with a large but finite momentum $\vb p$, which should be much larger than all other relevant momentum scales in the plasma. This plasma of quarks and gluons will be far from equilibrium during the initial stages in heavy-ion collisions, so we generalize the definition \eqref{eq:intro-qhat-definition} to account for momentum broadening in different directions,
\begin{align}
	\qhat^{ij}(p)=\int\dd[3]{\vb q}\,q^iq^j \frac{\dd\Gammael}{\dd[3]{\vb q}},\label{eq:def_qhat}
\end{align}
and, in particular, have also included the possibility of longitudinal momentum broadening, i.e., broadening along the direction of the propagation of the jet.

For the total transverse momentum broadening coefficient $\qhat$, we need to sum over the directions perpendicular to the jet direction. If we consider the jet to be moving in the $x$-direction, we obtain the usual jet quenching parameter
\begin{align}
	\label{eq:qhat_1122}
	\qhat = \qhat^{yy}+\qhat^{zz}\,,
\end{align}
which measures the average transverse momentum transfer squared to the jet parton per unit time. 
Additionally, we can also consider longitudinal momentum broadening, i.e., momentum broadening in the direction of the jet,
\begin{align}
    \hat q_L = \hat q^{xx}.
\end{align}

Let us recall the expression for the elastic scattering rate, Eq.~\eqref{eq:decay_rate}, which was derived in Section \ref{sec:decay-rate},
\begin{align}
    \begin{split}
	\Gammael&=\frac{1}{4p\nu_a}\sum_{bcd}\int_{\vec k\vec p'\vec k'}(2\pi)^4\delta^4(P+K-P'-K')\label{eq:decay_rate2}\\
	&\qquad\qquad\times\left|\mathcal M^{ab}_{cd}(\vecp, \vec k;\vec p',\vec k')\right|^2f_b(\vec k)\left[1\pm f_d(\vec k')\right]\left[1\pm f_c(\vec p')\right],
    \end{split}
\end{align}
and which we need as input for $\qhat$ using Eq.~\eqref{eq:def_qhat}.
The sum is over all possible scattering processes.
It describes the rate of a particle of species $a$ with momentum $\vb p$ being scattered out of its momentum state due to elastic collisions $p+k\to p'+k'$.
The expression is completely symmetric under the exchange of the outgoing particles $\vecp' \leftrightarrow \vec k'$ and $c\leftrightarrow d$, as it should be. However, including $q^iq^j$ as in Eq.~\eqref{eq:def_qhat} breaks this symmetry.
Effectively, we need to choose with respect to which outgoing particle we measure the transferred momentum.
We choose here to define the harder outgoing particle to be the jet particle and label it $c$ with momentum $p'$, which is consistent with other studies, e.g., \cite{Arnold:2008vd, Caron-Huot:2008zna, He:2015pra, Ghiglieri:2015ala}.
In particular, this definition arises naturally if one thinks of $\hat q$ as a diffusion coefficient (with small momentum transfer), and amounts to defining
\begin{align}
    \vb q=\vb p'-\vb p.\label{eq:introduce_q}
\end{align}
With this choice, we may relabel our momenta such that $p'>k'$, which leads to an additional factor of $2$.\footnote{
We use the identity
\begin{align*}
	\int_{\vb k'\vb p'} g(k',p')
	= 2\int_{\vb k'\vb p'}g(k',p')\Theta(p'-k'),
\end{align*}
for symmetric functions $g(p',k')=g(k',p')$.}
This will lead to 
more matrix elements than those from Table \ref{tab:amy_matrix_el}, since processes with the external outgoing particles switched are treated symmetrically there. For instance, $qg\leftrightarrow gq$ and $qg\leftrightarrow qg$ have the same matrix element in Ref.~\cite{Arnold:2002zm} while we need to explicitly distinguish them here. We will discuss this new complication in more detail and list the required matrix elements explicitly in Table \ref{tab:qhat_matrix_el} in Section \ref{sec:qhat-formula-finite-p}.

We then arrive at
\begin{align}
	\hat q^{ij} &= \frac{1}{2p\nu_a}\sum_{bcd}\int_{\substack{\,\,\vb k\vb p'\vb k'\\p'>k'}}q^iq^j(2\pi)^4\delta^4(P+K-P'-K')\nonumber\\
	&\qquad\times\left|\mathcal M_{cd}^{ab}(\vb p,\vb k;\vb p',\vb k')\right|^2 f_b(\vb k)\left[1\pm f_d(\vb k')\right]\left[1\pm f_c(\vb p')\right]. \label{eq:qhat_general}
\end{align}
Due to the requirement that $p'>k'$, if the jet energy $p$ is large enough, we may set $f(\vb p')\to 0$.

To simplify this integral, it is convenient to rewrite the integral measures in \eqref{eq:qhat_general} to
\begin{align}
\hat q^{ij} &= \frac{1}{2p\nu_a}\sum_{bcd}\int\frac{\dd[4]{K}\dd[4]{P'}\dd[4]{K'}}{(2\pi)^5}\qperp^i\qperp^j\delta^{(4)}(P+K-P'-K')\left|\mathcal M_{cd}^{ab}(\vb p,\vb k;\vb p',\vb k')\right|^2 \nonumber\\
& \qquad \qquad \times\delta(K^2)\delta(P'^2)\delta(K'^2)\Theta(K^0)\Theta(P'^0)\Theta(K'^0)f^b(\vb k)\left[1\pm f^d(\vb k')\right]\Theta(p'-k'). 
\end{align}
Using the delta function, we eliminate the $K'$ integral. For convenience, we 
will keep writing $K'$ or $\vb k'$ as a short notation for $P+K-P'$ or $\vb p+\vb k-\vb p'$, respectively.
Similarly to Ref.~\cite{Arnold:2003zc}, and in accordance with our discussion on how to introduce $\vb q$ (see Eq.~\eqref{eq:introduce_q}), we define $Q^\mu=(\omega, \vb q)^\mu$ as
\begin{align}
\begin{split}
Q=P'-P\quad \Leftrightarrow\quad & \vb q = \vb p' - \vb p = \vb k - \vb k',\\
& \omega = p' - p = k - k'. 
\end{split}
\end{align}
Note that unlike the external momenta $P,K, P'$ and $K'$, the transfer momentum $Q$ is not necessarily light-like,
i.e., $Q^2=\vb q^2-\omega^2\geq 0.$ 
Then we have 
$\dd[4] K\dd[4]P'=\dd[4]K\dd[4]Q$ and thus
\begin{align}
\begin{split}
	\hat q^{ij} &= \frac{1}{2p\nu_a}\sum_{bcd}\int\frac{\dd[4]{K}\dd[4]{Q}}{(2\pi)^5}q^iq^j\left|\mathcal M_{cd}^{ab}(\vb p,\vb k;\vb p',\vb k')\right|^2
	 f^b(\vb k)\left[1\pm f^d(\vb k')\right]\\
	&\qquad\times \Theta(p'-k')\Theta(K^0)\Theta(P^0 + \omega)\Theta(K^0 - \omega)\delta(K^2)\delta\left((P+Q)^2\right)\delta\left((K-Q)^2\right) 
	 .
    \end{split}
\end{align}
Using $P^2=K^2=0$, $P\cdot Q=-p\omega + pq\cos\thetaqp$ and $K\cdot Q=-k\omega + kq\cos\thetaqk$, where $\thetaqp$ is the angle between $\vb p$ and $\vb q$ (and $\thetaqk$ between $\vb k$ and $\vb q$), we can rewrite the last two delta functions as
\begin{align}
\begin{split}
	&\delta\left((P+Q)^2\right)\delta\left((K-Q)^2\right)\\
	&\quad=\frac{1}{4pkq^2}\delta\left(\cos\thetaqp-\frac{\omega}{q}-\frac{\omega^2-q^2}{2pq}\right)
	\delta\left(\cos\thetaqk-\frac{\omega}{q}+\frac{\omega^2-q^2}{2kq}\right).
\end{split}
\end{align}
This fixes the angles between $\vb q$ and $\vb k$, and between $\vb q$ and $\vb p$.
Additionally, the integration region is restricted to
\begin{align}
	|\omega| < q, && p >\frac{q-\omega}{2}, && k >\frac{q+\omega}{2}. \label{eq:phase_space_relation_omega_q_k}
\end{align}
Subsequently performing the $K^0$ integral yields
\begin{align}
\begin{split}
	\hat q^{ij} &= \frac{1}{16p^2\nu_a}\sum_{bcd}\int\frac{\dd[3]{\vb k}\dd[3]{\vb q}\dd{\omega}}{(2\pi)^5 q^2 k^2 }q^iq^j\left|\mathcal M_{cd}^{ab}(\vb p,\vb k;\vb p',\vb k')\right|^2 f^b(\vb k)\left[1\pm f^d(\vb k-\vec q)\right]\\
	&\times \Theta(p'-k')\Theta\left(p - \frac{q-\omega}{2}\right)\Theta\left(k - \frac{q+\omega}{2}\right)\Theta(q-|\omega|) \\
	&\times\delta\left(\cos\thetaqp-\frac{\omega}{q}-\frac{\omega^2-q^2}{2pq}\right)\delta\left(\cos\thetaqk-\frac{\omega}{q}+\frac{\omega^2-q^2}{2kq}\right)
	 .\label{eq:qhat_der_with_delta}
    \end{split}
\end{align}

\subsection{Coordinate systems}
\label{sec:coordinate_systems}
\begin{figure*}
	\centering
	\includegraphics[width=\linewidth]{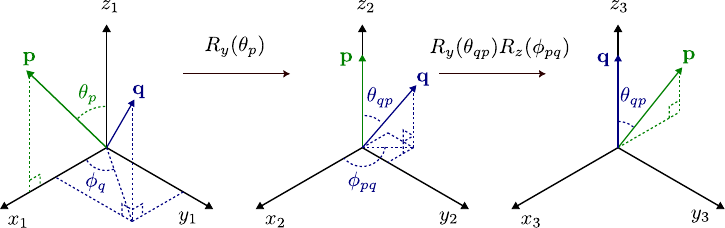}
	\caption{The integration frames. {\em (Left:)} `Lab frame'. The jet momentum $\vb p$ lies in the $x-z$ plane.
		{\em (Center:)} `$p$-frame', obtained by rotating the `lab frame' around the $y$-axis, such that $\vb p$ points in the $z$ direction. 
		{\em (Right:)} `$q$-frame'. Here, $\vb q$ points in the $z$ direction and $\vb p$ lies in the $x-z$ plane. Figure from \cite{Boguslavski:2023waw}.
	}
	\label{fig:frames}
\end{figure*}
We will now discuss how the integration variables and coordinate systems are chosen to perform this integral. We use a different choice here than when evaluating the elastic collision term in the kinetic theory simulations (see Appendix \ref{sec:coordinate-systems-appendix}).
In particular, since $f(\vb p)$ is stored on a grid in a specific coordinate system (let us call it \emph{lab frame}), we need to establish how the vectors $\vb k$ and $\vb k'$ in the lab frame depend on the integration variables.
In the \emph{lab frame}, we denote the vectors with lower index $1$,
\begin{subequations}
	\begin{align}
		\vb p_1&=p(\sin\theta_p,0,\cos\theta_p),\\
		\vb q_1&=q(\sin\theta_q\cos\phi_q,\sin\theta_q\sin\phi_q,\cos\theta_q),\\
		\vb k_1&=k(\sin\theta_k\cos\phi_k,\sin\theta_k\sin\phi_k,\cos\theta_k).\label{eq:labframe_k}
	\end{align}
\end{subequations}

In accordance with the discussion in Section \ref{sec:symmetries-f}, we assume azimuthal symmetry around the $z$-axis, so we can always rotate this frame such that $\vb p $ lies in the $x-z$ plane (effectively having the jet go into the $\vb x$ direction).
For choosing the integration frames, we follow Ref.~\cite{Arnold:2003zc} and use the jet momentum $\vb p$ as a distinct direction. 
In this second frame (\emph{$p$-frame}), denoted by a subscript 2, in which $\vb p$ points in the $z$ direction and that is obtained via a rotation of the \emph{lab frame} around the $y$-axis (see \fig\ref{fig:frames}),
\begin{subequations}
	\begin{align}
		\vb p_2&=p(0,0,1),\\
		\vb q_2&=q(\sin\thetaqp\cos\phi_{pq},\sin\thetaqp\sin\phi_{pq},\cos\thetaqp),\label{eq:pframe_q}\\
		\vb k_2&=k(\sin\theta_{pk}\cos\phi_{pk},\sin\theta_{pk}\sin\phi_{pk},\cos\theta_{pk}).
	\end{align}
\end{subequations}
Here, we perform the integral over $\vb q$.

The $\vb k$ integral is then performed in a third frame, in which $\vb q$ points in the $z$ direction and $\vb p$ lies in the $x-z$ plane. We call this the \emph{$q$-frame} and denote it by a subscript 3:
\begin{subequations}
	\begin{align}
		\vb p_3&=p(\sin\thetaqp,0,\cos\thetaqp),\label{eq:qframe_p}\\
		\vb q_3&=q(0,0,1), \label{eq:qframe_q}\\
		\vb k_3&=k(\sin\thetaqk\cos\phikq,\sin\thetaqk\sin\phikq,\cos\thetaqk). \label{eq:qframe_k}
	\end{align}
\end{subequations}

The components of the vectors transform between the frames according to the matrix relations
\begin{subequations}
	\begin{align}
		\vb v_2=A\vb v_1, && A&=R_y(\theta_p),\label{eq:def_A}\\
		\vb v_3 = B\vb v_2, && B&=R_y(\thetaqp)R_z(\phipq),\label{eq:def_B}
	\end{align}
\end{subequations}
where $R_y(\alpha)$ and $R_z(\alpha)$ denote the matrices corresponding to a rotation with angle $\alpha$ around the $y$- and $z$-axis, respectively.
The transformation matrices read
\begin{subequations}
	\begin{align}
		A&=\begin{pmatrix}
			\cos\theta_p&0&-\sin\theta_p\\
			0&1&0\\
			\sin\theta_p&0&\cos\theta_p
		\end{pmatrix},\\
		B&=\begin{pmatrix}
			\cos\thetaqp\cos\phipq&\cos\thetaqp\sin\phipq & -\sin\thetaqp\\
			-\sin\phipq&\cos\phipq&0\\
			\cos\phipq\sin\thetaqp&\sin\thetaqp\sin\phipq&\cos\thetaqp
		\end{pmatrix}.
	\end{align}
\end{subequations}
For the calculation of $\qhat^{ij}$ we use the components $\qperp^{i}$ of $\vec q$ in the $p$-frame,
\begin{subequations}
	\begin{align}
		q^1 &= (\vb q_2)^1 = q\sin\thetaqp\cos\phipq,\\
		q^2 &= (\vb q_2)^2 = q\sin\thetaqp\sin\phipq,\\
		q^3 &= (\vb q_2)^3 = q\cos\thetaqp\label{eq:qz_pframe}.
	\end{align}
\end{subequations}
In this way, the components $1$ and $2$ are perpendicular to $\vec p$ and quantify the momentum broadening transverse to the jet.

Having taken the Dirac delta functions in \eq\eqref{eq:qhat_general} into account, 
we choose $\phipq$, $\phikq$, $k$, $\omega$, and $q$ as independent integration variables. 
Therefore, we need to express all other quantities in terms of them. All other angles are then fixed. For example, the on-shell conditions $|\vec k'|^2=|\vec k-\vec q|^2=(k-\omega)^2$ and $|\vec k' +\vec q|^2=(k'+\omega)^2$ lead to
\begin{subequations}
	\begin{align}
		\cos\thetaqk&=\frac{\omega}{q}-\frac{\omega^2-q^2}{2kq},\\
		\cos\thetaqp&=\frac{\omega}{q}+\frac{\omega^2-q^2}{2pq},\label{eq:cos_thetapq}\\
		\cos\theta_{k'q}&=\frac{\omega}{q}+\frac{\omega^2-q^2}{2k'q}.
	\end{align}
\end{subequations}
Next, we discuss how to express the angles $\theta_k$ and $\theta_{k'}$ in terms of the integration variables needed for evaluating the distribution functions $f(\vb k)$ and $f(\vb k')$ in \eqref{eq:qhat_general} in the lab frame.
From Eqs.~\eqref{eq:def_A} and \eqref{eq:def_B}, we obtain $\vec k_1=A^TB^T\vec k_3$. From $(\vec k_1)_z$ we can read off 
\begin{subequations}\label{eq:cos-thetak_labframe}
\begin{align}
	&\cos\theta_k = \sin\phikq\sin\phipq\sin\thetaqk\sin\theta_p\\
	& \qquad\qquad -\cos\phikq\sin\thetaqk\left(\cos\phipq\cos\thetaqp\sin\theta_p+\cos\theta_p\sin\thetaqp\right)\nonumber\\
	& \qquad\qquad +\cos\thetaqk\left(\cos\theta_p\cos\thetaqp-\cos\phipq\sin\theta_p\sin\thetaqp\right),\nonumber
\end{align}
and a similar expression holds for $\cos\theta_{k'}$,
\begin{align}
	&\cos\theta_{k'} = \sin\phikq\sin\phipq\sin\theta_{qk'}\sin\theta_p\\
	& \qquad\qquad -\cos\phikq\sin\theta_{qk'}\left(\cos\phipq\cos\thetaqp\sin\theta_p+\cos\theta_p\sin\thetaqp\right)\nonumber\\
	& \qquad\qquad +\cos\theta_{qk'}\left(\cos\theta_p\cos\thetaqp-\cos\phipq\sin\theta_p\sin\thetaqp\right).\nonumber
\end{align}
\end{subequations}
The azimuthal angle is $\phi_{qk'}=\phikq$ because $\vec k'=\vec k -\vec q$ and $\vec q$ points in the $z$ direction in the \emph{$q$-frame}.

\subsection{Formula for the jet quenching parameter for finite jet energy\label{sec:qhat-formula-finite-p}}

We are now ready to give the formula for the components of $\qhat$,
\begin{align}
    \begin{split}
	\qhat^{ij}&=\frac{1}{2^9\pi^5\nu_a}\sum_{bcd}\int_0^{2\pi}\dd{\phipq}\int_0^{2\pi}\dd{\phikq}\int_0^\infty \dd{q} \int^q_{\max\left(-q,q-2p,\frac{q-2p}{3}\right)}\!\!\!\dd{\omega}\int_{\frac{q+\omega}{2}}^{p+2\omega}\!\!\!\dd{k} \\
    &\qquad\qquad\times
    q^iq^j\frac{|\mathcal M^{ab}_{cd}|^2}{p^2}f_b(\vb k)\left(1\pm f_d(\vb k')\right)\,. \label{eq:qhat_formula}
    \end{split}
\end{align}
Recall that $\nu_a = 2d_R$, where $d_R$ is the dimension of the representation of the jet particle.
The upper sign in Eq.~\eqref{eq:qhat_formula} is to be used when the $d$ particle is a boson (gluon), and the lower sign if it is a fermion (quark), see Eq.~\eqref{eq:boseenhancement-fermi-blocking}.
The components $q^i$ of \eq \eqref{eq:qhat_formula} in the \emph{$p$-frame} read
\begin{subequations}
	\begin{align}
		q^1 &= q\sin\thetaqp\cos\phipq\label{eq:qx_def},\\
		q^2 &= q\sin\thetaqp\sin\phipq\label{eq:qy_def},\\
		q^3 &= q\cos\thetaqp\label{eq:qz_def}.
	\end{align}
\end{subequations}
Although we started with the collision term \eqref{eq:c22_first} with the same matrix elements, our choice $p'>k'$ breaks the symmetry of exchanging the outgoing particles, $(abcd)\leftrightarrow(abdc)$ (see \eqref{eq:c22-matrixelements-symmetries}). This implies that we have to distinguish between, for example, `$q_1g\leftrightarrow q_1 g$' and `$q_1g \leftrightarrow g q_1$', and thus require more matrix elements.
Recall that they are conveniently given in terms of the Lorentz-invariant Mandelstam variables $s$, $t$, and $u$ (see Eq.~\eqref{eq:Mandelstam-definition}), which are defined with respect to the momenta corresponding to the particles with labels $a,b,c,d$, 
\begin{align}
	s=-(P+K)^2, && t=-(P'-P)^2, && u =-(K'-P)^2.\label{eq:def-mandelstam}
\end{align}
The additional matrix elements resulting from the reduced symmetry can be obtained from Table \ref{tab:amy_matrix_el} by relabeling $p'\leftrightarrow k'$ and $c\leftrightarrow d$, 
effectively switching 
\begin{align}
	s\to s, && t\to u, && u\to t.
\end{align}
These resulting matrix elements are still symmetric under $(abcd)\leftrightarrow(badc)$ and are listed in \tab\ref{tab:qhat_matrix_el}. 
\begin{table*}
	\begin{tabular}{ M{0.2\linewidth}  m{0.65\linewidth} }\toprule
		\vspace{5pt}
		$ab\leftrightarrow cd$ & $\left|\mathcal M^{ab}_{cd}\right|^2/g^4$ \\[5pt] 
		\hline 
		$q_1q_2\leftrightarrow q_1q_2,$ & \multirow{4}{*}{$8\frac{\dF ^2\CF^2}{\dA }\left(\frac{s^2+u^2}{\underline{t^2}}\right)$}\\ $q_1\bar q_2\leftrightarrow q_1\bar q_2,$ &\\ $\bar q_1 q_2\leftrightarrow \bar q_1 q_2,$ &\\ $\bar q_1\bar q_2\leftrightarrow \bar q_1\bar q_2$ &  \\[10pt] 
		
		$q_1q_2\leftrightarrow q_2q_1,$ & \multirow{4}{*}{$8\frac{\dF ^2\CF^2}{\dA }\left(\frac{s^2+t^2}{{u^2}}\right)$}\\ $q_1\bar q_2\leftrightarrow \bar q_2 q_1,$ &\\ $\bar q_1 q_2\leftrightarrow q_2\bar q_1,$ &\\ $\bar q_1\bar q_2\leftrightarrow \bar q_2\bar q_1$ &  \\[10pt] 
		
		$q_1q_1\leftrightarrow q_1q_1,$ & \multirow{2}{*}{$8\frac{\dF ^2\CF^2}{\dA }\left(\frac{s^2+u^2}{\underline{t^2}}+\frac{s^2+t^2}{{u^2}}\right)+16 \dF \CF\left(\CF-\frac{\CA  }{2}\right)\frac{s^2}{tu}$}\\ $\bar q_1 \bar q_1\leftrightarrow \bar q_1 \bar q_1$ & \\ [10pt]
		
		$q_1\bar q_1\leftrightarrow q_1\bar q_1$ &$8\frac{\dF ^2\CF^2}{\dA }\left(\frac{s^2+u^2}{\underline{t^2}}+\frac{t^2+u^2}{s^2}\right)+16 \dF \CF\left(\CF-\frac{\CA  }{2}\right)\frac{u^2}{st}$ \\[10pt]
		
		$q_1\bar q_1\leftrightarrow \bar q_1 q_1$ &$8\frac{\dF ^2\CF^2}{\dA }\left(\frac{s^2+t^2}{{u^2}}+\frac{u^2+t^2}{s^2}\right)+16 \dF \CF\left(\CF-\frac{\CA  }{2}\right)\frac{t^2}{su}$ \\[10pt]

		$q_1\bar q_1\leftrightarrow q_2\bar q_2,$ & \multirow{2}{*}{$8\frac{\dF ^2\CF^2}{\dA }\left(\frac{t^2+u^2}{s^2}\right)$}\\
		$q_1\bar q_1\leftrightarrow \bar q_2 q_2$ &\\[10pt]
		
		$q_1\bar q_1\leftrightarrow gg$ & $8 \dF  \CF^2\left(\frac{u}{\underline{\underline{t}}}+\frac{t}{{{u}}}\right) - 8\dF \CF \CA\left(\frac{t^2+u^2}{s^2}\right)$ \\[10pt] 
		
		$q_1 g\leftrightarrow q_1 g,$ & \multirow{2}{*}{$-8\dF \CF ^2\left(\frac{u}{s}+\frac{s}{{{u}}}\right)+8\dF \CF \CA  \left(\frac{s^2+u^2}{\underline{t^2}}\right)$}\\
		$\bar q_1 g\leftrightarrow \bar q_1 g$ & \\[10pt]
		
		$q_1 g\leftrightarrow gq_1 ,$ & \multirow{2}{*}{$-8\dF \CF ^2\left(\frac{t}{s}+\frac{s}{\underline{\underline{t}}}\right)+8\dF \CF \CA  \left(\frac{s^2+t^2}{{u^2}}\right)$}\\
		$\bar q_1 g\leftrightarrow g\bar q_1 $ & \\[10pt]
		
		$gg\leftrightarrow gg$ & $16\dA \CA  ^2\left(3-\frac{su}{\underline{t^2}}-\frac{st}{u^2}-\frac{tu}{s^2}\right)$\\[5pt] \bottomrule
	\end{tabular}
	\caption{Matrix elements for the jet quenching parameter $\hat q$, obtained from the matrix elements from Table \ref{tab:amy_matrix_el} by breaking the symmetry of exchanging outgoing particles. They are obtained by replacing $c\leftrightarrow d$ and $t\leftrightarrow u$. Singly-underlined denominators indicate infrared-sensitive contributions from soft gluon exchange and double-underlined denominators from soft fermion exchange.
		The group constants are given in Eq.~\eqref{eq:group-constants}. Table taken from \cite{Boguslavski:2023waw}.}
	\label{tab:qhat_matrix_el}
\end{table*}
In the underlined terms, medium effects have to be included, as we will discuss in Section \ref{sec:qhat_screening}.

The Mandelstam variables are given explicitly by
\begin{align}
    t &=\omega^2-q^2, \label{eq:mandelstam_t_qhat}\\
    \begin{split}
        s &= -\frac{t}{2q^2}\Big((p+p')(k+k')+q^2
        -\sqrt{(4pp'+t)(4k'k+t)}\cos\phikq\Big),
    \end{split} \label{eq:mandelstam_s_qhat}\\
    \begin{split}
        u &= \frac{t}{2q^2}\Big((p+p')(k+k')-q^2
        -\sqrt{(4pp'+t)(4k'k+t)}\cos\phikq\Big),
    \end{split}\label{eq:mandelstam_u_qhat}
\end{align}
with $ p'=p+\omega$ and $k'=k-\omega$.
The expressions for $s$ and $u$ as in Eqs.~\eqref{eq:mandelstam_s_qhat} and \eqref{eq:mandelstam_u_qhat} can also be found in \cite{Arnold:2003zc} and are similar to those used in the elastic collision terms, Eqs.~\eqref{eq:mandelstam_s_explicit-c22} and \eqref{eq:mandelstam_u_explicit-c22} (they only differ in the argument of the cosine).
Note that we defined the components $\qhat^{ij}$ with respect to the jet direction.
If the jet moves perpendicular to the beam-axis $z$ in the $x$-direction as in \fig\ref{fig:jet-quenching-illustration}, then $\cos\theta_p = 0$, and $\qhat^{yy}=\qhat^{22}$ is the momentum broadening in the $y$ direction and $\qhat^{zz}=\qhat^{11}$ is the momentum broadening in the beam direction, which sum to the usual $\hat q = \hat q^{11} + \hat q^{22}$. Longitudinal momentum broadening can be obtained from $\qhat_L=\qhat^{33}$. If we replace $\qhat^i\qhat^j$ by $\omega$ in \eqref{eq:qhat_formula}, we obtain elastic (collisional) energy loss.

\subsection{Symmetries of $\qhat^{ij}$}\label{sec:symmetries_of_qhat}
Let us now discuss the symmetries of the matrix $\qhat^{ij}$.
For a {\em spherically} symmetric distribution function $f(\vb k)=f(|\vb k|)$, it is easy to see that 
\begin{align}
	\qhat^{12}=\qhat^{13}=\qhat^{23}=0\,, \qquad \qhat^{11}=\qhat^{22}.
\end{align}
For a phase space density that is {\em azimuthally} symmetric around the $z$-axis (beam direction), i.e., $f(k,\cos\theta_k)$, we also find that
\begin{align}
	\qhat^{12}=\qhat^{23}=0\,.
\end{align}
For a jet moving in the $x$ or beam direction, we additionally find
\begin{align}
	\qhat^{13} = 0\,, \quad \text{if}~\cos\theta_p=0~\text{or}~\cos\theta_p=1\,.
\end{align}

The fact that $\qhat^{12}=\qhat^{23}=0$ can be seen by rewriting the angular integrals $\int_0^{2\pi}\dd{\phipq}\int_0^{2\pi}\dd{\phikq}g(\phipq,\phikq)=\int_{-\pi}^{\pi}\dd{\phipq}\int_{-\pi}^{\pi}\dd{\phikq}g(\phipq,\phikq)$ and then splitting the $\phipq$ integral into the integral from $(-\pi,0)$ and $(0,\pi)$ to arrive at
\begin{align*}
&\int_0^{2\pi}\dd{\phipq}\int_0^{2\pi}\dd{\phikq}g(\phipq,\phikq)
	=\int_0^{\pi}\dd{\phipq}\int_{-\pi}^{\pi}\dd{\phikq}\left[g(-\phipq,-\phikq)+g(\phipq,\phikq)\right]
\end{align*}
The angles $\theta_k$ and $\theta_{k'}$ depend on the angles $\phikq$ and $\phipq$, but are not changed by simultaneously replacing $\phikq\to-\phikq$ and $\phipq\to-\phipq$. In the matrix element, $\phikq$ appears in $s$ and $u$ in the cosine argument, which is an even function. The only change happens in $q_2\to -q_2$ in \eqref{eq:qy_def}, which results in $\qhat^{12}=\qhat^{23}=0$.

To show that $\qhat^{13}=0$ for $v_p=0$, we consider the phase shift $\phipq\to\phipq+\pi$ which, for $\cos\theta_p=0$, changes $\cos\theta_k\to-\cos\theta_k$. In our assumption \eqref{eq:distributionfunction-mirrorsymmetry}, the distribution function is symmetric around the $z=0$ plane, thus $f(k,-\cos\theta_k)=f(k,\cos\theta_k)$ and thus this only results in $q^1\to -q^1$. Thus we obtain $\qhat^{13}=0$.

\subsection{Medium screening effects}\label{sec:qhat_screening}
In Section \ref{sec:amy-screening-prescription}, we have already discussed how medium effects enter the kinetic theory description of QCD by effectively screening soft-momentum transfer in the matrix elements.

Similarly, also for the jet quenching parameter $\hat q$ screening becomes important when the Mandelstam variable $t$ becomes small $s \gg t \sim\mathcal O(m_D^2)$. This concerns the underlined terms with inverse powers of $t$ in the matrix elements listed in \tab \ref{tab:qhat_matrix_el}.
Because we enforced $p'>k'$, we could only have $|u|\ll s$  when $k\gg p$, which is highly suppressed by the fact that we are choosing $p$ to be hard and $k$ to be a medium particle.
Thus, unlike for the elastic collision term \eqref{eq:c22_first}, we only need to implement screening effects in terms with $t$ in the denominator.

Similar as in Section \ref{sec:amy-screening-prescription}, we include medium modifications by replacing
the singly underlined terms in the matrix elements in \tab \ref{tab:qhat_matrix_el} by
\begin{align}
	\label{eq:prescription_screening}
	M_0&=\frac{(s-u)^2}{t^2} 
	\to\left|G_R(P-P')_{\mu\nu}(P+P')^\mu (K+K')^\nu\right|^2 \equiv \Mscreen,
\end{align}
where $G_R$ denotes the retarded gluon propagator in the HTL approximation.
Recall that the retarded propagator is needed because we started from the decay rate \eqref{eq:decay_rate}, which includes the fully retarded amplitudes via the cutting rule \eqref{eq:cutting-rule}, thus the internal lines must be retarded propagators.

Furthermore, we will employ two different approximations, neglecting plasma instabilities as in the elastic collision kernel: First, we will use the isotropic hard thermal loop resummed gluon propagator, which we will refer to as \emph{isoHTL} screening.
In a second step, we further approximate and use the simple Debye-like screening prescription already introduced before (see \ref{sec:debye-like-screening}), but we will find that a different screening parameter $\xi_g^\perp$ is needed to account for the correct treatment of transverse momentum broadening.

In any case, the singly underlined terms in \tab\ref{tab:qhat_matrix_el} can be rewritten in terms of the unscreened (vacuum) expression $M_0$ using Eq.\eqref{eq:mandelstam_sum},
\begin{align}
	\label{eq:s2u2Overt2}
	\frac{s^2+u^2}{t^2} = \frac{1}{2}+\frac{1}{2}M_0, && \frac{su}{t^2}=\frac{1}{4}-\frac{1}{4}M_0.
\end{align}

Using the full isotropic HTL propagator (see \app \ref{app:full_htl_matrix_el} for details), the screened matrix element reads explicitly
\begin{align}
	\Mhtl=\frac{c_1^2}{A^2+B^2}+\frac{c_2^2}{C^2+D^2}-\frac{2c_1c_2(AC+BD)}{(A^2+B^2)(C^2+D^2)},\label{eq:full_HTL_finite_p_matrix_element_replacement}
\end{align}
where $A, B, C, D$ are obtained from the real and imaginary parts of the retarded HTL self-energies and are explicitly given by
\begin{subequations} \label{eq:parameters_for_full_HTL_matrix_element}
	\begin{align}
		A&=q^2+m_D^2\left(1+\frac{\omega}{2q}\ln\frac{q-\omega}{q+\omega}\right),&
		B&=-\frac{m_D^2\omega}{2q}\pi,\\
		C&=q^2-\omega^2+\frac{m_D^2}{2}\left(\frac{\omega^2}{q^2}+\left(\frac{\omega^2}{q^2}-1\right)\frac{\omega}{2q}\ln\frac{q-\omega}{q+\omega}\right),&
		D&=\frac{\pi m_D^2\omega}{4q}\left(1-\frac{\omega^2}{q^2}\right),
	\end{align}
\end{subequations}
and
\begin{subequations}\label{eq:parameters_c1c2_for_qhat_finitep}
	\begin{align}
		c_1&=(2p+\omega)(2k-\omega),&
        c_2&=4pk\sin\thetaqp\sin\thetaqk\cos\phikq.
	\end{align}
\end{subequations}
Note that for isotropic distributions, 
the last term  in \eq~\eqref{eq:full_HTL_finite_p_matrix_element_replacement} can be dropped since it is proportional to $\cos\phikq$ and will thus vanish in the angular integral needed for $\qhat$.

The doubly underlined terms in \tab\ref{tab:qhat_matrix_el} correspond to soft fermionic exchange. Here, we do not need to consider them explicitly because, as we will discuss in \se\ref{sec:pfinite}, they are subleading in $1/p$.

\begin{figure}
	\centering
	\includegraphics[width=0.5\linewidth]{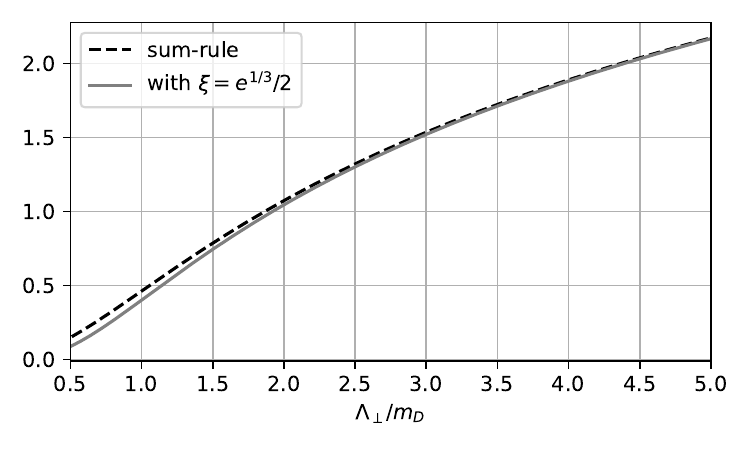}
	\caption{Shown are the HTL sum rule result on the left-hand side of \eq \eqref{eq:regularization_eq_to_solve} as a dashed curve and the values from the approximated expression on its right-hand side with the parameter $\xi$ given by \eqref{eq:xi_analytic} as a continuous line.
		The plot shows that for $\Lambda_\perp \gtrsim 4m_D$ the screening approximation of the full HTL matrix element provides accurate results. Figure taken from \cite{Boguslavski:2023waw}.}
	\label{fig:regularization}
\end{figure}

Next, we will consider the Debye-like screening prescription already mentioned in Section \ref{sec:debye-like-screening},
\begin{align}
	\Mhtl \to\Mxi=\frac{(s-u)^2}{t^2}\frac{q^4}{(q^2+(\xiscreenperp)^2m_D^2)^2}
	.
	\label{eq:isotropic_screening_approximation_xi}
\end{align}
As explained in Section \ref{sec:debye-like-screening}, this replacement can be justified when
we are not directly interested in the matrix element but in the (weighted) integral over it, as in computations of  $\qhat$ or $\Ctwotwo$.
The Debye-like screened matrix element $\Mxi$ agrees with $\Mhtl$ at large $q$, but behaves differently in the small $q$ region. It includes a constant $\xi_g^\perp$ that is fixed such that the integral over the Debye-like screened matrix element matches the result of the full isotropic HTL matrix element. For transverse momentum broadening, this integral needs to be taken in the high energy limit $p\to\infty$, be weighted with $\qperp^2$ and integrated over $\dd[2]{\vb \qperp}$ to obtain $\qhat$. 
Thus, similarly as fixing $\xiscreen$ in the elastic collision kernel using Eq.~\eqref{eq:ximatching-longitduinal-theory}, we fix $\xiscreenperp$ by requiring that
\begin{align} \label{eq:ximatching}
	\int_{0}^\infty\!\! \dd{\qperp}\qperp^3 \int_{-\infty}^\infty\!\! \frac{\dd{\omega}}{\sqrt{\qperp^2+\omega^2}}\int_0^{2\pi}\!\!\!\dd\phikq\left(\Mhtl - \Mxi\right)=0.
\end{align}
This matching is different than in the case needed for the time evolution (see Section \ref{sec:debye-like-screening}), where one is matching the longitudinal momentum transfer rather than the transverse one (replace $q_\perp^3\to\omega^2q_\perp$ in \eqref{eq:ximatching}), which leads to a different value $\xiscreen\neq\xiscreenperp$.

In the case considered here, we take both matrix elements in the limit $p\to\infty$, and additionally consider the soft limit $\omega\ll k$, $\qperp\ll k$. 
We then first integrate over $\omega$. For $\Mhtl$ this can, in the soft limit, be done analytically using a sum rule \cite{Aurenche:2002pd}, which is discussed in more detail in \app\ref{app:sum-rule}.
Then, the $\qperp$ integral is performed up to a cutoff $\Lambda_\perp$ to obtain the following condition 
\begin{align}
	&\frac{2}{3}\ln\left(1+\frac{\Lambda_\perp^2}{m_D^2}\right)\label{eq:regularization_eq_to_solve}\\
	&=4\ln\frac{\lperp}{2\,\xiscreenperp m_D}-\frac{\lperp^2}{(\xiscreenperp m_D)^2}-\frac{(\lperp^4+2\lperp^2(\xiscreenperp m_D)^2+4(\xiscreenperp m_D)^4)\ln\frac{\lperp}{\xiscreenperp m_D+\sqrt{\lperp^2+(\xiscreenperp m_D)^2}}}{(\xiscreenperp m_D)^3\sqrt{\lperp^2+(\xiscreenperp m_D)^2}}\nonumber
\end{align}
where the left-hand side stems from $\Mhtl$.
Expanding both sides of the equation for large cutoff $\Lambda_\perp\gg \xi m_D$ leads to
\begin{align}
	\xiscreenperp=\frac{e^{1/3}}{2}= 0.6978\dots \label{eq:xi_analytic}\,.
\end{align}
In \fig \ref{fig:regularization} we demonstrate the agreement of both sides of \eq \eqref{eq:regularization_eq_to_solve} as a function of the cutoff $\lperp$. For large momentum cutoffs $\Lambda_\perp \gtrsim 4m_D$, the approximation seems to work very well.
We note that the value $\xiscreenperp$ in \eq \eqref{eq:xi_analytic} entering the matrix element in $\qhat$ is slightly different from the one typically used obtained in section \ref{sec:debye-like-screening} for the elastic collision kernel $\Ctwotwo$, $\xiscreen$.

This approximation will be investigated in \se\ref{sec:qhat_special_cases} numerically by comparison with the HTL-screened results. 
There, it is found that the largest differences occur for a small cutoff $\Lambda_\perp$ or a large coupling $\lambda$. For physically motivated values $\lambda=10$ and $\Lambda_\perp=T$, the screening prescriptions differ by 30\%, showing that the choice of the screening prescription can be important for the evaluation of the jet quenching parameter $\qhat$.

\subsection{Towards the limit $p\to\infty$: NLO terms in $1/p$}
\label{sec:pfinite}

In the derivation of $\qhat$, we have considered the jet momentum $p$ to be much larger than all other momentum scales of the plasma. However, in the strict limit $p\to \infty$ the momentum diffusion coefficient $\qhat$ exhibits a logarithmic divergence, unless the exchanged momentum is limited by a cutoff. We will first, in this subsection, discuss the limit of $p$ being large, but not infinite. Then, in \se \ref{sec:pinf_formula}, we will introduce a cutoff on $\qperp$ and take $p\to \infty$.
In the limit of large $p$, only the terms $su/t^2$ and $(s^2+u^2)/t^2$ in the matrix elements (\tab\ref{tab:qhat_matrix_el}) are $\sim p^2$ and thus contribute.

For example, let us consider the screened gluonic matrix element,
\begin{align}
	\frac{\left|\mathcal M^{gg}_{gg}\right|^2}{p^2}&=16 \dA  \CA  ^2\frac{\left(2k-\omega-\sqrt{(2k-\omega)^2-q^2}\cos\phikq\right)^2}{(q^2+ \xiscreenperpsqr m_D^2)^2} \left(1+\frac{\omega}{p}+\mathcal{O}\left(\frac{1}{p^2}\right)\right).\label{eq:matrixelement_gluonic_limit}
\end{align}
Here, $k$ is a medium momentum scale (the collision integral is proportional to $f(\vb k)$) and we can thus assume that $k\ll p$ even if formally $k$ is integrated up to infinity.
Na\"ively, one might assume that $\frac{\omega}{p}=\mathcal O\left(\frac{1}{p}\right)$. However, $p$ appears in the boundaries of $\omega$, ($p>(q-\omega)/2$, see \eqref{eq:phase_space_relation_omega_q_k}), and, therefore, we need to carefully consider the term $\omega/p$. 
For positive $\omega$ we obtain $\omega < k \ll p$, such that it indeed becomes negligible. However, for negative $\omega$,
\begin{align}
	\left|\frac{\omega}{p}\right|=\frac{-\omega}{p}<\frac{p-k}{2p} = \frac{1}{2} - \frac{k}{2p},
\end{align}
which does not vanish for $p \to \infty$. 
A more careful calculation (carried out explicitly in  \app \ref{app:large-momentum-limits-qhat}) shows that the leading large $p$ contribution in \eq\eqref{eq:matrixelement_gluonic_limit}
diverges logarithmically $\sim\ln p$, whereas the $\mathcal{O}(\omega/p)$ contribution becomes constant in $p$. Thus, indeed, the leading behavior is obtained by considering the $p\gg \omega$ term(s) in the matrix element.

To summarize, for large jet energies $\Ejet =p$, the jet quenching parameter is given by
\begin{align}
	\qhat(p\gg \Teps) \simeq a_p\ln p + b_p, \label{eq:qhat_largep_behavior}
\end{align}
where $b_p$ is a (yet undetermined) coefficient depending on the precise form of the integrand.
Here, $\Teps$ is the characteristic momentum scale of plasma particles, e.g., the temperature in thermal equilibrium.
The coefficient $a_p$ for isotropic distributions is derived explicitly in  \app \ref{app:large-momentum-limits-qhat} as
\begin{align}
	\begin{split}
		a_p/\CR&=\frac{\CA g^4}{4\pi^3}\int_0^\infty \dd{k}k^2f_g(k)+\sum_{f}\frac{g^4}{8\pi^3}\int_0^\infty \dd{k} k^2 f_f(k),
	\end{split}\label{eq:qhat_ap_coefficient}
\end{align}
where $f_g$ is the gluon distribution function and the subscript $f$ in $f_f$ labels different quark species.

\subsection{Limit $p\to\infty$ with a momentum cutoff\label{sec:pinf_formula}}

\begin{figure}
    \centering
	\includegraphics[width=0.7\linewidth]{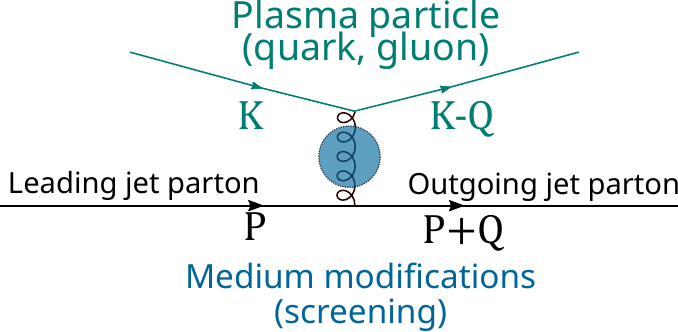}
	\caption{Feynman diagram for a $t$-channel gluon exchange process that dominates the matrix element for the jet quenching parameter $\qhat$ in the high-energy limit $p\to\infty$. In the internal gluon propagator, medium effects should be included as explained in \se\ref{sec:qhat_screening}. Figure reused from \cite{Boguslavski:2023waw}.}
	\label{fig:feynmandiag}
\end{figure}

We now introduce a cutoff $\lperp$ to restrict the maximum amount of transverse momentum transfer in the scattering, 
\begin{align}
	\qperp^2=q^2-\omega^2<\Lambda_\perp^2. \label{eq:qperp_cutoff}
\end{align} 

It is now possible to directly take the limit $p\to\infty$ in the matrix element without any potential divergences (as is shown in Appendix \ref{app:large-momentum-limits-qhat}),
which considerably simplifies the calculation. 
In this limit, there is also a reduced number of matrix elements, which are given explicitly in \tab \ref{tab:p-inf_matrix_el}. Only $t$-channel gluon exchange diagrams contribute, as depicted in \fig\ref{fig:feynmandiag}. 
Apart from that, a few changes need to be made to the formula of $\qhat$ presented in \se\ref{sec:qhat-formula-finite-p}. In Eq.~\eqref{eq:qhat_formula}, we need to change the integration limits to account for the transverse momentum cutoff condition \eqref{eq:qperp_cutoff},
\begin{align}
    \begin{split}
	\qhat^{ij}&=\frac{1}{2^9\pi^5\nu_a}\sum_{bcd}\int_0^{2\pi}\dd{\phipq}\int_0^{2\pi}\dd{\phikq}\\
    &\qquad\times\left(\int_0^{\lperp} \dd{q} \int^q_{-q}\!\!\!\dd{\omega}+\int_{\lperp}^\infty\dd{q}\left[\int_{-q}^{-\sqrt{q^2-\lperp^2}}\dd{\omega}+\int_{\sqrt{q^2-\lperp^2}^q}\dd{\omega}\right]\right)\\
    &\qquad\times\int_{\frac{q+\omega}{2}}^{\infty}\!\!\!\dd{k} 
    q^iq^j\lim_{p\to\infty}\frac{|\mathcal M^{ab}_{cd}|^2}{p^2}f_b(\vb k)\left(1\pm f_d(\vb k')\right)\,. 	\label{eq:qhat_formula_pinf}
    \end{split}
\end{align}
As before, the upper and lower signs in the term $(1 \pm f_d)$ denote bosonic particles (gluons) and fermionic particles (quarks), respectively, see Eq.~\eqref{eq:boseenhancement-fermi-blocking}.

In this limit, one needs to replace \eqref{eq:cos_thetapq} by $\cos\thetaqp = \omega / q$ (see \app\ref{app:qhat_large_p}) and the few nonvanishing matrix elements
for $\lim_{p\to\infty}\frac{|\mathcal M|^2}{p^2}$ are tabulated in \tab \ref{tab:p-inf_matrix_el}. In that limit, we do not need the explicit expressions \eqref{eq:mandelstam_s_qhat} and \eqref{eq:mandelstam_u_qhat} for $s,u$ in terms of our phase space integration variables. 
From the matrix elements in \tab\ref{tab:p-inf_matrix_el} and \eqref{eq:qhat_formula}, one naturally finds Casimir scaling
\begin{align}
	\frac{\qhat^{\mathrm{gluon}}}{\CA}=\frac{\qhat^{\mathrm{quark}}}{\CF},\label{eq:Casimir_scaling}
\end{align}
i.e., the jet quenching parameter for a quark and gluon jet are equivalent up to color factors.

\begin{table}
	\begin{tabular}{M{0.4\linewidth} M{0.6\linewidth}}\toprule
		$ab\leftrightarrow cd$ & $\lim_{p\to\infty}|\mathcal M^{ab}_{cd}|^2/(p^2g^4)$ \\ [10pt]
		\hline 
		$q_1q_i\leftrightarrow q_1q_i,$ & \multirow{4}{*}{$4\frac{\dF^2\CF ^2}{\dA }\tildeMscreen$}\\ $\bar q_1q_i\leftrightarrow \bar q_1 q_i,$ &\\ $q_1\bar q_i\leftrightarrow q_1\bar q_i,$ &\\ $\bar q_1\bar q_i\leftrightarrow \bar q_1\bar q_i$ &  \\ [10pt]
		$q_1g\leftrightarrow q_1g,$ & \multirow{2}{*}{$4 \dF\CF \CA   \tildeMscreen$}\\ $\bar q_1 g\leftrightarrow \bar q_1 g$ & \\ [10pt]
		$gg\leftrightarrow gg$ & $4\dA \CA^2 \tildeMscreen$\\ \bottomrule
	\end{tabular}
	\caption{The matrix elements for $\hat q$ as in \tab \ref{tab:qhat_matrix_el} in the limit $p\to\infty$. Here $\tildeMscreen=\lim_{p\to\infty}\Mscreen/p^2$ denotes the appropriate screening terms $\tildeMhtl$ or $ \tildeMxi$ as explained in \se\ref{sec:qhat_screening} and \app\ref{app:full_htl_matrix_el}, and quark flavors are labeled by the index $i$. 
	Table taken from \cite{Boguslavski:2023waw}.}
	\label{tab:p-inf_matrix_el}
\end{table}

The screened matrix element  $\tildeMscreen$ in \tab\ref{tab:p-inf_matrix_el} is implemented as detailed in \se\ref{sec:qhat_screening}. In the $p\to \infty$ limit, the isotropic HTL matrix element from Eq.~\eqref{eq:full_HTL_finite_p_matrix_element_replacement}  becomes
\begin{align}
	\tildeMhtl=\frac{\tilde c_1^2}{A^2+B^2}-\frac{\tilde c_2^2}{C^2+D^2}+\frac{2\tilde c_1\tilde c_2(AC+BD)}{(A^2+B^2)(C^2+D^2)},\label{eq:full_htl_matrix_element}
\end{align}
with the parameters $A,\, B,\, C,$ and $D$ given by \eqref{eq:parameters_for_full_HTL_matrix_element} 
and
\begin{subequations} \label{eq:c_parameters_for_pinf_matrix_element_HTL}
	\begin{align}
		\tilde c_1 &= \lim_{p\to\infty} \frac{c_1}{p}=2(2k-\omega),\label{eq:ctilde1}\\
		\tilde c_2 &= \lim_{p\to\infty}\frac{c_2}{p}=4k\sin\thetaqp\sin\thetaqk\cos\phikq. \label{eq:ctilde2}
	\end{align}
\end{subequations}
Again, for isotropic systems, we do not need to include the last term in \eqref{eq:full_htl_matrix_element}, since it vanishes in the angular integral when calculating $\qhat^{ij}$.

For the \emph{Debye-like} screening approximation, we obtain (c.f.~\eqref{eq:isotropic_screening_approximation_xi})
\begin{align}
	\tildeMxi = 4\frac{\left(2k-\omega -\sqrt{(2k-\omega)^2-q^2}\cos\phikq\right)^2}{(q^2+\xiscreenperpsqr m_D^2)^2},\label{eq:approximated_matrix_element}
\end{align}
with $\xiscreenperp=e^{1/3}/2$ as before.

\subsection{Limiting behavior for large cutoff}
\label{sec:limiting_behavior_large_cutoff}

The jet quenching parameter $\qhat$ exhibits a logarithmic behavior when the cutoff $\Lambda_\perp$  exceeds the typical hard momenta $\Teps$ 
of the plasma constituents.
\begin{equation}
	\qhat(\Lambda_\perp \gg \Teps) \simeq a_{\lperp}\ln\Lambda_\perp + b_{\lperp},\label{eq:qhat_behavior_large_cutoff}
\end{equation}
with (see Appendix \ref{app:large-momentum-limits-qhat})
\begin{align}
    a_{\lperp}/\CR=\frac{g^4}{\pi}\int\frac{\dd[3]{\vb k}}{(2\pi)^3}\left(\NC f_g(\vb k)+\frac{1}{2}\sum_{f}f_f(\vb k)\right).
\end{align}
For isotropic distributions, it reads
\begin{align}
	\begin{split}
		a_{\lperp}/\CR&\simeq\frac{\CA g^4}{2\pi^3}\int_0^\infty \dd{k}k^2f_g(k)+\sum_{f}\frac{g^4}{4\pi^3}\int_0^\infty \dd{k} k^2 f_f(k).
	\end{split}\label{eq:alperp}
\end{align}
This is the same logarithmic behavior as in Eq.~\eqref{eq:qhat_ap_coefficient}, keeping in mind that now the phase space is limited by $\Lambda_\perp^2$ rather than $p$, and thus $\ln p$ gets replaced by $2\ln \Lambda_\perp$.

For thermal equilibrium, this yields 
\begin{align}
	&\qhat^{\mathrm{therm}}(\Lambda_\perp \gg T) \simeq \frac{\CR}{\pi^3} g^4\zeta(3) T^3\left(\NC + \frac{3}{4}\nf\right)\ln\Lambda_\perp + \mathrm{const},
\end{align}
which is \eq \eqref{eq:qhat_a_coefficient_equilibrium} in \app \ref{sec:qhat_behavior_large_cutoff} and agrees 
with Ref.~\cite{Arnold:2008vd}, as we will later discuss around \eq \eqref{eq:qhat_hard_arnold}.

\subsection{Interpreting the momentum cutoff}
\label{sec:momentum_cutoffs}
A peculiar feature of the jet quenching parameter $\qhat$ is its dependence on a transverse momentum cutoff $\lperp$. Let us discuss here briefly where this dependence comes from and how to interpret this cutoff in physical terms. 

In a kinetic picture, the reason for the cutoff is taking the jet momentum $p\to\infty$. In this case, the jet particle can inject an unrestricted amount of transverse momentum into the collision, which leads to a logarithmic divergence. It has to be regulated by introducing a cutoff $\lperp$, which restricts transverse momentum transfer $\qperp<\lperp$. Practically all analytic calculations that rely on quasiparticles or hard-thermal loop frameworks, but even with different interaction potentials, need to employ this cutoff.

A simple way of setting the cutoff is to use the relation between the  
coefficient $a_{\lperp}$ for large cutoff $\Lambda_\perp$ and the coefficient $a_p$ for large (finite) jet energy $p$ (see Eqs.~\eqref{eq:qhat_ap_coefficient} and \eqref{eq:alperp}). Requiring that the dynamics of jet quenching calculated with a cutoff in the $p\to\infty$ approximation has the same leading logarithmic behavior as a kinematically more accurate one with a finite $p$, we should choose the cutoff such that
\begin{align}
	\Lambda_\perp^{\mathrm{kin}}\sim \sqrt{pT}, \label{eq:kinematic_cutoff}
\end{align}
where $p$ is the energy of the jet parton and $T$ is an additional dimensionful scale (e.g., the temperature in equilibrium).
This kinematic cutoff $\Lambda_\perp^{\mathrm{kin}}$ is widely used in the literature (see, e.g., \cite{Qin:2009gw, JET:2013cls, Xu:2014ica, He:2015pra, Cao:2021rpv, JETSCAPE:2021ehl, JETSCAPE:2022jer, Mehtar-Tani:2022zwf}).

While this is a straightforward result of our definition for $\qhat$ in \eq \eqref{eq:intro-qhat-definition}, it should be kept in mind that this parameter only encodes the momentum diffusion due to elastic $2\leftrightarrow 2$ scattering processes, and competing inelastic effects like splittings or gluon emissions are neglected. 
For radiative energy loss calculations, one can restrict the cutoff by considering the rate of momentum exchange processes and comparing it with the `life-time' of the leading parton under consideration. During an LPM splitting process (see also the discussion in section \ref{sec:lpm}), this corresponds to the \emph{formation time} $\tform$, Eq.~\eqref{eq:formation-time}.
Therefore, we are interested in the accumulated transverse momentum until a splitting occurs. To calculate radiative energy loss of the leading parton in the harmonic approximation, the jet quenching parameter $\qhat$ naturally appears in the expansion of the interaction potential in position space, Eq.~\eqref{eq:Cx_harmonic-approximation-first}.
In this approximation, it is sufficient to use a momentum cutoff $\lperp$ of the order of the typical total momentum transfer $Q_\perp$ during the formation time \cite{Arnold:2008iy}. By definition, it is given by $Q_\perp^2\sim \qhat \tform$, where for a small medium with length $L<\tform$ one should replace $\tform$ by $L$. The formation time of the splitting $p\to p_1+p_2$ can be estimated as $(\tform)^2\sim E_i/\qhat$, with $E_i$ being the energy of the emitted gluon, see Eq.~\eqref{eq:formation-time}. It has been argued \cite{Arnold:2008vd, Arnold:2008zu} that energy loss is dominated by processes in which both daughters share a similar energy fraction $p_1\sim p_2\sim p$, which enables us to use the leading-parton energy in the formation time estimate. With the parametric relation $\qhat\sim g^4 T^3$, we obtain for a large medium $L>\tform$ the expression
\begin{align}
	\lperp^{\mathrm{LPM}}\sim g\,(pT^3)^{1/4}. \label{eq:LPM_cutoff}
\end{align}
In order to present our results in a form that can be applied to different pictures of energy loss, we generally give our results as functions of $\lperp$.
In section \ref{sec:qhat_bottomup-cutoff-dependence}, we will use these cutoffs as a model to study the jet quenching parameter during the initial stages in heavy-ion collisions.

In Chapter \ref{sec:smalldistance-dipolecrosssection}, we will discuss how to obtain the small distance behavior of the dipole cross section $C(|\vb x|)$ in more detail, in particular without the need for a cutoff.

\section{Analytical results for the jet quenching parameter in special cases\label{sec:qhat_special_cases}}

Let us now calculate the jet quenching parameter $\qhat$ for some special cases. 
In Section \ref{sec:qhat-thermal}, we first review the derivation of $\qhat$ for quark-gluon plasmas in thermal equilibrium \cite{Aurenche:2002pd, Arnold:2008vd} and compare the results with numerical evaluations of Eq.~\eqref{eq:qhat_formula_pinf}. 
Additionally, an interpolation formula is provided that reproduces the numerically obtained values of the quenching parameter in thermal equilibrium $\qhattherm$ for different couplings and momentum cutoffs.

In Section \ref{sec:toy_models}, we then consider toy models for the bottom-up thermalization process in heavy-ion collisions (see Section \ref{sec:thermalization-expanding-systems}). First, an effectively two-dimensional distribution is studied to model the large momentum-space anisotropy encountered in the initial stages in heavy-ion collisions and then the thermal results of Section \ref{sec:qhat-thermal} are generalized to a scaled thermal distribution to model over- and under-occupied systems that are typically encountered in the bottom-up thermalization picture.

We also study the different contributions to $\qhat$ that are linear or quadratic in the distribution function by splitting the jet quenching parameter $\qhat$ into its individual components,
\begin{align}
    \qhat = \qhatf+\qhatff.\label{eq:qhat_f_ff_splitting}
\end{align}
Similarly as in Ref.~\cite{Kurkela:2021ctp}, we refer to $\qhatf$ as the \emph{classical} and $\qhatff$ as the \emph{Bose-enhanced} part of $\qhat$.\footnote{
This Bose-enhanced term can also be considered to be a \emph{classical field} contribution because it is dominant in 
highly occupied systems $f\gg 1$ that can be studied numerically using classical-statistical simulations. This can be seen in the 
limit of $\lambda\to 0$ with $\lambda f$ held constant, in which only $\qhatff$ survives.
}

\subsection{Thermal distribution\label{sec:qhat-thermal}}

The equilibrium form of the particle distributions is given by Eq.~\eqref{eq:thermal-distributionfunctions},
\begin{align}
f_{\pm}(k; T)=\frac{1}{\exp(k/T)\mp1},\label{eq:thermal_bose_fermi_combined}
\end{align}
where $T$ is the plasma temperature. The upper signs $f_+$ denote the Bose-Einstein distribution and $f_-$ is the Fermi-Dirac distribution.

In thermal equilibrium, $\qhat$ has already been calculated for the limiting cases of small and large cutoffs $\lperp$ in \cite{Arnold:2008vd, Caron-Huot:2008zna}, which is briefly summarized here.
In \se\ref{sec:scaled-thermal-distribution}, we will generalize this derivation to the case of a scaled thermal distribution, which is obtained by rescaling a thermal distribution.

For evaluating $\qhat$, we work in the $p\to \infty$ limit with a transverse momentum cutoff $\lperp$, as discussed in \se\ref{sec:pinf_formula}. Since the distribution function is spherically symmetric, one has $\qhat^{11}=\qhat^{22}$ and we can restrict the discussion to the sum $\qhat=\qhat^{11}+\qhat^{22}$. Our starting point is Eq.~\eqref{eq:qhat_formula_pinf}, where we integrate over the modulus of $\vec q=(\vec q_\perp, q^3)$. For $p\to\infty$, we obtain $q^3=\omega$ and thus
\begin{align}
q^2=q_\perp^2+\omega^2.
\end{align}

It will be useful to change the integration variables from $(q,\phi_{pq},\omega)$ to $(q^1,q^2,\omega)$, which yields a factor $q$ from the Jacobian,
\begin{align}
\begin{split}
&\int_0^{2\pi}\dd{\phipq}\int_0^\infty\dd{q}\int_{-q}^q\dd{\omega} \Theta(\Lambda_\perp^2 + \omega^2- q^2)=\int_{\qperp <\Lambda_\perp}\dd[2]{\vec q_\perp}\int_{-\infty}^\infty\frac{\dd{\omega}}{\sqrt{\qperp^2+\omega^2}}. \label{eq:coordinate_trafo_differentials_to_qperp}
\end{split}
\end{align}

In the limit $p\to\infty$, the matrix elements in \tab\ref{tab:p-inf_matrix_el} do not allow for identity-changing processes, which means that the leading parton $a$ and the outgoing parton $c$ are of the same type, $a=c$, and similarly $b=d$.
Therefore, we can scale out the Casimir factor of the jet $\CR$, and the prefactors in front of $\tildeMscreen$ in \tab\ref{tab:p-inf_matrix_el} neatly combine with $1/\nu_a$ for the degrees of freedom of the jet particle to 
\begin{subequations}
\label{eq:XiPlMin}
\begin{eqnarray}
    \Xi_+ \,& &= 2\NC, \\
    \Xi_- \,&= 4\nf\frac{\dF\CF}{\dA} &= 2\nf,
\end{eqnarray}
\end{subequations}
for scattering off a gluon and off a quark/anti-quark, respectively, which leads to Casimir scaling (see also \eq \eqref{eq:Casimir_scaling}).

There are two limiting cases in which an analytic result for $\qhat$ can be obtained: for small and large momentum cutoffs. We discuss both cases in the following.

\subsubsection{Small momentum cutoff}

For small momentum cutoffs $\lperp \ll T$, the expression for $\qhat$ in \eq \eqref{eq:qhat_formula_pinf} with \eqref{eq:coordinate_trafo_differentials_to_qperp} and the prefactors \eqref{eq:XiPlMin} becomes
\begin{align}
\label{eq:qhat_factorized}
&\hat q(\Lambda_\perp)=C_R\sum_{\pm}\Xi_\pm\frac{g^4}{2^9\pi^5}\!
\int_0^\infty\!\!\dd{k}\,f_\pm(k)\left(1\pm f_\pm(k)\right) \int_0^{\Lambda_\perp}\!\!\dd[2]{\vb q_\perp}\qperp^2\int_0^{2\pi}\!\dd{\phikq}\int_{-\infty}^\infty\!\!\frac{\dd{\omega}\tildeMhtl}{\sqrt{q_\perp^2+\omega^2}}.
\end{align}
We have extended the lower boundary\footnote{
The largest error of this approximation comes from the $f_+^2$ term. It can by estimated by $\int_0^{\frac{q+\omega}{2}}k^2f_+^2 < \frac{q+\omega}{2}\lim_{k\to 0}\left(k^2f_+^2\right)$, where the $k^2$ factor stems from $\tildeMhtl$ and we approximated the integral by the maximum value of the integrand at $k=0$. This yields the error estimate $\frac{T^2(q+\omega)}{2}$, which for $\qperp < \lperp \ll T$ is much smaller than the leading-order contribution $\int_0^{\infty}k^2f_+^2 = 2T^3(\zeta(2)-\zeta(3))$ (see \eq\eqref{eq:thermal_distributionfunctions_integrals})
for $q,\,\omega \ll T$.}
of the $k$-integral to $0$ and approximated $f(k-\omega)\approx f(k)$. This is appropriate because large values of $\omega$ are suppressed by the matrix element $\tildeMhtl$, as can be seen from Eq.~\eqref{eq:matrixelement_gluonic_limit}.

The last two integrals
can be evaluated analytically using a sum rule \cite{Aurenche:2002pd} as discussed in \app\ref{app:sum-rule},
\begin{align}
\begin{split}
\hat q(\Lambda_\perp)&=\frac{C_Rg^4}{(2\pi)^3}2\int_{\qperp < \Lambda_\perp}\frac{\dd[2]{\vb\qperp}}{2\pi}\qperp^2\frac{1}{\qperp^2(\qperp^2+m_D^2)}
\sum_{\pm}\Xi_{\pm}\int_0^\infty\dd{k}k^2\,f_\pm(k)\left(1\pm f_\pm(k)\right).
\end{split}\label{eq:qhat_soft_distribution_function_split_off}
\end{align}
Note that until now we have only assumed spherical symmetry, but have not used a specific form for the distribution function $f(k)$.
The thermal form of $\qhat$ for a small cutoff is then obtained by performing the integrals over the distribution function,
\begin{subequations}\label{eq:thermal_distributionfunctions_integrals}
    \begin{align}
        \int_0^\infty\dd{k}k^2f_\pm(k)&=2T^3\zeta_\pm(3),\\
        \int_0^\infty\dd{k}k^2 \left(f_\pm(k)\right)^2&=\pm 2T^3(\zeta_\pm(2)-\zeta_\pm(3)),
    \end{align}
\end{subequations}
where $\zeta_+(s)=\zeta(s)$ is the Riemann Zeta function and $\zeta_-(s)=(1-2^{1-s})\zeta(s)$ denotes its fermionic counterpart as in \re \cite{Arnold:2008vd}.
Using $\zeta(2)=\pi^2/6$, we obtain
\begin{align}
    \qhat(\lperp)=\int_{\qperp<\Lambda_\perp}\dd[2]{\vb \qperp} \qperp^2 \frac{g^2 C_R T }{(2\pi)^2}\frac{m_D^2}{ \qperp^2(\qperp^2+m_D^2)},\label{eq:qhat_from_collision_kernel_explicit}
\end{align}
from which we can read off the elastic scattering rate $\frac{\dd\Gammael}{\dd[2]{\qperp}}$.
This leads us to the thermal form of $\qhat$ for a small cutoff,\footnote{This form is actually valid in general for any isotropic distribution $f(k)$ with the replacement of 
$T\to T_\ast$ (see Eq.~\eqref{eq:tstar-definition})
and the more generally evaluated Debye mass $m_D$ as in \eq \eqref{eq:debyemass-general}.}
\begin{align}
    \qhattherm(\Lambda_\perp\ll T)=\frac{g^2}{4\pi} \CR T m_D^2\ln\left(1+\frac{\Lambda_\perp^2}{m_D^2}\right).\label{eq:qhat_thermal_equlibrium_soft}
\end{align}

For a thermal system, the terms containing $\zeta_\pm(3)$ cancel for the total $\qhat$, but are important if one considers the \emph{Bose-enhanced} part separately, as in \eqref{eq:qhat_f_ff_splitting}.
Splitting off the \emph{Bose-enhanced} term as in \eqref{eq:qhat_f_ff_splitting}, we obtain
\begin{subequations}
\begin{align}
	\qhatftherm(\Lambda_\perp\ll T)&= \zeta(3)\left(12\NC +9\nf\right)C_L\label{eq:qhat_thermal_f},\\
	\qhatfftherm(\Lambda_\perp\ll T)&=\left[2\NC(\pi^2-6\zeta(3))\right.  + \left.\nf(\pi^2-9\zeta(3))\right]C_L,\label{eq:qhat_thermal_ff}
\end{align}
\end{subequations}
with $C_L=\frac{g^4T^3 \CR}{24\pi^3}\ln\left(1+\frac{\Lambda_\perp^2}{m_D^2}\right)$ and the thermal Debye mass given by \eq \eqref{eq:equilibriumform-debyemass-tstar-meffs}.

\subsubsection{Large momentum cutoff\label{sec:thermal_qhat_large_momentum_cutoff}}

For large momentum cutoffs, the jet quenching parameter has been calculated in Ref.~\cite{Arnold:2008vd} for a thermal system. To later generalize this to a scaled thermal distribution, we briefly review the derivation here. It relies on constructing an interpolating formula for the elastic scattering rate
\begin{align}
    \frac{C(\qperp)}{(2\pi)^2}=\frac{\dd{\Gammael}}{\dd[2]{\qperp}}\simeq \CR\frac{g^4T^3 F(q_\perp/T)}{\qperp^2(\qperp^2+m_D^2)},\label{eq:def_F_function}
\end{align}
where the function $F(\qperp/T)$ interpolates between the known limits of this quantity (we will discuss both limits later in Chapter \ref{sec:collkern}, see Eq.~\eqref{eq:limits_general}) and can be calculated in the approximation $q\gg m_D$. It is then split into gluonic ($I_+$) and fermionic ($I_-$) contributions,\footnote{In principle,
we could take Eq.~\eqref{eq:qhat_factorized} instead and relax the assumption of small momentum transfer, i.e., keep $f_{\pm}(k)(1\pm f_\pm(k-\omega))$. However, the strategy employed in \re\cite{Arnold:2008vd} (scaling out this factor $F(\qperp/T)$ in \eqref{eq:def_F_function}) allows us to evaluate the expression analytically in the large $\qperp$ limit, where the matrix element does not need to be screened, and we can use the simpler form $su/t^2$ instead.}
\begin{align}
    F(\qperp/T)=\frac{1}{\pi^2}\left(\Xi_+ I_+(\qperp/T)+\Xi_- I_-(\qperp/T)\right).    
\end{align}
Following the notation in \re \cite{Arnold:2008vd}, we write these contributions to the collision kernel in the limit $p\to\infty$ and $\qperp\gg m_D$ as
\begin{align}
    \label{eq:Ipm}
    I_\pm\left(\frac{\qperp}{T}\right) =& \frac{\pi^2}{T^3}\int\frac{\dd{q_z}}{2\pi}\int\frac{\dd[3]{\vb k}}{(2\pi)^3} 2\pi\delta( q_z+|\vb k-{\vb q}|-k) \frac{(k-k_z)^2}{k|\vb k-{\vb{q}}|}f_\pm(\vb k)\left[1\pm f_\pm(\vb k - {\vb q})\right].
\end{align}
This formula follows directly from the t-channel matrix element in \tab\ref{tab:qhat_matrix_el}, i.e., $su/t^2$, with $t^2=\qperp^4$ being scaled out in \eqref{eq:def_F_function} and $s\simeq -u=2p(k-k_z)$.

\begin{figure}
    \centerline{
        \includegraphics[width=0.49\linewidth]{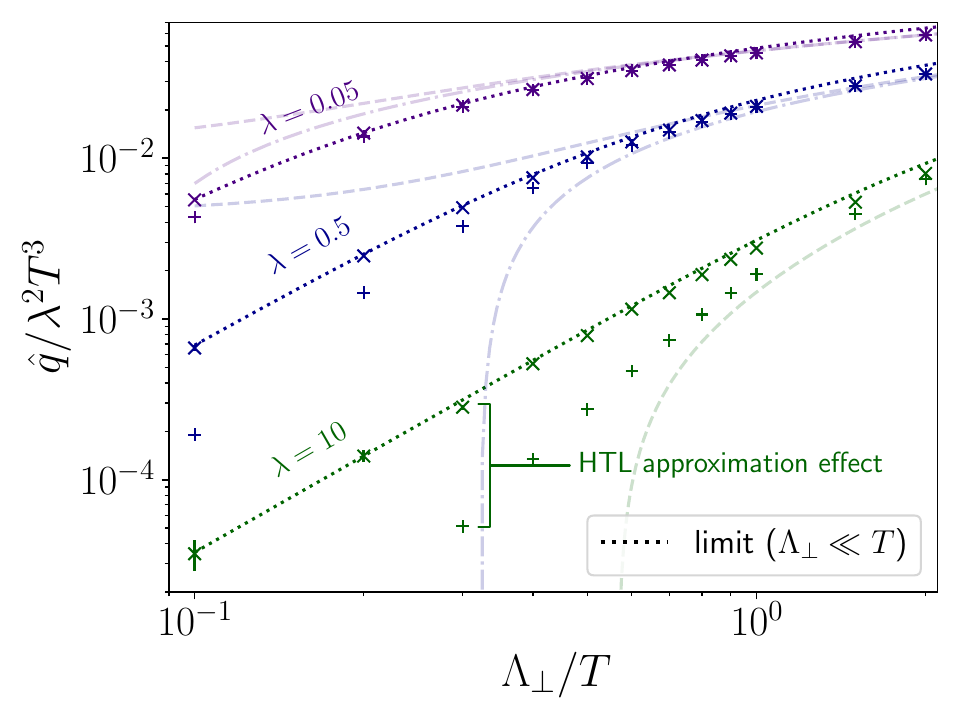}
    	\includegraphics[width=0.49\linewidth]{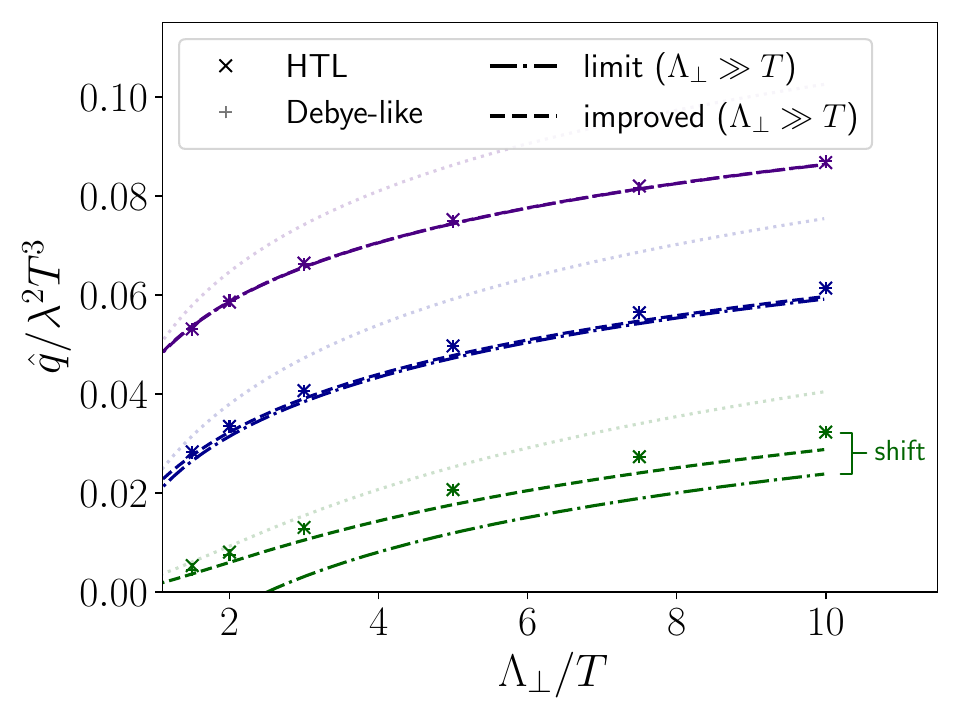}
    }
    \caption{The coefficient $\qhat$ for a quark jet in thermal equilibrium for different 't Hooft couplings $\lambda=g^2\NC$ and transverse momentum cutoffs $\Lambda_\perp$, with {\em (left:)}  small $\Lambda_\perp \lesssim T$ (logarithmically scaled axis), and {\em (right:)} large $\Lambda_\perp \gtrsim T$. The dotted curve labeled `limit ($\Lambda_\perp \ll T$)' shows the analytical limiting expression of \eqref{eq:qhat_thermal_equlibrium_soft}, the dash-dotted curve `limit ($\Lambda_\perp\gg T$)' illustrates \eqref{eq:qhat_hard_arnold}, and `improved ($\Lambda_\perp\gg T)$' denotes \eqref{eq:qhat_hard_arnold-improved}. The small `$+$'-symbols show the numerical results with the Debye-like screened matrix element \eqref{eq:approximated_matrix_element}, whereas the `$\times$'-symbols show the results with the full HTL matrix element \eqref{eq:full_htl_matrix_element}. The curves that are not valid in the respective limit are 
    displayed with lighter color
    but are still shown because \eqref{eq:qhat_thermal_equlibrium_soft} is often also used at large $\Lambda_\perp$ as an approximation. Figures adapted from~\cite{Boguslavski:2023waw}.
    }
	\label{fig:thermal_equilibrium}
\end{figure}

As in \eqref{eq:qhat_f_ff_splitting} we can identify the contributions coming from the $f$ and $f^2$ parts via
\begin{align}
    I_\pm(\qperp/T)=I_\pm^{\mathrm{f}}+I_\pm^{\mathrm{ff}}(\qperp/T),
\end{align}
where $I_\pm^{\mathrm{f}}$ will turn out to be a constant.
To evaluate them, the thermal functions are written as
\begin{align}
    f_\pm(p)=\sum_{m=1}^\infty\left(\pm1\right)^{m-1}e^{-m p/T}.
\end{align}
This can then be inserted into \eq \eqref{eq:Ipm} to rewrite the equation as a double sum,
\begin{align}
    I_\pm^{\mathrm{f}}\left(\frac{\qperp}{T}\right)&=\sum_{m=1}^\infty (\pm 1)^{m-1}
I_{m0}(\qperp/T),\\
    I_\pm^{\mathrm{ff}}\left(\frac{\qperp}{T}\right)&=\sum_{m=1}^\infty\sum_{n=1}^\infty(\pm1)^{m+n-1}I_{mn}(\qperp/T),
\end{align}
with
\begin{align}
    \label{eq:I_{mn}}
    I_{mn}\left(\frac{\qperp}{T}\right) =& \frac{\pi^2}{T^3}\int\frac{\dd{q_z}}{2\pi}\int\frac{\dd[3]{\vb k}}{(2\pi)^3}\, 2\pi\delta( q_z+|\vb k-{\vb{q}}|-k)\frac{(k-k_z)^2}{k|\vb k-{\vb{q}}|}e^{-m k/T}e^{-n|\vb k-\vb q|/T}\,.
\end{align}
In \cite{Arnold:2008vd}, $I_\pm$ was split in a similar way isolating the $n=0$ term $I_\pm(\infty)$, which is exactly the constant $I_\pm^{\mathrm{f}} = I_\pm(\infty)=\zeta_\pm(3)$. This is a consequence of 
the fact that for large momentum transfer only the $f$ part contributes, as discussed in  \se \ref{sec:limiting_behavior_large_cutoff}.
In Ref.~\cite{Arnold:2008vd}, the integrals in \eqref{eq:I_{mn}} are evaluated analytically,
\begin{align}
    I_{mn}\left(\frac{\qperp}{T}\right)=\frac{mn}{2(m+n)^3}\left(\frac{\qperp}{T}\right)^3K_2\left(\frac{q_\perp}{T}\sqrt{mn}\right),
\end{align}
where $K_\nu(z)$ is the modified Bessel function of the second kind. With that, we can write the collision kernel in thermal equilibrium as the infinite double sum
\begin{align}
    C(\qperp)=\frac{2\CR g^4 T^3}{\qperp^2(\qperp^2+m_D^2)\pi^2}\frac{q_\perp^2}{T^2}\sum_{m=1}^\infty\sum_{n=0}^\infty\left(\NC+\nf (-1)^{m+n-1}\right)\frac{mn}{2(m+n)^3}K_2\left(\frac{q_\perp}{T}\sqrt{mn}\right) .\label{eq:collision_kernel_analytic_thermal}
\end{align}

Performing the remaining integrals over $\vb\qperp$ as in Ref.~\cite{Arnold:2008vd} leads to the following expression for the jet quenching parameter $\qhat$ for large cutoffs $\Lambda_\perp\gg T$,
\begin{subequations}
\label{eq:qhat_thermal_hard_analytic}
\begin{align}
        \qhattherm(\Lambda_\perp\gg T)&= \CR \frac{g^4T^3}{\pi^2}\sum_\pm\Xi_\pm \mathcal I_\pm(\Lambda_\perp),\label{eq:qhat_hard_arnold}\\
	\mathcal I_\pm(\Lambda_\perp)&=\frac{\zeta_\pm(3)}{2\pi}\ln\left(\frac{\Lambda_\perp}{m_D}\right)+\Delta\mathcal I_\pm,\label{eq:qhat_thermal_hard_analytic_Ipm}\\
    \Delta\mathcal I_\pm &= \frac{\zeta_\pm(2)-\zeta_\pm(3)}{2\pi} \label{eq:qhat_thermal_hard_analytic_DeltaI}\left[\ln\left(\frac{T}{m_D}\right)+\frac{1}{2}-\gamma_E+\ln 2\right] -\frac{\sigma_\pm}{2\pi},\\
	\sigma_+&=0.386043817389949,\\
	\sigma_-&=0.011216764589789,
\end{align}
\end{subequations}
where $\gamma_E$ is the Euler-Mascheroni constant, and $\sigma_\pm=\sum_{k=1}^\infty\frac{(\pm 1)^{k-1}}{k^3}\ln[(k-1)!]$.

This formula \eqref{eq:qhat_thermal_hard_analytic}, as opposed to the one for small cutoffs \eqref{eq:qhat_thermal_equlibrium_soft}, has the (unphysical) feature that the logarithm $\ln T/m_D$ 
becomes negative for $m_D \geq T$.
Of course, in perturbation theory at weak couplings, one has $m_D\sim gT\ll T$.
However, to get an analytical expression that is well-behaved also for larger couplings, we add a constant to the argument of the logarithm, which still preserves the leading order accuracy at weak coupling.
To be explicit, we replace $2\ln x \to \ln(1+x^2)$ in both logarithms, and we will denote the resulting `improved' analytic expressions for $\qhat$ by $\qhatimproved$. Although the replacement does not change the result at leading order, we will see that this choice of regularization significantly improves the agreement with numerical evaluations of \eqref{eq:qhat_formula}, as we will discuss in the next subsection.
Moreover, the Bose-enhanced part $\qhatff$ of \eqref{eq:qhat_f_ff_splitting} is solely due to $\Delta\mathcal I_\pm$ in \eqref{eq:qhat_thermal_hard_analytic_Ipm}. With these replacements in the logarithm, the contribution $\qhatf$ has the same form as for small momentum cutoffs \eqref{eq:qhat_thermal_f}, $\qhatf(\Lambda_\perp\ll T)=\qhatf(\Lambda_\perp\gg T)$.

With this procedure, the improved version of Eq.~\eqref{eq:qhat_hard_arnold} becomes 
\begin{subequations}
\begin{align}
		\qhattherm_{\mathrm{im}}(\Lambda_\perp\gg T)&=
		\qhatftherm(\Lambda_\perp\ll T) + \qhatffthermimproved\label{eq:qhat_hard_arnold-improved}
\end{align}
with
\begin{align}
{\qhatffthermimproved}&={\CR g^4T^3}\sum_\pm\Xi_\pm \left\lbrace
 \frac{\zeta_\pm(2)-\zeta_\pm(3)}{4\pi^3}\right.  \left.\left[\ln\left(1+\frac{T^2}{m_D^2}\right) +1-2\gamma_E+2\ln 2\right] -\frac{\sigma_\pm}{2\pi^3}\right\rbrace . 
\end{align}
\end{subequations}

\subsubsection{Comparison with numerical results\label{sec:qhat_thermal_numerical}}

Let us now compare the analytical small and large cutoff limits of $\qhat$ given by \eqref{eq:qhat_thermal_equlibrium_soft} and \eqref{eq:qhat_hard_arnold} or the improved version \eqref{eq:qhat_hard_arnold-improved} to a numerical evaluation of $\qhat$ using \eqref{eq:qhat_formula_pinf}. For simplicity we consider a purely gluonic plasma, i.e., $\nf = 0$.
In particular, we want to study how well these analytic formulae 
describe the full numerical evaluation of the $\qhat$ integral, although being only valid for asymptotic regions of the cutoff $\Lambda_\perp$. We also want to compare the expressions using the full HTL screened matrix element (isoHTL) \eqref{eq:full_htl_matrix_element} with the simpler Debye-like screened matrix element \eqref{eq:approximated_matrix_element} and study the impact of different screening approximations on the jet quenching parameter $\hat q$.

\fig \ref{fig:thermal_equilibrium} shows $\qhat$ for various momentum cutoffs $\Lambda_\perp$ and different 't Hooft couplings $\lambda = g^2 \NC$. 
The prefactor $\lambda^2T^3$ is scaled out in the plots, leaving a nontrivial coupling dependence that enters via the Debye mass $m_D$ in the logarithms originating from the matrix element.
The curves show the analytic expressions for small cutoffs (dotted, \eq \eqref{eq:qhat_thermal_equlibrium_soft}), large cutoffs (dash-dotted, \eq \eqref{eq:qhat_hard_arnold}) and the improved large-cutoff version (dashed, \eqref{eq:qhat_hard_arnold-improved}), while the numerical evaluation of $\qhat$ is depicted by crosses for the HTL matrix elements \eqref{eq:full_htl_matrix_element} and plus signs for the Debye-like screened ones \eqref{eq:approximated_matrix_element}. 
In the left panel of \fig \ref{fig:thermal_equilibrium}, we observe that the small-cutoff form of $\qhat$ 
accurately agrees with the numerical evaluation using the full HTL matrix element 
in the corresponding region $\Lambda_\perp\ll T$, even for $\Lambda_\perp\to 0$. 
Note that the frequently employed form of $\qhat$ in this limit with the approximation $\ln\left(1+\frac{\Lambda_\perp^2}{m_D^2}\right)\to2\ln\frac{\Lambda_\perp}{m_D}$ (not shown in the figure)
would become negative at too small cutoffs $\lperp \sim m_D$, and is, thus, not well behaved in the limit $\lperp\to 0$.

In the region $\Lambda_\perp\gg T$ (right panel of \fig\ref{fig:thermal_equilibrium}), we observe that 
for small couplings $\lambda \sim 0.05$ both analytic large-cutoff expressions 
agree very well with the numerical values. However, they start to differ when increasing the coupling $\lambda \gtrsim 0.5$.
This is denoted as `shift' in \fig\ref{fig:thermal_equilibrium}.
We find that the values from the improved formula \eqref{eq:qhat_hard_arnold-improved} are closer to the numerical values 
than from the original formula \eqref{eq:qhat_hard_arnold}, i.e., the shift is smaller. However, for large couplings $\lambda \sim 10$, the improved analytic expression still seems to underestimate $\qhat$, with the difference being a constant. This is because the leading logarithmic behavior stems from the large $q_\perp$ behavior of the collision kernel (see Section \ref{sec:limiting_behavior_large_cutoff} and Appendix \ref{app:large-momentum-limits-qhat}), whereas for the constant, one requires the exact form of the collision kernel for all $q_\perp$ (in particular also for $q_\perp\approx T$), where the form \eqref{eq:collision_kernel_analytic_thermal} and approximations used in evaluating the $q_\perp$ integral over it \cite{Arnold:2008vd} are only valid for small couplings.

Turning now to a comparison of the matrix elements, 
we observe in \fig \ref{fig:thermal_equilibrium} that for small values of the coupling $\lambda \sim 0.05$ (left panel) as well as for large cutoffs $\Lambda_\perp \gg T$ (right panel), the results with the Debye-like screening approximation \eqref{eq:approximated_matrix_element} agree well with the full HTL matrix element \eqref{eq:full_htl_matrix_element}.
However, they start deviating with growing coupling
at small cutoffs $\Lambda_\perp \lesssim T$ (left panel). 
To guide the eye, for $\Lambda_\perp = 0.3 T$ this difference is denoted as `HTL approximation effect'.
For $\Lambda_\perp = T$ 
and $\lambda=10$
the deviation between the approximated 
and the HTL matrix elements is of the order of 
30\%.

\begin{table*}
\begin{tabular}{c c c c} \toprule
 $\lambda$ & $\tilde b$ & $\tilde d$ & $\tilde e$ \\ [0.5ex] 
 \hline
 $0.5$ & $0.0011944 \pm 0.0000020$ & $4.114 \pm 0.013$ & $-0.76919 \pm 0.00058$\\
$1.0$ & $0.0037772 \pm 0.0000062$ & $2.4910 \pm 0.0029$ & $-0.24707 \pm 0.00041$\\
$1.5$ & $0.007379 \pm 0.000013$ & $2.0956 \pm 0.0018$ & $0.03349 \pm 0.00032$\\
$2.0$ & $0.011905 \pm 0.000021$ & $1.9636 \pm 0.0014$ & $0.20498 \pm 0.00029$\\
$2.5$ & $0.017295 \pm 0.000031$ & $1.8987 \pm 0.0012$ & $0.32796 \pm 0.00028$\\
$3.0$ & $0.023563 \pm 0.000042$ & $1.8653 \pm 0.0010$ & $0.42226 \pm 0.00026$\\
$3.5$ & $0.030716 \pm 0.000054$ & $1.84570 \pm 0.00096$ & $0.49864 \pm 0.00025$\\
$4.0$ & $0.038770 \pm 0.000067$ & $1.83331 \pm 0.00088$ & $0.56271 \pm 0.00024$\\
$4.5$ & $0.047761 \pm 0.000082$ & $1.82484 \pm 0.00080$ & $0.61789 \pm 0.00023$\\
$5.0$ & $0.057714 \pm 0.000099$ & $1.81902 \pm 0.00075$ & $0.66626 \pm 0.00022$\\
$5.5$ & $0.06864 \pm 0.00012$ & $1.81444 \pm 0.00071$ & $0.70960 \pm 0.00021$\\
$6.0$ & $0.08061 \pm 0.00014$ & $1.81130 \pm 0.00069$ & $0.74868 \pm 0.00020$\\
$6.5$ & $0.09362 \pm 0.00015$ & $1.80845 \pm 0.00067$ & $0.78441 \pm 0.00020$\\
$7.0$ & $0.10772 \pm 0.00017$ & $1.80584 \pm 0.00066$ & $0.81733 \pm 0.00020$\\
$7.5$ & $0.12296 \pm 0.00020$ & $1.80380 \pm 0.00065$ & $0.84781 \pm 0.00019$\\
$8.0$ & $0.13933 \pm 0.00022$ & $1.80168 \pm 0.00064$ & $0.87635 \pm 0.00019$\\
$8.5$ & $0.15687 \pm 0.00024$ & $1.80026 \pm 0.00064$ & $0.90313 \pm 0.00019$\\
$9.0$ & $0.17562 \pm 0.00026$ & $1.79871 \pm 0.00064$ & $0.92836 \pm 0.00019$\\
$9.5$ & $0.19569 \pm 0.00029$ & $1.79776 \pm 0.00063$ & $0.95195 \pm 0.00019$\\
$10.0$ & $0.21701 \pm 0.00031$ & $1.79691 \pm 0.00063$ & $0.97442 \pm 0.00019$\\
$10.5$ & $0.23960 \pm 0.00034$ & $1.79628 \pm 0.00063$ & $0.99579 \pm 0.00019$\\
$11.0$ & $0.26361 \pm 0.00036$ & $1.79589 \pm 0.00063$ & $1.01605 \pm 0.00019$\\
$11.5$ & $0.28894 \pm 0.00039$ & $1.79532 \pm 0.00063$ & $1.03544 \pm 0.00019$\\
$12.0$ & $0.31570 \pm 0.00042$ & $1.79489 \pm 0.00063$ & $1.05399 \pm 0.00019$\\
$12.5$ & $0.34386 \pm 0.00045$ & $1.79432 \pm 0.00062$ & $1.07188 \pm 0.00019$\\
$13.0$ & $0.37349 \pm 0.00048$ & $1.79405 \pm 0.00062$ & $1.08902 \pm 0.00019$\\
$13.5$ & $0.40461 \pm 0.00052$ & $1.79343 \pm 0.00062$ & $1.10557 \pm 0.00019$\\
$14.0$ & $0.43722 \pm 0.00054$ & $1.79316 \pm 0.00062$ & $1.12149 \pm 0.00019$\\
$14.5$ & $0.47134 \pm 0.00058$ & $1.79241 \pm 0.00062$ & $1.13694 \pm 0.00019$\\
$15.0$ & $0.50704 \pm 0.00061$ & $1.79162 \pm 0.00061$ & $1.15192 \pm 0.00019$\\
$15.5$ & $0.54431 \pm 0.00066$ & $1.79053 \pm 0.00061$ & $1.16651 \pm 0.00019$\\
$16.0$ & $0.58324 \pm 0.00070$ & $1.78988 \pm 0.00060$ & $1.18054 \pm 0.00020$\\
$16.5$ & $0.62381 \pm 0.00074$ & $1.78933 \pm 0.00060$ & $1.19420 \pm 0.00019$\\
$17.0$ & $0.66613 \pm 0.00078$ & $1.78922 \pm 0.00060$ & $1.20734 \pm 0.00019$\\
$17.5$ & $0.71013 \pm 0.00081$ & $1.78864 \pm 0.00059$ & $1.22017 \pm 0.00019$\\
$18.0$ & $0.75590 \pm 0.00085$ & $1.78805 \pm 0.00059$ & $1.23280 \pm 0.00019$\\
$18.5$ & $0.80337 \pm 0.00090$ & $1.78734 \pm 0.00058$ & $1.24516 \pm 0.00019$\\
$19.0$ & $0.85257 \pm 0.00094$ & $1.78699 \pm 0.00058$ & $1.25721 \pm 0.00019$\\
$19.5$ & $0.90367 \pm 0.00099$ & $1.78674 \pm 0.00058$ & $1.26879 \pm 0.00019$\\
$20.0$ & $0.9565 \pm 0.0010$ & $1.78563 \pm 0.00058$ & $1.28029 \pm 0.00019$\\[1ex] \bottomrule
\end{tabular}
\caption{Fitted coefficients for the interpolation formula for $\qhat$ in \eq \eqref{eq:empirical_qhat3}. The values were obtained by numerically integrating \eqref{eq:qhat_formula_pinf} and the HTL-screened matrix element \eqref{eq:full_htl_matrix_element}, then numerically fitting the coefficient $\tilde b$ in the region $\Lambda_\perp \gg T$, and finally fitting $\tilde d$ and $\tilde e$ in the range $0.1 T < \Lambda_\perp < 15 T$.
Table from \cite{Boguslavski:2023waw}.
}
\label{tab:fitted_empirical3}

\end{table*}

\begin{figure*}
    \centerline{
    \includegraphics[width=0.49\linewidth]{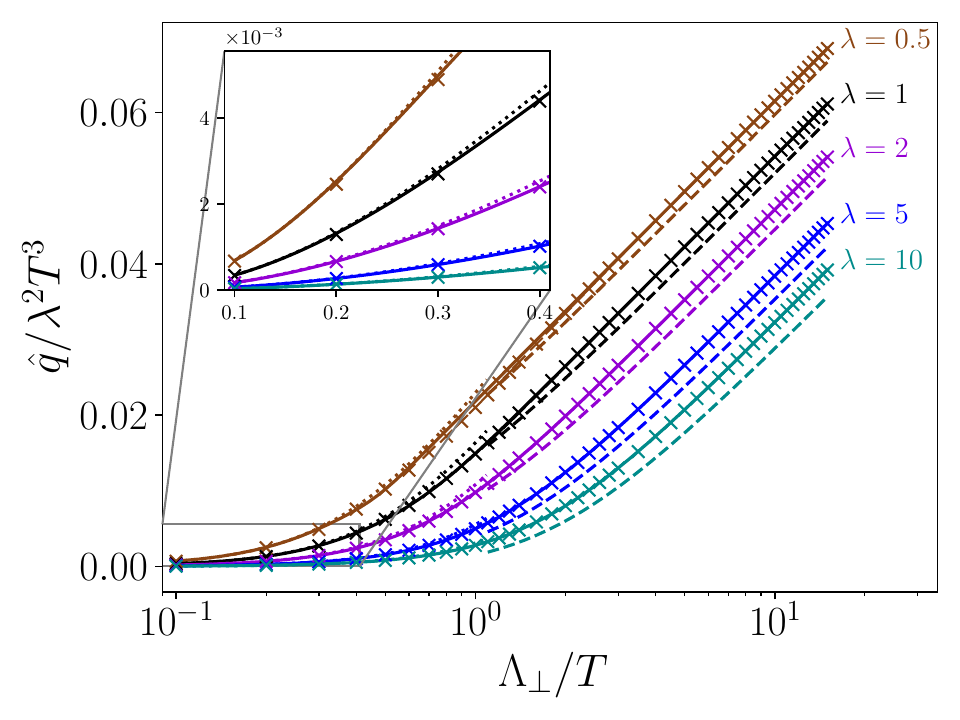}
    \includegraphics[width=0.49\linewidth]{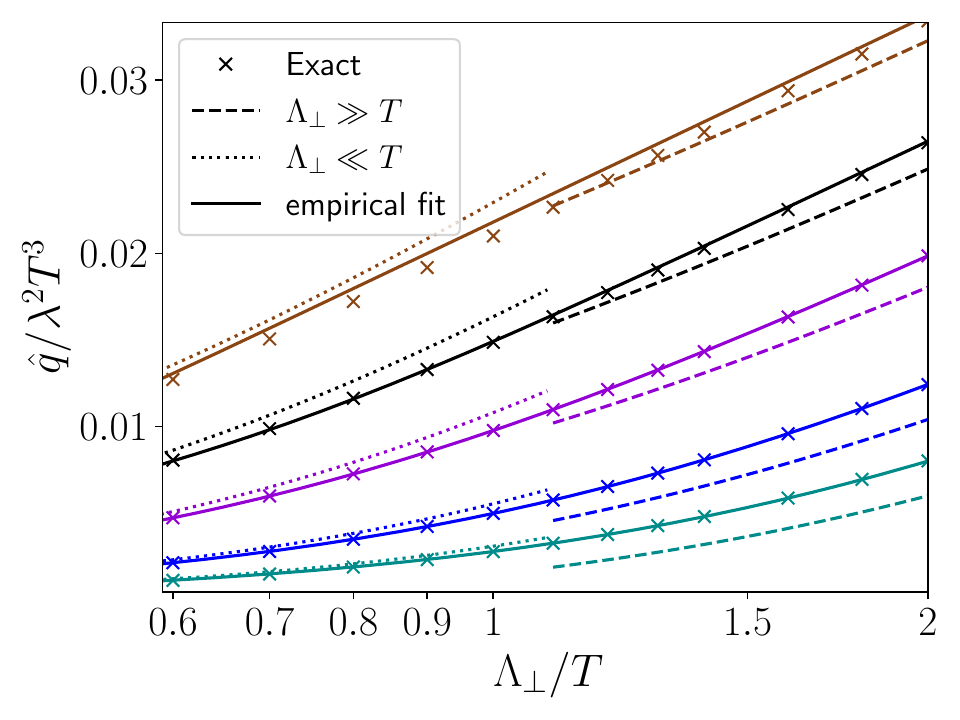}
    }
    \caption{{\em (Left:)} The interpolation formula \eqref{eq:empirical_qhat3} for the jet quenching parameter $\qhat$ with the fitted coefficients listed in \tab \ref{tab:fitted_empirical3} is shown as continuous lines for different couplings $\lambda$ (color-coded). The numerical evaluation of Eq.~\eqref{eq:qhat_formula_pinf} is shown as crosses, the limiting expressions for soft \eqref{eq:qhat_thermal_equlibrium_soft} and hard cutoffs \eqref{eq:qhat_hard_arnold} as dotted and dashed curves. The inset shows the behavior at small momentum cutoffs. 
    {\em (Right:)} Focus on the interpolation region $\Lambda_\perp \sim T$. For $\lambda \geq 1$ we see very good agreement with the numerical results. Plots adapted from \cite{Boguslavski:2023waw}.
    }\label{fig:qhat_empirical_fit3_combined}
\end{figure*}

\subsubsection{Interpolation formula for $\qhat$ in thermal equilibrium \label{sec:Empirical_fit_formula}}

We have now verified that the analytic expressions \eqref{eq:qhat_thermal_equlibrium_soft} and \eqref{eq:qhat_hard_arnold} describe $\qhat$ only in certain limits and \eq \eqref{eq:qhat_hard_arnold} only holds for small couplings $\lambda\lesssim 0.5$. For phenomenological calculations, a general simple formula (or parametrization) for $\qhat$ in thermal equilibrium
may be useful without the need to perform the high-dimensional integral \eqref{eq:qhat_formula} numerically for the required value of the coupling $\lambda$ and transverse momentum cutoff $\Lambda_\perp$.
Thus, we construct an interpolation formula that reproduces the analytical results in the limits $\Lambda_\perp \ll T$ and for $\Lambda_\perp\gg T$ and agrees with our numerical evaluation.

From \eqref{eq:qhat_thermal_equlibrium_soft} and \eqref{eq:qhat_hard_arnold} we know the behavior of $\qhat$ for small $\Lambda_\perp\ll T$ and large cutoffs $\Lambda_\perp\gg T$. As discussed before, \eqref{eq:qhat_hard_arnold} differs from the numerical evaluation of $\qhat$ by a constant shift for larger values of the coupling $\lambda \gtrsim 0.5$. Our strategy is to find an empirical fit function 
that smoothly interpolates in between,
\begin{align}
    \frac{\qhat^{\mathrm{emp}}}{\CR T^3}=
    \begin{cases}
        \tilde c\ln(1+\Lambda_\perp^2/m_D^2)\,, & \text{for } \Lambda_\perp \ll T \\
        \tilde a\ln(\Lambda_\perp/m_D) + \tilde b\,, & \text{for } \Lambda_\perp \gg T\,.
    \end{cases}\label{eq:qhat_empirical_fit_philosophy}
\end{align}
The switching between those two cases will be done using a hyperbole tangent that smears out a step function with width parameter $\tilde d$,
\begin{align}
    \Theta_{\tilde d}(x)=\frac{1+\tanh\left(\tilde d x\right)}{2}\,,
\end{align}
which approaches the usual step function for $\tilde d\to\infty$.

This leads to the following form for the fit formula
\begin{align}
\begin{split}
    \frac{\qhat^{\mathrm{emp}}}{\CR T^3}&=\tilde c\ln\left(1+\frac{\Lambda_\perp^2}{m_D^2}\right)\Theta_{\tilde d}\left(\tilde e-\ln\frac{\Lambda_\perp}{T}\right)+\left(\tilde a\ln\frac{\Lambda_\perp}{m_D} + \tilde b\right)\Theta_{\tilde d}\left(\ln\frac{\Lambda_\perp}{T}-\tilde e\right).\label{eq:empirical_qhat3}
\end{split}
\end{align}
For the coefficients $\tilde c$ and $\tilde a$, we use the prefactors of \eqref{eq:qhat_thermal_equlibrium_soft} and \eqref{eq:qhat_hard_arnold}, which 
have the following values for a gluonic plasma: 
\begin{align}
    \label{eq:tilde_ac_empfit}
    \tilde c = \frac{\lambda^2}{12\pi \NC}, && \tilde a =\frac{\lambda^2\zeta(3)}{\pi^3 \NC}\,.
\end{align}
This leaves only three fit parameters: The constant $\tilde b$ encodes the linear shift in the large $\Lambda_\perp/T$ region,
while $\tilde d$ and $\tilde e$ describe the width and position of the switching between the two limiting cases in \eqref{eq:qhat_empirical_fit_philosophy}.
First, the coefficient $\tilde b$ is obtained, such that it correctly reproduces $\qhat \simeq \tilde a\ln\Lambda_\perp/m_D + \tilde b$ in the large $\Lambda_\perp/T$ region. 
Then, the coefficients $\tilde d$ and $\tilde e$ are determined by fitting them to the numerical data. 

The results for the remaining fit parameters in \eq \eqref{eq:empirical_qhat3} are listed in \tab\ref{tab:fitted_empirical3} for the couplings $\lambda = 0.5 - 20$. The resulting $\qhat$ are shown
in \fig \ref{fig:qhat_empirical_fit3_combined} as continuous lines. For comparison, the numerically obtained values are included as crosses, and the limiting expressions for hard and soft cutoffs, Eqs.~\eqref{eq:qhat_hard_arnold} and \eqref{eq:qhat_thermal_equlibrium_soft}, respectively, as dashed and dotted lines.
Consistently with the construction of the fit formula, its values are seen to agree well with the numerically evaluated $\qhat$ in the left panel of \fig \ref{fig:qhat_empirical_fit3_combined} and in the inset showing the small cutoff behavior at $\Lambda_\perp \ll T$. 
The right panel of \fig \ref{fig:qhat_empirical_fit3_combined} shows the interpolation region $\Lambda_\perp \sim T$. We find a very good agreement with the numerics for $\lambda \geq 1$, while for smaller couplings $\lambda \lesssim 0.5$ deviations grow in this region. Note that the fit formula provides a smooth interpolating expression for $\qhat$ with improved accuracy in this region as compared to the previous limiting forms.

Thus, the expression \eqref{eq:empirical_qhat3} together with the coefficients in \tab \ref{tab:fitted_empirical3} can be used to obtain $\qhat$ in thermal equilibrium for any transverse momentum cutoff $\Lambda_\perp$ and the listed couplings $\lambda$ in the weak coupling limit at leading-order.

\subsection{Toy models for bottom-up thermalization}
\label{sec:toy_models}

As toy models for the bottom-up thermalization process that we discussed in Section \ref{sec:thermalization-expanding-systems}, we consider first an effectively two-dimensional 
distribution in \se \ref{sec:aniso_dist}. Then, $\qhat$ is computed analytically in \se \ref{sec:scaled-thermal-distribution} using an isotropic scaled thermal distribution, which models key features of the over- and under-occupied bottom-up stages.

\subsubsection{Effectively two-dimensional distribution}
\label{sec:aniso_dist}
As a toy model for the large anisotropies encountered at early times in the bottom-up thermalization scenario, we consider an effectively two-dimensional system and calculate the jet quenching parameter $\hat q$ for that system.
For that, we consider a distribution function with vanishing $k_z$ momentum,
\begin{align}
    f(\vb k)=B(k_x,k_y)\delta(k_z/Q), \label{eq:distribution_extremely_anisotropic}
\end{align}
where $B$ is (for the moment) an arbitrary function of $k_x$ and $k_y$, and $Q$ is an energy scale.
Due to its vanishing momentum in beam direction $k_z = 0$, such a state is similar in spirit to the Glasma, where the jet quenching parameter has also been studied \cite{Avramescu:2023qvv, Ipp:2020mjc, Ipp:2020nfu, Carrington:2021dvw, Carrington:2022bnv}.

First, let us focus on the \emph{Bose-enhanced} part $\qhatff$ in kinetic theory. This agrees with $\qhat$ in a classical-statistical framework since there is no $\qhatf$ contribution in the classical field limit.
By inserting the two-dimensional distribution \eqref{eq:distribution_extremely_anisotropic} into the $\qhat$ integral \eqref{eq:qhat_formula}, we immediately find
\begin{align}
\qhatff^{zz}=0,
\end{align}
due to its proportionality to $\int(q^z)^2\delta(k_z)\delta(k_z')$. 
Note that this is true regardless of the precise form of the matrix element or screening prescription. 
Thus, a purely two-dimensional momentum distribution remains two-dimensional 
in the classical field limit of kinetic theory.

Let us now consider a special case of \eqref{eq:distribution_extremely_anisotropic} for which we may obtain an analytic expression for the jet quenching parameter, where all particles have a specific momentum $\tilde k$,
\begin{align}
	f(\vec k)= A\,\delta\left(\frac{k_x^2+k_y^2-\tilde k^2}{Q^2}\right) \delta(k_z/Q)\,.\label{eq:special_distribution}
\end{align}

First, note that we do not need to enforce a momentum cutoff $\qperp<\lperp$, because $\qhatff$ is finite even for $p\to\infty$.
Hence, we consider here the case where the transverse momentum cutoff $\lperp$ is sufficiently large, at least $\lperp>2\tilde k$, as we will see below.

For $\qhatff^{yy}$ we start with Eq.~\eqref{eq:qhat_der_with_delta} and insert the special two-dimensional distribution $f(\vec k)$ from \eqref{eq:special_distribution}, obtaining
\begin{align}
	&\qhatff^{yy} = \frac{A^2Q^6}{16p^2\nu}\int\frac{\dd[3]{\vb k}\dd[3]{\vb q}\dd{\omega}}{(2\pi)^5 q^2 k^2 }q^yq^y\left|\mathcal M(\vb p,\vb k;\vb p',\vb k')\right|^2 \nonumber\\
	& \times \delta\left(\cos\thetaqp-\frac{\omega}{q}-\frac{\omega^2-q^2}{2pq}\right)\delta\left(\cos\thetaqk-\frac{\omega}{q}+\frac{\omega^2-q^2}{2kq}\right)\nonumber\\
	&\times\theta(p'-k')\theta\left(p - \frac{q-\omega}{2}\right)\Theta\left(k - \frac{q+\omega}{2}\right)\Theta(q-|\omega|)\nonumber\\
	&\times\delta(k_z)\delta(q_z)\,{\delta(k^2-\tilde k^2)}
	\,{\delta\left((\vb k-\vb q)^2-\tilde k^2\right)}
	.
\end{align}
We first integrate out $k_z$ and $q_z$.
Additionally, for $p\to\infty$ we may drop the theta functions containing $p$ and $p'$. 
We then rewrite the
delta functions as $\delta\left((\vb k-\vb q)^2-\tilde k^2\right)=\delta(\omega^2-2\omega\tilde k)=\frac{1}{2\tilde k}\left(\delta(\omega)+\delta(\omega-2\tilde k)\right)$, allowing us to integrate over $\omega$.
The $\delta(\omega-2k)$ term vanishes because then the third step function becomes $\Theta(-q/2)=0$. Integrating over $\omega$ then enforces $\omega=0$. Finally, in the large-$p$ limit, the first delta function can be simplified to $\delta(\cos\thetaqp)=q\delta(q_x)$, 
allowing us to perform the $q_x$ integral as well. We then arrive at
\begin{align}
	\qhatff^{yy} &= \frac{A^2Q^6}{16p^2\nu}\int\frac{\dd{k_x}\dd{k_y}\dd{q_y}}{(2\pi)^5 q_y k^2 }\left(q_y\right)^2\left|\mathcal M(\vb p,\vb k;\vb p',\vb k')\right|^2 \\
	&\qquad \times\delta\left(\cos\thetaqk-\frac{|q_y|}{2 \tilde k}\right)\Theta\left(\tilde k - \frac{|q_y|}{2}\right)\frac{1}{4\tilde k^2}\delta(k-\tilde k),\nonumber
\end{align}
where we have used $\delta(k^2-\tilde k^2)= \frac{1}{2\tilde k}\delta(k-\tilde k)$.
Effectively, $\vb q$ is parallel to the $y-$axis and $\vb k$ lies in the $x-y$ plane with length $\tilde k$ and $k_y=|q_y|/2$. For the matrix element we need $q=|q_y|$, $k=\tilde k$ and $\phikq$, which is the polar angle of $\vb k$ in a frame, in which $\vb q$ points in the $z$ direction and $\vb p$ lies in the $x-z$ plane, see \se\ref{sec:coordinate_systems}. In our case, $\vb q$ is orthogonal to $\vb p$, thus we perform the $\vb k$ integration in a frame, in which $\vb q = q \vb e_{z^3}$, $\vb p=-p\vb e_{x^3}$. Since $\vb k$ must lie in the $\vb p - \vb q$ plane, we obtain 
\begin{align}
	\phikq\in\left\{0,\pi\right\}.\label{eq:cond_phikq}
\end{align}
We get a factor $2$ from the symmetry $q_y\leftrightarrow -q_y$ and insert the gluonic matrix element from \tab \ref{tab:p-inf_matrix_el} with the Debye-like screening approximation \eqref{eq:approximated_matrix_element}, and sum over the possible values of $\cos\phikq$, 1 and -1. Thus we obtain, for a momentum cutoff $\lperp>2\tilde k$,
\begin{align}
	\qhatff^{yy}&=\frac{\dA \CA  ^2 A^2g^4Q^6}{2^6\pi^5 \nu \tilde k^3}\int_{0}^{2\tilde k}\dd{q}q
	\frac{\left(2\tilde k - \sqrt{4\tilde k ^2-q^2}\right)^2+\left(2\tilde k + \sqrt{4\tilde k ^2-q^2}\right)^2}{(q^2+\xi^2m_D^2)^2}\nonumber
     \\
	&=\frac{\dA \CA  ^2 A^2g^4Q^6}{(2\pi)^5 \nu \tilde k^3}\int_{0}^{2\tilde k}\dd{q}q
	\frac{8\tilde k^2-q^2}{(q^2+\xi^2m_D^2)^2}.
	\label{eq:special_case_to_integrate}
\end{align}
The integral over $q$ can be performed analytically, which yields 
\begin{align}
	\qhatff^{yy} &=\frac{\dA \CA^2 A^2 g^4}{2^7\pi^5d_R\tilde k^3}Q^6\Bigg\{4\tilde k^2\left(\frac{2}{\xi^2m_D^2}-\frac{1}{4\tilde k^2+\xi^2m_D^2}\right)+\ln\frac{\xi^2m_D^2}{4\tilde k^2+\xi^2m_D^2}\Bigg\},\label{eq:qhatffyy_special}
\end{align}
where $m_D^2=A\frac{g^2Q^3}{\pi^2\tilde k}$ according to Eq.~\eqref{eq:debyemass-general}.

Indeed, in Section \ref{sec:obtaining-qhat-between-glasma-and-hydro} we will observe that in the over-occupied and anisotropic earliest stage of bottom-up thermalization {one has $\qhat^{zz} < \qhat^{yy}$ and that this is due to $\qhatff^{zz}<\qhatff^{yy}$, which is consistent with the simple toy model presented here.

\subsubsection{Scaled thermal distribution\label{sec:scaled-thermal-distribution}}

Let us now study another aspect encountered during bottom-up thermalization: over and under-occupied systems. For simplicity, we use an isotropic toy model and consider a scaled thermal distribution, i.e., we scale the amplitude of the thermal distribution \eqref{eq:thermal_bose_fermi_combined} with $N_\pm$. Here $N_+$ denotes the scaling parameter of the Bose-Einstein distribution and $N_-$ the scaling parameter of the Fermi-Dirac distribution,
\begin{align}
f_{\pm}(k; T)=\frac{N_\pm}{\exp(k/T)\mp1}\,.\label{eq:scaled_bose_fermi_combined}
\end{align}
This allows us to easily
generalize the results obtained in \se\ref{sec:qhat-thermal} for the jet quenching parameter $\qhat$ in a thermal medium. We start with $\qhat$ given by \eq \eqref{eq:qhat_soft_distribution_function_split_off}. Splitting the $f$ and $ff$ contributions 
and using the integrals \eqref{eq:thermal_distributionfunctions_integrals} over thermal distributions, 
we obtain for small cutoff 
\begin{align}
	&\qhat(\Lambda_\perp\ll T, N_\pm)=\frac{g^4T^3\CR}{24\pi^3}\ln\left(1+\frac{\Lambda_\perp^2}{m_D^2}\right) \label{eq:qhat_scaled_soft}\\
	&\qquad \times\Big(\pi^2(2\NC(\NOg)^2  +\nf (\NOq)^2 )
	+\zeta(3)\left[9\nf \NOq(1-\NOq)-12\NC\NOg(\NOg-1)\right]\Big), \nonumber
\end{align}

which generalizes
the equilibrium ($N_\pm=1$) result in \eq \eqref{eq:qhat_thermal_equlibrium_soft}. Similarly, we can generalize the large cutoff 
formula \eqref{eq:qhat_hard_arnold} to
\begin{subequations}
\label{eq:qhat_thermal_hard_analytic_combined_notimproved}
\begin{align}
	\begin{split}
		\qhat(\Lambda_\perp\gg T, N_\pm)&=
		\CR \frac{g^4T^3}{\pi^2}\sum_\pm\Xi_\pm
		\mathcal I_\pm(\Lambda_\perp, N_\pm),\label{eq:qhat_hard_arnold_scaled_not_improved}
	\end{split}\\
	\mathcal I_\pm(\Lambda_\perp, N_\pm)&=\frac{N_\pm\zeta_\pm(3)}{2\pi}\ln\left(\frac{\Lambda_\perp}{m_D}\right)+(N_\pm)^2\Delta\mathcal I_\pm,\label{eq:qhat_thermal_hard_analytic_Ipm_scaled_notimproved}
\end{align}
with $\Delta\mathcal I_\pm$ given by \eq \eqref{eq:qhat_thermal_hard_analytic_DeltaI}, which entirely determines $\qhatff$,
\begin{align}
    \qhatff(\lperp\gg T, N_\pm)=\CR\frac{g^4T^3}{\pi^2}\sum_{\pm}\left(N_\pm\right)^2\Xi_\pm \Delta\mathcal I_\pm.\label{eq:scaled_qhatff_notimproved}
\end{align}
\end{subequations}

Furthermore, similarly to our discussion for the thermal result in \se \ref{sec:thermal_qhat_large_momentum_cutoff}, by replacing $2\ln (\Lambda_\perp/m_D) \to \ln(1+(\Lambda_\perp/m_D)^2)$ in \eqref{eq:qhat_thermal_hard_analytic_Ipm_scaled_notimproved},
we obtain an `improved' formula valid for large cutoffs that is finite even at small $\Lambda_\perp$ and generalizes
\eq \eqref{eq:qhat_hard_arnold-improved}.
Then we can again split off the Bose-enhanced contribution as in \eqref{eq:qhat_f_ff_splitting},
    $\qhat = \qhatf+\qhatff$.
We note that
$\qhatf(\lperp)$ has the same form for small and large cutoffs, 
\begin{subequations}
\label{eq:qhat_scaled_individual_components_combined}
\begin{align}
	\qhatf(\Lambda_\perp, N_\pm)&= \zeta(3)\left(12\NOg\NC +9\nf\NOq\right)C_L(\Lambda_\perp)\,, \label{eq:qhat_scaled_f}
\end{align}
with, as before, $C_L(\Lambda_\perp)=\frac{g^4T^3 \CR}{24\pi^3}\ln\left(1+\frac{\Lambda_\perp^2}{m_D^2}\right)$.
Again, the Bose-enhanced terms differ 
\begin{align}
	\qhatff(\lperp\ll T,N_\pm)&=\Big[2\NC(\NOg)^2(\pi^2-6\zeta(3))\label{eq:qhatff_scaled_soft}~+\nf(\NOq)^2(\pi^2-9\zeta(3))\Big]C_L(\Lambda_\perp),\\
    {\qhatffimproved(\lperp\gg T,N_\pm)}&={\CR g^4T^3}\sum_\pm\Xi_\pm(N_\pm)^2\label{eq:qhatff_hard_scaled_improved}\\
&\times \left\{
 \frac{\zeta_\pm(2)-\zeta_\pm(3)}{4\pi^3}\left[\ln\left(1+\frac{T^2}{m_D^2}\right) +1-2\gamma_E+2\ln 2\right]
    -\frac{\sigma_\pm}{2\pi^3}\right\}. \nonumber
\end{align}
\end{subequations}
The Debye mass entering these expressions for the scaled thermal distributions is given by (see \eq \eqref{eq:debyemass-general})
\begin{align}
	m_D^2=\frac{g^2T^2}{3}\left(\NOg\NC  +\frac{\NOq\nf }{2}\right). \label{eq:debye_mass_scaled_thermal}
\end{align}
Thus, $m_D$ scales with $\sqrt{\lambda N_\pm}$. The occupation of fermions $N_-$ cannot become large due to Pauli blocking. For large gluon occupancies $N_+$, this may pose a problem for the validity of perturbation theory that requires $m_D\ll T$ and that our arguments and the derivations in \cite{Arnold:2008vd} are based on.
 We can estimate the breakdown scale by requiring $m_D\ll T$, which leads to 
\begin{align}
    \NOg \ll \frac{1}{\NC}\left(\frac{3}{g^2}- \frac{\NOq\nf}{2}\right).\label{eq:qhat_scaled_hard_limitation}
\end{align}
This is, of course, in line with the usual limitations of perturbation theory, which breaks down at nonperturbatively large occupation numbers $f \gtrsim 1/g^2$.

\begin{figure*}
    \centerline{
        \includegraphics[width=0.49\linewidth]{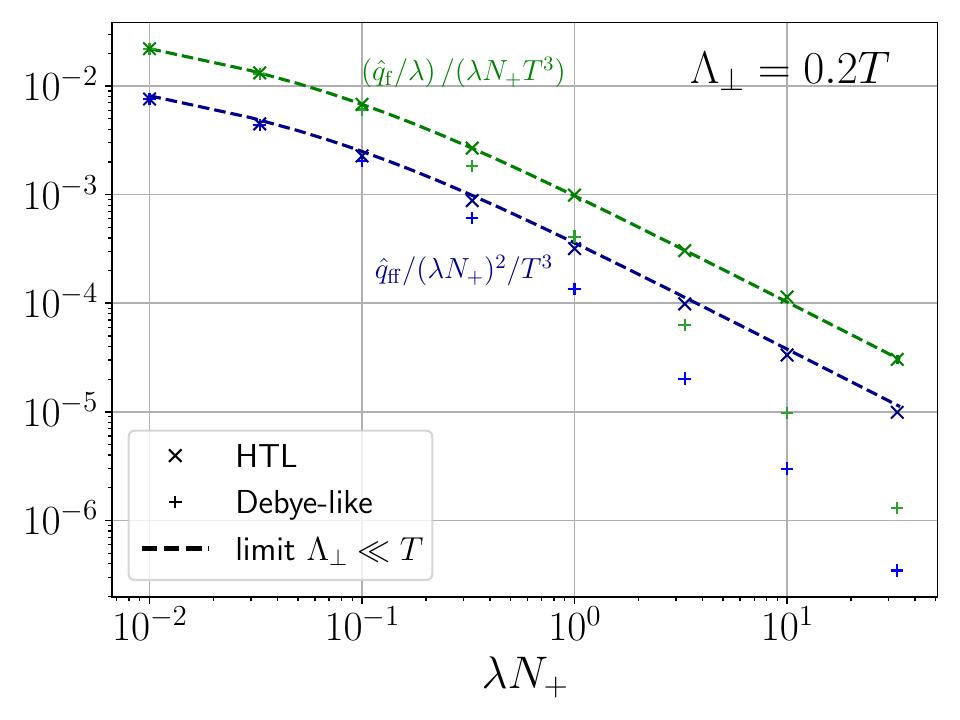}
        \includegraphics[width=0.49\linewidth]{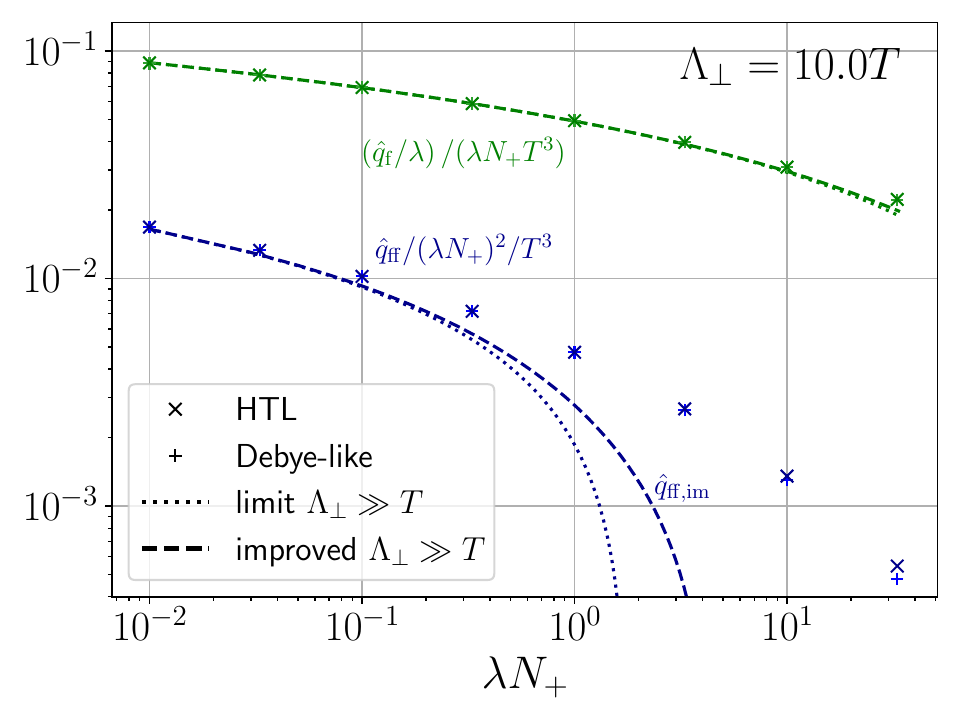}
    }
    \caption{\label{fig:qhat_components_combined}The individual components $\qhatf$ (green) and $\qhatff$ (blue) as defined in \eqref{eq:qhat_f_ff_splitting} and rescaled according to their parametric estimates in \eqref{eq:qhat_components_naive_behavior} as functions of their only argument $\lambda N_+$ for $\Lambda_\perp/T=0.2$ (left) and $\lperp/T=10$ (right). 
   The isoHTL screened results are shown with `$\times$'-symbols, and the Debye-like screened results are shown as `$+$'-symbols.
The left panel shows the small-cutoff form \eqref{eq:qhatff_scaled_soft} for $\qhatff$ and Eq.~\eqref{eq:qhat_scaled_f} for $\qhatf$ labeled as ``$\lperp\ll  T$''.
 The right panel depicts both large-cutoff expressions \eqref{eq:qhat_scaled_f} for $\qhatf$ and \eqref{eq:scaled_qhatff_notimproved} for $\qhatff$
 labeled ``limit $\lperp \gg T$''  and its improved version \eqref{eq:qhatff_hard_scaled_improved}
 labeled as ``improved $\lperp \gg T$''. Figures adapted from \cite{Boguslavski:2023waw}.
    }
\end{figure*}
\begin{figure*}
    \centerline{
        \includegraphics[width=0.49\linewidth]{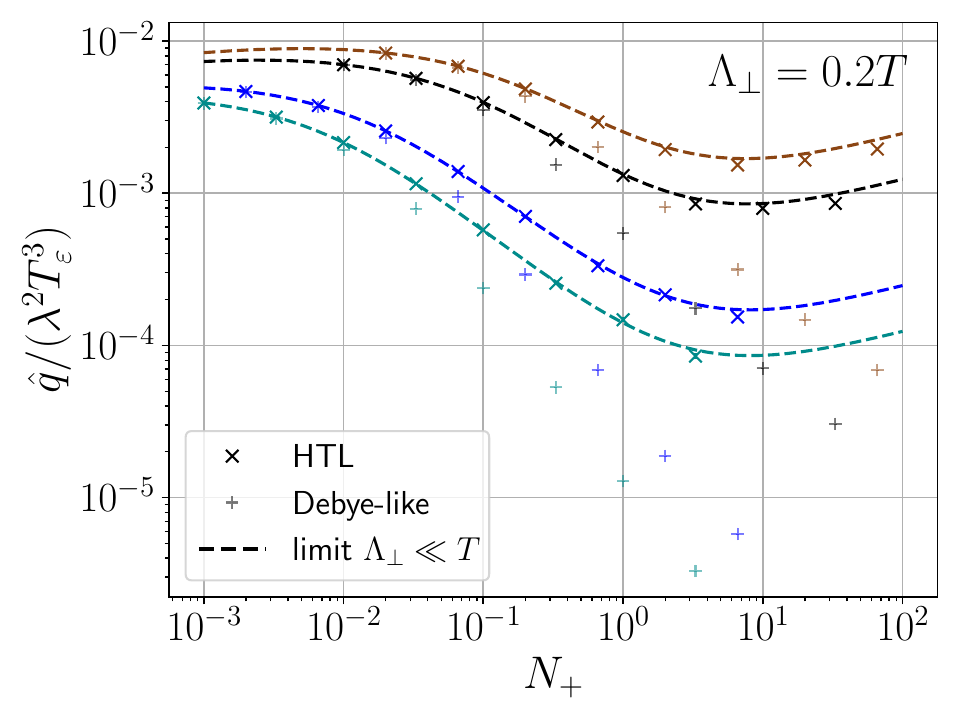}
        \includegraphics[width=0.49\linewidth]{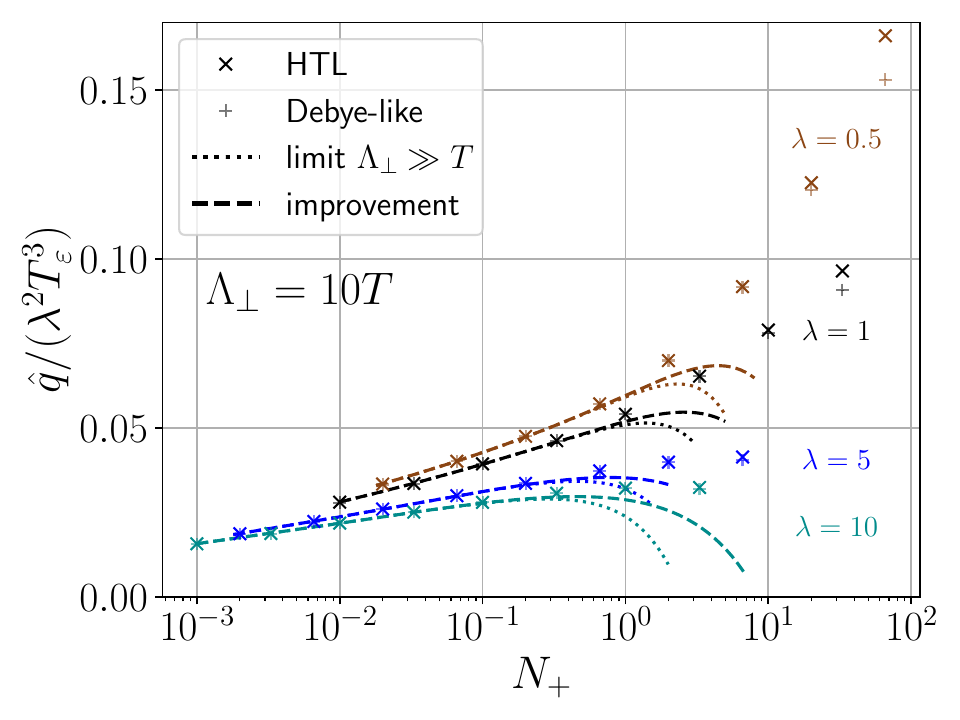}
    }
    \caption{\label{fig:qhat_scaled_combined}
The jet quenching parameter $\qhat$ for a scaled thermal distribution
\eqref{eq:scaled_bose_fermi_combined} as a function of $N_+$ for different couplings $\lambda$ for cutoff $\Lambda_\perp/T=0.2$ (left) and $\Lambda_\perp/T=10$ (right). As in Fig.~\ref{fig:qhat_components_combined}, the isoHTL results are shown with $\times$-symbols, and Debye-like screened results as $+$-symbols.
In the left panel, the small-cutoff form \eqref{eq:qhat_scaled_soft} is labeled as ``$\lperp\ll  T$''.
The right panel shows both the large-cutoff expression
Eq.~\eqref{eq:qhat_hard_arnold_scaled_not_improved} labeled as ``limit $\lperp \gg T$'' and the improved version obtained by summing Eqs.~\eqref{eq:qhat_scaled_f} and \eqref{eq:qhatff_hard_scaled_improved} labeled as ``improvement''. Figures adapted from \cite{Boguslavski:2023waw}. 
    }
\end{figure*}

Let us now compare these expressions with a numerical evaluation of $\qhat$ using \eq \eqref{eq:qhat_formula_pinf}.
For a purely gluonic system, the individual contributions $\qhatf/\lambda$ and $\qhatff$ (from Eqs.~\eqref{eq:qhat_scaled_individual_components_combined}) depend only on the product $\lambda N_+$, and not on the coupling $\lambda$ and occupation $N_+$ individually,\footnote{
The Debye mass entering the matrix element depends only on the product $\lambda N_+$ (see Eq.~\eqref{eq:debyemass-general}). The quantum part $\qhatf$ comes naturally with a factor $N_+$ from the single distribution function and a factor $\lambda^2$ from the matrix element, and thus $\qhatf/\lambda$ is only a function of the product $\lambda N_+$. The classical wave contribution $\qhatff$ comes naturally with a prefactor $\lambda^2 N_+^2$.
}
\begin{align}
    \qhat(\Lambda_\perp, N_+, \lambda) = \lambda\left(\frac{\qhatf}{\lambda}\right)(\Lambda_\perp,\lambda N_+)+ \qhatff (\Lambda_\perp,\lambda N_+). \label{eq:scaled_thermal_splitting}
\end{align}
These contributions are plotted in \fig \ref{fig:qhat_components_combined} for small (left) and large (right) cutoffs $\Lambda_\perp = 0.2 T$ and $10 T$, respectively, divided by the prefactor
\begin{align}
    \qhatf \sim \lambda (\lambda N_+) T^3, && \qhatff \sim (\lambda N_+)^2T^3.\label{eq:qhat_components_naive_behavior}
\end{align}
We observe that 
their values deviate significantly from the simple estimates in \eq \eqref{eq:qhat_components_naive_behavior}.
This is a consequence of 
screening effects and the scaling of the Debye mass. In particular, one finds for sufficiently small cutoffs $\Lambda_\perp \lesssim m_D, T$ and large occupancies that
\begin{align}
    \label{eq:invLNplus}
    \frac{\qhatf}{\lambda^2 N_+ T^3} \sim \frac{\qhatff}{\lambda^2 N_+^2 T^3} \sim \frac{\Lambda_\perp^2}{m_D^2} \sim (\lambda 
 N_+)^{-1}\,,
\end{align}
which is visible in the left panel of \fig \ref{fig:qhat_components_combined} for sufficiently large $\lambda N_+$. Note that for a sufficiently large $\lambda N_+ \gg 1$ the effective kinetic theory description used here ceases to be valid. 
Similarly to the equilibrium case discussed in \se \ref{sec:qhat-thermal} and particularly in \fig \ref{fig:thermal_equilibrium}, the expression for small cutoffs \eqref{eq:qhat_scaled_soft} nicely agrees with the numerical values in the small-cutoff asymptotic region, plotted in the left panel of \fig\ref{fig:qhat_components_combined}. In the right panel, for large cutoffs, 
we observe that the analytic form for $\qhatf$ in Eq.~\eqref{eq:qhat_scaled_f} remains a very good description coinciding with the numerical values, whereas the analytic estimate for $\qhatff$ in Eq.~\eqref{eq:scaled_qhatff_notimproved} (and its improvement Eq.~\eqref{eq:qhatff_hard_scaled_improved}) ceases to describe the data for nonperturbatively large occupancies $\lambda N_+ \gtrsim 1$. This is expected from the condition \eqref{eq:qhat_scaled_hard_limitation}, and we see sizable deviations already at $\lambda N_+ \gtrsim 0.1$.

The full HTL screening and the Debye-like screening \eqref{eq:approximated_matrix_element} nicely agree with each other at large cutoffs for the whole $\lambda N_+$ range despite the aforementioned limitations concerning $\qhatff$.
On the other hand, for small cutoffs (left panel), the Debye-like screening approximation shows large deviations from the full HTL screening, albeit in the large $\lambda N_+$  region which should be taken with caution, as discussed above. 
The resemblance to the thermal case here is, of course, no coincidence since by setting $N_+=1$, we recover the thermal results.

Recombining the contributions from $\qhatf$ and $\qhatff$, 
we show $\qhat$ in \fig\ref{fig:qhat_scaled_combined} for the couplings $\lambda = 0.5$, $1$, $2$, $5$ and $10$ as functions of the occupancy $N_+$, for the small cutoff $\Lambda_\perp/T=0.2$ in the left panel and the large cutoff $\Lambda_\perp/T=10$ in the right panel.
The values are shown scaled by the effective temperature $\Teps$ that represents the temperature of a thermal system with the same energy density, Eq.~\eqref{eq:Landau-matching-condition},
\begin{align}
    \Teps = (\NOg)^{1/4}T. \label{eq:effective_temperature}
\end{align}
For comparison, also 
the analytic expectations are shown for small \eqref{eq:qhat_scaled_soft} and large cutoffs \eqref{eq:qhat_thermal_hard_analytic_combined_notimproved} as well as its improved expression \eqref{eq:qhat_scaled_individual_components_combined}.
Similarly as for $\qhatf$ and $\qhatff$, we observe for $\qhat$ in \fig \ref{fig:qhat_scaled_combined} that the small-cutoff expression agrees well with our (HTL-)screened data points while the large-cutoff expressions describe the data points until $N_+ \lesssim 1/\lambda$. 
Moreover, the improved formula for large cutoffs increases the validity of the analytic result only to slightly larger occupancy $N_+$.
This plot emphasizes the importance of screening effects that prevent the na\"ive scaling with $N_+$ or $N_+^2$. 
We, therefore, should be cautious when using such analytic expressions to describe over-occupied systems with typical occupancies $N_+ \sim 1/\lambda$. 
Instead, transport coefficients in such systems can be studied using classical-statistical lattice simulations \cite{Boguslavski:2020tqz, Ipp:2020mjc, Ipp:2020nfu, Avramescu:2023qvv}. In particular, it has been shown \cite{Boguslavski:2020tqz} that nonperturbative corrections can be substantial.

Interestingly, as visible in \fig\ref{fig:qhat_scaled_combined}, increasing the occupancy $N_+$ does not appear to drastically increase the value of the jet quenching parameter $\qhat$. In particular, for small cutoff $\lperp/T=0.2$ visible in the left panel, we observe that the scaled $\qhat$ in fact decreases with increasing occupancy.
Even for large cutoffs (right panel), increasing the occupancy by several orders of magnitude only leads to a slight increase in the jet quenching parameter.
This behavior is due to a combination of two effects. 
The first effect is that the increasing occupation number also increases the Debye mass $m_D$. Thus, a detailed understanding of screening effects is particularly important for a quantitative analysis of $\qhat$.
The second effect comes from dividing the value of $\qhat$ by the third power of the effective temperature $\Teps\sim N_+^{1/4}$, which increases with the occupancy when the hard momentum scale $T$ is kept fixed.
\begin{figure}
	\centering
	\includegraphics[width=0.5\linewidth]{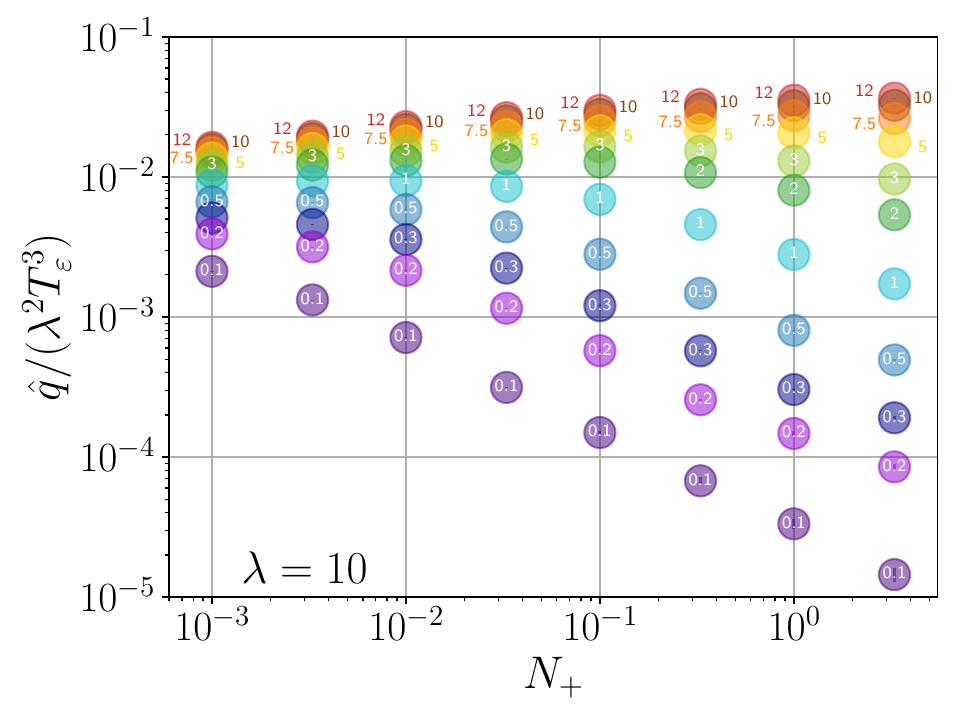}
	\caption{Jet quenching parameter $\qhat$ for a scaled Bose-Einstein distribution \eqref{eq:scaled_bose_fermi_combined} as a function of its amplitude $\NOg$ for different momentum cutoffs $\Lambda_\perp$ (numbers in circle markers) and coupling $\lambda=10$. Figure from \cite{Boguslavski:2023waw}.}
	\label{fig:qhat_scaled_all}
\end{figure}

\fig \ref{fig:qhat_scaled_all} provides an overview of the numerical values of $\qhat$ for the phenomenologically relevant coupling $\lambda = 10$ in heavy-ion collisions. 
Different values of the cutoff $\Lambda_\perp$
are color-coded and written in the circle markers in the figure. 
We observe the same behavior at small and large cutoffs that we have found in \fig \ref{fig:qhat_scaled_combined}. This involves a fast (power-law) decrease with growing occupancy $N_+$ at small cutoffs as $\qhat / (\lambda^2 \Teps^3) \sim N_+^{-3/4}$, and a slow growth at high cutoffs.
We additionally see how $\qhat$ interpolates smoothly between these two behaviors at small and large cutoffs. 
From a physical point of view, this confirms the observation that for small cutoffs, jet quenching in an over-occupied (isotropic) system similar to a scaled thermal distribution may be strongly suppressed. However, as stated below Eq.~\eqref{eq:invLNplus}, these parameters may lie beyond the range of applicability of the original integral formula for $\qhat$, Eq.~\eqref{eq:qhat_formula}.

\section{Obtaining $\qhat$ between Glasma and hydrodynamics\label{sec:obtaining-qhat-between-glasma-and-hydro}}
Finally, let us consider the jet quenching parameter $\hat q$ during a numerical simulation of the bottom-up thermalization process in heavy-ion collisions. This subsection is based on \cite{Boguslavski:2023alu}.

As already briefly discussed in the introduction, this is motivated by the large values of $\qhat$ reported during the Glasma stage \cite{Ipp:2020mjc, Ipp:2020nfu, Carrington:2020sww, Carrington:2021dvw, Carrington:2022bnv, Avramescu:2023qvv}, while other studies require it to be negligible at these earliest times \cite{Andres:2019eus}. Importantly, different jet quenching studies treat the earliest stages before hydrodynamics becomes applicable differently \cite{Andres:2019eus, Zigic:2019sth, Andres:2022bql, JETSCAPE:2023hqn} with partly different conclusions, showing the need for a better theoretical understanding of jet quenching during the initial stages.

\begin{figure}
    \centering
    \includegraphics[width=0.5\linewidth]{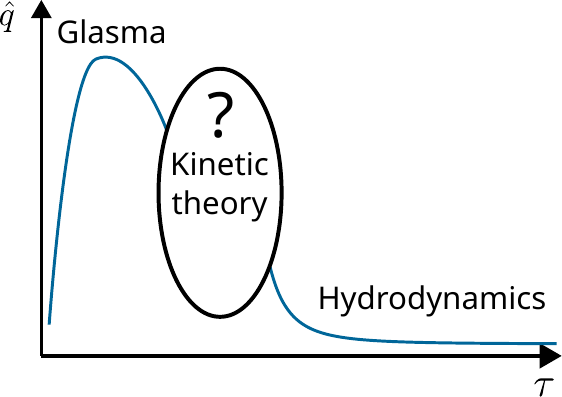}
    \caption{Sketch of the evolution of the jet quenching parameter $\hat q$ during the initial stages in heavy-ion collisions, depicting the goal of this section: To obtain the jet quenching parameter between the Glasma and hydrodynamic stage. Figure from \cite{Boguslavski:2023alu}.
    }
    \label{fig:qhat-schematic-glasma-hydro}
\end{figure}

The goal of this section is to obtain the jet quenching parameter $\qhat$ between the Glasma and hydrodynamic stage, as sketched in Fig.~\ref{fig:qhat-schematic-glasma-hydro}.
To do that, the Boltzmann equation \eqref{eq:actual-boltzmann-equation-to-solve-expanding-system}
\begin{align}
    -\pdv[f_{\vb p}]{\tau}+\frac{p_z}{\tau}\pdv[f_{\vb p}]{p_z}=\Conetwo[f_{\vb p}]+ \Ctwotwo [f_{\vb p}]\label{eq:boltzmann_equation_expanding}
\end{align}
is solved to numerically obtain the distribution function $f(\vb p,t)$ throughout the pre-equilibrium evolution, which is used as input for calculating the jet quenching parameter via Eq.~\eqref{eq:qhat_formula_pinf}. The distribution function is initialized with the initial conditions described in Section \ref{sec:boltzmann-equation-and-initial-condition} with anisotropy parameters $\xianiso\in\{4,10\}$. In the elastic collision kernel, soft-gluon exchanges are regulated using the Debye-like screening prescription as discussed in Section \ref{sec:debye-like-screening}.

\begin{figure}
\centerline{
\includegraphics[width=0.49\linewidth]{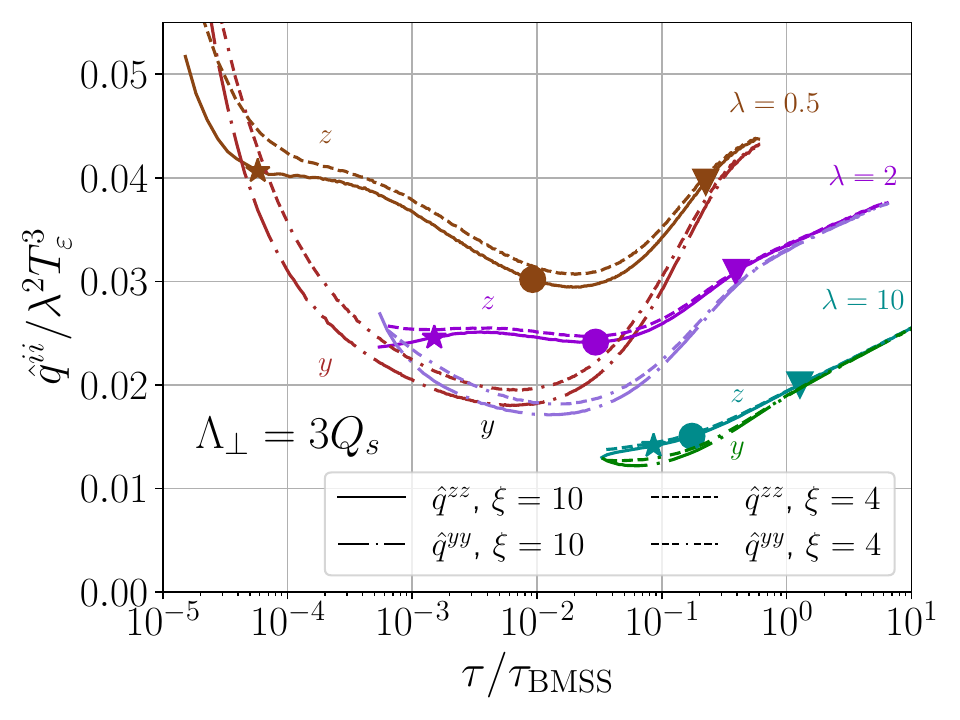}
}
\caption{\label{fig:qhat_couplings_aniso}     \label{fig:qhat_approach_to_thermal}
Jet quenching parameter in $y$ and $z$ directions rescaled by $\lambda^2 \Teps^3$ for momentum cutoff $\Lambda_\perp=3\Q$ for an expanding system for different couplings and initial conditions (solid: $\xi=10$, dashed: $\xi=4$).
Figure adapted from \cite{Boguslavski:2023alu}. 
}
\end{figure}
\begin{figure}
    \centering
    \centerline{
        \includegraphics[width=0.49\linewidth]{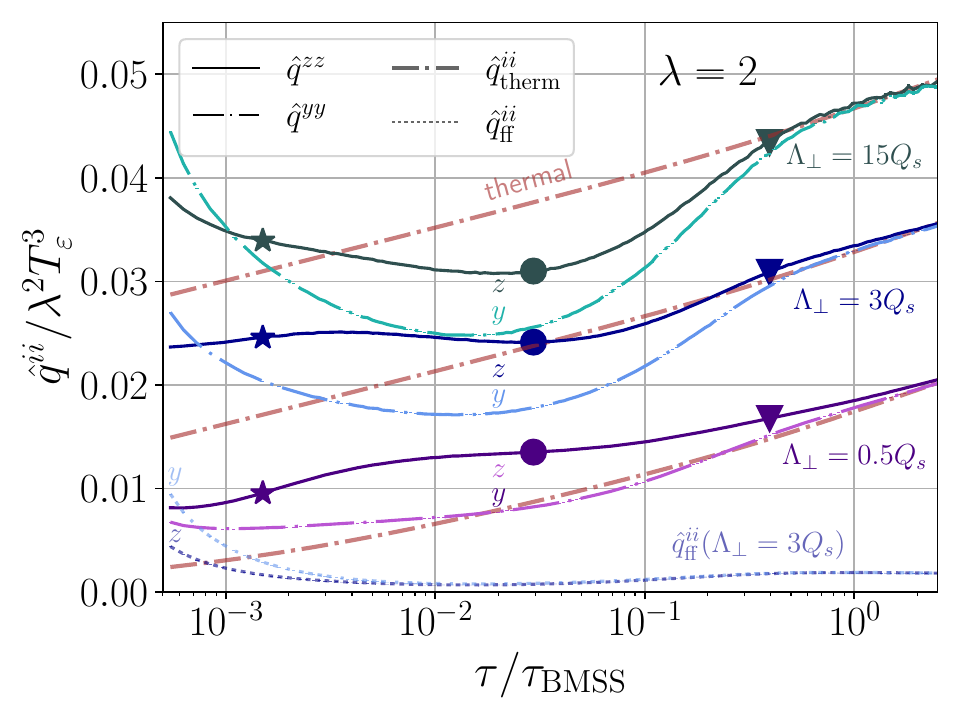}
        \includegraphics[width=0.49\linewidth]{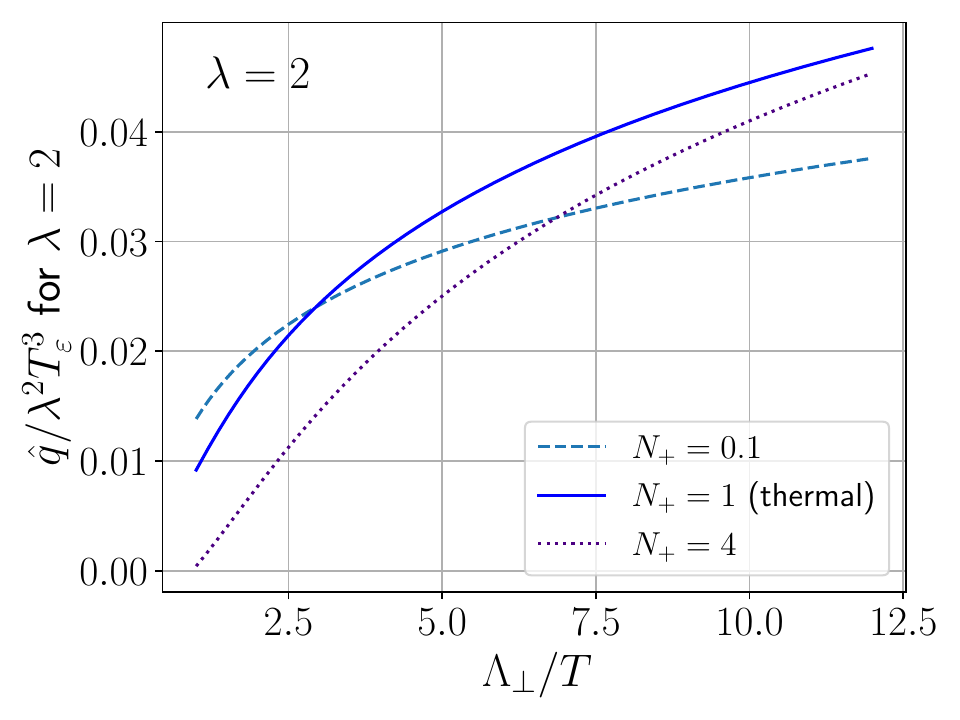}
    }
    \caption{
    (\emph{Left}): Evolution of $\qhat^{zz}$ and $\qhat^{yy}$ for a quark jet during bottom-up thermalization for $\lambda=2$ and different cutoffs $\Lambda_\perp$. The Bose enhanced contributions $\qhatff^{ii}$ for $\Lambda_\perp = 3\Q$ are added as dotted curves of the same color, visible at the bottom of the figure.
    For comparison, thermal curves for the same $\Teps(\tau)$ are shown as brown dash-dotted lines. Figure adapted from \cite{Boguslavski:2023alu}. 
    (\emph{Right}): Jet quenching parameter for a scaled thermal distribution with different scaling coefficients $N_+$ as a function of the transverse momentum cutoff $\lperp$. Figure from \cite{Boguslavski:2023waw}. 
    }
    \label{fig:different_cutoffs_comparison_thermal_qhat}
\end{figure}

\subsection{Results for a fixed momentum cutoff $\lperp$}
We start by discussing the resulting jet quenching parameter $\qhat$ at fixed transverse momentum cutoffs $\Lambda_\perp$ and later generalize this to more realistic models of evolving (time-dependent) momentum cutoffs. 
As discussed in Section \ref{sec:symmetries_of_qhat}, the mixed components $\qhat^{yz}=\qhat^{zy} = 0$ vanish due to symmetry arguments, so we need to focus here only on the diagonal components 
$\qhat^{zz}(\tau)$ and $\qhat^{yy}(\tau)$. 
They are plotted in \fig\ref{fig:qhat_couplings_aniso} for a cutoff $\Lambda_\perp = 3 \Q$ for different couplings and initial anisotropy parameters $\xianiso = 10$ (solid lines) and $\xianiso = 4$ (dashed lines). For these anisotropy parameters, we find little sensitivity regarding varying the initial conditions, less than $15\%$ for the considered parameters, and observe qualitatively similar behavior for different couplings. 

To further study the evolution of $\qhat^{ii}$, \fig \ref{fig:different_cutoffs_comparison_thermal_qhat} shows their values for different cutoffs $\Lambda_\perp$  for anisotropy parameter $\xianiso = 10$ and coupling $\lambda = 2$. Qualitatively similar effects can also be found for other couplings, and $\lambda=2$ is chosen for illustrative purposes.
The estimates for an energy-density matched (see Eq.~\eqref{eq:Landau-matching-condition}) thermal system $\qhat^{ii}_{\mathrm{therm}}=\frac{1}{2}\,\qhat_{\mathrm{therm}}$
are also shown as dash-dotted lines and are obtained by evaluating Eq.~\eqref{eq:qhat_formula_pinf} with a thermal distribution.
For that, the empirical formula \eqref{eq:empirical_qhat3} from Section \ref{sec:qhat-thermal} is used with the $\pmin$ corrected Debye mass \eqref{eq:debyemass-finitepmin} to account for discretization artifacts.
The contribution from the Bose-enhanced $\qhatff$ term is shown separately as dotted lines.
We observe that in general, the order of magnitude of  $\qhat^{ii}$ follows the energy-density matched thermal values. During the earliest stage of bottom-up thermalization, which is characterized by overoccupation and extreme anisotropy and consists of the time before the star markers, the results of $\qhat$ are above the energy-density matched thermal ones.

In the next stage, and in particular, when minimum occupancy is reached (marked by the circles), the values for large cutoffs $\Lambda_\perp$ undershoot the thermal ones, while those for a small cutoff overshoot them. 
This behavior can be understood by considering the scaled thermal distribution studied in the previous section \ref{sec:scaled-thermal-distribution}. The analytic expression for $\hat q$ for a scaled thermal distribution, Eq.~\eqref{eq:qhat_thermal_hard_analytic_combined_notimproved}, is shown for different occupancies in the right panel of Fig.~\ref{fig:different_cutoffs_comparison_thermal_qhat}. Similarly, as in the left panel, when increasing the cutoff $\lperp$, the thermal $\qhat$ first underpredicts the actual under-occupied $\qhat$, and this ordering reverses when the cutoff is increased. This simple toy model for an under-occupied distribution, thus, provides an intuitive explanation of the ordering of the thermal and nonthermal curves during the under-occupied regime in the bottom-up evolution.

Finally, approaching thermal equilibrium (signaled by the triangle markers), the values of $\qhat^{ii}$ also approach the thermal expectation. An interested reader might wonder why the thermal values seem to grow over time. This is because we consider $\hat q(\tau)$ at a fixed transverse momentum cutoff $\lperp$, but due to Bjorken expansion, all scales in the plasma continuously decrease. Thus, effectively, the cutoff increases and the rise observed in the jet quenching parameter comes from this increase.

For almost the entire evolution we find that momentum broadening in the beam direction is larger than transverse to it, $\qhat^{zz}>\qhat^{yy}$. 
This seems to be typical for anisotropic systems with occupancies up to order unity,
as has been found for transport coefficients in the context of kinetic theory \cite{Romatschke:2006bb, Dumitru:2007rp}. 
It should be emphasized that in our formulation, this ordering is a result of the anisotropic under-occupied distribution and, thus, does not stem from the matrix element for which an isotropic HTL screening prescription is used.
It leads to a sizable difference in the total momentum broadening in different directions. 
Moreover, a low momentum cutoff can be associated with the momentum broadening of the plasma constituents themselves. Thus the larger broadening in the $z$ direction for smaller $\Lambda_\perp$ is consistent with the isotropization dynamics in the bottom-up scenario. 
Remarkably, jet quenching studies in the Glasma \cite{Ipp:2020mjc, Ipp:2020nfu, Avramescu:2023qvv} have revealed a similar ordering $\qhat^{zz}>\qhat^{yy}$ as we find for most of the evolution of $\qhat$ in our kinetic simulations, although for a different reason. There the enhancement of $\qhat^{zz}$ seems to stem primarily from a slight asymmetry between the chromo-magnetic and -electric fields in the underlying classical-statistical description of the Glasma.

Interestingly, we find that for large cutoffs the ordering is reversed at early times before the star marker, leading to $\qhat^{zz}<\qhat^{yy}$. This mainly stems from the Bose enhancement of the over-occupied plasma phase at the beginning of the evolution. To see this, consider the Bose-enhancement part $\qhatff$ of $\qhat$ shown in \fig\ref{fig:qhat_approach_to_thermal} separately for the $y$ and $z$ directions. Note that $\qhatff$ is finite in the limit $\Lambda_\perp \to \infty$, and the value at $\Lambda_\perp=3\Q$ that is plotted is already close to that limit. While for the non-Bose enhanced term $\qhatf$, the anisotropy $p_z \ll p_\perp$ leads to $\qhatf^{zz}>\qhatf^{yy}$, for the Bose-enhanced term the effect is the opposite, $\qhatff^{zz}<\qhatff^{yy}$.
This ordering can be understood from studying the extremely anisotropic effectively two-dimensional distribution in Section \ref{sec:toy_models}. There, we found that in the limit of extreme anisotropy (no particles moving in the $z$ direction), $\qhat^{zz}_{\mathrm{ff}}=0$, which explains the ordering $\qhatff^{zz}< \qhatff^{yy}$ shown in Fig.~\ref{fig:different_cutoffs_comparison_thermal_qhat}.
\begin{figure}
    \centering
    \includegraphics[width=0.5\linewidth]{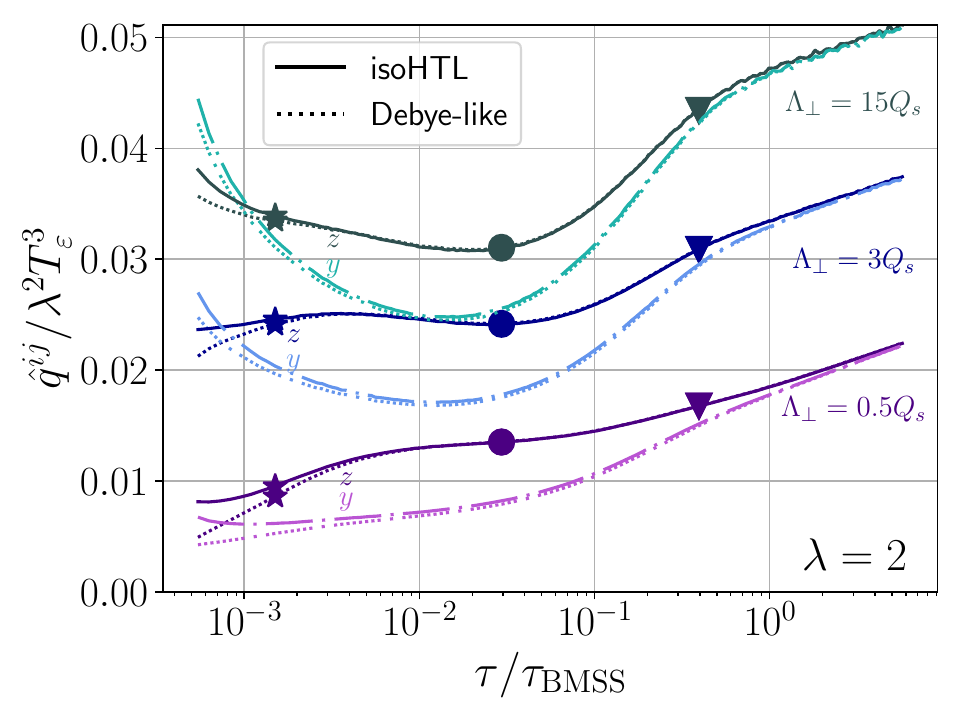}
    \caption{Components of the jet quenching parameter $\qhat^{ii}$ for differently screened matrix elements. Solid lines: isoHTL screening, dotted lines: Debye-like screening
    Figure adapted from Ref.~\cite{Boguslavski:2023alu}.
    }
    \label{fig:qhat_comparison_HTL}
\end{figure}

Finally, as in the previous sections, let us compare differently screened matrix elements for the evaluation of the jet quenching parameter $\qhat$. In Fig.~\ref{fig:qhat_comparison_HTL}, we show the components $\qhat^{ii}$ for both the isoHTL screened matrix element (\eqref{eq:full_htl_matrix_element}, solid lines) and the Debye-like screened matrix element (\eqref{eq:approximated_matrix_element}, dotted lines). We find that these two screening prescriptions only slightly differ at early times and for smaller cutoffs (as expected).

Note, however, that in the matrix element in the elastic collision term of the Boltzmann equation, always the Debye-like screened matrix element is used. We will study the effects of using the isoHTL screening for the time evolution in Chapter \ref{sec:improving-qcd-simulations}.


\subsection{Results for realistic cutoff dependence\label{sec:qhat_bottomup-cutoff-dependence}}
\subsubsection{Kinematic and LPM cutoff}
Until now, we have studied the jet quenching parameter $\qhat$ using a fixed cutoff $\Lambda_\perp$. This is 
unphysical since during the plasma expansion all characteristic energy scales decrease. To account for this, 
we choose 
cutoff models that depend on the jet energy $\Ejet$ and
effective
plasma temperature
$\Teps$, introduced and motivated in Section \ref{sec:momentum_cutoffs},
\begin{subequations}\label{eq:cutoff_combined}
\begin{align}
    \Lambda_\perp^{\mathrm{LPM}}(\Ejet,\Teps)&=\zeta^{\mathrm{LPM}}g\times(\Ejet \Teps^3)^{1/4},\label{eq:cutoff_LPM}\\
    \Lambda_\perp^{\mathrm{kin}}(\Ejet,\Teps)&=\zeta^{\mathrm{kin}}g\times(\Ejet \Teps)^{1/2}.\label{eq:cutoff_kinematic}
\end{align}
\end{subequations}
Recall that the first cutoff model $\Lambda_\perp^{\mathrm{LPM}}$ is a rough estimate of the accumulated transverse momentum during the formation time of a gluon emission during the LPM regime.
Variants of the kinematic
cutoff model $\Lambda_\perp^{\mathrm{kin}}$ have been widely used in the literature (see, e.g., \cite{Qin:2009gw, JET:2013cls, Xu:2014ica, He:2015pra, Cao:2021rpv, JETSCAPE:2021ehl,JETSCAPE:2022jer, Mehtar-Tani:2022zwf}) and take into account that the plasma particles the jet scatters off have momentum $k\sim T$.
These arguments only provide parametric estimates, so we have included a (so far unknown) proportionality constant $\zeta^{\dots}$, which will be determined by requiring that $\qhat$ reproduces some reference value.
It should be emphasized here that $\qhat$ is regarded as a medium property relevant for a jet with an appropriate fixed energy $\Ejet$ and we do not study the actual evolution of a jet. 

We do not expect any substantial differences between these two cutoff models \eqref{eq:cutoff_combined} since the dependence of $\qhat$ on the cutoff is only logarithmic for large $\Lambda_\perp$,
\begin{align}
    \qhat^{ii}(\Lambda_\perp\gg \Teps)\simeq a^i_{\lperp}\ln\frac{\Lambda_\perp}{\Q}+b_{\lperp}^i. \label{eq:qhat_form_large_cutoff}
\end{align}
In practice, we fit the coefficients $a_{\lperp}^i$ and $b_{\lperp}^i$ to the large cutoff behavior of the numerically obtained values for $\qhat$. Although, as discussed in Section \ref{sec:qhat_behavior_large_cutoff}, the coefficients $a^i_{\lperp}$ should be the same for different directions, we employ here a more general parametrization, allowing for differences in different directions. The resulting coefficients $a^y_{\lperp}$ and $a^z_{\lperp}$ differ only slightly.
The numerical values of $a^i_{\lperp}/\Q^3$ and $b^i_{\lperp}/\Q^3$ as functions of $\Q\tau$ for $\lambda=0.5, 1, 2, 5$ and $10$ for the initial conditions in \eq\eqref{eq:initial_cond} are publicly availabe in \cite{lindenbauer_2023_10419537}.

\subsubsection{Matching the Glasma: Problems and strategy employed here}
One of our goals is to make contact with the large $\qhat$ values reported in the Glasma \cite{Ipp:2020mjc, Ipp:2020nfu, Avramescu:2023qvv, Carrington:2021dvw}, and to assess whether these large values reported there are plausible from a kinetic theory perspective. 
To achieve that, ideally, we would take at some time, e.g., $\tau_0=1/\Q$, the full configuration obtained in a classical statistical simulation of the Glasma, extract a gluon distribution function (as, e.g., done in Ref.~\cite{Greif:2017bnr, Greif:2020rhi}) and use this as input for our kinetic theory simulation. Performing this matching seems feasible in principle but is very complicated in practice, particularly because the Glasma simulations that are used to obtain values for the jet quenching parameter $\qhat$ in Refs.~\cite{Ipp:2020nfu, Avramescu:2023qvv} employ a boost-invariant approximation (they are effectively 2+1D simulations). From that, we cannot immediately obtain any information about the $p_z$ dependence of the gluon distribution function $f(\vb p, \tau_0)$, which, physically, broadens due to plasma instabilities \cite{Mrowczynski:1993qm, Epelbaum:2013waa, Berges:2014yta}. 

As discussed in Section \ref{sec:boltzmann-equation-and-initial-condition}, as a first step,
a parameterization of the JIMWLK evolved (2+1D) Glasma result from Ref.~\cite{Lappi:2011ju} is used, which we also use as initial condition \eqref{eq:initial_cond}, similar to what is done in Ref.~\cite{Kurkela:2015qoa}.
There, the $p_z$ dependence is then obtained by ``de-''squeezing the transverse momentum $p_\perp \to \sqrt{p_\perp^2+\xianiso^2p_z^2}$.
To make a meaningful comparison to the Glasma results, the value of $\Q$ is chosen such that the energy density of the Glasma in Ref.~\cite{Ipp:2020nfu} is reproduced at the initial time $\Q\tau=1$ for coupling $\lambda=10$.
We allow for a different value of $\Q^{\mathrm{Glasma}}$ in Glasma simulations that might not correspond to the value of $\Q$ we use due to different definitions and conventions. 
For the Glasma simulation in Ref.~\cite{Ipp:2020nfu}, $\Q^{\mathrm{Glasma}}=\SI{2}{\giga\electronvolt}$ and $m/(g^2\mu)=0.1$ with $50$ color sheets have been used, where $g^2\mu$ and $\Q^{\mathrm{Glasma}}$ are related as in Ref.~\cite{Lappi:2007ku}.
Performing the matching of the energy density at $\Q\tau=1$ (which is an implicit equation requiring as input the energy density $\varepsilon^{\mathrm{Glasma}}(\tau)$ from Ref.~\cite{Ipp:2020nfu}) yields a value of $\Q=\SI{1.4}{\giga\electronvolt}$. Remarkably, this is the same value as obtained in Ref.~\cite{Keegan:2016cpi}, where it was found that precisely this value is needed 
for the EKT setup to be consistent with the later hydrodynamic evolution. This shows the consistency of both approaches.

To summarize, the matching is performed both in the form of the distribution function (parameterized and squeezed \cite{Kurkela:2015qoa}) and the energy density.

\subsubsection{Numerical results}
We fix the values of the parameters $\zeta^{\dots}$ in \eq\eqref{eq:cutoff_combined}
at the triangle marker close to thermal equilibrium. At this time, where the temperature is $\Teps=0.21\Q=\SI{295}{\mega\electronvolt}$ for a realistic coupling $\lambda=10$.
More concretely, we require the jet quenching parameter $\hat q$ to match the median value for $\qhat_{\mathrm{therm}}$ in the LBT parametrization of Ref.~\cite{JETSCAPE:2021ehl} by the JETSCAPE collaboration in order to be close to a traditional numerical estimate for a thermal distribution. For $\Ejet=\SI{100}{\giga\electronvolt}$, one finds the numerical values $\zeta^{\mathrm{LPM}}=0.70$ and $\zeta^{\mathrm{kin}}=0.16$, whereas for $\Ejet=\SI{20}{\giga\electronvolt}$ the values $\zeta^{\mathrm{LPM}}=1.14$ and $\zeta^{\mathrm{kin}}=0.40$ are obtained.
We will discuss different matching conditions in the next subsection, which, importantly, do not significantly change the results.

Fig.~\ref{fig:qhat_realisticv5a} shows the resulting $\qhat^{zz}$ and $\qhat^{yy}$ for the anisotropy parameter $\xianiso=10$ and jet energy $\Ejet = \SI{100}{\giga\electronvolt}$. Both cutoff parametrizations of \eq \eqref{eq:cutoff_combined} are shown for different values of the coupling $\lambda$ (color-coded).
We find that the evolution of $\qhat^{ii}$ is similar as for fixed cutoffs, with $\qhat^{zz}>\qhat^{yy}$ for most of the nonequilibrium evolution except for a short period of reversed ordering at the beginning for $\lambda = 2$. In particular, longitudinal momentum broadening is more efficient than expected from a thermal system, as indicated by the dashed thermal line for comparison, which we only show for the LPM cutoff \eqref{eq:cutoff_LPM}.
As anticipated, both cutoff models lead to similar results, with the LPM cutoff yielding systematically larger values than the kinematic cutoff model during the pre-hydrodynamic evolution. For $\lambda = 10$ they differ by less than $20\%$, while the variation of initial anisotropies has a much smaller impact, as we have seen in \fig\ref{fig:qhat_couplings_aniso}. 

\begin{figure}
  \centering
  \includegraphics[width=0.5\linewidth]{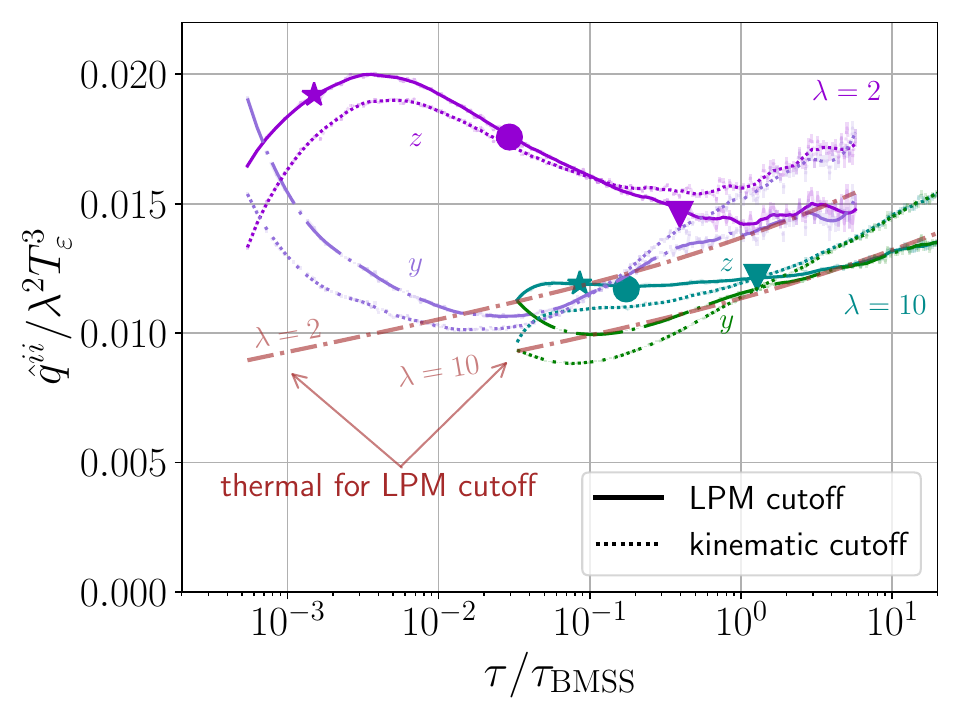}
  \caption{\label{fig:qhat_realisticv5a}
  Evolution of $\qhat^{zz}$ and $\qhat^{yy}$ for the cutoff models in \eqref{eq:cutoff_LPM} (solid) and \eqref{eq:cutoff_kinematic} (dashed) with jet energy $\Ejet=\SI{100}{\giga\electronvolt}$, $\Q=\SI{1.4}{\giga\electronvolt}$
  and $\Teps$ extracted from the plasma simulation for $\xi=10$. 
  The curves were smoothed using a Savitzky-Golay filter, while the original curves with estimated error bars are shown transparently beneath. 
  Thermal curves for the LPM cutoff model are included for comparison. Figure adapted from \cite{Boguslavski:2023alu}.
  }
\end{figure}

\begin{figure}
    \centering
    \includegraphics[width=0.49\linewidth]{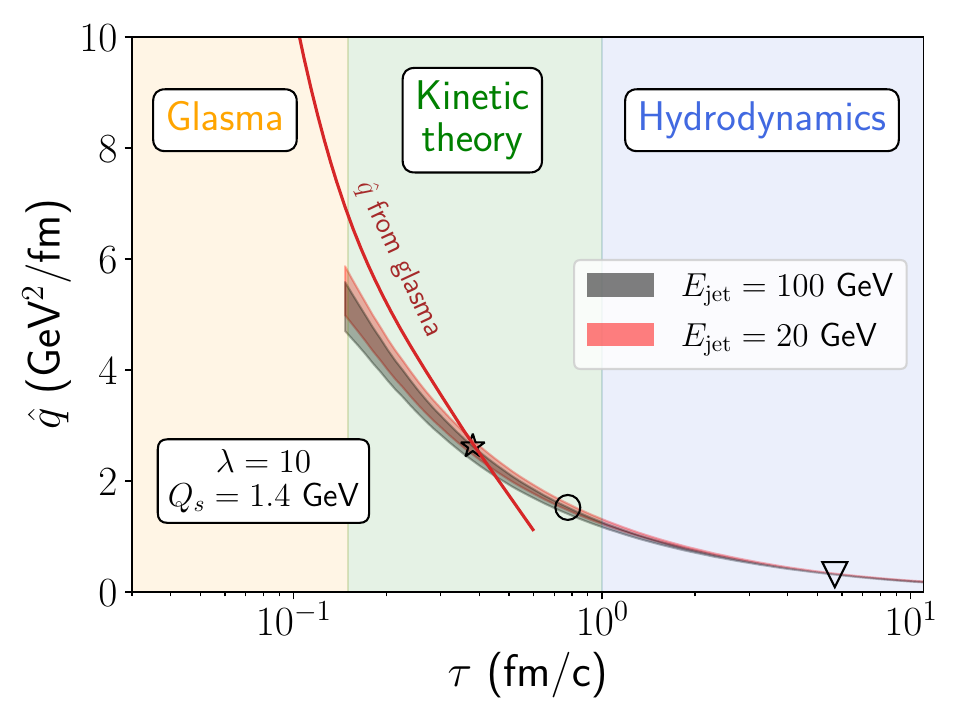}
    \includegraphics[width=0.49\linewidth]{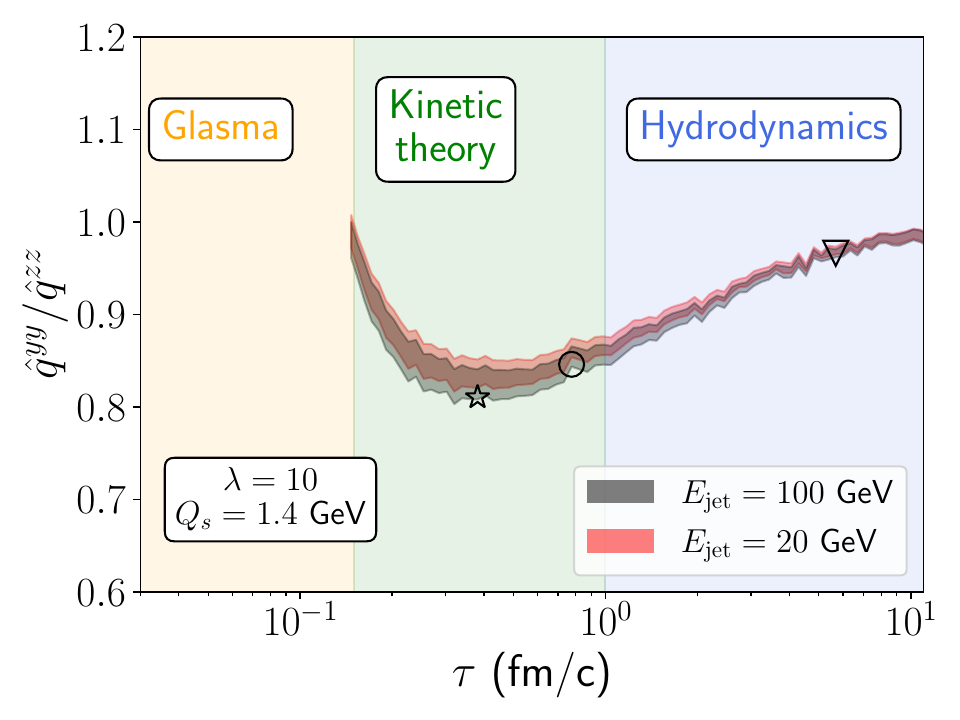}
    \caption{Jet quenching parameter $\hat q$ during the initial stages in heavy-ion collisions. Different regions with different effective descriptions are color-coded. (\emph{Left}): Jet quenching parameter for different jet energies and varying cutoff models (bands) and comparison to the Glasma results from \cite{Ipp:2020nfu}. (\emph{Right}): Anisotropy ratio $\qhat^{yy}/\qhat^{zz}$ during the initial stages. Figures reused from Ref.~\cite{Boguslavski:2023alu}.}
    \label{fig:qhat_stages}
\end{figure}

This relatively mild sensitivity to the initial parameters and cutoff models
enables us to obtain results for $\qhat$ throughout the pre-equilibrium stages, effectively extrapolating backward from a fit to a phenomenological extraction by the JETSCAPE collaboration at the triangle time marker close to equilibrium. 
The results of this procedure are shown in Fig.~\ref{fig:qhat_stages} for
jet energies $\Ejet = {\SI{20}{\giga\electronvolt}}$ and $\SI{100}{\giga\electronvolt}$. Variations in cutoff models and initial conditions are represented in the bands.
With the procedure described above to fix the cutoff, we find that the bands for different jet energies almost overlap.

Additionally, the numerical results of the Glasma calculation of Ref.~\cite{Ipp:2020nfu} are also included in Fig.~\ref{fig:qhat_stages}. We observe that at $\tau \sim \SI{0.2}{\femto\meter}/c$, the jet quenching parameter $\qhat \approx 4 - \SI{5}{\giga\electronvolt^2/\femto\meter}$, which is comparable (but not identical) to the values during the Glasma regime at this time. It should be noted that we do not precisely match the Glasma distribution function at our initialization time, and this matching should be regarded as a crude estimate. In particular, a similar matching procedure employed in Ref.~\cite{Boguslavski:2023fdm} by the same authors leads to a much worse matching for a closely related transport parameter, the \textit{heavy-quark diffusion coefficient} $\kappa$. This shows that the matching procedure here is perhaps too crude for a quantitative matching of the transport coefficients. However, this does not change the message of Ref.~\cite{Boguslavski:2023alu} that the seemingly large values of $\qhat$ during the Glasma phase are plausible from a kinetic theory perspective.

Finally, the right panel shows the ratio of the jet broadening coefficients $\qhat^{yy}/\qhat^{zz}$ for these times, showing a suppression of up to 20\% between the Glasma and the hydrodynamic regimes.

Note that the boundary between kinetic theory and hydrodynamics in Fig.~\ref{fig:qhat_stages} is---somewhat arbitrarily---chosen to be $\tau=\SI{1}{\femto\meter/\c}$ to give an estimate of where typical hydrodynamic evolution is expected to be valid. The actual numerical values shown in the plots are all obtained using kinetic theory, as discussed in Section \ref{sec:performing-qcd-kinetic-theory-simulations}, 
and it should be noted that at late times QCD kinetic theory simulations are indistinguishable from hydrodynamics (see Fig.~\ref{fig:pressureratio-approach-hydro}).

\subsubsection{Matching different cutoffs}
\begin{figure}
    \centering
    \centerline{
        \includegraphics[width=0.49\linewidth]{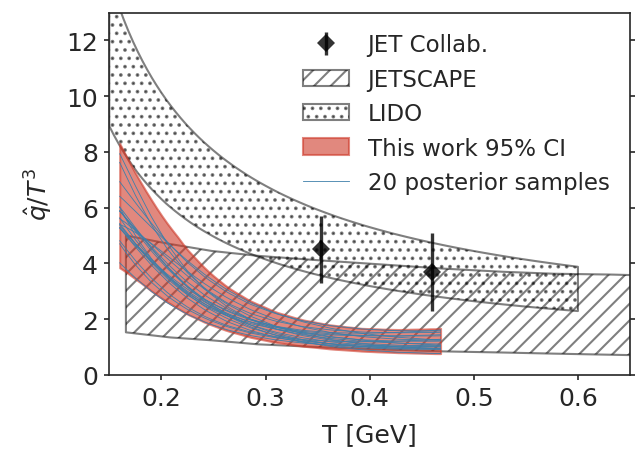}
        \includegraphics[width=0.49\linewidth]{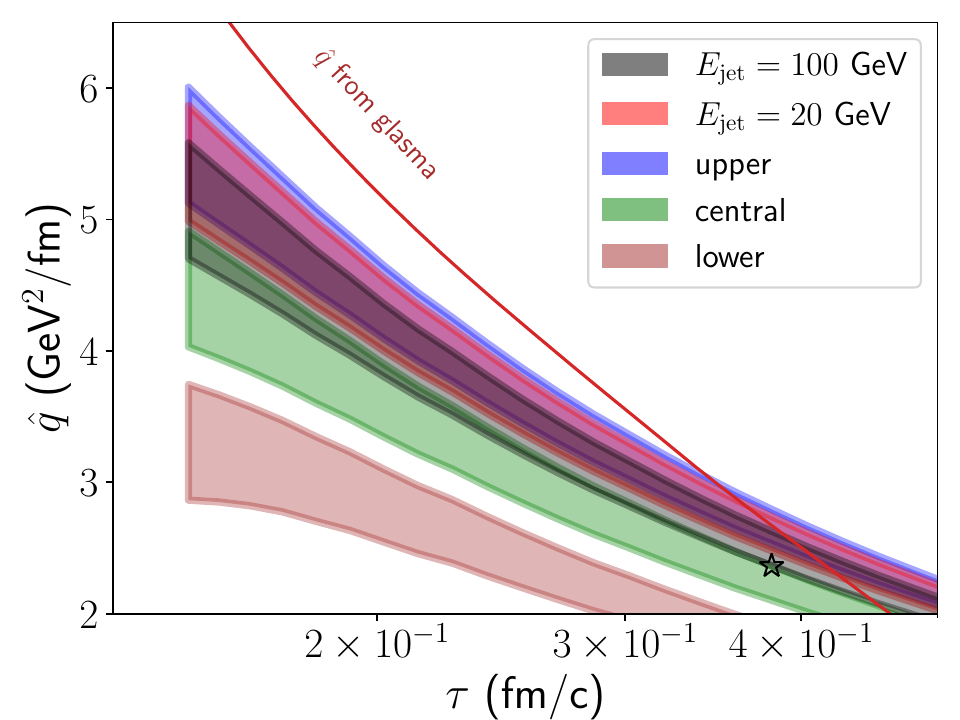}
    }
    \caption{
    (\emph{Left}): Temperature dependence of $\qhat$ resulting from an information field based Bayesion inference. Figure from \cite{Xie:2022ght}.
    (\emph{Right}): Early-time behavior of $\hat q$ (jet energy $\SI{50}{\giga\electronvolt}$) for different cutoff matchings from Ref.~{\cite{Xie:2022ght}}, as explained in the text. Figure from \cite{Boguslavski:2023alu}.}
    \label{fig:different_parametrization_included}
\end{figure}
A perhaps unsatisfying aspect of the previous discussion is that in order to present numerical results, we were required to fix the proportionality constants introduced in \eqref{eq:cutoff_combined} to reproduce a reference value of the jet quenching parameter $\qhat$. For that, we chose to use the JETSCAPE result \cite{JETSCAPE:2021ehl} obtained from Bayesian inference, because a convenient and simple parameterization is provided there. However, it is an interesting question to which extent the results would change had we chosen to fix the cutoff in a different way.
Here, we explore matching to the values from Ref.~{\cite{Xie:2022ght}} for comparison, which are depicted in the left panel of Fig.~\ref{fig:different_parametrization_included}. At the matching temperature $\Teps={\SI{295}{\mega\electronvolt}}$, their curves are consistent with the LBT model employed in the previous matching but exhibit a larger spread.
To quantify the uncertainty in the extraction, we choose upper ($\qhat = 2.5 T^3$), central ($\qhat = 2 T^3$), and lower ($\qhat = 1.5 T^3$) parts of the error band in the left panel of Fig.~\ref{fig:different_parametrization_included}. 
While these values are independent of the jet energy, we choose $\Ejet={\SI{50}{\giga\electronvolt}}$ for Eqs.~{\eqref{eq:cutoff_combined}} to make a concrete comparison.
The results (for the early times) are shown in the right panel of Fig.~\ref{fig:different_parametrization_included}, where we observe that the upper and central values are compatible with the JETSCAPE parametrization and also consistent with the Glasma values, whereas the lower bound yields slightly smaller values.


\section{Concluding remarks}
In this chapter, we studied how the jet quenching parameter $\qhat$ can be obtained within QCD kinetic theory, both for finite jet energy and for the limit of an infinite jet energy with a transverse momentum cutoff $\lperp$. We generalized earlier analytic calculations of this parameter to toy models relevant to the bottom-up thermalization process in heavy-ion collisions. In particular, we considered a scaled thermal distribution to model an under-occupied stage, and an effectively two-dimensional distribution, modeling the initial large anisotropy in heavy-ion collisions. Comparing with numerical results, we found that different screening prescriptions lead to different results for small couplings or cutoffs. In particular, we investigated HTL and Debye-like screening in the matrix element needed for the evaluation of $\qhat$. Furthermore, we found that the previous analytic expressions, which are derived in the weakly coupled limit, do not accurately describe the jet quenching parameter when extrapolated to larger values of the coupling, for which we provided a convenient parameterization.

Finally, we went on to study the jet quenching parameter $\qhat$ during the Glasma and hydrodynamic phase at the early stages in heavy-ion collisions.
We find that the numerical results for the jet quenching parameter $\qhat$ follow roughly Landau-matched thermal estimates while exhibiting a sizable anisotropy $\qhat^{zz}>\qhat^{yy}$ for most of the pre-equilibrium evolution. This emphasizes the importance of going beyond approximating $\qhat$ with its equilibrium value, which seems to systematically underestimate $\qhat$ and is isotropic by construction.

The results of $\qhat$ during the kinetic regime could be used in the future to extend current frameworks that employ a hydrodynamic medium evolution to extract $\qhat$ from experimental data \cite{JET:2013cls, JETSCAPE:2021ehl, Xie:2022ght}. 
Although based on scattering processes with on-shell partons, the extracted values for $\qhat$ can also enter jet evolution models in order to include medium effects during the initial large virtuality phase \cite{JETSCAPE:2023hqn}.

The anisotropy $\qhat^{zz} > \qhat^{yy}$ remains during most of the pre-hydrodynamic evolution including the Glasma and kinetic stages, and, thus, may leave imprints on experimental observables. In particular, it has been pointed out that such anisotropies in the jet quenching parameter lead to jet hadron polarization \cite{Hauksson:2023tze}, and may be probed experimentally by studying spin-polarized and azimuthal jet observables \cite{Barata:2024bqp}.

In Chapter \ref{sec:collkern}, we will consider the collision kernel $C(\vb q_\perp)$, which can be thought of as a generalization of the jet quenching parameter, during the initial stages in heavy-ion collisions.

%% file: 470_limiting_attractors.tex
As discussed in the introduction, it has been found that the time evolution of the nonequilibrium QCD plasma generated in heavy-ion collision quickly follows a universal curve, the hydrodynamic attractor. This was first seen in the context of hydrodynamics, where the pressure anisotropy for various different initial conditions quickly collapses on the universal hydrodynamic attractor curve. Moreover, it has been found that a similar attractor exists for kinetic theory simulations at different couplings, when the time is rescaled with the relaxation time $\tauR$ from Eq.~\eqref{eq:relaxation-time}, $\tauR=4\pi\eta/s/T$ \cite{Keegan:2015avk, Kurkela:2018wud}. However, as discussed in Section \ref{sec:thermalization-expanding-systems}, there is another time scale involved when studying the bottom-up equilibration in heavy-ion collisions from kinetic theory. There, the equilibration time scales parametrically as $\tauT\sim \alpha_s^{-13/5}$. In this chapter, we discuss the relevance of these different time scales for the pressure ratio, jet quenching parameter and the \emph{heavy-quark diffusion coefficient}, which is a related quantity describing the momentum diffusion of heavy-quarks in the quark-gluon plasma.

This chapter is based on Ref.~\cite{Boguslavski:2023jvg}.

\section{Time scales and initial conditions}
In Section \ref{sec:thermalization-expanding-systems}, we discussed how, in a weak-coupling picture, the out-of-equilibrium plasma created in heavy-ion collisions reaches equilibrium according to the bottom-up hydrodynamization picture \cite{Baier:2000sb}.
Within this picture, the time scale for thermalization is given by the parametric estimate $\tauT$ from Eq.~\eqref{eq:bottom-up-timescale} (restated below), while first-order hydrodynamics is governed by the time scale $\tauR$, which is related to the shear viscosity over entropy ratio $\eta/s$ (see Eq.~\eqref{eq:relaxation-time}),
\begin{align}
	\tauT(\lambda) = \alpha_s^{-13/5}/Q_s, && \tauR(\lambda, \tau) =  \dfrac{4 \pi \,\eta/s(\lambda)}{\Teps(\tau)}.
	\label{eq:timescales-attractors}
\end{align} 
Recall the relation for the coupling constant $\alpha_s = \lambda /(4\pi \NC)$. 
The numerical values of\footnote{In this thesis, only the ratio of the shear viscosity over entropy density $\eta/s$ is considered, and this ratio depends on the coupling $\lambda$.} $\eta/s(\lambda)$ can be obtained by using a functional basis \cite{Arnold:2003zc} or by the late-time behavior of the pressure ratio (e.g., in \cite{Kurkela:2018vqr}). We will discuss how to extract $\eta/s$ in more detail in Chapter \ref{sec:improving-qcd-simulations}. For now, the values from Ref.~\cite{Keegan:2015avk} are used (given for $1\leq \lambda \leq 10$). For $\lambda=0.5$ and $\lambda=20$, the values used here have been extracted\footnote{The values used here slightly differ from the values extracted in Section \ref{sec:extract_etas}. This is because the current chapter is based on Ref.~\cite{Boguslavski:2023jvg}, which appeared earlier than Ref.~\cite{Boguslavski:2024kbd}, upon which Chapter \ref{sec:improving-qcd-simulations} is based, where $\eta/s(\lambda=0.5)\approx 77$ and $\eta/s(\lambda=20)\approx 0.2$ is found using a more rigorous analysis.} using the methods described in Chapter \ref{sec:improving-qcd-simulations},
\begin{align}
    \frac{\eta}{s}(\lambda{=}0.5) = 80\,, \qquad \frac{\eta}{s}(\lambda{=}20) = 0.22\,.
\end{align}

The equilibration process is simulated using QCD kinetic theory, as already employed in the previous Chapter \ref{sec:momentum-broadening-of-jets}. More details on the exact setup are detailed in Section \ref{sec:performing-qcd-kinetic-theory-simulations}. For that, the Boltzmann equation \eqref{eq:actual-boltzmann-equation-to-solve-expanding-system}
\begin{align}
    \left(\pdv{\tau}-\frac{p_z}{\tau}\pdv{p_z}\right)f(\tau,\vb p)=-\Conetwo[f(\tau,\vb p)]-\Ctwotwo[f(\tau,\vb p)],\label{eq:actual-boltzmann-equation-to-solve-expanding-system-limmiting-attractors}
\end{align}
is numerically solved to obtain the time evolution of the gluon distribution function $f(\vb p,\tau)$. Only gluons are considered, which are the dominant degrees of freedom for equilibration and hydrodynamization. As initial conditions, we employ Eq.~\eqref{eq:initial_cond}, with $\xianiso\in\{4,10\}$. In the plots, also the time markers introduced in Section \ref{sec:time-markers-and-scales} are added: The star marker is placed when the occupancy drops below unity, the circle marker at minimum occupancy and the triangle marker at the time where the pressure ratio $P_L/P_T=0.5$.

\begin{figure}
	\centering
    \centerline{
	   \includegraphics[width=0.5\linewidth]{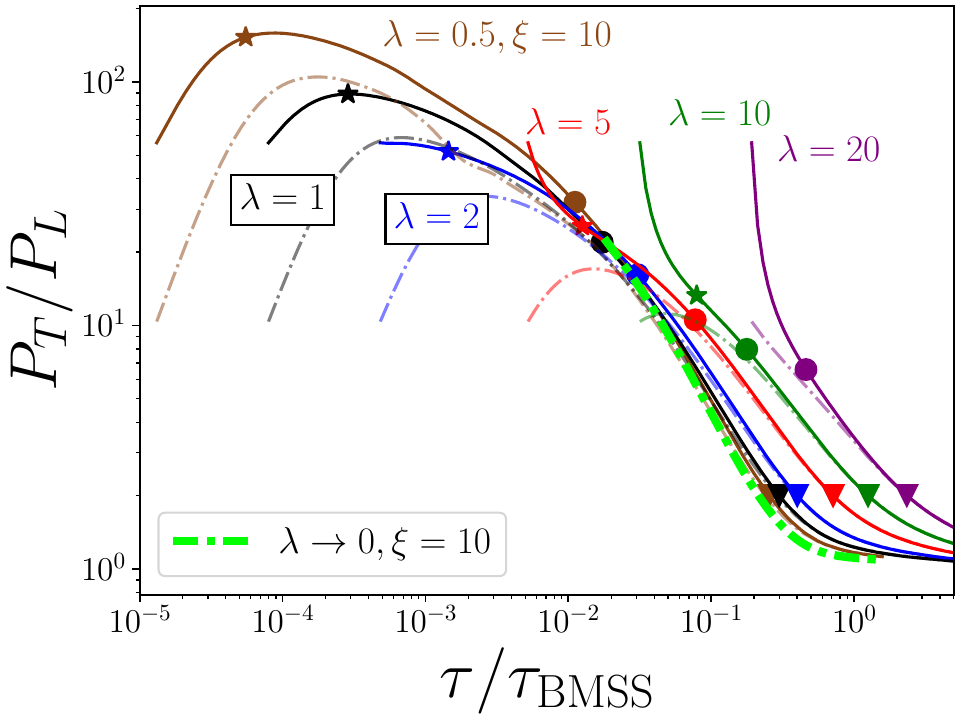}
	   \includegraphics[width=0.5\linewidth]{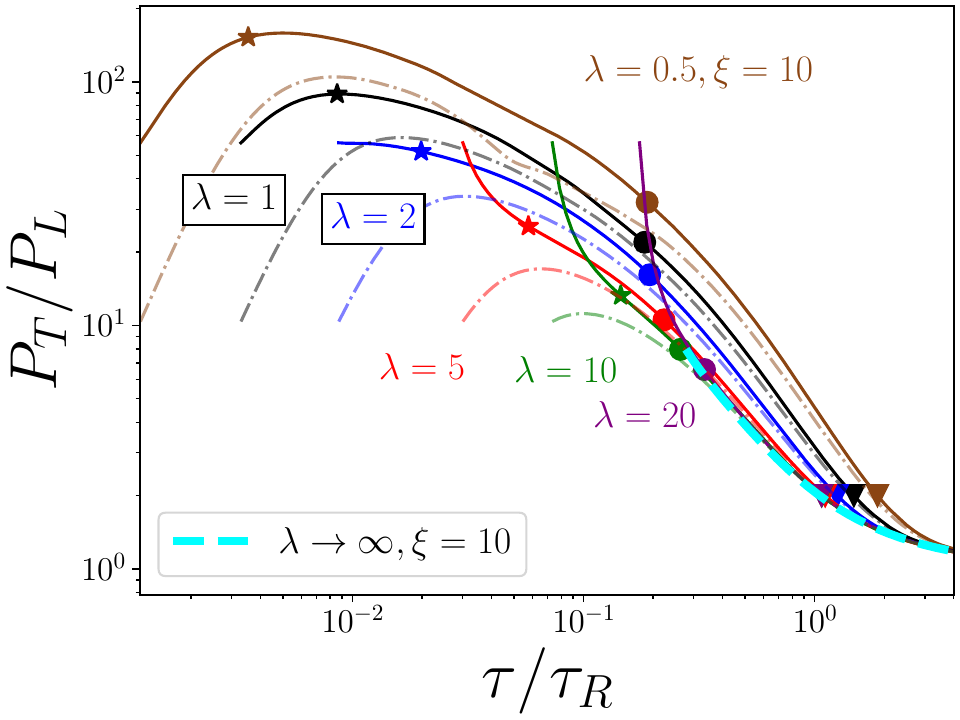}
    }
	\caption{Pressure ratio as functions of $\tau/\tauT$ (left) and $\tau/\tauR$ (right). 
		The extrapolations to vanishing and infinite coupling (the limiting attractors) are performed for each value of $\tau$ as demonstrated in \fig \ref{fig:PTPLratioAndFits}, and are denoted as thick dashed lines. Figures from Ref.~\cite{Boguslavski:2023jvg}.
	}
	\label{fig:pressureRatioVsToverTauR}
\end{figure}

\begin{figure}
	\centering
    \centerline{
	\includegraphics[width=0.5\linewidth]{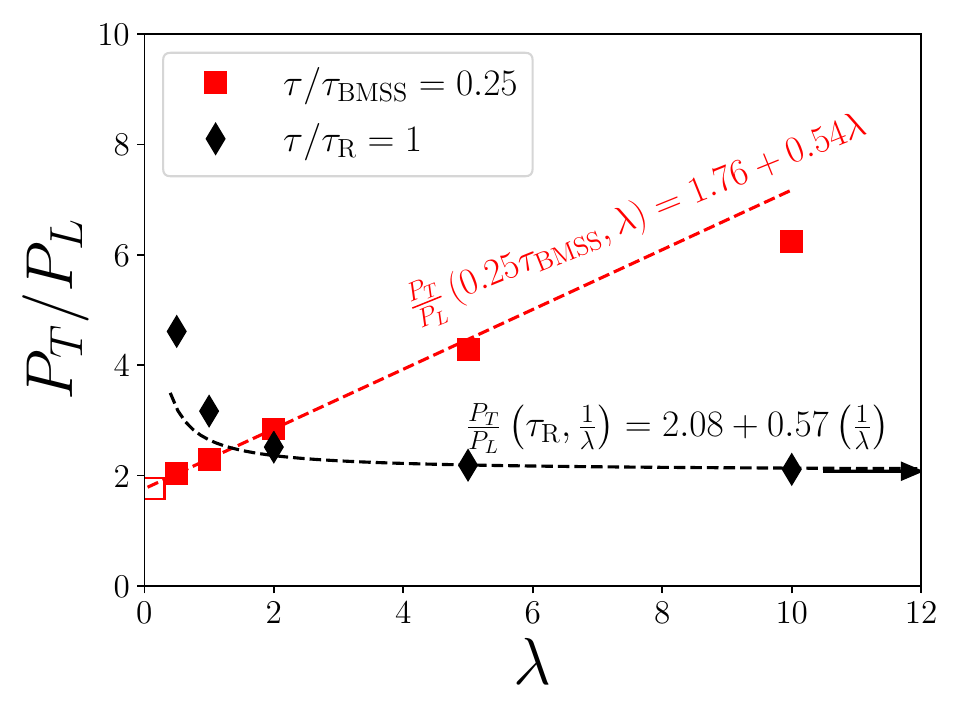}
    \includegraphics[width=0.5\linewidth]{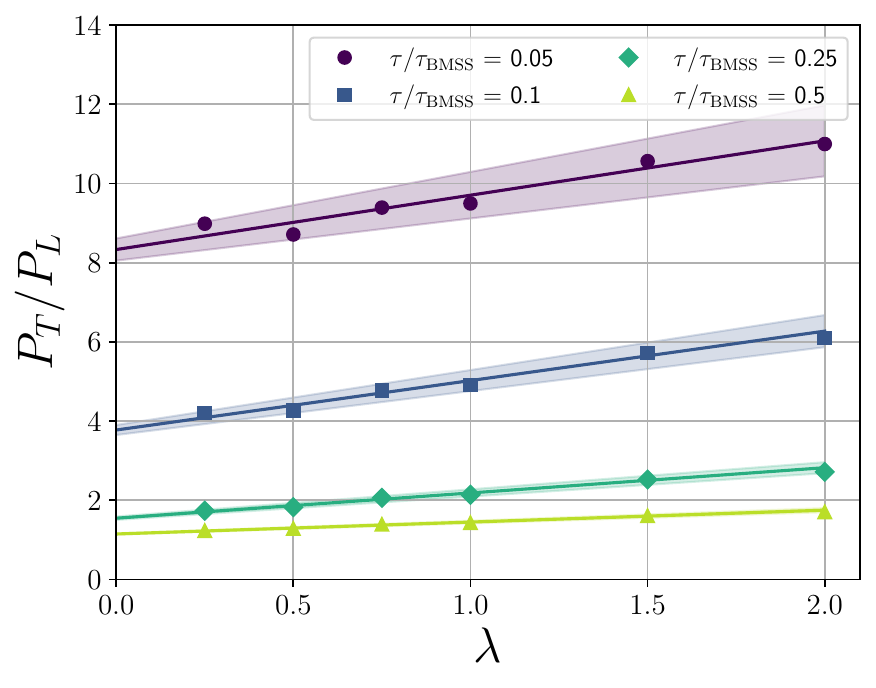}
    }
	\caption{(Left): Pressure ratio at a fixed time as a function of the coupling $\lambda$ in the units of both time scales. We also show the fits and their parametrizations, which are used to extrapolate to vanishing coupling (empty square) or infinite coupling (black arrow). Figure taken from \cite{Boguslavski:2023jvg}.
    (Right): Extrapolation procedure for the bottom-up limiting attractor for different times. Additional data points for $\lambda\in\{0.25, 0.75, 1.5\}$.
	}
	\label{fig:PTPLratioAndFits}
\end{figure}

\section{Limiting attractors in the pressure ratio}
\label{sec:occupnumberpressratio}
First, let us consider the pressure ratio $P_T/P_L$, obtained from the diagonal components of the energy-momentum tensor $T^{\mu\nu}$ from Eq.~\eqref{eq:energy-momentum-tensor-from-f}.
Figure \ref{fig:pressureRatioVsToverTauR} shows the pressure ratio in units of both time scales: in the left panel as a function of $\tau/\tauT$
and in the right panel as a function of $\tau/\tauR$.
We observe that for each coupling, the curves from different initial conditions with $\xianiso = 4$ and $10$ approach each other.
This collapse of curves for various different initial conditions for a single coupling has already been extensively studied in the literature \cite{Heller:2015dha, Almaalol:2020rnu, Du:2022bel}.
In the context of this chapter, it will be more interesting to compare simulations at different values of the coupling.

Let us start our discussion with the pressure ratio in the right panel of Fig.~\ref{fig:pressureRatioVsToverTauR}, where the time is rescaled with the relaxation time $\tauR$.
For sufficiently large values of the coupling (consider $\lambda\geq 5$), the curves become indistinguishable after $\tau/\tauR\gtrsim 0.3$.
This serves as motivation to define a \emph{limiting attractor} for large couplings, which can be obtained by extrapolating $\lambda\to\infty$ linearly in $1/\lambda$, i.e.,
\begin{align}
    \frac{P_T}{P_L}(\tau/\tauR)= a(\tau/\tauR)+b(\tau/\tauR)\frac{1}{\lambda}, \label{eq:extrapolate_infinite_coupling}
\end{align}
where $a$ and $b$ are functions of the rescaled time $\tau/\tauR$.
We define the value of the limiting large coupling (hydrodynamic) attractor as the value of the coefficient $a(\tau/\tauR)$, which is shown as a light-blue dashed line in Fig.~\ref{fig:pressureRatioVsToverTauR}. 
How quickly this hydrodynamic limiting attractor is approached, depends on the value of the coupling.
For large values $\lambda\geq 5$, the approach occurs already 
close to the circle markers.
In contrast, for weaker couplings $\lambda \leq 2$, we find that the hydrodynamic limiting attractor is approached at a significantly later time, even after the triangle marker.

In contrast to this,
we show the pressure ratio plotted as a function of the time variable rescaled with the bottom-up time scale $\tauT$ in the left panel of \fig\ref{fig:pressureRatioVsToverTauR}.
There, we observe
a rather opposite picture, where the curves corresponding to smaller couplings seem to approach each other, even when the system is still far from equilibrium. Similarly as in the extrapolation to infinite coupling \eqref{eq:extrapolate_infinite_coupling}, we define a limiting weak-coupling \emph{bottom-up} attractor obtained by extrapolating to vanishing coupling,
\begin{align}
        \frac{P_T}{P_L}(\tau/\tauT)= a(\tau/\tauT)+b(\tau/\tauT)\times{\lambda} ,\label{eq:extrapolate_zero_coupling}
\end{align}
with two different functions $a$ and $b$ of the rescaled time $\tau/\tauT$.
Similarly, as before, the coefficient $a(\tau/\tauT)$ gives the numerical value of the bottom-up limiting attractor and is shown as a green dash-dotted line in the figure.

We explicitly illustrate the extrapolation procedure that constructs the limiting attractors visible in \fig \ref{fig:PTPLratioAndFits}. 
The left panel shows for each coupling $\lambda$ the corresponding value of the pressure ratio $P_T/P_L$ at a fixed rescaled time $\tau/\tauT = 0.25$. 
Performing the linear fit from \eqref{eq:extrapolate_zero_coupling} allows extrapolating to the limiting value of $P_T/P_L$ for $\lambda \to 0$. Similarly, for fixed $\tau/\tauR=1$, the Eq.~\eqref{eq:extrapolate_infinite_coupling} leading to the hydrodynamic limiting attractor is shown as a black dashed curve in \fig\ref{fig:PTPLratioAndFits}.
One might argue that here a linear fit is performed using only a few points. To study the bottom-up limiting attractor in more detail, several additional\footnote{These additional simulations were performed about two years after those used in Ref.~\cite{Boguslavski:2023jvg}. They use the same initial condition, but the improved adaptive time step employed in Ref.~\cite{Boguslavski:2024kbd} and Chapter \ref{sec:improving-qcd-simulations}, and better discretization parameters. It can be clearly seen that this leads to a slight systematic shift in Fig.~\ref{fig:PTPLratioAndFits} (right panel) in the form that the newer runs with $\lambda\in\{0.25,0.75,1.5\}$ lie slightly above the straight-line fit, whereas the older results lie slightly below. Despite this small bias due to better accuracy, these additional runs further corroborate the results of Ref.~\cite{Boguslavski:2023jvg} presented in the present chapter.} kinetic theory simulations at weak couplings $\lambda\in\{0.25, 0.75, 1.5\}$ were performed, and used to obtain the bottom-up limiting attractor in the right panel of Fig.~\ref{fig:PTPLratioAndFits}. There, we observe that for $\lambda\leq 2$, the values lie approximately on a straight line. The error bar shown in the figure is an estimate of the fit uncertainty from a linear regression procedure.

To summarize, we find that the pressure ratio at weak coupling between the circle and the triangle markers is better described using the bottom-up limiting attractor, while the hydrodynamic limiting attractor provides a better description for strong couplings after the triangle marker.
However, it should be noted that at sufficiently late times, even for weak couplings, all curves converge to the hydrodynamic attractor, which is required by the fact that hydrodynamics emerges from kinetic theory close to equilibrium.

\section{Limiting attractors for transport coefficients of hard probes}
\label{subsec:diffusionquenching}
Having identified these limiting attractors in the pressure ratio, we may wonder about other quantities exhibiting similar behavior. In the context of this thesis, it is natural to consider the \emph{jet quenching parameter} $\qhat$, which we have discussed extensively in the previous Chapter \ref{sec:momentum-broadening-of-jets}. Additionally, we will also consider the closely related \emph{heavy-quark diffusion coefficient}.

\subsection{Jet quenching parameter and heavy-quark diffusion coefficient}
As discussed thoroughly in the previous Chapter \ref{sec:momentum-broadening-of-jets}, the \emph{jet quenching parameter} $\qhat$ quantifies the momentum broadening of jets, (see Eq.~\eqref{eq:intro-qhat-definition}),
\begin{align}
    \qhat = \dv[\langle p_\perp^2\rangle]{t}.
\end{align}
For the derivation in the previous Chapter \ref{sec:momentum-broadening-of-jets}, we have taken the idealistic case of a massless jet particle with momentum much larger than all plasma scales. On the other hand, we may also consider the opposite case: a heavy quark. It can be modeled as a particle at rest (in the plasma rest frame) with a mass much larger than all scales in the plasma. This heavy quark will receive momentum kicks from the plasma, which leads to diffusive behavior.
Physically, the momentum of the heavy quark can then be modeled by the Langevin equation \cite{Moore:2004tg}
\begin{align}
    \dv[p_i]{t}=\xi_i(t)-\eta_Dp_i, && \langle \xi_i(t)\xi_j(t')\rangle=\kappa_i\delta_{ij}\delta(t-t').\label{eq:langevin_heavyquark}
\end{align}
Here, $\eta_D$ is a drag coefficient, and $\xi_i(t)$ encodes random momentum kicks. The \emph{heavy quark diffusion coefficient} $\kappa$ quantifies the mean squared momentum transfer per time,
\begin{align}
    \dv[\langle (\Delta p)^2\rangle]{t}=\sum_i\kappa_i \overset{\substack{\text{rotational}\\\text{symmetry}}}{=} 3\kappa. \label{eq:kappa_momentumderivative}
\end{align}
In the Langevin equation \eqref{eq:langevin_heavyquark}, we have allowed for the possibility of having different diffusion coefficients $\kappa_i$ in different directions in an anisotropic plasma. 

\begin{table}
    \centering
    \begin{tabular}{cccccc}\toprule
        Particle & Momentum & Mass & $\dv[\langle \Delta p_x^2\rangle]{t}$ & $\dv[\langle \Delta p_y^2\rangle]{t}$ & $\dv[\langle \Delta p_z^2\rangle]{t}$\\ [0.5em]
         \hline

        Jets &
            $\begin{aligned}
                p&\to\infty\\
                \hat{\vb {p}}&=\hat{\vb {e}}_x
            \end{aligned}$
            & $m\to 0$   & $\qhat^{xx}=\qhat_L$ & $\qhat^{yy}$ & $\qhat^{zz}$\\ [2em]
        Heavy quarks& $p\to 0$ & $m\to\infty$ & $\kappa_T$ & $\kappa_T$ & $\kappa_z$\\ [0.5em] \bottomrule
    \end{tabular}
    \caption{Comparison of momentum diffusion coefficients for jets and heavy quarks in an idealized scenario. The top row shows the properties of an idealized jet particle and corresponding transport coefficients $\qhat^{ii}$, while the bottom row lists properties and transport coefficients for a heavy quark.}
    \label{tab:momentum-diffusion-coefficients}
\end{table}

We compare the different definitions in these idealized scenarios in Table \ref{tab:momentum-diffusion-coefficients}. In principle, these coefficients can be thought of as the limits of a more general transport coefficient which depends both on the mass and momentum of the particle.

For the jet quenching parameter, we use Eq.~\eqref{eq:qhat_formula_pinf}, which can be written as 
\begin{align}
	\hat q^{ij} &=  \frac{1}{4 \nu_g}\lim_{p\to\infty}\int_{\substack{\vb k\vb k'\vb p'\\q_\perp < \Lambda_\perp}} q_\perp^i q_\perp^j (2\pi)^4\delta^4(P+K-P'-K') \,\frac{\left|\mathcal M_{gg}^{gg}\right|^2}{p} f(\vb k)\left(1+ f(\vb k')\right).
\end{align}
In a similar way, the heavy quark momentum diffusion coefficient $\kappa$ can be obtained via \cite{Moore:2004tg}
\begin{align}
	\label{eq:kappa_master_formula}
	\kappa_i &= \frac{1}{2 M}\int_{\vb{k} \vb{k^\prime} \vb{p^\prime}} \left(2 \pi \right)^3 \delta^3\left( \vb{p} +\vb{k} - \vb{p^\prime} - \vb{k^\prime} \right)  2 \pi \delta \left(k^\prime - k \right) q_i^2 
	\left| \mathcal{M}_g \right|^2 f(\vb{k}) (1+f(\vb{k^\prime})).
\end{align}
Here, $M$ is the mass of the heavy quark and is considered to be much larger than all other relevant scales.
Similarly to the jet quenching parameter, $\vb{p}, \vb{p^\prime}$ are the incoming and outgoing heavy quark momenta, $\vb{k}, \vb{k^\prime}$ are the incoming and outgoing momenta of the plasma particles. The transferred momentum $q_i^2 \in \left\{q_z^2, q_T^2=q_x^2=q_y^2\right\}$ distinguishes longitudinal or transverse momentum transfer, and will be used for the longitudinal ($\kappa_z$) or transverse ($\kappa_T$) diffusion coefficient, respectively. With the convention employed here (and also in Eqs.~\eqref{eq:langevin_heavyquark} and \eqref{eq:kappa_momentumderivative}), we obtain $3 \kappa = 2 \kappa_T + \kappa_z$, and the ratio $\kappa_T / \kappa_z=1 $ for an isotropic system. For the integration measure \eqref{eq:integration-measure}, we use $K^0=M$ for the heavy quark and $k^0 = \left| \vb{k} \right|$ for plasma particles. 

At leading order in the coupling and inverse heavy quark mass, $\kappa$ is dominated by the t-channel gluon exchange matrix element \cite{Moore:2004tg}, which we use in an isotropic approximation,
\begin{align}
	\label{eq:MatrixElement}
	\left|\mathcal{M}_g \right|^2 = \NC \CR g^4 16 M^2 \,\frac{k^2 (1+ \cos^2 \theta_{\vb{k} \vb{k^\prime} } )}{\left(q^2 + m_D^2 \right)^2}\,.
\end{align}
More details on the formula and implementation can be found in Ref.~\cite{Boguslavski:2023fdm}.

\subsection{Numerical results}
\begin{figure*}
	\centerline{
		\includegraphics[width=0.47\linewidth]{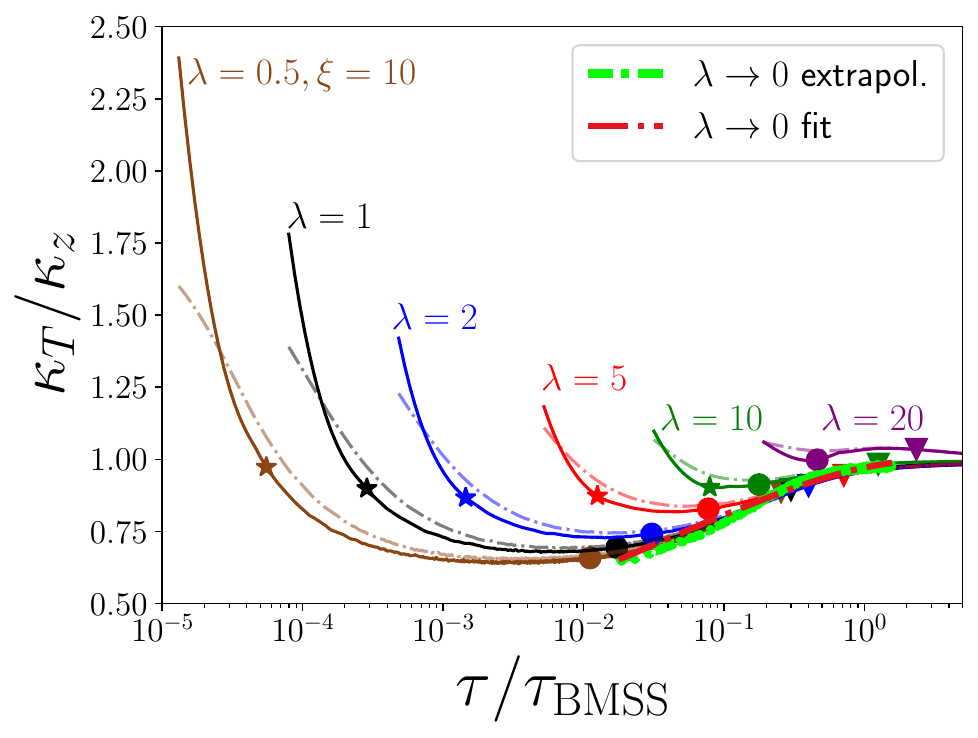}
		\includegraphics[width=0.47\linewidth]{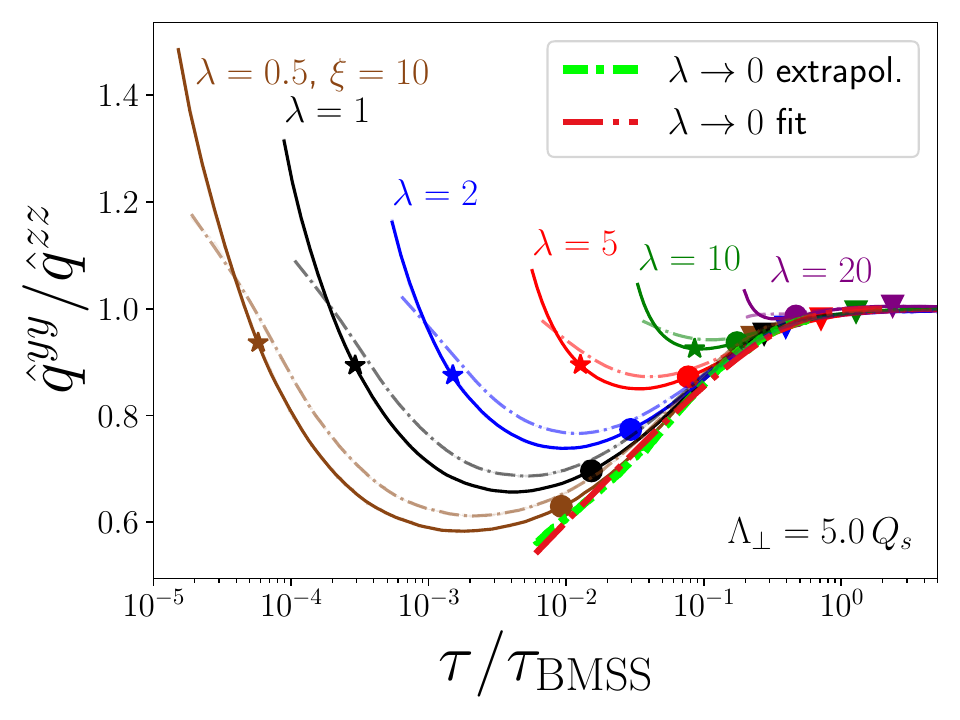}
	}
	\centerline{
		\includegraphics[width=0.47\linewidth]{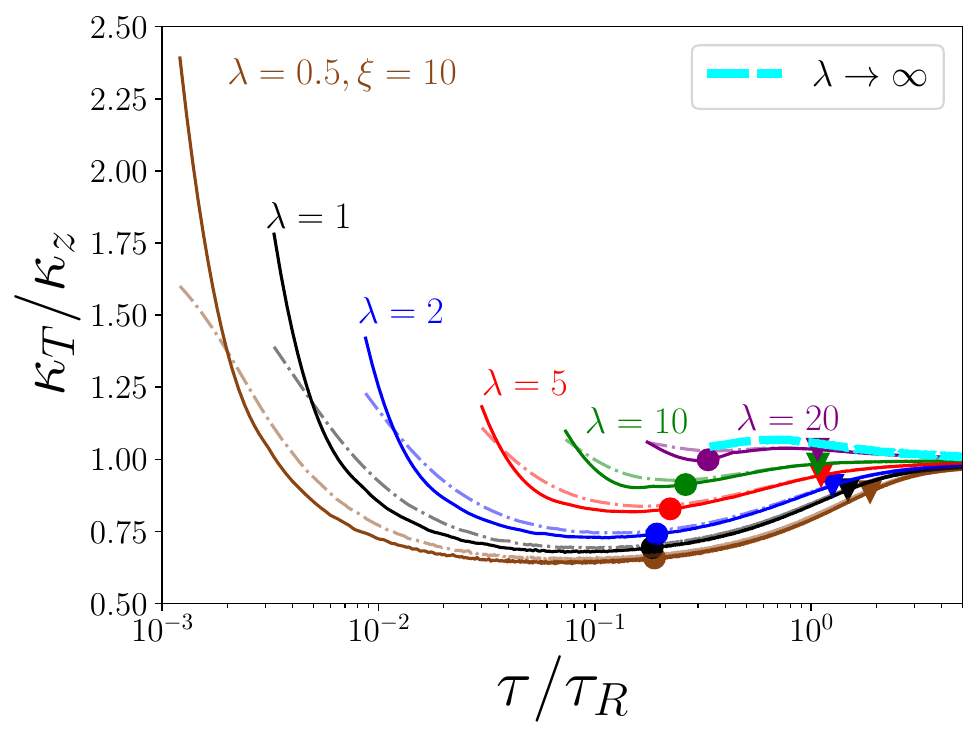}
		\includegraphics[width=0.47\linewidth]{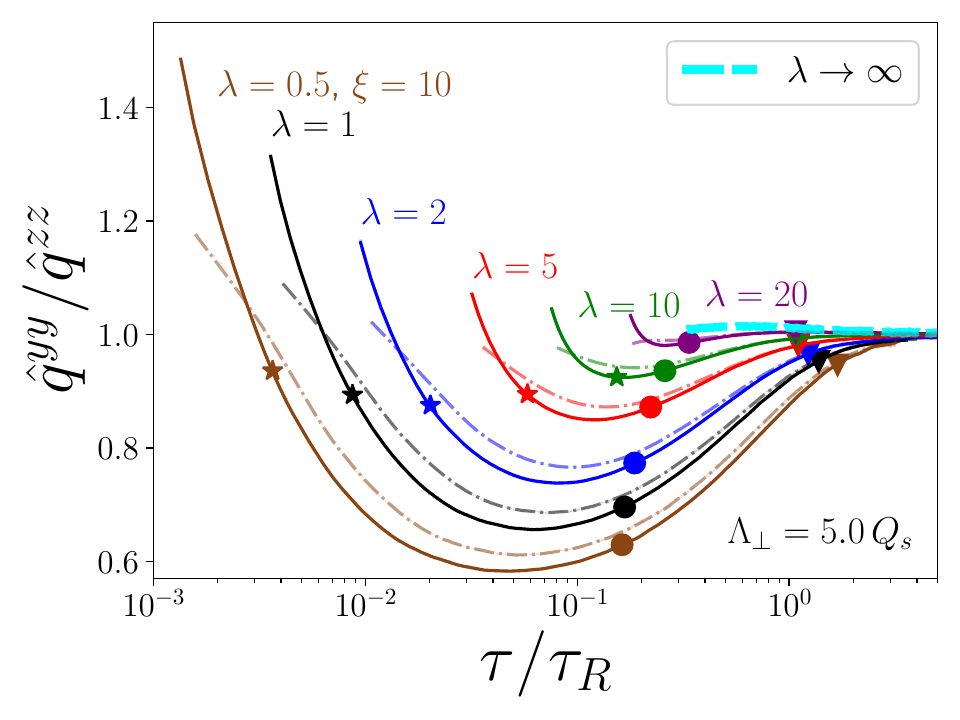}
	}
	\centerline{
		\includegraphics[width=0.47\linewidth]{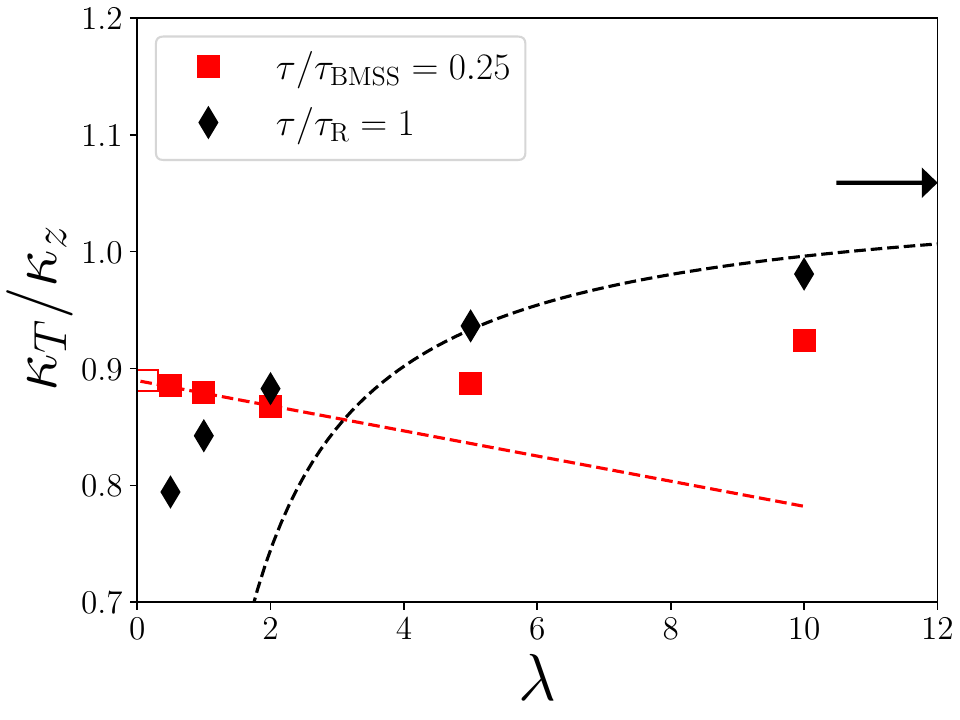}
		\includegraphics[width=0.485\linewidth]{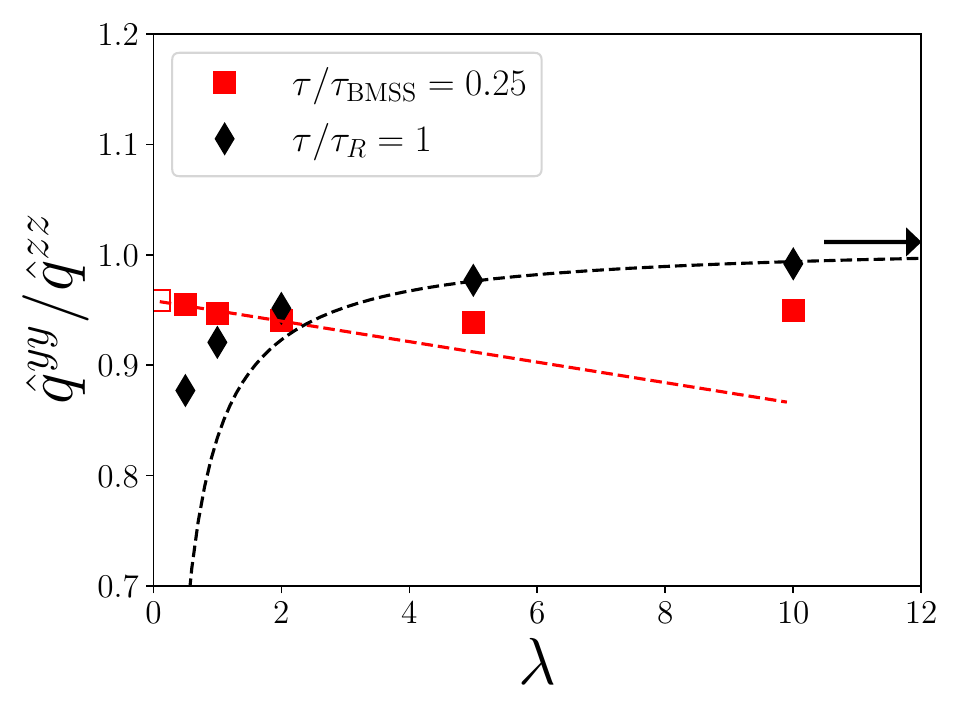}
	}
	\caption{Ratios of transverse and longitudinal momentum broadening coefficients of hard probes using different time scalings. Full and dashed lines correspond to anisotropy parameters $\xianiso=10$ and $\xianiso=4$, respectively. The left column depicts the ratio of the heavy quark diffusion $\kappa_T/\kappa_z$ where additionally a Savitzky-Golay filter \cite{doi:10.1021/ac60214a047} is applied to the curves to smoothen the data. 
		Similarly, the right column shows the ratio of jet quenching parameters $\qhat^{yy}/\qhat^{zz}$ for fixed $q_\perp$ cutoff $\Lambda_\perp=5\,Q_s$.
		In the top row, time is rescaled with $\tauT$, and in the center row with $\tauR$. The bottom row illustrates the extrapolation procedure to vanishing coupling (empty square) or infinite coupling (black arrow), with the latter 
		performed on the three largest couplings including $\lambda=20$ not shown in the plots. Plots from Ref.~\cite{Boguslavski:2023jvg}.
		\label{fig:kappa_qhat_combined}
	}
\end{figure*}

\begin{figure}
    \centering
    \centerline{
    \includegraphics[width=0.5\linewidth]{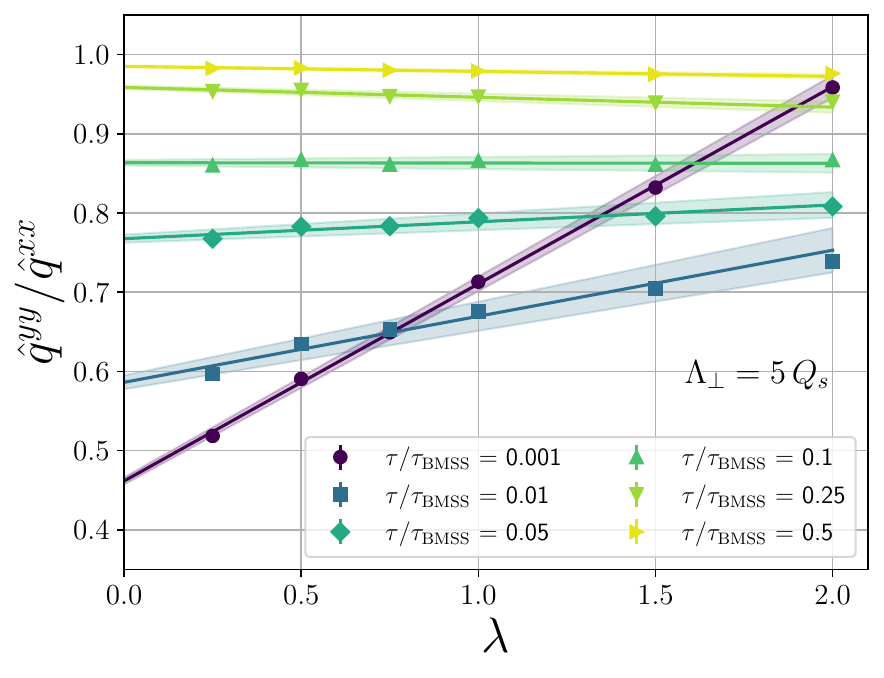}
    \includegraphics[width=0.5\linewidth]{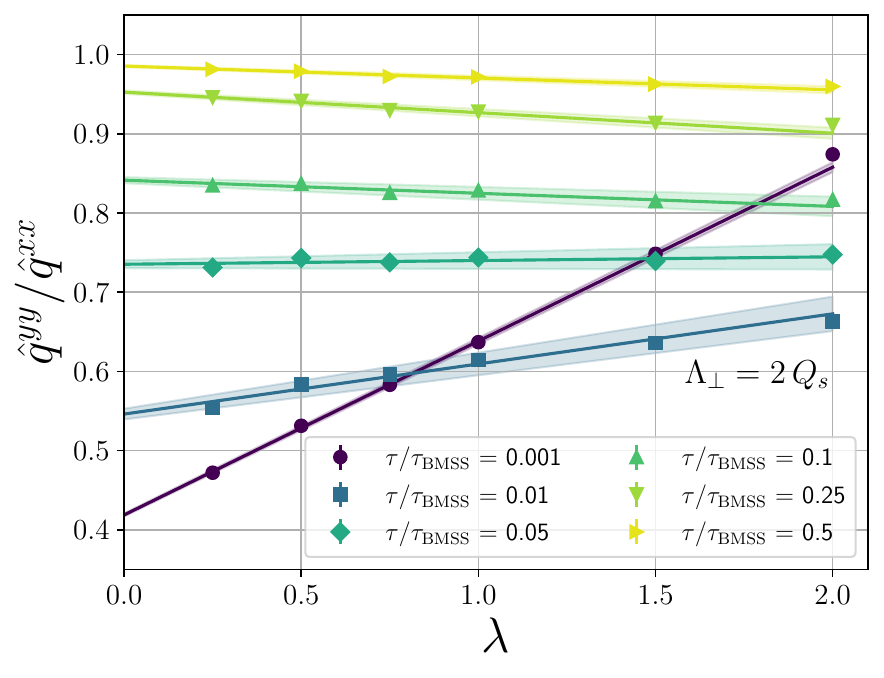}
    }
    \caption{Additional examples for extrapolating to the bottom-up limiting attractor for the ratio of the jet quenching parameters $\qhat^{yy}/\qhat^{zz}$. Shown are values for various rescaled times $\tau/\tauT$ for different couplings $\lambda$, including $\lambda\in\{0.25, 0.75, 1.5\}$. In the different panels, I show different transverse momentum cutoffs $\lperp$.
    }
    \label{fig:qhat-extrapolation-example}
\end{figure}

We have identified limiting attractors in the pressure anisotropy ratio in the previous section. Let us now consider anisotropy ratios of transport parameters of hard probes, i.e., consider the ratio of the jet quenching parameter in $y$ and $z$ direction, $\qhat^{yy} / \qhat^{zz}$, for a large transverse momentum cutoff in \eq \eqref{eq:qhat_general} and of the transverse and longitudinal heavy-quark diffusion coefficient $\kappa_T/\kappa_z$ in \eq \eqref{eq:kappa_master_formula}. 

These ratios are presented in \fig \ref{fig:kappa_qhat_combined}, with $\kappa_T/\kappa_z$ in the left and $\qhat^{yy} / \qhat^{zz}$ for cutoff $\Lambda_{\perp} = 5 \qs$ in the right column, both for a wide range of couplings $\lambda = 0.5$ to $20$. The top row depicts them as functions of time scaled by the bottom-up time scale $\tauT$. There, we observe a remarkable qualitative similarity in the evolution of both anisotropy ratios.
We find that after the circle marker, the curves corresponding to different couplings quickly approach each other, and even the curves for larger couplings seem to approach this universal curve.
As before, the extrapolation to zero coupling according to Eq.~\eqref{eq:extrapolate_zero_coupling} is shown as a light green curve.
Note that curves for weaker couplings approach this attractor earlier, and, in particular, at a time when the system is still far from equilibrium.

For possibly later convenience, these bottom-up limiting attractor curves can be fit to the simple form
\begin{align}
	\label{eq:fit1}
	R_{\qhat, \kappa}(\tau) = 1 + c_1^{\qhat,\kappa}\ln\left(1 - e^{-c_2^{\qhat,\kappa} \tau/\tauT}\right).
\end{align}
For the jet quenching ratio $R_{\qhat}(\tau) \approx \qhat^{yy}/\qhat^{zz}$, this yields $c_1^{\qhat}=0.12$ and $c_2^{\qhat}=3.45$ while the ratio of the heavy quark diffusion coefficient $R_{\kappa}(\tau) \approx \kappa_T/\kappa_z$ leads to $c_1^{\kappa}=0.093$ and $c_2^{\kappa}=1.33$. 
For both anisotropy ratios, the fits \eqref{eq:fit1} are included in the top panels of \fig \ref{fig:kappa_qhat_combined} as dash-dotted lines, labeled ``$\lambda\to 0$ fit''. 

Note that the parametrization in \eq \eqref{eq:fit1} should be taken with caution. While its advantage 
lies in its simplicity and small number of fit parameters, the approach toward unity may not be captured completely by this simple ad hoc form. In particular, upon closer inspection, we observe that it provides a more accurate description for the bottom-up limiting attractor of the $\qhat$ ratio but shows more pronounced deviations for the $\kappa$ ratio.
Moreover, since the functional form 
can become negative at early times while the $\qhat$ and $\kappa$ anisotropies are always positive in kinetic theory, we may expect the fits to deviate substantially from the bottom-up limiting attractors at very early times $\tau \ll 0.01\, \tauT$. Irrespective of the exact parametrization, we emphasize that the bottom-up limiting attractors are well-defined at early times.

In contrast, the hydrodynamic limiting attractors for these anisotropy ratios 
seem to offer less predictive power.
This is shown in the central panels of \fig \ref{fig:kappa_qhat_combined} where the ratios are depicted as functions of time scaled with $\tauR$. As before, the hydrodynamic limiting attractors is obtained by extrapolating to $\lambda\to\infty$ at fixed $\tau/\tauR$ according to Eq.~\eqref{eq:extrapolate_infinite_coupling}. 
The resulting attractors (light blue curves) predict a ratio close to unity long before the system reaches isotropy signaled by the triangle marker.
Additionally, the functional form of this attractor is different from the curves for finite couplings and is approached at much later times, even after the triangle marker.
Therefore, the bottom-up limiting attractors provide a much more accurate description of these ratio observables for modeling the pre-equilibrium behavior of hard probes.

The bottom panels demonstrate the validity of the fit forms \eqref{eq:extrapolate_infinite_coupling} and \eqref{eq:extrapolate_zero_coupling} at fixed times $\tau/\tauT=0.25$ and $\tau/\tauR=1$ for the smallest and largest couplings, respectively.

Furthermore, in Fig.~\ref{fig:qhat-extrapolation-example}, the extrapolation procedure for the bottom-up limiting attractor is further demonstrated at various different times, including results for a different cutoff $\lperp=2\Q$ as well. For times between $0.001\leq \tau/\tauT\leq 0.5$, the values of $\qhat^{yy}/\qhat^{zz}$ are seen to fall nicely on a straight line, which demonstrates that the extrapolation procedure and the bottom-up limiting attractor is well-defined for these times. Remarkably, the extrapolation seems to work very well even at the very early time $\tau/\tauT=0.001$.

\section{Note on the time scales}
\begin{figure}
	\centering
	\includegraphics[width=0.5\linewidth]{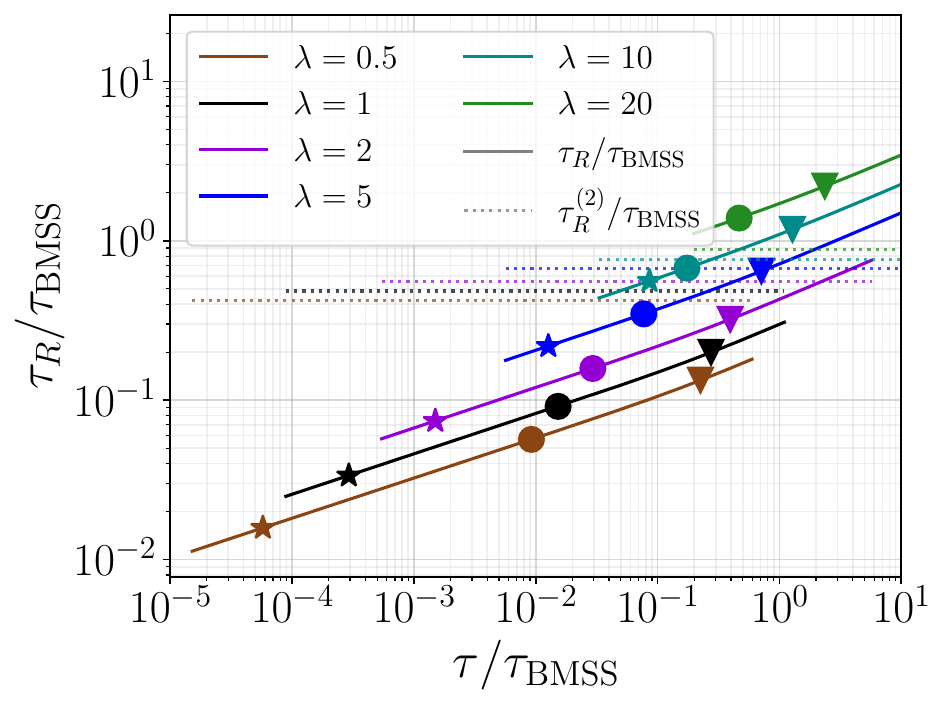}
	\caption{Ratio of the two different time scales $\tauR/\tauT$ during the simulation for different couplings. Dotted lines show the simple estimate \eqref{eq:relaxation-time-coupling} for the relaxation time $\tauR^{(2)}$. Plot adapted from Ref.~\cite{Boguslavski:2023jvg}.}
	\label{fig:time_comparison}
\end{figure}

\label{app:time_scales}
One might wonder how different the time scales $\tauT$ and $\tauR$ in Eq.~\eqref{eq:timescales-attractors} actually are in practice. However, it is not straightforward to compare them since the bottom-up time scale $\tauT$ depends only on the coupling $\lambda$ and is therefore constant in time, while the kinetic relaxation time $\tauR$ also
includes the effective (Landau-matched \eqref{eq:Landau-matching-condition}) temperature, which decreases throughout the time evolution.

To facilitate comparing these time scales, we may try to express $\tauR$ in terms of the coupling $\lambda$. This can be done by utilizing the parametric estimate for the maximum temperature obtained in the bottom-up picture \cite{Baier:2000sb}, $T_{\mathrm{max}}\sim \alpha_s^{2/5}$ and using the shear viscosity over entropy density \cite{Arnold:2003zc} $\eta/s\sim \alpha_s^{-2}$ (neglecting the logarithmic dependence). In summary, we would then obtain
\begin{align}
    \tau_R=\frac{4\pi\eta/s}{T} \overset{?}{\sim} \alpha_s^{-12/5}=\tauR^{(2)}, \label{eq:relaxation-time-coupling}
\end{align}
which is to be compared with $\tauT\sim\alpha_s^{-13/5}$. Thus, these time scales seem to be very similar. However, the limiting hydrodynamic attractor seems to work, especially for larger couplings $\lambda\gtrsim 5$, where perturbative estimates such as those used in \eqref{eq:relaxation-time-coupling} are questionable. Indeed, the values of $\eta/s$ used when employing the rescaling with the relaxation time \eqref{eq:timescales-attractors} are extracted from numeric simulations and do not follow the weakly coupling estimate $\eta/s\sim \alpha_s^{-2}$. Thus, Eq.~\eqref{eq:relaxation-time-coupling} does not accurately represent the parametric dependence of $\tauR$ on the coupling $\lambda$.

Numerically, it is, of course, straightforward to compare both time scales $\tauT$ and $\tauR$ in Eq.~\eqref{eq:timescales-attractors} directly. Their ratio is plotted in \fig\ref{fig:time_comparison}, with $\tauR$ as solid lines and $\tauR^{(2)}$ as dotted lines.
For small couplings $\lambda \lesssim 2$, the kinetic relaxation time $\tauR$ is much smaller than the bottom-up thermalization estimate $\tauT$ for the entire simulation, as visible in \fig\ref{fig:time_comparison}. In contrast, for larger values of $\lambda$, the relaxation time is comparable to and even becomes larger than the bottom-up estimate. In particular, both time scales are approximately identical at the triangle marker ($P_T/P_L=2$) for $\lambda = 10$. Comparing $\tauR^{(2)}$ to $\tauT$ leads to the same ordering as $\tauR$, but the timescales are more similar, as is already evident from their definition. However, as discussed before, the time scale $\tauR^{(2)}$ does not accurately describe the time-varying $\tauR$ in the region of interest (moderate to large couplings).

For weakly coupled systems, the observation that $\tauT\gg\tauR$ is consistent with the fact that the bottom-up picture dominates the equilibration process, which can also be seen by the emergence of the bottom-up limiting attractor for small couplings. It can be understood by the fact that hydrodynamization is dominated by its longest time scale. On the other hand, for larger couplings, we have $\tauR\gtrsim\tauT$, which is in line with the hydrodynamic limiting attractor becoming more dominant for larger couplings.
This provides a simple explanation for the observed behavior in this chapter.

\section{Concluding remarks}
\label{sec:conc}
In this chapter, we have focused on the universal features of a set of observables during the bottom-up thermalization process.
We have established the concept of limiting attractors, which can be constructed according to Eqs.~\eqref{eq:extrapolate_infinite_coupling} and \eqref{eq:extrapolate_zero_coupling} by extrapolating to vanishing or infinite coupling at fixed rescaled times. The bottom-up (weak-coupling) limiting attractor can be obtained using the characteristic bottom-up time scale $\tauT$, while for the hydrodynamic (strong-coupling) limiting attractor the relaxation time $\tauR$ is used.

It should be noted that these two limiting attractors are not contradictory, but should be rather thought of as complementary. For instance, certain observables might be more sensitive to the hydrodynamic limiting attractor (for example the pressure ratio), and others to the bottom-up limiting attractor (for instance the hard probes transport coefficients ratios considered here).
Perhaps surprisingly, the often-used hydrodynamic (limiting) attractor seems not to be very useful for the anisotropy ratios of the hard probe transport coefficients considered here, as it predicts only very small deviations from unity while the ratios for finite couplings differ from it both in shape and magnitude.

In this chapter, we consider a purely gluonic system, but the inclusion of quarks would be straightforward. Although qualitative changes to the picture described here would not be expected (because gluons are the dominant degree of freedom during hydrodynamization), it would be important to study whether including quarks enhances or diminishes these limiting attractors. 

Furthermore, it would be interesting to perform a broader study to identify other quantities and observables that are sensitive to the bottom-up limiting attractor. This may open the possibility of identifying observables in which both attractors are clearly visible, allowing a detailed study of the interplay of these two limiting attractors.

%% file: 500_improving_QCD_kinetictheory.tex
In Chapter \ref{sec:qcd-kinetic-theory}, we have introduced and discussed QCD kinetic theory \cite{Arnold:2002zm}, and how medium-effects are included in the Boltzmann equations for the quark and gluon distribution functions $f_s(\vb p,t)$. These medium effects have been included in a simple way, which we have referred to as \emph{Debye-like screening}. In this chapter, we discuss how to do better; by including the fully resummed hard thermal loop propagator in the matrix elements of the elastic collision term. To quantify the effects of these changes, 
we will study how the thermalization of QCD is affected by this different screening choice. We will start by considering isotropic systems, and conclude this section with simulations of Bjorken expanding systems now using the full (isotropic) hard-thermal loop matrix element.

This chapter is based on Ref.~\cite{Boguslavski:2024kbd}.

\section{Recapitulate: Debye-like screening in QCD kinetic theory}
As already discussed in Section \ref{sec:debye-like-screening}, medium-effects are included in QCD kinetic theory in the matrix elements for the elastic collision term using the simple replacement \eqref{eq:amy_replacement},
\begin{align}
	{\frac{(s-u)^2}{\underline{t^2}}}&\to \left|\Gret_{\mu\nu}(P-P')\;(P+P')^\mu(K+K')^\nu\right|^2\label{eq:amy_replacement_htlsection},
\end{align}
where $\Gret$ is the retarded HTL propagator. 
In the Debye-like screening approximation, this propagator is approximated as \eqref{eq:simple-isotropic-screening}
\begin{align}
	G_{\mu\nu}=\frac{g_{\mu\nu}}{Q^2} \, \frac{q^2}{q^2+\xiscreensqr m_D^2},
	\label{eq:simple-isotropic-screening_htlsection}
\end{align}
effectively replacing
\begin{align}
    \frac{(s-u)^2}{t^2}\to \frac{(s-u)^2}{t^2}\frac{q^4}{(q^2+\xiscreensqr m_D^2)^2} \label{eq:simple_replacement_debyelike_htlsection}
\end{align}
in the underlined terms in Tab.~\ref{tab:amy_matrix_el}. Note that, because of the symmetry in exchanging the outgoing particles (exchanging $u$ and $t$ channel), the $u$ channel divergences that would require screening can always be mapped to a $t$ channel divergence and be screened via the replacement in Eq.~\eqref{eq:simple_replacement_debyelike_htlsection}.

In this chapter, we will discuss that other replacements, such as replacing $su/t^2$, are equally valid and use the full isotropic hard-thermal-loop propagator. We will begin by considering this propagator in a general (isotropic linear) gauge.

\section{Hard thermal loop propagator}
\subsection{Isotropic HTL gluon propagator in a linear gauge\label{sec:isoHTLgeneral}}

While the HTL gluon propagator is gauge-dependent, calculating several physical quantities has been shown to be gauge-independent \cite{Kobes:1990dc, Kobes:1990xf, Rebhan:1992ca}. In particular, in an isotropic system, the HTL gluon self-energy can always be written in terms of two functions $\Pi_a(Q)$ and $\Pi_b(Q)$.

In this section, we will discuss the most general form of the gluon propagator with HTL self-energy corrections in a linear gauge that does not break the rotational symmetry in the plasma rest frame. 
We will show that all gauge-dependent parts of the propagator are proportional to the exchange momentum $Q$ and will vanish when contracted with the external momenta in \eq \eqref{eq:amy_replacement_htlsection},
confirming that the replacement \eqref{eq:amy_replacement_htlsection} is gauge-independent.

The reason for restricting to not breaking the rotational symmetry is that this leaves only two relevant directions for the self-energy: The plasma rest frame, which is defined by the vector $\tilde n^\mu=(1,\vb 0)$ and momentum of the propagating gluon
$Q^\mu$.
This will allow for a 4-dimensional basis for the self-energy and propagator, as, e.g., in Ref.~\cite{Bellac:2011kqa}. If we included another direction, we would need a higher-dimensional basis, e.g., the one employed in Ref.~\cite{Carrington:2021bnk}.

In our case, any tensor quantity in this system can be constructed from $\tilde n^\mu$, $Q^\mu$, and the metric $g^{\mu\nu}$. For simplicity, we construct our tensor basis with the vector $n^\mu$ that is orthogonal to $Q^\mu$,
\begin{align}
	n_\mu &= P_{\mu\nu}\tilde n^\nu, & P_{\mu\nu}=g_{\mu\nu}-\frac{Q_\mu Q_\nu}{Q^2}.
\end{align}

In momentum space, we can represent any linear gauge condition \cite{Baier:1992mg, Bellac:2011kqa} $f^\mu A_\mu=0$
by a general vector $f^\mu(Q)$. If it further does not break rotational invariance (which we assumed), we may decompose $f_\mu$ into a part parallel and transverse to $Q$,
\begin{align}
	f_\mu &= a(Q)Q_\mu + b(Q)n_\mu,
\end{align}
This general linear gauge includes the Lorenz (covariant) gauge, $f_\mu = Q_\mu$, and the Coulomb gauge, where $f_0 = 0$ and $f_i=q_i$, as well as the temporal axial gauge $f_0=\Lambda$ and $f_i=0$, where $\Lambda$ is a constant scale needed for dimensional reasons.
A symmetric basis for the gluon propagator is given by \cite{Bellac:2011kqa}
\begin{subequations}
	\begin{align}
		B_{\mu\nu}&=\frac{n_\mu n_\nu}{n^2}, &	C_{\mu\nu}&=n_\mu Q_\nu + n_\nu Q_\mu, \\
		E_{\mu\nu}&=\frac{Q_\mu Q_\nu}{Q^2}, &
		A_{\mu\nu}&=P_{\mu\nu}-B_{\mu\nu}.
	\end{align}
\end{subequations}
The free propagator then reads \cite{Baier:1992mg, Bellac:2011kqa}
\begin{align}
	G^0_{\mu\nu}=\frac{P_{\mu\nu}}{Q^2}-\frac{1}{f_e Q^2}\left(f_c C_{\mu\nu}+(f_b+ Q^2)E_{\mu\nu}\right),
\end{align}
with
\begin{align}
	f_b=\frac{b^2n^2}{\xigauge},&& f_c=\frac{ab}{\xigauge}, && f_e=\frac{a^2Q^2}{\xigauge},\label{eq:f_relations}
\end{align}
where $\xigauge$ is the gauge-fixing parameter introduced in the Fadeev-Popov procedure \cite{Bellac:2011kqa}.
Including the self-energy (see \eqref{eq:self-energy-schematically}), we obtain the dressed propagator,
\begin{align}
	G_{\mu\nu}&=\frac{A_{\mu\nu}}{Q^2+\Pi_a}+\frac{B_{\mu\nu}}{\tilde b(Q)}-\frac{f_c+\Pi_c}{\tilde b(Q)(f_e+\Pi_e)}C_{\mu\nu}
	+\frac{Q^2+\Pi_b+f_b}{\tilde b(Q)(f_e+\Pi_e)}E_{\mu\nu},
\end{align}
with
\begin{align}
	\tilde b(Q)&=Q^2+\Pi_b+f_b-n^2Q^2\frac{(f_c+\Pi_c)^2}{f_e+\Pi_e},
\end{align}
and where we have similarly decomposed the self-energy in the same tensor basis,
\begin{align}
	\Pi_{\mu\nu} = \Pi_a A_{\mu\nu} + \Pi_b B_{\mu\nu} + \Pi_c C_{\mu\nu} + \Pi_e E_{\mu\nu}.
\end{align}
The Ward identity forces the HTL self-energy to be transverse, $Q^\mu \Pi_{\mu\nu}=0$ (see, e.g., \cite{Blaizot:2001nr}).
In that case, one finds that $\Pi_c=\Pi_e=0$, which simplifies $\tilde b(Q)=Q^2+\Pi_b$. All dependence from the gauge choice is now in the parameters $f_b$, $f_c$, and $f_e$, which always appear together with factors of $Q_\mu$.
A closer inspection reveals that all terms proportional to $Q_\mu$ or $Q_\nu$ yield zero when contracted with the external momenta in Eq.~\eqref{eq:amy_replacement_htlsection}. This follows from
\begin{align}
	Q \cdot (P+P')&=(P'-P)\cdot(P+P')=P'^2-P^2=0,
\end{align}
and similarly with $P\leftrightarrow K$.
Therefore, the screening prescription in Eq.~\eqref{eq:amy_replacement_htlsection} is gauge invariant for a general linear gauge.

\subsection{IsoHTL screening}
Let us now consider including the full isotropic HTL retarded propagator in Eq.~\eqref{eq:amy_replacement_htlsection} for the gluonic matrix element,
\begin{align}
	&\frac{|\Mhtl|^2}{16d_AC_A^2g^4}=\frac{1}{4}\bigg(9+\frac{(t-u)^2}{s^2}\label{eq:full_isotropic_HTL_matrixelement}
	 +2\left|\GretHTL_{\mu\nu}(P-P')(P+P')^\mu(K+K')^\nu\right|^2\bigg).
\end{align}
As before, we will refer to this screening prescription as \emph{isoHTL}. Note that, as we have discussed in \se\ref{sec:isoHTLgeneral}, Eq.~\eqref{eq:full_isotropic_HTL_matrixelement} is gauge invariant.
We have already considered a very similar contraction with the external momenta as in Eq.~\eqref{eq:full_isotropic_HTL_matrixelement} for obtaining the isoHTL screened matrix element for the jet quenching parameter in Eq.~\eqref{eq:full_htl_matrix_element}. The steps to obtain the isoHTL screened matrix element here are almost identical (see Appendix \ref{app:isoHTL-screening-for-c22}), leading to
\begin{align} 	
	&\left|\GretHTL_{\mu\nu}(P-P')(P+P')^\mu (K+K')^\nu\right|^2 \label{eq:HTL_propagator_explicit_expression}
	=\frac{c_1^2}{A^2+B^2}+\frac{c_2^2}{C^2+D^2}-\frac{2c_1c_2(AC+BD)}{(A^2+B^2)(C^2+D^2)},
\end{align}
with $A,\,B,\,C,\,D$ and $c_1$ given by Eqs.~\eqref{eq:parameters_for_full_HTL_matrix_element} and \eqref{eq:parameters_c1c2_for_qhat_finitep}, and the only difference being $c_2$, which now reads
\begin{align}
		c_2&=4pk\sin\thetaqp\sin\thetaqk\cos(\phiqk-\phi_{qp}).\label{eq:constant_c2}
\end{align}
More details and a more elaborate discussion on the differences to the isoHTL screening for the jet quenching parameter are discussed in Appendix \ref{app:isoHTL-screening-for-c22}.

\subsection{Debye-like screening as an approximation to isoHTL}
\label{sec:debye_as_approx}
As already briefly discussed in Section \ref{sec:debye-like-screening}, the Debye-like screening prescription \eqref{eq:simple_replacement_debyelike_htlsection} can be understood as a simple approximation to the full isotropic HTL matrix element \eqref{eq:HTL_propagator_explicit_expression}, which we briefly reiterate here. In the original work \cite{AbraaoYork:2014hbk}, the screening parameter
$\xiscreen$ was chosen to reproduce (longitudinal) soft momentum transfer in elastic collisions. This can be seen as follows:
For isotropic distributions, mimicking the steps to obtain the jet quenching parameter from Section \ref{sec:obtaining_qhat_from_kinetic_theory}, we may write the collision kernel as \cite{AbraaoYork:2014hbk}%
\begin{align}
	\begin{split}
		\Ctwotwo&=\frac{1}{2^9\pi^5\nu}\int_0^\infty\dd{k}\int_0^{2\pi}\dd{\phiqp}\\
		&\!\!\!\times\int_{-p}^k\!\!\!\dd{\omega}\!\!\left\{f(p)f(k)(1+f(p+\omega))(1+f(k-\omega)) - f(p+\omega)f(k-\omega)(1+f(p))(1+f(k))\right\}\\
		&\!\!\!\times\int_{|\omega|}^{\min(2k-\omega,2p+\omega)}\dd{q}\int_0^{2\pi}\dd{\phiqk}\frac{\left|\mathcal M\right|^2}{p^2}.
	\end{split}
\end{align}
Screening effects are only important for soft internal momenta, $q,\,\omega\ll k,\,p$. When expanding the distribution functions for small $\omega$, the first nonvanishing term is quadratic
in $\omega$ since the matrix element is even. One therefore requires that in this limit
\begin{align}
	\int_{-\infty}^\infty\dd{\omega}\omega^2\int_{|\omega|}^\infty\dd{q}\int_0^{2\pi}\dd\phi \left(\left|\Mhtl\right|^2-\left|\Mdebyeone\right|^2\right)=0, \label{eq:ximatching-longitduinal}
\end{align}
which fixes the constant $\xiscreen=e^{5/6}/\sqrt{8}$ \cite{AbraaoYork:2014hbk}.

\begin{figure}
    \centerline{
        \includegraphics[width=0.5\linewidth]{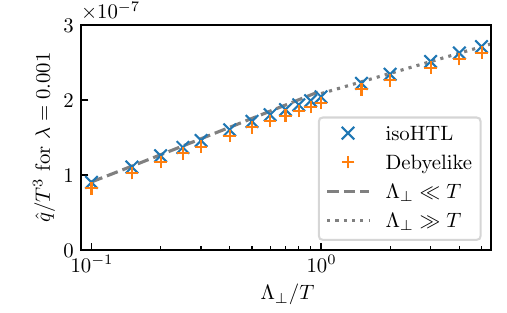}
        \includegraphics[width=0.5\linewidth]{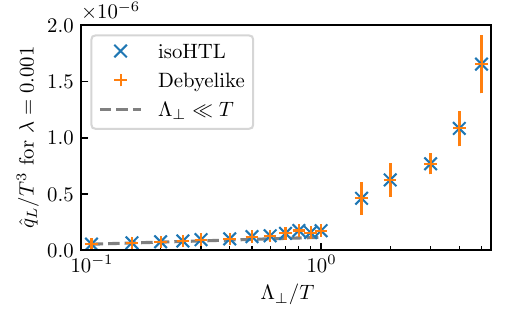}
    }
    \caption{Transverse (left) and longitudinal (right) jet quenching parameter in a thermal plasma using both the Debye-like screening prescription with $\xiscreen=e^{5/6}/\sqrt{8}$ (usually employed in the elastic collision term) and the isoHTL screened Matrix element. Figure taken from an upcoming publication \cite{Boguslavski:2025a}.}
    \label{fig:qhat-vs-qhatL-debyelike-vs-isohtl}
\end{figure}
In Section \ref{sec:qhat_screening}, we have already seen that using the Debye-like screening prescription for transverse momentum broadening, we obtain a different constant $\xiscreenperp=e^{1/3}/2$. That is, to match transverse momentum broadening in Section \ref{sec:qhat_screening} using \eqref{eq:ximatching}, we needed to include $q_\perp^2$ in the matching, whereas here we include $\omega^2$ in Eq.~\eqref{eq:ximatching-longitduinal}. For a highly energetic jet, we have $q_z=\omega$ from \eqref{eq:vpq_ptoinf}, and, hence, we match longitudinal broadening here, whereas in Section \ref{sec:qhat_screening}, we matched transverse momentum broadening.
Consequently, the Debye-like screening prescription cannot simultaneously approximate both longitudinal and transverse momentum diffusion, even in isotropic systems, while the isoHTL prescription can. 
Therefore, the isoHTL screening prescription is more general and should be used when both processes are important. This is illustrated in Fig.~\ref{fig:qhat-vs-qhatL-debyelike-vs-isohtl}, where the transverse and longitudinal jet quenching parameter is plotted with both the Debye-like screened matrix element (with screening constant $\xiscreen$) and the isoHTL screened matrix element. 
Also the analytic limits of $\qhat$ from Eqs.~\eqref{eq:qhat_thermal_equlibrium_soft} and \eqref{eq:qhat_hard_arnold} are shown, and the corresponding analytic form of $\qhat_L$ from Ref.~\cite{Ghiglieri:2015ala} valid for small cutoffs.
We observe that only the isoHTL-screened matrix element reproduces the analytic limit in both transverse and longitudinal momentum broadening, whereas the Debye-like screened matrix element can only reproduce one of them. In this case, by choosing $\xiscreen$, the longitudinal broadening is reproduced.

It should be emphasized, again, that both of these screening prescriptions neglect the effect of plasma instabilities, which are generically seen to occur in anisotropic systems \cite{Mrowczynski:1988dz, Mrowczynski:1993qm, Arnold:2003rq, Romatschke:2003ms, Romatschke:2004jh, Romatschke:2006bb, Kurkela:2011ub, Hauksson:2021okc}. However, numerical evidence indicates that these instabilities do not play a dominant role at the time scales of interest for kinetic theory simulations \cite{Berges:2013eia, Berges:2013fga} and when a quasiparticle picture becomes applicable \cite{Boguslavski:2018beu, Boguslavski:2021buh, Boguslavski:2021kdd}.

\subsection{Different Debye-like screening prescriptions\label{sec:different_debyelike_screening_prescriptions}}

As discussed in \se\ref{sec:amy-screening-prescription}, screening effects need only be included for $|t|\ll s \approx -u$, where the screening prescription screens the divergencies for $t\to 0$. However, in this limit,
\begin{align}
	I_1=\frac{(s-u)^2}{4t^2}, && I_2 = -\frac{su}{t^2}, && I_3=\frac{s^2}{t^2},\label{eq:possible-screening-mandelstam-bare}
\end{align}
are equivalent up to $\mathcal O(|t|/s)$, in particular, $I_3/I_1=1+\mathcal \mathcal O(|t|/s)$ and $I_2/I_1=1+\mathcal O(t^2/s^2)$. 

Instead of screening the first term in \eqref{eq:possible-screening-mandelstam-bare}, $(s-u)^2/(4t^2)$, we can also apply the screening prescription \eqref{eq:simple_replacement_debyelike_htlsection} to any of the other terms in \eqref{eq:possible-screening-mandelstam-bare}.
The gluonic matrix element can be rewritten in several forms, where these terms appear explicitly,
\begin{subequations}\label{eq:equivalent-matrixelements}
	\begin{align}
		\frac{|\mathcal M|^2}{16d_AC_A^2g^4}
		&=\frac{1}{4}\left(9+2\frac{(s-u)^2}{t^2}+\frac{(t-u)^2}{s^2}\right)\\
		&=3-2\frac{su}{t^2}-\frac{tu}{s^2}\\
		&=3+2\frac{s^2}{t^2}+2\frac{s}{t}-\frac{tu}{s^2}\,.
	\end{align}
\end{subequations}
These vacuum matrix elements \eqref{eq:equivalent-matrixelements} are equivalent because of the condition for the Mandelstam variables \eqref{eq:mandelstam_sum}, $s+t+u=0$. To leading-order (or up to $\mathcal O(|t|/s)$), the expressions in \eqref{eq:possible-screening-mandelstam-bare} coincide, and we thus have a particular freedom of how to implement the screening. The choice mentioned in Ref.~\cite{Arnold:2002zm} is to implement the screening as in the scalar quark case, Eq.~\eqref{eq:scalar_quark_result}, which makes the gauge-independence manifest.

In this chapter, we study including the simple Debye-like isotropic screening \eqref{eq:simple_replacement} in the different equivalent expressions \eqref{eq:possible-screening-mandelstam-bare} in \eqref{eq:equivalent-matrixelements}, by using the matrix elements,
\begin{subequations}\label{eq:different_regularization_matrix_element}\begin{align}
		\frac{|\Mdebyeone|^2}{16d_AC_A^2g^4}&=\frac{1}{4}\left(9+2\frac{(s-u)^2}{t^2} \, \frac{q^4}{(q^2+\xiscreensqr m_D^2)^2} +\frac{(t-u)^2}{s^2}\right),\label{eq:usual_screened_matrix_element}\\
		\frac{|\Mdebyetwo|^2}{16d_AC_A^2g^4}&=3-2\frac{su}{t^2}\, \frac{q^4}{(q^2+\xiscreensqr m_D^2)^2} -\frac{tu}{s^2},
		\label{eq:Debye2_screened_matrix_element}\\
		\frac{|\Mdebyethree|^2}{16d_AC_A^2g^4}&=3+2\left(\frac{s^2}{t^2}+\frac{s}{t}\right) \frac{q^4}{(q^2+\xiscreensqr m_D^2)^2} -\frac{tu}{s^2}\,.
		\label{eq:Debye3_screened_matrix_element}
	\end{align}
\end{subequations}
Because of Eq.~\eqref{eq:mandelstam_sum}, the second and third matrix elements exactly coincide, $\Mdebyetwo\equiv\Mdebyethree$, and we only need to consider differences between $\Mdebyeone\neq \Mdebyetwo$.

Note that for the region $|t|\ll s$, 
the agreement of the matrix elements \eqref{eq:different_regularization_matrix_element}
is independent of the values of $q$ and $m_D$. On the other hand, for sufficiently large $q \gg m_D$ and independent of the values of $s,t$, these matrix elements only differ up to factors of $\mathcal O\left(m_D^2/q^2\right)$.
Since screening effects are only important for these regions of small $q\sim m_D\sim gT$ or small $|t|\ll s$, the matrix elements \eqref{eq:different_regularization_matrix_element} are leading-order equivalent.

\section{Thermalization of isotropic systems\label{sec:thermalization_isotropic}}

\begin{figure*}
	\centerline{
		\includegraphics[width=0.47\linewidth]{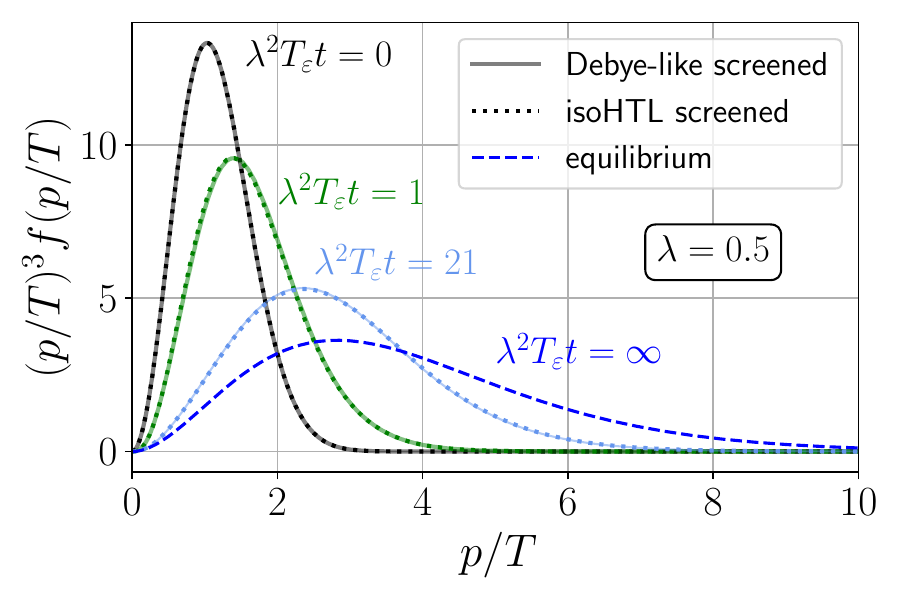}
		\includegraphics[width=0.47\linewidth]{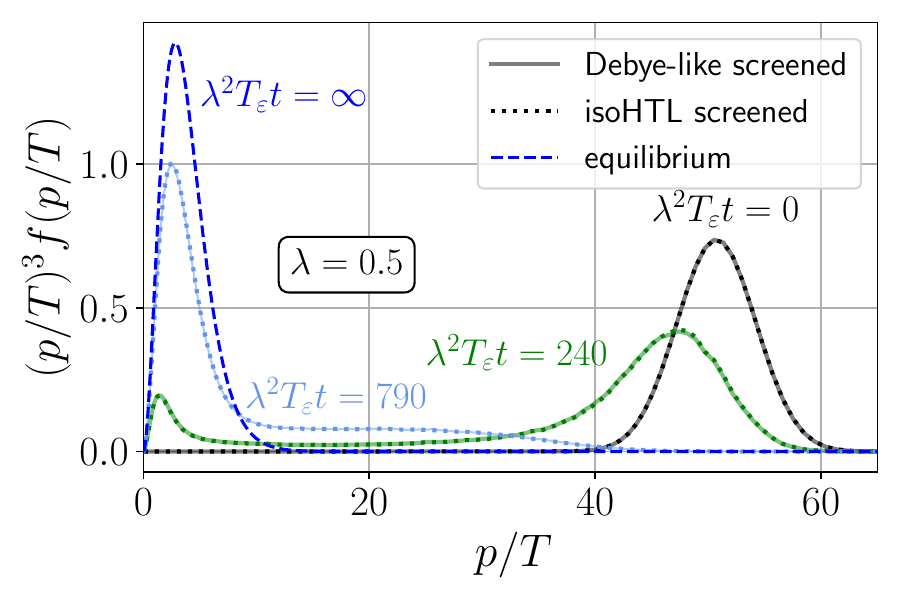}
	}
	\centerline{
		\includegraphics[width=0.47\linewidth]{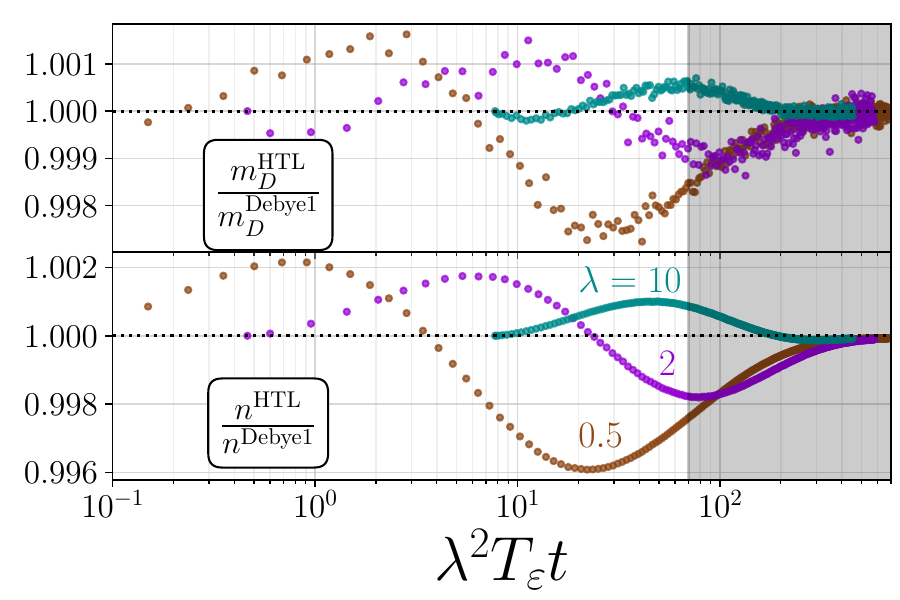}
		\includegraphics[width=0.47\linewidth]{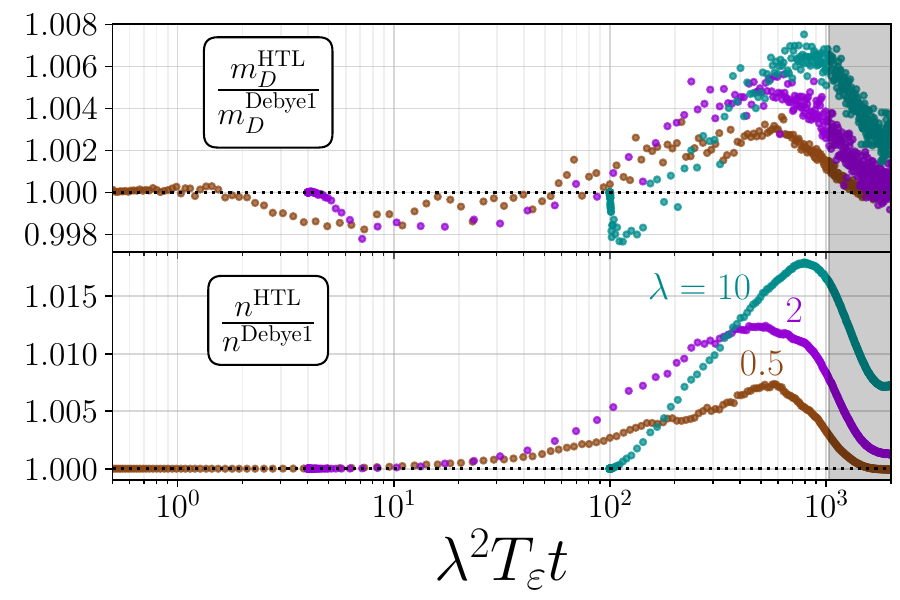}
	}
	
	\caption{Thermalization of an initially over-occupied (left) and under-occupied (right) gluonic plasma. The top row shows the distribution function $f$ scaled with $p^3$ corresponding to the contribution of $f$ to the energy density.
    The bottom row shows ratios of observables (Debye mass $m_D$ and particle number $n$) calculated with isoHTL screening \eqref{eq:full_isotropic_HTL_matrixelement} over Debye-like screening (variant 1, \eqref{eq:usual_screened_matrix_element}) as functions of the rescaled time $\lambda^2 \Teps t$. Figures from \cite{Boguslavski:2024kbd}.
	}
	\label{fig:iso_thermalization}
\end{figure*}

We have already discussed how isotropic systems thermalize in Section \ref{sec:thermalization_isotropic}.
In this section, we will now study how the screening choice modifies QCD thermalization in isotropic systems. For that, numerical simulations using the isoHTL screened matrix element \eqref{eq:full_isotropic_HTL_matrixelement} are performed and compared with Debye-like screened simulations. In the isotropic case, we consider non-expanding isotropic systems where the longitudinal expansion term in \eq \eqref{eq:actual-boltzmann-equation-to-solve-expanding-system} is omitted.
Similarly to \cite{Kurkela:2014tea, Fu:2021jhl}, we start with both an under- and over-occupied initial distribution. The under-occupied simulations are initialized via
\begin{align}
	f(p)=A\exp\left(-(p-Q)^2/(Q/10)^2\right),
\end{align}
with $A$ chosen\footnote{In the simulation, all quantities are given in dimensionless units. There, we first fix $Q=50$, then adjust $A$ such that the energy density is that of a thermal system with $T=1$.} such that $Q=50 T$, where $T$ is the temperature of the equilibrium system after thermalization.
As an initial condition for the over-occupied system, the parametrization of the system's self-similar scaling solution from \cite{AbraaoYork:2014hbk}
\begin{align}
	f(p)=\frac{(Qt)^{-4/7}}{\lambda \tilde p}\left(0.22 e^{-13.3\tilde p}+2 e^{-0.92 \tilde p^2}\right),
\end{align}
is used, with $\tilde p=(p/Q)(Qt)^{-1/7}$ and initial time $Qt=0.12$.

The Boltzmann equation \eqref{eq:actual-boltzmann-equation-to-solve-expanding-system} is solved numerically until equilibrium is reached, using a Debye-like $\Mdebyeone$ or isoHTL screened matrix element $\Mhtl$.
The system's time evolution can be fully described by the one-dimensional distribution function, which is depicted in the top row of \fig\ref{fig:iso_thermalization} rescaled with $p^3 f(p,t)$ for initially over- (left) and under-occupied systems (right).
We observe that the curves corresponding to the Debye-like (full lines) and isoHTL screening (dotted lines) almost coincide. Hence, for isotropic systems, the Debye-like screening prescription provides a good quantitative approximation of the full HTL matrix element. 

Next, as an attempt to better quantify deviations between the different screening prescriptions, we consider a measure for the thermalization time $\ttherm$.
We define this time scale implicitly in terms of the effective temperatures following \cite{Fu:2021jhl}
\begin{align}\left(T_0(\ttherm)/T_1(\ttherm)\right)^{\pm 4}=0.9,\label{eq:def_thermalizationtime}
\end{align}
with $+$ for under- and $-$ for overoccupied systems. The effective temperatures are defined via
\begin{align}
	T_\alpha=\left[\frac{2\pi^2}{\Gamma(\alpha+3)\zeta(\alpha+3)}\int\frac{\dd[3]{\vb p}}{(2\pi)^3}p^\alpha f(p)\right]^{1/(\alpha+3)}.
\end{align}
The first moment $T_1=\Teps$ has the physical interpretation of 
the temperature of a thermal system with the same energy density (Landau matching, see Eq.~\eqref{eq:Landau-matching-condition}),
\begin{align}\Teps(\tau)=\left(\frac{30\varepsilon(\tau)}{\pi^2\nu_g}\right)^{1/4}.\label{eq:Teps}
\end{align}
Due to energy conservation in non-expanding systems, it is constant throughout the evolution and corresponds to the temperature of the thermal system after equilibration. The other effective temperature, $T_0$, is related to the particle number density $n$, via the relation 
$T_0=\left(\pi^2 n/(\zeta(3)\nu_g)\right)^{1/3}$.
The thermalization times are listed in \tab \ref{tab:IR_UV_thermalizationtimes}. We observe that the full HTL matrix element leads to only slightly smaller thermalization times.

\begin{table}
    \centering
	\begin{tabular}{c c c c c} \toprule
        & \multicolumn{2}{c}{Overoccupied} & \multicolumn{2}{c}{Underoccupied}\\
		$\lambda$ & Debye-like: $\lambda^2 T \ttherm$ & isoHTL: $\lambda^2T \ttherm$ & Debye-like: $\lambda^2 T \ttherm$ & isoHTL: $\lambda^2T \ttherm$ \\ [0.7ex]
		\hline
		$0.5$ &$69.7$ & $67.0$ & $1029$ & $1022$\\
		$2$ & $86.0$ & $84.4$ &  $1127$ & $1112$\\
		$10$ & $93$ & $93$  & $1245$ & $1216$
		\\[0.5ex]
        \bottomrule
	\end{tabular}
	\caption{Thermalization times for initially over- and under-occupied systems. Table adapted from Ref.~\cite{Boguslavski:2024kbd}.
	}
	\label{tab:IR_UV_thermalizationtimes}
\end{table}

To also quantify differences in the dynamics, in the bottom row of \fig\ref{fig:iso_thermalization}, the evolution of the Debye mass \eqref{eq:debyemass-general} and number density $n$ are depicted as ratios of a simulation with isoHTL screening over one with Debye-like screening. 
The gray area indicates when the system is close to equilibrium, i.e., after the thermalization time defined in \eqref{eq:def_thermalizationtime}.
We observe that both the Debye mass and the number density differ only at a sub-percent level between the simulations.

Finally, the excellent agreement for isotropic systems among the screening prescriptions might not come as a big surprise.
In fact, as detailed in \se \ref{sec:debye_as_approx}, the Debye-like screening prescription was introduced specifically in the isotropic non-expanding case to approximate (isotropic) HTL screening.

\section{Results\label{sec:results} with longitudinal expansion}

We now turn to systems undergoing Bjorken expansion, relevant for the initial stages in heavy-ion collisions. For that, we need to include the additional expansion term in the Boltzmann equation \eqref{eq:actual-boltzmann-equation-to-solve-expanding-system} that was absent for the isotropic case in the previous subsection.

\subsection{Initial conditions, time markers and time scales}

We use the initial conditions discussed in Section \ref{sec:boltzmann-equation-and-initial-condition} with $\xianiso=10$, and the
time markers introduced in section \ref{sec:time-markers-and-scales}. Recall that
the star and circle markers are related to the occupancy $\langle pf \rangle/\langle p\rangle$ (star at occupancy $1/\lambda$ and circle at minimum), and the triangle marker is placed when the pressure ratio $P_T/P_L=2$, indicating that the system has almost isotropized.

We will also use the time scales introduced in Section \ref{sec:time-markers-and-scales} and used extensively in Chapter \ref{sec:limiting_attractors}, both the relaxation time $\tauR$ and the bottom-up time scale $\tauT$,
\begin{align}
    \tauR=\frac{4\pi\eta/s}{\Teps}, && \taubmss = \left(\frac{\lambda}{4\pi\NC}\right)^{-13/5}/Q_s\,. \label{eq:timescales-improvingqcd}
\end{align}

\subsection{Comparison of Debye-like screening prescriptions}
\begin{figure}
	\centering
	\centerline{ 
		\includegraphics[width=0.33\linewidth]{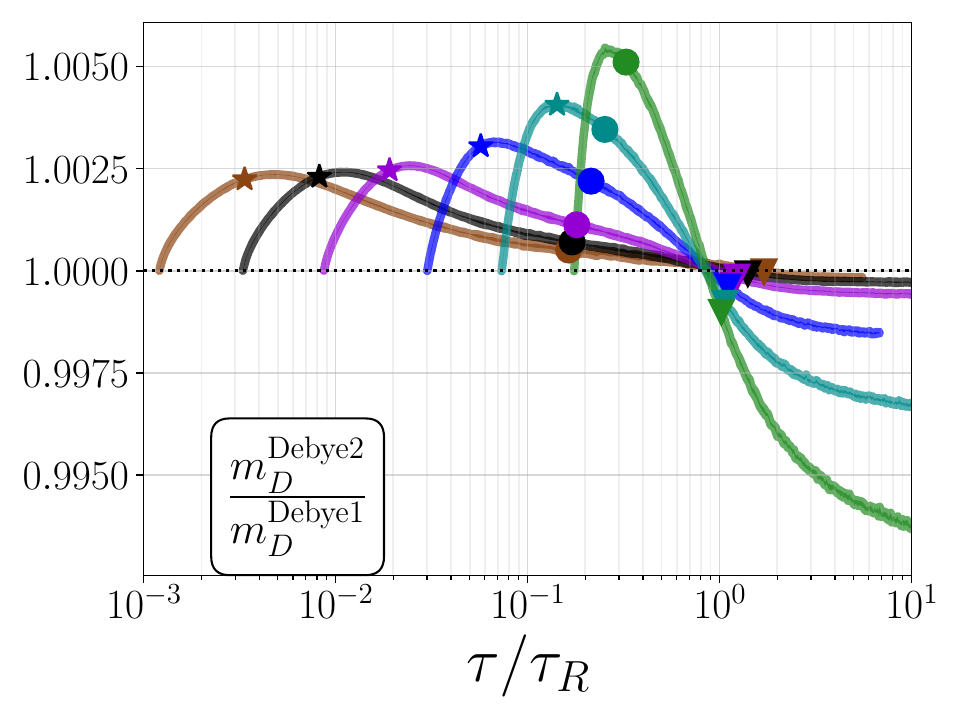}
		\includegraphics[width=0.33\linewidth]{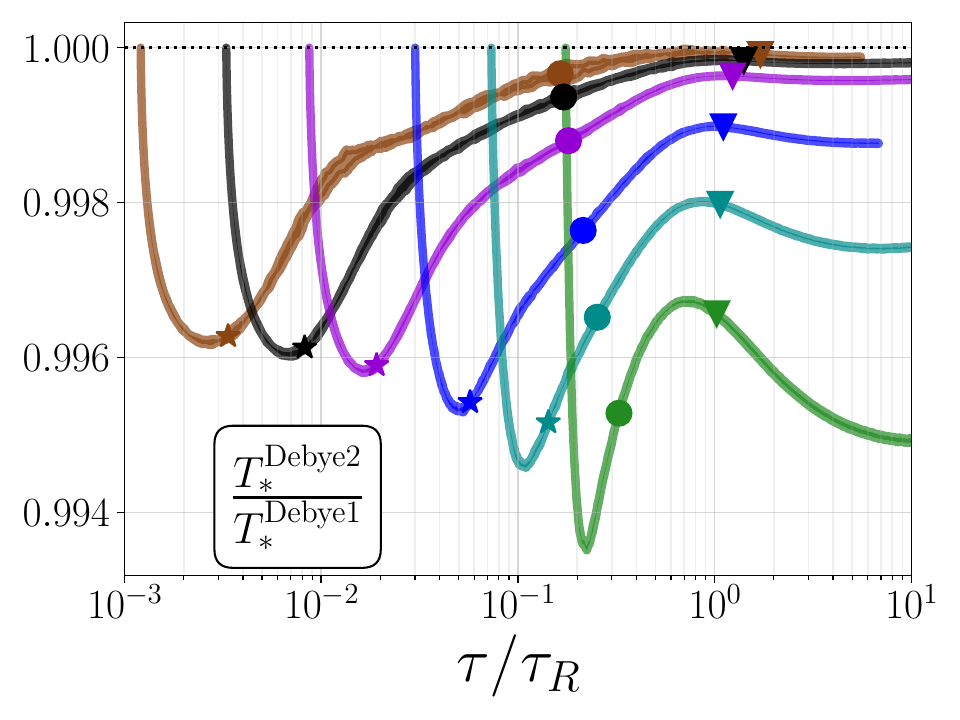}
		\includegraphics[width=0.33\linewidth]{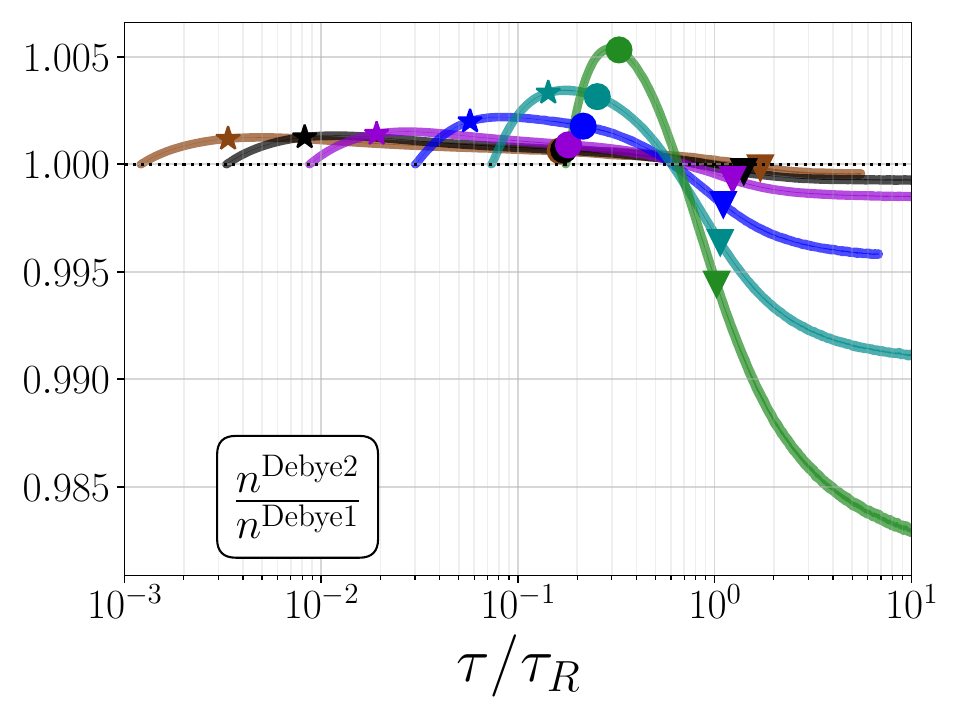}
	}
	\centerline{ 
		\includegraphics[width=0.33\linewidth]{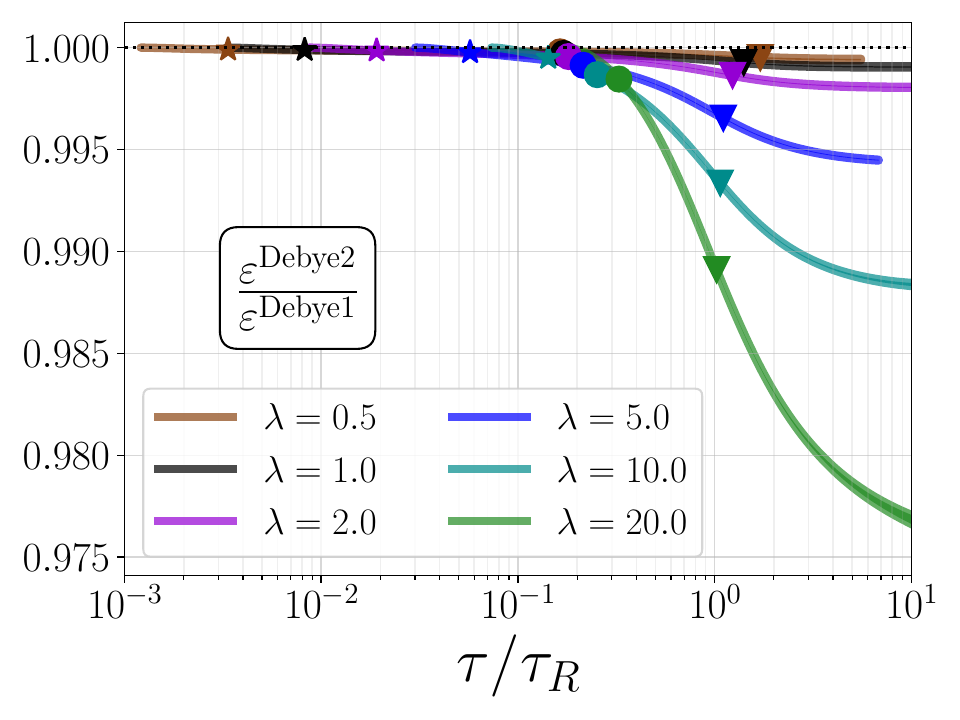}
		\includegraphics[width=0.33\linewidth]{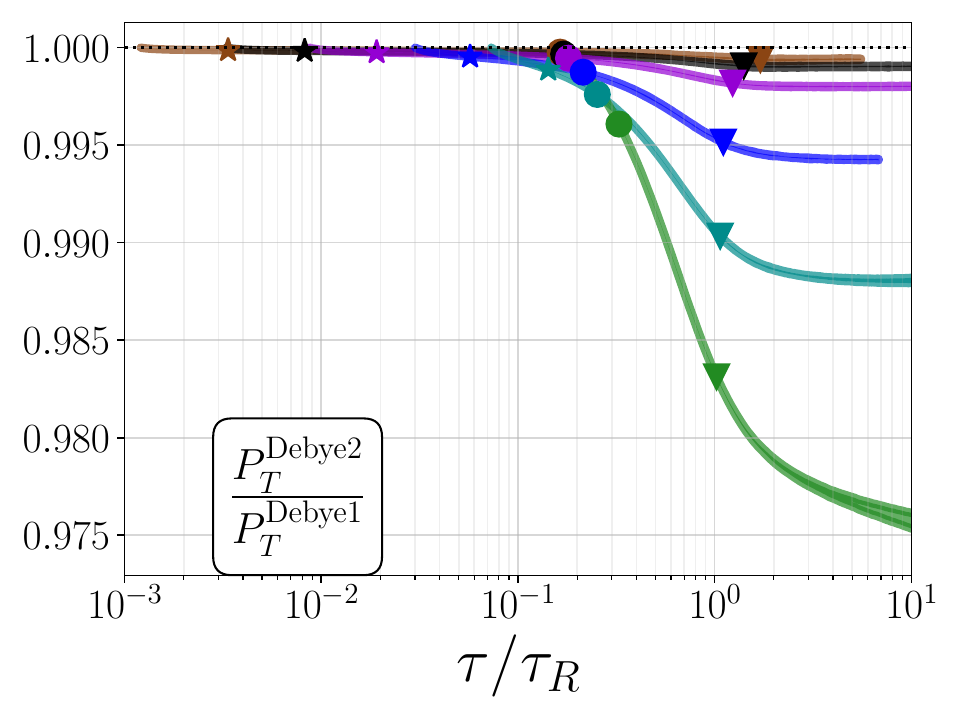}
		\includegraphics[width=0.33\linewidth]{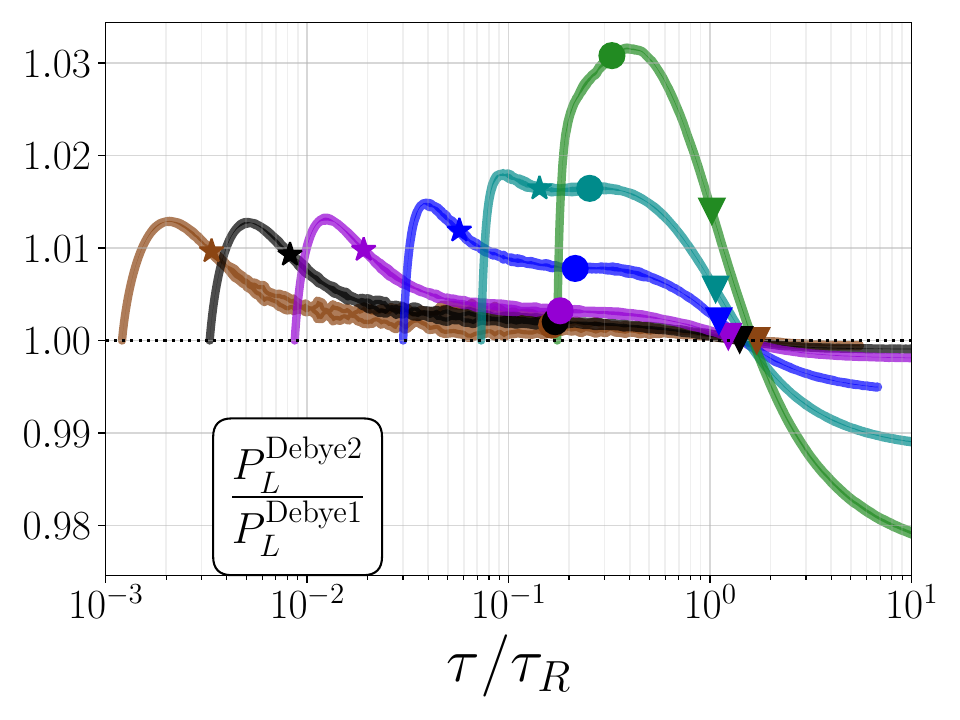}
	}
	\centerline{ 
		\includegraphics[width=0.33\linewidth]{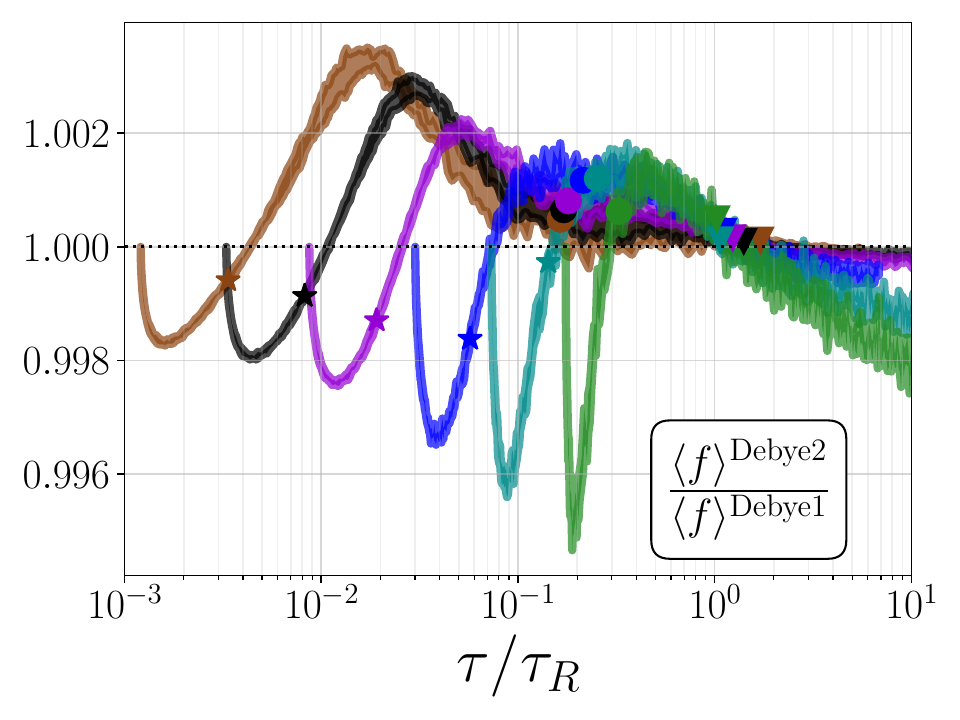}
		\includegraphics[width=0.33\linewidth]{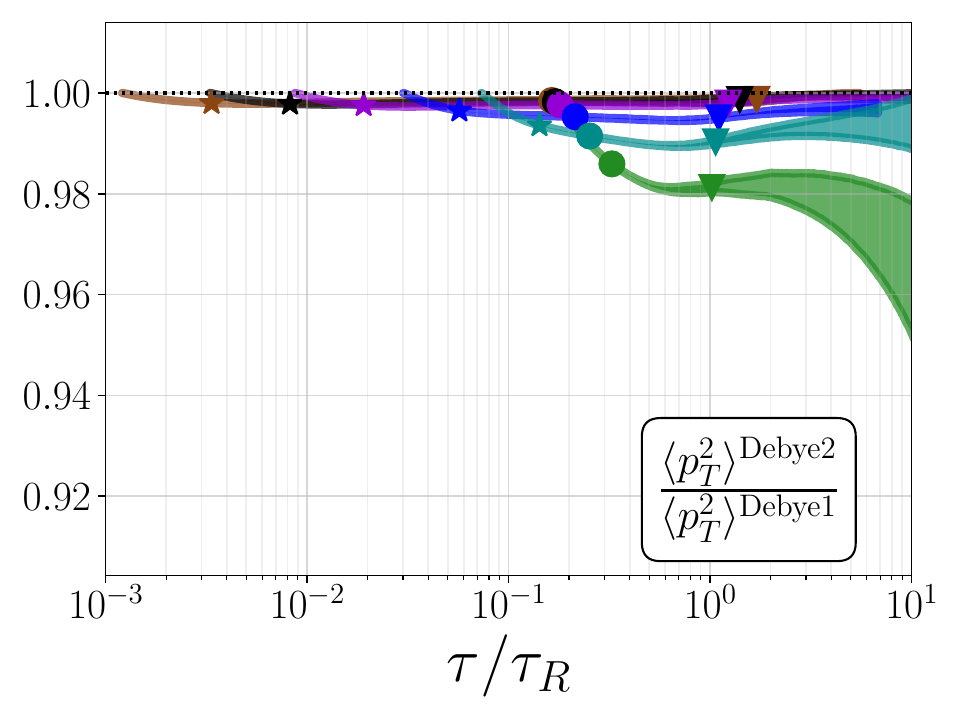}
		\includegraphics[width=0.33\linewidth]{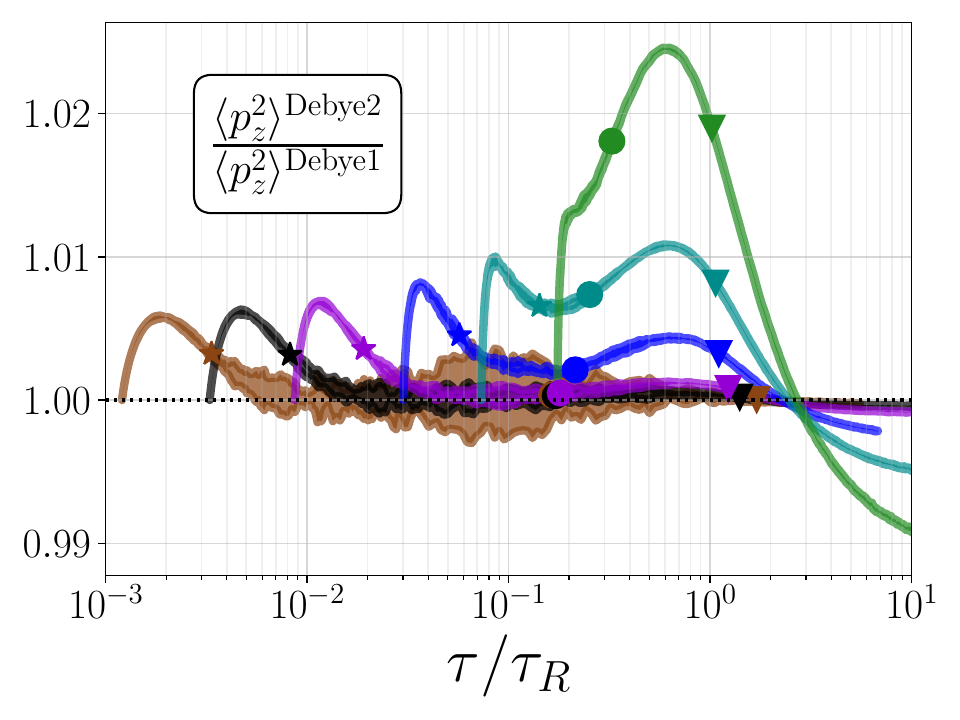}
	}
	\caption{Ratios of various observables of simulations with different Debye-like screening prescriptions from \eq \eqref{eq:different_regularization_matrix_element}. 
		The bands denote estimates for the standard error, obtained from statistically averaging over five independent simulations. Plots from Ref.~\cite{Boguslavski:2024kbd}.
	}
	\label{fig:debye-like-comparison}
\end{figure}

First, let us consider differences arising from the different variants of implementing the Debye-like screening prescription as written in \eqref{eq:different_regularization_matrix_element}.
The numerical results for simulations with identical initial conditions but different Debye-like screening prescriptions are presented in Fig.~\ref{fig:debye-like-comparison}, where the ratio of various observables is plotted.
These observables consist of various moments of the distribution function, such as components of the energy-momentum tensor $T^{\mu\nu}$ from Eq.~\eqref{eq:energy-momentum-tensor-from-f}, and Debye mass $m_D$ from Eq.~\eqref{eq:debyemass-general}, and the effective infrared temperature $T_\ast$ given by \eq\eqref{eq:tstar-definition}.
We observe only very small changes, mostly on a sub-percent level. To verify that these small changes are not caused by random fluctuations or noise from the Monte Carlo evaluation of the collision kernels, several (four to five) simulations with identical parameters but different seeds for the random number generator were performed. The error bands in the figure are estimates of the standard error of the results, see Appendix \ref{sec:statistical-averages}.

As can be seen in \fig \ref{fig:debye-like-comparison}, the different Debye-like screening prescriptions differ less for smaller couplings. This is expected because the different screening prescriptions are leading-order equivalent, as we discussed in section \ref{sec:different_debyelike_screening_prescriptions}.
Perhaps surprisingly then, also for larger values of the coupling (even for $\lambda=20$), the considered observables differ only at a percent level.
This is because these matrix elements \eqref{eq:different_regularization_matrix_element} are not only leading-order equivalent but also equivalent in the small angle scattering region where $|t|\ll s$, in accordance with our discussion in Section \ref{sec:different_debyelike_screening_prescriptions}.
In fact, it is well-known that the dynamics is mostly dominated by small-angle scatterings, which is used also as the basis for other QCD studies using the Boltzmann equation in diffusion approximation (BEDA) 
\cite{Blaizot:2013lga, Brewer:2019oha, Brewer:2022vkq, Rajagopal:2024lou, BarreraCabodevila:2022jhi, Cabodevila:2023htm, BarreraCabodevila:2025vir}. 
Indeed, it can be checked that processes with $s\gg |t|$ are dominant for QCD kinetic theory simulations, which is presented in more detail in Appendix \app\ref{app:validity_screening}.

\subsection{Comparison with isoHTL: Pressure ratio}
We now turn to the isoHTL screening prescription \eqref{eq:HTL_propagator_explicit_expression}.
Numerically, employing the isoHTL prescription leads to larger noise and a more unstable numerical evaluation of the differential equation \eqref{eq:actual-boltzmann-equation-to-solve-expanding-system}, and requires an improved step size algorithm which is given in detail in Appendix \ref{app:adative_stepsize}.
We will first discuss the results for the pressure ratio and then move on to other observables.
\begin{figure}
	\centerline{
	\includegraphics[width=0.5\linewidth]{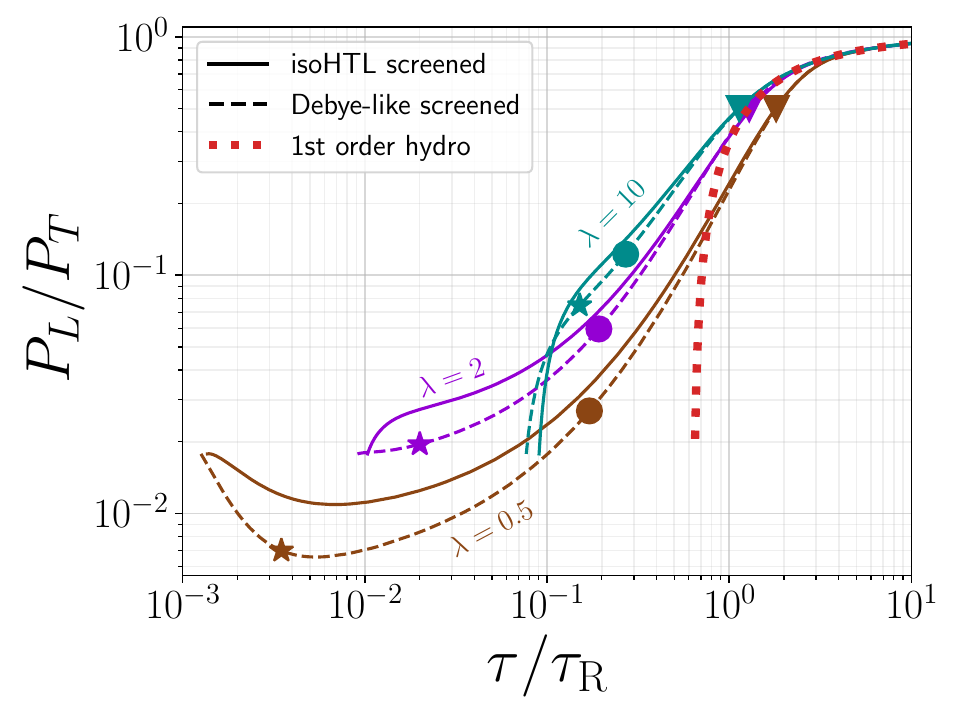}
	\includegraphics[width=0.5\linewidth]{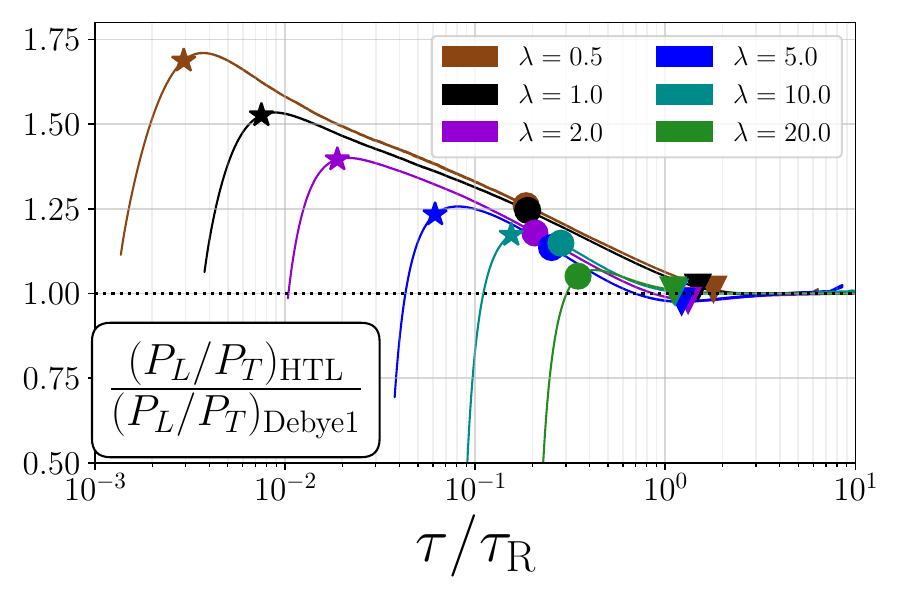}}
	\caption{Pressure ratio as a function of time for different couplings (colors). In the left panel, the results using the isoHTL-screened matrix element are shown as solid lines, and for the Debye-like screened matrix element as dashed lines. The first-order hydrodynamic estimate is shown as a thick red dotted line. The right panel shows the pressure ratio of the isoHTL-screened simulations normalized to the pressure ratio with Debye-like screening. Time is rescaled with the kinetic relaxation time $\tau_R$. Figures adapted from \cite{Boguslavski:2024kbd}. 
	}
	\label{fig:pressure-ratio}
\end{figure}

Fig.~\ref{fig:pressure-ratio} shows the results for the pressure ratio for different couplings (color-coded) in the left panel, and the ratio between the isoHTL and Debye-like pressure ratio in the right panel. Before discussing these results in more detail, a note about the time scales is in order.
As already discussed in Section \ref{sec:time-markers-and-scales}, curves of the pressure ratio nicely collapse at late times when the time is rescaled with the relaxation time $\tauR=4\pi\eta/s /\Teps$.,
\begin{align}
	\frac{P_L}{P_T}=1-\frac{2}{\pi}\frac{\tau_R}{\tau}, \label{eq:PL_over_PT_first_order_hydro}
\end{align}
which is shown in the left panel of \fig\ref{fig:pressure-ratio} as a thick red dotted line.

Importantly, the shear viscosity $\eta$ depends on the screening prescription, and for the curves to collapse in Fig.~\ref{fig:pressure-ratio}, one needs to use different values of $\eta/s$ for different screening prescriptions. We will discuss how to extract this transport parameter in more detail in the next section.

Coming back to the pressure ratio shown in Fig.~\ref{fig:pressure-ratio}, we find that the maximum pressure anisotropy is almost halved by employing the isoHTL screening prescription, i.e., the lowest value of $P_L/P_T$ is larger.
While both the isoHTL-screened (solid line) and Debye-like screened curves (dashed line, corresponding to \eqref{eq:usual_screened_matrix_element}) start at the same value of the pressure ratio by using the same initial condition, the time evolution with the different screening prescriptions shows clear differences: Simulations with Debye-like screening become more anisotropic as compared to isoHTL screening, which is more pronounced at small values of the coupling $\lambda$.

This can be understood from the observation that while the Debye-like screening prescription \eqref{eq:usual_screened_matrix_element} approximates well the longitudinal momentum transfer, it underestimates transverse momentum broadening as encoded in the jet quenching parameter $\qhat$, see Fig.~\ref{fig:qhat-vs-qhatL-debyelike-vs-isohtl}. 
However, transverse momentum broadening
is an essential ingredient in the bottom-up equilibration process \cite{Baier:2000sb}. The Debye-like screening prescription, therefore, leads to less efficient transverse momentum broadening and to a larger anisotropy.

In contrast, the late-time evolution is less sensitive to the screening prescription if the relaxation time $\tauR$ in \eq \eqref{eq:timescales-attractors} is adjusted accordingly. The required parameter $\eta/s$ 
is obtained such that the curves follow this late-time first-order hydrodynamic estimate \eqref{eq:PL_over_PT_first_order_hydro}. We will study the approach to hydrodynamics and thermalization first and then discuss the extraction procedure of $\eta/s$ together with the numerical results in \se \ref{sec:extract_etas}.

Note also that the curves in the right panel of Fig.~\ref{fig:pressure-ratio} do not start at $1$ despite the simulations being initialized with the same initial condition. This is because due to the different values of $\eta/s$ for different screening prescriptions, the system is initialized at a different $\tau/\tauR$ (as can be seen in the left panel).

\subsection{Numerical extraction of the specific shear viscosity $\eta/s$\label{sec:extract_etas}}

As input for the relaxation time $\tauR$ in \eqref{eq:timescales-attractors}, the value of the transport parameter $\eta/s$ for every coupling and screening prescription is needed. This first-order hydrodynamic parameter quantifies the late-time approach to isotropy.
It has been obtained perturbatively in Ref.~\cite{Arnold:2000dr, Arnold:2003zc}.

\subsubsection{General strategy}
The general strategy employed here is to compare the late-time evolution of the numerical simulations with a first-order conformal hydrodynamic system, in which the shear viscosity over entropy density $\eta/s$ is the only medium parameter and uniquely governs the relaxation to isotropy.
For a Bjorken expanding system, the pressure ratio is given by Eq.~\eqref{eq:PL_over_PT_first_order_hydro}.
However, other ways to extract $\eta/s$ have been used in the literature. For example, the authors of Ref.~\cite{Kurkela:2018vqr} fit to the pressure difference
\begin{align}
	\frac{P_T-P_L}{\varepsilon+P}=2\left(\frac{\eta/s}{\tau \Teps}\right)+4C_2/3\left(\frac{\eta/s}{\tau \Teps}\right)^2,\label{eq:pressure_difference_formula}
\end{align}
which originates from a second-order hydrodynamic formulation. Here, $C_2$ is an additional fitting constant, and the (isotropic) pressure $P$ is related to the energy density $\varepsilon$ via the conformal relation $P=\varepsilon/3$.
Both expressions \eqref{eq:PL_over_PT_first_order_hydro} and \eqref{eq:pressure_difference_formula} will be used here to obtain the value of $\eta/s$ for a given coupling $\lambda$ and screening prescription. While both equations give similar results, they are not identical, and their difference is taken as an estimate of the systematic uncertainty.

\begin{figure}
    \centering
    \includegraphics[width=\linewidth]{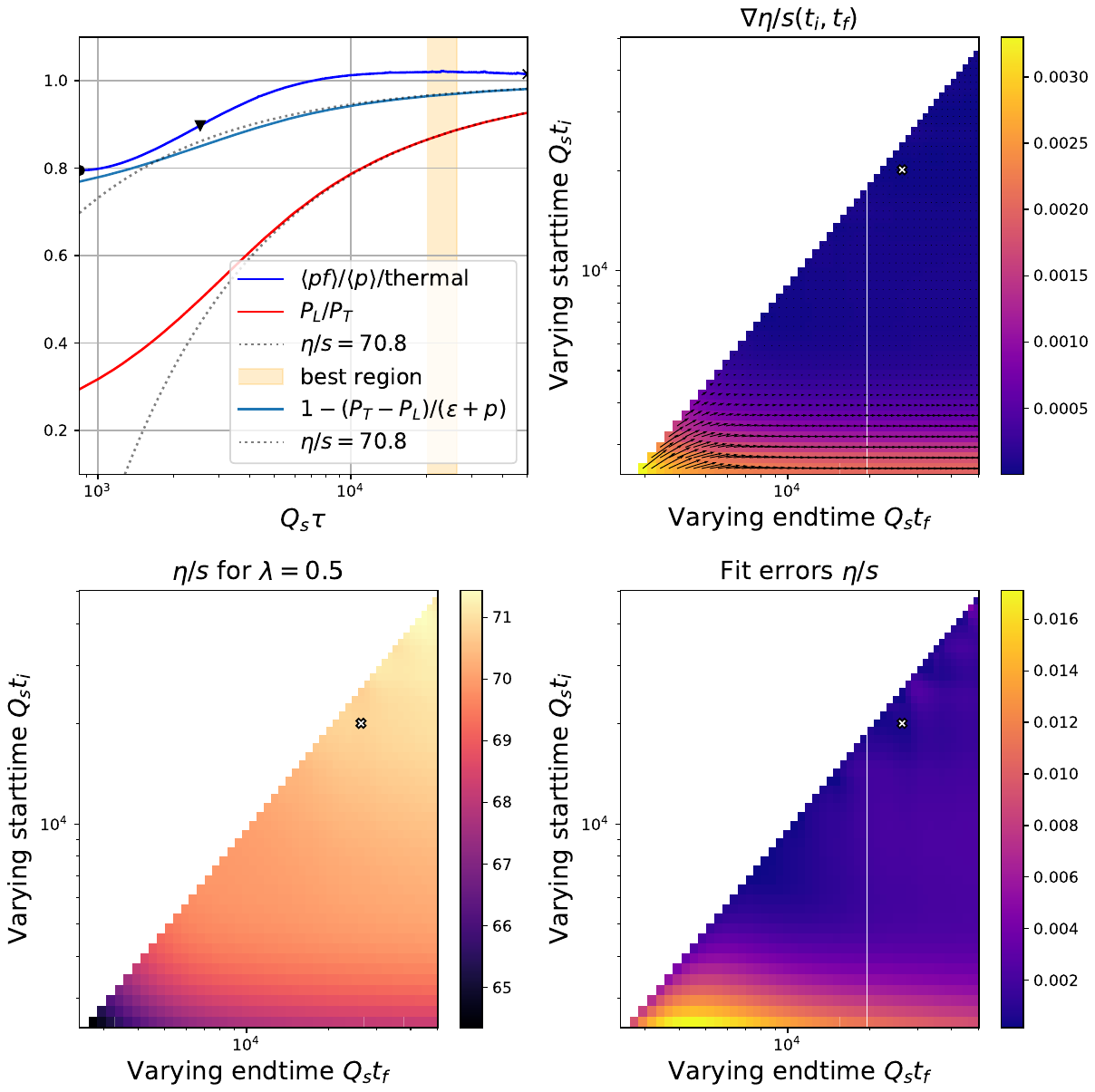}
    \caption{One example of how $\eta/s$ is fitted and its dependence on the start and end of the fit time, for $\lambda=0.5$ and isoHTL screening.
    (\emph{Top left}): Pressure ratio and pressure anisotropy as in \eqref{eq:PL_over_PT_first_order_hydro} and \eqref{eq:pressure_difference_formula}. Shown are also the fits with the best value of $\eta/s$.
    (\emph{Bottom left}): Fitted values of $\eta/s$ as a function of the start and end time.
    (\emph{Top right}): Gradient of the fitted values of $\eta/s$ with respect to varying start and end time.
    (\emph{Bottom right}): Fit errors from fitting $\eta/s$ to Eq.~\eqref{eq:PL_over_PT_first_order_hydro}. The cross is the best value obtained as explained in the text.
    }
    \label{fig:etas_fit}
\end{figure}

\subsubsection{Detailed fitting procedure}
More concretely, the parameter $\eta/s$ is extracted by fitting Eq.~\eqref{eq:PL_over_PT_first_order_hydro}
at some late time interval $(t_i,t_f)$ to the pressure anisotropy $P_L/P_T$ of the simulation.
There are two main causes of inaccuracy in this procedure: On the one hand, any fixed-order hydrodynamic formula only describes the system's behavior sufficiently close to thermal equilibrium (in the present case: at sufficiently late times).
Before that, corrections to the first-order form are expected, which will make the obtained values less reliable when decreasing the initial fit time $t_i$. On the other hand, the numerical simulations performed here are plagued with discretization artifacts that become worse over time (see Section \ref{sec:discretization-artifacts}).
Hence, a too large final time $t_f$ may worsen the fit.

Therefore, to obtain the best fit, the start ($t_i$) and end times ($t_f$) of the fitting process are varied. The earliest time for $t_i$ is chosen as the time when 
the pressure anisotropy is $P_L/P_T=0.5$ (triangle marker). 
For each $(t_i, t_f)$ pair, the value of $\eta/s$ and its fit error is recorded. Fig.~\ref{fig:etas_fit} illustrates this procedure for one particular example. The top left panel shows the pressure ratio to which Eq.~\eqref{eq:PL_over_PT_first_order_hydro} is fitted. The bottom left panel shows the values of $\eta/s$ obtained for different pairs of initial and final times for the fit, $(t_i, t_f)$.
Calculating also the gradients $\nabla_{t_i,t_f}\eta/s$ (Top right panel) allows extracting the value where the fit-error (bottom right panel) times the gradient is the smallest.
This fitting procedure is then performed for several simulations with different random number generator seeds 
and the results are averaged. From that, the statistical uncertainty is obtained by using a simple error of the mean estimate (see Appendix \ref{sec:statistical-averages}).
The procedure described here is not meant to account for all possible systematic errors. The goal here is to obtain a robust procedure to obtain $\eta/s$ and investigate whether there are \emph{systematic differences} in the value of $\eta/s$ for \emph{different screening prescriptions}.

\subsubsection{Initial conditions}
The shear viscosity over entropy density $\eta/s$ is a medium parameter and as such, independent of the specific initial condition. For $\lambda\geq 1$, the same initial distribution \eqref{eq:initial_cond} as for all expanding simulations (with $\xianiso=10$) is used, for $\lambda=0.5$ late-time discretization artifacts are reduced by choosing the initialization time $Q_s\tau=50$ and initial anisotropy $\xianiso=2$ with the suitable initial amplitude $A(\xianiso{=}2)=0.96789$, see Table \ref{tab:initconds}.

\subsubsection{Results for $\eta/s$}
The results of this procedure are summarized in \tab\ref{tab:etas_values}, where also the statistical error estimate for the fit procedure is given.

We find that both Debye-like screening prescriptions and both fit formulas \eqref{eq:PL_over_PT_first_order_hydro} and \eqref{eq:pressure_difference_formula} lead to similar values of $\eta/s$ which are consistent with earlier numerical extractions \cite{Keegan:2015avk}.
However, the isoHTL values are, in general, about 10\% - 20\% smaller than those from Debye-like screening. Additionally, they are much closer to the perturbative estimates at next-to-leading-log accuracy \cite{Arnold:2003zc}, which are labeled as ``pQCD NLL'' and given by
\begin{align}
	\label{eq:pQCD_NLL}
	\left.\frac{\eta}{s}\right|_{\mathrm{NLL\,pQCD}}=\frac{34.784}{\lambda^2\ln\left[4.789/\sqrt{\lambda}\right]}.
\end{align}
This formula is based on an expansion in inverse logarithms and is thus only valid at small couplings at next-to-leading logarithmic (NLL) accuracy, and there is good agreement for $\lambda \lesssim 1$ between this formula and the isoHTL values.
This may indicate that there are no large systematic biases in the extraction procedure performed here. In general, any such systematic bias would be similar for all screening prescriptions considered in this paper, and thus the decrease of the value of $\eta/s$ for the isoHTL screening prescription can be seen as a robust statement.

\begin{table}
		\begin{tabular}{c c c c c c c c c}\toprule
            $\lambda$&\multicolumn{2}{c}{Debye1} & \multicolumn{2}{c}{Debye2}& \multicolumn{2}{c}{isoHTL} & pQCD \\
			 & Eq.~\eqref{eq:PL_over_PT_first_order_hydro} &  Eq.~\eqref{eq:pressure_difference_formula} & Eq.~\eqref{eq:PL_over_PT_first_order_hydro} &  Eq.~\eqref{eq:pressure_difference_formula} & Eq.~\eqref{eq:PL_over_PT_first_order_hydro} &  Eq.~\eqref{eq:pressure_difference_formula} & NLL \\ [0.5ex]
			\hline
			
			$0.5$ & $76.6(9)$ & $78.2(1)$ & $77.0(2)$ & $78.2(1)$ & $70.6(9)$ & $71(2)$ & 72.7\\
			$1.0$ & $23.249(6)$ & $25.0(3)$ & $23.228(6)$ & $25.0(3)$ & $21.55(6)$ & $21.6(7)$ & 22.2\\
			$2.0$ & $7.50(4)$ & $7.55(4)$ & $7.49(4)$ & $7.517(10)$ & $6.59(9)$ & $6.3(4)$ & 7.13\\
			$5.0$ & $1.7037(5)$ & $1.731(5)$ & $1.6909(4)$ & $1.717(1)$ & $1.44(2)$ & $1.48(3)$ & 1.83\\
			$10.0$ & $0.5940(6)$ & $0.60(2)$ & $0.5830(7)$ & $0.55(3)$ & $0.505(2)$ & $0.513(8)$ & 0.838\\
			$20.0$ & $0.2120(5)$ & $0.20(1)$ & $0.2027(6)$ & $0.20(1)$ & $0.1759(3)$ & $0.180(2)$ & 1.27\\[1ex]
            \bottomrule
		\end{tabular}
	\caption{Extraction of $\eta/s$ values for various couplings $\lambda$ using the screening prescriptions Debye1 \eqref{eq:usual_screened_matrix_element}, Debye2 \eqref{eq:Debye2_screened_matrix_element}, isoHTL \eqref{eq:full_isotropic_HTL_matrixelement} and the extraction procedures \eqref{eq:PL_over_PT_first_order_hydro} and \eqref{eq:pressure_difference_formula}. The error estimate from our fitting procedure is explained in the text. 
		NLL pQCD denotes the weak coupling perturbative expression \eqref{eq:pQCD_NLL} from \cite{Arnold:2003zc}. Table adapted from \cite{Boguslavski:2024kbd}.
	}
	\label{tab:etas_values}
\end{table}
\begin{figure}
	\centering
	\includegraphics[width=0.5\linewidth]{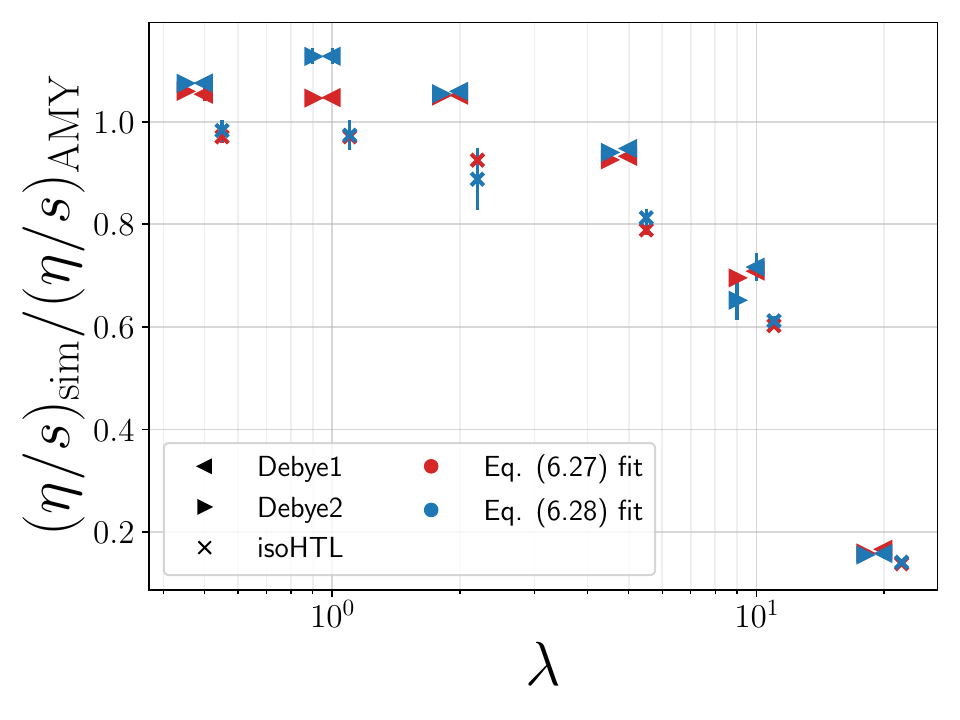}
	\caption{Numerically extracted values $\eta/s$ from \tab\ref{tab:etas_values}. The simulation and fit results for the different screening methods and fitting formulas are normalized by the NLL pQCD results stated in the table. For visual ease, the markers are slightly shifted around the value of $\lambda$. Adapted from \cite{Boguslavski:2024kbd}.
	}
	\label{fig:etas_figure2}
\end{figure}

\fig\ref{fig:etas_figure2} shows the ratio of the extracted values of $\eta/s$ over the perturbative NLL results from Ref.~\cite{Arnold:2003zc}. As mentioned before, for small couplings $\lambda \lesssim 1$, the values of $\eta/s$ for the isoHTL matrix element are very close to the pQCD values, whereas the Debye-like screened matrix elements consistently lead to larger values of $\eta/s$.
At larger couplings, we see increasing discrepancies between the perturbative NLL values and the extracted results. Nevertheless, the isoHTL screening continues to yield consistently smaller values than the Debye-like prescriptions.
At such large coupling strengths, we should recall that
the inverse log expansion from Ref.~\cite{Arnold:2003zc} breaks down, and also kinetic theory becomes less accurate.
However, it is interesting and encouraging to note that both screening prescriptions lead to rather similar values even at large couplings.

\begin{table}
    \centering
	\begin{tabular}{c c c} \toprule
		$\lambda$ & Debye1, \eqref{eq:PL_over_PT_first_order_hydro} &  Debye1, \eqref{eq:pressure_difference_formula} \\ [0.5ex]
		\hline
		$0.53$ & $70.49(3)$ & $72.4(8)$\\
		$2.2$ & $6.39(3)$ & $6.417(4)$\\
		$11.2$ & $0.5012(8)$ & $0.507(8)$
		\\[1ex]
        \bottomrule
	\end{tabular}
	\caption{Extraction of $\eta/s$ values for different couplings $\lambda$ and the Debye-like screening \eqref{eq:usual_screened_matrix_element} for the $\eta/s$ extraction procedures \eqref{eq:PL_over_PT_first_order_hydro} and \eqref{eq:pressure_difference_formula}.  Table from Ref.~\cite{Boguslavski:2024kbd}.
	}
	\label{tab:etas_values2}
\end{table}

\subsection{Simulations at the same specific shear viscosity $\eta/s$}
\begin{figure}
	\centering\includegraphics[width=0.5\linewidth]{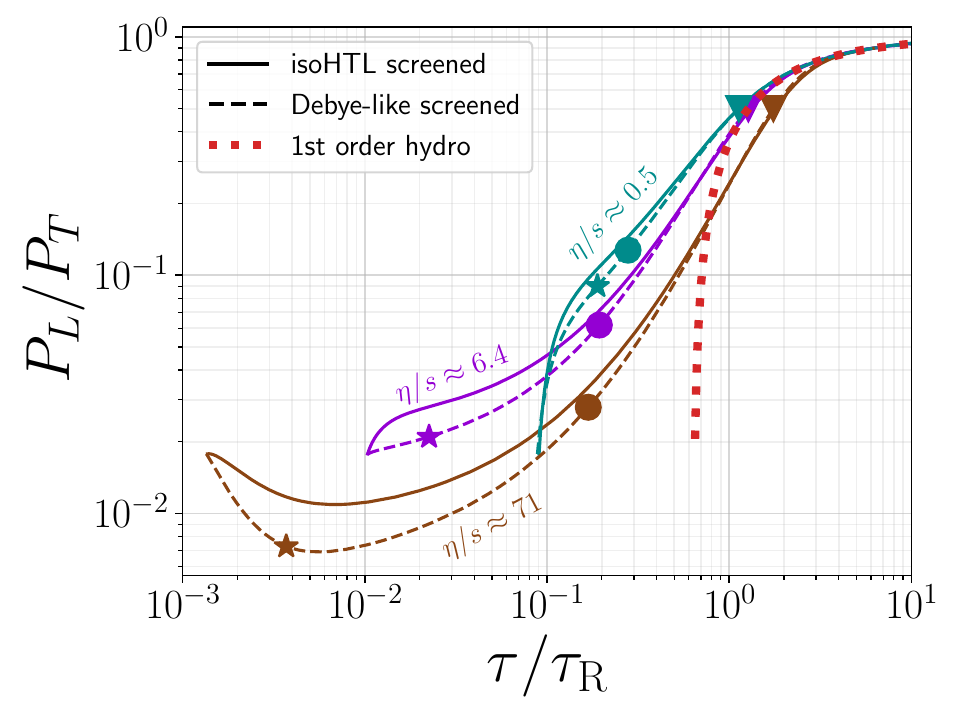}
	\caption{
		Pressure ratio for systems with similar $\eta/s$. The values of $\lambda$ are adjusted between the isoHTL and Debye-like screened simulations to yield a similar value of $\eta/s$. Figure from \cite{Boguslavski:2024kbd}.
	}
	\label{fig:pressure_ratio_etas}
\end{figure}
In the previous sections, we have compared the results using Debye-like and isoHTL screening for the same value of the coupling $\lambda$. But then we showed that shear viscosity $\eta$ entering the relaxation time $\tauR$ via Eq.~\eqref{eq:timescales-improvingqcd} depends on the screening prescription. Therefore, one might ask whether the changes in the pressure ratio for different screening prescriptions (shown in Fig.~\ref{fig:pressure-ratio}) are due to the different values of $\eta/s$. We will see here that this is not the case.
To do that, instead of comparing simulations with the same value of $\lambda$, we compare at the same value of $\eta/s$.

In particular, the isoHTL screened simulations at couplings $\lambda\in\{0.5,2,10\}$ are compared with Debye-like screened simulations with couplings $\lambda\in\{0.53,2.2,11.2\}$, which lead to similar values of $\eta/s$ (see \tab\ref{tab:etas_values2}).
This comparison is illustrated in \fig \ref{fig:pressure_ratio_etas} for the pressure ratio $P_L/P_T$ as a function of the rescaled time $\tau/\tauR$. Again, at late times, the curves coincide and fall on the universal hydrodynamic curve \eqref{eq:PL_over_PT_first_order_hydro} (red dotted line) as expected from an approach to a hydrodynamical evolution. 
In contrast, as before, we find that the evolution at early times differs substantially for the different screening prescriptions. 
The overall behavior is very similar to the comparison between screening prescriptions at the same value of the coupling in \fig\ref{fig:pressure-ratio}.
Thus, the early-time evolution of the pressure ratio (and similarly of other quantities) is significantly different for the isoHTL screening prescription, both when compared at the same value of the coupling $\lambda$ or shear viscosity over entropy density $\eta/s$.

\subsection{Thermalization and hydrodynamization time}
\begin{figure}
	\centerline{
	\includegraphics[width=0.5\linewidth]{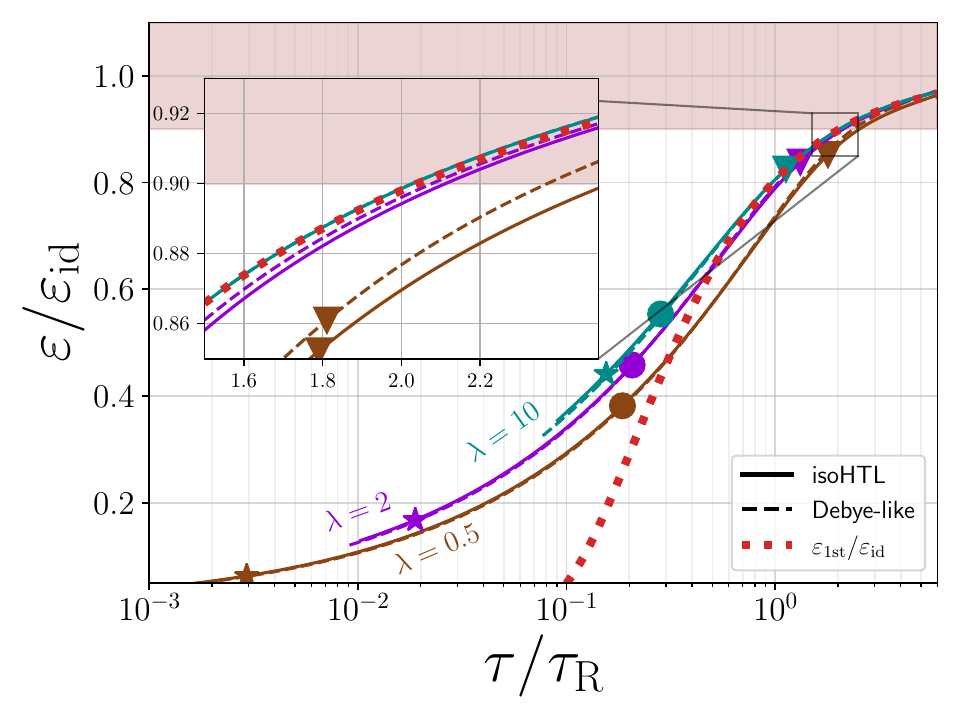}
	\includegraphics[width=0.5\linewidth]{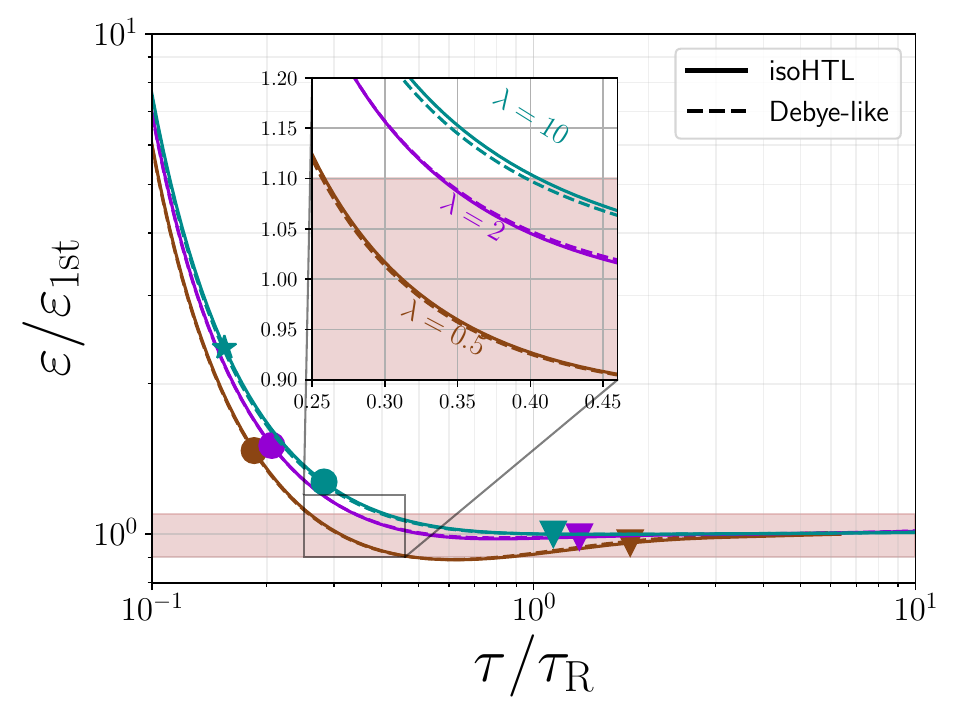}
    }
	\caption{Energy density of the nonequilibrium simulation for different couplings $\lambda$ (color coded) normalized to its ideal (left) and first-order hydrodynamic estimate (right). The results using the isoHTL-screened matrix element are shown as solid lines, for the Debye-like screened matrix element as dashed lines. The top panel also depicts the first-order hydrodynamic estimate as a thick red dotted line. The shaded region indicates a less than $10$~\% deviation from unity. Figure from Ref.~\cite{Boguslavski:2024kbd}.
	}
	\label{fig:thermalization_and_hydrodynamization}
\end{figure}
Next we move on to study how quickly the system thermalizes, and how quickly a hydrodynamic description becomes applicable.
In Ref.~\cite{Kurkela:2018xxd}, the thermalization time is defined as the time when the energy density of the nonequilibrium kinetic theory simulation first gets within $10\%$ of the ideal hydrodynamic estimate,
\begin{align}
	\left|1-\frac{\varepsilon(\tautherm)}{\varepsilon_{\mathrm{id}}(\tautherm)}\right|)=\left|1-\frac{\Teps^4(\tautherm)}{\Tid^4(\tautherm)}\right|
	=0.1\, .
\end{align}
Similarly, the hydrodynamization time is defined as the time when the full kinetic theory result gets within $10\%$ of the first-order hydrodynamic estimate,
\begin{align}
	\left|1-\frac{\varepsilon(\tauhydro)}{\varepsilon_{\mathrm{1st}}(\tauhydro)}\right|=
    \left|1-\frac{\Teps^4(\tauhydro)}{\Tfirst^4(\tauhydro)}\right|=
    0.1\,.
\end{align}
In ideal and first-order hydrodynamics, the temperature evolution is given by
\begin{align}   
	\Tid(\tau)&=\frac{c_1}{\tau^{1/3}},\\
	\frac{\Tfirst(\tau)}{\Tid(\tau)}&=1-\frac{1}{6\pi}\frac{\tauR}{\tau},
\end{align}
with a constant $c_1$, which has to be fitted at late times from the late-time behavior of the nonequilibrium energy density (or rather the effective temperature $\Teps$ from Eq.~\eqref{eq:Landau-matching-condition}).

The left panel of \fig \ref{fig:thermalization_and_hydrodynamization} shows the energy density of the simulation normalized to the energy density of ideal hydrodynamics for different values of the coupling $\lambda$. We observe that both the Debye-like screening results (dashed lines, corresponding to Eq.~\eqref{eq:usual_screened_matrix_element}) and the lines associated with simulations using the isotropic HTL matrix element (solid lines, corresponding to Eq.~\eqref{eq:full_isotropic_HTL_matrixelement}) lie almost on top of each other. This is due to the rescaling of time using the relaxation time $\tauR$, in which the different values for the shear viscosity $\eta/s$ are already included. 
In particular, we find that thermalization occurs roughly at the same time, at $\tautherm\approx 2\,\tauR$ for $\lambda \gtrsim 2$.

The right panel of \fig\ref{fig:thermalization_and_hydrodynamization} shows the energy density normalized to the first-order hydrodynamic estimate. Similar to the thermalization case, we find that hydrodynamization occurs roughly at the same time when comparing both screening approximations.
The results for different values of the coupling $\lambda$ differ more than the different screening prescriptions themselves.
Hydrodynamization occurs roughly at $\tauhydro\approx 0.35\,\tauR$.

These results indicate that the impact of changing the screening prescription on the approach to hydrodynamics and to local thermal equilibrium is mainly captured by the values of the coefficient $\eta/s$.

\subsection{Impact of screening prescriptions on early-time dynamics}

\begin{figure}
	\centering
	\includegraphics[width=0.5\linewidth]{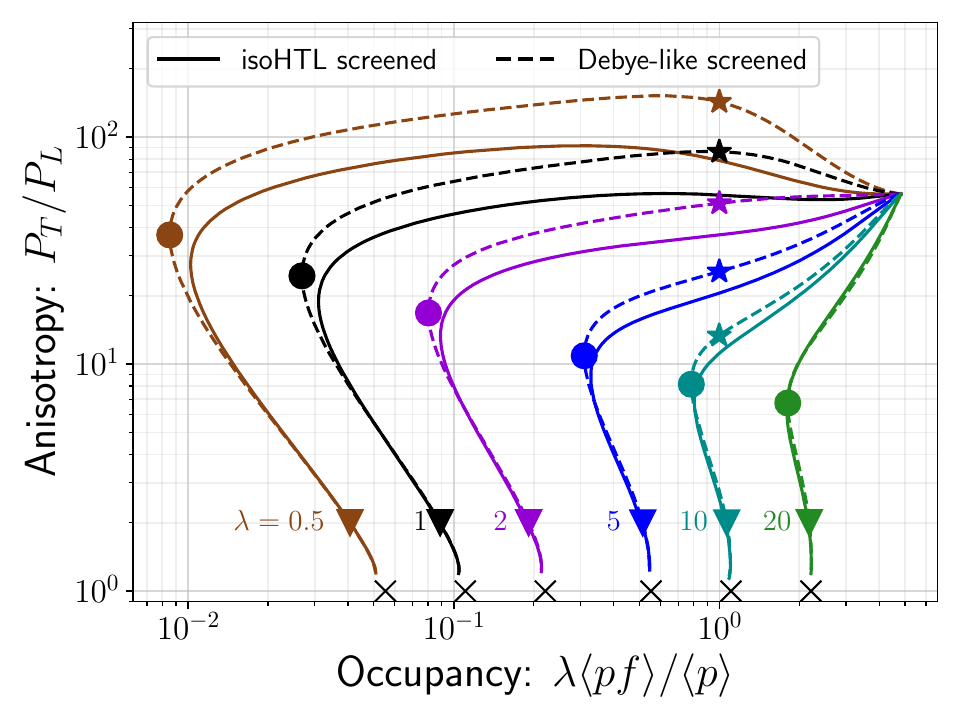}
	\caption{Comparison of simulations with isoHTL (solid) and Debye-like (dashed lines) screening prescriptions in the anisotropy (measured by the pressure ratio $P_T/P_L$) - occupancy (measured by $\lambda \langle pf \rangle/\langle p \rangle$) plane. Figure from \cite{Boguslavski:2024kbd}.
	}
	\label{fig:overview-curves}
\end{figure}
\begin{figure*}
	\centerline{
		\includegraphics[width=0.33\linewidth]{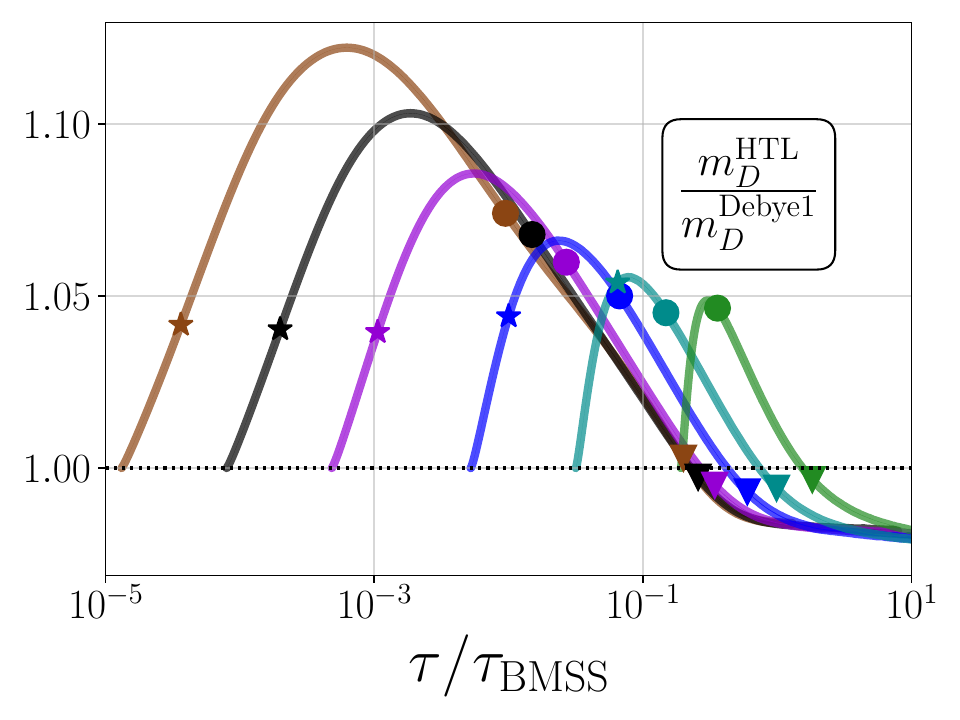}
		\includegraphics[width=0.33\linewidth]{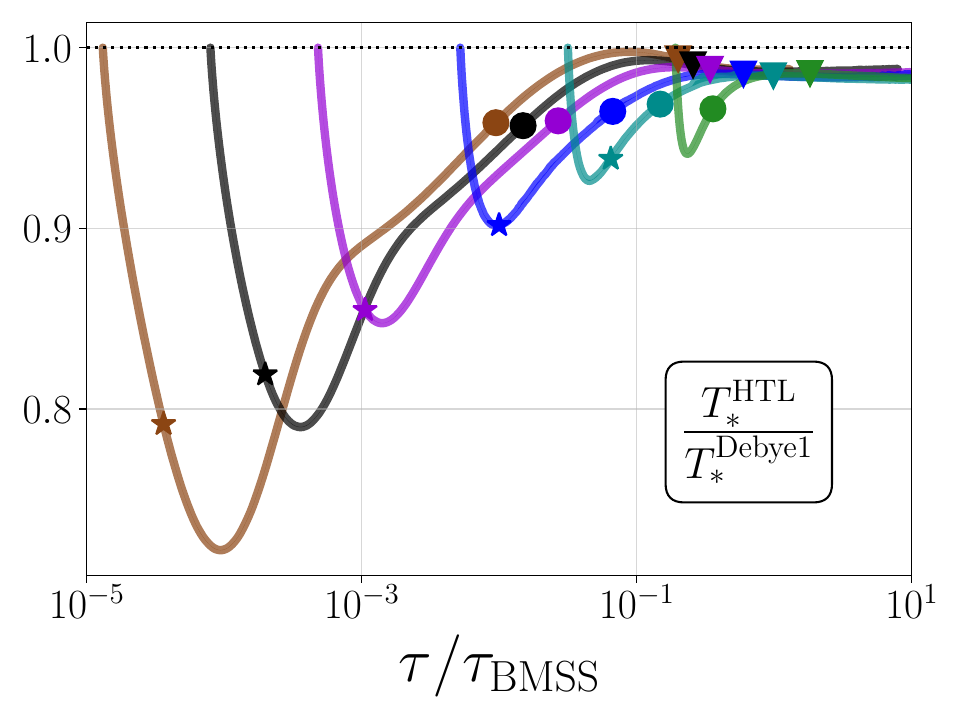}
		\includegraphics[width=0.33\linewidth]{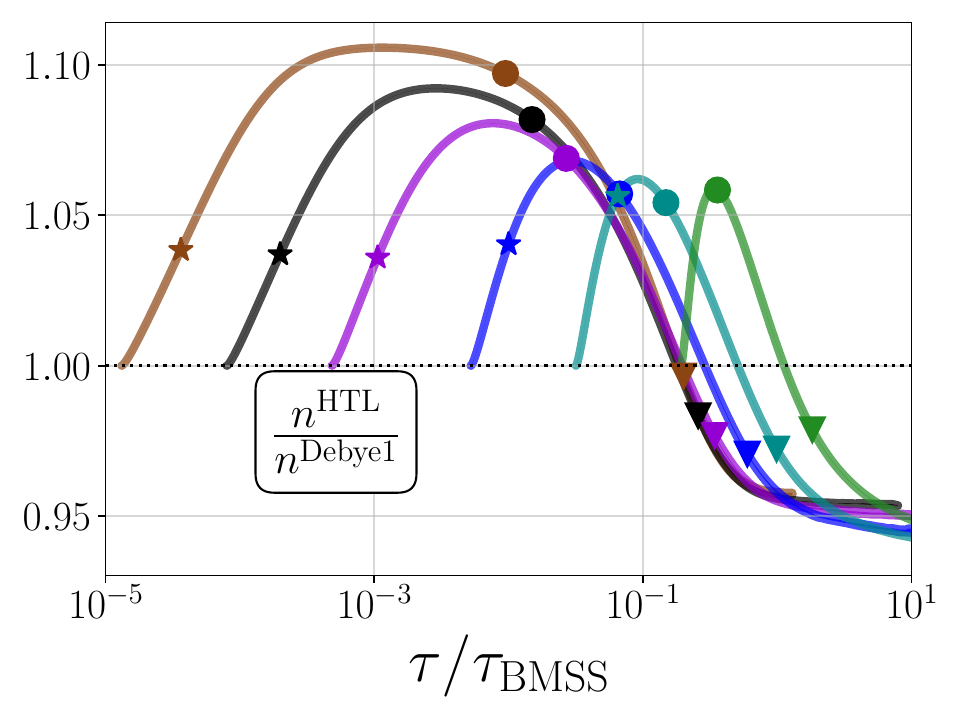}
	}
	\centerline{ 
		\includegraphics[width=0.33\linewidth]{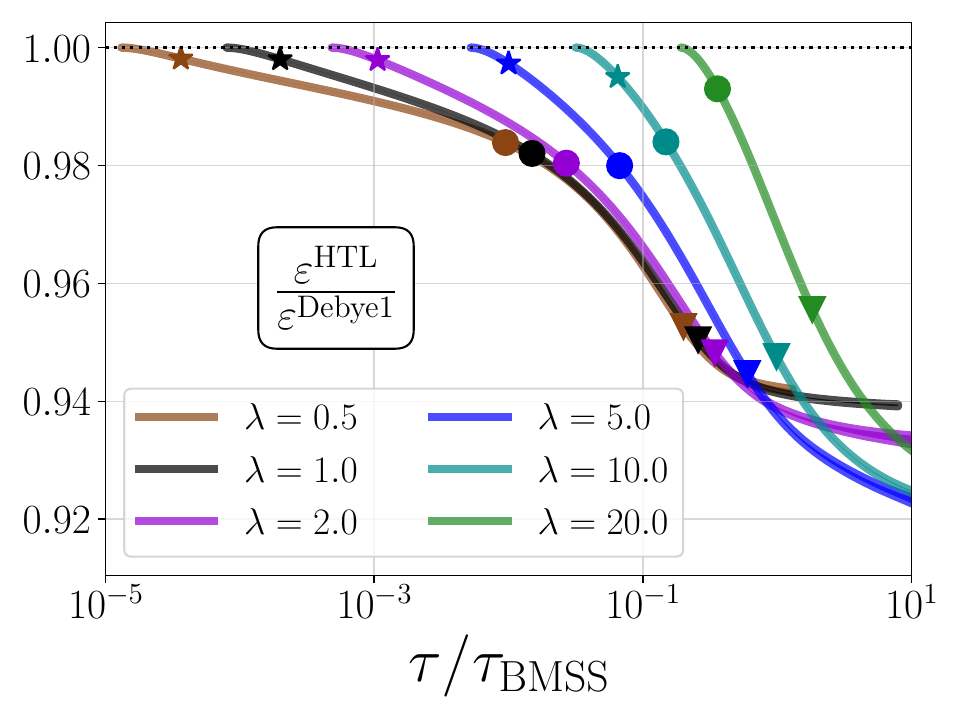}
		\includegraphics[width=0.33\linewidth]{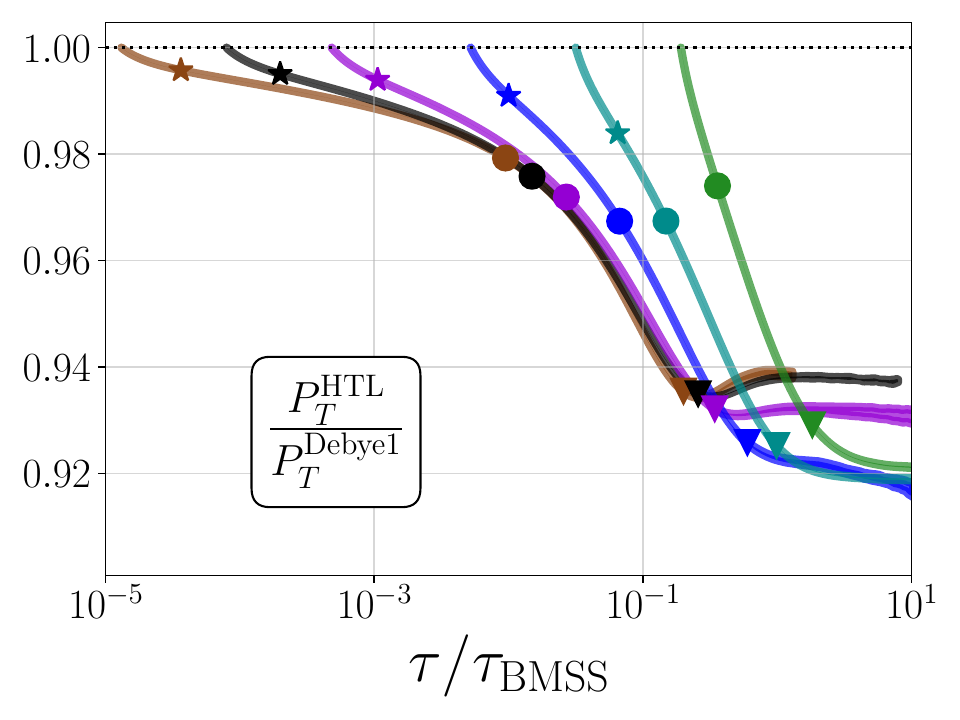}
		\includegraphics[width=0.33\linewidth]{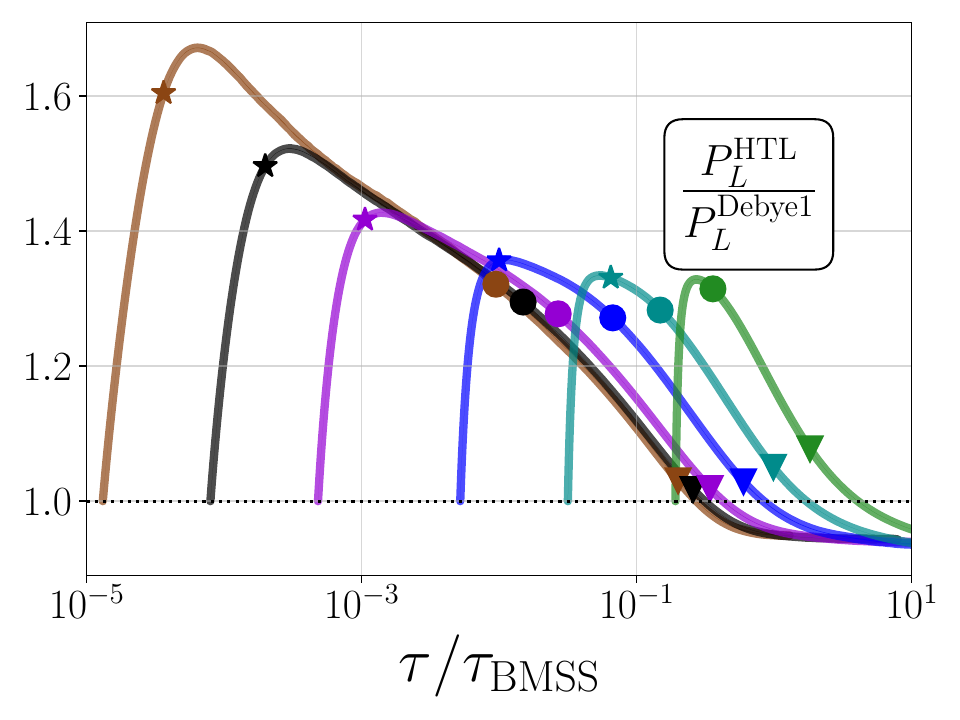}
	}
	\centerline{ 
		\includegraphics[width=0.33\linewidth]{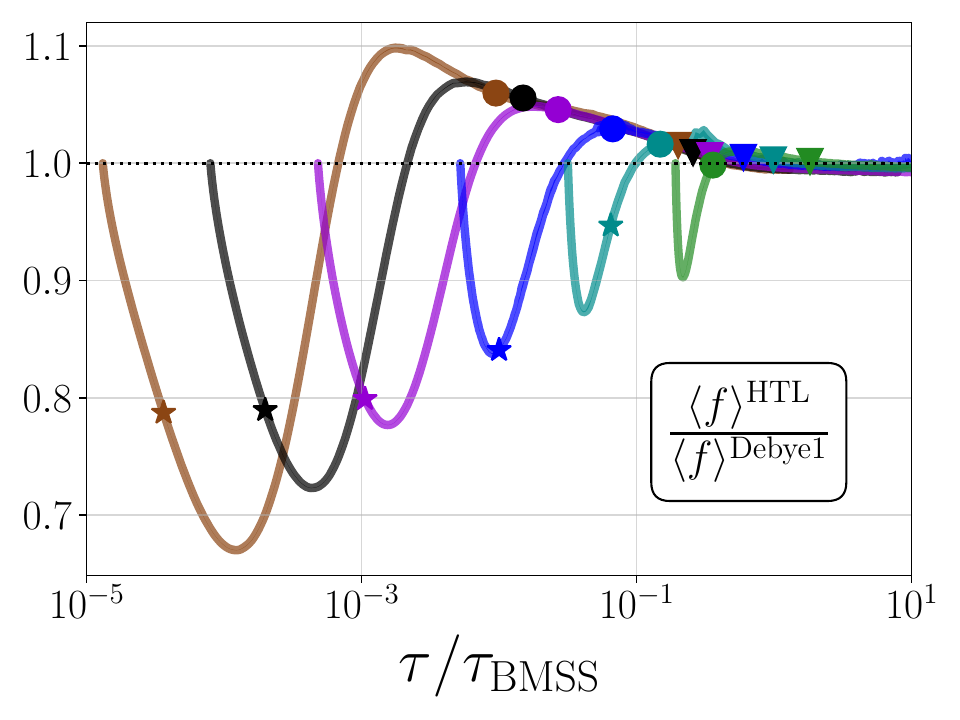}
		\includegraphics[width=0.33\linewidth]{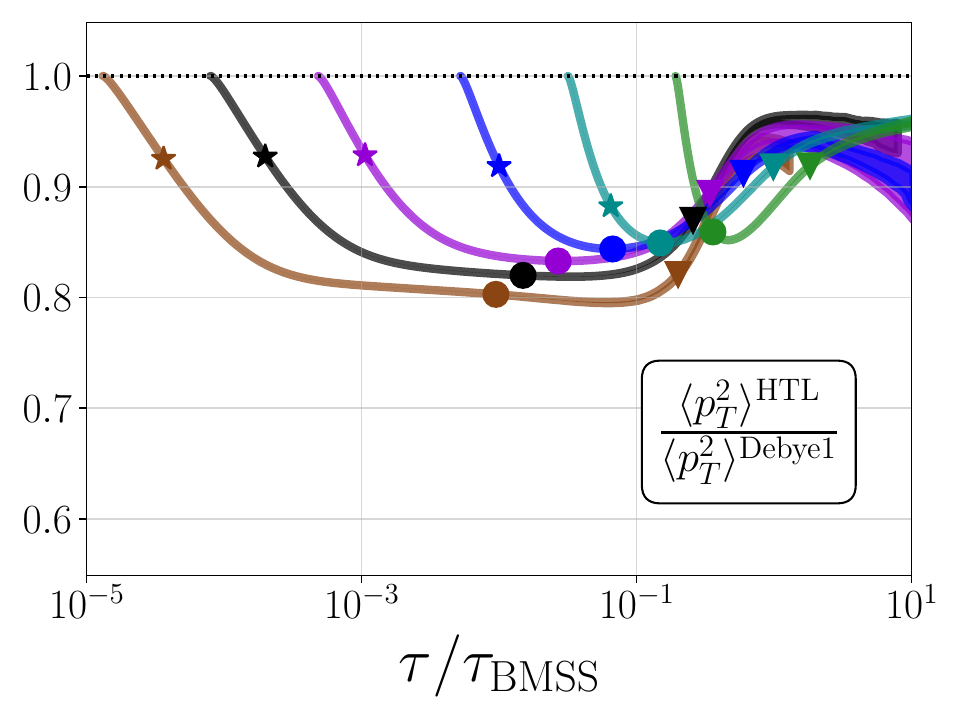}
		\includegraphics[width=0.33\linewidth]{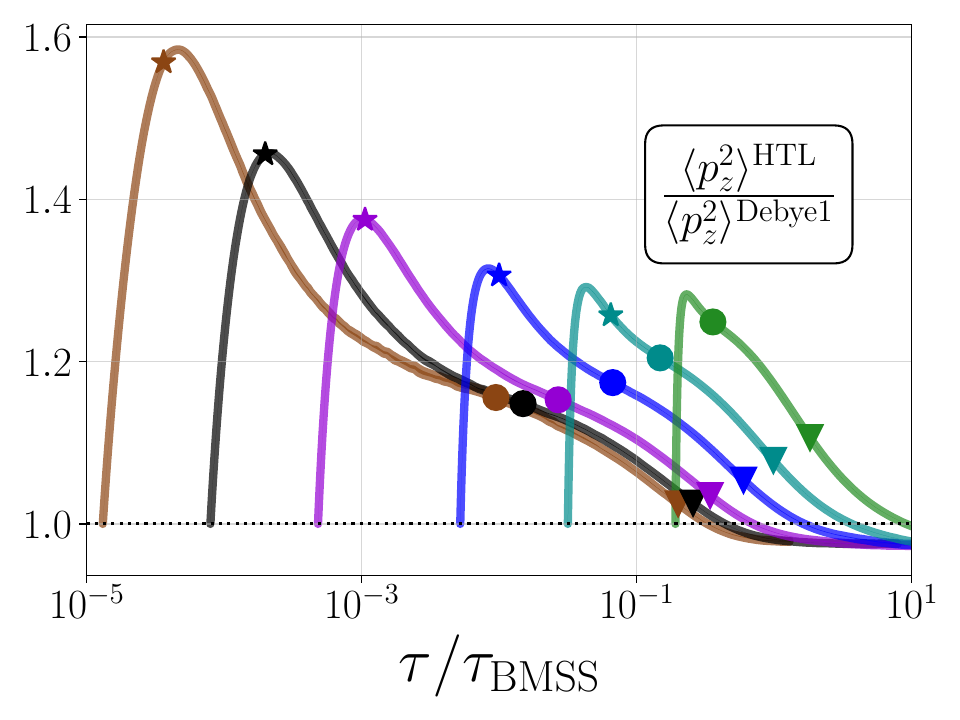}
	}
	\caption{\label{fig:obs_ratios}Observable ratios obtained from simulations with isoHTL screening normalized by those with the Debye-like screening prescription. 
		Time is rescaled with the bottom-up thermalization time \eqref{eq:taubmss}. The band corresponds to an error estimate using a statistical average of 5 simulations. Figures from Ref.~\cite{Boguslavski:2024kbd}.
	}
\end{figure*}
We have depicted the bottom-up thermalization process \cite{Baier:2000sb} already in Fig.~\ref{fig:bottomup-overview} in Section \ref{sec:thermalization-expanding-systems} (see also Fig.~\ref{fig:overview-curves-appendix} in Appendix \ref{sec:additional-results-bottomup}). We will now discuss how different screening prescriptions impact the curves shown in these figures.
\fig\ref{fig:overview-curves} depicts the system's time evolution in the occupancy-anisotropy plane, both for Debye-like screened matrix elements (dashed lines) and the isoHTL screening (solid lines.
All simulations start with the same initial condition given by \eq \eqref{eq:initial_cond}, shown in the upper-right part of the plot, and then evolve towards thermal equilibrium (black crosses).
The isoHTL matrix element leads to a visibly different evolution, especially during the far-from-equilibrium initial stages before minimal anisotropy is reached (circle marker). These differences are seen throughout a wide range of occupation numbers. For instance, for weak couplings, the system reaches a significantly smaller anisotropy in terms of $P_T/P_L$, as we have seen before in Fig.~\ref{fig:pressure-ratio}.
Additionally, the minimal occupancy reached is larger than for Debye-like screening. Nonetheless, we do not find significant qualitative changes to the bottom-up picture of thermalization \cite{Baier:2000sb}.

We now move on to study a wide range of different moments of the distribution function, the same moments that were already shown in Fig.~\ref{fig:debye-like-comparison}:
the Debye mass $m_D$, effective temperature $T_\ast$, particle density $n$, energy density and pressures $\varepsilon$, $P_T$, $P_L$, and moments of the distribution function $\langle f \rangle$, $\langle p_T^2 \rangle$ and $\langle p_z^2 \rangle$. 
\fig \ref{fig:obs_ratios} shows the ratio of these moments obtained from simulations with the isoHTL screening prescription over the same quantities from a Debye-like screened simulation. 
Time is rescaled using the bottom-up timescale $\tauT$, because then all ratios start at unity, which is not the case when using the relaxation time $\tauR$ because of the different shear viscosities for the different screening prescriptions, as can be seen, e.g., in Fig.~\ref{fig:pressure-ratio}.
Different couplings are indicated with different colors. The bands indicate an estimate of the statistical uncertainty from the Monte Carlo evaluation of the collision terms. They are obtained by performing five simulations for every set of initial conditions with different random number seed and calculating the standard error (see Appendix \ref{sec:statistical-averages}).
We observe that most considered quantities show larger deviations between the screening prescriptions for smaller values of the coupling.
This is in line with our expectation that for larger values of the coupling the screening prescriptions differ less, as discussed in \se \ref{sec:ekt_extrapolation_large_couplings}.
Nonetheless, sizable corrections can be found even at large couplings.

For instance, the Debye mass in the isoHTL simulation is enhanced by almost $15\%$ for $\lambda=0.5$ between the star and circle marker, and by about $5\%$ for $\lambda = 20$. The average occupancy $\langle f \rangle$ 
is reduced 
in the isoHTL simulations by over $30 \%$ at early times for weak couplings, but even for $\lambda=20$ we find a reduction of more than $10\%$. 
In contrast, the particle density is enhanced by $5 - 10 \%$.

However, the largest deviations concern measures of the bulk anisotropy of the plasma.
At late times, the components of the energy-momentum tensor from the isoHTL simulations are about $10\%$ smaller than for the Debye-like screened simulations,
which implies a smaller temperature in the subsequent equilibrium plasma.
In contrast, the longitudinal pressure increases by more than $60\%$ for $\lambda=0.5$, and by more than $30\%$ for $\lambda=10$ and $\lambda=20$ before eventually also decreasing to values similar to $P_T$.
This effect results from the longitudinal pressure being dominated by particles close to the transverse plane that undergo transverse momentum broadening, i.e., in $p_z$ direction. Since isoHTL leads to a more accurate and more efficient transverse momentum broadening than the simple Debye-like screening prescription (see, e.g., Fig.~\ref{fig:thermal_equilibrium}), one arrives at the observed higher values of $P_L$. 
Similar effects can be seen in the moments of $f$. Indeed, the anisotropy ratio $\langle p_z^2 \rangle / \langle p_T^2 \rangle$ is also considerably larger for isoHTL simulations than for Debye-like screening.
In particular, $\langle p_z^2 \rangle$ overshoots and $\langle p_T^2 \rangle$ is lower in the HTL cases at early times. This significant reduction of anisotropy at early times is, hence, one of the main qualitative differences between these different screening prescriptions. 

\subsection{Evolution of the jet quenching parameter}
\begin{figure}
	\centering
	\includegraphics[width=0.5\linewidth]{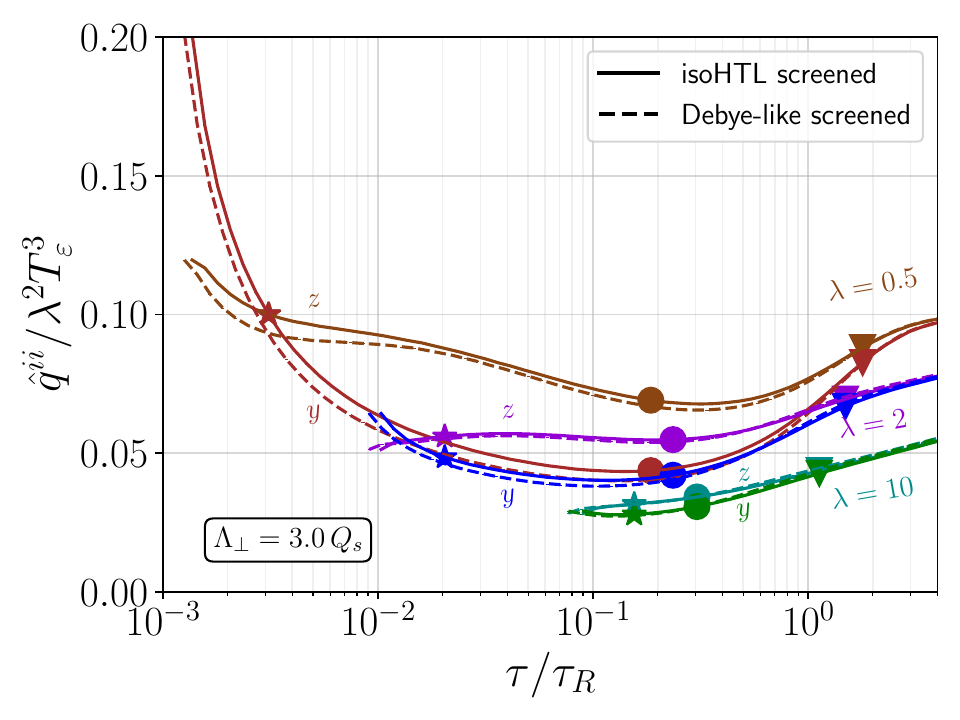}
	\caption{Evolution of the jet quenching parameter $\qhat$ for couplings $\lambda\in\{0.5, 2, 10\}$. The legend refers to the screening prescriptions used in the time evolution of the plasma background. Figure from \cite{Boguslavski:2024kbd}.
	}
	\label{fig:qhat}
\end{figure}
We have studied the jet quenching parameter with various screening prescriptions already in Chapter \ref{sec:momentum-broadening-of-jets}. However, in the previous chapters, the simulations from which $f(\vb p,\tau)$ was taken as input to calculate the jet quenching parameter were all performed using the Debye-like screening prescription in the matrix element. Here, we show that the jet quenching parameter $\hat q$ is only mildly influenced by the choice of screening in the medium evolution.

As in the previous chapters, Eq.~\eqref{eq:qhat_formula_pinf} is used to obtain the jet quenching parameter for a fixed transverse momentum cutoff $\lperp$.

Fig.~\ref{fig:qhat} shows the time evolution of the jet quenching parameter along the beam axis $\qhat^{zz}$ and in the transverse direction $\qhat^{yy}$ for different couplings and screening prescriptions of the plasma.
The results are quantitatively close to those presented in the previous chapters, and the curves for isoHTL and Debye-like screened evolutions only mildly differ. Thus, the choice of either Debye-like or isoHTL screening prescription has only a negligible influence on the time evolution and value of the jet quenching parameter $\qhat$.

\section{Concluding remarks}
In this chapter, we have discussed results for QCD kinetic theory simulations with the isoHTL screening prescription in the elastic collision term, and compared with simulations using the Debye-like screening prescription.
We first reviewed the Debye-like and isoHTL screening prescriptions and discussed, in particular, their gauge invariance.
For isotropic systems, we find only minor differences, but for a Bjorken expanding system relevant for the initial nonequilibrium stages in heavy-ion collisions, we find that the maximum anisotropy is drastically reduced when using the isoHTL screening prescription. Changes can also be found in the transport parameter $\eta/s$, the shear viscosity over entropy density, which determines the approach to isotropy.
Employing the isoHTL screening prescription leads to a reduction of $\eta/s$ at the same value of the coupling $\lambda$, and for small couplings, $\eta/s$ agrees with the perturbative expression.
Finally, we studied whether the jet quenching parameter is influenced by the different screening prescriptions used in the background medium evolution. We find that this is not the case, the value of the jet quenching parameter is almost insensitive to the screening prescription used in the QCD kinetic theory evolution of the background.

%% file: 700_collision_kernel.tex
In 
Chapter \ref{sec:momentum-broadening-of-jets}, we considered the jet quenching parameter $\qhat$, which characterizes the small-distance behavior of the dipole cross section.
In this chapter, we go beyond the small distance form and obtain the full dipole cross section, as the Fourier transform of the elastic collision kernel. We will then use this dipole cross section to calculate the gluon emission rates in the AMY formalism. This chapter is based on a paper in preparation \cite{Altenburger:2025a}.

\section{Going beyond the jet quenching parameter: Elastic collision kernel}
As discussed in the introduction in Section \ref{sec:introduction}, the jet quenching parameter $\qhat$ can be obtained as a moment of the elastic collision kernel (see Eq.~\eqref{eq:intro-qhat-definition}),
\begin{align}
    \qhat = \int\frac{\dd[2]{\vb q_\perp}}{(2\pi)^2}q_\perp^2 C(\vb q_\perp),
\end{align}
which is directly related to the elastic scattering rate
\begin{align}
    C(\vb q_\perp)=(2\pi)^2\frac{\dd \Gammael}{\dd[2]{\vb q_\perp}}. \label{eq:collkern-relation-scatteringrate}
\end{align}
While the jet quenching parameter quantifies the rate of change of transverse momentum squared, $\qhat = \dv[\langle \Delta p_\perp^2\rangle]{t}$, the collision kernel $C(\vb q_\perp)$ encodes the rate for the jet parton to receive a transverse momentum kick with momentum $\vb q_\perp$.
Additionally, this elastic collision kernel $C(\vb q_\perp)$ is needed as input to calculate the rate of inelastic gluon emissions (as discussed in Chapter \ref{sec:jet-energy-loss}), making it an essential quantity that encodes all the relevant medium properties.
For solving the AMY rate equations \eqref{eq:amy-integralequation-long} relevant for gluon emission, it is convenient to transform to impact parameter space, where the relevant quantity is then the dipole cross section
\begin{align}
    C(\vb x)=\int\frac{\dd[2]{\vb q_\perp}}{(2\pi)^2}(1-e^{i\vb x\cdot \vb q_\perp})C(\vb q_\perp). \label{eq:chapter-collkern-fouriertrafo}
\end{align}

\subsection{Generalizing the jet quenching parameter to obtain the collision kernel}
In Chapter \ref{sec:momentum-broadening-of-jets}, we have discussed how to obtain the jet quenching parameter $\qhat$ in QCD kinetic theory.
The collision kernel can be obtained in a very similar way, which we will briefly discuss here.

We start again with the expression for the elastic scattering rate \eqref{eq:decay_rate},
\begin{align}
    \Gammael &= \frac{1}{4p\nu_a}\sum_{bcd}\int\frac{\dd[3]{\vb k}}{(2\pi)^32k}\frac{\dd[3]{\vb {p'}}}{(2\pi)^32p'}\frac{\dd[3]{\vb k'}}{(2\pi)^3 2k'}|\mathcal M^{ab}_{cd}(\vb p, \vb k; \vb p', \vb k')|^2\nonumber\\
    &\times(2\pi)^4\delta^4(P+K-P'-K')
     f(\vb k)(1\pm f(\vb k'))(1\pm f(\vb p')), \label{eq:decay_rate3}
\end{align}
As noted before, the scattering rate \eqref{eq:decay_rate3} is symmetric under the exchange of outgoing particles $p'\leftrightarrow k'$ (switching $u$ and $t$ channel), so we can always enforce $p'>k'$, which leads to an additional factor $2$.
We first integrate over $\vb k'$ using the delta function and then perform a variable substitution $\vb p'=\vb p+\vb q$. We pull out the integral over $\vb q_\perp$ and use Eq.~\eqref{eq:collkern-relation-scatteringrate} to arrive at 
\begin{align}
    \begin{split}
    C(\vb q_\perp)&=\frac{1}{2p\nu_a}\sum_{bcd}\int\frac{\dd[3]{\vb k}}{(2\pi)^3}\int\dd {q_z} \frac{|\mathcal M^{ab}_{cd}(\vb p, \vb k; \vb p+\vb q, \vb k-\vb q)|^2}{8|\vb k||\vb k-\vb q||\vb p+\vb q|} \\ &\times f(\vb k)(1+f(\vb k-\vb q))\delta(p+|\vb k|-|\vb p+\vb q|-|\vb k-\vb q|).\label{eq:formula_Cq_appendix}
    \end{split}
\end{align}
In the limit $p\to\infty$, as already was the case for the jet quenching parameter $\qhat$ in Section \ref{sec:pinf_formula}, only the matrix elements in Table \ref{tab:p-inf_matrix_el} remain, which can be written in terms of the screened matrix element $\tildeMscreen$. For instance, the matrix element for gluon-gluon scattering reduces to
\begin{align}
    \lim_{p\to\infty}\frac{|\mathcal M^{gg}_{gg}|^2}{p^2}=4 d_A C_A^2\left|G_R(P-P')_{\mu\nu}(P+P')^\mu(K+K')^\nu\right|^2,
\end{align}
and we can use the isotropic retarded hard-thermal loop propagator $G_R$ for the internal soft gluon propagator (isoHTL screening), Eq.~\eqref{eq:full_htl_matrix_element}.

The delta function can be rewritten as \cite{Arnold:2000dr}
\begin{align}
        \delta(q_z-k+k')&=\frac{k'}{kq}\delta\left(\cos\thetaqk-\frac{q_z}{q}+\frac{t}{2kq}\right)\Theta(k-q_z),
\end{align}
and only yields a contribution if $k>\frac{q_z+q}{2}$. 
We then arrive at
\begin{align}
\begin{split}
    C(\vb q_\perp)&=\frac{1}{16p\nu_a(2\pi)^3}\sum_{bcd}\int_0^{2\pi}\dd{\phiqk}\int_{-\infty}^\infty\frac{\dd{q_z}}{q}\int_{\frac{q+q_z}{2}}^\infty\dd{k}\\
    &\qquad\times\lim_{p\to\infty}\frac{|\mathcal M^{ab}_{cd}(\vb p, \vb k; \vb p+\vb q, \vb k-\vb q)|^2}{p^2}f(\vb k)(1+f(\vb k-\vb q)). \label{eq:Cq_final}
    \end{split}
\end{align}

Note that in Chapter \ref{sec:momentum-broadening-of-jets}, we have introduced $\omega=p'-p$, which is equivalent to $q_z\equiv\omega$ in the limit $p\to\infty$. We proceed similarly as in Chapter \ref{sec:momentum-broadening-of-jets}, and perform the $\vb k$ integral in a frame in which $\vb q$ points in the $z$ direction, and the $\vb q$ integral is performed in a frame in which $\vb p$ points in the $z$ direction. 
In practice for an implementation, we need the angles $\theta_k$ and $\theta_{k'}$, which we obtain as in Chapter \ref{sec:momentum-broadening-of-jets} via Eq.~\eqref{eq:cos-thetak_labframe}. Note that the collision kernel $C(\vb q_\perp)$ is a function of the two-dimensional vector $\vb q_\perp$ in the transverse plane to the jet. Only the component of $\vb q$ parallel to the jet is integrated over (see also Eq.~\eqref{eq:collisionkernel-formula-from-correlator}).
The remaining two-dimensional vector $\vb q_\perp$ can be parameterized as
\begin{align}
    \vb q_\perp=
    q_\perp\begin{pmatrix}
        \cos\phi\\ \sin\phi
    \end{pmatrix}
    =
    q_\perp\begin{pmatrix}
        \cos\phi_{pq}\\ \sin\phi_{pq}
    \end{pmatrix}=\begin{pmatrix}
        q_x\\q_y
    \end{pmatrix}=\begin{pmatrix}
        -q^{(1)}_z\\ q^{(1)}_y
    \end{pmatrix},\label{eq:qperp-coordinate-systems}
\end{align}
and $q=\sqrt{q_\perp^2+q_z^2}$, all in accordance with Eq.~\eqref{eq:pframe_q} in Section \ref{sec:coordinate_systems}. For a jet moving in the $x$ direction, $\phi_{pq}=0$ corresponds to momentum broadening along the beam direction, while $\phi_{pq}=\pi/2$ to momentum broadening transverse to the beam direction and to the jet direction.

A note about coordinate systems is in order. The upper index $(1)$ denotes the components in the coordinate system, in which the beam axis defines the $z$-axis, which is the coordinate system that we called \emph{lab frame} in Section \ref{sec:coordinate_systems}. However, since $\vb q_\perp$ is a two-dimensional vector in the transverse plane to the jet, it is more natural to use a coordinate system, where these two components are labeled $q_x$ and $q_y$. This is the coordinate system, in which the jet points in the $z$ direction, which we called \emph{$p$ frame} in Section \ref{sec:coordinate_systems}. For our purposes here, it will be more natural to work in this frame, and to label the transverse components $\vb q_\perp=(q_x,q_y)$, as in Eq.~\eqref{eq:qperp-coordinate-systems}. The spatial coordinate in the transverse plane is $\vb x = (x,y)$. Furthermore, we will frequently abbreviate the angle of $\vb q_\perp$ in the transverse plane to the jet as $\phi_{pq}=\phi$.

Eq.~\eqref{eq:Cq_final} is the final result of this subsection and is the equation that is implemented and evaluated numerically.

\subsection{Symmetries of the collision kernel}
Next, let us discuss the symmetries of the collision kernel $C(\vb q_\perp)$, and of the dipole cross section $C(\vb x)$. In momentum space, we can adopt the notation $C(\vb q_\perp)=C(q_x, q_y)$. Since the distribution function $f(\vb k)$ is symmetric around mid-rapidity (mirror-symmetry around the $z=0$ plane, see Eq.~\eqref{eq:distributionfunction-mirrorsymmetry}), $f(k,\cos\theta_k)=f(k,-\cos\theta_k)$, we obtain $C(q_z^{(1)},q_y^{(1)})=C(-q_z^{(1)}, q_y^{(1)})$, and, hence, $C(q_x,q_y)=C(-q_x,q_y)$. Additionally, the distribution function is symmetric under rotations in the transverse plane (see Eq.~\eqref{eq:distributionfunction-rotationsymmetry}), which implies $C(q_x,q_y)=C(q_x,-q_y)$. Combining this, we obtain
\begin{align}
    C(q_x, q_y)=C(\pm q_x, \pm q_y). \label{eq:symmetry_Cq}
\end{align}
With that, the Fourier transform \eqref{eq:chapter-collkern-fouriertrafo} becomes real,
\begin{align}
    C(x,y)&= \int\frac{\dd[2]{\vb q_\perp}}{(2\pi)^2}(1-e^{iq_x x+iq_yy})C(q_x,q_y)=
    \int\frac{\dd[2]{\vb q_\perp}}{(2\pi)^2}(1-\cos(\vb q_\perp\cdot \vb x))C(q_x,q_y) \label{eq:fouriertrafo_real}
\end{align}
From that, we can infer the symmetry properties of the dipole cross section
\begin{align}
    C(x,y)=C(\pm x,\pm y).
\end{align}
Therefore, we can map all angles $\phi > \pi/2$ into the first quadrant,
\begin{align}
    C(|\vb x|,\phi)=\begin{cases}
        C(|\vb x|,\phi), & 0 <\phi <\pi/2\\
        C(|\vb x|,\pi-\phi), & \pi/2 < \phi < \pi\\
        C(|\vb x|,\phi-\pi), & \pi < \phi <3/2 \pi\\
        C(|\vb x|,2\pi-\phi), & 3/2 \pi < \phi < 2\pi
    \end{cases}\label{eq:Cb_symmetry}
\end{align}
It will be convenient to note here how calculating the coefficients of a Fourier series is simplified under these symmetries. For a function $D(\phi)$ obeying the symmetries \eqref{eq:Cb_symmetry},
\begin{align}
    \int_0^{2\pi}\dd{\phi}e^{-im\phi}D(\phi)
    &=\int_0^{\pi/2}\dd{\phi}D(\phi)
    \left(e^{-im\phi}+e^{-im(\pi-\phi)}+e^{-im(\phi-\pi)}+e^{-im(2\pi-\phi)}\right)\nonumber\\
    &=\begin{cases}
        0, & m \text{ odd}\\
        4\int_0^{\pi/2}\dd\phi \cos m\phi \,D(\phi), & m\text{ even}
    \end{cases}
\end{align}
\subsection{Analytic limits: Small and large momentum transfers\label{sec:collkern-limits-analytic}}
Next, we discuss the analytic limits of the collision kernel $C(\vb q_\perp)$ for small and large momentum transfers. Both limits have implicitly already been discussed before for the jet quenching parameter, but it is insightful to consider these limits here specifically for the collision kernel.

First, we consider large momentum transfer, $|\vb q_\perp|\to\infty$, for which it is convenient to start with Eq.~\eqref{eq:formula_Cq_appendix}. We consider the case of pure gluons, but including quarks is considered in Appendix \ref{sec:qhat_behavior_large_cutoff}. For $p\to\infty$, $|\vb p+\vb q|\to p + q_z$, when $\vb p=(0,0,|\vb p|)$, and then the delta function can be rewritten to constrain $q_z$,
\begin{align}
    \delta(k-q_z-|\vb k-\vb q|)\to \frac{q_\perp^2}{2(k-k_z)^2}\delta\left(q_z-\frac{\vb q_\perp\cdot \vb k_\perp-q_\perp^2/2}{k-k_z}\right),
\end{align}
where in the normalization we have already taken the leading term in the limit $q_\perp\to\infty$. Thus, $q_z\approx -q_\perp^2/(2(k-k_z))$, and $q\to\infty$. In this limit, screening effects can be neglected in the matrix element, and we may take the vacuum form of the matrix element 
\begin{align}
    |\mathcal M^{gg}_{gg}|^2=-16 g^4d_A C_A^2\frac{su}{t^2}=16 g^4d_AC_A^2\frac{4p^2(k-k_z)^2}{q_\perp^4}.
\end{align}
Inserting this, we obtain
\begin{align}
    C(\vb q_\perp)
    &=\frac{4d_AC_A^2g^4}{\nu_g}\frac{1}{q_\perp^4}\int\frac{\dd[3]{\vb k}}{(2\pi)^3}\frac{k-k_z}{k}f(\vb k).
\end{align}
Therefore, the broadening kernel is proportional to the number density $J^0=n=\nu_g\int\frac{\dd[3]{\vb p}}{(2\pi)^3}f(\vb p)$ and the number current in the jet direction $J^z=\nu_g\int\frac{\dd[3]{\vb p}}{(2\pi)^3}f(\vb p) p_z/|\vb p|$, which is zero for a distribution that obeys the symmetry condition \eqref{eq:distributionfunction-mirrorsymmetry}.
Using that, we obtain 
\begin{align}
    \lim_{|\vb q_\perp|\to\infty}C(\vb q_\perp)\to\frac{g^4\CR\mathcal N}{q_\perp^4},
\end{align}
where $\mathcal N$, in the general case also including quarks (see Appendix \ref{sec:qhat_behavior_large_cutoff}), is given by
\begin{align}
    \mathcal N = \int\frac{\dd[3]{\vb k}}{(2\pi)^3}\left(2\NC f_g(\vb k)+\sum_f f_f(\vb k)\right).
\end{align}
Importantly, this does not depend on the angle of $\vb q_\perp$ for an arbitrary\footnote{As long as the assumptions needed for the elastic scattering rate \eqref{eq:decay_rate3} are valid.} distribution function $f(\vb k)$.

The opposite limit, $|\vb q_\perp|\to 0$ is more difficult to take because in this limit screening effects become important. 
While for isotropic systems, the $q_z$ integral can be performed analytically using a sum rule (see Appendix \ref{app:sum-rule}), in the anisotropic case, this is more difficult. Here, it will be enough to demonstrate that the leading behavior is $1/q_\perp^2$.
It is not apriori clear that this should be the case since screening effects would naively change the propagators from $1/q^2\to 1/(q^2+\tilde m^2)$, where $\tilde m$ is some effective screening mass, thus becoming finite at low momenta. Thus, we need to consider the HTL self-energy correction instead and reevaluate the integral more cautiously.

To do that, we start with \eqref{eq:Cq_final}, and, following the steps in Appendix \ref{app:sum-rule}, we write the matrix element as (without assuming isotropic distributions),
\begin{align}
    \lim_{p\to\infty}\frac{|\mathcal M^{gg}_{gg}|^2}{4\dA\CA^2}&=\tilde c_1^2|G_L|^2+\tilde c_2^2|G_T|^2-2\tilde c_1\tilde c_2|G_L|^2|G_T|^2(AC+BD)\label{eq:htl-matrix-el-general}
\end{align}
It is enough to focus on the transverse propagator, (with $x=q_z/q)$
\begin{align}
    |G_T|^2=-\frac{4q}{q_z m_D^2\pi(1-x^2)}\frac{\IM \Pi^{T}(x)}{(q^2+\RE\Pi^{T}(x))^2+(\IM \Pi^{T}(x))^2},
\end{align}
which, when integrated over $q_z$ using the sum rule from Appendix \ref{app:sum-rule} gives
\begin{align}
    \int_{-\infty}^\infty\frac{\dd q_z}{q}|G_T|^2=-\frac{4}{m_D^2}\left[\frac{1}{q_\perp^2+\RE\Pi^T(\infty)}-\frac{1}{q_\perp^2+\RE\Pi^T(0)}\right]=-\frac{4}{m_D^2}\left[\frac{1}{q_\perp^2+m_D^2/3}-\frac{1}{q_\perp^2}\right].
\end{align}
This gives exactly the $1/q_\perp^2$ contribution that we wanted to show and
proves that the leading contribution goes like $\sim 1/q_\perp^2$ for small $q_\perp$, even in an anisotropic system.\footnote{A more careful analysis potentially revealing more of the angular information of the coefficient $a(\phi)$ in Eq.~\eqref{eq:limits_qperp} may be possible, but since we deal with an isotropic screening prescription, it seems going beyond a simple scaling needed for the extrapolations used here, is not needed and perhaps not useful.}

Thus, to summarize, the limiting cases are
\begin{subequations}\label{eq:limits_general}
\begin{align}
    C(\vb \qperp)\to \begin{cases}
        \frac{a(\phi)}{\qperp^2}, & q_\perp \to 0\\
        \frac{g^4\CR \mathcal N}{\qperp^4}, & q_\perp \to \infty
    \end{cases}\label{eq:limits_qperp}
\end{align}
where the coefficient $a(\phi)$ may depend on the direction of $\vb q_\perp$.

For an isotropic system, the coefficient $a$ for the small $q_\perp$ case can be obtained analytically using exactly the same rule mentioned above. It leads to 
\begin{align}
\lim_{q_\perp\to 0}C(q_\perp) \to 
g^2C_RT_\ast\,\frac{m_D^2}{q_\perp^2(q_\perp^2+m_D^2)}, 
\label{eq:analytic_collision_kernel}
\end{align}
\end{subequations}
and in equilibrium we have $T_\ast\to T$ and the equilibrium form of the Debye mass \eqref{eq:equilibriumform-debyemass-tstar-meffs}.

\subsection{Numerical results}
Similar to the jet quenching parameter $\qhat$ in Chapter \ref{sec:momentum-broadening-of-jets}, Eq.~\eqref{eq:Cq_final} is evaluated using Monte Carlo integration. We consider a purely gluonic system with a gluon jet, but in the limit $p\to\infty$, the collision kernel for a quark jet or parton can be obtained using Casimir scaling \eqref{eq:Casimir_scaling},
\begin{align}
    \frac{C^{\mathrm{gluon}}(\vb q_\perp)}{\CA}=\frac{C^{\mathrm{quark}}(\vb q_\perp)}{\CF}.
\end{align}

\subsubsection{Thermal equilibrium}
\begin{figure}
    \centering
    \includegraphics[width=0.5\linewidth]{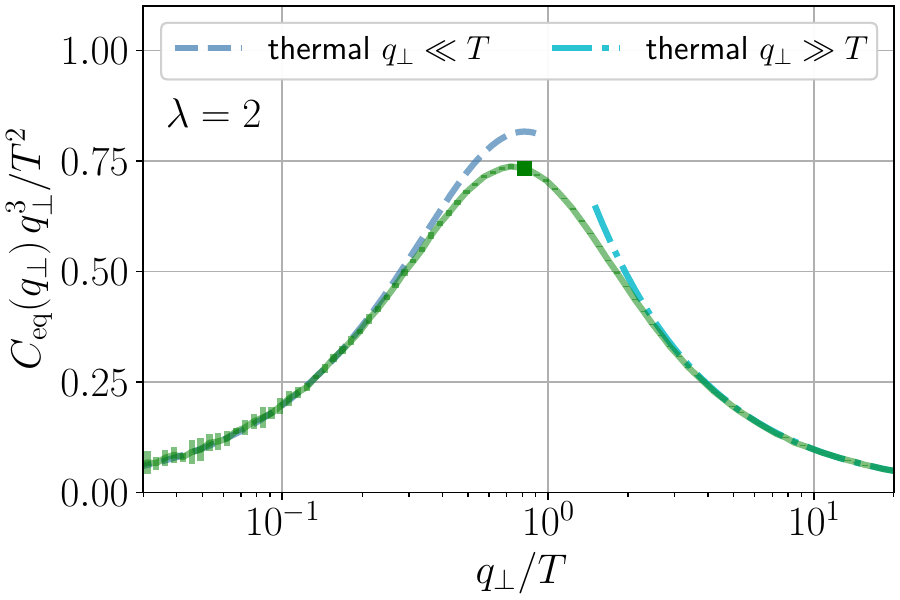}
    \caption{Collision kernel $C(q_\perp)$ for a thermal gluonic system multiplied with $q_\perp^3$ in units of the plasma temperature $T$ for coupling $\lambda=2$. This is the integrand of the jet quenching parameter $\qhat$. Shown are also the analytic small and large $q_\perp$ limits \eqref{eq:limits_general}, as well as the Debye mass $m_D$ as a square on the curve. 
    }
    \label{fig:collision-kernel-thermal}
\end{figure}
First, we consider a thermal system, i.e., take $f(\vb k)$ to be the Bose-Einstein distribution $f_+(k)$, Eq.~\eqref{eq:thermal-distributionfunctions}. The numerical results for the collision kernel multiplied with $q_\perp^3$ are shown in Fig.~\ref{fig:collision-kernel-thermal} as a solid green curve. the analytic limits \eqref{eq:limits_general} are added as dashed and dash-dotted curves. We see that they nicely agree in their respective region of applicability. The collision kernel $\Ceq(q_\perp)$ in thermal equilibrium is approximately peaked at the Debye mass, which is indicated by a square on the curve.

We will frequently use the thermal form $\Ceq(q_\perp)$ to normalize the nonequilibrium kernel $C(\vb q_\perp)$, in order to highlight the deviations from the thermal form. For that, always the numerically evaluated $\Ceq(q_\perp)$ is used, and computed with the corresponding discretization parameters as in the nonequilibrium case.

\subsubsection{Bjorken expanding systems}
We now consider gluonic plasmas undergoing Bjorken expansion, motivated by the initial stages in heavy-ion collisions. We take the initial conditions described in more detail in Section \ref{sec:boltzmann-equation-and-initial-condition} with $\xianiso=10$, and use the couplings $\lambda=2$ and $\lambda=10$. We use $\lambda=2$, because for this coupling the qualitative and quantitative features of the bottom-up thermalization (see Chapter \ref{sec:thermalization-expanding-systems}) are still clearly visible, and the system is still moderately weakly coupled, and choose $\lambda=10$ as a more phenomenologically relevant value.

\begin{figure}
    \centering
    \centerline{
    \includegraphics[width=0.33\linewidth]{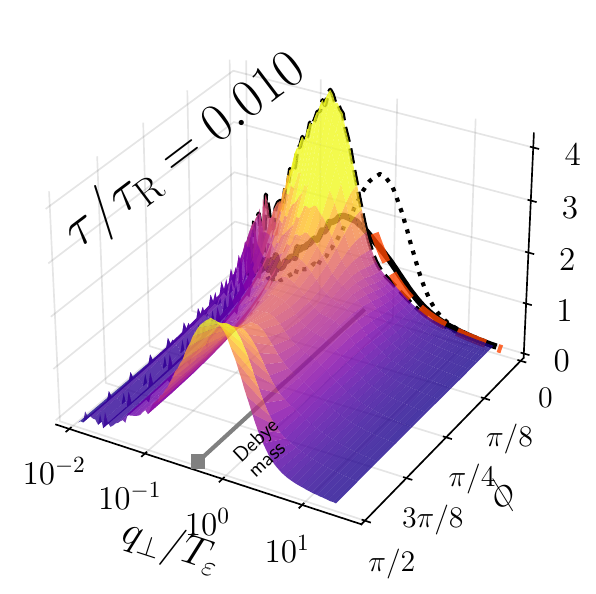}
    \includegraphics[width=0.33\linewidth]{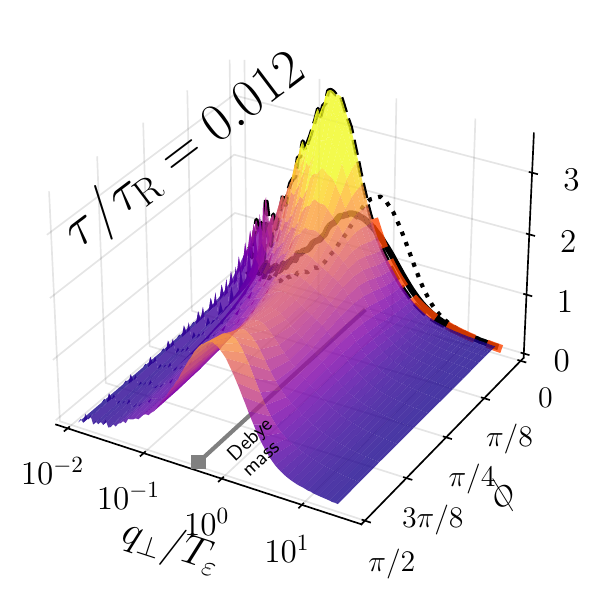}    
    \includegraphics[width=0.33\linewidth]{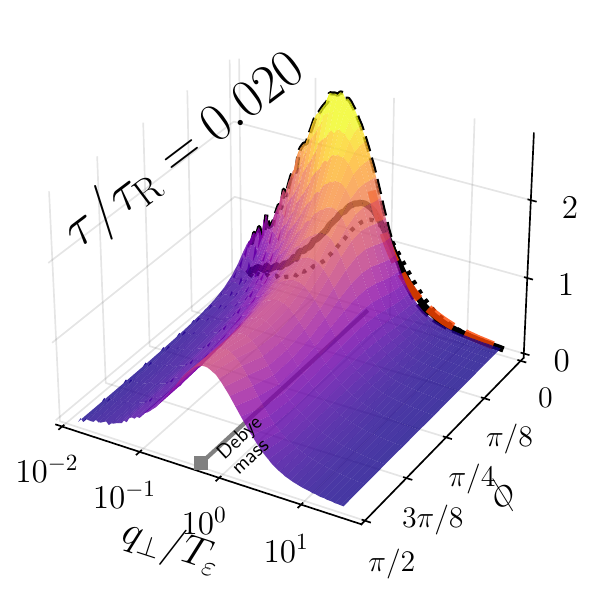}
    }
    \centerline{
    \includegraphics[width=0.33\linewidth]{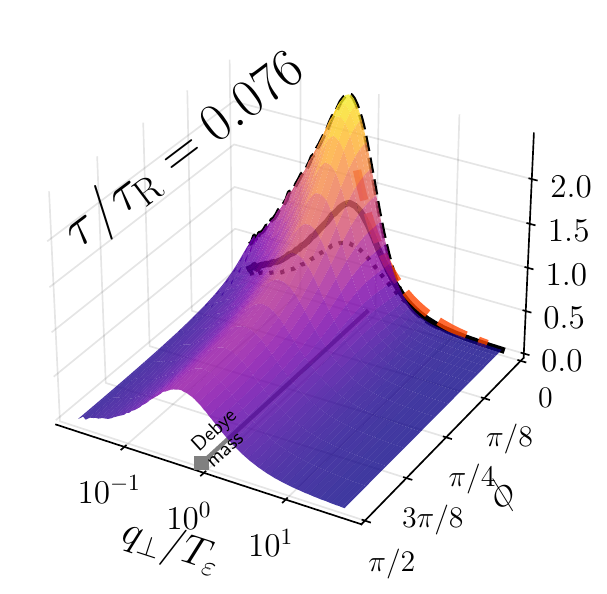}
    \includegraphics[width=0.33\linewidth]{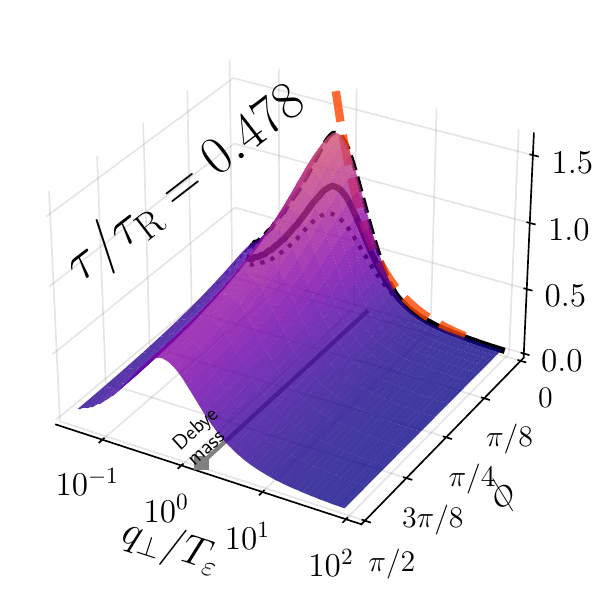}
    \includegraphics[width=0.33\linewidth]{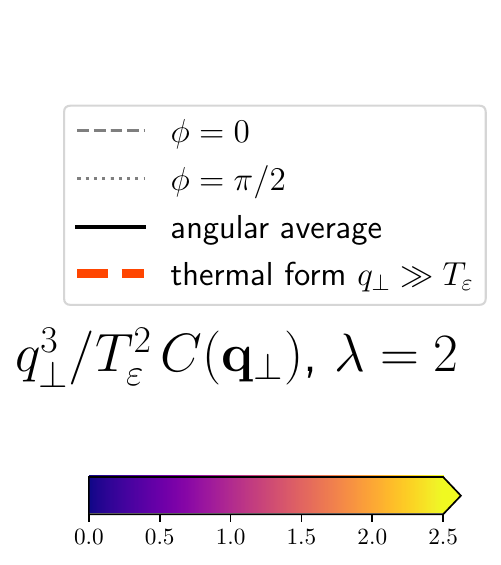}
    }
    \caption{Collision kernel $C(\vb q_\perp)$ of a gluonic plasma undergoing bottom-up thermalization for various times during the evolution for coupling $\lambda=2$. The back plane shows the projection of the angles $\phi=0$ and $\phi=\pi/2$, the angular average $\langle C(q_\perp)\rangle_\phi$ and the large $q_\perp$ result \eqref{eq:limits_qperp} in thermal equilibrium with the same energy density.
    }
    \label{fig:3dplots-lambda2}
\end{figure}
\begin{figure}
    \centering
    \centerline{
    \includegraphics[width=0.33\linewidth]{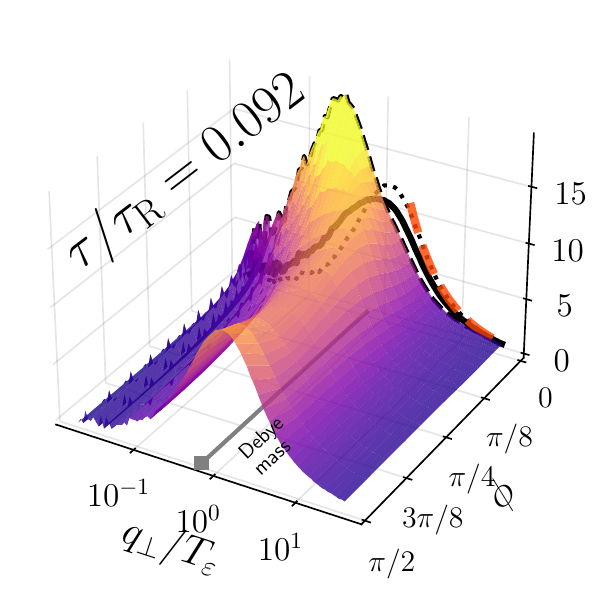}
    \includegraphics[width=0.33\linewidth]{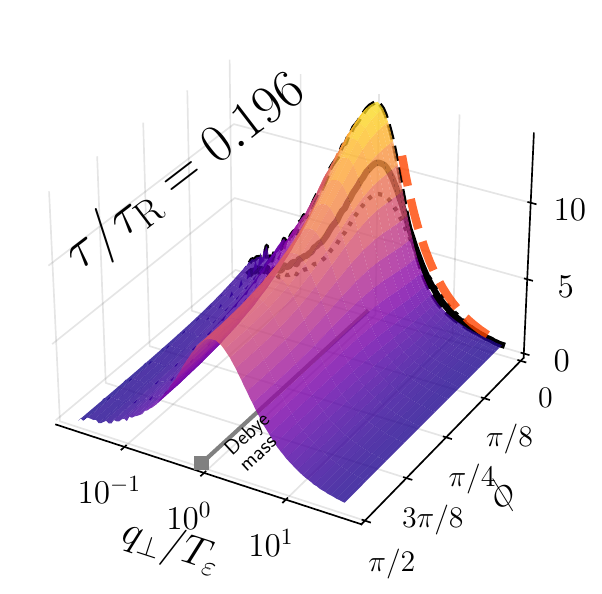}
    \includegraphics[width=0.33\linewidth]{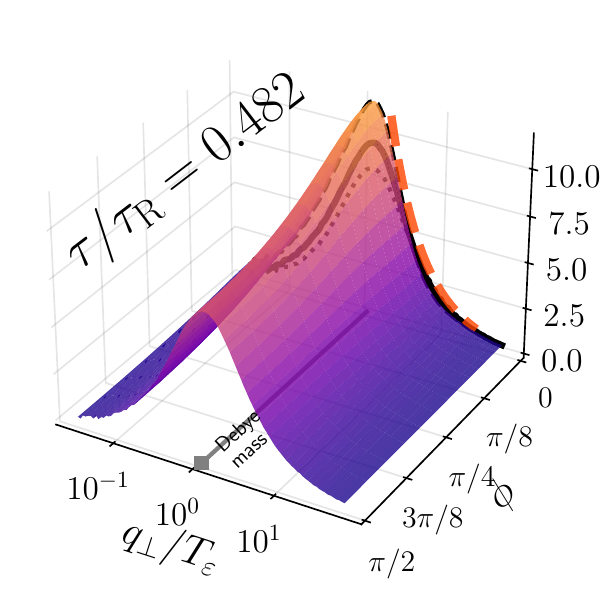}
    }
    \centerline{
    \includegraphics[width=0.33\linewidth]{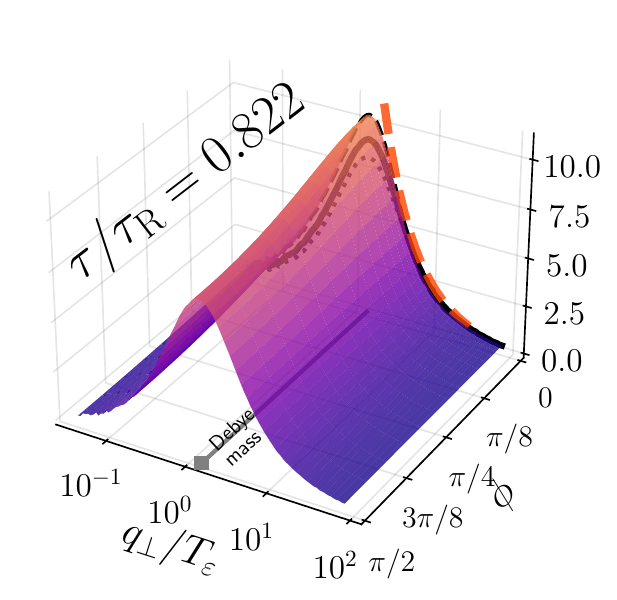}
    \includegraphics[width=0.33\linewidth]{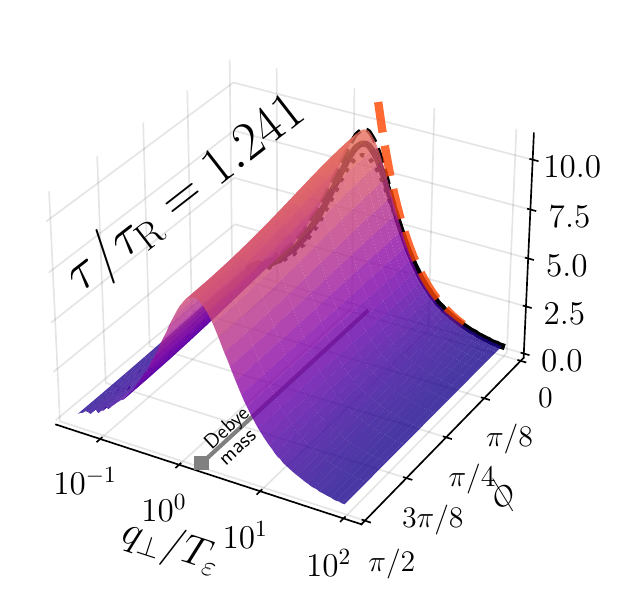}
    \includegraphics[width=0.33\linewidth]{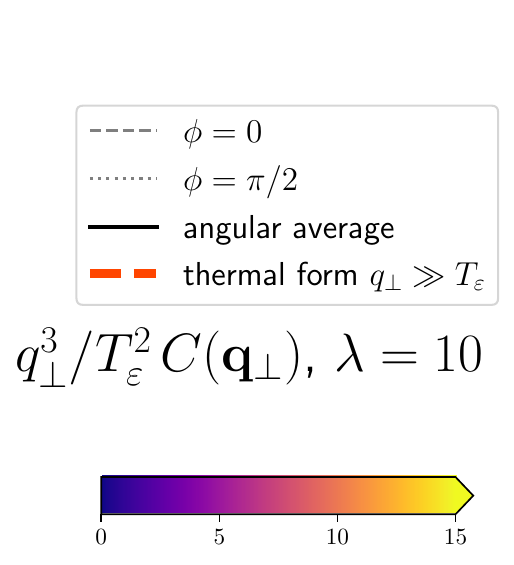}
    }
    \caption{Collision kernel $C(\vb q_\perp)$ of a gluonic plasma undergoing bottom-up thermalization for various times during the evolution for coupling $\lambda=10$. The back plane shows the projection of the angles $\phi=0$ and $\phi=\pi/2$, the angular average $\langle C(q_\perp)\rangle_\phi$ and the large $q_\perp$ result \eqref{eq:limits_qperp} in thermal equilibrium with the same energy density.
    }
    \label{fig:3dplots-lambda10}
\end{figure}
Figs.~\ref{fig:3dplots-lambda2} and \ref{fig:3dplots-lambda10} show the results for the anisotropic collision kernel $C(\vb q_\perp)$ for various times and the two couplings $\lambda=2 $ (Fig.~\ref{fig:3dplots-lambda2}) and $\lambda=10$ (Fig.~\ref{fig:3dplots-lambda10}). The anisotropic kernel is shown as a surface plot with the height (and color) representing the value of the collision kernel, rescaled with $q_\perp^3/\Teps^2$, representing the integrand of the jet quenching parameter $\qhat$. Recall that $\Teps$ is an effective temperature of the nonequilibrium system obtained via the Landau matching condition \eqref{eq:Landau-matching-condition}.
The kernel for $\phi=0$ (broadening along beam axis), $\phi=\pi/2$ (broadening transverse to beam) and the angular averaged collision kernel
\begin{align}
    \langle C(q_\perp)\rangle_\phi=\frac{1}{2\pi}\int_0^{2\pi}\dd\phi C(q_\perp, \phi)\label{eq:collkern-angular-average}
\end{align}
are shown as black lines on the back plane, where also the large $q_\perp$ result \eqref{eq:limits_qperp} for a thermal system with temperature $\Teps$ are shown as an orange dashed line. Additionally, the Debye mass is marked with a gray square (and line) on the $x$-axis (where $q_\perp/\Teps$ is plotted). The time variable is given as a ratio over the relaxation time $\tauR$ (see Eq.~\eqref{eq:relaxation-time}).

At late times, the kernel is seen to be isotropic (flat in the $\phi$ direction), while at early times it is most anisotropic, with a large peak at $\phi=0$ and a smaller one at $\phi=\pi/2$. The latter is only visible at very early times and is more pronounced for $\lambda=2$. The peak itself is at the Debye mass for later times and is shifted towards smaller momenta at early times and small angles.
Clearly visible is also the noise from the Monte Carlo evaluation of the integral at small momenta and early times, see, e.g., the top left panel in Fig.~\ref{fig:3dplots-lambda2}.

\begin{figure}
    \centering
    \centerline{
    \includegraphics[width=0.5\linewidth]{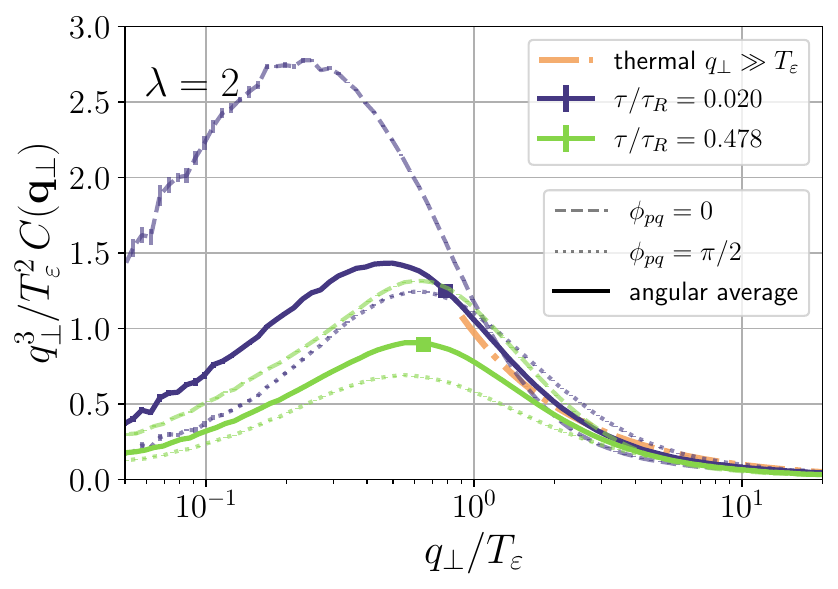}
    \includegraphics[width=0.5\linewidth]{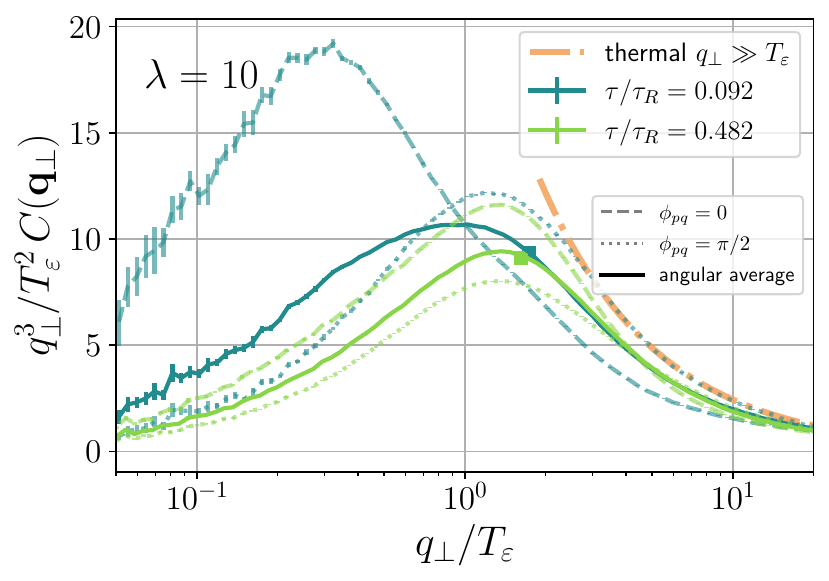}
    }
    \centerline{
    \includegraphics[width=0.5\linewidth]{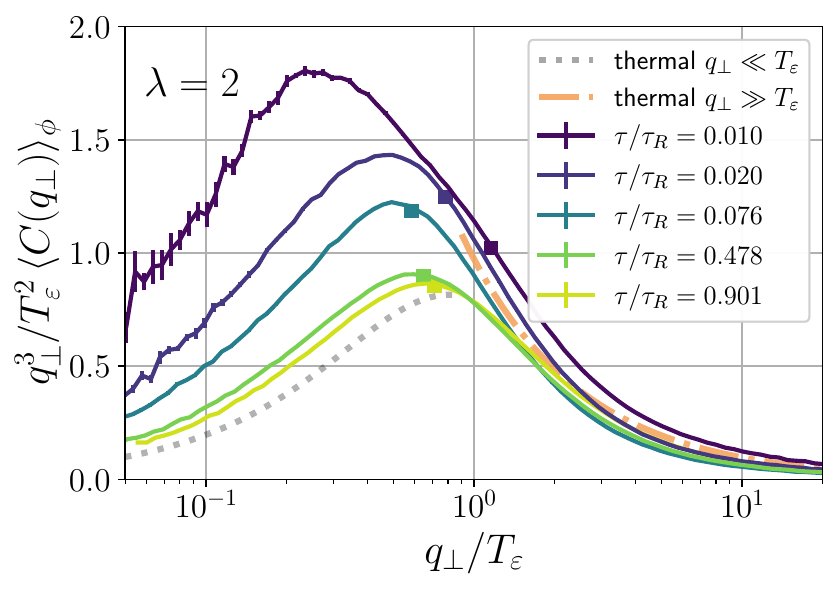}
    \includegraphics[width=0.5\linewidth]{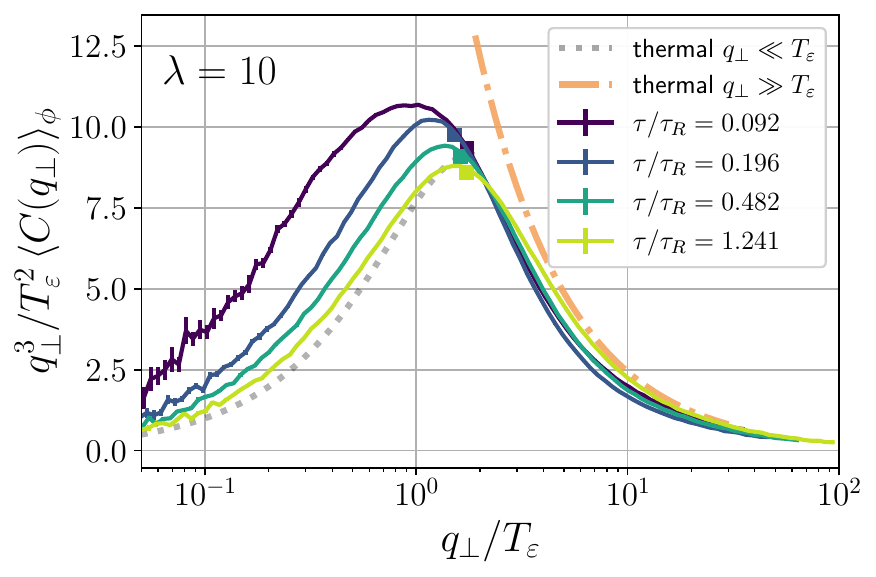}
    }
    \centerline{
    \includegraphics[width=0.5\linewidth]{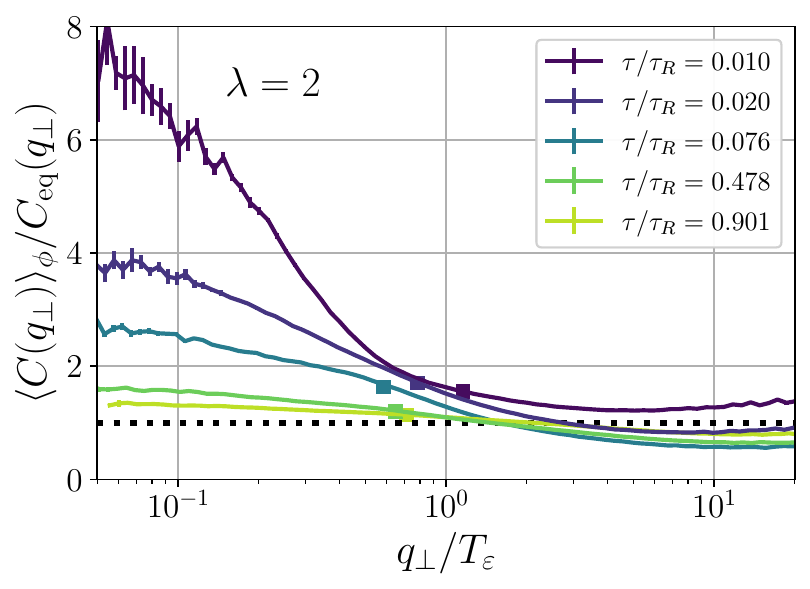}
    \includegraphics[width=0.5\linewidth]{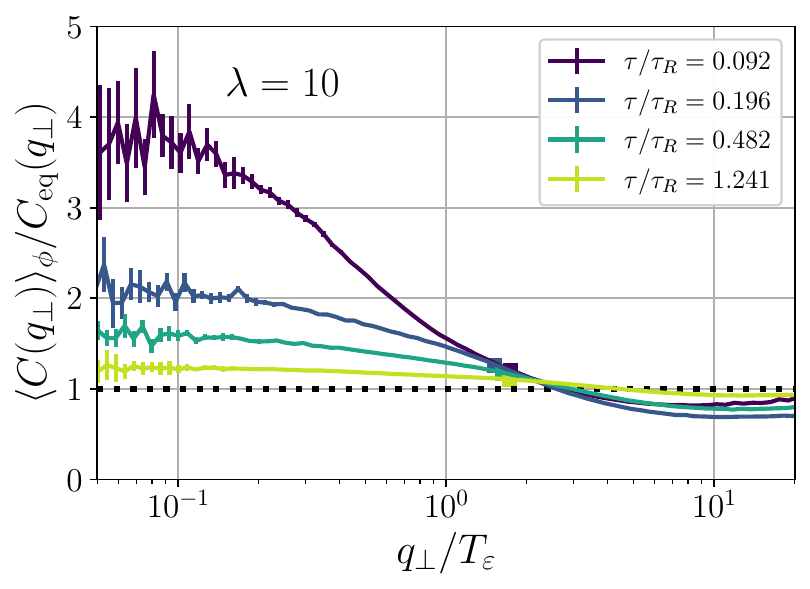}
    }
    \caption{Collision kernel $C(\vb q_\perp)$ for different times as a function of the momentum transfer $q_\perp$.
    Runs with coupling $\lambda=2$ are depicted in the left column and those with $\lambda=10$ in the right column. The box indicates the Debye mass.
    (\emph{Top}): Collision kernel for two angles and angular average at an early and late time.
    (\emph{Center}): Angular averaged collision kernel \eqref{eq:collkern-angular-average} for different times (color-coded). Shown are also the limiting analytic expressions for a thermal system \eqref{eq:limits_general}.
    (\emph{Bottom}): Angular averaged collision kernel normalized to the thermal one for the Landau-matched temperature $\Teps$ for different times.
    }
    \label{fig:Cq_variants}
\end{figure}

The angular resolution can be studied in more detail in the top panels of Fig.~\ref{fig:Cq_variants}. Here, as already shown on the back sides of the three-dimensional plots in Figs.~\ref{fig:3dplots-lambda2} and \ref{fig:3dplots-lambda10}, both the angular averaged kernel \eqref{eq:collkern-angular-average}, and the two angles $\phi=0$ and $\phi=\pi/2$ are shown for two distinct times. It can be clearly seen that the anisotropy decreases at later times. The box marks the Debye mass. As already mentioned before, for very early times, the nonequilibrium collision kernel is peaked at momenta below the Debye mass, and the peak is shifted even more to lower momenta for smaller angles $\phi$. This might suggest that at early times, the effective screening scale is angle-dependent, and considerably lower than the Debye mass at small angles.

This is further corroborated by the central and bottom panels in Fig.~\ref{fig:Cq_variants}. The central panels show the angular averaged collision kernel for different times. We clearly observe the evolution towards its equilibrium form and its peak shifting from the left towards the Debye mass, which is again indicated by the squares. In the bottom panels, the collision kernel is normalized to its equilibrium form. There, the enhancement of small momentum exchanges for early times is most visible. For instance, for $\lambda=2$ and $\tau/\tauR\leq 0.08$, small momentum exchange processes are enhanced by more than a factor $2$, which is similar for $\lambda=10$ and $\tau/\tauR\leq 0.2$.
This indicates that small momentum exchange processes are more likely than in thermal equilibrium and that using a thermal form for the collision kernel drastically underestimates the momentum transfer at small momenta $q_\perp$. Thus, describing the momentum broadening during the splitting process as a diffusion process in transverse momentum space, as required for the multiple-soft-scattering approximation (see Chapter \ref{sec:jet-energy-loss}) is a good approximation in this region.

Since the enhancement gets larger for smaller angles $\phi$, scatterings with momentum transfer along the beam axis are more likely. This counteracts the highly anisotropic plasma with small longitudinal momentum extent and contributes to a broadening of the distribution.

At late times, the collision kernel coincides with its equilibrium form, as we can see explicitly in the bottom panel in Fig.~\ref{fig:Cq_variants}, where at late times the ratio of the nonequilibrium kernel to its equilibrium counterpart is close to unity for all momenta.

\begin{figure}
    \centering
    \includegraphics[width=0.5\linewidth]{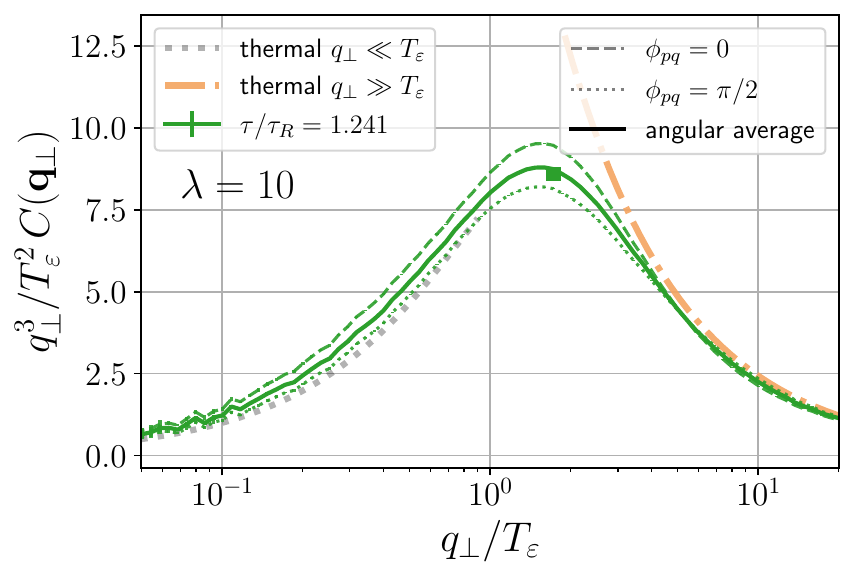}
    \caption{Collision kernel for two angles and angular average for $\lambda=10$, showing the anisotropy present even at a relatively late time.
    }
    \label{fig:Cq_lambda10_latetimes}
\end{figure}
Finally, due to the Bjorken expansion, the plasma never fully isotropizes. For instance, the pressure ratio only approaches but never reaches unity (up to numerical uncertainties). In particular, also the collision kernel is expected to remain anisotropic, even at late times. Fig.~\ref{fig:Cq_lambda10_latetimes} shows the results for the nonequilibrium kernel for $\lambda=10$, where we observe a clearly visible anisotropy in the kernel even at rather late times $\tau/\tauR=1.2$, where a hydrodynamic description starts becoming applicable (see Fig.~\ref{fig:pressureratio-approach-hydro}). This underlines that understanding which physical effects and observables are sensitive to the anisotropy in the collision kernel (and related jet quenching parameter) is an important step to understanding the nonequilibrium stages in heavy-ion collisions.

\section{Dipole cross section}
We have already discussed in Section \ref{sec:amy-rate-equation} that the splitting rate is obtained by transforming the collision kernel to  impact parameter space \eqref{eq:fouriertrafo_real},
\begin{align}
    C(\vb x)=\int\frac{\dd[2]{\vb q_\perp}}{(2\pi)^2}(1-\cos(\vb q_\perp\cdot \vb x))C(\vb q_\perp)\label{eq:fouriertrafo-real-dipolecrosssection}
\end{align}
to obtain the dipole cross section.
We will now study the form of this dipole cross section $C(\vb x)$ in expanding systems.
The Fourier transform \eqref{eq:fouriertrafo_real} is performed numerically. As $C(\vb q_\perp)$ is obtained on a finite grid $q_\perp^{\mathrm{min}}<q_\perp < q_\perp^{\mathrm{max}}$, smaller and larger values of $q_\perp$ are obtained by extrapolation using the limiting forms \eqref{eq:limits_qperp}. In practice, the small $q_\perp$ behavior is fitted to $a_1(\phi)/q_\perp^2$, and the large $q_\perp$ behavior to $a_2(\phi)/q_\perp^4$.

The integral in Eq.~\eqref{eq:fouriertrafo-real-dipolecrosssection} can be done analytically for the small $q_\perp$ form of the collision kernel for isotropic distributions \eqref{eq:analytic_collision_kernel},
\begin{align}
    C(q_\perp)=\frac{\CR g^2m_D^2 T_\ast}{q_\perp^2(q_\perp^2+m_D^2)},\label{eq:small-qperp-form-collkern-dipolecrosssectionsection}
\end{align}
which yields (see Appendix \ref{app:analytic-result-cross-section})
\begin{align}
    \Cisoappr(|\vb x|)&=\frac{C_Rg_s^2T_\ast}{2\pi}\left(\gamma_E+K_0(|\vb x| m_D)+\log\frac{|\vb x| m_D}{2}\right),\label{eq:Cb_iso_appr}
\end{align}
with $T_\ast$ and $m_D$ given by Eqs.~\eqref{eq:tstar-definition} and \eqref{eq:debyemass-general}.
In equilibrium, this then becomes
\begin{align}
    \Ceqappr(|\vb x|)&=\frac{C_Rg_s^2T}{2\pi}\left(\gamma_E+K_0(|\vb x| m_D)+\log\frac{|\vb x| m_D}{2}\right),\label{eq:Cb_eq_appr}
\end{align}
with the equilibrium Debye mass \eqref{eq:equilibriumform-debyemass-tstar-meffs}.
This form \eqref{eq:Cb_eq_appr} or also the small $q_\perp$ expression \eqref{eq:small-qperp-form-collkern-dipolecrosssectionsection} is also often used in the literature (see, e.g., \cite{Andres:2020vxs, Barata:2021wuf, AbraaoYork:2014hbk, Caron-Huot:2010qjx}). For instance, it is used in QCD kinetic theory simulations to calculate the splitting rates $\gamma^{a}_{bc}$ as discussed in Section \ref{sec:equations-of-qcd-kinetic-theory}, or also used as a medium model in jet quenching studies. While Eq.~\eqref{eq:Cb_eq_appr} is convenient because of its simple analytic form, it should be emphasized that it is not the correct form of the dipole cross section in thermal equilibrium. For that, not only the small $q_\perp$ form \eqref{eq:analytic_collision_kernel}, but the whole possibly numerically evaluated collision kernel needs to be Fourier transformed using Eq.~\eqref{eq:fouriertrafo-real-dipolecrosssection}. Therefore, we label it $\Ceqappr$, because it constitutes an analytic approximation to the true thermal form of the dipole cross section.

\subsection{Small distance limit\label{sec:smalldistance-dipolecrosssection}}
As explained before, for highly energetic particles (see Chapter \ref{sec:jet-energy-loss}), the small distance form of the dipole cross section is important. Here, we discuss the analytic limit of the small distance behavior of the dipole cross section and show that it can be expressed using the jet quenching parameter $\qhat(\lperp)$ with a transverse momentum cutoff.

In an isotropic system (when $C(\vb q_\perp,t)$ does not depend on the angle $\phi$), the angular integral in \eqref{eq:fouriertrafo-real-dipolecrosssection} can be done analytically,
\begin{align}
	C(|\vb x|)=\int_0^\infty \frac{\dd{q_\perp}}{2\pi}q_\perp (1-J_0(|\vb x| q_\perp))C(q_\perp). \label{eq:fouriertrafo-isotropic}
\end{align}

First, let us consider the isotropic case of thermal equilibrium. There, to find the small distance $|\vb x|\ll 1/T$ behavior of the dipole cross section $C(|\vb x|)$, one might na\"ively expect it to be sufficient to expand the integrand in powers of $|\vb x|$, ($1-J_0(|\vb x| q_\perp))=\vb x^2q_\perp^2/4 +\mathcal O(\vb x^4q_\perp^4)$). However, in this expansion
\begin{align}
    C(|\vb x)\sim \vb x^2+\mathcal O(\vb x^4), \label{eq:Cx-naive-expansion}
\end{align}
the first nonvanishing term for the dipole cross section $C(x)$ is given by the jet quenching parameter $\qhat$,
\begin{align}
	\qhat \sim \int_0^\infty\frac{\dd{q_\perp}}{2\pi}q_\perp^3 C(q_\perp) \to \infty,
\end{align}
and is divergent (without introducing a UV cutoff), as we have discussed before extensively in Section \ref{sec:pinf_formula}.
Clearly, the problem is the highly oscillatory function $J_0(|\vb x| q_\perp)$ at large $q_\perp$. The result could be made finite by introducing again a UV cutoff $\lperp$, but it is not clear at this stage how this would appear naturally in Eq.~\eqref{eq:fouriertrafo-isotropic}.

And even worse, for the analytically solvable case, \eqref{eq:Cb_eq_appr}, the small-$|\vb x|$ behavior is
not given by \eqref{eq:Cx-naive-expansion} but
\begin{align}
	C(|\vb x|\ll 1/T)=\vb x^2(a_1+b_1\log |\vb x|)+\dots
\end{align}
In particular, the logarithmic dependence on $|\vb x|$ will be important for small-$|\vb x|$.
It is now also clear that the na\"ive expansion in \eqref{eq:Cx-naive-expansion} was doomed to fail since it would have never led to a logarithm by just expanding the integrand for small-$|\vb x|$.

A better way to do it in full generality is to introduce a cutoff $\lperp$ in the integral, and to split the integral in a part below and above the cutoff,
\begin{align}
	C(\vb x)=\int_{q_\perp <\lperp} \frac{\dd[2]{\vb q_\perp}}{(2\pi)^2}(1-e^{i\vb q_\perp\cdot \vb x})C(\vb q_\perp)+\int_{\lperp}^\infty \frac{\dd{q_\perp}}{2\pi}q_\perp C(q_\perp)(1-J_0(bq_\perp)). \label{eq:cx-splitting}
\end{align}
Here, we have already used that for a general distribution function, the collision kernel becomes isotropic for sufficiently large $q_\perp$, and is given by (see Eq.~\eqref{eq:limits_qperp})
\begin{align}
    C(q_\perp)=\frac{g^4\CR\mathcal N}{q_\perp^4}.\label{eq:Cq-largeqperpform}
\end{align}

In the first integral in Eq.~\eqref{eq:cx-splitting}, we can expand the exponential for small $\vb x$, which yields exactly the jet quenching parameter $\qhat$ with a transverse momentum cutoff $\lperp$,
\begin{align}
    \int_{q_\perp <\lperp} \frac{\dd[2]{\vb q_\perp}}{(2\pi)^2}(1-e^{i\vb q_\perp\cdot \vb x})C(\vb q_\perp)&\approx \frac{1}{2}\int_{q_\perp <\lperp} \frac{\dd[2]{\vb q_\perp}}{(2\pi)^2}(x^2 q_x^2+y^2q_y^2)C(\vb q_\perp)\\
    &=\frac{1}{2}\left(x^2\qhat^{xx}(\lperp)+ y^2\qhat^{yy}(\lperp)\right),
\end{align}
where we have parameterized the vector $\vb x = (x,y)$. For large enough cutoffs $\lperp$, we have seen before in Section \ref{sec:limiting_behavior_large_cutoff} (see Eq.~\eqref{eq:qhat_behavior_large_cutoff}) that the jet quenching parameter $\hat q$ has the simple form
\begin{align}
    \qhat^{xx}=a\log\lperp/Q_s+b_x, && \qhat^{yy}=a\log\lperp/Q_s+b_y\label{eq:qhat-logarithmic-dependence}
\end{align}

The second integral in Eq.~\eqref{eq:cx-splitting} can be solved analytically for the isotropic large $q_\perp$ form \eqref{eq:Cq-largeqperpform} (which is also valid for an anisotropic distribution function $f(\vb p)$),
\begin{align}
    &\int_{\lperp}^\infty \frac{\dd q_\perp}{2\pi}\frac{\CR g^4\mathcal N}{q_\perp^3}(1-J_0(|\vb x| q_\perp))\\
    &=\frac{\CR g^4 \mathcal N}{256\pi}\left(\lperp^2 \vb x^4\, {}_2 F_3\left(1,1;\,2,3,3;\, -\frac{\lperp^2 \vb x^2}{4}\right)-32x^2\left(-1+\gamma_E+\log\frac{\lperp |\vb x|}{2}\right)\right),\nonumber\\
    &=-\frac{\CR g^4 \mathcal N}{8\pi} \vb x^2\left(\gamma_E-1+\log\frac{\lperp |\vb x|}{2}\right)+\mathcal O(\vb x^4)\label{eq:large-qperpform-collkern-integrated}
\end{align}
where ${}_p F_q$ is the generalized hypergeometric function and $\gamma_E$ is the Euler Mascheroni constant.

Importantly, since the sum of the two integrals in \eqref{eq:cx-splitting} cannot depend on $\lperp$ for small $|\vb x|$, the terms $\log\lperp$ must exactly cancel in the sum. In particular, because the large $q_\perp$ form \eqref{eq:large-qperpform-collkern-integrated} is isotropic, so must be the cutoff dependence in Eq.~\eqref{eq:qhat-logarithmic-dependence}. This provides another argument for why the coefficient $a$ in \eqref{eq:qhat-logarithmic-dependence} cannot depend on the direction.

Adding these two expressions, we finally obtain the small $|\vb x|$ form of the dipole cross section,

\begin{align}
    C(\vb x,t)\approx\frac{1}{2}(x^2 b_x(t)+y^2b_y(t))+\frac{\vb x^2\hat q_0(t)}{2}\left(1-\gamma_E-\log\frac{|\vb x|Q_s}{2}\right), \label{eq:smallxform-general}
\end{align}
where we have redefined $\hat q_0(t)=a(t)$, and $a$, $b_x$ and $b_y$ come from the large cutoff parameterization of the jet quenching parameter \eqref{eq:qhat-logarithmic-dependence}, which leads to
\begin{align}
    \qhat_0(t)=a(t)=\frac{g^4\CR\mathcal N(t)}{4\pi}.
\end{align}

More conventionally (see, e.g., \cite{Barata:2021wuf}), the small-$|\vb x|$ form is often written as
\begin{align}
    C(\vb x, t)\approx\frac{1}{4}\hat q_0(t) \vb x^2\log\frac{1}{\vb x^2\mu_\ast^2(t)}
\end{align}
which, for an anisotropic medium, can be generalized to
\begin{align}
    C(\vb x, t)\approx\frac{1}{4}\hat q_0(t)\left(\vb x^2\log\frac{1}{\vb x^2\mu_{\ast0}^2}+x^2\log\frac{Q_s^2}{\mu_x^2}+y^2\log\frac{Q_s^2}{\mu_y^2}\right),
\end{align}
where we need to identify
\begin{align}
	\mu_i^2=e^{-\frac{2b_i}{\hat q_0}}, && \mu_{\ast 0}^2=\frac{1}{4}e^{2\gamma_E-2}. && \hat q_0=a.
\end{align}

For the special case of thermal equilibrium (and pure gluons), we can use the result for $\qhat$ from Eq.~\eqref{eq:qhat_hard_arnold} to obtain
\begin{align}
	\Ceq(\vb x)&\approx\frac{\vb x^2}{4}\frac{C_RN_C g^4 T^3}{\pi^3}\\
	&\times\left[\zeta(3)\left(1-\gamma_E+\ln\frac{2}{|\vb x| m_D}\right)
	+(\zeta(2)-\zeta(3))\left\{\log\frac{T}{m_D}+\frac{1}{2}-\gamma_E+\log 2\right\}-\frac{\sigma_+}{2\pi}
	\right]
\end{align}

from which we may deduce in equilibrium
\begin{subequations}
\begin{align}
	\hat q_0&=\frac{\CR \NC g^4T^3\zeta(3)}{2\pi^3},\\ \mu_\ast^2&=\frac{m_D^2}{4}\left(\frac{m_D}{2T}\right)^{2\zeta(2)/\zeta(3)-2}\exp\left[2\gamma_E-2-\left(\frac{\zeta(2)}{\zeta(3)}-1\right)(1-2\gamma_E)+\frac{\sigma_+}{\pi\zeta(3)}\right].
\end{align}
\end{subequations}

\subsection{Numerical results}
\begin{figure}
    \centering
    \centerline{
        \includegraphics[width=0.5\linewidth]{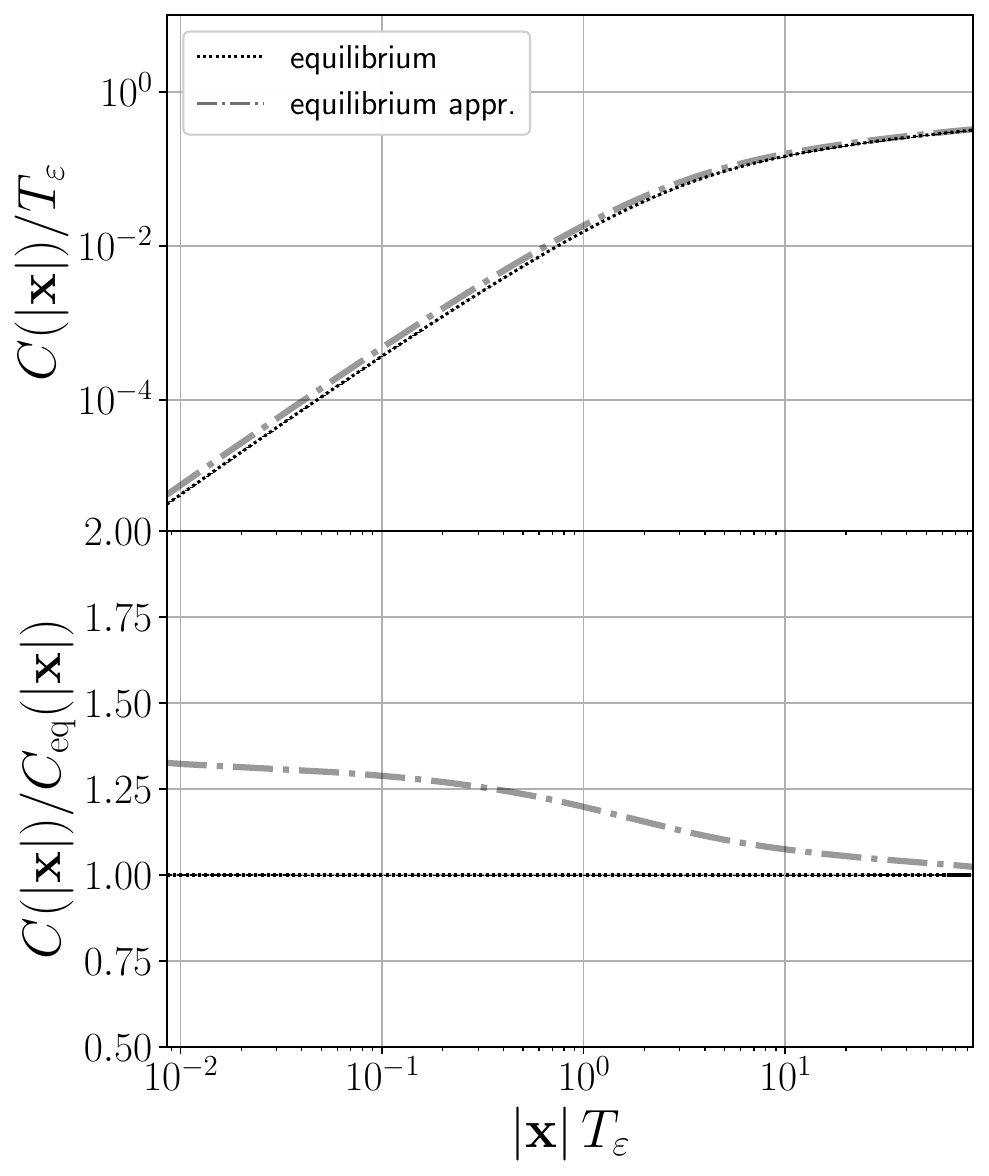}
        }        
    \caption{Dipole cross section $C(\vb x)$ in thermal equilibrium, compared with an often used analytic approximation \eqref{eq:Cb_eq_appr} for $\lambda=2$. 
    }
    \label{fig:dipole-crosssection-thermalcomparison}
\end{figure}

Let us now move on to discuss the numerical results for the dipole cross section, which we first discuss in thermal equilibrium.
Fig.~\ref{fig:dipole-crosssection-thermalcomparison} compares the thermal form for the dipole cross section $\Ceq(|\vb x|)$ with the analytic approximation $\Ceqappr(|\vb x|)$ from Eq.~\eqref{eq:Cb_eq_appr}. While the upper panel shows both curves on a logarithmic scale, where only little differences are visible, the lower panel depicts their ratio. We find that at small $|\vb x|$, which is the region relevant for highly energetic partons, these two expressions differ by more than $25\%$, which implies that the small-$|\vb x|$ behavior of $\Ceqappr$ significantly overestimates the true small-$|\vb x|$ behavior.

\begin{figure}
    \centering
    \centerline{
        \includegraphics[width=0.5\linewidth]{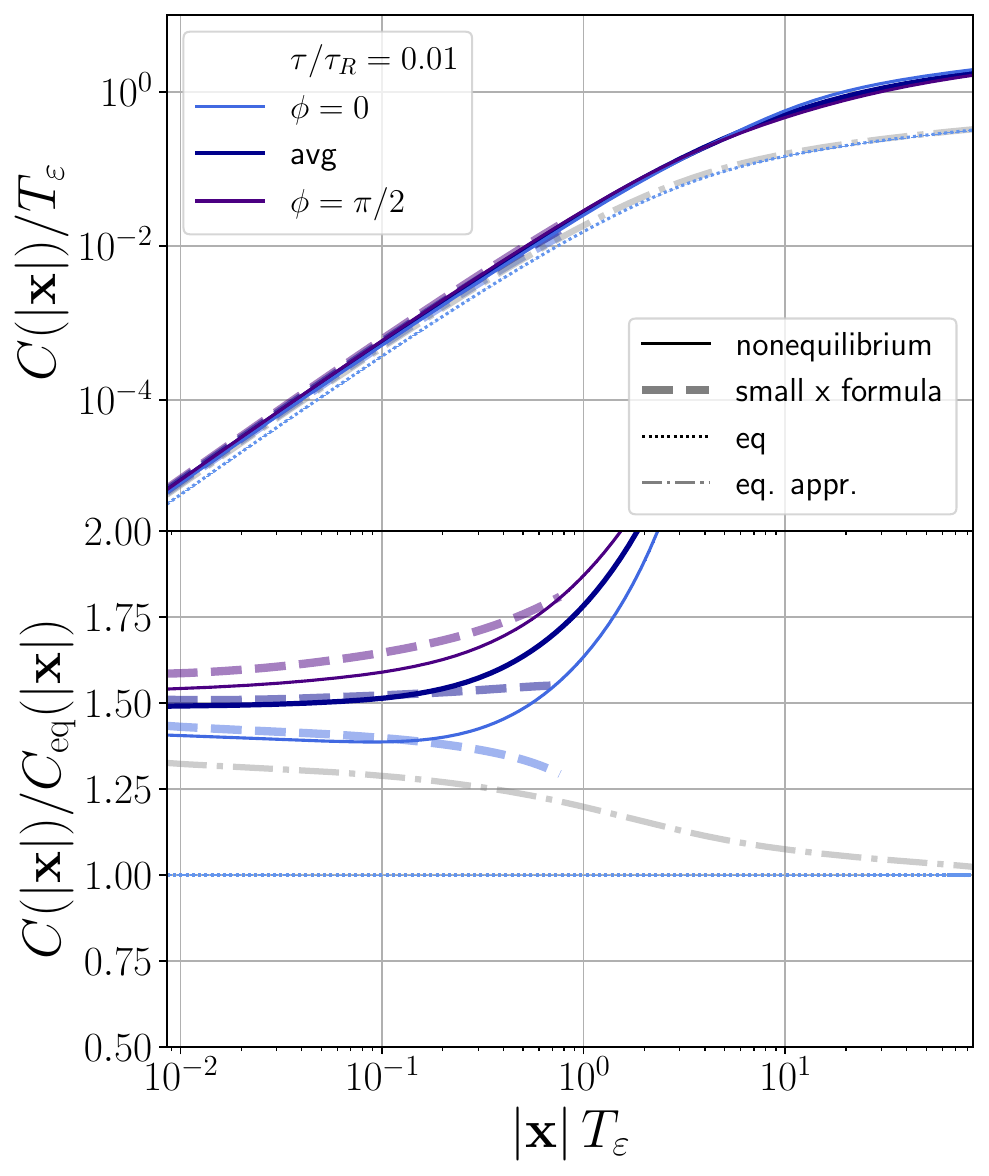}
        \includegraphics[width=0.5\linewidth]{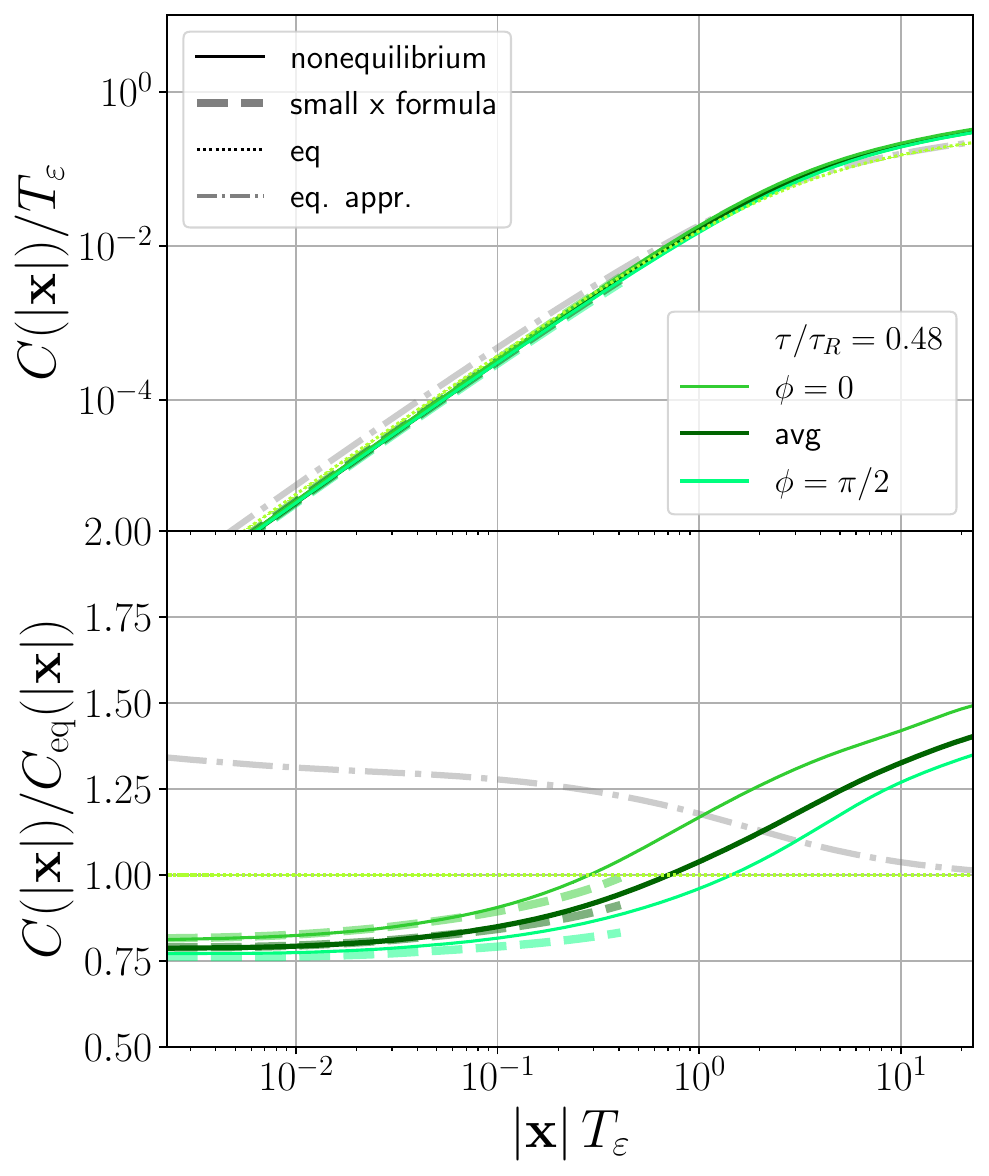}
        }        
    \caption{Dipole cross section $C(\vb x)$ for two angles and angular average for two distinct times (left and right panels). In the top panels, the dipole cross section itself is shown, in the bottom panels it is normalized to its thermal value. We also show the small-$|\vb x|$ formula \eqref{eq:smallxform-general}.
    }
    \label{fig:dipolecrosssection}
\end{figure}
Next, we consider the dipole cross section of the expanding plasma for $\lambda=2$ and two early times in Fig.~\ref{fig:dipolecrosssection}. In particular, we consider the times $\tau/\tauR=0.01$ (left panels) and $\tau/\tauR=0.48$ (right panels). In the upper panels, the dipole cross section is plotted in units of the temperature $\Teps$; in the lower panels, it is normalized to its equilibrium form. The two angles $\phi=0$ and $\phi=\pi/2$, and the angular averaged dipole cross section $\langle C(x)\rangle_\phi$ are shown as solid lines. We compare the full dipole cross section with the small $x$ result \eqref{eq:smallxform-general}, which is shown as dashed lines, and agrees very well for both times and angles.

This angular averaged dipole cross section can be obtained from the angular averaged collision kernel, as we show now.
For that, consider the angular averaged dipole cross section,
\begin{align}
    \langle C(|\vb x|)\rangle_\phi&=\frac{1}{2\pi}\int_0^{2\pi}\!\!\dd\phi C(|\vb x|,\phi)=\frac{1}{2\pi}\int_0^{2\pi}\dd\phi_x\int\!\frac{\dd[2]{\vb q_\perp}}{(2\pi)^2}(1-\cos(\vb q_\perp\cdot\vb x))C(\vb q_\perp)\\
    &=\frac{1}{2\pi}\int_0^{2\pi}\!\!\dd\phi_x\frac{1}{2\pi}\int_0^{2\pi}\!\dd\phi_q\int\!\frac{\dd q_\perp q_\perp}{2\pi}(1-\cos(q_\perp |\vb x| \cos(\phi_x-\phi_q)))C(q_\perp, \phi_q)
\end{align}
We can now swap the $\phi_x$ and $\phi_q$ integrations and define a new integration variable $\tilde\phi_x=\phi_x-\phi_q$, such that
\begin{align}
    \langle C(|\vb x|)\rangle_\phi&=\frac{1}{2\pi}\int_0^{2\pi}\dd\phi_q\frac{1}{2\pi}\int_{-\phi_q}^{2\pi-\phi_q}\dd{\tilde \phi_x}\int\frac{\dd q_\perp q_\perp}{2\pi}(1-\cos(q_\perp |\vb x| \cos\tilde\phi_x))C(q_\perp,\phi_q).
\end{align}
Because we integrate $\tilde \phi_x$ over the whole period of the cosine, we may shift it freely, and then have
\begin{align}
    \langle C(|\vb x|)\rangle_\phi&=\frac{1}{2\pi}\int_0^{2\pi}\dd{\tilde\phi_x}\int\frac{\dd{q_\perp}q_\perp}{2\pi}(1-\cos(q_\perp |\vb x| \cos\tilde\phi_x))\frac{1}{2\pi}\int_0^{2\pi}\dd\phi_q C(q_\perp,\phi)\\
    &=\int\frac{\dd[2]{\vb q_\perp}}{(2\pi)^2}(1-\cos(q_\perp |\vb x| \cos\phi_q))\langle C(q_\perp)\rangle_\phi,
\end{align}
which proves that to obtain the angular average of the dipole cross section $\langle C(|\vb x|)\rangle_\phi$, it is enough to use the isotropic already angular averaged collision kernel $\langle C(q_\perp)\rangle_\phi$ (Eq. \eqref{eq:collkern-angular-average}) in the Fourier transform \eqref{eq:fouriertrafo-real-dipolecrosssection}.

\begin{figure}
    \centering
    \centerline{
        \includegraphics[width=0.5\linewidth]{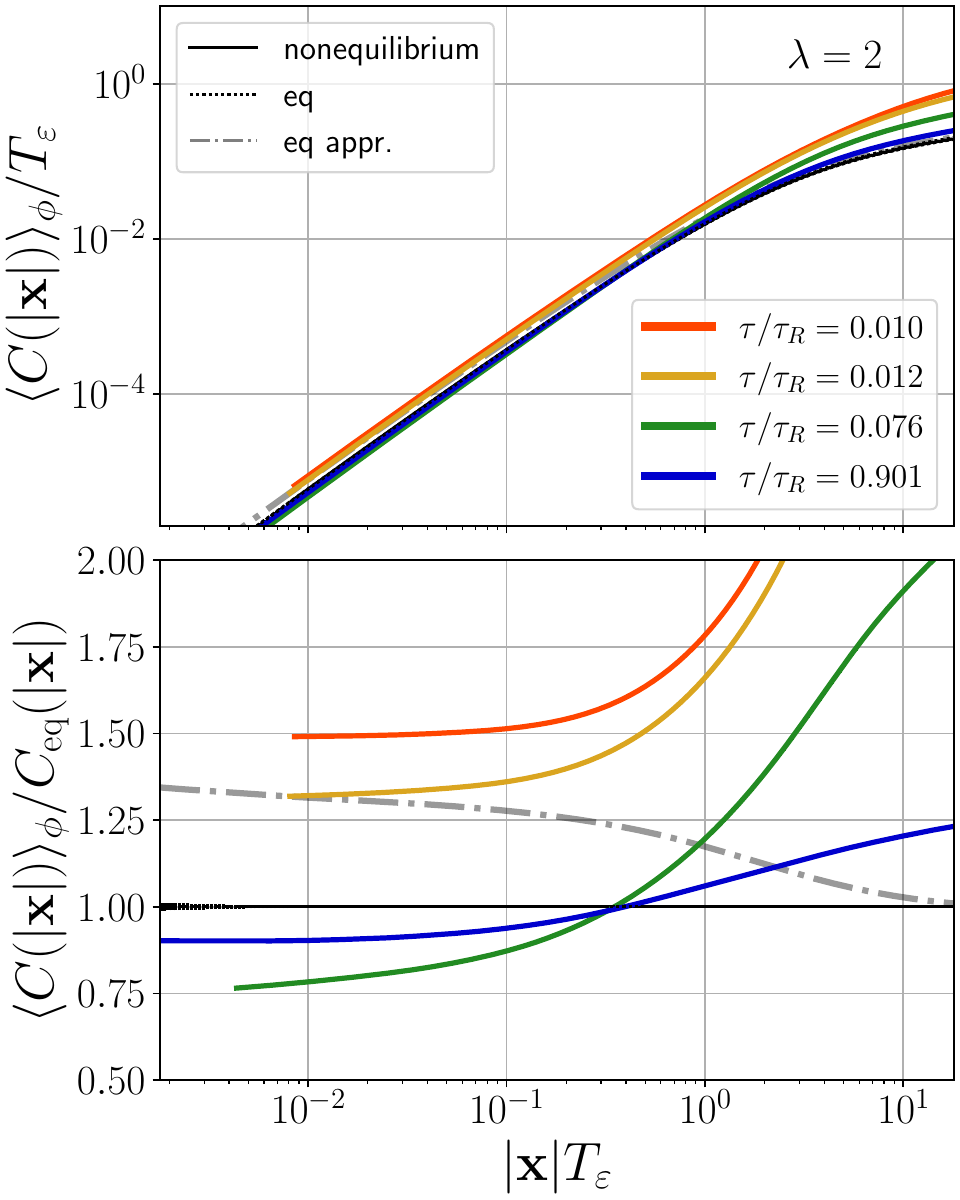}
        \includegraphics[width=0.5\linewidth]{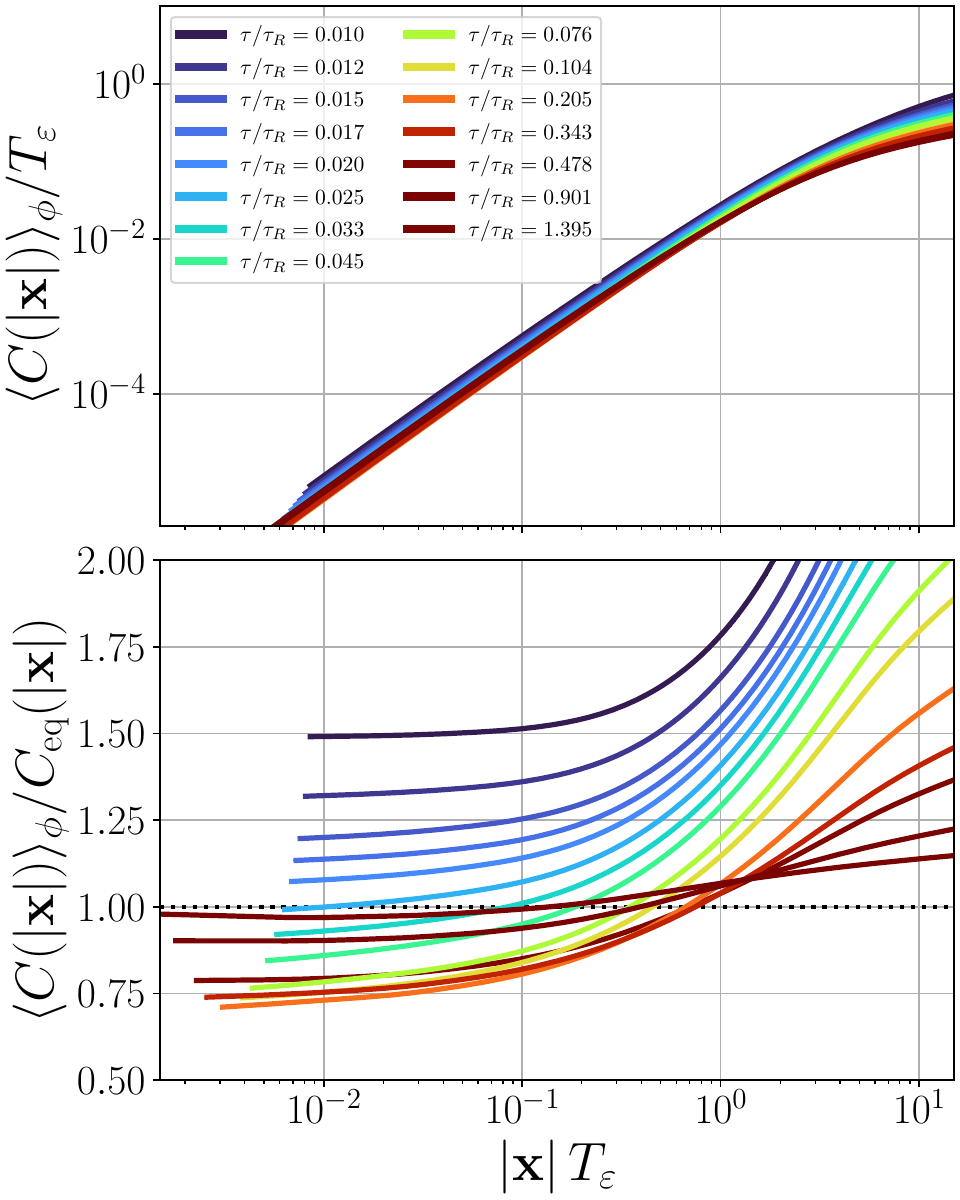}        
        }        
    \caption{Dipole cross section $\langle C( x)\rangle_\phi$ for different times and $\lambda=2$. In the left panel, four times are chosen to illustrate the general behavior, whereas in the right panel, more times are shown to show the continuous behavior. 
    }
    \label{fig:dipolecrosssection-angularaverages-lambda2}
\end{figure}
\begin{figure}
    \centering
    \centerline{
        \includegraphics[width=0.5\linewidth]{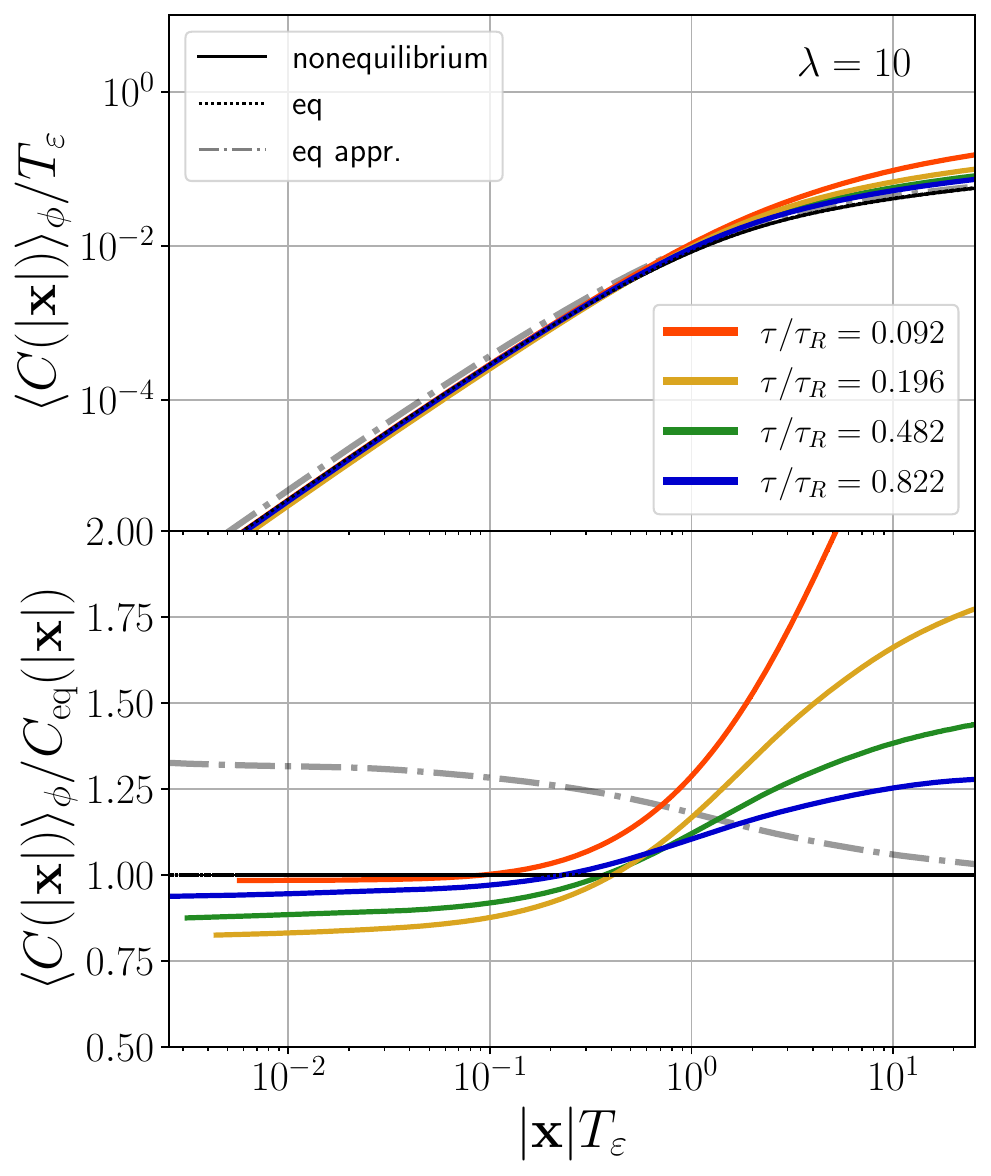}
        \includegraphics[width=0.5\linewidth]{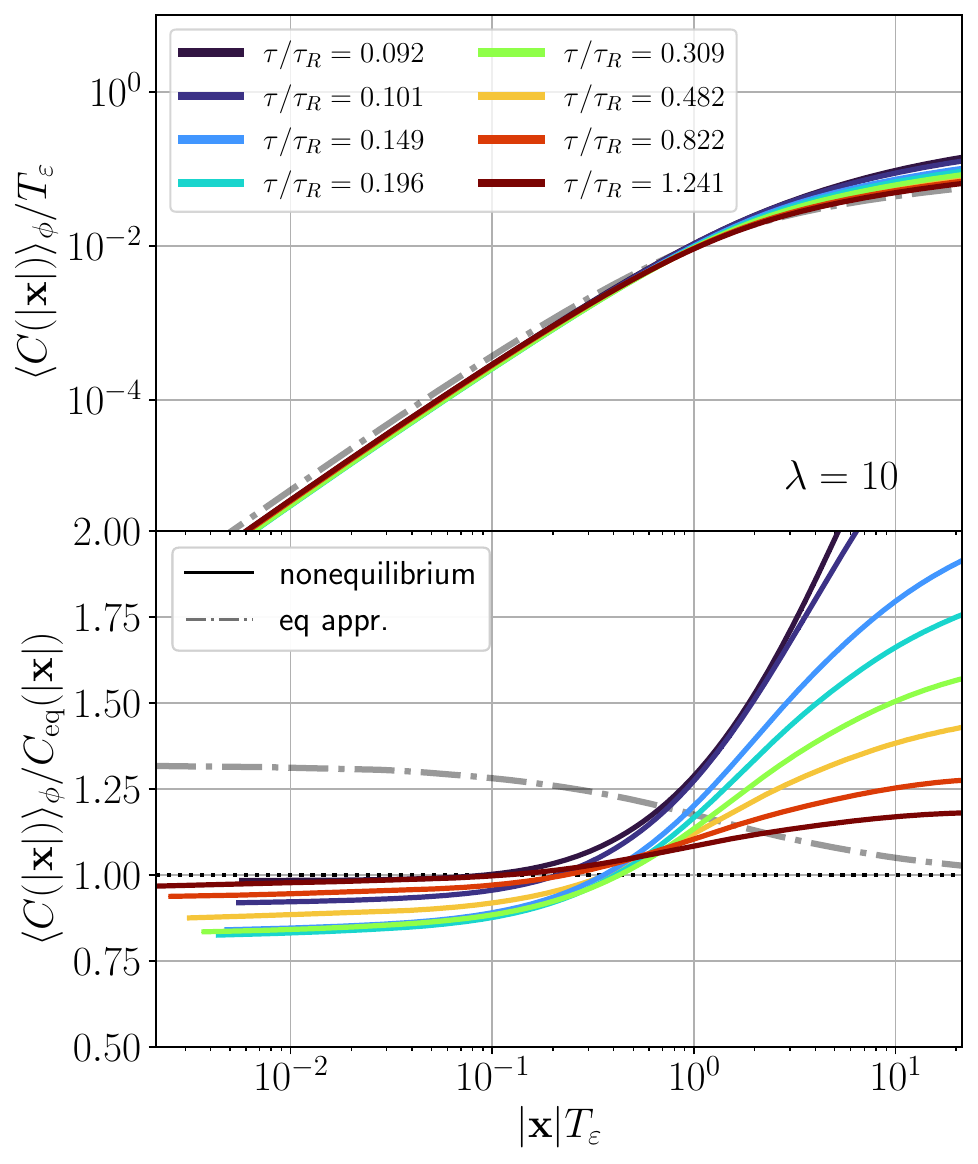}        
        }        
    \caption{Dipole cross section $\langle C( x)\rangle_\phi$ for different times and $\lambda=10$. In the left panel, four times are chosen to illustrate the general behavior, whereas in the right panel, more times are shown to show the continuous behavior.
    }
    \label{fig:dipolecrosssection-angularaverages-lambda10}
\end{figure}

\begin{figure}
    \centering
    \centerline{
        \includegraphics[width=0.5\linewidth]{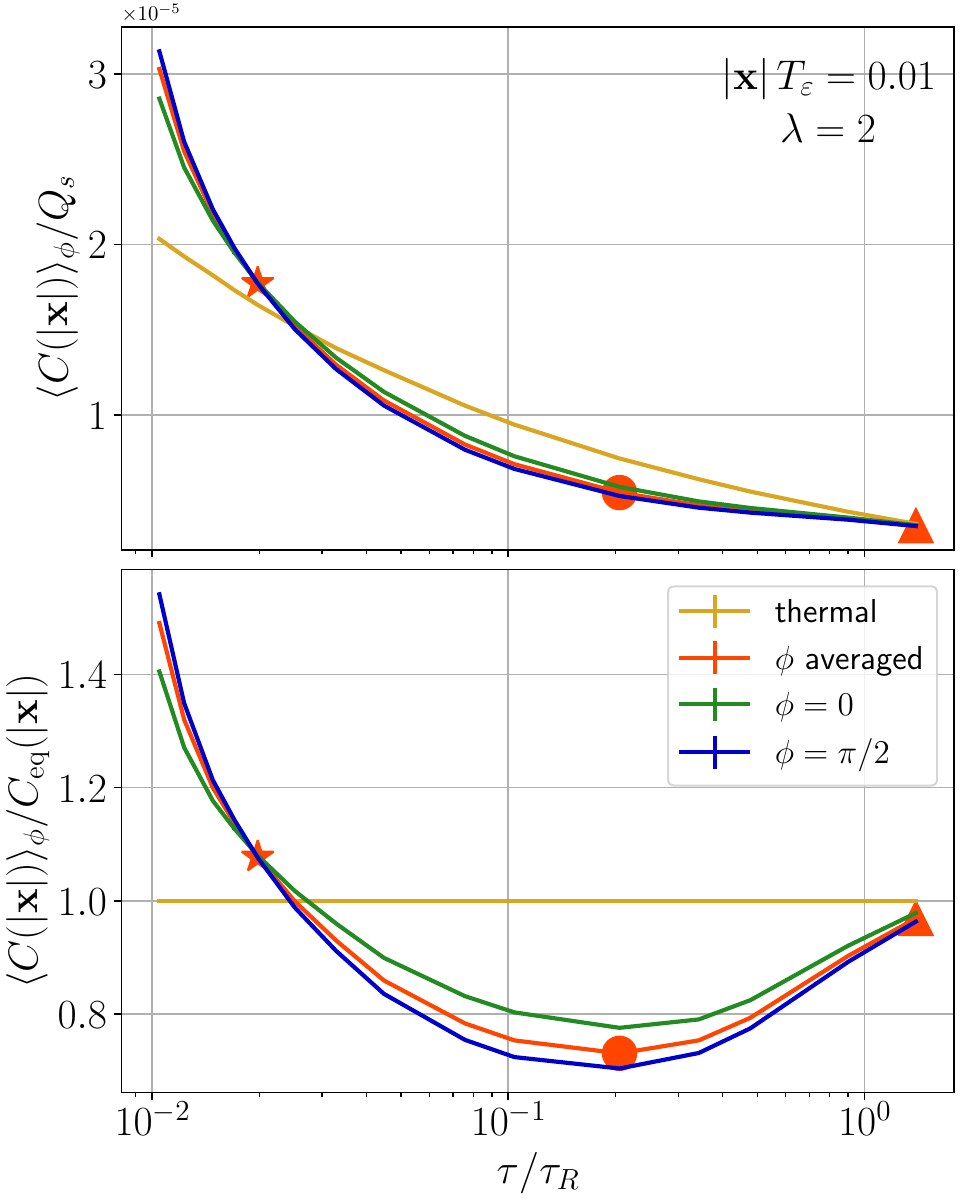}
        \includegraphics[width=0.5\linewidth]{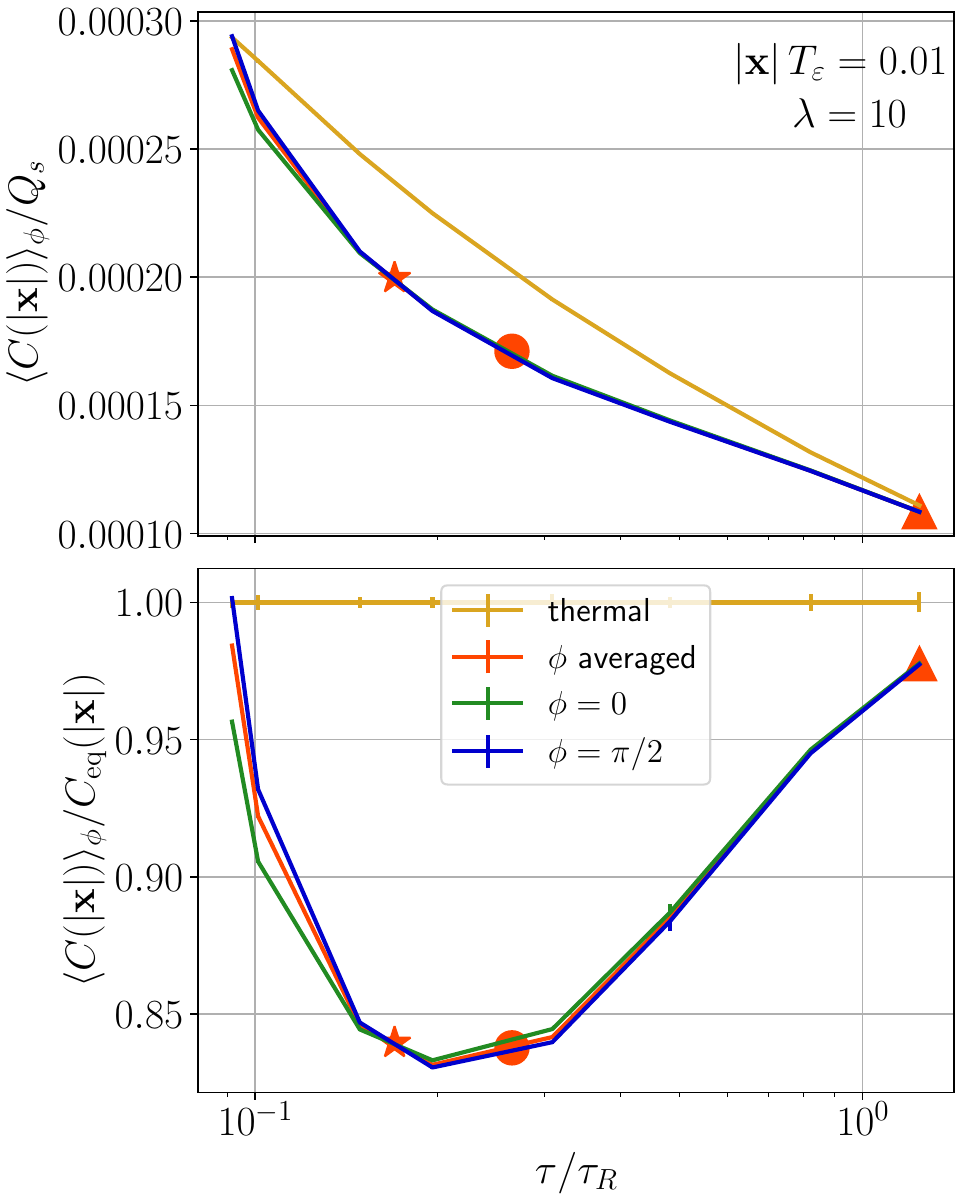}
        }        
    \caption{
    Dipole cross section $C(\vb x)$ for a fixed $|\vb x|\Teps = 0.01$ as a function of time. Shown is both the angular averaged cross section $\langle C(|\vb x|\rangle_\phi$ as a red curve, and the cross section for angles $\phi=0$ and $\phi=\pi/2$. The upper panels show the cross section normalized to the saturation momentum $Q_s$, the lower panels to the thermal cross section. Shown are couplings $\lambda=2$ (left) and $\lambda=10$ (right).
    }
    \label{fig:dipole-cross-section-smallx}
\end{figure}

We will now move on to discuss the time evolution of the dipole cross section.
The angular averaged dipole cross section $\langle C(x)\rangle_\phi$ is depicted for different times for couplings $\lambda=2$ in Fig.~\ref{fig:dipolecrosssection-angularaverages-lambda2} and $\lambda=10$ in Fig.~\ref{fig:dipolecrosssection-angularaverages-lambda10}. In the top panels, the dipole cross section is normalized by the effective temperature $\Teps$ from Eq.~\eqref{eq:Landau-matching-condition}, and in the lower panels by its equilibrium value. In the left panels, four distinct times are shown, whereas the right panels show more times to illustrate the continuous evolution.
We observe that the dipole cross section as a function of time behaves qualitatively differently for small and large $|\vb x|$, as compared to its thermal form. For all couplings and times, at large $\vb x$, the dipole cross section significantly exceeds its equilibrium values and continuously decreases towards equilibrium throughout the time evolution. In contrast, at small $|\vb x|$, which is the relevant region for highly energetic partons (jet quenching), the evolution is not monotonous, and we consider this region in more detail in Fig.~\ref{fig:dipole-cross-section-smallx}.
There,
the small-$|\vb x|$ behavior is shown
at $|\vb x|\Teps=0.01$ as a function of time.
In particular, for $\lambda=2$, the dipole cross section is initially about $50\%$ larger than in equilibrium, then quickly drops below its equilibrium value at around $\tau/\tauR\approx0.025$, and then approaches its equilibrium value from below. Therefore, for times $\tau/\tauR \gtrsim 0.025$, gluon radiation and thus jet quenching is suppressed as compared to thermal equilibrium. Na\"ively, in the infinite static medium, the rate/spectrum would go as $\sqrt{\qhat}$ (see Eq.~\eqref{eq:rate-lpm}), and $\qhat$ determines the small distance behavior of $\langle C(|\vb x|)\rangle_\phi$. Thus, the suppression in the spectrum is expected to be less dramatic than the suppression of the dipole cross section. The smallest value is at about $0.75$ of its thermal value, and we would therefore expect the suppression to be of $\sqrt{0.75}\approx 0.87$, only of about $10\%$. This na\"ive estimate is, of course, very simplistic; in practice, one needs to integrate over the whole splitting process.

An interesting feature of Fig.~\ref{fig:dipole-cross-section-smallx} is the angular ordering, i.e. that the dipole cross section for $\phi=0$ is initially suppressed until the star marker and then enhanced as compared to the $\phi=\pi/2$ cross section. This is similar to the behavior of the jet quenching parameter which was observed in Chapter \ref{sec:momentum-broadening-of-jets}.

For $\lambda=10$, the qualitative behavior, i.e., that the dipole cross section at small $|\vb x|$ first drops and then rises again as compared to its equilibrium value stays the same, but for the whole evolution, it is below the thermal value, which indicates suppression of jet quenching during the whole evolution.
Although this is qualitatively in agreement with findings of phenomenological studies that find jet quenching needs to be suppressed before $\tau=\qty{0.6}{\femto\meter/\c}$ \cite{Andres:2019eus}, the effect coming from the rather mild suppression of the dipole cross section at small $|\vb x|$ is rather too small to explain this. However, this result provides a hint that the nonequilibrium dynamics may influence the collision kernel and dipole cross section and can lead to an effective suppression of jet quenching during the initial stages.

\section{Gluon radiation in an anisotropic medium}
As explained in Chapter \ref{sec:jet-energy-loss}, the collision kernel is the key input for calculating the rate of inelastic gluon emissions. It is used as the medium input for jet quenching calculations, where one considers a highly energetic parton emitting a gluon. In that case, the dominant contribution comes from the small distance (small-$|\vb x|$) behavior of the dipole cross section.

For less energetic partons, the whole form of the collision kernel becomes important. This is the case needed for calculating the splitting and merging rates $\gamma^{a}_{bc}$ included in QCD kinetic theory simulations as discussed in Section \ref{sec:equations-of-qcd-kinetic-theory}.
In current QCD kinetic theory implementations (see, e.g., \cite{Kurkela:2014tea, Kurkela:2015qoa, Kurkela:2018oqw, Kurkela:2018xxd, Du:2020dvp, Du:2020zqg, AbraaoYork:2014hbk, kurkela_2023_10409474}) only the approximated thermal dipole cross section \eqref{eq:Cb_iso_appr} is used, valid for isotropic systems, effectively taking the analytic Fourier transform \eqref{eq:fouriertrafo_real} of the small $q_\perp$ limit \eqref{eq:analytic_collision_kernel}.

\subsection{AMY rate equation}
In this section, we will consider the full form of the dipole cross section $C(\vb x)$ to calculate the rate in the formalism of Arnold, Moore, and Yaffe.
Recall from Section \ref{sec:amy-rate-equation} that the rate for the process $g\to gg$
\cite{Arnold:2002ja, Arnold:2002zm} is given by 
\begin{align}
    \gamma=\frac{p^4+p'^4+k'^4}{p^3p'^3k'^3}\frac{d_A\alpha_s}{2(2\pi)^3}\int\frac{\dd[2]{\vb h}}{(2\pi)^2}2\vb h\cdot \mathrm{Re} \vb F, \label{eq:gammarate}
\end{align}
where $\vb F$ is the solution to the integral equation
\begin{align}
    2\vb h &= i\delta E(\vb h)\vb F(\vb h)+\frac{1}{2}\int\frac{\dd[2]{\vb q_\perp}}{(2\pi)^2}C(\vb q_\perp)\label{eq:integralequation-amy}\\
    &\times \left[(3\vb F(\vb h)-\vb F(\vb h-p\vb q_\perp)-\vb F(\vb h-k\vb q_\perp)-\vb F(\vb h+p'\vb q_\perp)\right],\nonumber
\end{align}
with $\delta E(\vb h)=m_D^2/4\times (1/k+1/p-1/p')+h^2/(2pkp')$.
The expression for this rate was derived with the assumption of an infinite medium and that the collision kernel $C(\vb q_\perp)$ does not significantly change (i.e., is constant) during the formation time ($\tform\sim\sqrt{\omega/\hat q}$, see Eq.~\eqref{eq:formation-time}) of a splitting process, but it is valid for all emitted gluon energies.
As discussed before, despite these approximations, these rates are used in kinetic theory simulations with the simplified isotropic form \eqref{eq:analytic_collision_kernel} of the collision kernel \cite{Arnold:2002zm, AbraaoYork:2014hbk, Kurkela:2014tea, Kurkela:2015qoa, Kurkela:2018xxd, Du:2020zqg, kurkela_2023_10409474,
Kurkela:2018oqw, Kurkela:2018xxd, Du:2020dvp}.

In this thesis, this rate equation is solved numerically for a general collision kernel $C(\vb q_\perp)$ satisfying the symmetry conditions \eqref{eq:symmetry_Cq} (but it is easy to generalize the method discussed here for a generic kernel). Previously in the literature, numerical evaluations of this rate only considered isotropic collision kernels (or, equivalently, isotropic dipole cross sections) \cite{Andres:2020vxs, Andres:2023jao, Caron-Huot:2010qjx, Moore:2021jwe, Schlichting:2021idr, Yazdi:2022bru, Modarresi-Yazdi:2024vfh}, or solved the rate equation \eqref{eq:gammarate} perturbatively around isotropy \cite{Hauksson:2023dwh}. To the best of my knowledge, I present here the first method to obtain the rate \eqref{eq:gammarate} for an anisotropic collision kernel $C(\vb q_\perp)$.

\subsection{Solving the AMY rate equation numerically}
We follow (and generalize) the method outlined in Ref.~\cite{Aurenche:2002wq} and solve Eq.~\eqref{eq:integralequation-amy} in impact parameter space, where the equation reduces to
\begin{align}
    (A-D(z,\vb x)-B\nabla^2)\vb F(\vb x)=-2i\nabla \delta(\vb x)\label{eq:impactparameterspaceequation}
\end{align}
with 
\begin{align}
    A&=i\frac{m_D^2}{4p}  \left(\frac{1}{z} + \frac{1}{1-z}-1\right),\\
    B&=\frac{i}{2pz(1-z)},\\
    D(z,\vb x)&=-\frac{1}{2} \left(C(\vb x) + C(z\vb x)+C((1-z)\vb x)\right),
\end{align}
and $z$ is the energy fraction of the emitted gluon, i.e., $p\to zp + (1-z)p$.
By going from the original integral equation \eqref{eq:integralequation-amy} to impact parameter space \eqref{eq:impactparameterspaceequation}, we go from the collision kernel $C(\vb q_\perp)$ to the dipole cross section $C(\vb x)$, see Appendix \ref{app:amyrates-details}.

In the isotropic case, $\vb F(\vb x)\sim \vb x F(|\vb x|)$, and Eq.~\eqref{eq:impactparameterspaceequation} for small $|\vb x|$ only has two linearly independent solutions.
For the general case, one can decompose the angular information in Fourier modes, and find that there are two linearly independent solutions for every Fourier mode, effectively summing up to infinitely many different independent solutions.
The boundary conditions are dictated for small $\vb x$ by the delta function in Eq.~\eqref{eq:impactparameterspaceequation}, and by requiring that $F(\vb x)\to 0$ for $|\vb x|\to \infty$, as well as that the rate \eqref{eq:gammarate} is finite.

The rate is solved for 7 and 11 Fourier modes to ensure that the results do not depend on the truncation of the Fourier series.
While in the isotropic case, it is enough to solve two independent ordinary differential equations, in the anisotropic case, $n_{\mathrm{fourier}}+3$ different systems of $n_{\mathrm{fourier}}$ coupled ordinary differential equations need to be solved.
More details on the numerical method is provided in Appendix \ref{app:amyrates-details}.

For solving the differential equation, the dipole cross section $C(\vb x)$ is needed for possibly arbitrary small and large $x$. In practice, this is achieved by considering the analytic limits of $C(\vb x)$. We have already extensively discussed the small-$|\vb x|$ limit in Section \ref{sec:smalldistance-dipolecrosssection}, where we found that $C(|\vb x|)\sim \vb x^2\log |\vb x|$. 
We will now briefly also discuss the opposite limit $|\vb x|\to\infty$. The Fourier transform \eqref{eq:chapter-collkern-fouriertrafo} features a rapidly oscillating function, and the integral is dominated by the region where the exponent is approximately unity, i.e., $q_\perp\sim 1/|\vb x|$. In the limit $|\vb x|\to\infty$, we, therefore, need to consider the $q_\perp\to 0$ limit of the collision kernel, which we have already worked out in Eq.~\eqref{eq:limits_qperp}. In this limit, $C(\qperp)\sim1/q_\perp^2$, which leads to a logarithmic behavior when performing the integral \eqref{eq:chapter-collkern-fouriertrafo}.

To summarize, the dipole cross section exhibits the following asymptotic behavior
\begin{align}
    C(|\vb x|,\phi)\to\begin{cases}
        \vb x^2(a_1(\phi)\log |\vb x|+a_2(\phi)), & |\vb x|\to 0\\
        a_3(\phi)\log |\vb x| + a_4(\phi), & |\vb x|\to\infty,
    \end{cases}\label{eq:dipolecrosssection-limits}
\end{align}
with angular dependent coefficients $a_i(\phi)$. In the numerical method, Eq.~\eqref{eq:dipolecrosssection-limits} is used to extrapolate the numerical results for the dipole cross section to arbitrarily small and large values of $|\vb x|$.

\subsection{Numerical results}
\begin{figure}
    \centering
    \centerline{
    \includegraphics[width=0.5\linewidth]{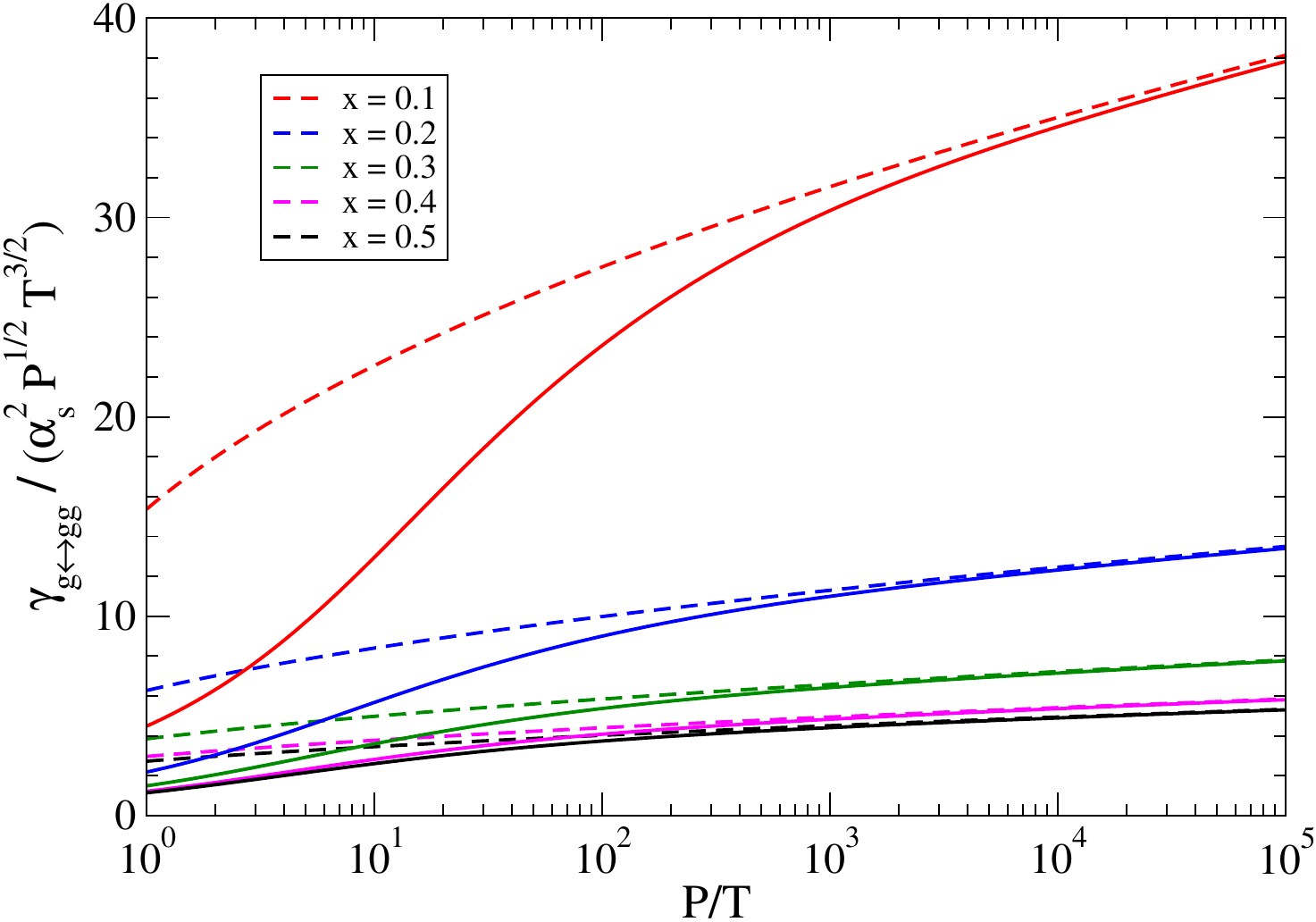}
    \includegraphics[width=0.5\linewidth]{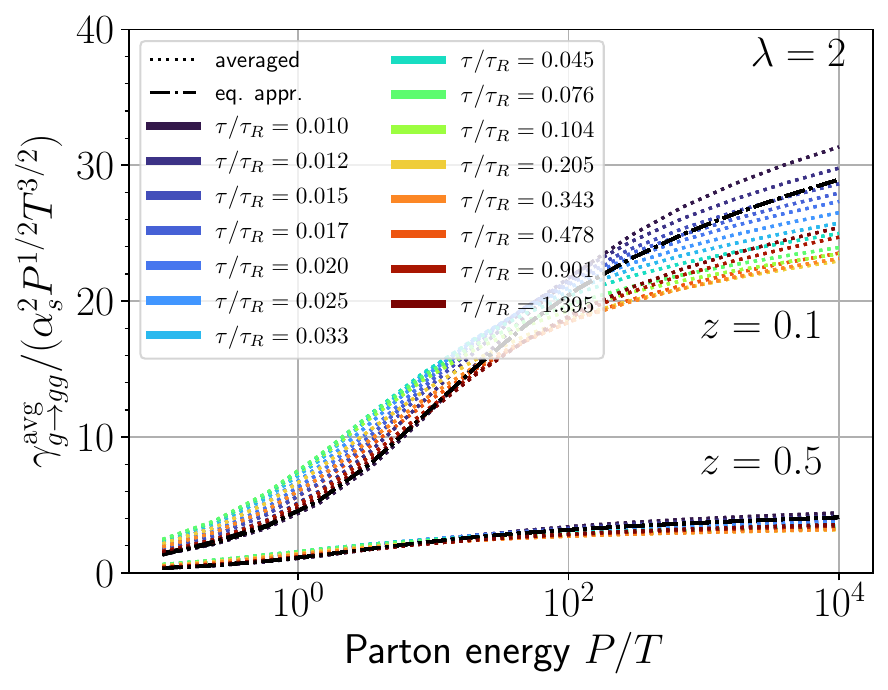}
    }     
    \caption{Gluon splitting rate $\gamma$ for various parton energies $P$. The axis are scaled such that in equilibrium all coupling and temperature dependence cancels.
    (\emph{Left}): Plot from Ref.~\cite{Arnold:2008zu} showing the splitting rate in thermal equilibrium for the approximated dipole cross section \eqref{eq:Cb_eq_appr} (dashed lines) and its leading-log approximation (solid lines).
    (\emph{Right}): 
    Splitting rate as computed for collision kernels at different times for Bjorken expanding system at coupling $\lambda=2$. Shown are, for simplicity, only the rates obtained from the averaged kernel (dotted lines) and from the approximated dipole cross section \eqref{eq:Cb_eq_appr}.
    }
    \label{fig:scaled-rate}
\end{figure}
We now move on to discuss the numerical results for the splitting rate.
As in the whole thesis, we consider only gluons, i.e., the process of inelastic gluon radiation $g\to gg$.
First, to validate the numerical approach, we compare the splitting rate obtained from the numerical approach with previously known results for the splitting rate in thermal equilibrium \cite{Arnold:2008zu}, which are shown in the left panel of Fig.~\ref{fig:scaled-rate}. The axes are scaled in a way to make all curves fall on top of each other in thermal equilibrium for different temperatures $T$ and couplings $\lambda$. The dashed lines denote the numerically obtained rate from Ref.~\cite{Arnold:2008zu} using the small $q_\perp$ form of the collision kernel \eqref{eq:analytic_collision_kernel}, which is equivalent to using the approximated analytic dipole cross section \eqref{eq:Cb_eq_appr} in impact parameter space. The dashed lines represent the next-leading-log solution of Ref.~\cite{Arnold:2008zu} valid at large jet momenta. This corresponds to an expansion in logarithms, and is, thus, similar to the improved opacity expansion \cite{Mehtar-Tani:2019tvy, Mehtar-Tani:2019ygg, Barata:2021wuf}.
Different colors denote different splitting fractions $z$ (labeled in the left plot as $x$). The right panel shows the results of using the numerical method, applied for angular averaged dipole cross sections $\langle C(|\vb x|)\rangle_\phi$, and also for the approximated thermal cross section \eqref{eq:Cb_eq_appr} as black dash-dotted lines. These latter lines should be compared with the dashed curves in the left panel, and show excellent agreement, validating the numerical method used here. The colored dotted lines correspond to the rates obtained from the dipole cross sections at different times and roughly follow the thermal estimates in magnitude.

\begin{figure}
    \centering
    \centerline{
        \includegraphics[width=0.5\linewidth]{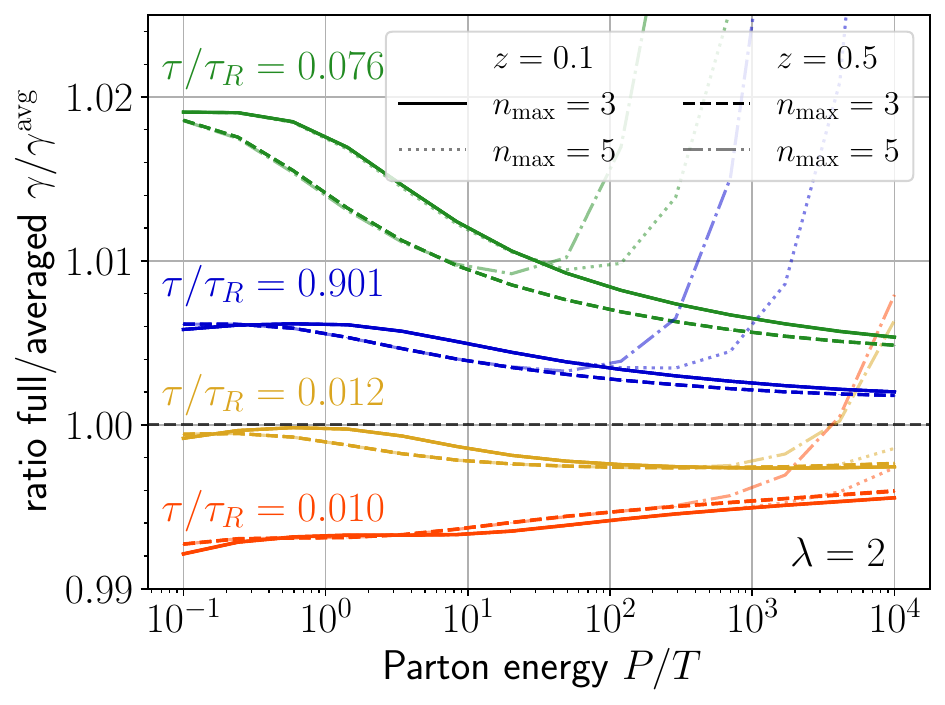}
        \includegraphics[width=0.5\linewidth]{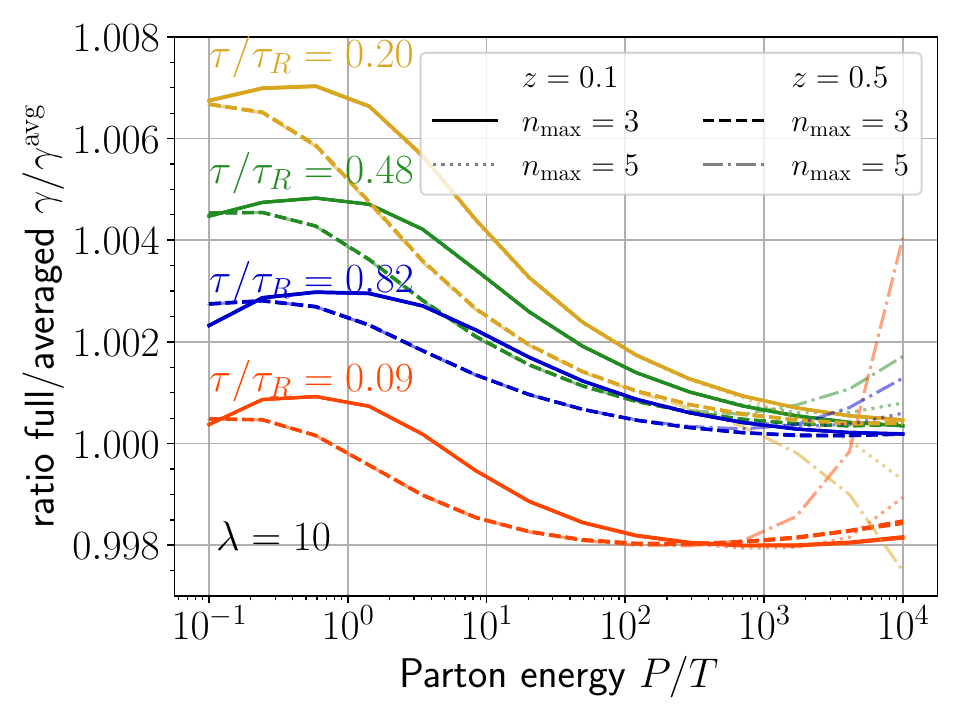}
        }     
    \centerline{
        \includegraphics[width=0.5\linewidth]{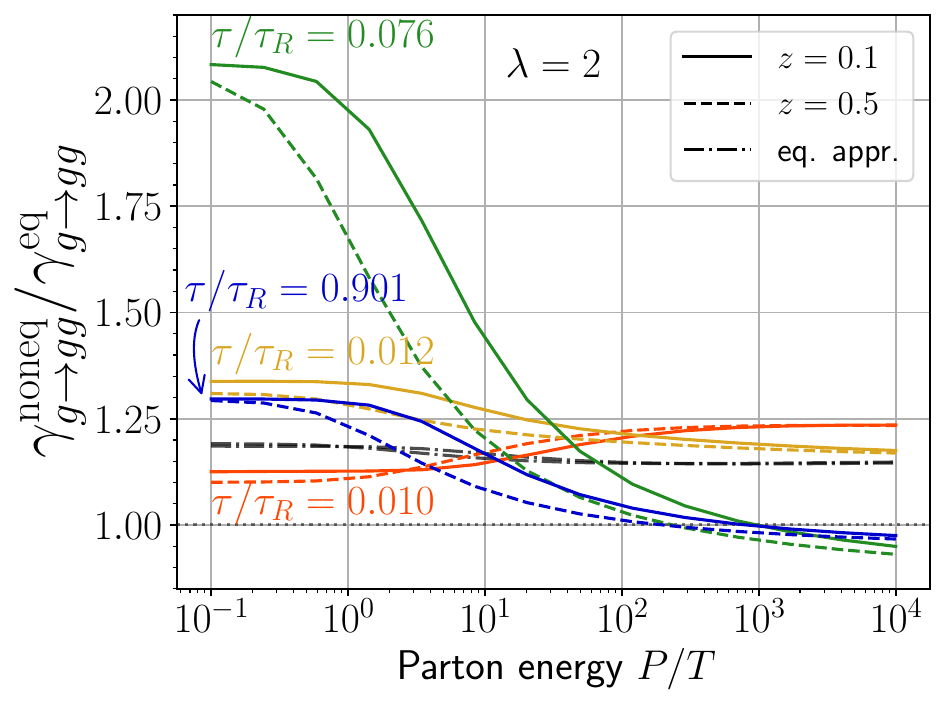}
        \includegraphics[width=0.5\linewidth]{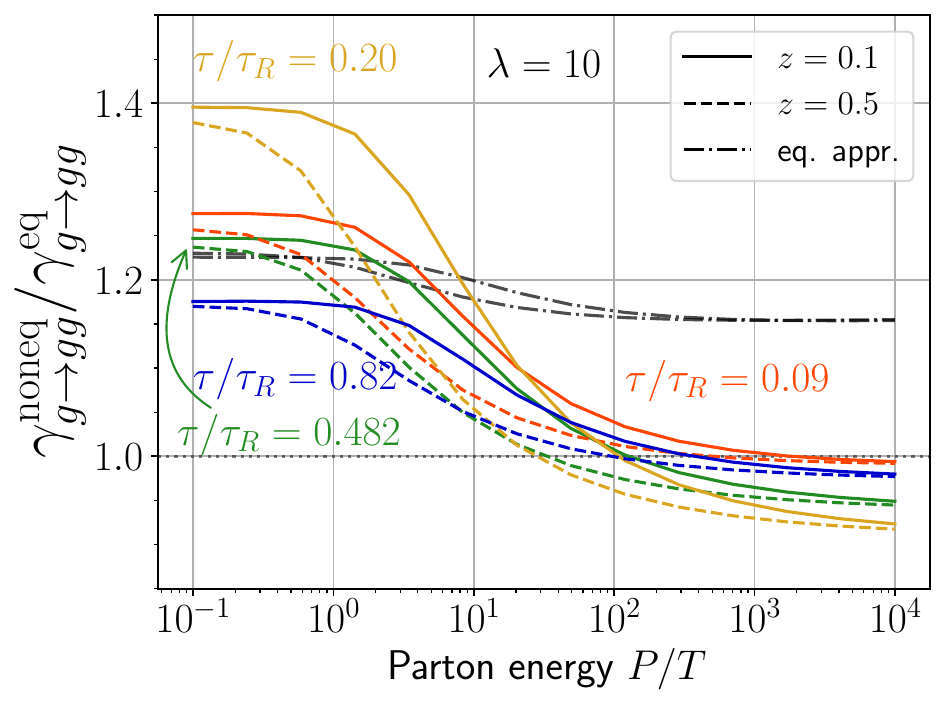}
        }        
         \centerline{
        \includegraphics[width=0.5\linewidth]{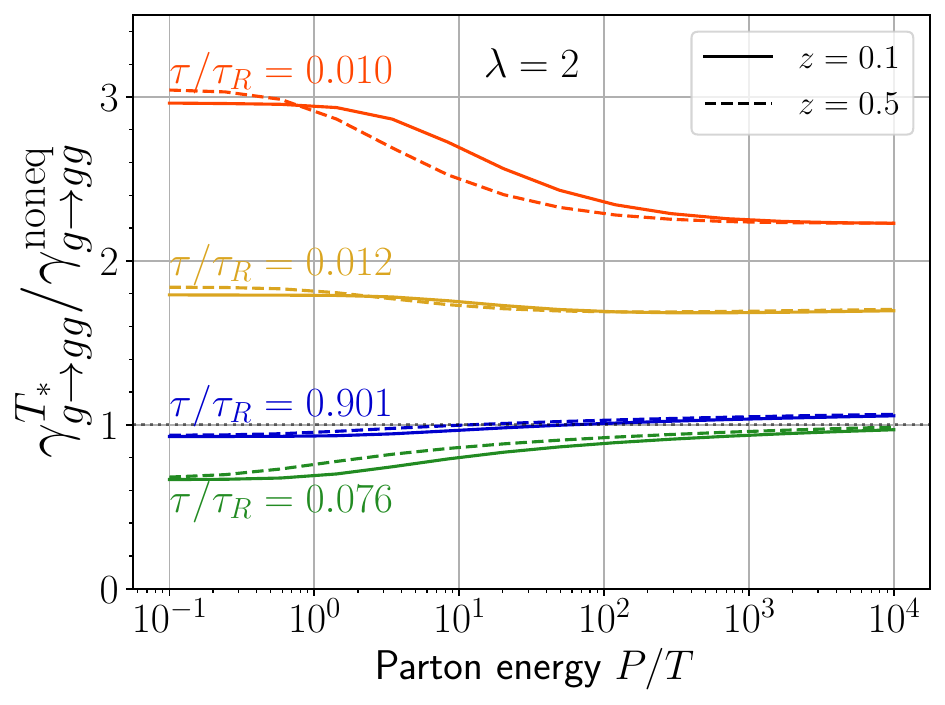}
        \includegraphics[width=0.5\linewidth]{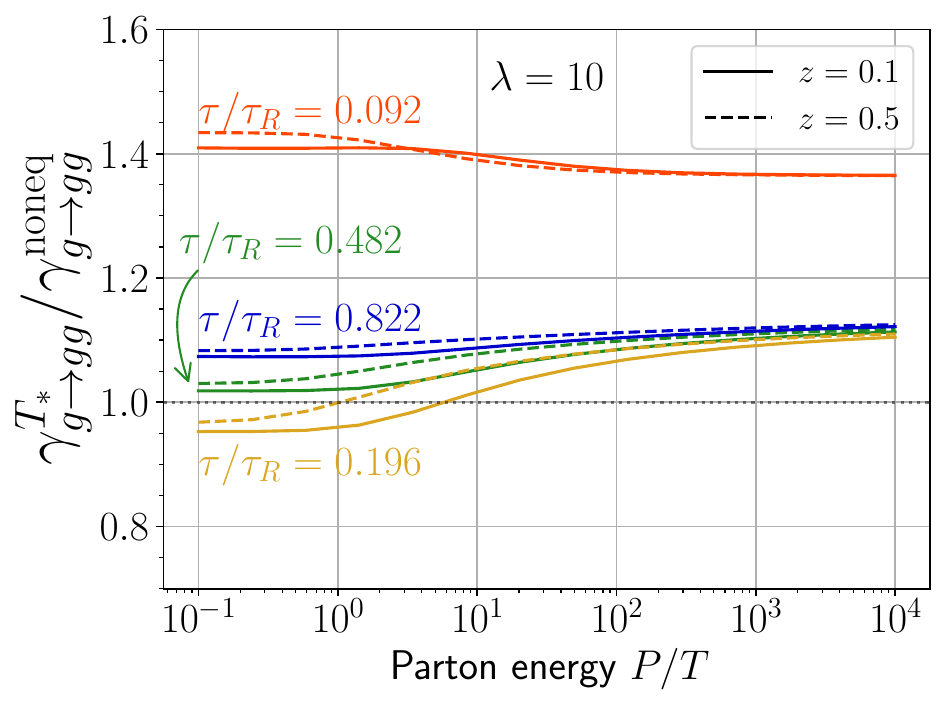}
        }        
    \caption{
    Gluon splitting rate for $\lambda=2$ (left column) and $\lambda=10$ (right column) in a Bjorken expanding gluonic plasma at different times (color-coded). The parameter $z$ is the energy fraction of one of the emitted gluons.
    (\emph{Top}): Ratio of the rate for the anisotropic collision kernel $C(\vb q_\perp)$ over the rate obtained from the angular averaged one $\langle C(q_\perp)\rangle_\phi$. 
    (\emph{Middle}): Rate for the nonequilibrium kernel over the Landau-matched equilibrium one. Additionally shown is the rate obtained from the approximated thermal dipole cross section \eqref{eq:Cb_eq_appr} as a black dash-dotted line.
    (\emph{Bottom}): 
    Rate using the $T_\ast$ approximated dipole cross section as input \eqref{eq:Cb_iso_appr} to the thermal one.
    }
    \label{fig:rates-manypanels}
\end{figure}

We now turn to study the rates obtained from the nonequilibrium kernel in more detail.
Fig.~\ref{fig:rates-manypanels} shows the gluon splitting rate $\gamma$ obtained from the nonequilibrium dipole cross section $C(\vb x,\tau)$ at several times for a Bjorken expanding plasma for couplings $\lambda=2$ (left column) and $\lambda=10$ (right column). The different times are color-coded, with a consistent color scheme for the same coupling.

In the upper row, the rate for the anisotropic nonequilibrium collision kernel $C(\vb q_\perp,\tau)$ is plotted over the rate from its angular average $\langle C(q_\perp,\tau)\rangle_\phi$. Remarkably, this ratio is close to unity (less than $2\%$ deviations) for both considered couplings and all considered times. This implies that obtaining the rate for the averaged collision kernel $\langle C(q_\perp)\rangle_\phi$ provides already a very good estimate for the rate obtained from the anisotropic kernel $C(\vb q_\perp)$. In the latter case, the maximum number of Fourier modes is varied ($\nmax = 3$ and $\nmax = 5$), where we observe that $\nmax=3$ coincides with $\nmax=5$ for almost the entire parton energy range, and, thus, already provides a very good description. For larger parton energies $P\geq 50 \Teps$, using more Fourier modes leads to numerical problems (e.g., finding the unique linear combination of solutions that satisfies all the boundary conditions). This is seen in the plots by the diverging lines at larger parton energies $P$. These numerical problems will be discussed in more detail in Appendix \ref{app:amyrates-details}.

Let us now move on to the central panels, where the rate obtained from the nonequilibrium kernel $C(\vb q_\perp)$ is compared to the Landau-matched thermal rate $\gamma^{\mathrm{eq}}$. Here, we observe an enhancement of the rate at small parton energies and a suppressed rate at large energies. This suppression aligns with the discussion in the last section regarding the reduction of the small-$|\vb x|$ behavior of the dipole cross section. However, interpreting the rate at large momenta $P$ is complicated by the fact that an underlying assumption for the rate equation \eqref{eq:gammarate} is that the collision kernel $C(\vb q_\perp)$ does not change during the emission process. This is only true for processes with small formation time \eqref{eq:formation-time},
\begin{align}
    \tform \sim \sqrt{\frac{\omega}{\qhat}}=\sqrt{\frac{zP}{\qhat}}. \label{eq:formationtime-gluonenergy-relation}
\end{align}
Let us provide a simple estimate of until which parton energies we may trust the results in Fig.~\ref{fig:rates-manypanels}.
For example, consider $\lambda=2$ (left panels), where in the middle panels, the rate (or rather ratio) seems to be approximately constant between $0.076 < \tau/\tauR < 0.9$ (corresponding to $Q_s\Delta\tau\approx 400$). If we take as average temperature $\Teps \approx 0.3 Q_s$ from Tab.~\ref{tab:simulation-lambda2-isoHTL}, and $\qhat\approx 0.16\Teps^3\approx 0.004Q_s^3$ from Fig.~\ref{fig:different_cutoffs_comparison_thermal_qhat}, this leads to $\omega\approx 640 Q_s\approx 2000 \Teps$, for which the formation time is of this order. 

For $\lambda=10$, we can consider the right center panel of Fig.~\ref{fig:rates-manypanels}. Taking the time between the red and yellow curve corresponding to $Q_s\Delta\tau\approx 2$ with $\Teps/Q_s\approx 0.5$ from Tab.~\ref{tab:simulation-lambda10-isoHTL}, and taking $\qhat\approx 2 \Teps^3\approx 0.25 Q_s^3$ from Fig.~\ref{fig:qhat_realisticv5a}, we can estimate $\omega \lesssim Q_s\approx 2\Teps$, which implies that for $z=0.1$, the formation time is smaller or of that order for $P/T\lesssim 20$.

If we are less restrictive and consider the rate (or rather ratio) between the green ($\tau/\tauR=0.48$) and blue curve ($\tau/\tauR=0.82$) to be almost constant, this corresponds to roughly $\Delta\tau\approx10/Q_s$, and take $\Teps=0.2 Q_s$ and $\qhat\approx 2 T^3\approx 0.08 Q_s^3$, we obtain $\omega = 8 Q_s\approx 40\Teps $, leading to a maximum parton energy of $P \approx 400 \Teps $, where the rate is still larger than its thermal equilibrium counterpart.

These simple estimates verify that the rates cannot be used for very energetic particles at all times, but can be used for particles with momenta of the order of the effective plasma temperature $\Teps$, for example, in kinetic theory simulations.

In the bottom row, we compare the rate $\gamma^{T_\ast}$ obtained from the isotropic approximation to the dipole cross section $\Cisoappr$ in \eqref{eq:Cb_iso_appr} to the nonequilibrium one. We find that at early times the rate $\gamma^{T_\ast}$ significantly overestimates the nonequilibrium rate. In particular for $\lambda=2$, we find that initially the rate is overestimated by more than a factor of two. Even for $\lambda=10$, $\gamma^{T_\ast}$ is about $40\%$ larger than the nonequilibrium rate.

\begin{figure}
    \centering
    \centerline{
        \includegraphics[width=0.5\linewidth]{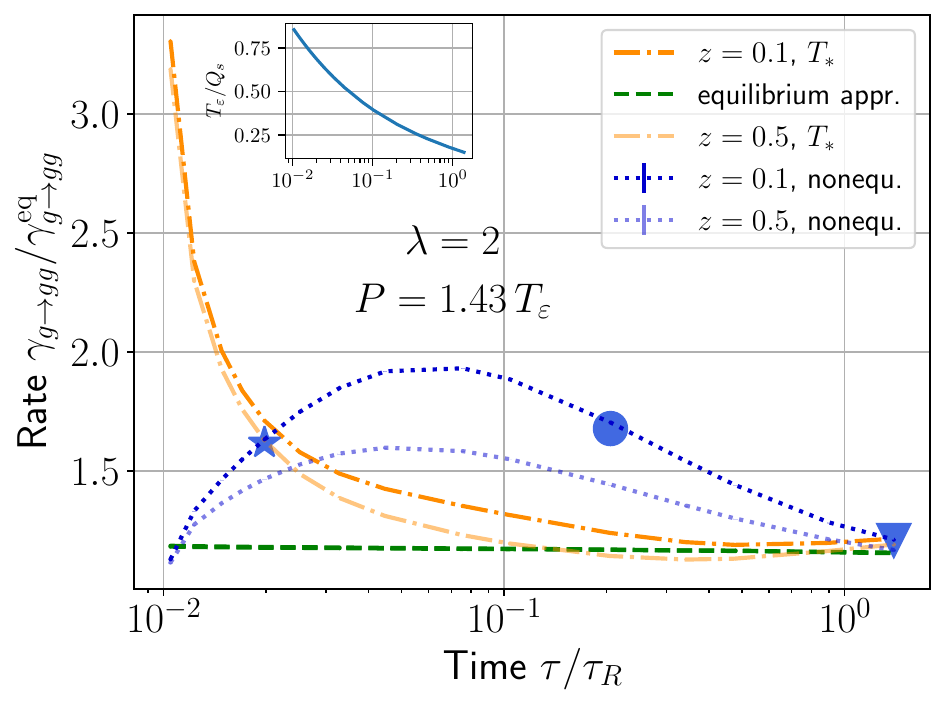}
        \includegraphics[width=0.5\linewidth]{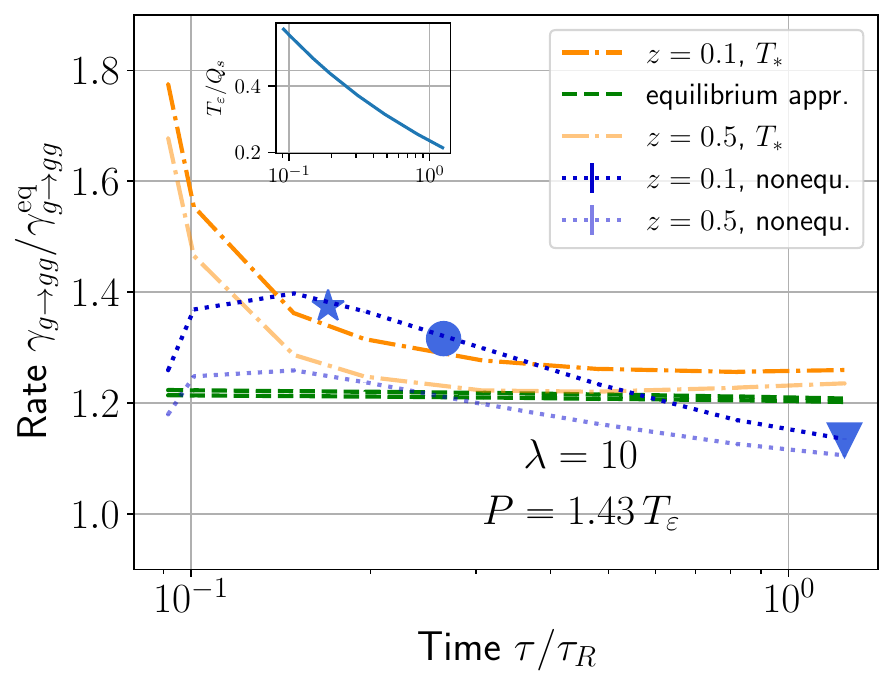}
        }        
    \caption{
    Comparison of the rate $\gamma$ obtained from the nonequilibrium collision kernel $C(\vb q_\perp)$ (blue dotted line), the equilibrium approximated cross section \eqref{eq:Cb_eq_appr} (green dashed line) and the isotropic approximated cross section \eqref{eq:Cb_iso_appr} (orange dash-dotted line) for a parton with energy $P=1.25\Teps$ as a function of time $\tau$. The left panel shows the results for coupling $\lambda=2$, the right panel for $\lambda=10$. 
    }
    \label{fig:rate-time-evolution}
\end{figure}

Finally, in Fig.~\ref{fig:rate-time-evolution}, we compare the splitting rate $\gamma$ at a specific constant parton energy over effective temperature $P/\Teps$, obtained from the full anisotropic nonequilibrium collision kernel $C(\vb q_\perp,\tau)$ (blue dotted line), and for various approximations: The rate from the approximated thermal cross section \eqref{eq:Cb_eq_appr} is shown as a green dashed curve, and the rate from the approximated isotropic cross section \eqref{eq:Cb_iso_appr} is shown as an orange dash-dotted line. All of the values are normalized to the thermal values. The left panel shows the results for $\lambda=2$, the right panel for $\lambda=10$. In both cases, all of the curves are above unity, implying that the splitting rate for this parton energy is larger than expected from a Landau-matched thermal equilibrium. The thermal approximated rate in green is constant, at a value of $20\%$ above the thermal rate. The nonequilibrium rate (blue dotted curves) initially grows up to $20\%$ above the thermal rate ($50\%$ for $\lambda=2$) and then decreases, but at the triangle marker, the differences to thermal equilibrium are still larger than $10\%$. Surprisingly, the rate for the isotropic approximation \eqref{eq:Cb_iso_appr} that is used in QCD kinetic theory simulations exhibits a qualitatively different behavior. It peaks at initial times with up to $60\%$ above the thermal values ($300\%$ in the case of $\lambda=2$), then drops below the nonthermal rate, crossing it approximately at the star marker, where the system becomes under-occupied. It then approaches the approximated thermal rate (which it should approach at late times). The failure to exactly approach the thermal rate at the very end is due to discretization artifacts.

This qualitative and quantitative different behavior of the rate $\gamma^{T_\ast}$ obtained from the isotropic cross section \eqref{eq:Cb_iso_appr} poses several questions regarding QCD kinetic theory simulations, in which this rate is commonly applied. In particular, at weak couplings, the deviations seem to become even larger. Since the bottom-up thermalization process relies crucially on the splitting rate, through which a soft thermal bath is formed, a modification of this rate in simulations can have sizable consequences. The rate obtained from the nonequilibrium kernel seems to be initially smaller during the over-occupied stage, which implies that fewer soft gluons are initially emitted. In the under-occupied stage (after the star marker), for $\lambda=2$, the nonequilibrium rate becomes larger than $\gamma^{T_\ast}$, possibly affecting the second and third stage in the bottom-up equilibration simulations.
Clearly, only further simulations using this nonequilibrium rate can clarify whether the equilibration and hydrodynamization process is substantially modified, which will be an exciting opportunity in the future.

\section{Concluding remarks}
In this chapter, we studied the elastic collision kernel $C(\vb q_\perp,\tau)$, which generalizes the jet quenching parameter. The results obtained from numerical QCD kinetic theory simulations of the bottom-up equilibration process
reveal an effectively angle-dependent screening scale and
indicate that the jet quenching parameter $\qhat$ receives larger contributions from small-momentum exchanges than in a corresponding thermal equilibrium, particularly in regions below the Debye mass. The anisotropy and magnitude of the kernel are consistent with the evaluation of the jet quenching parameter $\qhat$ in Chapter \ref{sec:momentum-broadening-of-jets}.

We then moved on to study the dipole cross section, which can be obtained from the collision kernel using a (one-subtracted) Fourier transform \eqref{eq:chapter-collkern-fouriertrafo}. Its small distance behavior is relevant for calculating the gluon emission probability of a highly energetic parton, and, thus, the relevant region for jet quenching. This small distance behavior of the dipole cross section can be well described by the analytic formula \eqref{eq:smallxform-general}, which needs as medium input only the jet quenching parameter as a function of the cutoff $\qhat(\lperp)$. We found that this formula also reproduces very well the angular information. Furthermore, for calculating gluon emission rates of softer partons, the whole form of the dipole cross section is important. We have discussed the analytic limits of this quantity and compared it to its equilibrium estimates at various times. We find that while the nonequilibrium dipole cross section at large $|\vb x|$ is significantly larger than in equilibrium, at small distances $|\vb x|$, it is slightly smaller, indicating a suppression of jet quenching.

In the final part of this chapter, we considered the AMY gluon splitting rates obtained from the dipole cross section. For the first time, a numerical method is developed to obtain this quantity for an anisotropic dipole cross section. Remarkably, the rate obtained from the anisotropic cross section is well approximated by the rate obtained from the angular averaged cross section. However, it is still substantially different from the rate for an equilibrium collision kernel, especially at early times (up to $50\%$ for $\lambda=2$).
Additionally, we have discussed that the approximation for the collision kernel used in QCD kinetic theory simulations leads to significantly larger rates at early times (up to $300\%$ for $\lambda=2$), which potentially impacts QCD kinetic theory simulations of the bottom-up equilibration and hydrodynamization process in heavy-ion collisions.

%% file: 800_summary.tex
This thesis focuses on the initial nonequilibrium stages in heavy-ion collisions, which can be described and modeled using QCD kinetic theory. Potential experimental probes of these initial stages are high-energy particles which are measured as jets in the detectors. Their energy loss can---in the harmonic approximation---be described by a single medium parameter, the jet quenching parameter $\qhat$.

In Chapter \ref{sec:momentum-broadening-of-jets}, we discussed how this jet quenching parameter $\qhat$ can be obtained for a nonequilibrium QCD plasma using QCD kinetic theory, particularly for a Bjorken expanding system relevant for the initial stages in heavy-ion collisions.
We studied this parameter for several toy models of different parts of these initial stages, and then extracted and obtained its value numerically using QCD kinetic theory for such expanding systems. The parameter is found to be similar in magnitude to calculations and simulations from the earlier Glasma stage, as well as in its qualitative properties, including increased broadening along the beam axis for a jet moving perpendicular to it. 
This first extraction using QCD kinetic theory marks an important step in understanding pre-equilibrium jet quenching.

We also discussed the jet quenching parameter and the related \emph{heavy-quark diffusion coefficient} in the context of hydrodynamic attractors in Chapter \ref{sec:limiting_attractors}. While for their anisotropy ratio, the commonly used time scaling associated with the hydrodynamic attractor seems to be less useful than for other observables, we identified a new feature in these transport coefficients. When time is rescaled with the parametric estimate for the bottom-up thermalization of weakly-coupled expanding systems, we observe that these anisotropy ratios admit an extrapolation to vanishing couplings, which we refer to as \emph{limiting attractors}. We also observe this weak-coupling bottom-up limiting attractor in the pressure ratio, which additionally allows an extrapolation to the strong-coupling hydrodynamic limiting attractor. While the hydrodynamic limiting attractor is also visible for the jet quenching parameter and heavy-quark diffusion coefficient ratio, it offers less predictive power since even moderate values of the couplings quantitatively and qualitatively deviate from this attractor.

In Chapter \ref{sec:improving-qcd-simulations}, we then moved on to discuss in detail screening approximations which are typically employed in QCD kinetic theory simulations. In particular, we investigated how including the fully resummed HTL propagator in the elastic collision term modifies previous results on QCD thermalization obtained with a simpler Debye-like screening prescription. While for isotropic systems the effects are negligible, large and significant deviations can be observed in several quantities for expanding systems, which are relevant for the initial stages in heavy-ion collisions. For instance, the maximum pressure anisotropy is significantly reduced when employing HTL screening.
The approach to hydrodynamics is less affected when the different screening prescriptions are accounted for with different numerical values of the hydrodynamic transport parameter $\eta/s$. For this parameter, we found that the HTL screening prescription consistently leads to smaller values, which are closer to their perturbative estimate than when employing Debye-like screening. Furthermore, we studied the impact of this screening prescription on the jet quenching parameter $\qhat$, which was found to be only mildly influenced by different screening prescriptions. However, it should be noted that similar to all other previous treatments in QCD kinetic theory, we still lack a proper understanding of the effect of plasma instabilities in these simulations, which are currently neglected by employing an isotropic screening approximation.

Finally, we generalized the jet quenching parameter to obtain the elastic collision kernel $C(\vb q_\perp,\tau)$ in Chapter \ref{sec:collkern}. We found that it is anisotropic and the contribution to the jet quenching parameter is peaked at the Debye mass for late times and for broadening transverse to the beam axis. For early times and broadening along the beam axis, the peak is shifted to smaller momentum exchanges, which could be interpreted as an emerging angle-dependent effective screening mass. This contribution to $\qhat$ is significantly enhanced at early times for small momenta and suppressed for large momenta when compared to a corresponding thermal system, while at late times, it approaches its thermal form. We then moved on to obtain the dipole cross section as a Fourier transform of the collision kernel. We first verified that its small distance behavior is accurately described by the formula obtained in Section \ref{sec:smalldistance-dipolecrosssection}, and then studied and compared its nonequilibrium and angularly averaged form to the thermal form. There, we found that the dipole cross section is smaller than in thermal equilibrium for small distances throughout the whole evolution for larger couplings (and most of the evolution for coupling $\lambda=2$). In contrast, for large distances, it significantly exceeds its thermal form and continuously decreases to thermal equilibrium throughout the hydrodynamization process. Finally, we also considered the dipole cross section as input to calculate the gluon splitting rates used in QCD kinetic theory simulations. There, we found that the rates obtained from an anisotropic collision kernel can be well approximated by the rate from an angular averaged kernel. However, these nonequilibrium rates significantly differ from those obtained in a corresponding thermal system. In particular, they are $20-50\%$ enhanced. Moreover, the rate obtained from an isotropic approximated collision kernel employed in QCD kinetic theory simulations exhibits a significantly different behavior than the actual nonequilibrium rate, both in magnitude and functional time dependence, with possible and as of yet unexplored consequences for QCD kinetic theory simulations.

This first extraction of both the jet quenching parameter $\qhat$ and the collision kernel during the initial stages in heavy-ion collisions using QCD kinetic theory marks an important step in increasing our understanding of pre-equilibrium jet quenching in heavy-ion collisions. In particular, many features found in these quantities, such as anisotropies and deviations from equilibrium, are unaccounted for in current simulations of heavy-ion collisions and will be even more important in the upcoming collisions of light ions at the LHC.

While this thesis thus constitutes an important theoretical and conceptual improvement of medium properties relevant for jet quenching studies during the initial stages, further studies are needed to quantify the effect of these initial nonequilibrium stages. In particular, the effect of using the isotropic approximated kernel for obtaining the nonequilibrium splitting rates in QCD kinetic theory simulations will need to be explored in further simulations, to investigate if qualitative or quantitative changes arise from this approximation.
Moreover, it will be interesting to study the new concept of limiting attractors proposed in this thesis, in particular, to identify other observables that exhibit a weak-coupling bottom-up limiting attractor. Studying how this bottom-up limiting attractor interferes and interplays with the hydrodynamic limiting attractor may improve our understanding of the hydrodynamization and equilibration process of QCD in heavy-ion collisions and beyond.
Furthermore, the anisotropy and evolution of the jet quenching parameter and collision kernel during the nonequilibrium evolution should be included in the phenomenological modeling of heavy-ion collisions to assess their impact on jet quenching. This will enable finding, identifying, and proposing new experimental observables that may be sensitive to the initial stages and, thus, may offer the exciting opportunity of probing the nonequilibrium evolution of the QCD plasma in heavy-ion collisions.

%% file: 850_qcd.tex
In this appendix, we discuss details about QCD, correlation functions and nonequilibrium field theory.
We start in Section \ref{sec:qcd-lagrangian} by discussing the QCD Lagrangian and its building blocks. In Section \ref{app:closed-timepath}, we discuss aspects of nonthermal quantum field theory such as the closed time path and different types of propagators and correlation functions. In Section \ref{sec:perturbation-theory-self-energy}, we discuss how the self-energy arises, and in \ref{app:correlator-relations}, the various propagators and correlation functions are listed, together with some convenient properties.
In Section \ref{app:htl}, we discuss the hard thermal loop propagators, how they can be used in the isoHTL screening for the jet quenching parameter and elastic collision term, and in particular how to use the sum rule from Ref.~\cite{Aurenche:2002pd} to obtain an analytic result for the collision kernel in isotropic systems. In Section \ref{app:scattering-soft-momentum-exchange}, we verify that for soft-gluon exchange, the AMY screening prescription \eqref{eq:amy_replacement} is leading-order accurate by considering explicitly quark and gluon scatterings with soft momentum transfer.

\section{QCD Lagrangian and notation\label{sec:qcd-lagrangian}}
Quantum chromodynamics is the quantum field theory describing the strong interaction. It is a nonabelian gauge theory with gauge group $\mathrm{SU}(\NC)$ with $\NC=3$ for QCD (but we can leave it arbitrary for the moment since it does not further complicate the discussion).
This implies that the Lagrangian
\begin{align}
    \mathcal L=\sum_f\bar\psi_i^f\left(i\gamma^\mu D_\mu{}_{,ij}-m_f\right)\psi_j^f-\frac{1}{2}\Tr\left(F^{\mu\nu}F_{\mu\nu}\right)
\end{align}
is invariant under the gauge transformation
\begin{align}
    \psi_i^f(X)\to U_{ij}(X)\psi_j^f(X), && A_\mu(X)\to U(X)A_\mu(X)U^\dagger(X)+\frac{i}{g}U(X)\partial_\mu U^\dagger(X),
\end{align}
where $U \in \mathrm{SU}(\NC)$ with components $U_{ij}$ is a unitary $\NC\times\NC$ matrix with unit determinant, $\det U(X)=1$.
The Dirac field $\psi_{i,\alpha}^f(X)$ is a spinor field with color ($i$), flavor ($f$) and spinor $(\alpha)$ index. The gamma matrices $(\gamma^\mu)_{\alpha\beta}$ are matrices in spinor space (i.e., they act on the spinor index of $\psi$) and satisfy the Clifford algebra relation
\begin{align}
    \{\gamma^\mu,\gamma^\nu\}=\gamma^\mu\gamma^\nu+\gamma^\nu\gamma^\mu=-2\eta^{\mu\nu}.
\end{align}
The adjoint spinor can be obtained via $\bar\psi=\psi^\dagger \gamma^0$.

The index $f$ runs over the number of $\nf$ quark flavors with mass $m_f$. For this thesis, the typical momenta of particles will be much larger than their rest mass (for the light flavors), making $m_f=0$ a good approximation.

The gauge field $A_\mu(X)\in\mathfrak{su}(\NC)$ is an element of the Lie Algebra $\mathfrak{su}(\NC)$, and as such, can be represented by a traceless hermitian matrix. If we want to explicitly expand in the degrees of freedom, we can expand it in terms of the $\dA=\NC^2-1$ basis vectors (or \emph{generators}) $t_a$,
\begin{align}
    A_\mu(X)=A_\mu^a(X) t^a.\label{eq:definition-generators}
\end{align}
It enters the covariant derivative
\begin{align}
    D_\mu=\partial_\mu-igA_\mu(X),
\end{align}
describing how a vector rotates in color space when it is parallel transported.

The \emph{generators} $t_a$ fulfill the commutation relations
\begin{align}
    [t^a,t^b]=i f^{abc}t^c,
\end{align}
and can be represented as $\dR\times\dR$ matrices, where $\mathrm{R}$ labels different representations. For our purposes, the \emph{fundamental} ($\mathrm{R}=\mathrm{F}$) and \emph{adjoint} ($\mathrm{R}=\mathrm{A}$) representation are of importance. Their dimensions are
\begin{subequations}\label{eq:group-constants}
\begin{align}
    \dF=\NC, && \dA=\NC^2-1.
\end{align}
Furthermore, in every representation, there is the invariant object $t_at_a=\sum_a t_at_a=\CR\mathbb I$, where $\CR$ is called the \emph{quadratic Casimir}. Their values for different representations are given by
\begin{align}
    \CF = \frac{\NC^2-1}{2\NC}, && \CA=\NC.
\end{align}
\end{subequations}

The coupling $g$ often appears combined with $\NC$ as the 't Hooft coupling $\lambda$, or $\alpha_s$, 
\begin{align}
    \lambda=g^2\NC, && \alpha_s=\frac{g^2}{4\pi}.
\end{align}

The field strength tensor $F_{\mu\nu}$ is a generalization of the field strength tensor of classical electromagnetism, and is given by
\begin{align}
    F_{\mu\nu}=\frac{i}{g}[D_\mu,D_\nu]=\partial_\mu A_\nu-\partial_\nu A_\nu -ig[A_\mu,A_\nu].
\end{align}
\section{Correlation functions and closed time path\label{app:closed-timepath}}
In this subsection, we discuss aspects of quantum field theories out of equilibrium. The presentation here follows Ref.~\cite{Ghiglieri:2020dpq} but with the conventions from Ref.~\cite{Blaizot:2001nr}.
In a quantum field theory at finite temperature and out of equilibrium, it is useful to define several correlation functions. For a general quantum system, the expectation value of an operator $\mathcal O$ is given by
\begin{align}
    \langle \mathcal O\rangle=\Tr\left(\hat \rho \mathcal O\right),\label{eq:general-trace}
\end{align}
where $\hat \rho$ is the density matrix ($\hat\rho= e^{-\beta H}/Z$ in thermal equilibrium). If $\hat\rho(t_0)$ is known at a given time $t_0$ and the operator $\mathcal O$ has several time arguments but is not time ordered, e.g., $\mathcal O=\mathcal O(t_1)\mathcal O(t_2)$, suitable time translation operators $U(t_f,t_i)$ need to be inserted in \eqref{eq:general-trace}.
Representing them as a path integral is more tricky than for time-ordered operators since a path integral naturally introduces a time ordering along its path. For a product of operators of two different times $t_1\neq t_2$, this can be achieved effectively by introducing two sets of fields, one set living on a forward while the other on the backward time path.

To illustrate this, take the correlation function
\begin{align}
    \langle \mathcal O(t_1)\mathcal O(t_2)\rangle&=\Tr\left(\hat \rho(t_0)\mathcal O(t_1)\mathcal O(t_2)\right)\\
    &=\sum_{ijklmn}\bra{\phi_i}\hat \rho(t_0)\ket{\phi_j}\bra{\phi_j}U(t_0,t_1)\ket{\phi_k}\bra{\phi_k}\mathcal O(t_1)\ket{\phi_l}\bra{\phi_l}U(t_1,t_2)\ket{\phi_m}\nonumber\\
    &\qquad\qquad\times\bra{\phi_m}\mathcal O(t_2)\ket{\phi_n}\bra{\phi_n}U(t_2,t_0)\ket{\phi_i}.\label{eq:twopointfunction}
\end{align}
Next, note that the time evolution operator can be represented as a path integral
\begin{align}
    \bra{\phi_i}U(t_1,t_0)\ket{\phi_i}=\bra{\phi_i}e^{-i\hat H(t_1-t_0)}\ket{\phi_j}=\int_{\phi_1(t_0)=\phi_j}^{\phi_1(t_1)=\phi_i}\mathcal D\phi_1(t)e^{i S(\phi_1)},\label{eq:timeevolutionoperator}
\end{align}
representing the time evolution from initial time $t_0$ to $t_1$. It naturally satisfies the composition rule $U(t_0,t_1)U(t_1,t_2)=U(t_0,t_2)$. Eq.~\eqref{eq:twopointfunction} can then be written as a path integral (assuming $t_0<t_1<t_2$)
\begin{align}
\begin{split}
     \langle \mathcal O(t_1)\mathcal O(t_2)\rangle&=\sum_{ijklmn}\bra{\phi_i}\hat \rho(t_0)\ket{\phi_j}\int^{\phi_2(t_1)=\phi_k}_{\phi_2(t_0)=\phi_j}\mathcal D\phi_2(t)e^{-iS(\phi_2)}\bra{\phi_k}\mathcal O(t_1)\ket{\phi_l}\\
     &\quad \times \int^{\phi_2(t_2)=\phi_m}_{\phi_2(t_1)=\phi_l}\mathcal D\phi_2(t)e^{-iS(\phi_2)}\bra{\phi_m}\mathcal O(t_2)\ket{\phi_n}\\
     &\quad\times \int_{\phi_1(t_0)=\phi_i}^{\phi_1(t_2)=\phi_n}\mathcal D\phi_1(t)e^{iS(\phi_1)}\label{eq:twopointfunction-pathintegral}
     \end{split}
\end{align}
\begin{figure}
    \centering
    \centerline{
    \includegraphics[width=0.5\linewidth]{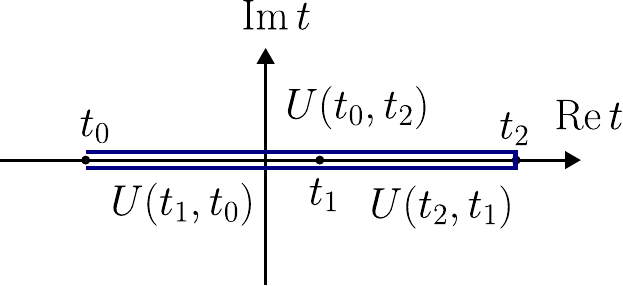}
    \includegraphics[width=0.5\linewidth]{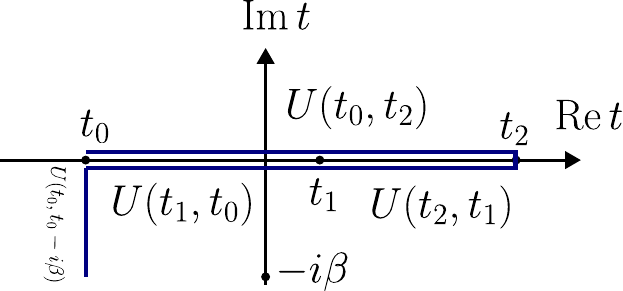}
    }
    \caption{The closed time path of the path integral \eqref{eq:twopointfunction-pathintegral}.}
    \label{fig:closed-timepath}
\end{figure}
and represented graphically in Fig.~\ref{fig:closed-timepath}.
Effectively, we first evolve forward in time to reach the maximum time extent $t_2$ (using the fields on the forward path $\phi_1$) and then backward to reach $t_1$ and back to $t_0$ using the fields on the backward path $\phi_2$. Going back to the initial time $t_0$ is a consequence of the trace in Eq.~\eqref{eq:general-trace}. Note that the action in the exponent of Eq.~\eqref{eq:twopointfunction-pathintegral} in the path integral involving the fields $\phi_2$ has a negative sign, which is a consequence of the backward time evolution.
For a thermal system, the density operator $\hat \rho=e^{-\beta H}/Z$ can be interpreted as a time evolution operator in imaginary time up to time extent $\beta=1/T$, together with a periodicity condition at the boundary $\phi_i(t_0)=\phi_i(t_0-i\beta)$. The thermal case is depicted in the left panel of Fig.~\ref{fig:closed-timepath}.

Considering the fields on the forward ($\phi_1$) and backward ($\phi_2$) to be different objects allows for a path integral representation of correlation functions, which are not time ordered. These types of correlation functions appear in thermal and nonthermal situations quite naturally. For instance, the Wightman functions are the (not time-ordered) correlators,
\begin{align}
    G^>(t_1,t_0)=\langle \phi(t_1)\phi(t_0)\rangle, && G^<(t_1,t_0)=\langle\phi(t_0)\phi(t_1)\rangle \label{eq:wightman}
\end{align}
They can be obtained more generally by introducing the generating functional using the closed path integral using the two fields $\phi_1$ and $\phi_2$ and corresponding sources,
\begin{align}
    Z[J_1,J_2]=\sum_{ij}\bra{\phi_i}\hat \rho\ket{\phi_j}\int_{\phi_i}^{\phi_j}\mathcal D\phi_1\mathcal D\phi_2 e^{iS(\phi_1)-iS(\phi_2)-\int\dd[4]{x}(J_1(x)\phi_1(x)-J_2(x)\phi_2(x))}.\label{eq:partition-function}
\end{align}
Using this, the various correlation functions can be obtained using functional derivatives,
\begin{align}
    \vb D_{ij}=\left.\frac{\delta}{\delta J_i}\frac{\delta }{\delta J_j}Z[J_1,J_2]\right|_{J=0}=\begin{pmatrix}
        \langle \phi_1\phi_1\rangle& \langle\phi_1\phi_2\rangle\\
        \langle\phi_2\phi_1\rangle& \langle\phi_2\phi_2\rangle
    \end{pmatrix}=\begin{pmatrix}
        G&G^<\\
        G^> &G^{\bar{\mathrm{F}}}
    \end{pmatrix},
\end{align}
where $G^>$ and $G^<$ are the Wightman functions \eqref{eq:wightman}, $D$ is the time-ordered (Feynman) propagator, 
\begin{align}
    G(t_1,t_0)=\Theta(t_1-t_0)G^>(t_1,t_0)+\Theta(t_0-t_1)G^<(t_1,t_0), \label{eq:Feynman-propagator-in-terms-of-wightman-functions}
\end{align}
and $G^{\bar{\mathrm{F}}}$ is the anti time-ordered (Dyson) propagator.

It is often useful to perform a basis transformation to the $r/a$ basis, which is obtained by the linear transformation
\begin{align}
    \phi_r=\frac{1}{2}(\phi_1+\phi_2), && \phi_a=\phi_1-\phi_2, && \phi_1=\phi_r+\frac{1}{2}\phi_a, && \phi_2=\phi_r-\frac{1}{2}\phi_a,\label{eq:12-ar-transformation}
\end{align}
in which the propagator matrix has the form
\begin{align}
    \vb G=\begin{pmatrix}
        \langle\phi_r\phi_r\rangle&\langle\phi_r\phi_a\rangle\\
        \langle\phi_a\phi_r\rangle & \langle\phi_a\phi_a\rangle
    \end{pmatrix}=
    \begin{pmatrix}
        G^{rr}& -iG^R\\
        -iG^A& 0
    \end{pmatrix},
\end{align}
where $G^{rr}=G^>+G^<$ is called the statistical two-point function or Hadamard propagator, and $D^R$ and $D^A$ are the retarded and advanced propagators\footnote{We follow here the conventions of \cite{Blaizot:2001nr}, which differ by a factor of $i$ from the one in Ref.~\cite{Ghiglieri:2020dpq} for the retarded and advanced propagator.},
\begin{align}
    G^R(t_1,t_0)=i\Theta(t_1-t_0)\rho(t_1,t_0), && G^A(t_1,t_0)=-i\Theta(t_0-t_1)\rho(t_1,t_0), \label{eq:retarded-advanced-propagator}
\end{align}
where the spectral function or Jordan propagator $\rho$ is given by 
\begin{align}
    \rho(t_1,t_0)=G^>-G^<=-i(G^R-G^A).
\end{align}
An important feature of the r/a basis is that the correlation function of two $a$ fields vanishes $\langle\phi_a\phi_a\rangle=0$, which simplifies calculations. Additionally, because of the step functions in \eqref{eq:retarded-advanced-propagator}, any closed loop containing only retarded or advanced propagators vanishes as well.

Physically, the retarded propagator represents causality flow, and we represent them pictorially by an arrow that points towards the r-field,
\begin{align}
G_{ra}(t_0,t_1)=G^R(t_0,t_1)=\raisebox{-0.5\height}{\begin{tikzpicture}
	\begin{feynman}
		\vertex[label=below: \(r\)] (a);
		\vertex[right=2cm of a, label=below: \(a\)] (b);
		\diagram* {
			(a) -- [anti fermion] (b),
		};
	\end{feynman}
\end{tikzpicture}}.
\end{align}
In this basis, the vertices can be obtained from Eq.~\eqref{eq:partition-function} by taking the difference of the interaction part of the action with $\phi_1$ fields and $\phi_2$ fields and using the transformation \eqref{eq:12-ar-transformation},
\begin{align}
    S_I(\phi_1)-S_I(\phi_2)=S_I\left(\phi_r+\phi_a/2\right)-S_I\left(\phi_r-\phi_a/2\right),
\end{align}
which implies that there are only vertices with an odd number of $a$ fields.

The propagators and correlation functions are often used in momentum space. More generally, if the system is not translational invariant (e.g., inhomogeneous), one may instead perform a Wigner transform and expand in gradients for slowly varying fields. This is sketched in Section \ref{sec:kinetic-theory-derivation} to obtain the Boltzmann equation from a quantum field theory.

The Wigner transform of a function $f(X,Y)$ is given by
\begin{align}
    \tilde f(K,\bar X)=\int\dd[4]{S}e^{-iK\cdot S}f\left(X(s,\bar X),Y(s,\bar X)\right),\label{eq:Wigner-trafo}
\end{align}
with the difference $S$ and central coordinates $X$ given by
\begin{align}
    S^\mu=X^\mu-Y^\mu, && \bar X^\mu=\frac{X^\mu+Y^\mu}{2}.
\end{align}
Note that the Wigner transform \eqref{eq:Wigner-trafo} reduces to an ordinary Fourier transform when the system is translational invariant (does not depend on $\bar X$).

A more detailed review of nonequilibrium quantum field theory can be found, e.g., in Refs.~\cite{Berges:2004yj, Calzetta_Hu_2008}.

\section{Perturbation theory and self-energy\label{sec:perturbation-theory-self-energy}}
In perturbation theory, one separates the action into an action for a free theory, which is typically analytically solvable, and an interaction part, which is suppressed by a small dimensionless parameter $g\ll 1$ (which for QCD is satisfied at high energies due to \emph{asymptotic freedom}. The full propagators are obtained from Eq.~\eqref{eq:general-trace} with the full action, while the free propagators (indicated by an additional index $0$) are obtained using only the free action. The interaction part gives rise to vertices, while the free theory part gives rise to the free propagators $G_0$.
Any quantities of interest are then expanded in terms of vertices and free propagators $G_0$ to obtain corrections proportional to the coupling $g$, $g^2$, \dots 
This is typically represented graphically as an expansion in loops.

An important concept is the concept of a self energy, which determines the difference between the full and free propagator,\footnote{Different conventions exist that differ in signs or whether or not to also put an imaginary unit $i$ there.}
\begin{align}
    G^{-1}(X,Y)=G_0^{-1}(X,Y)+\Pi(X,Y),\label{eq:self-energy-schematically}
\end{align}
where the inverse propagators are defined via
\begin{align}
    \int\dd[4]{Z} G^{-1}(X,Z)G(Z,Y)=\delta ^4 (X-Y), && \int\dd[4]{Z} G_0^{-1}(X,Z)G_0(Z,Y)=\delta ^4(X-Y).
\end{align}

Similar to the Wightman functions and their relation to the propagator in Eq.~\eqref{eq:Feynman-propagator-in-terms-of-wightman-functions}, one may also decompose the self-energy into
\begin{align}
    \Pi(t_1,t_0)=\Theta(t_1-t_0)\Pi^>(t_1,t_0)+\Theta(t_0-t_1)\Pi^<(t_1,t_0)-i\delta(t_1-t_0)\Pi^\delta(t_1),\label{eq:self-energy-decomposition}
\end{align}
where we have included the possibility of a singular term $\Pi^\delta$, see, e.g., \cite{Blaizot:2001nr}.
Similar as with the propagators \eqref{eq:retarded-advanced-propagator}, we may introduce a retarded self-energy
\begin{align}
    \Pi^R(t_1,t_0)=-i\Theta(t_1-t_0)\Gamma(t_1,t_0), && \Pi^A(t_1,t_0)=i\Theta(t_0-t_1)\Gamma(t_1,t_0), \label{eq:retarded-advanced-self-energy}
\end{align}
with 
\begin{align}
    \Gamma = -(\Pi^>-\Pi^<).
\end{align}

\section{Correlator relations and properties \label{app:correlator-relations}}
To summarize, the different propagators and correlators are given by
\begin{subequations}\label{eq:propagators}
\begin{align}
    G^>{}_{\mu\nu}^{ab}(X,Y)&=\langle A^a_\mu(X)A_\nu^b(Y)\rangle,\label{eq:wightman-correlator-gluons}\\
    G^<{}_{\mu\nu}^{ab}(X,Y)&=\langle A^b_\nu(Y)A_\mu^a(X)\rangle = G^>{}_{\nu\mu}^{ba}(Y,X),\\
    G{}_{\mu\nu}^{ab}(X,Y)&=\Theta(X^0-Y^0)G^>{}^{ab}_{\mu\nu}(X,Y)+\Theta(Y^0-X^0)G^<{}^{ab}_{\mu\nu}(X,Y)\label{eq:wightman-functions-decomposition}\\
    \rho^{ab}_{\mu\nu}(X,Y)&=\langle [A^a_\mu(X),A_\nu^b(Y)]\rangle
    \\
    G^R{}^{ab}_{\mu\nu}(X,Y)&=i\Theta(X^0-Y^0)\rho^{ab}_{\mu\nu}(X,Y),\\
    G^A{}^{ab}_{\mu\nu}(X,Y)&=-i\Theta(Y^0-X^0)\rho^{ab}_{\mu\nu}(X,Y) = G^R{}^{ba}_{\nu\mu}(Y,X)
\end{align}
Here, $\Theta(x)$ is the usual step function,
\begin{align}
    \Theta(x)=\begin{cases}
        1, & x \geq 0,\\
        0, & x < 0
    \end{cases}
\end{align}
\end{subequations}
Several useful properties of these correlators and the self-energies are listed in the following:
\begin{align}
    \left(G^{\substack{>\\<}}(Y,X)\right)^*&=G^{\substack{>\\<}}(X,Y)\\
    G^A(X,Y)&=G^R(Y,X), & \Pi^R(X,Y)&=\Pi^A(Y,X)\\
    \left(\tilde G^A(K,X)\right)^*&=\tilde G^R(K,X)\\
    \tilde \rho(K,X)&=2\IM \tilde G^R(K,X), & \tilde \Gamma(K,X)&=-2\IM \tilde\Pi^R(K,X) \label{eq:spectral-function-retarded-propagator-imag}\\
    \tilde G^{\substack{>\\<}}(-K,X)&=\tilde G^{\substack{<\\>}}(K,X)\label{eq:wightmanfunction-negativek}\\
    \tilde\rho(-K,X)&=-\tilde\rho(K,X)\label{eq:rho-antisymmetric}\\
    G^R(-K,X)&=G^A(K,X) \label{eq:retarded-propagator-with-negative-momentum}\\
    \rho=G^>-G^<&=-i(G^R-G^A), & -\Gamma = \Pi^>-\Pi^<&=-i(\Pi^R-\Pi^A)
\end{align}

The definitions for the correlation functions \eqref{eq:propagators} hold generically for any bosonic fields. For fermions, additional minus signs have to be considered, and the commutators are replaced by anticommutators.

\section{Hard thermal loops\label{app:htl}}
It is well-known in thermal field theory that to cure infrared problems, the propagators for the soft modes have to be resummed \cite{Bellac:2011kqa, Laine:2016hma, Kapusta:2006pm}.
At a fixed order in the coupling $g$, it turns out that more diagrams contribute than what would be expected from a na\"ive loop expansion. In fact, for diagrams with soft external momenta, it is needed to integrate out the hard loop momenta, which is formalized in the hard thermal loop (HTL) theory \cite{Braaten:1989mz, Braaten:1990az, Frenkel:1989br}.

Parts of the following discussion are based on Appendix B in \cite{Boguslavski:2023waw}.

\subsection{Hard thermal loop resummed gluon propagator}
The HTL retarded propagator in strict Coulomb gauge\footnote{By Coulomb gauge we mean using $\partial_i A^i$ as the gauge function and by strict we mean enforcing it strictly, i.e. $\partial_iA^i=0$, which amounts to setting $\xi=0$ in the Faddeev-Popov procedure \cite{Bellac:2011kqa}.} is given by\footnote{We take the explicit expression from Ref.~\cite{Blaizot:2001nr} (BI), but write it in the form of \cite{Ghiglieri:2020dpq} (GKSV). As mentioned before, different authors use different conventions of factors of $\pm i$. The conventions used by these authors are related by $G_R^{\mathrm{BI}}=iG_R^{\mathrm{GKSV}}$ for the retarded (and advanced) propagators.} \cite{Blaizot:2001nr, Ghiglieri:2020dpq}
\begin{subequations}\label{eq:HTL-propagators}
\begin{align}
    \GLR(Q)&=\frac{-1}{q^2+\PiLR(\omega/q)},\\
    G^{ij}_R(Q)=\left(\delta^{ij}-\frac{q^iq^j}{q^2}\right)\GTR(Q)&=\frac{\delta^{ij}-\frac{q^iq^j}{q^2}}{q^2-\omega^2+\PiTR(\omega/q)}
\end{align}
\end{subequations}
with $x=\omega/q$ and the self-energies
\begin{subequations}\label{eq:htl-selfenergies}
\begin{align}
    \RE\PiLR(x)&=m_D^2\left(1-\frac{x}{2}\ln\left|\frac{x+1}{x-1}\right|\right), &
    \IM\PiLR(x)&=\frac{xm_D^2\pi}{2}\Theta(1-|x|)\\
    \RE \PiTR(x)&=\frac{m_D^2}{2}-\frac{1}{2}(1-x^2)\RE\PiLR, &
    \IM\PiTR(x)&=-\frac{1}{2}(1-x^2)\IM\PiLR.
\end{align}
\end{subequations}
The Debye mass can be obtained via the integral \eqref{eq:debyemass-general} and in equilibrium has the value \eqref{eq:equilibriumform-debyemass-tstar-meffs},
\begin{align}
    m_D^2=g^2T^2\left(\frac{\NC}{3}+\frac{\nf}{6}\right).
\end{align}
For later use, let us also list here the spectral function $\rho(Q)=2\IM G^R(Q)$ (see Eq.~\eqref{eq:spectral-function-retarded-propagator-imag}),
\begin{subequations}\label{eq:HTL-spectral-functions}
\begin{align}
    \rho^{00}(Q)&=\frac{2\IM \Pi^{00}(Q)}{(q^2+\RE\Pi^{00}(Q))^2+(\IM\Pi^{00}(Q))^2}, \\
    \rho^T(Q)&=\frac{-2\IM \Pi^T(Q)}{(q^2-\omega^2+\RE\Pi^T(Q))^2+(\IM\Pi^T(Q))^2}.
\end{align}
\end{subequations}
In the next subsections, we will use the propagators to obtain explicitly the isoHTL screened matrix element used for evaluating the jet quenching parameter and the elastic collision term. Additionally, we will discuss how to perform a certain integral over these propagators, where the spectral function will appear.

\subsection{IsoHTL screening in the jet quenching parameter}
\label{app:full_htl_matrix_el}
Here, we want to explicitly derive the expression (including all the kinematic contractions) for the full isotropic HTL matrix element \eqref{eq:full_htl_matrix_element} needed for the jet quenching parameter $\qhat$ in Chapter \ref{sec:momentum-broadening-of-jets}.
We start with the AMY screening prescription Eq.~\eqref{eq:prescription_screening} (see also Eq.~\eqref{eq:amy_replacement}), 
\begin{align}
	\Mhtl = \left|G_R(P-P')_{\mu\nu}(P+P')^\mu (K+K')^\nu\right|^2, \label{eq:prescription_screening_HTL}
\end{align}
Due to the kinematic constraint $|\omega| < q$ (see \eqref{eq:phase_space_relation_omega_q_k}), the variable $x$ appearing in the self-energies \eqref{eq:htl-selfenergies} is always $|x|<1$, and thus the imaginary parts of the self energies \eqref{eq:htl-selfenergies} always contribute. Note that  $G_R(-Q)$ corresponds to the advanced propagator (see Eq.~\eqref{eq:retarded-propagator-with-negative-momentum}), which has a different imaginary part in the self-energy, $\IM\Pi_R(-Q)=-\IM\Pi_R(Q)$.
Let us further abbreviate 
\begin{align}
    &\GLR(-Q)=: z_L=\frac{-1}{A+Bi}, && \GTR(-Q)=:z_T=\frac{1}{C+Di},\\
    &A = q^2+\RE\PiLR(x), && B =\IM\PiLR(-x),\\
    &C = q ^2-\omega^2+\RE\PiTR(x), && D =  \IM\PiTR(-x).
\end{align}
It will turn out that $B$ and $D$ only appear quadratically or as a product, i.e., we can consider the internal propagator to have momentum $Q$ or $(-Q)$. Thus, we do not need to distinguish them from $\IM\Pi_R(x)$. 
We can now split the retarded propagator in \eqref{eq:prescription_screening_HTL} into its temporal and spatial parts and use the expressions for $\vb p$, $\vb q$, and $\vb k$ in the $q$-frame, i.e., using their parametrizations \eqref{eq:qframe_p}, \eqref{eq:qframe_q} and \eqref{eq:qframe_k},
\begin{align}
    Q&=P'-P=\begin{pmatrix}
        \omega\\0\\0\\q
    \end{pmatrix}, &
    P&=p\begin{pmatrix}
        1\\\sin\thetaqp\\0 \\ \cos\thetaqp  
    \end{pmatrix}, &
    K&=k\begin{pmatrix}
        1\\ \sin\thetaqk\cos\phiqk \\ \sin\thetaqk \sin\phiqk \\ \cos\thetaqk
    \end{pmatrix}.\label{eq:external_momenta-qhat}
\end{align}

to obtain 
\begin{align}
\begin{split}
    \Mhtl&=\left|c_1z_L+c_2z_T\right|^2=c_1^2|z_L|^2+c_2^2|z_T|^2+c_1c_2(z_L\bar z_T+\bar z_L z_T),
\end{split}
\end{align}
where $\bar z$ means taking the complex conjugate of $z$ and $c_1=(2p+\omega)(2k-\omega)$ and $c_2=4pk\sin\thetaqp\sin\thetaqk\cos\phikq$.

This leads to $|z_L|^2=|\GLR(Q)|^2=(A^2+B^2)^{-1}$, $|z_T|^2=|\GTR(Q)|^2=(C^2+D^2)^{-1}$ and
\begin{align}
	\bar z_L z_T+ z_L\bar z_T&=-2(AC+BD)|z_L|^2|z_T|^2, 
\end{align}
and eventually we obtain
\begin{align}
 \Mhtl=\frac{c_1^2}{A^2+B^2}+\frac{c_2^2}{C^2+D^2}-\frac{2c_1c_2(AC+BD)}{(A^2+B^2)(C^2+D^2)}.\label{eq:full_htl_matrix_element_APP_qhat_finitep}
\end{align}
The last term is proportional to $\cos\phikq$ and may, therefore, be dropped for isotropic distributions $f(k)$.

The rescaled matrix element $\tilde M = \lim_{p\to\infty}\Mhtl/p^2$ in the limit $p\to\infty$ can be obtained easily by scaling out $p$ 
(see \eq \eqref{eq:c_parameters_for_pinf_matrix_element_HTL})
\begin{align}
    \tilde c_1 = \lim_{p\to\infty}c_1/p=2(2k-\omega),&&
    \tilde c_2 = \lim_{p\to\infty}c_2/p=4k\sin\thetaqp\sin\thetaqk\cos\phikq, 
\end{align}
which yields
\begin{align}
    \label{eq:full_htl_matrix_element_APP}    
    \tildeMhtl=\frac{\tilde c_1^2}{A^2+B^2}+\frac{\tilde c_2^2}{C^2+D^2}-\frac{2\tilde c_1\tilde c_2(AC+BD)}{(A^2+B^2)(C^2+D^2)}.
\end{align}
Similarly, as before, for isotropic distributions, $f(k)$, the last term does not contribute and may be dropped.

\subsection{IsoHTL screening in the elastic collision kernel\label{app:isoHTL-screening-for-c22}}
In Chapter \ref{sec:improving-qcd-simulations}, we consider isoHTL screening in the elastic collision term $\Ctwotwo$. The difference to the previous Section \ref{app:full_htl_matrix_el} is that Eq.~\eqref{eq:external_momenta-qhat} is replaced by \eqref{eq:external_momenta}, in which the momentum $\vb p$ is parametrized differently, including an additional angle $\phi_{qp}$. This is because for $\qhat$, as explained in Section \ref{sec:coordinate_systems}, the incoming parton specifies a specific direction $\vb p$, and one performs the $\vb q$ integral in a frame, in which $\vb p$ points in the $z$ direction and the $\vb k$ integration in a frame in which $\vb q$ points in the $z$ direction.
For QCD kinetic theory simulations, however, one needs to solve the more general integration measure \eqref{eq:integration-measure}, performing the $\vb q$ integration in the ``lab frame'' (in which also the distribution function is stored), and both the $\vb p$ and $\vb k$ integrations in the frame in which $\vb q$ points in the $z$ direction. This leads to the different parameterizations \eqref{eq:external_momenta-qhat} and \eqref{eq:external_momenta}.

Effectively, the only change is in the coefficient $c_2$ in \eqref{eq:full_htl_matrix_element_APP_qhat_finitep}, with 
\begin{align}
    c_2=4pk\sin\thetaqp\sin\thetaqk\cos(\phiqk-\phiqp).
\end{align}

\subsection{Sum rule\label{app:sum-rule}}

In this appendix, we show that one can analytically perform the integral over $\omega$ over the HTL matrix element \eqref{eq:full_htl_matrix_element_APP},
\begin{align}
    \int_{-\infty}^\infty\frac{\dd{\omega}}{q}\tildeMhtl,
\end{align}
using the sum rule from \re \cite{Aurenche:2002pd}. This sum rule simplifies the evaluation of an integral over a ``spectral function''
\begin{align}
\begin{split}
    \int_0^1\frac{\dd{x}}{x}\frac{2\IM\Pi(x)}{(z+\RE\Pi(x))^2+(\IM \Pi(x))^2}
    =\pi\left[\frac{1}{z+\RE\Pi(\infty)}-\frac{1}{z+\RE\Pi(0)}\right],
    \end{split}\label{eq:AGZ-sumrule}
\end{align}
provided that the function $\Pi(x)$ fulfills the conditions $\IM\Pi(0)=0$, $\IM\Pi(x)=0$ for $x\geq 1$ and $\RE\Pi(x)\geq 0 $ for $x\geq 1$.

Let us now calculate the collision kernel $C(\vb q_\perp)$ for small $q_\perp \ll k\ll p$. Thus, effectively, we can take the formalism developed in Section \ref{sec:momentum-broadening-of-jets} for the jet quenching parameter $\qhat$ (related to the collision kernel via Eq.~\eqref{eq:intro-qhat-definition}), and employ the approximations $q_\perp \ll k \ll p$. Effectively, we can thus work in the limit $p\to \infty$ discussed in Section \ref{sec:pinf_formula}. We consider isotropic systems and may thus drop the last term in the isoHTL screened matrix element \eqref{eq:full_htl_matrix_element}. Recall that in this limit $\cos\thetaqp=\cos\thetaqk=\omega/q$, and then \eqref{eq:full_htl_matrix_element} reduces to

\begin{align}
    \langle\tildeMhtl\rangle_\phi=\int_0^{2\pi}\frac{\dd\phiqk}{2\pi}\tildeMhtl &= 16k^2\left(\left|\GLR\right|^2+\frac{1}{2}\left(1-\frac{\omega^2}{q^2}\right)^2\left|\GTR\right|^2\right),
\end{align}
where we have inserted back the more physical quantities $|G^{00}(Q)|^2=(A^2+B^2)^{-1}$ and $|G^T(Q)|^2=(C^2+D^2)^{-1}$, and already integrated out the angle $\phiqk$. Importantly, the collision kernel then reads (see Eq.~\eqref{eq:qhat_factorized})
\begin{align}
    C(\vb q_\perp)=\left(\CR\sum_{\pm}\Xi_{\pm}\frac{g^4}{2^8\pi^4}\int_0^\infty \dd{k} k^2f_\pm(k)(1\pm f_\pm(k))\right)\int_0^{2\pi}\dd{\phiqk}\int_{-\infty}^\infty\frac{\dd\omega \tilde M_{\mathrm{HTL}}}{k^2q},
\end{align}
where, importantly, we have factored out the $k$-dependence and the distribution functions. As we will see, the integral over $\omega$ can now be performed using the sum rule \eqref{eq:AGZ-sumrule},

for which we write $|G_R|^2$ in terms of the self-energy $\Pi_R$, and expand the fraction with the imaginary part of the self-energy,
\begin{subequations}
\begin{align}
    \left|\GLR\right|^2&=\frac{2q}{\omega m_D^2\pi}\frac{\IM\PiLR(x)}{\left(q^2+\RE\PiLR\right)^2+\left(\IM\PiLR\right)^2}\\
    \left|\GTR\right|^2&=-\frac{4q}{\omega m_D^2\pi(1-x^2)}\frac{\IM\PiTR(x)}{\left(\qperp^2+\RE\PiTR\right)^2+\left(\IM\PiTR\right)^2},
\end{align}
\end{subequations}
where we can immediately recognise the spectral functions \eqref{eq:HTL-spectral-functions},
\begin{align}
    \langle\tildeMhtl\rangle_\phi= \frac{16k^2 q}{\omega m_D^2\pi}\left(\rho^{00}(x)+\rho^{33}(x)\right)=\frac{16k^2 q}{\omega m_D^2\pi}\left(\rho^{00}(x)+(1-x^2)\rho^T(x)\right).\label{eq:sumrule-spectral-density}
\end{align}
A similar trick is also used in \cite{Ghiglieri:2015ala}.

Together with 
the substitution $\frac{\dd{\omega}}{\omega}=\frac{\dd{x}}{x(1-x^2)}$, this results in
\begin{align}
    &\int_{-\infty}^\infty\frac{\dd{\omega}}{q} \langle\tildeMhtl\rangle_\phi =2\int_{0}^\infty\frac{\dd{\omega}}{q} \langle\tildeMhtl\rangle_\phi 
    = \frac{32 k^2}{m_D^2}\left[\frac{1}{\qperp^2+\frac{m_D^2}{3}}-\frac{1}{\qperp^2+m_D^2}-\frac{1}{\qperp^2+\frac{m_D^2}{3}}+\frac{1}{\qperp^2}\right]\label{eq:sumrule-result}.
\end{align}
For the longitudinal propagator $|\GLR|^2$ the factor $(1-x^2)$ from the coordinate transformation needs to be absorbed into the self-energy $\tildePiLR(x)=(1-x^2)\PiLR(x)$.
The relevant limits read
\begin{subequations}
\begin{align}
    \RE\PiTR(0)&=0,&\RE\tildePiLR(0)&=m_D^2,\\
    \RE\PiTR(\infty)&=\frac{m_D^2}{3},&\RE\tildePiLR(\infty)&=\frac{m_D^2}{3}.
\end{align}
\end{subequations}
This leads to the familiar result
\begin{align}
\frac{m_D^2}{32k^2}\int_0^{2\pi}\frac{\dd{\phikq}}{2\pi}\int_{-\infty}^\infty\frac{\dd{\omega}}{q}\tildeMhtl = \frac{m_D^2}{\qperp^2(\qperp^2+m_D^2)}
    \label{eq:sumrule_result_final}
\end{align}
A similar result is obtained in \cite{Aurenche:2002pd, Caron-Huot:2008zna} in thermal equilibrium, where also the integral over the distribution functions $f(k)(1+f(k))$ is automatically included. Here, we have shown explicitly that the matrix element itself gives rise to the form \eqref{eq:sumrule_result_final}. In the soft limit, this enables us to perform the integral over the distribution function separately, allowing a straightforward generalization to nonequilibrium systems.
Finally, we note that this sum rule is used in \se \ref{sec:qhat_screening} to set the parameter $\xiscreenperp$ in the Debye-like screening prescription.

An interested reader might wonder what this has to do with the Wightman function appearing in the integral equation \eqref{eq:amy-integralequation-long}.
Another interested reader might wonder if the approach taken here was perhaps overly complicated. Perhaps, using a different approach, the spectral function \eqref{eq:spectral-function-retarded-propagator-imag} would appear more naturally. Indeed, both of these questions are answered by the following short discussion following the introduction of Ref.~\cite{Hauksson:2021okc}.

Let us consider a quark moving through the plasma with large momentum $K^\mu=(k,\vb k)$, which has the decay rate \cite{Bellac:2011kqa} (or scattering rate)
\begin{align}
    \Gamma=\frac{1}{4k}\Tr[\gamma^\mu K_\mu \Pi^>(K)]
\end{align}

\begin{figure}
    \centering
    \includegraphics[width=0.5\linewidth]{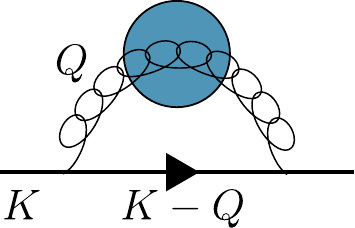}
    \caption{Self-energy of an energetic quark, representing Eq.~\eqref{eq:self-energy-fermion-appendix}.}
    \label{fig:self-energy-fermion-appendix}
\end{figure}

Calculating the self-energy via (see Fig.~\ref{fig:self-energy-fermion-appendix})
\begin{align}
    \Pi^>(K)=g^2\CF\int\frac{\dd[4]{Q}}{(2\pi)^4}G_{\mathrm{gluon}}^{>,\mu\nu}(Q)\gamma_\mu \gamma_\nu  G_{\mathrm{quark}}^>(K-Q),\label{eq:self-energy-fermion-appendix}
\end{align}
where we used the vertex factors $(ig)\times (-ig)=g^2$, and have the additional minus sign for the $2$ vertex. We may now use Eq.~\eqref{eq:kadanoff-baym-ansatz} to write the Wightman functions $G^>(K)=\rho(K)(1\pm N(K))$ as spectral density $\rho$ and off-shell distribution function $N$. For the quark, since it is highly energetic $K\gg Q$, we may use the quasiparticle approximation, representing the spectral function as a delta function \cite{Bellac:2011kqa}

\begin{align}
    \rho_{\mathrm{quark}}(P)=2\pi\sign(P^0)\gamma^\mu P_\mu\delta(P^2).
\end{align}
With the trace identity
\begin{align}
    K_\alpha (K-Q)_\beta \Tr[\gamma^\alpha\gamma^\mu\gamma^\beta\gamma^\nu]=4\left(K^\mu (K-Q)^\nu + K^\nu (K-Q)^\mu - K\cdot (K-Q)\eta^{\mu\nu}\right)\approx 8K^\mu K^\nu-4K^2 \eta^{\mu\nu},
\end{align}
where we used the symmetry of the propagator $G^>{}^{\mu\nu}$, we obtain
\begin{align}
    \Gamma=\frac{\pi}{2k}g^2\CF\int\frac{\dd[4]{Q}}{(2\pi)^4}G_{\mathrm{gluon}}^>{}^{\mu\nu}(Q)\left(8K_\mu K_\nu-4K^2\eta_{\mu\nu}\right) \sign(K^0-Q^0)\delta\left((K-Q)^2\right)(1-N(K-Q)).
\end{align}
The delta function sets the distribution function on-shell $N(K-Q)=f(\vb k-\vb q)$ (see Eq.~\eqref{eq:distributionfunction-onshell}), and setting the self-energy on-shell (as in Eq.~\eqref{eq:self-energy-onshell}), we obtain
\begin{align}
    \Gamma=2\pi g^2\CF\int\frac{\dd[4]{Q}}{(2\pi)^4}G^{>\mu\nu}_{\mathrm{gluon}}(Q)v_\mu v_\nu \delta(v\cdot Q)(1-f(\vb k-\vb q)),
\end{align}
where we have neglected $Q^2\ll 2K\cdot Q$.
Since the scale separation $k\gg \text{medium scale} \gg q$, we can set $f(\vb k-\vb q)\to 0$. Therefore, the differential scattering rate is given by
\begin{align}
    \frac{\dd{\Gamma}}{\dd[4]{Q}}=\frac{2\pi g^2\CF}{(2\pi)^4} G^{>\mu\nu}_{\mathrm{gluon}}(Q)v_\mu v_\nu \delta (v\cdot Q)
\end{align}
This expression is exactly what enters in the AMY integral equation \eqref{eq:amy-integralequation-long}. If we take the particle to be moving in the $z$ direction, we have $v^\mu=(1,0,0,1)$, and the delta function enforces $Q^0=q_z$. We may now use the relation of the Wightman function to the spectral function \eqref{eq:kadanoff-baym-ansatz},
\begin{align}
    G_{\mathrm{gluon}}^{>,\mu\nu}(Q)=\rho^{\mu\nu}(Q)\left(1+f(Q^0)\right)\approx \frac{T_\ast}{Q^0}\rho^{\mu\nu}(Q),
\end{align}
where we have approximated the distribution function for very soft momenta using the infrared temperature $T_\ast$. Contracting this with $v^\mu v^\nu$, we obtain 
\begin{align}
    \frac{\dd{\Gamma}}{\dd[2]{\vb q_\perp}}=\frac{g^2\CF T_\ast}{(2\pi)^3}\int \frac{\dd \omega}{\omega}\left(\rho^{00}+\rho^{33}\right)=\frac{\CF g^2 T_\ast }{(2\pi)^2}\frac{m_D^2}{q_\perp^2(q_\perp^2+m_D^2)},
\end{align}
where we used the sum rule result \eqref{eq:sumrule-result}.
This is exactly, up to prefactors, the collision kernel
\begin{align}
    C(\vb q_\perp)=(2\pi)^2\frac{\dd\Gamma}{\dd[2]{\vb q_\perp}}=g^2\CF \int\frac{\dd Q^0\dd Q^z}{(2\pi)^2}2\pi\delta (v\cdot Q)G^{>\mu\nu}_{\mathrm{gluon}}(Q)v_\mu v_\nu\overset{\substack{\text{isotropic}\\\text{soft}}}{=}\frac{\CF g^2T_\ast m_D^2}{\qperp^2(\qperp^2+m_D^2)}\label{eq:isotropic-system-collision-kernel}
\end{align}
Thus, we have seen that the collision kernel can be represented via a Wightman function, and how the spectral function appears.

\subsection{An analytic results for the dipole cross section\label{app:analytic-result-cross-section}}
Here we show how to obtain the dipole cross section \eqref{eq:intro-dipole-crosssection-fouriertrafo},
\begin{align}
    C(\vb x)=\int\frac{\dd[2]{\vb q_\perp}}{(2\pi)^2}C(\vb q_\perp)\left(1-e^{i\vb x\cdot\vb q_\perp}\right), \label{eq:intro-dipole-crosssection-fouriertrafo-appendix}
\end{align}
using the small $q_\perp$ form of the collision kernel for an isotropic system \eqref{eq:isotropic-system-collision-kernel},
\begin{align}
    C(\vb q_\perp)=\frac{\CR g^2T_\ast m_D^2}{\qperp^2(\qperp^2+m_D^2)}.
\end{align}
First, the $\vb q_\perp$ integral is split into modulus and angle, and the angular integral can be performed to yield the Bessel function $J_0$,
\begin{align}
    C(\vb x)&=\frac{\CR g^2T_\ast}{(2\pi)^2}\int_0^\infty\dd{q_\perp}q_\perp\int_0^{2\pi}\dd{\phi}\left(1-e^{i q |\vb x|\cos\phi}\right)\frac{m_D^2}{q_\perp^2(q_\perp^2+m_D^2)}\\
    &=\CR g^2T_\ast \int_0^\infty\frac{\dd{q_\perp}}{2\pi}(1-J_0(|\vb x| q_\perp))\frac{m_D^2}{q_\perp(q_\perp^2+m_D^2)}.
\end{align}
Using $\dv[J_0(|\vb x|q)]{q}=-|\vb x| J_1(|\vb x| q)$ and $\int\dd{q}\frac{m_D^2}{q(q^2+m_D^2)}=\frac{1}{2}\log\frac{q^2}{q^2+m_D^2}$, we obtain
by partial integration (the boundary terms vanish because $J_0(0)=1$)
\begin{align}
    C(\vb x)&=-\CR g^2T_\ast\frac{|\vb x|}{2}\int_0^\infty\frac{\dd{q_\perp}}{2\pi}J_1(|\vb x|q_\perp)\log\frac{q_\perp^2}{q_\perp^2+m_D^2}\\
    &=\frac{\CR g^2T_\ast}{2\pi}\left(\gamma_E+K_0(|\vb x|m_D)+\log\frac{|\vb x| m_D}{2}\right),\label{eq:analytic-result-dipole-cross-section-appendix}
\end{align}
where $\gamma_E$ is the Euler-Mascheroni constant, and $K_0$ is the modified Bessel function of the second kind. This expression can be expanded for small $|\vb x|$,
\begin{align}
    C(\vb x)=\frac{\CR g^2 T_\ast m_D^2}{8\pi}\vb x^2\left(1-\gamma_E-\log\frac{|\vb x|m_D}{2}\right)+\mathcal O(\vb x^4, \vb xx^4\log|\vb x|).
\end{align}
It should be noted that Eq.~\eqref{eq:analytic-result-dipole-cross-section-appendix} is not the correct form of the dipole cross section in thermal equilibrium, because we started off with only the small $q_\perp$ form of the collision kernel.

\section{Scattering with soft gluon exchange\label{app:scattering-soft-momentum-exchange}}
\begin{figure}
	\centering
	\includegraphics[width=0.7\linewidth]{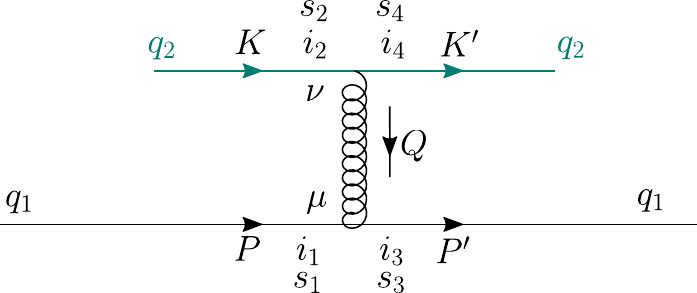}
	\caption{Feynman diagram of a quark with momentum $q_1$ scattering off another quark ($q_2$) with two distinct quark flavors. The color, spin, and Lorentz indices are shown explicitly for illustration. Figure from \cite{Boguslavski:2024kbd}.}
	\label{fig:quarkscattering}
\end{figure}
In this Appendix, we verify that quark or gluon scatterings with soft-gluon exchange are equivalent to scalar scatterings as in \eq \eqref{eq:scalar_quark_result} up to $\mathcal O(Q/P)$. Or, put differently, that for soft-gluon exchange the matrix elements are at leading order independent of the spin of the hard particles.

This presentation follows Ref.~\cite{Boguslavski:2024kbd}.

\subsection{Quark scattering}
First, we calculate the square of the matrix element for quark scattering with a general internal gluon propagator.
The amplitude for quark scattering of two different flavors with incoming momenta $P$ and $K$ (spin states $s_1$, $s_2$ and colors $i_1$, $i_2$) to the outgoing momenta $P'$ and $K'$ (spin states $s_3$ and $s_4$, colors $i_3$ and $i_4$), as depicted in \fig\ref{fig:quarkscattering}, is given by%
\footnote{The Feynman rules can be found in any standard QFT textbook, e.g. see \cite{Srednicki:2007qs} for the mostly plus metric convention.}
\begin{align}
	i\mathcal M^{s_1i_1,s_2i_2}_{s_3i_3,s_4i_4} &=
	-g^2\left(\bar u^{s_3}(P')\gamma^\mu u^{s_1}(P)\right)\left(\bar u^{s_4}(K')\gamma^\nu u^{s_2}(K)\right)  \left(-iG_{\mu\nu}^{ab}(Q)\right)t^a_{i_3i_1}t^b_{i_4i_2}\,.
\end{align}
To calculate the square $\left|\mathcal M\right|^2 = \sum_{s_j, i_j} \left(\mathcal M^{s_1i_1,s_2i_2}_{s_3i_3,s_4i_4}\right)^\ast \mathcal M^{s_1i_1,s_2i_2}_{s_3i_3,s_4i_4}$, we need to sum over the color indices and use 
\begin{align} \left(t^a_{i_3i_1}t^b_{i_4i_2}\right)^\ast t^c_{i_3i_1}t^d_{i_4i_2}=\Tr(t^at^c)\Tr(t^bt^d)=n_F^2\delta^{ac}\delta^{bd},
\end{align}
where $n_F=1/2=\CF \NC/\dA$ is the index of the fundamental representation of $\mathrm{SU}(\NC)$. Summing over all spins, we obtain
\begin{align}
	\begin{split}
		\left|\mathcal M\right|^2 &=\sum_{s_1s_2s_3s_4} n_F^2\delta^{ac}\delta^{bd}g^4G_{\mu\nu}^{ab}(Q)\left(G_{\rho\sigma}^{cd}(Q)\right)^\ast \\
		&\times \left(\bar u^{s_3}(P')\gamma^\mu u^{s_1}(P)\right)\left(\bar u^{s_1}(P)\gamma^\rho u^{s_3}(P')\right)\\
		&\times \left(\bar u^{s_4}(K')\gamma^\nu u^{s_2}(K)\right)\left(\bar u^{s_2}(K)\gamma^\sigma u^{s_4}(K')\right)
	\end{split}
\end{align}
Using the identity $\sum_{s_1}u_i^{s_1}(P)\bar u^{s_1}_j(P)=-P_\mu(\gamma^\mu)_{ij}=-\slashed{P}_{da}$, we obtain
\begin{align}
	\left|\mathcal M\right|^2
	&=n_F^2g^4G^{ab}_{\mu\nu}(Q)\left(G^{ab}_{\rho\sigma}(Q)\right)^\ast\Tr(\gamma^\mu\slashed{P}\gamma^\rho\slashed{P'})\Tr(\gamma^\nu\slashed{K}\gamma^\sigma\slashed{K'})\nonumber.
\end{align}
With
\begin{align}
	\Tr(\slashed{P}\gamma^\mu\slashed{P'}\gamma^\nu)=4(P^\mu P'{}^\nu + P^\nu P'{}^\mu - g^{\mu\nu}P\cdot P'),
\end{align}
we finally arrive at
\begin{align}
	\begin{split}
		\left|\mathcal M\right|^2&=16 n_F^2g^4 G^{ab}_{\mu\nu}(Q)\left(G^{ab}_{\rho\sigma}\right)^\ast(Q)
		\left[P^\mu P'{}^\rho + P^\rho P'{}^\mu - g^{\mu\rho}P\cdot P'\right]
		\left[K^\nu K'{}^\sigma + K^\sigma K'{}^\nu - g^{\nu\sigma}K\cdot K'\right].
		\label{eq:quark_scattering_general_matrix_element}
	\end{split}
\end{align}
This expression is independent of the precise form of the gluon propagator $G$. 
Note that all terms in the propagator $G_{\mu\nu}$ proportional to $Q_\mu$ or $Q_\nu$ (which we argue in \se\ref{sec:isoHTLgeneral} are the terms that depend on the specific gauge choice) do not contribute, i.e.,
\begin{align}
	&Q_\mu\left[P^\mu(P+Q)^\nu+ P^\nu(P+Q)^\mu-g^{\mu\nu} P\cdot Q\right]
	=P^\nu \left(2(Q\cdot P)+Q^2\right)=P^\nu(P+Q)^2=P^\nu P'^2= 0.\nonumber
\end{align}

To expand for small $Q$, we use $P'=P+Q$, $K'=K-Q$. 
To lowest order,
the only $Q$-dependence remains in the propagator and we obtain
\begin{align}
	|\mathcal M|^2&=16 n_F^2g^4G_{\mu\nu}G^\ast_{\rho\sigma}\, 4P^\mu P^\rho K^\nu K^\sigma\left(1+\mathcal O\left(\frac{Q}{P}\right)\right).
\end{align}
To the lowest order in $Q$, this is equivalent to 
\begin{align}
	|\mathcal M|^2&=4 n_F^2g^4G_{\mu\nu}G^\ast_{\rho\sigma}\times(P+P')^\mu (P+P')^\rho (K+K')^\nu (K+K') ^\sigma\times\left(1 + \mathcal O(Q/P)\right)\nonumber,
\end{align}
which at the same order
is the same as the scalar quark result \eqref{eq:scalar_quark_result}.

\subsection{Gluon scattering}
After having considered the case of quark scattering, we turn to gluon scattering. In this subsection, we show that in the soft limit, medium effects in gluon scattering enter in the same way.

\begin{figure}
	\centering
	\includegraphics[width=0.4\linewidth]{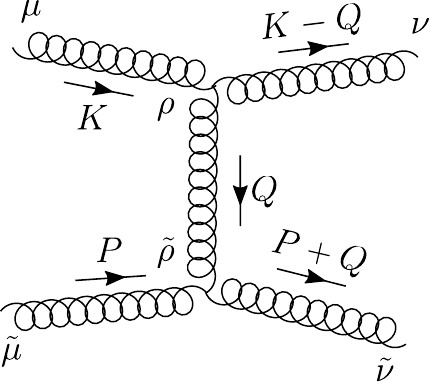}
	\caption{t-channel diagram for gluon scattering. Figure from \cite{Boguslavski:2024kbd}.
	}
	\label{fig:gluonscattering}
\end{figure}
Gluon-gluon scattering is depicted pictorially in \fig\ref{fig:gluonscattering}. The dominant process for screening effects is the $t$-channel, where $|t|\ll s$, or---equivalently---the $u$-channel, which can be readily obtained by exchanging $u\leftrightarrow t$ in the collision kernel.

Using the Feynman rules, we can write (suppressing color factors)
\begin{align}
	\mathcal M^{\mu\tilde\mu\nu\tilde\nu}&\propto G_{\rho\tilde\rho}(Q)\left[g^{\mu\rho}(K+Q)^\nu+g^{\rho\nu}(K-2Q)^\mu+g^{\nu\mu}(Q-2K)^\rho\right]\nonumber\\
	&\qquad\quad\,\,\,\times \left[g^{\tilde\mu\tilde\nu}(2P+Q)^{\tilde\rho}+g^{\tilde\nu\tilde\rho}(-P-2Q)^{\tilde\mu}+g^{\tilde\rho\tilde\mu}(Q-P)^{\tilde\nu}\right].\label{eq:gluonscattering-general}
\end{align}
In the soft limit, for small $Q$, we can, to a first approximation, neglect all $Q$ dependence except for the propagator. Additionally, when summing over polarizations, we need to contract the ``same'' external lines $\left(\mathcal M^{\mu\tilde\mu\nu\tilde\nu}\right)^\ast \mathcal M_{\mu\tilde\mu\nu\tilde\nu}$, and then the first line simplifies to
\begin{align}
	\begin{split}
		\left[g^{\mu\rho}K^\nu+g^{\rho\nu}K^\mu-2g^{\nu\mu}K^\rho\right]\left[g_{\mu\rho'}K_\nu+g_{\rho'\nu}K_\mu-2g_{\nu\mu}K_{\rho'}\right]
		=2K^2\delta_{\rho'}^\rho+(4D-6)K^\rho K_{\rho'},
	\end{split}
\end{align}
where $D=g^{\mu}_\mu$ is the space-time dimension.
With $P^2=K^2=0$, we obtain
\begin{align}
	|\mathcal M|^2&\propto K^\rho K_{\rho'}P^{\tilde\rho}P_{\tilde\rho'}G_{\rho\tilde\rho}(Q)\bar{G}^{\rho'\tilde\rho'}(Q) \times\left(1 + \mathcal O\left(\frac{Q}{K},\frac{Q}{P}\right)\right)\\
	&=|(K+K')^\rho (P+P')^{\tilde\rho}G_{\rho\tilde\rho}(Q)|^2\times\left(1 + \mathcal O\left(\frac{Q}{K},\frac{Q}{P}\right)\right).
	\label{eq:leadingorderinQgluon}
\end{align}
Again, to the lowest order in $Q$, this is the same as the scalar quark result \eqref{eq:scalar_quark_result}.

%% file: 950_numerical_details.tex
In this appendix, we discuss several numerical and kinematic aspects relevant for QCD kinetic theory simulations. In particular, we discuss how the distribution function is discretized, and how the collision terms are rewritten in a symmetrized form to solve the Boltzmann equation for the discretized distribution function. We discuss the coordinate systems that are used for the integral measure of the elastic collision term and discuss how the collision terms are evaluated using Monte Carlo sampling. We also discuss the adaptive step size algorithm used for the isoHTL-screened simulations.

\section{Discretization\label{app:discretization}}
In kinetic theory, all medium information is encoded in the distribution function $f(t, \vb p)$. In our case, for a given time $t$ (or proper time $\tau$), it depends only on two parameters (see the discussion in section \ref{sec:symmetries-f}), the magnitude of the momentum $p$ and its polar angle $\theta_p$.
How the distribution function $f(\vb p)$ is stored on a finite grid follows the \emph{discrete momentum} method of Ref.~\cite{AbraaoYork:2014hbk},
where we store the number density per bin,
\begin{align}
	n_{ij}=\lambda\int\frac{\dd[3]{\vb p}}{(2\pi)^3}f(\vb p) w_i(p)  \tilde w_j(\cos\theta),\label{eq:def_nij}
\end{align}
with the piecewise linear wedge functions $w$, defined for the momentum case as
\begin{align}
	w_i(p)=\begin{cases}
		\frac{p-p_{i-1}}{p_i-p_{i-1}}, & p_{i-1} < p < p_i\\
		\frac{p_{i+1}-p}{p_{i+1}-p_{i}}, & p_{i} < p < p_{i+1}\\
		0& \text{else,}
	\end{cases}
\end{align}
and for the polar angle similarly with the grid points substituted by $p_i \to \cos\theta_i$. The distribution function $f(\vb p)$ can be (approximately) recovered from the moments $n_{ij}$ by linear interpolation between the grid points, using the simple relation
\begin{align}
	f(p_i,\cos\theta_j)=\frac{(2\pi)^3n_{ij}}{\lambda p_i^2 4\pi \Delta V^p_i\Delta V^\theta_j},
\end{align}
with the volume factors defined via
\begin{align}
	\Delta V_i^p=\int_{-\infty}^\infty\dd{x}w_i(x), && \Delta V_i^\theta=\int_{-1}^1\dd{\cos\theta}\tilde w_i(\cos\theta).
\end{align}
This simplifies obtaining observables because the integral for an observable \eqref{eq:observables} can be computed via
\begin{align}
    n\langle O(\vb p)\rangle=\int\frac{\dd[3]{\vb p}}{(2\pi)^3}\mathcal O(\vb p)f(\vb p)\approx\sum_{ij}n_{ij}\mathcal O(\vb p_{ij}),
\end{align}
which is exact for the number and energy density
\begin{align}
    n=\sum_{ij}n_{ij}, && \varepsilon = \sum_{ij}n_{ij}p_i,
\end{align}
because we can insert \eqref{eq:def_nij} and use
\begin{align}
    \sum_i w_i(p)=1, && \sum_iw_i(p)p_i=p.
\end{align}
See also Appendix A of Ref.~\cite{Du:2020dvp} for a more detailed discussion about the discretization.

The discretization parameters used for simulations in this thesis are summarized in Table~\ref{tab:discretization-params}.

\begin{table}
    \centering
    \begin{tabular}{l c c c c c}\toprule
        Description & Chapter(s)& $N_p$ & $N_\theta$ & $\pmin$ & $\pmax$\\
        \hline
        Runs with Bjorken expansion for $\qhat$ & \ref{sec:momentum-broadening-of-jets}, \ref{sec:limiting_attractors} & $100$& $180$ & $0.03$ & $6$ \\     
        Runs extracting $\kappa$ &\ref{sec:limiting_attractors} & $100$ & $180$ & $0.05$ & $6$\\
        Additional runs with $\lambda\in\{0.25, 0.75, 1.5\}$ &\ref{sec:limiting_attractors}  & $180$ & $250$ & $0.03$ & $6$\\
        Bjorken expanding isoHTL/Debye-like & \ref{sec:improving-qcd-simulations} & $200$ & $180$ & $0.03$ & $8$\\
        Over-occupied, with $\lambda=0.5$ &\ref{sec:improving-qcd-simulations} & $300$ & $1$ & $0.01$ & $25$\\
        Over-occupied, with $\lambda=2$ &\ref{sec:improving-qcd-simulations}& $200$ & $1$ & $0.01$ & $12$ \\
        Over-occupied, with $\lambda=10$ &\ref{sec:improving-qcd-simulations}& $200$ & $1$ & $0.01$ & $10$\\
        Under-occupied runs &\ref{sec:improving-qcd-simulations} & $500$ & $1$ & $0.03$ & $100$\\
        Runs with Bjorken expansion &\ref{sec:collkern} & $200$ & $180$ &$0.03$ & $8$\\
        \bottomrule
    \end{tabular}
    \caption{Discretization parameters for the simulations used in this thesis.}
    \label{tab:discretization-params}
\end{table}

\section{Symmetrization of the collision terms\label{sec:symmetrization-collkernels}}
For the numerical evaluation, it is convenient to rewrite Eq.~\eqref{eq:c22_first} in a symmetric form. For pure gluons, it then reads
\begin{align}
\label{eq:C22-symmetrized}
\begin{split}
\Ctwotwo&[f(\tilde{\vb{p}})] = \frac{\left( 2 \pi \right)^3}{4 \pi \tilde{p}^2} \dfrac{1}{8 \nu_g} \int \dd{\Gamma_{\mathrm{PS}}} \left|\mathcal M(\vb p,\vb k;\vb p', \vb k') \right|^2  \\
& \times  \Big( f(\vb{p}) f(\vb{k}) (1 + f(\vb{p^\prime})) ( 1 + f(\vb{k^\prime}) - f(\vb{p^\prime}) f(\vb{k^\prime}) (1 + f(\vb{p}) (1 + f(\vb{k}) \Big)  \\
& \times \left( \delta^3(\vb{\tilde{p}} - \vb{p})  + \delta^3(\vb{\tilde{p}} - \vb{k}) - \delta^3 (\vb{\tilde{p}} - \vb{p^\prime}) - \delta^3(\vb{\tilde{p}} - \vb{k^\prime})\right),
\end{split}
\end{align}
with the integration measure
\begin{align}\label{eq:integration-measure-c22}
 \int \dd{\Gamma_{\mathrm{PS}}}  &=  \int_{\vb{p k p' k'}}
 \left(2 \pi \right)^4 \delta^4\left( P + K - P^\prime - K^\prime \right) \\
 & = \frac{1}{2^{12} \pi^8} \int_0^\infty \dd q \int_{-q}^q \dd \omega \int_{\frac{q-\omega}{2}}^\infty \dd p  \int_{\frac{q+\omega}{2}}^\infty \dd k \int_{-1}^1 \dd \cos\theta_q\int_0^{2\pi}\dd{\phi_q} \int_0^{2 \pi} \dd \phi_{pq} \int_0^{2 \pi} \dd\phiqk. \nonumber
 \end{align}
We will discuss the integration variables in more detail in the next section \ref{sec:coordinate-systems-appendix}.

As discussed in Section \ref{sec:amy-screening-prescription}, the integrand is symmetric under the exchange of the outgoing particles with momenta $k'\leftrightarrow p'$, which exchanges the $u-$ and $t-$channel. This shifts the dominating integration region from small $|u|$ and small $|t|$ to the small $|t|$ region only, which greatly simplifies the screening prescription and importance sampling.

The inelastic collision term \eqref{eq:c12} accounts for collinear splitting and merging. In its symmetrized form, and for pure gluons, it is given by
\begin{align}
    \begin{split}\label{eq:c12-symmetrized}
        \Conetwo[f(\vb {\tilde p})]&=
        \frac{(2\pi)^3}{4\pi\tilde p^2}\frac{1}{\nu}\int_0^\infty \dd{p}\int_0^{p/2}\dd{k'}4\pi\gamma^{p}_{p',k'}\\
        &\quad\times\Big\{f(\vb p)(1+f(p'\vbphat))(1+f(k'\vbphat))\\
        &\qquad\quad-f(p'\vbphat)f(k'\vbphat)(1+f(\vb p))\Big\}\\
        &\times\left[\delta(\tilde p-p)-\delta(\tilde p-p')-\delta(\tilde p-k')\right]
    \end{split}
\end{align}

\section{Coordinate system and Mandelstam variables\label{sec:coordinate-systems-appendix}}
The elastic collision term, after symmetrization \eqref{eq:C22-symmetrized} features an integration over $\vb p$, $\vb k$, and $\vb q$. Similarly to Ref.~\cite{Arnold:2003zc}, the integrals over $\vb p$ and $\vb k$ are performed in a frame, in which $\vb q$ defines the $z$-direction with the original $z$-axis lying in the $xz$ plane. However, contrary to Ref.~\cite{Arnold:2003zc}, $\vb p$ is not required to be in the $xz$ plane, since it is also integrated. This is also different to how the jet quenching parameter $\qhat$ is obtained in Section \ref{sec:coordinate_systems}, where the jet direction $\vb p$ was fixed.
As integration variables it is convenient to choose the exchange momentum $\vb q$, parametrized by its magnitude $q$, polar and azimuthal angle $\cos\theta_q$, and $\phi_q$,
the exchange energy $\omega$, the magnitude of the vectors $p$ and $k$, and the azimuthal angles of $\vb p$ and $\vb k$ in a frame (\emph{$q$-frame}), in which $\vb q$ points in the $z$ direction and the original $z$-axis lies in the $xz$ plane. This fixes the kinematics completely. We denote these azimuthal angles by $\thetaqp$ and $\thetaqk$. Energy conservation results in the kinematic conditions
\begin{align}
	|\omega| < q, && p > \frac{q-\omega}{2}, && k > \frac{q+\omega}{2},
\end{align}
and fixes the polar angles $\thetaqp$ and $\thetaqk$ of $\vb p$ and $\vb k$ in the \emph{$q$-frame} (see, e.g., \cite{Arnold:2003zc} or the discussion in Section \ref{sec:coordinate_systems}),
\begin{align}
    \cos\thetaqp=\frac{\omega}{q}+\frac{\omega^2-q^2}{2pq}, && \cos\thetaqk=\frac{\omega}{q}-\frac{\omega^2-q^2}{2kq}.
\end{align}

In terms of these integration variables, we parameterize the relevant vectors (in the \emph{$q$-frame}) using these variables as
\begin{align}
    Q&=P'-P=\begin{pmatrix}
        \omega\\0\\0\\q
    \end{pmatrix}, &
    P&=p\begin{pmatrix}
        1\\\sin\thetaqp\cos\phiqp\\ \sin\thetaqp\sin\phiqp \\ \cos\thetaqp  
    \end{pmatrix}, &
    K&=k\begin{pmatrix}
        1\\ \sin\thetaqk\cos\phiqk \\ \sin\thetaqk \sin\phiqk \\ \cos\thetaqk
    \end{pmatrix}.\label{eq:external_momenta}
\end{align}
In terms of the integration variables, the Mandelstam variables needed for the matrix elements are given by
\begin{align}
	t &=\omega^2-q^2, \label{eq:mandelstam_t_explicit-c22}\\
	\begin{split}
		s &= -\frac{t}{2q^2}\Big((p+p')(k+k')+q^2-\sqrt{(4pp'+t)(4k'k+t)}\cos(\phiqk-\phiqp)\Big),
	\end{split} \label{eq:mandelstam_s_explicit-c22}\\
	\begin{split}
		u &= \frac{t}{2q^2}\Big((p+p')(k+k')-q^2-\sqrt{(4pp'+t)(4k'k+t)}\cos(\phiqk-\phiqp)\Big).
	\end{split}\label{eq:mandelstam_u_explicit-c22}
\end{align}

\section{Inelastic collision term}
For the inelastic collision term, the splitting rate $\gamma^g_{gg}$ from Section \ref{sec:amy-rate-equation} has to be evaluated. This is done using the method described in Ref.~\cite{AbraaoYork:2014hbk} following Ref.~\cite{Aurenche:2002wq}. The integral equation \eqref{eq:amy-integralequation-long} is solved in impact parameter space, where the dipole cross section \eqref{eq:analytic-result-dipole-cross-section-appendix} is used. We do not discuss this method here in detail, but discuss a general method for obtaining the splitting rate for an arbitrary collision kernel $C(\vb q_\perp)$ in Appendix \ref{app:amyrates-details}.

\section{Monte Carlo evaluation of the collision terms}
To derive the differential equations for the moments $n_{ij}$, one needs to rewrite the Boltzmann equation \eqref{eq:boltzmann_equation} in terms of $n_{ij}$,
\begin{align}
    v^\mu\partial_\mu n_{ijk}&=\lambda\int\frac{\dd[3]{\vb {\tilde p}}}{(2\pi)^3}w_{i}({\tilde p})\tilde w_j(\cos\tilde\theta)v^\mu\partial_\mu f(\vb{\tilde p}) =-\lambda\int\frac{\dd[3]{\vb {\tilde p}}}{(2\pi)^3}w_{i}({\vb p})\tilde w_j(\cos\tilde\theta)\mathcal C[f(\vb{\tilde p})].
\end{align}
This replaces the delta functions in the collision terms \eqref{eq:C22-symmetrized} and \eqref{eq:c12-symmetrized} by the wedge functions $w_{i}$ and $\tilde w_j$. The collision terms are then computed using Monte Carlo importance sampling.

For the elastic collision term $\Ctwotwo$, points from the integration measure \eqref{eq:integration-measure-c22} are sampled using Monte Carlo importance sampling. For finite grid boundaries, we need to require that all momenta sampled lie within the grid.
For the $q$, $\omega$, $p$, and $k$ integration, the grid boundaries $\pmin$ and $\pmax$ have to be considered, and
these integrals become
\begin{align}
    &\int_0^{\pmax}\dd{q}\int_{\max(-q,\pmin-\pmax)}^{\min(q,\pmax-\pmin)}\dd{\omega}\int_{\max\left(\frac{q-\omega}{2},\pmin,\pmin-\omega\right)}^{\min(\pmax,\pmax-\omega)}\dd{p}\int_{\max\left(\frac{q+\omega}{2},\pmin,\pmin+\omega\right)}^{\min(\pmax,\pmax+\omega)}\dd{k}\nonumber,
\end{align}
which ensures that all momenta $k$, $k'=k-\omega$, $p$, $p'=p+\omega$ lie within the grid.
The second boundaries for $\omega$ come from the requirement that $\pmax>\pmin+\omega$ and $\pmax+\omega > \pmin$ in the $k$ integral.

For the $q$-integral, we sample from a $\dd{q}/(q+\xi m_D)^4$ distribution, with $\xiscreen=e^{5/6}/\sqrt{8}$ (see section \ref{sec:debye-like-screening}), although its precise value is not important for the sampling process. While $\omega$ is sampled uniformly, both $k$ and $p$ are sampled from a $\dd{k}/k$ distribution.

\section{Adaptive step size\label{app:adative_stepsize}}
Here, the adaptive step size is described, which is used for simulations with the isoHTL screened matrix element in Chapter\footnote{The presentation here follows Ref.~\cite{Boguslavski:2024kbd}.} \ref{sec:improving-qcd-simulations}.
The philosophy of the adaptive timestep is to set it such that the relative change of specific observables is smaller than a predefined constant $\epsgoal$. This is similar to the stepsize employed in Ref.~\cite{Du:2020dvp}, but the method here goes beyond linearization by calculating the exact change in the considered observables. It is found that this approach is better stabilizes simulations with isoHTL screening.

To be more concrete, in every timestep $t_{k+1}=t_{k}+\Delta t$, the stored values of $n_{ij}$ are changed (defined in \eqref{eq:def_nij}) such that
\begin{align}
	n_{ij}(t_{k+1})=n_{ij}(t_{k})+\Delta t \, c_{ij},
\end{align}
and $c_{ij}$ are numerical estimates (from Monte Carlo integrals) of the sum of all collision terms in Eq.~\eqref{eq:boltzmann_equation}.

With the discretization choice \eqref{eq:def_nij}, several quantities can be easily calculated,
\begin{subequations}\label{eq:observables-timestep}
	\begin{align}
		\varepsilon&=\int\frac{\dd[3]{\vb p}}{(2\pi)^3}p f(\vb p)=\sum_{ij} \frac{n_{ij} p}{\lambda}\\
		n&=\int\frac{\dd[3]{\vb p}}{(2\pi)^3} f(\vb p)=\sum_{ij} \frac{n_{ij} }{\lambda}\\
		n\langle p_z^2\rangle&=\int\frac{\dd[3]{\vb p}}{(2\pi)^3}p^2\cos^2\theta f(\vb p)\approx\sum_{ij} \frac{n_{ij}}{\lambda} p^2\cos^2\theta\\
		n\langle f\rangle &=\int\frac{\dd[3]{\vb p}}{(2\pi)^3}f^2(\vb p)\approx\sum_{ij} \frac{n_{ij}^2}{\lambda^2}\frac{(2\pi)^3}{\Delta V_i^p\Delta V_j^\theta p_i^2 4\pi}\\
		n\langle pf\rangle&=\int\frac{\dd[3]{\vb p}}{(2\pi)^3}pf^2(\vb p)\approx\sum_{ij} \frac{n_{ij}^2}{\lambda^2} p_i\frac{(2\pi)^3}{\Delta V_i^p\Delta V_j^\theta p_i^2 4\pi}\\
		m^2&=2\lambda\int\frac{\dd[3]{\vb p}}{(2\pi)^3} \frac{f(\vb p)}{p}\approx2\sum_{ij} \frac{n_{ij}} {p_i}\,.
	\end{align}
\end{subequations}
Note that, except for $n\langle f\rangle$ and $n\langle pf\rangle$, these quantities are linear in $n_{ij}$.
After having evaluated the collision terms $c_{ij}$, we can accurately predict how these observables change in a single timestep and adjust the timestep such that their relative change does not exceed the parameter $\epsgoal$.

For quantities that are linear in $n_{ij}$ (e.g., $\varepsilon$, $m^2$, $n$, $n\langle p_z^2\rangle$), their exact change per unit time can be obtained by replacing $n_{ij}\to c_{ij}$ in Eq.~\eqref{eq:observables-timestep}. For instance, the energy density $\varepsilon$ changes by
\begin{align}
	\delta\varepsilon:=\frac{\Delta \varepsilon}{\Delta t}&=\sum_{ij} c_{ij} p_i.
\end{align}
With the relative change  
$\frac{\Delta \varepsilon}{\varepsilon}=\frac{\delta\varepsilon \Delta t}{\varepsilon},$
the new time step can then be set according to
\begin{align}
	\Delta t = \frac{\epsgoal}{\left|\frac{\delta\varepsilon}{\varepsilon}\right|}\, .
\end{align}

The change of the quadratic quantities (e.g., $n\langle f\rangle$ and $n\langle pf\rangle$), could be approximated linearly, e.g., for $\langle f\rangle$,
\begin{align}
	\Delta (n\langle f\rangle) \approx \sum_{ij}2 n_{ij} c_{ij} V_{ij}
\end{align}
with 
\begin{align}
	V_{ij}=\frac{(2\pi)^3}{\Delta V_i^p\Delta V_j^\theta p_i^2 4\pi}
\end{align}
but this is only a crude estimate of the change and it turns out that using the correct change leads to a more stable evolution. The exact change $\Delta n\langle f\rangle = (n\langle f\rangle)_{i+1} - (n\langle f\rangle)_i$ can easily be obtained via
\begin{align}
	\Delta n\langle f\rangle &=\sum_{ij}V_{ij}\left( (n_{ij}+c_{ij}\Delta t)^2- n_{ij}^2\right)\\
	&=\sum_{ij}V_{ij}\left(2n_{ij}c_{ij}\Delta t + c_{ij}^2(\Delta t)^2\right).
\end{align}
Enforcing a maximum relative change $|\Delta n\langle f\rangle/n\langle f\rangle| \leq  \epsgoal$ 
leads to a quadratic equation in $\Delta t$,
\begin{align}
	|a\Delta t+b\Delta t^2|=\epsgoal,\label{eq:quadratic_equation}
\end{align}
with $b > 0$ and
\begin{align}
	a=\frac{\sum_{ij}V_{ij}2 n_{ij}c_{ij}}{n\langle f\rangle}, && b = \frac{\sum_{ij}V_{ij}(c_{ij})^2}{n\langle f \rangle}.
\end{align}

To solve this quadratic equation for the most restrictive (i.e., smallest) $\Delta t$, we need to consider the two cases:

\begin{enumerate}
	\item \emph{ $a > 0$}:
	
	Here, $a\Delta t+b \Delta t^2$ monotonically increases and is always positive, 
	thus the timestep is given by
	\begin{align}
		\Delta t = -\frac{a}{2b}+\sqrt{\frac{a^2}{4b^2}+\frac{\epsgoal}{b}}.
	\end{align}
	
	\begin{figure}[t!]
		\centering
		\includegraphics[width=0.5\linewidth]{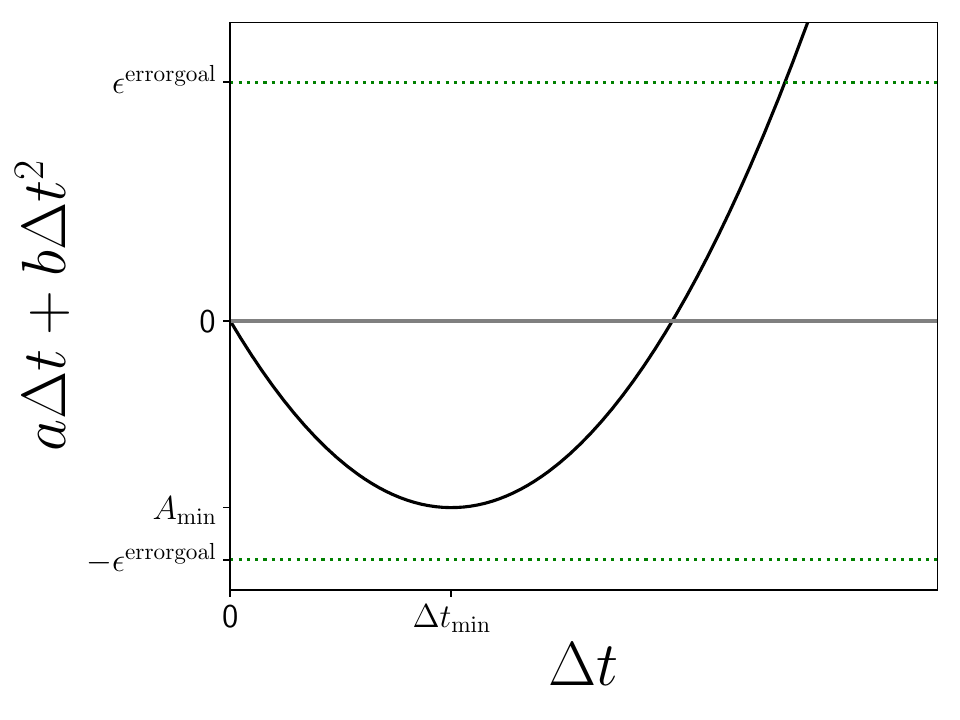}
		\caption{Sketch of the quadratic equation \eqref{eq:quadratic_equation} for $a<0$. Figure from \cite{Boguslavski:2024kbd}.
        }
		\label{fig:parabola}
	\end{figure}
	
	\item For \emph{ $a < 0$},
	the parabola \eqref{eq:quadratic_equation} is sketched in Fig.~\ref{fig:parabola}. Its minimum is at $\Delta t_{\mathrm{min}}=-a/(2b)$, with the minimum value $A_{\mathrm{min}}=-a^2/(4b)$.
	
	When $|A_{\mathrm{min}}|<\epsgoal$ as in Fig.~\ref{fig:parabola}, we need to solve for positive $\epsilon$, which yields the same result as above (again, we need the positive square root, as can be easily seen from the figure)
	\begin{align}
		\Delta t = -\frac{a}{2b}+\sqrt{\frac{a^2}{4b^2}+\frac{\epsgoal}{b}}.
	\end{align}
	For $|A_{\mathrm{min}}|>\epsgoal$, we need to equate to negative $\epsgoal$, i.e., solve
	$a\Delta t+ b\Delta t^2=-\epsgoal,$
	which yields
	\begin{align}
		\Delta t=-\frac{a}{2b}-\sqrt{\frac{a^2}{4b^2}-\frac{\epsgoal}{b}}.
	\end{align}
\end{enumerate}

The new goal for the timestep is then set by determining the minimum of these 
\begin{align}
	\Delta t^{\mathrm{goal}}=\min_{i\in \{\varepsilon, n, n\langle p_z^2\rangle, n\langle f\rangle, n\langle pf\rangle, m^2\}}\Delta t_i
\end{align}
and then adjusting it smoothly to the old one via
\begin{align}
	\Delta t^{\mathrm{new}}=\sqrt{(\Delta t^{\mathrm{goal}})^{2\alpha}(\Delta t^{\mathrm{old}})^{2-2\alpha}},
\end{align}
where $\alpha$ is a constant $0 <\alpha < 1$ that 
ensures an interpolation between the previous value and the goal time step. While $\alpha=1$ corresponds to always choosing the goal timestep, $\alpha=0$ corresponds to a constant time step (i.e., never choosing the goal time step). Choosing $\alpha$ between these limits prevents large variations of this time step due to statistical fluctuations. In practice, a value of $\alpha=0.9$ is used for the simulations performed here.

\section{Kinematic considerations\label{app:kinematic-considerations}}
In this section, we discuss why $s$ is always the largest Mandelstam variable, and that if $Q^2$ is small, also $q,\omega$ are necessarily small. This section follows Ref.~\cite{Boguslavski:2024kbd}.

\subsection*{Why $s$ is always the largest Mandelstam variable}
Here, we want to show that $s$ is always the largest Mandelstam variable. This has the consequence that no $s$-channel processes needs to be screened in Section \ref{sec:amy-screening-prescription}.

The Mandelstam variable $s$ corresponds to the square of the center-of-mass energy. Since the Mandelstam variables are Lorentz invariant, we may compute them in any frame.

For simplicity, let us choose the center-of-mass frame, where
\begin{subequations}
	\begin{align}
		P&=(p,p,0,0),& K&=(p,-p,0,0),\\
		P'&=(p, p\,\vb e_x + \vb q), & K'&=(p,-p\,\vb e_x-\vb q),
	\end{align}
\end{subequations}
and rotate the frame such that $q_z=0$. Note that in this frame, the energy of each particle is conserved separately in an elastic scattering process.

From $P'^2=0$ we can compute $2pq_x+q^2=2pq_x+q_x^2+q_y^2=0$, and thus $q_x=-p\pm \sqrt{p^2-q_y^2}$.
Therefore, $|q_y| \leq p$, and $|q_x| \leq 2p$. Additionally,
$q_x=-\frac{q^2}{2p}$.

The Mandelstam variable $t$ is given by $t=-q^2$, which is thus bounded by $|t| \leq 4p^2=s$.
Therefore, $s$ is always the largest Mandelstam variable, and no $s$-channel process will need to be screened.

\subsection*{Kinematic considerations: From $Q^2$ small follows $q,\omega$ small}

Here, we show that the condition $0<-t\ll s$, or $Q^2\ll -(P+K)^2$ implies that all components of $Q$, i.e., $|\omega|$ and $q$, are small.
For that, we consider a plasma with typical excitations at momentum $T$,
\begin{subequations}
	\begin{align}
		P&=T(1,1,0,0), & K&=T(1, \vec n), \\
		P'&=(T+\omega,T+q_x,q_y,q_z), & K'&=(T-\omega,T\vec n - \vec q),
	\end{align}
\end{subequations}
with $\vb n=(n_x,n_y,0)$ being a unit vector in the $x-y$ plane, i.e., $n_x^2+n_y^2=1$.
The condition $|t|=|Q^2|\ll s$ is equivalent to
\begin{align}
	q^2-\omega^2\ll 2T^2(1-n_x),\label{eq:mandelstam_requirement}
\end{align}
where we used that $q^2 > \omega^2$.

We show in the following that Eq.~\eqref{eq:mandelstam_requirement} implies $|\vb q| \gtrsim |\omega| \ll T$.

Let us assume 
\begin{align}
|\vb q| \approx T \label{eq:assumption-kin}
\end{align}
and show that this leads to a contradiction. Eq.~\eqref{eq:assumption-kin} requires $|\omega| \approx T$ for Eq.~\eqref{eq:mandelstam_requirement} to be fulfilled. We will show in the following that this is kinematically forbidden, and thus $|\vb q|\ll T$.

We thus assume $|\omega|\approx T$ and introduce a parameter $\alpha \ll 1-n_x$ to rewrite Eq.~\eqref{eq:mandelstam_requirement} to $q^2-\omega^2=\mathcal O(\alpha T^2)$.
From $P'^2=0$ we can 
obtain\footnote{Using $P'^2=q^2-\omega^2-2T(\omega-q_x)$} an expression for $q_x = \omega + \mathcal O(\alpha^2 T)$.
Similarly, from $K'^2=0$, we obtain\footnote{Using $K'^2=(K-Q)^2=q^2-\omega^2-2T[-\omega+q_xn_y+q_yn_y]$ and then using $q_x=\omega - \frac{q^2-\omega^2}{2T}$} $q_y=\frac{1-n_x}{n_y}\omega + \mathcal O(\alpha T)$. Inserting back into Eq.~\eqref{eq:mandelstam_requirement} and using $\frac{(1-n_x)^2}{1-n_x^2}=\frac{1-n_x}{1+n_x}$ yields an equation for $q_z$,
\begin{align}
	q^2-\omega^2 = \frac{1-n_x}{1+n_x}\omega^2+q_z^2+\mathcal O(\alpha  T^2)=0,
\end{align}
which has no solution for which $q_z^2 > 0$. Therefore, the assumption $q^2\approx\omega^2\approx T^2$ leads to a contradiction, and we have shown that $|\vb q|, |\omega| < q \ll T$.

\section{Dominance of small angle scatterings\label{app:validity_screening}}

\begin{figure*}
	\includegraphics[width=0.95\linewidth]{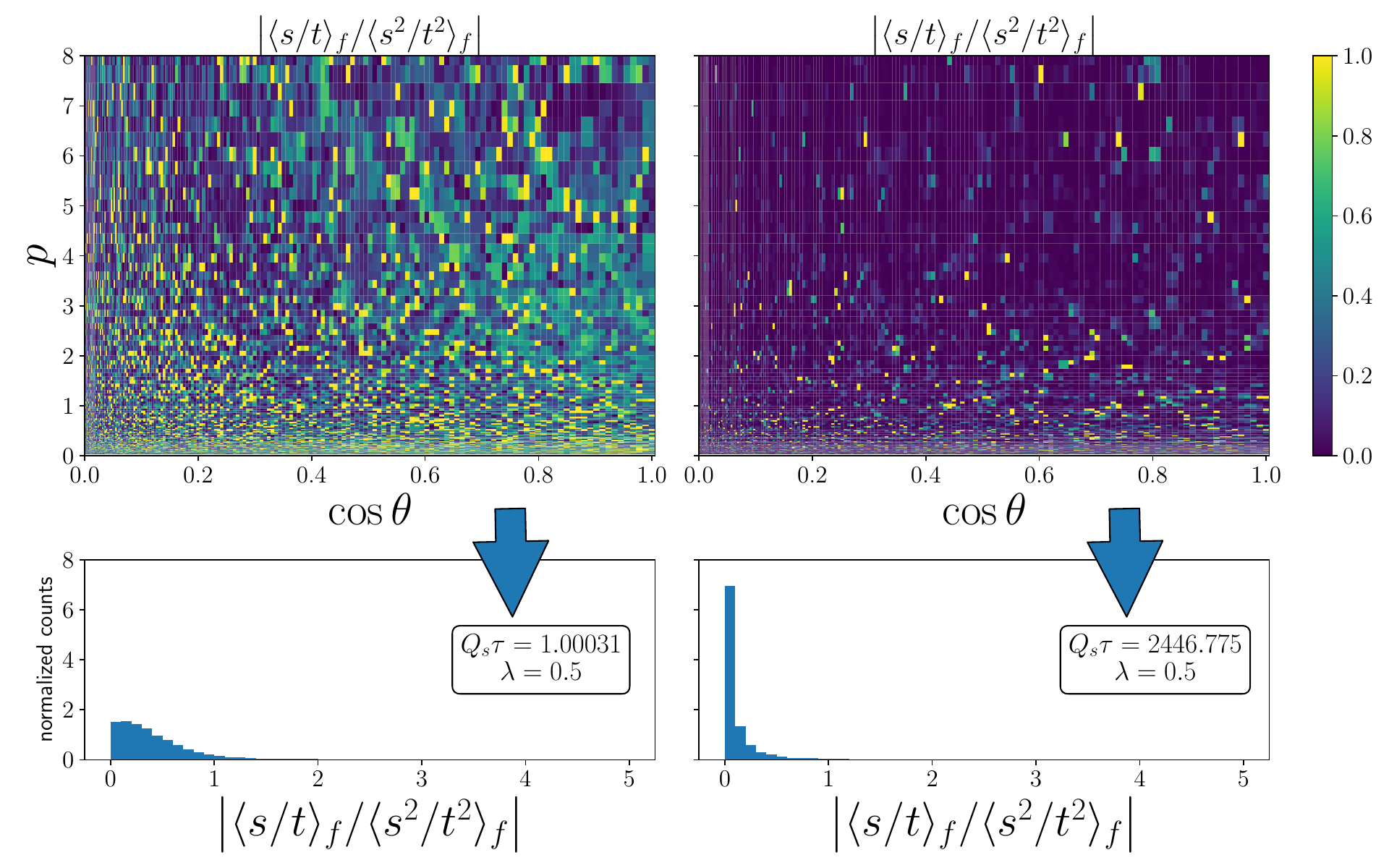}
	\includegraphics[width=0.95\linewidth]{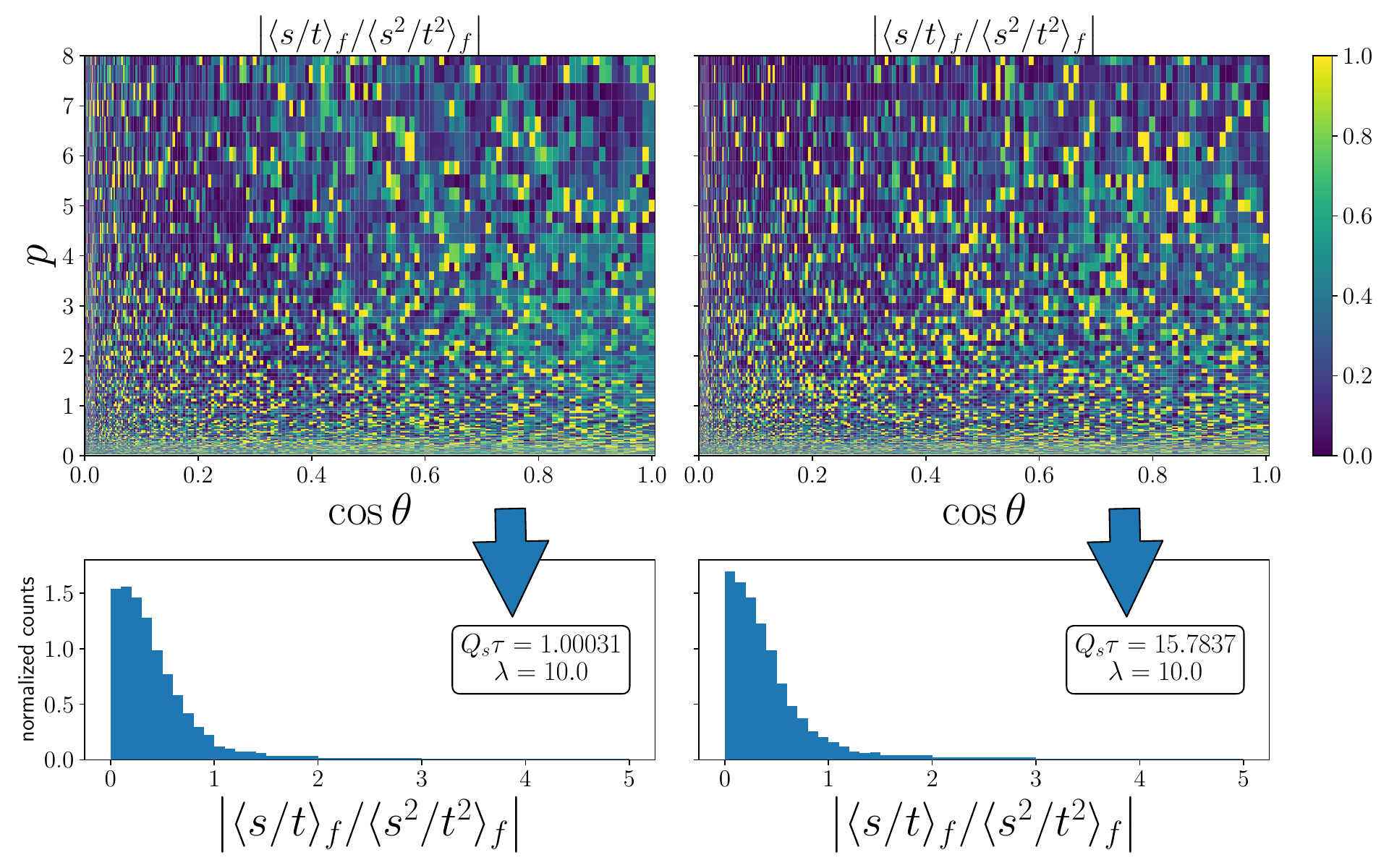}
	\caption{Ratio of contributions of the elastic collision term with only $s/t$ as matrix element over $s^2/t^2$, denoted $\left|{\langle s/t\rangle_f}/{\langle s^2/t^2\rangle_f}\right|$, which is defined in Eq.~\eqref{eq:rate-of-change-dist-funct-el-col}. All four panels 
		show $\left|{\langle s/t\rangle_f}/{\langle s^2/t^2\rangle_f}\right|$ as a function of the momentum $\vb p$, parametrized by its length $p$ and angle $\theta_p$ with respect to the beam axis. The histograms show the distribution of values in the plots above.
		The left column corresponds to the initial time and the right column to a later time for the couplings $\lambda=0.5$ {\em (top)} and $\lambda=10$ {\em (bottom)}. Figure from \cite{Boguslavski:2024kbd}.
	}
	\label{fig:change_05_10}
\end{figure*}

Here, it is demonstrated that $|t|\ll |s|\sim |u|$ is the dominant regime for elastic processes in the collision term. This implies that the different screening prescriptions that we discussed in \eq \eqref{eq:different_regularization_matrix_element} in Chapter \ref{sec:improving-qcd-simulations} agree for $|t|/s \ll 1$ and $|t/u| \ll 1$.

To show this, typical distribution functions $f(\vb p)$ are considered during simulations of the equilibration process in expanding systems, and the ``expectation'' value of $s/t$ and $s^2/t^2$ is calculated by integrating over phase-space and the statistical factors. More concretely, 
kinetic theory simulations using the Debye-like screened matrix element \eqref{eq:usual_screened_matrix_element} are performed, and the distribution function $f(\vb p)$ is extracted at two distinct times.
Then the contribution to the collision term from the small $|t|\ll s$ contributions are calculated,
\begin{align}
	&\left|\frac{\left.\pdv[f(\vb p)]{\tau}\right|_{\text{el. coll. with $s/t$}}}{\left.\pdv[f(\vb p)]{\tau}\right|_{\text{el. coll. with $s^2/t^2$}}}\right| =\left|
	\frac{\int\dd\Gamma\frac{s}{t}\, F^{2 \leftrightarrow 2}}{\int\dd\Gamma\frac{s^2}{t^2}\, F^{2 \leftrightarrow 2}}
	\right|=\left|\frac{\langle s/t\rangle_f}{\langle s^2/t^2\rangle_f}\right|,\label{eq:rate-of-change-dist-funct-el-col}
\end{align}
where we have introduced the short notation $\langle \dots\rangle_f$ to denote the phase-space average including the distribution functions $F^{2 \leftrightarrow 2} = f_pf_k(1+f_{p'})(1+f_{k'})-f_{p'}f_{k'}(1+f_{p})(1+f_{k})$. Note that the result still depends on the vector $\vb p$, which is the argument of the initial distribution function.

The results of this procedure are presented for two distinct times in \fig \ref{fig:change_05_10} 
for couplings $\lambda=0.5$ and $\lambda=10$. We observe that the values of $|\langle s/t\rangle/\langle s^2/t^2\rangle|$ fluctuate depending on the momentum bin, coupling, and time. However, when plotted within a histogram below (indicated by blue arrows), we find consistently dominating values below $1$. Therefore, we conclude that indeed $\langle s/t\rangle$ terms are suppressed as compared to $\langle s^2/t^2\rangle$ terms, and the results with the different Debye-like screening prescriptions coincide
not only for weak coupling, as argued in Chapter \ref{sec:improving-qcd-simulations}.

\section{Statistical averages\label{sec:statistical-averages}}
The method for obtaining the collision integrals described here is based on Monte Carlo integration, which is based on random numbers. There, the value of an integral is estimated based on random samples in the integration region. Thus, the resulting estimate of the integral is inherently stochastic.
In Chapter \ref{sec:improving-qcd-simulations}, different simulations with different seeds for the random number generators are performed. Every simulation, thus, will yield slightly different results. An estimate for the error of the mean, or the standard error \cite{Dowdy_Weardon_Chilko_2004}
can be given by taking $n$ samples of a considered quantity, $\mathcal O_i$, and calculating the sample average $\bar O$ and sample variance $s^2$,
\begin{align}
    \bar O=\frac{1}{n}\sum_{i=1}^n \mathcal O_i, && s^2=\frac{1}{n-1}\sum_{i=1}^n\left(\mathcal O_i-\bar O\right)^2.
\end{align}
The standard error $\bar \sigma$ is then calculated as
\begin{align}
    \bar\sigma=\sqrt{\frac{s^2}{n}}.
\end{align}
This is the method for calculating the error bars in the time evolution in Chapter \ref{sec:improving-qcd-simulations}. The same method is also used to estimate the error from the Monte Carlo evaluation of the integral for the jet quenching parameter $\qhat$ in Chapter \ref{sec:momentum-broadening-of-jets}.

%% file: 908_large_momentum_limits.tex
In this appendix, we study the behavior of the jet quenching parameter $\qhat$ for large jet momentum $p$ and, in particular, how to correctly perform the limit $p\to\infty$. We verify that taking the term which is leading order in $1/p$ in the integrand of $\qhat^{ij}$ is indeed sufficient to obtain the correct leading-order contribution. 
This is not trivial since $p$ also appears in the integration boundaries. 
Thus, there are two possible sources for large-$p$ contributions to $\qhat$: the integrand and the integral boundaries. Our strategy here is to expand the integrand in orders of $1/p$ and then perform the integrals. This Appendix is based on Appendix C in \cite{Boguslavski:2023waw}.

We illustrate the large $p$ behavior of $\qhat$ using the gluonic matrix element, 
\begin{align}
    \hat q &\sim \int_0^\infty\dd{k}\!\!\int_{-\frac{p-k}{2}}^k\!\!\!\dd{\omega}\!\!\int_{|\omega|}^{\text{min}(p+p',k+k')}\!\!\!\!\!\!\!\dd{q} \label{eq:qhat_largep_integrand}
    q^2(1-\cos^2\thetaqp)\frac{\left|\mathcal M^{gg}_{gg}\right|^2}{p^2}f_b(\vb k)\left(1\pm f_d(\vb k')\right).
\end{align}
The distribution function $f(\vb k)$ provides a natural upper limit for the momentum of the plasma constituent $k$, which we assume to be much smaller than the jet momentum $p$, thus $k \ll p$. Hence, the minimum of $(p+p',k+k') = (2p+\omega, 2k-\omega)$ is always $2k-\omega$, because 
$2k-\omega < 2p+\omega$ for $\omega > k-p$, which
is always fulfilled due to the lower boundary of the $\omega$-integral, $\omega > \frac{k-p}{2}$. Then the only $p$-dependence in the integration boundaries of Eq.~\eqref{eq:qhat_largep_integrand} comes from the lower limit of the $\omega$ integral.

Therefore, we will be interested in the region $\omega < 0$, where $|\omega|$ is very large. In particular, we will assume $|\omega|>\lxi\gg k$ with a new scale $\lxi$, which will lead to simplifications in the matrix element. Additionally, the \emph{Bose-enhanced} term $\qhatff$ that includes $f(\vb k) f(\vb k')$ does not contribute in this limit, since $f(\vb k')\approx 0$. Focusing only on relevant terms, i.e., disregarding the $k$-integral since it cannot contribute to any large $p$ behavior, we analyze
\begin{align}
    \hat q &\sim \int_{-\frac{p-k}{2}}^{-\lxi}\dd{\omega}\int_{|\omega|}^{2k-\omega}\dd{q} \label{eq:qhat_largep_integrand2}
    q^2(1-\cos^2\thetaqp)\frac{\left|\mathcal M^{gg}_{gg}\right|^2}{p^2}.
\end{align}

\section{The integrand of $\qhat$ for large $p$}

Now let us expand the integrand in \eqref{eq:qhat_largep_integrand2} for large $p$ explicitly.

\subsection{The large $p$ limit of $\cos\thetaqp$}

First, we consider $\cos\thetaqp$ in the limit $p\to\infty$. Our starting point is Eq.~\eqref{eq:cos_thetapq},
\begin{align}
\cos\thetaqp=\frac{1}{q}\left(\omega+\frac{\omega^2-q^2}{2p}\right).
\label{eq:vpq_v2}
\end{align}
Using \eq \eqref{eq:phase_space_relation_omega_q_k}, i.e., $q<2k-\omega$, one has
\begin{align}
    \frac{q^2-\omega^2}{2p}< \frac{4k(k-\omega)}{2p}\ll |\omega|,
\end{align}
which leads
for $p\to\infty$ to,
\begin{align}
\cos\thetaqp=\frac{\omega}{q}, \label{eq:vpq_ptoinf}
\end{align}
where the neglected terms are $\mathcal O(k/p)$.
However, considering the term $1-\cos^2\thetaqp$ that appears in \eqref{eq:qhat_largep_integrand2} is more subtle, because the seemingly leading term in a $1/p$ expansion, $\omega/q$, can get close to unity, $|\omega|<q$. This leads to the corrections
\begin{align}
    1-\cos^2\thetaqp&=1-\frac{\omega^2}{q^2}+\frac{\omega(q^2-\omega^2)}{pq^2}+\dots \nonumber \\
    &=\left(1-\frac{\omega^2}{q^2}\right)\left(1+\frac{\omega}{p}+\dots\right),\label{eq:one_minus_vpq_limit}
\end{align}
and the correction term $\omega/p$ could possibly become large at the lower boundary of the $\omega$ integral, $\omega > -(p-k)/2$. Thus, in the limit $k/p\to 0$, $\cos\thetaqp\to\omega/q$, but $1-\cos^2\thetaqp\nrightarrow 1-\omega^2/q^2$.

\subsection{Matrix element for large $|\omega|$}

For large $|\omega|$, and therefore also large $q$ (from Eq.~\eqref{eq:qhat_largep_integrand2}), we do not need to take into account screening effects $\mathcal{O}(m_D)$ in the matrix element, such that \eqref{eq:full_HTL_finite_p_matrix_element_replacement} reduces to $M_{\mathrm{screen}} \approx M_0=(s-u)^2/t^2$. The contribution from the transverse propagator in the sum in \eqref{eq:full_HTL_finite_p_matrix_element_replacement} is negligible for large $|\omega|$, and we are left with
\begin{align}
    \frac{\left|\mathcal M^{gg}_{gg}\right|^2}{g^4p^2}=16 \dA\CA^2 \frac{\omega^2}{q^4}\left(1+\frac{\omega}{p}+\dots\right).
    \label{eq:limit_gluonic_matrix_element}
\end{align}
Collecting the pieces, we can rewrite the relevant integrand in \eqref{eq:qhat_largep_integrand2} as
\begin{align}
    &q^2(1-\cos^2\thetaqp)\frac{\left|\mathcal M^{gg}_{gg}\right|^2}{g^4p^2} 
    = 16\dA\CA^2 \frac{(q^2-\omega^2)\omega^2}{q^4}\left(1 + \frac{2\omega}{p}+\dots\right).\label{eq:qhat_largep_integrand_explicit}
\end{align}

\subsection{Integral over a more generic integrand}
The integrand, which is relevant for determining the large $p$ dominant behavior for $\qhat$, is a sum of terms $q^n\omega^m$, as can be seen from \eqref{eq:qhat_largep_integrand_explicit}.
Therefore, let us analyze a general integrand of this form and define the integral $I_{nm}$ (which should not be confused with Eq.~\eqref{eq:I_{mn}} in Section \ref{sec:qhat-thermal}),
\begin{align}
I_{nm}(p)=\int_{-\frac{p-k}{2}}^{-\lxi}\dd{\omega}\int_{-\omega}^{2k-\omega}\dd{q}\,q^n\omega^m. \label{eq:analysis_integrand_1}
\end{align}
Although $|\omega|\gg 2k$, we cannot neglect $2k$ in the upper integration boundary of the $q$-integral (because it will be important later on), which additionally complicates the analysis.
Using $\otild=-\omega>0$ to get rid of additional minus signs, we obtain for $n\neq -1$
\begin{align}
I_{nm}(p)&=(-1)^m\int_{\lxi}^{\frac{p-k}{2}}\dd{\otild}\,\int_{\otild}^{2k+\otild}\dd{q}\, q^nx^m,\\
&=\frac{(-1)^m}{n+1}\int_{\lxi}^{\frac{p-k}{2}}\dd{\otild}\,x^m\left[(2k+\otild)^{n+1}-\otild^{n+1}\right].
\end{align}
We expand the first term in a power series using the Binomial series 
\begin{align}
(x+y)^r=\sum_{k=0}^\infty\begin{pmatrix}
r \\ k
\end{pmatrix} x^{r-k}y^k,\quad x,y\in\mathbb R, \label{eq:binomial_series}
\end{align}
with $ |x| > |y|$ and $  r\in\mathbb C$.
We thus obtain
\begin{align}
I_{nm}&=\frac{(-1)^m}{n+1}\int_{\lxi}^{\frac{p-k}{2}}\dd{\otild}\otild^m\left[\sum_{j=0}^\infty \begin{pmatrix}n+1 \\ j\end{pmatrix}\otild^{n+1-j}(2k)^j-\otild^{n+1}\right] \nonumber \\
&=\frac{(-1)^m}{n+1}\!\!\int_{\lxi}^{\frac{p-k}{2}}\dd{\otild}\!\sum_{j=1}^\infty \!\begin{pmatrix}n+1 \\ j\end{pmatrix}\! \otild^{n+m+1-j}(2k)^j \\
&=\!\frac{(-1)^m}{n+1}\!\Bigg(\!\!\!\sum_{\substack{j=1\\j\neq n+m+2}}^\infty \!\!\!\!\!\!\!\begin{pmatrix}n+1 \\ j\end{pmatrix}\!\left.\frac{\otild^{n+m+2-j}}{n+m+2-j}\right|_{\otild=\lxi}^{\frac{p-k}{2}}(2k)^j
+\begin{pmatrix}n+1 \\ n+m+2\end{pmatrix}\ln\left(\frac{p-k}{2\lxi}\right)(2k)^{n+m+2}\Bigg).
\end{align}
Since we are interested in the behavior at large $p$, we drop the lower boundary $\otild=\lxi$ and take only the leading-order (LO) terms with the largest powers of $p$ into account. 
Those are obtained for $j=1$, for which the (generalized) binomial coefficient yields $n+1$. It will be useful to also consider the next-to-leading order (NLO) terms in $p$. We obtain, up to an additive constant,
\begin{subequations} 
\begin{align}
\ILO_{nm}&\simeq\begin{cases}\frac{(-1)^m (2k)}{n+m+1}\left(\frac{p-k}{2}\right)^{n+m+1}, & n+m+1 \neq 0\\
(-1)^m(2k)\ln(p), & n+m+1 = 0\end{cases} \label{eq:ILO}\\
\INLO_{nm}&\simeq\begin{cases}\frac{(-1)^m (2k)^2}{(n+1)(n+m)}\begin{pmatrix}n+1 \\ 2\end{pmatrix}\left(\frac{p-k}{2}\right)^{n+m}, & n+m \neq 0\\
\frac{(-1)^m}{n+1}\begin{pmatrix}n+1 \\ 2\end{pmatrix}(2k)^2\ln(p), & n+m = 0.\end{cases}\label{eq:INLO}
\end{align}
\end{subequations}
Note that the inclusion of $k$ in $(p-k)^{n+m+1}$ in the LO term is because it will contribute at NLO.

\section{Behavior of $\qhat$ for large $p$\label{app:qhat_large_p}}

Gathering the results of the previous sections and applying them to the integral of $\qhat$ in \eqref{eq:qhat_largep_integrand2} with the integrand \eqref{eq:qhat_largep_integrand_explicit}, we obtain
\begin{subequations}
\begin{align}
    \qhatLO &\sim I_{-2,2} - I_{-4,4} = (2k)^2\ln p + \text{ const}, \label{eq:qhat_LO_in_p}\\
    \qhatNLO &\sim \frac{1}{p}\left(I_{-2,3}-I_{-4,5}\right) = \text{const}+\mathcal O\left(\frac{1}{p}\right).
\end{align}
\end{subequations}

Note that here $\mathrm{NLO}$ denotes the terms proportional to $1/p$ in the integrand. For both cases, $n+m$ is constant, thus the leading terms \eqref{eq:ILO} cancel and we need the next-to-leading terms \eqref{eq:INLO}. We observe that, due to the logarithmic enhancement of the leading-order contributions $\qhatLO$, the next-to-leading order contributions $\qhatNLO$ become negligible\footnote{This is not a trivial statement: for an integrand $q^n\omega^m(1+\omega/p)$ with $n+m>0$, we would obtain $\frac{\qhatNLO}{\qhatLO}=\frac{\frac{1}{p}\left(a p^{n+m+2} + b\right)}{cp^{n+m+1}+e} \sim \frac{a}{c} + \mathcal O\left(\frac{1}{p}\right)$, thus the ratio $\text{NLO}/\text{LO}$ does not tend to $0$ for $p\to\infty$. This implies that multiplying the LO term with $\omega/p$ and integrating over it yields a term of the same order as the LO term.} 
for sufficiently large $p$, and $\qhat$ can be written in the form of Eq.~\eqref{eq:qhat_largep_behavior},
\begin{align}
\qhat(p\gg Q) \simeq a_p\ln p + b_p. 
\end{align}
Therefore, for sufficiently large jet momenta $p$, it is in principle enough to expand the matrix element and the integrand for large $p$ and take only the leading-order contribution in $p$. Note, however, that to obtain the constant term it is not enough to use the leading large $p$ behavior, but one must use the full matrix element.

Let us now calculate the coefficient of the logarithm, $a_p$.
Until now, we have not considered the exact form of the distribution function $f(\vb k)$ and merely used that it provides an upper cutoff for the $k$ integral. The numerical value of the coefficient $a_p$ will depend on the exact form of $f(\vb k)$.
 
Let us consider a gluon jet scattering off a gluon in the plasma and start with Eq.~\eqref{eq:qhat_formula},
\begin{align}
\begin{split}
    \qhat&\simeq\frac{16g^4\CA^2}{2^{10}\pi^5}\int_0^{2\pi}\dd\phipq\int_0^{2\pi}\dd\phikq\int_0^\infty\dd{k}\int_{-\frac{p-k}{2}}^k\dd{\omega}\int_{|\omega|}^{2k-\omega}\dd{q}
     f(\vb k)\frac{\omega^2(q^2-\omega^2)}{q^4},
\end{split} \label{eq:qhat_calculate_ap1}
\end{align}
where we have taken the leading term in the large $p$ integrand \eqref{eq:qhat_largep_integrand_explicit} that leads to the logarithmic behavior. Additionally, as explained below Eq.~\eqref{eq:qhat_largep_integrand}, it is sufficient to use $2k-\omega$ as the upper boundary of the $q$-integral.

In comparison to the general integrand we analysed in Eq.~\eqref{eq:analysis_integrand_1}, there appears also the distribution function $f(\vb k)=f(k,\cos\theta_k)$, and the angle depends on $\cos\thetaqk$ and $\cos\thetaqp$ as well, which are functions of $\omega$ and $q$. For the large $p$ behavior, we are interested in the region $|\omega|\gg k$ and $q\sim |\omega|$, which renders $v_{pq}\to -1$. For $v_{kq}$, however, we cannot make a definite statement, since it changes from $-1$ to $1$ when $q$ varies between its integration boundaries $|\omega|<q<2k-\omega$.
Therefore, we will restrict to isotropic distributions $f(k)$ here that only depend on the magnitude of $\vb k$. Then, the $\omega$ and $q$ integrations in Eq.~\eqref{eq:qhat_calculate_ap1} are given by Eq.~\eqref{eq:qhat_LO_in_p}.

For an isotropic plasma consisting of quarks and gluons with distributions $f_q$ and $f_g$, the coefficient $a_p$ is then given by Eq.~\eqref{eq:qhat_ap_coefficient},
\begin{align}
\begin{split}
    a_p/\CR&=\frac{\CA g^4}{4\pi^3}\int_0^\infty \dd{k}k^2f_g(k)+\sum_{f}\frac{g^4}{8\pi^3}\int_0^\infty \dd{k} k^2 f_q(k).
    \end{split}
\end{align}
In thermal equilibrium, this reduces to
\begin{align}
    \frac{a_p^{\mrm{eq}}}{\CR}=\frac{g^4\zeta(3)T^3}{2\pi^3}\left(\NC + \frac{3}{4}\nf\right).\label{eq:qhat_ap_coefficient_equilibrium}
\end{align}

This logarithmic behavior of $\qhatLO$ implies that the limit $p\to\infty$ requires a UV cutoff to render $\qhat$ finite. This is typically done by restricting the transverse momentum transfer, $\qperp < \Lambda_\perp$. We will verify in the next section that in this limit, $\qhat$ is finite, and we only need to consider the leading-order term in $p$ in the integrand.

\subsection{Large $p$ behavior combined with a transverse momentum cutoff}

Using a transverse momentum cutoff, $\qperp < \Lambda_\perp \ll p$, we can retrace the steps in the previous sections. For large $p$, apart from factors $\mathcal O(1/p)$, this amounts to
$\qperp^2 = q^2-\omega^2<\Lambda_\perp^2$ (c.f., \eqref{eq:qperp_cutoff}), and thus we modify $I_{nm}$ to
\begin{align}
I_{nm}(p)=\int_{-\frac{p-k}{2}}^{-\lxi}\!\!\dd{\omega}\int_{-\omega}^{\sqrt{\omega^2+\Lambda_\perp^2}}\!\!\!\dd{q}\,q^n\omega^m.\label{eq:analysis_integrand_2v2}
\end{align}
The upper limit in the $q$ integral replaces $2k-\omega$ in Eq.~\eqref{eq:analysis_integrand_1} for sufficiently large $\lxi$ because $\sqrt{\omega^2+\Lambda_\perp^2}<2k-\omega$ for $-\omega > \frac{\Lambda_\perp^2}{4k} - k$,
which, since $-\omega > \lxi$, can always be fulfilled by choosing 
\begin{align}
\lxi >\frac{\Lambda_\perp^2}{4k}-k. \label{eq:qperp_limit_xi_condition}
\end{align}
Similarly as before, with $\otild=-\omega$ and for $n\neq -1$, we obtain
\begin{align}
I_{nm}(p)&=\frac{(-1)^m}{n+1}\!\int_{\lxi}^{\frac{p-k}{2}}\!\!\!\!\dd{\otild} \otild^m\left[\left(\otild^2+\Lambda_\perp^2\right)^{\frac{n+1}{2}}-\otild^{n+1}\right].
\end{align}
For the convergence of the Binomial series \eqref{eq:binomial_series}, we need to check that $\Lambda_\perp^2 < \otild^2$, which follows from \eqref{eq:qperp_limit_xi_condition},\footnote{We know that $\Lambda_\perp^2 < 4k\lxi\left(1+\frac{k}{\lxi}\right)\approx 4k\lxi$ and thus $\Lambda_\perp^2/\lxi^2 < \frac{4k}{\lxi}\ll 1$, which makes $\Lambda_\perp < \lxi$ and thus $\Lambda_\perp < \otild$.}
and thus,
\begin{align}
&I_{nm}\!=\!\frac{(-1)^m}{n+1}\!\int_{\lxi}^{\frac{p-k}{2}}\!\!\!\!\!\dd{\otild}\!\sum_{j=1}^\infty \!\begin{pmatrix}(n+1)/2 \\ j\end{pmatrix}\!\otild^{n+m+1-2j}(\Lambda_\perp)^{2j}\\
&~=\frac{(-1)^m}{n+1}\Bigg[\!\!\!\!\!\!\!\!\!\sum_{\substack{j=1\\j\neq (n+m+2)/2}}^\infty \!\!\!\!\!\!\!\!\!\!\!\begin{pmatrix}(n+1)/2 \\ j\end{pmatrix}\!\!\left.\frac{\otild^{n+m+2-2j}}{n+m+2-2j}\right|_{\otild=\lxi}^{\frac{p-k}{2}}\!\!\Lambda_\perp^{2j}+\begin{pmatrix}(n+1)/2 \\ (n+m+2)/2\end{pmatrix}\ln\left(\frac{p-k}{2\lxi}\right)\Lambda_\perp^{n+m+2}\Bigg].
\end{align}
We now obtain, up to an additive constant (in $p$), 
\begin{align}
I^{\mathrm{LO}}_{nm}&=\begin{cases}\frac{(-1)^m \Lambda_\perp^2}{2(n+m)}\left(\frac{p-k}{2}\right)^{n+m}, & n+m \neq 0\\
(-1)^m\Lambda_\perp^2\ln(p) , & n+m = 0\end{cases} \\
I^{\mathrm{NLO}}_{nm}&=\begin{cases}\frac{(-1)^m \Lambda_\perp^4B}{(n+1)(n+m-2)} \left(\frac{p-k}{2}\right)^{n+m-2}, & n+m-2 \neq 0\\
\frac{(-1)^m}{n+1}B\Lambda_\perp^4\ln(p) , & n+m -2 = 0,\end{cases} 
\end{align}
where $B=\begin{pmatrix}(n+1)/2 \\ 2\end{pmatrix}$.
In this case, denoting again with $\mathrm{NLO}$ the terms proportional to $1/p$ in the integrand, we obtain
\begin{subequations}
    \begin{align}
        \qhatLO &\sim I_{-2,2} - I_{-4,4} \sim\,\text{const}\,+ \mathcal O\left(\frac{1}{p^{2}}\right) , \\
        \qhatNLO &\sim \frac{1}{p}\left(I_{-2,3}-I_{-4,5}\right) \sim \frac{\text{const}}{p}+\mathcal O\left(\frac{1}{p^2}\right).
    \end{align}
\end{subequations}
Thus, with a transverse momentum cutoff $\Lambda_\perp$ in place, we can explicitly take the limit of infinite jet momentum, $p\to\infty$, and obtain a finite jet quenching parameter $\qhat$. Moreover, it is then sufficient to take the leading-order terms in the integrand (in particular the matrix elements) in an expansion in $1/p$.

\subsection{Large momentum cutoff $\Lambda_\perp$ behavior}
\label{sec:qhat_behavior_large_cutoff}

Removing the momentum cutoff, i.e. taking $\Lambda_\perp\to\infty$, leads to a divergent jet quenching parameter $\qhat$. It is reasonable to assume that this divergence will be logarithmic, as it is for finite but large jet momentum. We will now show this explicitly and calculate the coefficient of the logarithm.

We continue working in the limit $p\to\infty$, and thus take only the leading-order terms in $p$ in the integrand into account. For that we start with Eq.~\eqref{eq:qhat_der_with_delta} with explicit step-functions and perform the coordinate transformation in Eq.~\eqref{eq:coordinate_trafo_differentials_to_qperp} to integrate over $\vb \qperp$,
\begin{align}
\begin{split}
    \qhat &\sim \int_0^\infty\dd{k}\int_{q_\perp < \Lambda_\perp}\dd[2]{\vb{\qperp}}\int_{-\infty}^{k-\frac{\qperp^2}{4k}}\dd{\omega} \qperp^2 f(\vb k)\left(1\pm f(\vb k')\right)\frac{\left|\mathcal M^{gg}_{gg}\right|^2}{p^2 \sqrt{\qperp^2+\omega^2}},
\end{split}
\end{align}
where we have not included the angular integrals.

The behavior for large $\Lambda_\perp$ is of course dominated by the upper integration boundary of the $\qperp$ integral. For large $\Lambda_\perp$, we now split the integral into a part $0 < q_\perp < \lxi$ and $\lxi < \qperp < \Lambda_\perp$. We can choose the scale $\lxi \gg k$, such that $\qperp\gg k$, since the momentum $k$ has a natural upper cutoff coming from the distribution function $f(\vb k)$. Then, also $|\omega|>\frac{\qperp^2}{4k}-k$ must be very large and thus $f(\vb k')\approx 0$. Hence we arrive at the scale separation
\begin{align}
    |\omega|\gg \qperp\gg k. \label{eq:large_lperp_scale_separation}
\end{align}
Using this for the integrand \eqref{eq:qhat_largep_integrand_explicit}, we obtain
\begin{align}
\begin{split}
    \qhat &\sim \int_\lxi^{\Lambda_\perp}\dd{\qperp}q_\perp\int_{-\infty}^{k-\frac{\qperp^2}{4k}}\dd{\omega}\qperp^2 
   \frac{1}{|\omega|^3}\approx 8k^2\int_\lxi^{\Lambda_\perp}\dd{\qperp}\frac{q_\perp^3}{q_\perp^4}= 8k^2\ln\Lambda_\perp+\text{const}\label{eq:qhat_integral_to_make_log_lambda}
\end{split}
\end{align}
We have thus shown that we can write $\qhat$ in that limit as in Eq.~\eqref{eq:qhat_behavior_large_cutoff},
\begin{align}
    \qhat(\Lambda_\perp \gg Q) \simeq a_{\lperp}\ln\Lambda_\perp +b_{\lperp}.
\end{align}

It is also possible to determine the coefficient of $\ln\Lambda_\perp$, similarly as in \app\ref{app:qhat_large_p}.
While in Ref.~\cite{Boguslavski:2023waw}, this was only done for isotropic systems, it is actually quite straightforward to generalize this to anisotropic system if one starts from the collision kernel, which we consider in more detail in Chapter \ref{sec:collkern}.

Starting from the general expression \eqref{eq:formula_Cq_appendix},
\begin{align}
    \begin{split}
    C(\vb q_\perp)&=\frac{1}{2p\nu}\sum_{bcd}\int\frac{\dd[3]{\vb k}}{(2\pi)^3}\dd {q_z} \frac{|\mathcal M^{ab}_{cd}(\vb p, \vb k; \vb p+\vb q, \vb k-\vb q)|^2}{8|\vb k||\vb k-\vb q||\vb p+\vb q|} \\ &\times f(\vb k)(1+f(\vb k-\vb q))\delta(p+|\vb k|-|\vb p+\vb q|-|\vb k-\vb q|),
    \end{split}
\end{align}
we take the limit of large $q_\perp$ following Section \ref{sec:collkern-limits-analytic}. Using
\begin{align}
    \delta(k-q_z-|\vb k-\vb q|)\to \frac{q_\perp^2}{2(k-k_z)^2}\delta(q_z-\frac{\vb q_\perp\cdot \vb k_\perp-q_\perp^2/2}{k-k_z}),
\end{align}
and the form for the matrix elements
\begin{align}
    \left|\mathcal M^{gg}_{gg}\right|^2=16 g^4 \dA\CA^2\frac{4p^2(k-k_z)^2}{\qperp^4}, && \left|\mathcal M^{gq}_{gq}\right|^2=16 g^4 \dF\CF\CA\frac{4p^2(k-k_z)^2}{\qperp^4},
\end{align}
we obtain
\begin{align}
    C(q_\perp\gg \Teps)=\frac{2\CR g^4}{q_\perp^4}\int\frac{\dd[3]{\vb k}}{(2\pi)^3}\left(1-\frac{k_z}{k}\right)\left(\NC f_g (\vb k)+\frac{1}{2}\sum_f f_f(\vb k)\right),
\end{align}
where, according to our symmetry arguments \eqref{eq:distributionfunction-mirrorsymmetry}, the term proportional to $k_z/k$ vanishes.
Integrating this to obtain $\qhat$, we obtain
\begin{align}
    \qhat(\lperp\gg \Teps)\simeq \frac{\CR g^4}{\pi}\int\frac{\dd[3]{\vb k}}{(2\pi)^3}\left(\NC f_g(\vb k)+\frac{1}{2}\sum_f f_f(\vb k)\right)\log\lperp + \mathrm{const}
\end{align}
For a plasma consisting of quarks and gluons with distributions $f_q$ and $f_g$.

For isotropic systems, we obtain
\begin{align}
\begin{split}
    \frac{a_{\lperp}}{\CR}&=\frac{\NC g^4}{2\pi^3}\int_0^\infty \dd{k}k^2f_g(k)+\sum_{f}\frac{g^4}{4\pi^3}\int_0^\infty \dd{k} k^2 f_f(k),
    \end{split}
\end{align}
which reduces for thermal distributions to
\begin{align}
    \frac{a_{\lperp}^{\mrm{eq}}}{\CR}=\frac{g^4\zeta(3)T^3}{\pi^3}\left(\NC + \frac{3}{4}\nf\right).\label{eq:qhat_a_coefficient_equilibrium}
\end{align}
This nicely agrees with Eq.~\eqref{eq:qhat_hard_arnold} that stems from \cite{Arnold:2008vd}.

%% file: 960_amyrates_numericaldetails.tex
In this Appendix, more details are provided on the numerical method used to solve the integral equation \eqref{eq:integralequation-amy},
\begin{align}
    2\vb h &= i\delta E(\vb h)\vb F(\vb h)+\frac{1}{2}\int\frac{\dd[2]{\vb q_\perp}}{(2\pi)^2}C(\vb q_\perp)\label{eq:app-integralequation-amy}\\
    &\times \left[(3\vb F(\vb h)-\vb F(\vb h-p\vb q_\perp)-\vb F(\vb h-k\vb q_\perp)-\vb F(\vb h+p'\vb q_\perp)\right],\nonumber
\end{align}
with $\delta E(\vb h)=m_D^2/4\times (1/k+1/p-1/p')+h^2/(2pkp')$.
This needs to be solved for the function $F(\vb h)$, which enters the rate via Eq.~\eqref{eq:gammarate}
\begin{align}
    \gamma=\frac{p^4+p'^4+k'^4}{p^3p'^3k'^3}\frac{d_A\alpha_s}{2(2\pi)^3}\int\frac{\dd[2]{\vb h}}{(2\pi)^2}2\vb h\cdot \mathrm{Re} \vb F, \label{eq:appendix-gammarate}
\end{align}
While in this thesis, the purely gluonic case is considered, quarks can be included by taking different color factors in \eqref{eq:app-integralequation-amy} and different vacuum splitting functions in \eqref{eq:appendix-gammarate}.

\section{Equations and boundary conditions}
First, we transform the integral equation to impact parameter space using the Fourier transformed quantities,
\begin{align}
    \vb F(\vb x)=\int\frac{\dd[2]{\vb h}}{(2\pi)^2}e^{i\vb x\cdot\vb h}\vb F(\vb h), && \vb F(\vb h)=\int\dd[2]{\vb x}e^{-i\vb x\cdot\vb h}\vb F(\vb x), \label{eq:app-fouriertransform-2d}\\
    \tilde C(\vb x)=\int\frac{\dd[2]{\vb q_\perp}}{(2\pi)^2}e^{i\vb x\cdot\vb q_\perp}C(\vb q_\perp).
\end{align}
Here, I have introduced a tilde on the Fourier-transformed collision kernel because it is more convenient to think of the object
\begin{align}
    C(\vb x)=\tilde C(0)-\tilde C(\vb x)=\int\frac{\dd[2]{\vb q_\perp}}{(2\pi)^2}(1-e^{i\vb x \cdot \vb q_\perp})C(\vb q_\perp).
\end{align}
With the abbreviations
\begin{align}
    A=im(z,P)=i\left(\frac{m_{\infty,z}^2}{2zP}+\frac{m_{\infty,(1-z)}^2}{2(1-z)P}-\frac{m_{\infty,1}^2}{2P}\right), && B=\frac{i}{2Pz(1-z)}
\end{align}
we may rewrite the integral equation \eqref{eq:app-integralequation-amy} into the shorter form 
\begin{align}
    2\vb h=\vb F(\vb h)(A+B\vb h^2)+\frac{1}{2}\int\frac{\dd[2]{\vb q}}{(2\pi)^2}C(\vb q)
    \left\{3\vb F(\vb h)-\vb F(\vb h-\vb q)-\vb F(\vb h-z\vb q)-\vb F(\vb h-(1-z)\vb q)\right\}.
\end{align}
Note that $A, B \in\mathbb C$ but purely imaginary such that $iA \in \mathbb R$, $A/B\in\mathbb R$.

Inserting now the Fourier transforms \eqref{eq:app-fouriertransform-2d}, we obtain Eq.~\eqref{eq:impactparameterspaceequation} in impact parameter space,
\begin{align}
    (A-D(z,\vb x)-B\nabla^2)\vb F(\vb x)=-2i\nabla \delta(\vb x),\label{eq:app-impactparameterspaceequation2}
\end{align}
where we have introduced the function
\begin{align}
    D(z,\vb x)=-\frac{1}{2}(C(\vb x)+C((1-z)\vb x)+C((1-z)\vb x)).
\end{align}

Methods to solve this equation for isotropic $D(z, b)$ have been described in Refs.~\cite{Aurenche:2002wq, Moore:2021jwe}. Building on these previous methods, I describe here the method used in this thesis to solve this equation for any anisotropic $D(\vb b)=D(b,\phi)$.

First, note that the delta function can be seen as imposing a boundary condition on $\vb F$, which can be seen by considering the most singular terms
\begin{align}
    -2i\nabla\delta(\vb x)=-B\nabla^2 F(\vb x).\label{eq:differentialequation-delta}
\end{align}
This is solved by going to Fourier space, where we can use
\begin{align}
    \delta (\vb x)=\int\frac{\dd[2]{\vb q_\perp}}{(2\pi)^2}e^{i\vb q_\perp\cdot \vb x},
\end{align}
such that $\vb F(q_\perp) = 2\frac{\vb q_\perp}{B q_\perp^2}$. The backward Fourier transform can be done using the parameterization $\vb x = |\vb x|(\cos\phi_x,\sin\phi_x)$, $\vb q_\perp=q_\perp(\cos\phi_q,\sin\phi_q)$,
\begin{align}
    \int\dd[2]{\vb q_\perp}\frac{\vb q_\perp}{q_\perp^2}e^{i\vb x\cdot\vb q_\perp}&=|\vb x| \int_0^{2\pi}\dd\phi_q\int_0^\infty\dd q_\perp \begin{pmatrix}
        \cos\phi_q\\\sin\phi_q
    \end{pmatrix} e^{i|\vb x| q_\perp \cos(\phi_q-\phi_x)}\\
&=|\vb x|\int_0^{2\pi}\dd{\tilde\phi_q}\int_0^\infty\dd{q_\perp}\begin{pmatrix}
        \cos(\tilde \phi_q+\phi_x)\\ \sin(\tilde\phi_q + \phi_x)
    \end{pmatrix} e^{i|\vb x| q_\perp \cos\tilde\phi_q}
    \\&=|\vb x|\int_0^{2\pi}\dd{\tilde\phi_q}\int_0^\infty\dd{q_\perp}\begin{pmatrix}
        \cos\phi_x\cos\tilde\phi_q-\sin\phi_x\sin\tilde\phi_q\\ \sin\phi_x\cos\tilde\phi_q+\cos\phi_x\sin\tilde\phi_q
    \end{pmatrix} e^{i|\vb x| q_\perp \cos\tilde\phi_q},\\
    &=2\pi i \frac{\vb x}{\vb x^2}
\end{align}
where we used
\begin{align}
    \int_0^{2\pi}\dd\phi \cos\phi e^{i |\vb x| q_\perp \cos\phi}=2\pi i J_1(q_\perp |\vb x|), && \int_0^{2\pi}\dd\phi \sin\phi e^{i |\vb x| q_\perp \cos\phi}=0.
\end{align}

Eq.~\eqref{eq:differentialequation-delta} (and the full differential equation \eqref{eq:app-impactparameterspaceequation2}) is thus solved by
\begin{align}
	\lim_{|\vb x|\to 0}\vb F(\vb x)=\frac{i}{B\pi}\frac{\vb x}{\vb b^2}.\label{eq:boundarycondition}
\end{align}
This we can take as a boundary condition and solve Eq.~\eqref{eq:app-impactparameterspaceequation2} for $|\vb x|>0$.

Next, we redefine $\vb F=|\vb x| {\vb g }=|\vb x|(g_x, g_y)$, for which the differential equation becomes
\begin{align}
    \left[\frac{A-D(|\vb x|,\phi)}{B}-\partial_{|\vb x|}^2-\frac{3}{|\vb x|}\partial_{|\vb x|}-\frac{1}{\vb x^2}-\frac{1}{\vb x^2}\partial_\phi^2\right]\vb g(|\vb x|,\phi)=0.
\end{align}
This redefinition has the advantage that the real part of the integral appearing in \eqref{eq:appendix-gammarate} can be obtained by the function $\vb g$ evaluated at $\vb x\to 0$.
\begin{align}
    \int\frac{\dd[2]{\vb h}}{(2\pi)^2}\RE\,\left(\vb h\cdot \vb F(\vb h)\right)&=\RE\,\int\frac{\dd[2]{\vb h}}{(2\pi)^2}\int\dd[2]{\vb x}e^{-i\vb x\cdot\vb h}\vb F(\vb x)\cdot\vb h\\
    &=\RE\,\int\frac{\dd[2]{\vb h}}{(2\pi)^2}\int\dd[2]{\vb x}(i\vb \nabla_x)e^{-i\vb x\cdot\vb h}\cdot \vb F(\vb b)\\
    &=\RE\,\int\frac{\dd[2]{\vb h}}{(2\pi)^2}\int\dd[2]{\vb x}e^{-i\vb x\cdot\vb h}(-i\nabla\cdot\vb F(\vb x))\\
    &=\RE\,\left(-i\nabla\cdot\vb F(\vb x)\right)_{\vb x=0}\\
    &=\IM\,\left(\nabla\cdot\vb F(\vb x)\right)_{\vb x=0}\\
    &=\IM\,\left[\frac{g_x(0)}{\cos\phi}+\frac{g_y(0)}{\sin\phi}\right].\label{eq:app-whatwewanttocompute}
\end{align}

For the angular information, we decompose $\vb g$ in its Fourier modes,
\begin{align}
    \vb g(|\vb x|,\phi)&=\sum_n\vb g_n(|\vb x|)e^{in\phi}, \\
    D(|\vb x|,\phi)&=\sum_m D_m(|\vb x|)e^{im\phi},\\
    D_m(|\vb x|)&=\frac{1}{2\pi}\int_0^{2\pi}\dd{\phi}e^{-im\phi}D(|\vb x|,\phi).
\end{align}
In terms of these Fourier modes $\vb g_n$, the differential equation becomes
\begin{align}
    \left[\frac{A}{B}-\partial_b^2-\frac{3}{|\vb x|}\partial_b+\frac{n^2-1}{\vb x^2}\right]\vb g_n(|\vb x|)=\sum_m\frac{D_m(|\vb x|)}{B}\vb g_{n-m}(|\vb x|)\label{eq:equation_gn}
\end{align}
Note that the boundary condition \eqref{eq:boundarycondition} only affects the modes $n=\pm1$, which 
can be seen by rewriting Eq.~\eqref{eq:boundarycondition} using exponential functions,
\begin{align}
	\lim_{|\vb x|\to 0}\vb g(|\vb x|,\phi)=\frac{1}{2B\pi}\frac{1}{\vb x^2}\begin{pmatrix}
	    i(e^{i\phi}+e^{-i\phi})\\e^{i\phi}-e^{-i\phi}
	\end{pmatrix}.
    \label{eq:boundarycondition-exponentials}
\end{align}
This fixes the small $|\vb x|$ limit of $\vb g_{\pm 1}$.
In the isotropic limit, $D_m\sim \delta_{m0}$ and does not couple different Fourier modes. Therefore, in this isotropic limit, only the modes for $m=\pm 1$ need to be solved for.

Let us now consider the case of small $|\vb x|$, where $D\sim \vb x^2\log |\vb x|$ can be neglected against $A$, i.e., $|D_m(|\vb x|)|\ll |A|$. In this region, Eq.~\eqref{eq:equation_gn} can be solved analytically,
\begin{align}
	\vb g_n(|\vb x|)=\frac{\vb c_1 I_{|n|}(|\vb x|\sqrt{A/B}) + \vb c_2 K_{|n|}(|\vb x|\sqrt{A/B})}{|\vb x|},
\end{align}
where $I_n$ and $K_n$ are the modified Bessel functions of the first and second kind.
Thus, at small $x$, the general solution is given by the linear combination
\begin{align}
	\vb g(|\vb x|,\phi)=\sum_n e^{in\phi} \frac{\vb c_I^n I_{|n|}(|\vb x|\sqrt{A/B}) + \vb c_K^n K_{|n|}(|\vb x|\sqrt{A/B})}{|\vb x|}.\label{eq:general-solution}
\end{align}
In practice, we will need to truncate the series, and only consider modes with $-\nmax \leq  n \leq \nmax$, leading to $n_{\mathrm{fourier}}=2n_{\mathrm{max}}+1$ different Fourier modes for every component of $\vb g = (g_x, g_y)$. We then have $2(2n_{\mathrm{max}}+1)$ linearly independent solutions at small $|\vb x|$. 
For a general solution, we thus need $2(2\nmax +1)$ boundary conditions.

Eq.~\eqref{eq:equation_gn} is a coupled system of $2n_{\mathrm{max}}+1$ ordinary second-order differential equations, for which we need $2(2n_{\mathrm{max}}+1)$ boundary conditions, which fix all the $c_I^n$ and $c_K^n$ uniquely. One natural boundary condition to impose is regularity at infinity \cite{Aurenche:2002wq},
\begin{align}
	\lim_{|\vb x|\to\infty}\vb g_n(|\vb x|)=0,\label{eq:boundary_cond_infinity}
\end{align} 
which yields $2n_{\mathrm{max}}+1$ conditions for every component of $\vb g$.
We also have the boundary condition \eqref{eq:boundarycondition-exponentials} at small $|\vb x|$. To achieve that, we expand the Bessel functions for small $|\vb x|$, 
\begin{align}
 I_n(|\vb x|)/|\vb x| \sim |\vb x|^{n-1}(1+\mathcal O(|\vb x|)), && K_n(|\vb x|)/|\vb x|\sim |\vb x|^{n-1}(1+\mathcal O(|\vb x|)) + \# |\vb x|^{-n-1}(1+\mathcal O(|\vb x|))   .
\end{align}
Comparing with \eqref{eq:boundarycondition}, we conclude that this fixes $c_K^{\pm 1}$ (2 additional conditions per component of $\vb g$). Furthermore, no function may diverge more quickly than $1/\vb x^2$ at the origin, which fixes $\vb c_K^m=0$ for $m\geq 2$, resulting in $2(n_{\mathrm{max}}-1)$ additional conditions per component of $\vb g$.
With that we are missing one (complex) boundary condition to determine the system completely\footnote{An interested reader might ask if excluding the $n=0$ modes would solve the problem: This would reduce the number of independent solutions by $2$, and the system would then be overdetermined.}. Since we are interested in the imaginary part of the constant to which $\vb g_n$ converges (see Eq.~\eqref{eq:app-whatwewanttocompute}) for $|\vb x|\to 0$, and since both $K_0/|\vb x|$ and $I_0/|\vb x|$ diverge in that limit, we can use one complex (or two real) boundary conditions to set $\IM \,\vb c_I^0=\IM\, \vb c_K^0=0$. To summarize, the boundary conditions are given by
\begin{subequations}\label{eq:app-boundaryconditions-combined}
\begin{align}
    \lim_{|\vb x|\to\infty}\vb g_n(\vb x)&=\vb 0,\label{eq:boundary_cond_infinity2}\\
	\vb c_K^1&=\frac{\sqrt{A/B}}{2B\pi}\begin{pmatrix}
	    i\\1
	\end{pmatrix},\\
    \vb c_K^{-1}&=\frac{\sqrt{A/B}}{2B\pi}\begin{pmatrix}
	    i\\-1
	\end{pmatrix},\\
    \RE\,\vb c_I^0&=\vb 0,\\
    \RE\,\vb c_K^0 &= \vb 0,
\end{align}
\end{subequations}
where the $\vb c_I$ and $\vb c_K$ coefficients are given by the small $x$ approximation of the full solution.

To solve the differential equation with these boundary conditions, we make use of the fact that since it is a linear homogeneous equation, any linear combination also solves the equation. We thus solve $m$ linearly independent systems $\{\vb g_n^{(m)}\}$ with $m$ linearly independent initial conditions
\begin{align}
	\vb c_K^n&=0, & |n| \geq 2, \label{eq:initialcondition-enforcement}
\end{align}
and initialize every system with exactly one nonzero coefficient.
This leads to $2(2n_{\mathrm{max}}+1) - 2(n_{\mathrm{max}}-2)=2\nmax + 6$ linearly independent sets of solutions $\{\vb g_n^{(m)}\}$. The full solution may be obtained by superimposing
\begin{align}
	\vb g(|\vb x|,\phi)=\sum_{m}\vb a_m  g^{(m)}(|\vb x|,\phi),
\end{align}
by choosing the coefficients $\vb a_m$ such that the boundary condition \eqref{eq:app-boundaryconditions-combined} is fulfilled.
We have now put the information of the vector components of $\vb g$ into the coefficients $\vb a_m$. The advantage of that is that we only need to solve the $m$ systems once, and can then obtain two different sets of coefficients $\{a_m^x, a_m^y\}$.
In practice, this leads to
a linear system for $\vb a_m$,
\begin{align}
    \sum_m \vb a_m\lim_{|\vb x|\to\infty} g_n^{(m)}(|\vb x|)=0,\label{eq:linearsystem}
\end{align}
which can be solved for a sufficiently large $\xmax$, and supplemented by \eqref{eq:app-boundaryconditions-combined},
\begin{subequations}\label{eq:linearsystem-complete}
\begin{align}
    \sum_m \vb a_m g_n^{(m)}(\xmax)&=0\\
    \sum_m \vb a_m c_K^1{}^{(m)}&=\frac{\sqrt{A/B}}{2B\pi}\begin{pmatrix}
	    i\\1
	\end{pmatrix},\label{eq:actual_boundary-condition1}\\
    \sum_m\vb a_m  c_K^{-1}{}^{(m)}&=\frac{\sqrt{A/B}}{2B\pi}\begin{pmatrix}
	    i\\-1
	\end{pmatrix},\label{eq:actual_boundary-condition2}\\
    \sum_m  \RE\,\vb a_m c_I^0{}^{(m)}&=\vb 0,\\
    \sum_m \RE\,\vb a_m c_K^0{}^{(m)} &= \vb 0, 
\end{align}
\end{subequations}
I have checked explicitly that the results are independent of the choice of $\xmax$.

Every system with index $(m)$ is initialized at a value $x=x_{\mathrm{min}}$, chosen such that $D(x_{\mathrm{min}},\phi)/A<0.00001$ with exactly one coefficient $c_{I,K}^n$ (taken to be a scalar not a vector) in Eq.~\eqref{eq:general-solution} nonzero (except for those in Eq.~\eqref{eq:initialcondition-enforcement}). For every system $(m)$, the system \eqref{eq:equation_gn} is then integrated outwards using a fourth-fifth order Runge Kutta with time stepping until the absolute value of one of the solutions becomes larger than a threshold.
Of all the systems, we then take the smallest of the maximum $x$, and solve the linear system \eqref{eq:linearsystem-complete}. For larger $\nmax$, the linear system becomes increasingly difficult to solve numerically, as the \texttt{scipy}\footnote{\cite{2020SciPy-NMeth}} routine used to solve the linear system encounters an almost singular matrix.
Finally, the solution can be written as
\begin{align}
	\vb g(|\vb x|,\phi)=\sum_m \vb a_m\sum_n e^{in\phi} \frac{c_I^n{}^{(m)} I_{|n|}(|\vb x|\sqrt{A/B}) + c_K^n{}^{(m)} K_{|n|}(|\vb x|\sqrt{A/B})}{|\vb x|},\label{eq:general-solution_finally}
\end{align}
where the $c_{I,K}^n{}^{(m)}$ are the initial conditions.

Eventually, the prefactor of the $I_1$ solution is needed, because in the limit $|\vb x|\to 0$, only the $I_1(|\vb x|\sqrt{A/B})=\frac{\sqrt{A/B}}{2}+\mathcal O(\vb x^2)$ contribute. Thus, at small $|\vb x|$,
\begin{align}
    \vb g(|\vb x|,\phi) = \sum_m\vb a_m \frac{\sqrt{A/B}}{2}\left(c_I^{1(m)}e^{i\phi}+c_I^{-1(m)}e^{-i\phi}\right).
\end{align}
Comparing with \eqref{eq:app-whatwewanttocompute}, this clearly induces additional requirements\footnote{These requirements amount to the whole function $g(\phi)$ to be odd or even.} on the coefficients $(a_m)_x$ and $(a_m)_y$, which we take as an uncertainty measure for our method. We thus obtain
\begin{subequations}\label{eq:actually-getting-the-coefficients}
\begin{align}
    \IM \, g_x(0)/\cos\phi&=\sqrt{A/B}\,\IM\sum_m (a_m)_x c_I^{1(m)}=\sqrt{A/B}\,\IM\,\sum_m(a_m)_xc_I^{-1(m)}\\
    \IM \, g_y(0)/\cos\phi&=\sqrt{A/B}\,\IM\left(i\sum_m (a_m)_x c_I^{1(m)}\right)=-\sqrt{A/B}\,\IM \left(i\sum_m(a_m)_xc_I^{-1(m)}\right).
\end{align}
\end{subequations}
Now I want to briefly discuss that these additional requirements on the coefficients are actually not an additional equation but follow from the system \eqref{eq:linearsystem-complete}. To do that, we consider as a toy model the simplified case of first having only 4 sets and only considering the modes $n=\pm1$. With that assumption, the function $g^{(m)}$ for one set of initial conditions $c_i^{(m)}$ can be written as
\begin{align}
    g^{(m)}=e^{i\phi}(c_1^{(m)}f_1(|\vb x|)+c_2^{(m)}f_2(|\vb x|))+e^{-i\phi}(c_3^{(m)}f_1(|\vb x|)+c_4^{(m)}f_2(|\vb x|)).
\end{align}
As initial conditions, we assume that we use $c_i^{(m)}=\delta_i^m$.
In that simplified case, the requirement of vanishing at infinity leads to
\begin{align}
    \begin{split}\label{eq:linearsystem-sample}
    c_1 a_1+c_2 a_2+c_3 a_3+c_4 a_4&=0\\
    c_3 a_1 + c_4 a_2 + c_1 a_3 + c_2 a_4&=0, 
    \end{split}
\end{align}
where we have used that $g^{(1)}$ and $g^{(3)}$ are similar because they are both initialized with the $f_1$ function. As an additional boundary condition, we need to have $a_1=\pm a_3$, enforcing symmetric or antisymmetric boundary condition (corresponding to the cosine or sine in \eqref{eq:boundarycondition-exponentials}, and additionally fixing the value of one coefficient, e.g., $a_1=1$. Thus, we have the additional conditions (mimicking \eqref{eq:actual_boundary-condition1} and \eqref{eq:actual_boundary-condition2})
\begin{align}
    a_3=\pm a_1, && a_1 = 1. \label{eq:toymodel-symmetricondition}
\end{align}
Inserting this into \eqref{eq:linearsystem-sample}, we can subtract (or add) those two equations to obtain
\begin{align}
    a_2(c_2 \mp c_4)+a_4(c_4\mp c_2)=0,
\end{align}
which leads to $a_2=\pm a_4$, i.e., the symmetry condition for \eqref{eq:toymodel-symmetricondition} enforces the same symmetry on the other coefficients $a_2$ and $a_4$.

While this holds for only considering two modes, $n=\pm 1$, this is also true for adding higher modes. For instance, let us consider now adding additionally $n=\pm 2$, such that Eq.~\eqref{eq:linearsystem-sample} then reads
\begin{subequations}
\begin{align}
    c_1 a_1 + c_2 a_2 + c_3 a_3+c_4 a_4 + c_5 a_5 + c_6 a_6 &= 0\\
    c_3 a_1 + c_4 a_2 + c_1 a_3 + c_2 a_4 + c_6 a_5 + c_5 a_6 &=0\\
    c_ 7 a_1 + c_8 a_2 + c_9 a_3+c_{10}a_4+c_{11}a_5+c_{12}a_6&=0\\
    c_9 a_1 + c_{10}a_2 + c_7 a_3 + c_8 a_4+c_{12} a_5 + c_{11} a_6&=0
\end{align}
\end{subequations}
which, with the condition \eqref{eq:toymodel-symmetricondition} leads to
\begin{align}
    0&=a_2(c_2\mp c_4)+a_4(c_4\mp c_2)+a_5(c_5\mp c_6)+a_6(c_6\mp c_5)\\
    0&=a_2(c_8\mp c_{10})+a_4(c_{10}\mp c_8)+a_5(c_{11}\mp c_{12})+a_6 (c_{12}\mp c_ {11}),
\end{align}
which, similar than before, upon eliminating $a_5$ and $a_6$, leads to
\begin{align}
    0&=a_2\left(\frac{c_2\mp c_4}{c_5\mp c_6}-\frac{c_8\mp c_{10}}{c_{11}\mp c_{12}}\right)+a_4\left(\frac{c_4\mp c_2}{c_5\mp c_6}-\frac{c_{10}\mp c_8}{c_{11}\mp c_{12}}\right).
\end{align}
Similar as before, this leads to $a_2=\pm a_4$, and thus to the same symmetry condition as for the other coefficients $a_2$ and $a_4$.

Thus, the symmetry between $c_I^{1(m)}$ and $c_I^{-1(m)}$ in Eq.~\eqref{eq:actually-getting-the-coefficients} is not an additional input but a consequence of the linear system \eqref{eq:linearsystem-complete}.

\section{Numerical cross-checks}
\begin{figure}
    \centering
    \centerline{
    \includegraphics[width=0.5\linewidth]{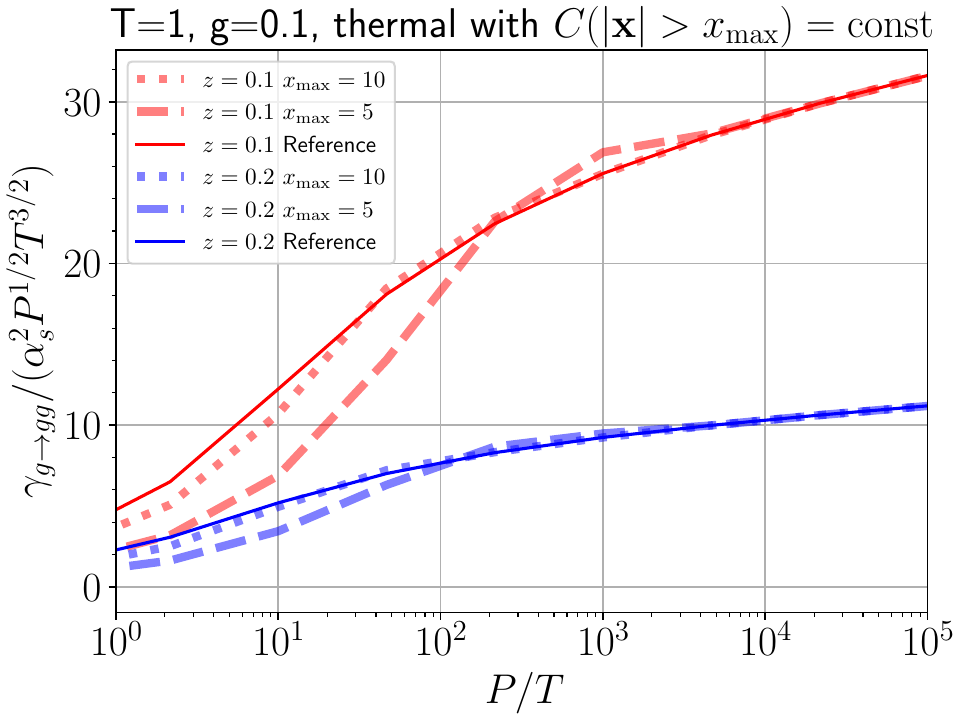}
    \includegraphics[width=0.5\linewidth]{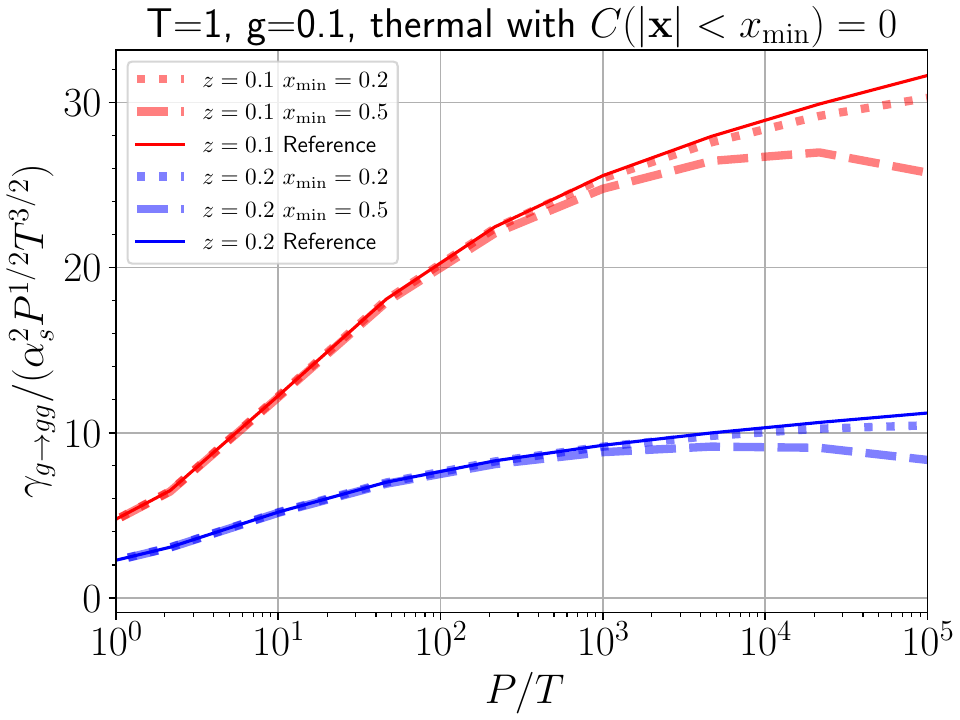}
    }
    \caption{Numerically obtaining the rate $\gamma$ for Eq.~\eqref{eq:Cb_eq_appr_appendix} with small $|\vb x|$ (left) or large $|\vb x|$ (right) behavior modified.}
    \label{fig:rate_isotropic_crosscheck}
\end{figure}
As a numerical crosscheck, the rate is calculated for an isotropic dipole cross section \eqref{eq:Cb_eq_appr},
\begin{align}
    \Ceqappr(|\vb x|)=\frac{\CA g^2 T}{2\pi}\left(\gamma_E+K_0(m_D|\vb x|)+\log\frac{m_D|\vb x|}{2}\right).\label{eq:Cb_eq_appr_appendix}
\end{align}
This is compared with a variation in the small or large $|\vb x|$ region. In the left panel of \fig\ref{fig:rate_isotropic_crosscheck}, the large $|\vb x|$ behavior is modified such that it remains constant for $|\vb x|>x_{\mathrm{max}}$. In the right panel the small $|\vb x|$ behavior is modified, such that $C(|\vb x|<x_{\mathrm{min}})=0$. We find that modifying the small $|\vb x|$ behavior changes the rate for large parton energies $P$, while changing the large $|\vb x|$ behavior results in modifications for small parton energies.

\begin{figure}
    \centering
    \includegraphics[width=0.5\linewidth]{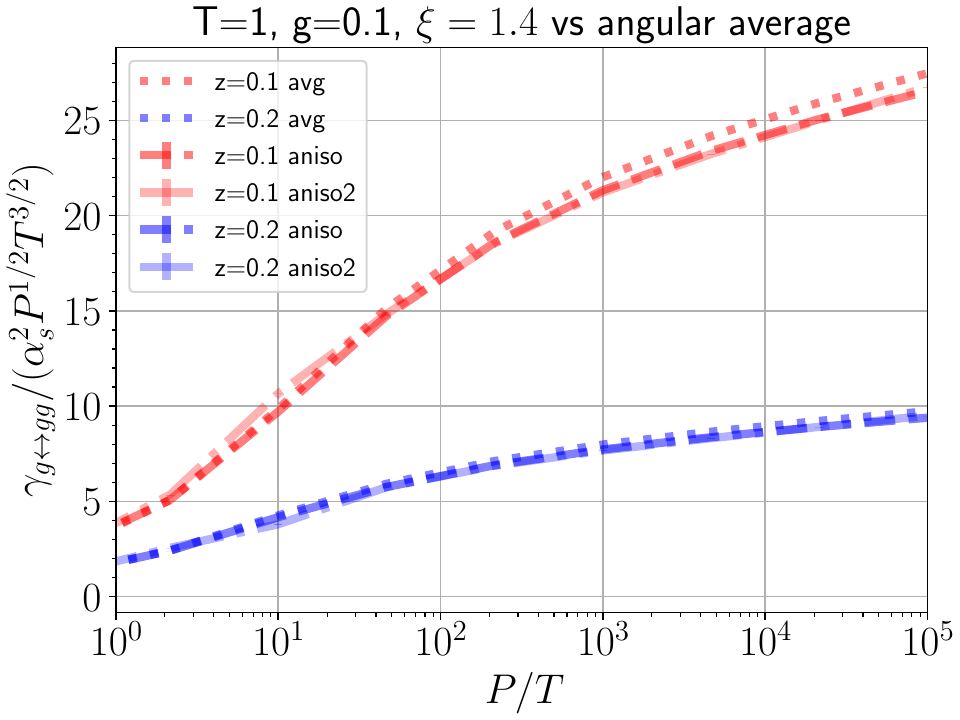}
    \caption{Splitting rate for an anisotropic dipole cross section \eqref{eq:Cb_eq_appr_appendix} with \eqref{eq:debyemass-anisotropic-model} and comparison to the rate from the angular averaged kernel.}
    \label{fig:rate_anisotropic_crosscheck}
\end{figure}
In Fig.~\ref{fig:rate_anisotropic_crosscheck}, the anisotropic method is tested using \eqref{eq:Cb_eq_appr_appendix}, but an anisotropic Debye mass
\begin{align}
    m_D^2(\phi)=\left(1-\frac{2\xi}{3}+\xi\cos^2\phi\right)m_D,\label{eq:debyemass-anisotropic-model}
\end{align}
also used in Ref.~\cite{Hauksson:2023dwh} based on Ref.~\cite{Romatschke:2003ms}. This model is used as a numerical cross-check. Its physical origin stems from a simple anisotropic plasma model using a squeezed thermal distribution, where this relation for the dominant mass $m_+$ is found in Ref.~\cite{Romatschke:2003ms} in the static limit for small anisotropies. Using this model, $T$ does no longer represent the actual effective temperature $\Teps$ from \eqref{eq:Landau-matching-condition}, but it is still used here as a dimensionful scale. The results for the rather large anisotropy $\xi=1.4$ are shown in Fig.~\ref{fig:rate_anisotropic_crosscheck}, which we compare with the rate from the angular averaged kernel \eqref{eq:collkern-angular-average}. Two different sets of parameters are used for the number of Fourier modes, and grid values $|\vb x_i|$ where the Fourier modes are calculated, and are shown as different line styles in Fig.~\ref{fig:rate_anisotropic_crosscheck}.
Both anisotropic evaluations agree very well.
As in the main text in Chapter \ref{sec:collkern}, we find that the angular averaged kernel provides a good approximation of the anisotropic kernel when considering the rate. Note the small deviations for $z=0.1$ despite the extremely anisotropic kernel with $\xi=1.4$.

%% file: 900-additional-bottomup-things.tex
In this Appendix, further results and plots for the simulations performed for this thesis are provided.
\begin{figure}
    \centering
    \includegraphics[width=0.5\linewidth]{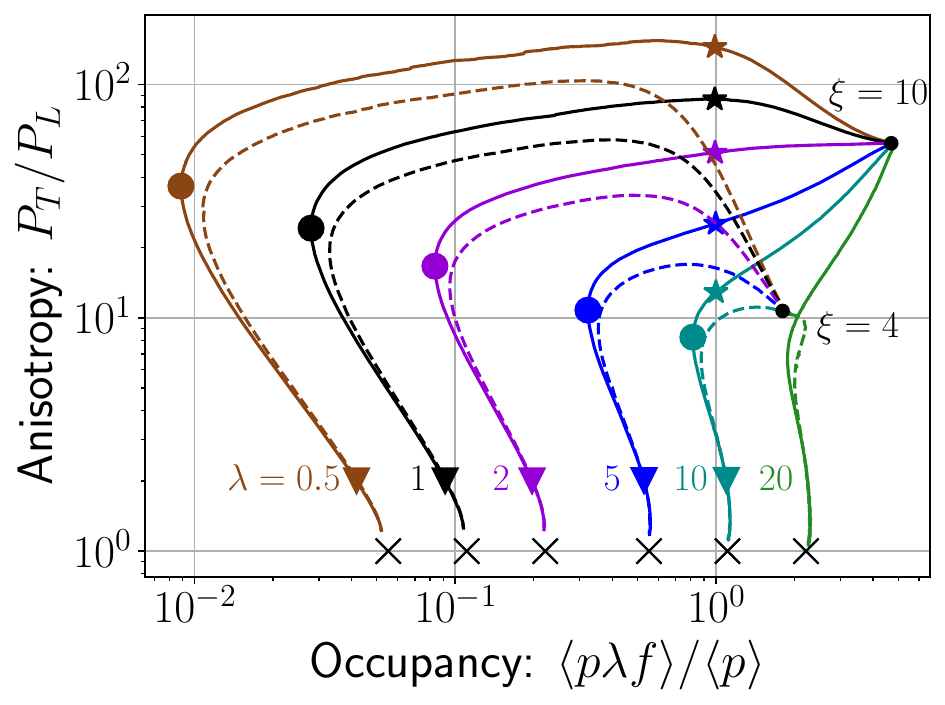}
    \caption{Time evolution of expanding systems in the anisotropy-occupancy plane for various couplings for a Debye-like screening prescription.}
    \label{fig:overview-curves-appendix}
\end{figure}
\begin{figure}
    \centering
    \centerline{
    \includegraphics[width=0.5\linewidth]{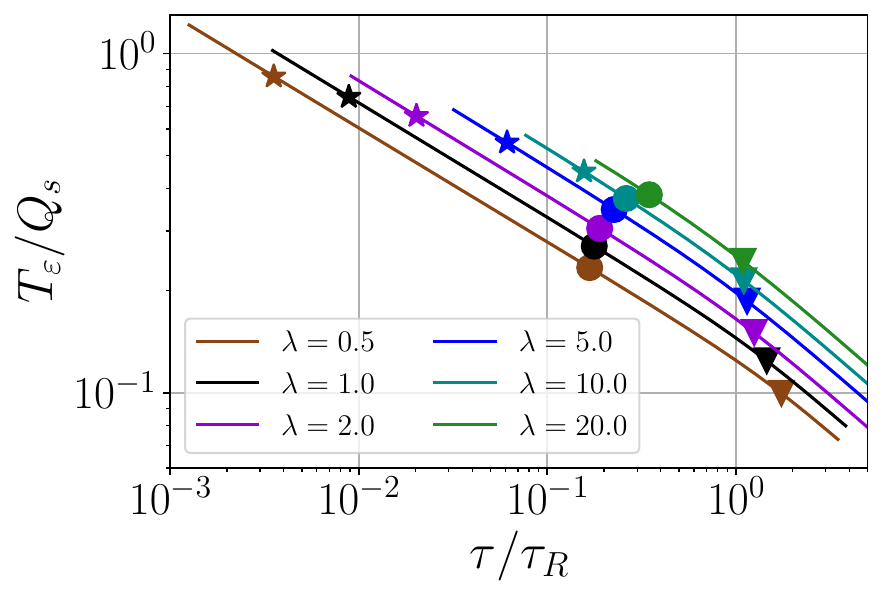}
    \includegraphics[width=0.5\linewidth]{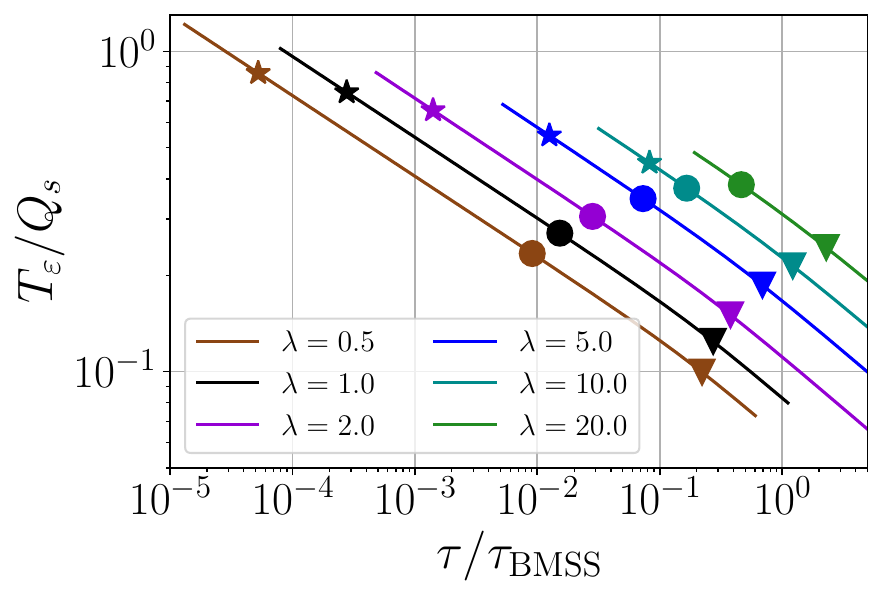}
    }
    \caption{Time evolution of the effective temperature $\Teps$ from Eq.~\eqref{eq:Landau-matching-condition} of an expanding system. (\emph{Left}: Time is rescaled with the relaxation time \eqref{eq:relaxation-time}. (\emph{Right}): Time is rescaled with the bottom-up timescale \eqref{eq:bottom-up-timescale}.}
    \label{fig:temperature-evolution}
\end{figure}
\begin{table}[]
    \centering
    \include{tables/table_export_xi10.0_m1_L0.5.tex}
    \caption{Simulation results of an expanding system with $\lambda=0.5$ and Debye-like screening prescription.}
    \label{tab:simulation-lambda05-debyelike}
\end{table}
\begin{table}[]
    \centering
    \include{tables/table_export_xi10.0_m1_L2.0.tex}
    \caption{Simulation results of an expanding system with $\lambda=2$ and Debye-like screening prescription.}
    \label{tab:simulation-lambda2-debyelike}
\end{table}
\begin{table}[]
    \centering
    \include{tables/table_export_xi10.0_m1_L10.0_GeV}
    \caption{Simulation results of an expanding system with $\lambda=10$ and Debye-like screening prescription. Values in \unit{\giga\electronvolt} and \unit{\femto\meter/\c} are obtained by using $Q_s=\qty{1.4}{\giga\electronvolt}$.}
    \label{tab:simulation-lambda10-debyelike}
\end{table}
\begin{table}[]
    \centering
    \include{tables/table_export_xi10.0_m101_L0.5.tex}
    \caption{Simulation results of an expanding system with $\lambda=0.5$ and isoHTL screening prescription.}
    \label{tab:simulation-lambda05-isoHTL}
\end{table}
\begin{table}[]
    \centering
    \include{tables/table_export_xi10.0_m101_L2.0.tex}
    \caption{Simulation results of an expanding system with $\lambda=2$ and isoHTL screening prescription.}
    \label{tab:simulation-lambda2-isoHTL}
\end{table}
\begin{table}[]
    \centering
    \include{tables/table_export_xi10.0_m101_L10.0_GeV}
    \caption{Simulation results of an expanding system with $\lambda=10$ and isoHTL screening prescription. Values in \unit{\giga\electronvolt} and \unit{\femto\meter/\c} are obtained by using $Q_s=\qty{1.4}{\giga\electronvolt}$.}
    \label{tab:simulation-lambda10-isoHTL}
\end{table}

Fig.~\ref{fig:overview-curves-appendix} shows the time evolution of an expanding system in the anisotropy-occupancy plane, similar to Fig.~\ref{fig:bottomup-overview}, but with more couplings and initial conditions. Fig.~\ref{fig:temperature-evolution} depicts the temperature of an expanding system with the Debye-like screening prescription.
For couplings $\lambda\in\{0.5,2,10\}$, several results are also tabulated in Tabs.~\ref{tab:simulation-lambda05-debyelike} - \ref{tab:simulation-lambda10-isoHTL}. There, one may easily obtain numerical values for the effective temperature $\Teps$ (Eq.~\eqref{eq:Landau-matching-condition}), infrared temperature $T_\ast$ (Eq.~\eqref{eq:tstar-definition}), screening mass \eqref{eq:debyemass-general}, occupancy $\langle pf\rangle/\langle p\rangle$ and pressure ratio $P_L/P_T$. The energy density $\varepsilon$ can be obtained from the effective temperature $\Teps$ via Eq.~\eqref{eq:Landau-matching-condition}. For the relaxation time $\tauR$, the values of $\eta/s$ from Tab.~\ref{tab:etas_values} are used. In the left column, the times corresponding to the star, circle and triangle marker are indicated with the corresponding symbol. For $\lambda=10$, the time is additionally given in \unit{\femto\meter/\c} and the effective temperature in \unit{\giga\electronvolt} using $Q_s=\qty{1.4}{\giga\electronvolt}$.

%% file: tables/table_export_xi10.0_m1_L0.5.tex
\begin{tabular}{c c c c c c c c c} \toprule
& \multicolumn{3}{c}{$\lambda=0.5$,} & \multicolumn{5}{c}{Screening: Debye-like}\\
 & $Q_s\tau$ & $\tau/\tau_{\mathrm{R}}$ & $\tau/\tau_{\mathrm{BMSS}}$ & $T_\varepsilon/Q_s$ & $T_*/Q_s$ &$m_D/Q_s$  & $\langle pf \rangle/\langle p\rangle$ & $P_L/P_T$\\\hline
 & $1.0$ & $0.0013$ & $0.000013$ & $1.2$ &  $9.5$ & $0.52$ & $9.7$ & $0.018$\\
 & $1.3$ & $0.0015$ & $0.000017$ & $1.1$ &  $5.7$ & $0.38$ & $5.5$ & $0.014$\\
 & $1.7$ & $0.0019$ & $0.000023$ & $1.1$ &  $4.5$ & $0.31$ & $4.1$ & $0.011$\\
 & $2.3$ & $0.0024$ & $0.000030$ & $0.98$ &  $3.6$ & $0.26$ & $3.1$ & $0.0086$\\
 & $3.0$ & $0.0029$ & $0.000040$ & $0.92$ &  $3.1$ & $0.22$ & $2.5$ & $0.0075$\\
$\filledstar$ & $4.0$ & $0.0035$ & $0.000052$ & $0.86$ &  $2.6$ & $0.18$ & $2$ & $0.0069$\\
 & $5.2$ & $0.0043$ & $0.000068$ & $0.8$ &  $2.3$ & $0.16$ & $1.6$ & $0.0066$\\
 & $6.8$ & $0.0053$ & $0.000090$ & $0.75$ &  $2$ & $0.14$ & $1.2$ & $0.0065$\\
 & $8.9$ & $0.0065$ & $0.00012$ & $0.7$ &  $1.7$ & $0.12$ & $0.96$ & $0.0066$\\
 & $11.6$ & $0.0079$ & $0.00015$ & $0.65$ &  $1.4$ & $0.1$ & $0.72$ & $0.0068$\\
 & $15.2$ & $0.0097$ & $0.00020$ & $0.61$ &  $1.2$ & $0.089$ & $0.53$ & $0.0069$\\
 & $19.8$ & $0.012$ & $0.00026$ & $0.57$ &  $1.1$ & $0.079$ & $0.38$ & $0.0072$\\
 & $25.8$ & $0.014$ & $0.00034$ & $0.54$ &  $0.92$ & $0.07$ & $0.26$ & $0.0077$\\
 & $33.6$ & $0.017$ & $0.00044$ & $0.5$ &  $0.81$ & $0.062$ & $0.18$ & $0.008$\\
 & $43.7$ & $0.021$ & $0.00058$ & $0.47$ &  $0.73$ & $0.056$ & $0.13$ & $0.0086$\\
 & $56.9$ & $0.026$ & $0.00075$ & $0.44$ &  $0.67$ & $0.051$ & $0.09$ & $0.0092$\\
 & $74.3$ & $0.032$ & $0.00098$ & $0.41$ &  $0.61$ & $0.046$ & $0.066$ & $0.0099$\\
 & $96.6$ & $0.039$ & $0.0013$ & $0.38$ &  $0.56$ & $0.042$ & $0.05$ & $0.011$\\
 & $125.7$ & $0.047$ & $0.0017$ & $0.36$ &  $0.51$ & $0.039$ & $0.039$ & $0.012$\\
 & $164.1$ & $0.057$ & $0.0022$ & $0.34$ &  $0.47$ & $0.036$ & $0.031$ & $0.013$\\
 & $213.8$ & $0.07$ & $0.0028$ & $0.31$ &  $0.42$ & $0.034$ & $0.026$ & $0.014$\\
 & $278.9$ & $0.085$ & $0.0037$ & $0.29$ &  $0.38$ & $0.031$ & $0.022$ & $0.016$\\
 & $363.1$ & $0.1$ & $0.0048$ & $0.28$ &  $0.35$ & $0.03$ & $0.02$ & $0.019$\\
 & $472.9$ & $0.13$ & $0.0062$ & $0.26$ &  $0.31$ & $0.028$ & $0.018$ & $0.022$\\
 & $615.8$ & $0.15$ & $0.0081$ & $0.24$ &  $0.28$ & $0.027$ & $0.018$ & $0.025$\\
$\circ$ & $801.2$ & $0.19$ & $0.011$ & $0.23$ &  $0.26$ & $0.025$ & $0.018$ & $0.03$\\
 & $1043.9$ & $0.23$ & $0.014$ & $0.21$ &  $0.23$ & $0.024$ & $0.018$ & $0.036$\\
 & $1360.3$ & $0.28$ & $0.018$ & $0.2$ &  $0.21$ & $0.024$ & $0.02$ & $0.045$\\
 & $1768.4$ & $0.34$ & $0.023$ & $0.18$ &  $0.2$ & $0.023$ & $0.022$ & $0.056$\\
 & $2302.8$ & $0.41$ & $0.030$ & $0.17$ &  $0.18$ & $0.022$ & $0.024$ & $0.071$\\
 & $2999.8$ & $0.5$ & $0.039$ & $0.16$ &  $0.17$ & $0.021$ & $0.028$ & $0.091$\\
 & $3904.5$ & $0.61$ & $0.051$ & $0.15$ &  $0.15$ & $0.021$ & $0.033$ & $0.12$\\
 & $5085.0$ & $0.74$ & $0.067$ & $0.14$ &  $0.14$ & $0.02$ & $0.039$ & $0.15$\\
 & $6630.1$ & $0.9$ & $0.087$ & $0.13$ &  $0.13$ & $0.02$ & $0.047$ & $0.2$\\
 & $8630.3$ & $1.1$ & $0.11$ & $0.12$ &  $0.13$ & $0.019$ & $0.056$ & $0.27$\\
 & $11250.5$ & $1.3$ & $0.15$ & $0.11$ &  $0.12$ & $0.019$ & $0.067$ & $0.35$\\
 & $14675.9$ & $1.6$ & $0.19$ & $0.1$ &  $0.11$ & $0.018$ & $0.078$ & $0.45$\\
$\triangledown$ & $19127.1$ & $1.9$ & $0.25$ & $0.096$ &  $0.1$ & $0.017$ & $0.088$ & $0.55$\\
 & $24886.6$ & $2.3$ & $0.33$ & $0.088$ &  $0.097$ & $0.016$ & $0.096$ & $0.65$\\
 & $32404.6$ & $2.7$ & $0.43$ & $0.081$ &  $0.091$ & $0.015$ & $0.1$ & $0.74$\\
 & $42160.0$ & $3.3$ & $0.55$ & $0.075$ &  $0.084$ & $0.014$ & $0.1$ & $0.8$\\
\bottomrule
\end{tabular}

%% file: tables/table_export_xi10.0_m1_L2.0.tex
\begin{tabular}{c c c c c c c c c} \toprule
& \multicolumn{3}{c}{$\lambda=2.0$,} & \multicolumn{5}{c}{Screening: Debye-like}\\
 & $Q_s\tau$ & $\tau/\tau_{\mathrm{R}}$ & $\tau/\tau_{\mathrm{BMSS}}$ & $T_\varepsilon/Q_s$ & $T_*/Q_s$ &$m_D/Q_s$  & $\langle pf \rangle/\langle p\rangle$ & $P_L/P_T$\\\hline
 & $1.0$ & $0.0091$ & $0.00048$ & $0.86$ &  $2.7$ & $0.52$ & $2.4$ & $0.018$\\
 & $1.3$ & $0.011$ & $0.00064$ & $0.8$ &  $1.7$ & $0.38$ & $1.3$ & $0.018$\\
 & $1.7$ & $0.014$ & $0.00084$ & $0.75$ &  $1.4$ & $0.31$ & $0.89$ & $0.018$\\
 & $2.3$ & $0.017$ & $0.0011$ & $0.7$ &  $1.2$ & $0.26$ & $0.64$ & $0.019$\\
$\filledstar$ & $3.0$ & $0.021$ & $0.0015$ & $0.65$ &  $1.1$ & $0.23$ & $0.48$ & $0.02$\\
 & $3.9$ & $0.025$ & $0.0019$ & $0.61$ &  $0.94$ & $0.2$ & $0.35$ & $0.021$\\
 & $5.2$ & $0.031$ & $0.0025$ & $0.57$ &  $0.83$ & $0.17$ & $0.26$ & $0.022$\\
 & $6.8$ & $0.038$ & $0.0033$ & $0.53$ &  $0.75$ & $0.15$ & $0.18$ & $0.023$\\
 & $8.8$ & $0.046$ & $0.0043$ & $0.5$ &  $0.67$ & $0.14$ & $0.13$ & $0.025$\\
 & $11.5$ & $0.056$ & $0.0055$ & $0.46$ &  $0.61$ & $0.13$ & $0.099$ & $0.027$\\
 & $15.0$ & $0.069$ & $0.0073$ & $0.43$ &  $0.55$ & $0.12$ & $0.075$ & $0.03$\\
 & $19.6$ & $0.084$ & $0.0095$ & $0.4$ &  $0.5$ & $0.11$ & $0.061$ & $0.033$\\
 & $25.4$ & $0.1$ & $0.012$ & $0.38$ &  $0.46$ & $0.1$ & $0.051$ & $0.037$\\
 & $33.3$ & $0.12$ & $0.016$ & $0.35$ &  $0.42$ & $0.095$ & $0.046$ & $0.043$\\
 & $43.4$ & $0.15$ & $0.021$ & $0.33$ &  $0.38$ & $0.09$ & $0.043$ & $0.05$\\
 & $56.6$ & $0.19$ & $0.027$ & $0.31$ &  $0.34$ & $0.086$ & $0.042$ & $0.059$\\
$\circ$ & $73.8$ & $0.23$ & $0.036$ & $0.29$ &  $0.31$ & $0.082$ & $0.042$ & $0.07$\\
 & $96.5$ & $0.27$ & $0.047$ & $0.27$ &  $0.29$ & $0.079$ & $0.044$ & $0.086$\\
 & $125.8$ & $0.33$ & $0.061$ & $0.25$ &  $0.26$ & $0.076$ & $0.048$ & $0.11$\\
 & $164.5$ & $0.41$ & $0.079$ & $0.23$ &  $0.24$ & $0.073$ & $0.053$ & $0.13$\\
 & $214.5$ & $0.49$ & $0.10$ & $0.22$ &  $0.23$ & $0.071$ & $0.059$ & $0.17$\\
 & $279.3$ & $0.6$ & $0.13$ & $0.2$ &  $0.21$ & $0.068$ & $0.066$ & $0.21$\\
 & $363.7$ & $0.73$ & $0.18$ & $0.19$ &  $0.2$ & $0.065$ & $0.075$ & $0.27$\\
 & $474.3$ & $0.88$ & $0.23$ & $0.17$ &  $0.18$ & $0.062$ & $0.083$ & $0.34$\\
 & $617.8$ & $1.1$ & $0.30$ & $0.16$ &  $0.17$ & $0.059$ & $0.092$ & $0.42$\\
$\triangledown$ & $803.6$ & $1.3$ & $0.39$ & $0.15$ &  $0.16$ & $0.056$ & $0.099$ & $0.51$\\
 & $1046.4$ & $1.5$ & $0.51$ & $0.14$ &  $0.15$ & $0.052$ & $0.1$ & $0.59$\\
 & $1360.6$ & $1.8$ & $0.66$ & $0.13$ &  $0.14$ & $0.048$ & $0.11$ & $0.66$\\
 & $1779.4$ & $2.2$ & $0.86$ & $0.12$ &  $0.13$ & $0.044$ & $0.11$ & $0.73$\\
 & $2328.5$ & $2.6$ & $1.1$ & $0.11$ &  $0.12$ & $0.04$ & $0.11$ & $0.77$\\
 & $3034.0$ & $3.2$ & $1.5$ & $0.098$ &  $0.11$ & $0.037$ & $0.11$ & $0.81$\\
 & $3959.8$ & $3.8$ & $1.9$ & $0.09$ &  $0.1$ & $0.034$ & $0.11$ & $0.84$\\
 & $5171.6$ & $4.5$ & $2.5$ & $0.083$ &  $0.092$ & $0.031$ & $0.11$ & $0.87$\\
 & $6741.4$ & $5.4$ & $3.3$ & $0.076$ &  $0.085$ & $0.028$ & $0.11$ & $0.89$\\
 & $8767.8$ & $6.5$ & $4.2$ & $0.07$ &  $0.079$ & $0.025$ & $0.11$ & $0.91$\\
 & $11433.4$ & $7.8$ & $5.5$ & $0.064$ &  $0.074$ & $0.023$ & $0.1$ & $0.93$\\
\bottomrule
\end{tabular}

%% file: tables/table_export_xi10.0_m1_L10.0_GeV.tex
\begin{tabular}{c c c c c c c c c c c} \toprule
& \multicolumn{3}{c}{$\lambda=10.0$,} & \multicolumn{5}{c}{Screening: Debye-like}\\
 & $Q_s\tau$ & $\frac{\tau}{\mathrm{fm/c}}$  &$\tau/\tau_{\mathrm{R}}$ & $\frac{\tau}{\tau_{\mathrm{BMSS}}}$ & $T_\varepsilon/Q_s$ & $\frac{T_\varepsilon}{\mathrm{GeV}}$ & $T_*/Q_s$ &$m_D/Q_s$  & $\langle pf \rangle/\langle p\rangle$ & $P_L/P_T$\\\hline
 & $1.0$ & $0.14$ & $0.077$ & $0.032$ & $0.57$ & $0.8$ & $0.89$ & $0.52$ & $0.48$ & $0.018$\\
 & $1.3$ & $0.18$ & $0.094$ & $0.042$ & $0.54$ & $0.75$ & $0.71$ & $0.41$ & $0.23$ & $0.041$\\
 & $1.7$ & $0.24$ & $0.12$ & $0.054$ & $0.5$ & $0.7$ & $0.62$ & $0.37$ & $0.15$ & $0.057$\\
 & $2.2$ & $0.32$ & $0.14$ & $0.071$ & $0.47$ & $0.65$ & $0.55$ & $0.34$ & $0.11$ & $0.071$\\
$\filledstar$ & $2.9$ & $0.41$ & $0.17$ & $0.093$ & $0.44$ & $0.61$ & $0.5$ & $0.32$ & $0.093$ & $0.084$\\
 & $3.8$ & $0.54$ & $0.21$ & $0.12$ & $0.41$ & $0.57$ & $0.45$ & $0.3$ & $0.085$ & $0.099$\\
 & $5.0$ & $0.7$ & $0.25$ & $0.16$ & $0.38$ & $0.53$ & $0.41$ & $0.28$ & $0.082$ & $0.12$\\
$\circ$ & $6.5$ & $0.92$ & $0.31$ & $0.21$ & $0.35$ & $0.49$ & $0.38$ & $0.27$ & $0.082$ & $0.14$\\
 & $8.5$ & $1.2$ & $0.37$ & $0.27$ & $0.33$ & $0.46$ & $0.35$ & $0.26$ & $0.085$ & $0.17$\\
 & $11.1$ & $1.6$ & $0.45$ & $0.35$ & $0.31$ & $0.43$ & $0.32$ & $0.25$ & $0.089$ & $0.21$\\
 & $14.4$ & $2$ & $0.55$ & $0.46$ & $0.28$ & $0.4$ & $0.3$ & $0.23$ & $0.095$ & $0.26$\\
 & $18.8$ & $2.6$ & $0.66$ & $0.60$ & $0.26$ & $0.37$ & $0.28$ & $0.22$ & $0.1$ & $0.31$\\
 & $24.6$ & $3.5$ & $0.8$ & $0.78$ & $0.24$ & $0.34$ & $0.26$ & $0.21$ & $0.1$ & $0.38$\\
 & $32.2$ & $4.5$ & $0.97$ & $1.0$ & $0.23$ & $0.32$ & $0.24$ & $0.19$ & $0.11$ & $0.45$\\
$\triangledown$ & $42.2$ & $5.9$ & $1.2$ & $1.3$ & $0.21$ & $0.29$ & $0.22$ & $0.18$ & $0.11$ & $0.52$\\
 & $54.9$ & $7.7$ & $1.4$ & $1.7$ & $0.19$ & $0.27$ & $0.2$ & $0.17$ & $0.11$ & $0.59$\\
 & $71.8$ & $10$ & $1.7$ & $2.3$ & $0.18$ & $0.25$ & $0.19$ & $0.15$ & $0.11$ & $0.65$\\
 & $93.9$ & $13$ & $2$ & $3.0$ & $0.16$ & $0.23$ & $0.17$ & $0.14$ & $0.11$ & $0.71$\\
 & $122.2$ & $17$ & $2.4$ & $3.9$ & $0.15$ & $0.21$ & $0.16$ & $0.13$ & $0.11$ & $0.75$\\
 & $159.2$ & $22$ & $2.9$ & $5.1$ & $0.14$ & $0.19$ & $0.15$ & $0.12$ & $0.11$ & $0.79$\\
 & $207.7$ & $29$ & $3.5$ & $6.6$ & $0.13$ & $0.18$ & $0.14$ & $0.11$ & $0.11$ & $0.83$\\
 & $271.0$ & $38$ & $4.2$ & $8.6$ & $0.12$ & $0.16$ & $0.13$ & $0.099$ & $0.11$ & $0.86$\\
\bottomrule
\end{tabular}

%% file: tables/table_export_xi10.0_m101_L0.5.tex
\begin{tabular}{c c c c c c c c c} \toprule
& \multicolumn{3}{c}{$\lambda=0.5$,} & \multicolumn{5}{c}{Screening: isoHTL}\\
 & $Q_s\tau$ & $\tau/\tau_{\mathrm{R}}$ & $\tau/\tau_{\mathrm{BMSS}}$ & $T_\varepsilon/Q_s$ & $T_*/Q_s$ &$m_D/Q_s$  & $\langle pf \rangle/\langle p\rangle$ & $P_L/P_T$\\\hline
 & $1.0$ & $0.0014$ & $0.000013$ & $1.2$ &  $9.5$ & $0.52$ & $9.7$ & $0.018$\\
 & $1.3$ & $0.0017$ & $0.000017$ & $1.1$ &  $5.1$ & $0.39$ & $5.1$ & $0.017$\\
 & $1.7$ & $0.002$ & $0.000022$ & $1.1$ &  $3.8$ & $0.32$ & $3.5$ & $0.015$\\
 & $2.2$ & $0.0025$ & $0.000029$ & $1$ &  $3.1$ & $0.28$ & $2.6$ & $0.014$\\
$\filledstar$ & $2.9$ & $0.003$ & $0.000038$ & $0.93$ &  $2.5$ & $0.24$ & $1.9$ & $0.013$\\
 & $3.7$ & $0.0037$ & $0.000049$ & $0.87$ &  $2.1$ & $0.2$ & $1.5$ & $0.012$\\
 & $4.8$ & $0.0044$ & $0.000064$ & $0.82$ &  $1.7$ & $0.18$ & $1.1$ & $0.011$\\
 & $6.3$ & $0.0054$ & $0.000083$ & $0.76$ &  $1.5$ & $0.15$ & $0.84$ & $0.011$\\
 & $8.2$ & $0.0066$ & $0.00011$ & $0.72$ &  $1.3$ & $0.13$ & $0.63$ & $0.011$\\
 & $10.6$ & $0.008$ & $0.00014$ & $0.67$ &  $1.1$ & $0.12$ & $0.48$ & $0.011$\\
 & $13.8$ & $0.0098$ & $0.00018$ & $0.63$ &  $0.97$ & $0.1$ & $0.36$ & $0.011$\\
 & $17.9$ & $0.012$ & $0.00024$ & $0.59$ &  $0.87$ & $0.092$ & $0.27$ & $0.011$\\
 & $23.3$ & $0.014$ & $0.00031$ & $0.55$ &  $0.78$ & $0.082$ & $0.21$ & $0.012$\\
 & $30.3$ & $0.018$ & $0.00040$ & $0.51$ &  $0.72$ & $0.073$ & $0.16$ & $0.012$\\
 & $39.5$ & $0.021$ & $0.00052$ & $0.48$ &  $0.66$ & $0.065$ & $0.12$ & $0.013$\\
 & $51.3$ & $0.026$ & $0.00067$ & $0.45$ &  $0.61$ & $0.059$ & $0.092$ & $0.013$\\
 & $66.7$ & $0.032$ & $0.00088$ & $0.42$ &  $0.56$ & $0.053$ & $0.071$ & $0.014$\\
 & $86.7$ & $0.039$ & $0.0011$ & $0.4$ &  $0.52$ & $0.049$ & $0.056$ & $0.015$\\
 & $112.8$ & $0.047$ & $0.0015$ & $0.37$ &  $0.48$ & $0.045$ & $0.045$ & $0.016$\\
 & $146.6$ & $0.057$ & $0.0019$ & $0.35$ &  $0.44$ & $0.041$ & $0.037$ & $0.018$\\
 & $190.7$ & $0.07$ & $0.0025$ & $0.32$ &  $0.41$ & $0.038$ & $0.031$ & $0.019$\\
 & $247.9$ & $0.085$ & $0.0033$ & $0.3$ &  $0.37$ & $0.035$ & $0.027$ & $0.021$\\
 & $322.5$ & $0.1$ & $0.0042$ & $0.28$ &  $0.34$ & $0.033$ & $0.024$ & $0.024$\\
 & $419.5$ & $0.13$ & $0.0055$ & $0.27$ &  $0.31$ & $0.031$ & $0.022$ & $0.027$\\
 & $545.4$ & $0.15$ & $0.0072$ & $0.25$ &  $0.28$ & $0.029$ & $0.021$ & $0.031$\\
$\circ$ & $709.0$ & $0.19$ & $0.0093$ & $0.23$ &  $0.26$ & $0.028$ & $0.02$ & $0.037$\\
 & $922.0$ & $0.23$ & $0.012$ & $0.22$ &  $0.24$ & $0.026$ & $0.021$ & $0.043$\\
 & $1199.3$ & $0.27$ & $0.016$ & $0.2$ &  $0.22$ & $0.025$ & $0.022$ & $0.052$\\
 & $1559.9$ & $0.33$ & $0.021$ & $0.19$ &  $0.2$ & $0.024$ & $0.023$ & $0.063$\\
 & $2028.4$ & $0.41$ & $0.027$ & $0.18$ &  $0.18$ & $0.023$ & $0.026$ & $0.078$\\
 & $2637.5$ & $0.49$ & $0.035$ & $0.17$ &  $0.17$ & $0.023$ & $0.029$ & $0.098$\\
 & $3429.2$ & $0.6$ & $0.045$ & $0.15$ &  $0.16$ & $0.022$ & $0.034$ & $0.12$\\
 & $4458.0$ & $0.72$ & $0.059$ & $0.14$ &  $0.15$ & $0.021$ & $0.039$ & $0.16$\\
 & $5795.5$ & $0.88$ & $0.076$ & $0.13$ &  $0.14$ & $0.021$ & $0.046$ & $0.2$\\
 & $7534.5$ & $1.1$ & $0.099$ & $0.12$ &  $0.13$ & $0.02$ & $0.054$ & $0.26$\\
 & $9799.1$ & $1.3$ & $0.13$ & $0.12$ &  $0.12$ & $0.019$ & $0.064$ & $0.34$\\
 & $12745.3$ & $1.5$ & $0.17$ & $0.11$ &  $0.11$ & $0.018$ & $0.074$ & $0.42$\\
$\triangledown$ & $16573.2$ & $1.9$ & $0.22$ & $0.099$ &  $0.11$ & $0.017$ & $0.083$ & $0.52$\\
 & $21555.5$ & $2.2$ & $0.28$ & $0.092$ &  $0.1$ & $0.016$ & $0.092$ & $0.62$\\
 & $28024.8$ & $2.7$ & $0.37$ & $0.085$ &  $0.094$ & $0.015$ & $0.097$ & $0.7$\\
 & $36435.8$ & $3.2$ & $0.48$ & $0.078$ &  $0.087$ & $0.014$ & $0.1$ & $0.77$\\
 & $47372.4$ & $3.8$ & $0.62$ & $0.072$ &  $0.081$ & $0.013$ & $0.1$ & $0.82$\\
 & $61601.2$ & $4.6$ & $0.81$ & $0.066$ &  $0.075$ & $0.012$ & $0.1$ & $0.86$\\
 & $80090.0$ & $5.5$ & $1.1$ & $0.061$ &  $0.07$ & $0.011$ & $0.1$ & $0.88$\\
\bottomrule
\end{tabular}

%% file: tables/table_export_xi10.0_m101_L2.0.tex
\begin{tabular}{c c c c c c c c c} \toprule
& \multicolumn{3}{c}{$\lambda=2.0$,} & \multicolumn{5}{c}{Screening: isoHTL}\\
 & $Q_s\tau$ & $\tau/\tau_{\mathrm{R}}$ & $\tau/\tau_{\mathrm{BMSS}}$ & $T_\varepsilon/Q_s$ & $T_*/Q_s$ &$m_D/Q_s$  & $\langle pf \rangle/\langle p\rangle$ & $P_L/P_T$\\\hline
 & $1.0$ & $0.01$ & $0.00048$ & $0.86$ &  $2.7$ & $0.52$ & $2.4$ & $0.018$\\
 & $1.3$ & $0.013$ & $0.00063$ & $0.8$ &  $1.6$ & $0.39$ & $1.2$ & $0.023$\\
 & $1.7$ & $0.015$ & $0.00082$ & $0.75$ &  $1.3$ & $0.33$ & $0.74$ & $0.025$\\
 & $2.2$ & $0.019$ & $0.0011$ & $0.7$ &  $1.1$ & $0.28$ & $0.51$ & $0.027$\\
$\filledstar$ & $2.9$ & $0.023$ & $0.0014$ & $0.66$ &  $0.92$ & $0.25$ & $0.36$ & $0.028$\\
 & $3.7$ & $0.028$ & $0.0018$ & $0.62$ &  $0.82$ & $0.22$ & $0.26$ & $0.029$\\
 & $4.8$ & $0.034$ & $0.0023$ & $0.58$ &  $0.74$ & $0.19$ & $0.19$ & $0.031$\\
 & $6.3$ & $0.041$ & $0.0030$ & $0.54$ &  $0.67$ & $0.17$ & $0.15$ & $0.032$\\
 & $8.2$ & $0.05$ & $0.0039$ & $0.5$ &  $0.62$ & $0.16$ & $0.11$ & $0.034$\\
 & $10.6$ & $0.06$ & $0.0051$ & $0.47$ &  $0.57$ & $0.14$ & $0.09$ & $0.037$\\
 & $13.8$ & $0.073$ & $0.0067$ & $0.44$ &  $0.52$ & $0.13$ & $0.073$ & $0.04$\\
 & $18.0$ & $0.089$ & $0.0087$ & $0.41$ &  $0.48$ & $0.12$ & $0.062$ & $0.044$\\
 & $23.4$ & $0.11$ & $0.011$ & $0.39$ &  $0.44$ & $0.11$ & $0.054$ & $0.048$\\
 & $30.4$ & $0.13$ & $0.015$ & $0.36$ &  $0.41$ & $0.1$ & $0.049$ & $0.054$\\
 & $39.5$ & $0.16$ & $0.019$ & $0.34$ &  $0.37$ & $0.098$ & $0.046$ & $0.062$\\
 & $51.4$ & $0.2$ & $0.025$ & $0.32$ &  $0.34$ & $0.092$ & $0.045$ & $0.071$\\
$\circ$ & $66.8$ & $0.24$ & $0.032$ & $0.29$ &  $0.31$ & $0.088$ & $0.045$ & $0.084$\\
 & $86.9$ & $0.29$ & $0.042$ & $0.27$ &  $0.29$ & $0.083$ & $0.046$ & $0.1$\\
 & $113.0$ & $0.35$ & $0.055$ & $0.26$ &  $0.27$ & $0.08$ & $0.049$ & $0.12$\\
 & $146.9$ & $0.42$ & $0.071$ & $0.24$ &  $0.25$ & $0.076$ & $0.053$ & $0.15$\\
 & $191.0$ & $0.51$ & $0.092$ & $0.22$ &  $0.23$ & $0.073$ & $0.059$ & $0.18$\\
 & $248.4$ & $0.62$ & $0.12$ & $0.21$ &  $0.21$ & $0.07$ & $0.065$ & $0.22$\\
 & $323.0$ & $0.75$ & $0.16$ & $0.19$ &  $0.2$ & $0.067$ & $0.072$ & $0.28$\\
 & $420.0$ & $0.91$ & $0.20$ & $0.18$ &  $0.19$ & $0.064$ & $0.08$ & $0.34$\\
 & $546.2$ & $1.1$ & $0.26$ & $0.17$ &  $0.17$ & $0.06$ & $0.088$ & $0.42$\\
 & $710.3$ & $1.3$ & $0.34$ & $0.15$ &  $0.16$ & $0.057$ & $0.095$ & $0.5$\\
$\triangledown$ & $923.4$ & $1.6$ & $0.45$ & $0.14$ &  $0.15$ & $0.053$ & $0.1$ & $0.58$\\
 & $1200.7$ & $1.9$ & $0.58$ & $0.13$ &  $0.14$ & $0.049$ & $0.1$ & $0.65$\\
 & $1561.0$ & $2.3$ & $0.75$ & $0.12$ &  $0.13$ & $0.045$ & $0.11$ & $0.72$\\
 & $2030.2$ & $2.7$ & $0.98$ & $0.11$ &  $0.12$ & $0.042$ & $0.11$ & $0.77$\\
 & $2639.6$ & $3.2$ & $1.3$ & $0.1$ &  $0.11$ & $0.038$ & $0.11$ & $0.81$\\
 & $3431.9$ & $3.9$ & $1.7$ & $0.094$ &  $0.1$ & $0.035$ & $0.11$ & $0.84$\\
 & $4462.6$ & $4.6$ & $2.2$ & $0.086$ &  $0.096$ & $0.032$ & $0.11$ & $0.87$\\
 & $5802.8$ & $5.5$ & $2.8$ & $0.079$ &  $0.089$ & $0.029$ & $0.11$ & $0.89$\\
 & $7545.7$ & $6.6$ & $3.6$ & $0.073$ &  $0.082$ & $0.026$ & $0.1$ & $0.91$\\
 & $9811.5$ & $7.9$ & $4.7$ & $0.067$ &  $0.076$ & $0.024$ & $0.1$ & $0.93$\\
 & $12758.2$ & $9.4$ & $6.2$ & $0.061$ &  $0.071$ & $0.022$ & $0.1$ & $0.94$\\
 & $16587.0$ & $11$ & $8.0$ & $0.056$ &  $0.066$ & $0.02$ & $0.099$ & $0.95$\\
\bottomrule
\end{tabular}

%% file: tables/table_export_xi10.0_m101_L10.0_GeV.tex
\begin{tabular}{c c c c c c c c c c c} \toprule
& \multicolumn{3}{c}{$\lambda=10.0$,} & \multicolumn{5}{c}{Screening: isoHTL}\\
 & $Q_s\tau$ & $\frac{\tau}{\mathrm{fm/c}}$  &$\tau/\tau_{\mathrm{R}}$ & $\frac{\tau}{\tau_{\mathrm{BMSS}}}$ & $T_\varepsilon/Q_s$ & $\frac{T_\varepsilon}{\mathrm{GeV}}$ & $T_*/Q_s$ &$m_D/Q_s$  & $\langle pf \rangle/\langle p\rangle$ & $P_L/P_T$\\\hline
 & $1.0$ & $0.14$ & $0.091$ & $0.032$ & $0.57$ & $0.8$ & $0.89$ & $0.52$ & $0.48$ & $0.018$\\
 & $1.3$ & $0.18$ & $0.11$ & $0.041$ & $0.54$ & $0.75$ & $0.66$ & $0.42$ & $0.18$ & $0.054$\\
 & $1.7$ & $0.24$ & $0.13$ & $0.054$ & $0.5$ & $0.7$ & $0.58$ & $0.38$ & $0.12$ & $0.076$\\
$\filledstar$ & $2.2$ & $0.31$ & $0.16$ & $0.070$ & $0.47$ & $0.66$ & $0.52$ & $0.36$ & $0.096$ & $0.094$\\
 & $2.9$ & $0.4$ & $0.2$ & $0.091$ & $0.44$ & $0.61$ & $0.48$ & $0.33$ & $0.086$ & $0.11$\\
 & $3.7$ & $0.52$ & $0.24$ & $0.12$ & $0.41$ & $0.57$ & $0.44$ & $0.31$ & $0.081$ & $0.13$\\
$\circ$ & $4.9$ & $0.68$ & $0.29$ & $0.15$ & $0.38$ & $0.53$ & $0.4$ & $0.29$ & $0.081$ & $0.15$\\
 & $6.3$ & $0.89$ & $0.35$ & $0.20$ & $0.35$ & $0.5$ & $0.37$ & $0.28$ & $0.082$ & $0.18$\\
 & $8.2$ & $1.2$ & $0.43$ & $0.26$ & $0.33$ & $0.46$ & $0.34$ & $0.26$ & $0.085$ & $0.21$\\
 & $10.7$ & $1.5$ & $0.52$ & $0.34$ & $0.31$ & $0.43$ & $0.32$ & $0.25$ & $0.089$ & $0.25$\\
 & $13.9$ & $2$ & $0.62$ & $0.44$ & $0.28$ & $0.4$ & $0.3$ & $0.24$ & $0.093$ & $0.3$\\
 & $18.1$ & $2.6$ & $0.75$ & $0.58$ & $0.26$ & $0.37$ & $0.27$ & $0.22$ & $0.098$ & $0.36$\\
 & $23.6$ & $3.3$ & $0.91$ & $0.75$ & $0.24$ & $0.34$ & $0.25$ & $0.21$ & $0.1$ & $0.42$\\
 & $30.7$ & $4.3$ & $1.1$ & $0.97$ & $0.23$ & $0.32$ & $0.24$ & $0.19$ & $0.1$ & $0.49$\\
$\triangledown$ & $40.0$ & $5.6$ & $1.3$ & $1.3$ & $0.21$ & $0.29$ & $0.22$ & $0.18$ & $0.11$ & $0.56$\\
 & $51.9$ & $7.3$ & $1.6$ & $1.6$ & $0.19$ & $0.27$ & $0.2$ & $0.17$ & $0.11$ & $0.62$\\
 & $67.5$ & $9.5$ & $1.9$ & $2.1$ & $0.18$ & $0.25$ & $0.19$ & $0.15$ & $0.11$ & $0.68$\\
 & $87.9$ & $12$ & $2.3$ & $2.8$ & $0.16$ & $0.23$ & $0.17$ & $0.14$ & $0.11$ & $0.73$\\
 & $114.4$ & $16$ & $2.7$ & $3.6$ & $0.15$ & $0.21$ & $0.16$ & $0.13$ & $0.11$ & $0.77$\\
 & $148.9$ & $21$ & $3.2$ & $4.7$ & $0.14$ & $0.19$ & $0.15$ & $0.12$ & $0.11$ & $0.8$\\
 & $193.7$ & $27$ & $3.9$ & $6.1$ & $0.13$ & $0.18$ & $0.14$ & $0.11$ & $0.11$ & $0.84$\\
 & $251.9$ & $35$ & $4.6$ & $8.0$ & $0.12$ & $0.16$ & $0.13$ & $0.099$ & $0.11$ & $0.86$\\
\bottomrule
\end{tabular}

%% file: 980_different_notations.tex
\begin{itemize}
    \item $v_{\dots}$ in Ref.~\cite{Boguslavski:2023waw} is written as $\cos\theta_{\dots}$ here.
    
    In Ref.~\cite{Boguslavski:2023waw}, the cosine of angles is usually abbreviated with the letter $v_{\dots}=\cos\theta_{\dots}$. In this thesis, I choose to always write the cosine explicitly. For the angle between $\vb p$ and $\vb q$, $\thetaqp$, I write here for the cosine $\cos\thetaqp=v_{pq}$.

    \item The parameter $\xi$ used in Ref.~\cite{Boguslavski:2023alu, Boguslavski:2023jvg} labels the anisotropy in the initial condition \eqref{eq:initial_cond}, which we denote as $\xianiso$ here. The parameter $\xi$ is used in this thesis to denote the gauge parameter in the Fadeev-Popov procedure. Furthermore, the same letter $\xi$ was used in Ref.~\cite{Boguslavski:2023waw} to denote the constant used in the Debye-like screening prescription. We denote this here by $\xiscreen$ for Debye-like screening in the elastic collision term \eqref{eq:c22_first}, and $\xiscreenperp$ for Debye-like screening for the jet quenching parameter $\qhat$.

    \item The convention for the HTL propagators \eqref{eq:HTL-propagators} differs from the one used in Ref.~\cite{Boguslavski:2023waw, Boguslavski:2024kbd} by a factor $i$ and correspond here to the convention used by Blaizot and Iancu \cite{Blaizot:2001nr} to be consistent with the propagators defined throughout the thesis. In both references, the propagators enter only via the AMY replacement \eqref{eq:amy_replacement} using the absolute value, so there are no other differences.

    \item The angle $\phi_{kq}$ from Ref.~\cite{Boguslavski:2023waw} is written $\phi_{qk}$ here, because it measures the azimuthal angle of $\vb k$ in a frame which is defined by $\vb q$.

\end{itemize}

%% file: 985_tools_used.tex
While working on my thesis, I used several scientific tools and libraries which I would like to acknowledge. 
I would like to acknowledge the use of Mathematica \cite{Mathematica}, Scipy \cite{2020SciPy-NMeth}, Numpy \cite{harris2020array}, Matplotlib \cite{Hunter:2007}, mpmath \cite{mpmath}, sympy \cite{10.7717/peerj-cs.103}, and, of course, various large language model (LLM) tools such as \verb|ChatGPT|\footnote{\url{https://chatgpt.com/}}, \verb|Mistral|\footnote{\url{https://mistral.ai/}}, \verb|copilot|\footnote{\url{https://copilot.microsoft.com}} or others, which I mainly used when getting stuck in a coding question and to help me write code more efficiently.